\documentclass[%
  a4paper,
  12pt,
  twoside,
  ]
  {report}

\newcommand*{\myName} {Guido Walter Pettinari}

\newcommand*{\myUni} {University of Portsmouth}
\newcommand*{\myFaculty} {Institute of Cosmology and Gravitation}

\newcommand*{\myTitle} {The intrinsic bispectrum of the Cosmic Microwave Background}

\usepackage[dvipsnames]{xcolor} 

\newcommand*{\runinhead}{\paragraph}

\let\saveboldmath\boldmath
\usepackage{mathptmx} 
\let\boldmath\saveboldmath
\usepackage{helvet} 
\usepackage{courier} 
\usepackage{type1cm} 

\usepackage{pdfpages} 
\usepackage{lmodern} 
\usepackage{relsize} 
\usepackage{bm} 
\usepackage{tabu} 
\usepackage{longtable}
\usepackage{esdiff} 
\usepackage{xspace} 
\usepackage{aas_macros} 
\usepackage{makeidx}  
\usepackage[reqno]{amsmath} 
\usepackage{amssymb}
\usepackage[tight,nice]{units} 
\usepackage{subfig} 
\usepackage{fixltx2e} 
\usepackage{graphicx} 
\usepackage{epstopdf} 
\usepackage{rotating} 
\usepackage{multirow} 
\usepackage{caption} 
\usepackage{xargs} 
\usepackage{setspace} 
\usepackage{multicol} 
\usepackage[bottom]{footmisc} 
\usepackage{hyperref} 
\hypersetup{%
    colorlinks=true, linktocpage=true, pdfstartpage=3, pdfstartview=FitV,%
    breaklinks=true, pdfpagemode=UseNone,%
    pageanchor=true,%
    pdfpagemode=UseOutlines, plainpages=false, bookmarksnumbered,%
    bookmarksopen=true, bookmarksopenlevel=1, hypertexnames=true,%
    pdfhighlight=/O,
    urlcolor=webbrown, linkcolor=RoyalBlue, citecolor=webgreen, 
    urlcolor=Black, linkcolor=Black, citecolor=Black,
    pagebackref=False,
    pdftitle={\myTitle},%
    pdfauthor={\myName, \myUni, \myFaculty},%
    pdfsubject={},%
    pdfkeywords={},%
    pdfcreator={},%
    pdfproducer={}%
}

\usepackage{chapterbib} 
\newcommand{\chapterbib}{
  \bibliographystyle{spbasic}
  \bibliography{phd_thesis}
}
\usepackage[square,numbers,sort]{natbib} 

\def\eprinttmp@#1arXiv:#2 [#3]#4@{
\ifthenelse{\equal{#3}{x}}{\href{http://arxiv.org/abs/#1}{#1}}{\href{http://
arxiv.org/abs/#2}{arXiv:#2} [#3]}}

\providecommand{\eprint}[1]{\eprinttmp@#1arXiv: [x]@}
\newcommand{\adsurl}[1]{\href{#1}{ADS}}
\providecommand{\bibinfo}[2]{\ifthenelse{\equal{#1}{isbn}}{
\href{http://cosmologist.info/ISBN/#2}{#2}}{#2}}

\usepackage{lettrine}
\usepackage[palatino]{quotchap}
\definecolor{chaptergrey}{rgb}{0.422,0.648,0.801}

\usepackage
[
        textwidth=16.3cm,
        outer=2.5cm,
        tmargin=30mm,
        bmargin=27mm,
        a4paper
]
{geometry}

\DeclareSymbolFont{cmsymbols}{OMS}{cmsy}{m}{n}
\SetSymbolFont{cmsymbols}{bold}{OMS}{cmsy}{b}{n}
\DeclareSymbolFontAlphabet{\mathcal}{cmsymbols}

\allowdisplaybreaks

\usepackage{etoolbox}
\preto\align{\par\nobreak\small\noindent}
\expandafter\preto\csname align*\endcsname{\par\nobreak\small\noindent}
\preto\multline{\par\nobreak\small\noindent}
\expandafter\preto\csname align*\endcsname{\par\nobreak\small\noindent}
\preto\flalign{\par\nobreak\small\noindent}
\expandafter\preto\csname align*\endcsname{\par\nobreak\small\noindent}
\preto\eqnarray{\par\nobreak\small\noindent}
\expandafter\preto\csname align*\endcsname{\par\nobreak\small\noindent}

\newcommand*{\msk}{\\[0.25cm]} 
\newcommand*{\nmsk}{\notag\msk} 
\newcommand*{\lmsk}{\\[0.4cm]} 
\newcommand*{\llmsk}{\\[0.6cm]} 

\makeatother

\renewcommand{\min}{\text{min}}
\renewcommand{\max}{\text{max}}
\newcommand*{\sci} [2] {\ensuremath{{#1} \times 10^{#2}}} 
\newcommand*{\abs} [1] {\left|{#1}\right|}
\newcommand*{\avg} [1] {\ensuremath{\left\langle\,{#1}\,\right\rangle}\xspace}
\newcommand*{\avgc} [1] {\ensuremath{{\left\langle\,{#1}\,\right\rangle}_c}\xspace}
\newcommand*{\avgbig} [1] {\ensuremath{\bigl\langle\,{#1}\,\bigr\rangle}\xspace}

\newcommand*{\ie} {\emph{i.\,\!e.}\xspace}
\newcommand*{\eg} {\emph{e.\,\!g.}\xspace}
\newcommand*{\etc} {\emph{etc.}\xspace}

\newcommand*{\LCDM} {$ \Lambda \text{CDM} $\xspace}
\newcommand*{\FLRW} {\emph{FLRW}\xspace}

\newcommand{\dd}{\textrm{d}}
\newcommand{\threev}[3]{\ensuremath{\begin{pmatrix}{#1}\\{#2}\\{#3}\\\end{pmatrix}}\xspace}
\newcommand{\sub}[1]{\ensuremath{\text{\smaller #1}}}

\newcommand{\Hc}{\ensuremath{\mathcal{H}}\xspace}
\newcommand*{\T} {\ensuremath{\mathcal{T}}\xspace}
\newcommand*{\R} {\ensuremath{\mathcal{R}}\xspace}
\renewcommand*{\O} {\ensuremath{\mathcal{O}}\xspace}

\newcommand*{\pfrac} [2] {\ensuremath{\frac{\partial {#1}}{\partial {#2}}}}
\newcommand*{\ppfrac} [2] {\ensuremath{\frac{\partial^2 {#1}}{\partial {#2}^{2}}}}

\newcommand*{\D}[2]{\ensuremath{{#1}_{#2}}\xspace}
\newcommand*{\U}[2]{\ensuremath{{#1}^{#2}}\xspace}
\newcommand*{\DD}[3]{\ensuremath{{#1}_{{#2}{#3}}}\xspace}
\newcommand*{\UD}[3]{\ensuremath{{{#1}^{#2}}_{#3}}\xspace}
\newcommand*{\UU}[3]{\ensuremath{{#1}^{{#2}{#3}}}\xspace}
\newcommand*{\KronUD} [2] {\UD{\delta}{\,#1}{#2}}

\newcommand*{\TODO}[2]{
  \ifthenelse{\equal{#1}{1}}{\textcolor{red}{\textbf{{TO DO: #2}}}}{}
  \ifthenelse{\equal{#1}{2}}{\textbf{{TO DO: #2}}}{}
  \ifthenelse{\equal{#1}{3}}{\emph{{TO DO: #2}}}{}
}

\renewcommand{\vec}[1]{{\boldsymbol {#1}}} 

\DeclareMathSymbol{\GammaIt}{\mathalpha}{letters}{"00}

\newcommand*{\keyword} [2] [\keywordentry] {%
\def\keywordentry{#2}%
\emph{#2}\index{#1}%
}
\newcommand*{\indexword} [2] [\indexwordentry] {%
\def\indexwordentry{#2}%
{#2}\index{#1}%
}
\newcommand*{\tref} [1] {Table \ref{#1}\xspace}

\newcommand*{\sref} [1] {Sec.\ \ref{#1}\xspace}

\newcommand*{\eref} [1] {Eq.\ \ref{#1}\xspace}

\newcommand*{\fref} [1] {Figure \ref{#1}\xspace}

\newcommand*{\cref} [1] {Chapter~\ref{#1}\xspace}

\newcommand*{\annotate}[1]{}

\def\Nicefrac#1#2{ \raise.5ex\hbox{$#1$}%
    \kern-.1em/\kern-.15em
    \lower.25ex\hbox{$#2$}}

\newcommand*{\SONG}{\textsf{SONG}\xspace}
\newcommand*{\tauz}{\ensuremath{\tau_\sub{0}}\xspace}

\newcommand{\rhoc} {\ensuremath{\rho_\text{crit}}\xspace}
\newcommand{\onesigma} {$ 68\% $ confidence level\xspace }
\newcommand{\twosigma} {$ 95\% $ confidence level\xspace }
\newcommand{\christoffel} [3] { {\Gamma^{#1}}_{{#2}{#3}} }
\newcommand{\ems} {\text{ems}} 
\newcommand{\obs} {\text{obs}} 
\newcommand{\rec} {\text{rec}} 

\newcommand{\Psid} {\dot{\Psi}} 
\newcommand{\Phid} {\dot{\Phi}} 
\newcommand{\Phidd} {\ddot{\Phi}} 

\newcommand{\g} {\gamma}
\newcommand{\G} {\ensuremath{\mathcal{G}}\xspace}
\newcommandx{\inlinereaction} [5] [3=\middlesymbol] {
\def\middlesymbol{\rightarrow}%
#1\;+\;#2\;#3\;#4\;+\;#5%
}
\newcommandx{\reaction} [5] [3=\middlesymbol] {%
\def\middlesymbol{\rightarrow}%
#1\;+\;#2\quad#3\quad#4\;+\;#5%
}

\newcommand{\krec}{\ensuremath{k_\text{rec}}\xspace}
\newcommand{\taurec}{\ensuremath{\tau_\text{rec}}\xspace}

\newcommand{\pert}[2]{\ensuremath{#1^{(#2)}}\xspace}
\newcommand{\PsiFO}{\pert{\Psi}{1}}
\newcommand{\PsiSO}{\pert{\Psi}{2}}
\newcommand{\PhiFO}{\pert{\Phi}{1}}
\newcommand{\PhiSO}{\pert{\Phi}{2}}
\newcommand{\omegaSO}{\ensuremath{\pert{\omega}{2}}\xspace}
\newcommand{\gammaSO}{\ensuremath{\pert{\gamma}{2}}\xspace}
\newcommand{\omegaSOm}[1]{\ensuremath{\pert{\omega}{2}_{[#1]}}\xspace}
\newcommand{\gammaSOm}[1]{\ensuremath{\pert{\gamma}{2}_{[#1]}}\xspace}
\newcommand{\rhoz}{\ensuremath{\bar\rho}\xspace}
\newcommand{\Pz}{\ensuremath{\bar P}\xspace}

\newcommand{\e}[2]{\ensuremath{{\tD{e}{#1}}^{#2}}\xspace}
\newcommand{\eInv}[2]{{\ensuremath{{\tU{e}{#1}}_{#2}}}\xspace}
\newcommand{\tU}[2]{{\ensuremath{#1^{\underline{#2}}}}\xspace}
\newcommand{\tD}[2]{{\ensuremath{#1_{\underline{#2}}}}\xspace}
\newcommand{\tUD}[3]{{\ensuremath{{#1^{\underline{#2}}}_{\underline{#3}}}}\xspace}
\newcommand{\tUU}[3]{{\ensuremath{#1^{\underline{#2}\underline{#3}}}}\xspace}
\newcommand{\tDD}[3]{{\ensuremath{#1_{\underline{#2}\underline{#3}}}}\xspace}

\newcommand*{\col}[2]{
  \ifthenelse{\equal{#1}{0}}{\ensuremath{\textcolor{Green}{#2}}}{}
  \ifthenelse{\equal{#1}{1}}{\ensuremath{\textcolor{Red}{#2}}}{}
  \ifthenelse{\equal{#1}{2}}{\ensuremath{\textcolor{RoyalBlue}{#2}}}{}
}

\newcommand*{\I}[1]{\col{1}{#1}}
\newcommand*{\II}[1]{\col{2}{#1}}

\newcommand{\tx}{\ensuremath{t,\vec{x}}}
\newcommand{\taux}{\ensuremath{\tau,\vec{x}}}
\newcommand{\tauk}{\ensuremath{\tau,\vec{k}}}

\newcommand{\tauini}{\ensuremath{\tau_\text{in}}\xspace}
\newcommand{\tauend}{\ensuremath{\tau_\text{end}}\xspace}
\newcommand{\xdep}[1]{\ensuremath{\vec{x_1},\dotsc,\vec{x_{#1}}}\xspace}
\newcommand{\kudt}[1]{\ensuremath{\vec{k_1},\vec{k_2},\vec{k_3}}\xspace}

\newcommand{\dirac}[1]{\ensuremath{\delta\left(#1\right)}\xspace}

\newcommand{\diracMinus}{\dirac{\vec{k}-\vec{k_1}-\vec{k_2}}}
\newcommand{\K}[1]{\ensuremath{\mathcal{K}\,\left\{\,{#1}\,\right\}}\xspace}

\newcommand{\probab}{\ensuremath{\mathcal{P}}}

\newcommand*{\con} [1] {\avgc{#1}}
\newcommand*{\ddR} {\ensuremath{\mathcal{D}[\R(\vecx)]}\xspace}
\newcommand*{\ddRhat} {\ensuremath{\mathcal{D}[\hat\R(\vecx)]}\xspace}

\newcommand{\vecv}{\ensuremath{\vec{v}}\xspace}
\newcommand{\vecp}{\ensuremath{\vec{p}}\xspace}

\newcommand{\ppp}{\ensuremath{\vec{p},\vec{p}'}\xspace}
\newcommand{\vecq}{\ensuremath{\vec{q}}\xspace}

\newcommand{\vecr}{\ensuremath{\vec{r}}\xspace}
\newcommand{\vecs}{\ensuremath{\vec{s}}\xspace}
\newcommand{\vect}{\ensuremath{\vec{t}}\xspace}
\newcommand{\vecx}{\ensuremath{\vec{x}}\xspace}
\newcommand{\vecy}{\ensuremath{\vec{y}}\xspace}
\newcommand{\vecz}{\ensuremath{\vec{z}}\xspace}
\newcommand{\vecxp}{\ensuremath{\vec{x}^\prime}\xspace}
\newcommand{\xone}{\ensuremath{\vec{x_1}}\xspace}
\newcommand{\xtwo}{\ensuremath{\vec{x_2}}\xspace}
\newcommand{\xtre}{\ensuremath{\vec{x_3}}\xspace}
\newcommand{\veck}{\ensuremath{\vec{k}}\xspace}
\renewcommand{\k}{\veck}

\newcommand{\kone}{\ensuremath{\vec{k_1}}\xspace}
\newcommand{\ktwo}{\ensuremath{\vec{k_2}}\xspace}
\newcommand{\ktre}{\ensuremath{\vec{k_3}}\xspace}
\newcommand{\kn}{\ensuremath{\vec{k_n}}\xspace}
\newcommand{\konep}{\ensuremath{\vec{k_1}^{\prime}}\xspace}
\newcommand{\ktwop}{\ensuremath{\vec{k_2}^{\prime}}\xspace}
\newcommand{\ktrep}{\ensuremath{\vec{k_3}^{\prime}}\xspace}

\newcommand{\n}{\ensuremath{\vec{n}}\xspace}
\newcommand{\vece}{\ensuremath{\vec{e}}\xspace}

\newcommand{\fz}{\bar{f}_p}
\newcommand{\fpz}{\bar{f}_{p'}}
\newcommand{\f}[1]{{f^{(#1)}}_{\vecp}}
\newcommand{\fps}[1]{{f^{(#1)}}_{\vecp'}}

\newcommand{\QTT} {\ensuremath{{Q}_{\text{\color{Black} {TT}}}}\xspace}
\newcommand{\QST} {\ensuremath{{Q}_{\text{\color{Black} {ST}}}}\xspace}
\newcommand{\QTR} {\ensuremath{{Q}_{\text{\color{Black} {TR}}}}\xspace}
\newcommand{\QSS} {\ensuremath{{Q}_{\text{\color{Black} {SS}}}}\xspace}

\newcommand{\chimatrix}[3]{\ensuremath{\chi_{\,{#1},[{#2}]}^{\,#3}}}
\newcommand{\xivector}[2]{\ensuremath{\xi_{\,[{#1}]}^{\,#2}}}
\newcommand{\koneM}[1]{\ensuremath{{k_1}_{[#1]}}}

\newcommand{\scalarP}[2]{{\ensuremath{\vec{#1}\nobreak\cdot\nobreak\vec{#2}}}\xspace}
\newcommand{\scalarp}[2]{{\ensuremath{\vec{#1}\nobreak\cdot\nobreak\vec{#2}}}\xspace}
\newcommand{\tensorP}[3]{\ensuremath{({#2}\nobreak{#3})_{[#1]}}\xspace}

\newcommand{\kokt}{\scalarP{k_1}{k_2}}

\newcommand*{\PONE}{P2009 \cite{pitrou:2009a}\xspace}
\newcommand*{\BFONE}{BF2010 \cite{beneke:2010a}\xspace}
\newcommand*{\BFTWO}{BFK2011 \cite{beneke:2011a}\xspace}

\newcommand{\FST}{\ensuremath{\mathcal{L}_\text{\,\tiny FS}}}
\newcommand{\RT}{\ensuremath{\mathcal{L}_\text{\,\tiny R}}}
\newcommand{\LT}{\ensuremath{\mathcal{L}_\text{\,\tiny L}}}
\newcommand{\Lcut}{\ensuremath{L_\text{cut}}\xspace}
\renewcommand{\S}{\ensuremath{\mathcal{S}}\xspace}

\newcommand{\threej}[6]{%
\mbox{\footnotesize$\begin{pmatrix}{#1\!}&{#2\!}&{#3}\\{#4\!}&{#5\!}&{#6}\end{pmatrix}$}}
\newcommand{\sixj}[6]{%
\mbox{\footnotesize$\begin{Bmatrix}{#1\!}&{#2\!}&{#3}\\{#4\!}&{#5\!}&{#6}\end{Bmatrix}$}}
\newcommand{\gaunt}[6]{%
\mathcal{G}\,^{#1#2#3}_{#4#5#6}
}
\newcommand{\GAUNT}[6]{%
\threej{#1}{#2}{#3}{0}{0}{0}%
\threej{#1}{#2}{#3}{#4}{#5}{#6}
}
\renewcommand{\L}{\ensuremath{\ell}\xspace}
\newcommand{\lmin}{{\ensuremath{\ell_\text{min}}}\xspace}
\newcommand{\lmax}{{\ensuremath{\ell_\text{max}}}\xspace}
\newcommand{\Lmax}{\ensuremath{L_\text{max}}\xspace}
\newcommand{\kmin}{\ensuremath{k_\text{min}}\xspace}
\newcommand{\kmax}{\ensuremath{k_\text{max}}\xspace}
\newcommand{\lone}{\ensuremath{{\ell_1}}\xspace}
\newcommand{\ltwo}{\ensuremath{{\ell_2}}\xspace}
\newcommand{\ltre}{\ensuremath{{\ell_3}}\xspace}
\newcommand{\ldep}{\ensuremath{{\lone\ltwo\ltre}}\xspace}
\newcommand{\lm}{{\ensuremath{\L m}}\xspace}
\newcommand{\lmp}{{\ensuremath{\L' m'}}\xspace}
\newcommand{\lmone}{{\ensuremath{\lone m_1}}\xspace}
\newcommand{\lmtwo}{{\ensuremath{\ltwo m_2}}\xspace}
\newcommand{\lmtre}{{\ensuremath{\ltre m_3}}\xspace}
\newcommand{\coll}{\ensuremath{\mathfrak{C}}\xspace}
\newcommand{\betaop}[2]{\ensuremath{\beta_{\,#2}\left[\,{#1}\,\right]}\xspace}
\newcommand{\DeltaT}{{\ensuremath{\tilde{\Delta}}}\xspace}

\newcommand{\INs}{\ensuremath{\mathcal{I}}\xspace}
\newcommand{\EMs}{\ensuremath{\mathcal{E}}\xspace}
\newcommand{\BMs}{\ensuremath{\mathcal{B}}\xspace}
\newcommand{\BAs}{\ensuremath{b\,}\xspace}
\newcommand{\CDs}{\ensuremath{c\,}\xspace}
\newcommand{\NEs}{\ensuremath{\mathcal{N}}\xspace}

\newcommand{\IN}[2]{\ensuremath{\INs^{\,\!#1}_{\,\!#2}}\xspace}
\newcommand{\EM}[2]{\ensuremath{\EMs^{\,\!#1}_{\,\!#2}}\xspace}
\newcommand{\BM}[2]{\ensuremath{\BMs^{\,\!#1}_{\,\!#2}}\xspace}
\newcommand{\NE}[2]{\ensuremath{\NEs^{\,\!#1}_{\,\!#2}}\xspace}
\newcommand{\BA}[3]{\ensuremath{\,{}_{#1}\BAs^{\,\!#2}_{\,\!#3}}\xspace}
\newcommand{\CD}[3]{\ensuremath{\,{}_{#1}\CDs^{\,\!#2}_{\,\!#3}}\xspace}

\newcommand{\bmult}[4]{\ensuremath{\,{}_{#2}#1_{#3#4}}\xspace}

\newcommand{\M}{\ensuremath{\mathcal{M}}\xspace}
\newcommand{\Q}[1]{\ensuremath{\mathcal{Q}^#1}\xspace}
\newcommand{\QLb}{{Q}^{L}_\BAs}
\newcommand{\QLI}{{Q}^{L}_\INs}
\newcommand{\QLE}{{Q}^{L}_\EMs}
\newcommand{\QLB}{{Q}^{L}_\BMs}
\newcommand{\QLN}{{Q}^{L}_\NEs}

\newcommand{\QCI}{{Q}^\mathfrak{C}_\INs}
\newcommand{\QCE}{{Q}^\mathfrak{C}_\EMs}
\newcommand{\QCB}{{Q}^\mathfrak{C}_\BMs}

\newcommand*{\intr}{\text{intr}}
\newcommand*{\lin}{\text{lin}}
\newcommand*{\trisp}{\text{trisp}}
\newcommand*{\local}{\text{local}}
\newcommand*{\equil}{\text{eq}}
\newcommand*{\orth}{\text{orth}}
\newcommand*{\SN} {\ensuremath{S/N}\xspace}
\newcommand*{\F} [2] {\ensuremath{F^{{#1},{#2}}}\xspace}

\newcommand*{\fnl}{\ensuremath{f_\sub{NL}}\xspace}
\newcommand*{\fnlcon}{\ensuremath{f_\sub{NL}^\text{intr}}\xspace}
\newcommand*{\taunl}{\ensuremath{\tau{_{NL}}}\xspace}
\newcommand*{\gnl}{\ensuremath{g{_{NL}}}\xspace}

\defcitealias{burrage:2009a}{BDS}

\newboolean{catalogswithnumbers}
\setboolean{catalogswithnumbers}{false} 
\ifthenelse{\boolean{catalogswithnumbers}}{

}{ 

}

\makeindex

\begin{document}



\includepdf{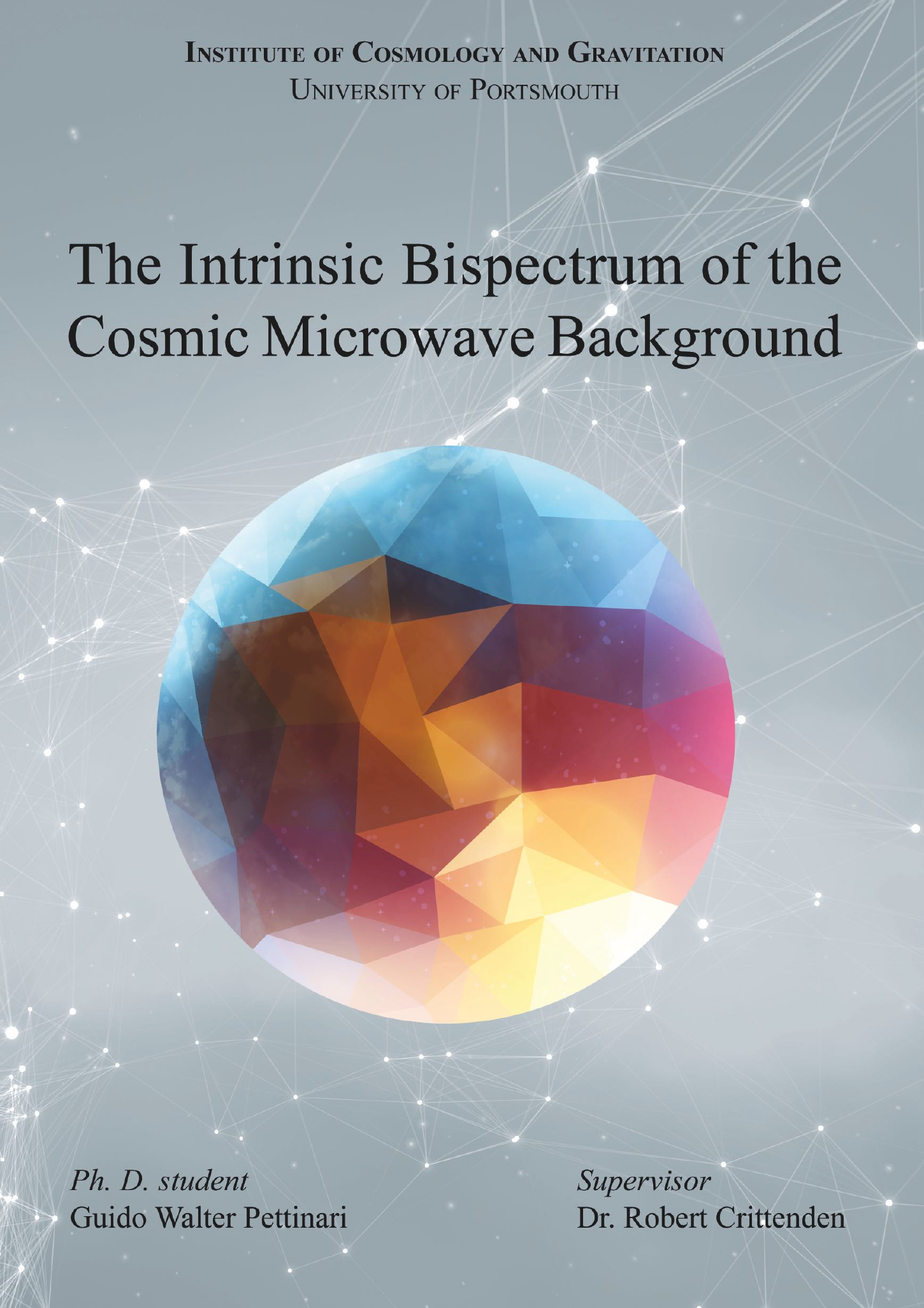}

\cleardoublepage

%
%
%
 
\chapter*{Preface}

\pagenumbering{gobble}
\pagenumbering{roman}

Cosmology, intended as the study of the origin and evolution of the Universe and its components, has advanced from being a philosophical discipline to a data-driven science.
Much of this progress was achieved in the last few decades thanks to the wealth of cosmological data from Earth and space-based experiments.
The abundance of observational constraints has considerably narrowed the space for theoretical speculation, to the point that now most of the cosmological community agrees on a \emph{standard model of cosmology}.

A crucial assumption of this model is that the structure observed in the Universe, such as planets, stars and galaxies, can be ultimately traced back to tiny density perturbations in the early Universe.
Therefore, a huge theoretical and experimental effort is being made by cosmologists and particle physicists to gain insight of the mechanism of generation of these primordial fluctuations, which remains still largely unknown.
The bispectrum of the cosmic microwave background (CMB) has been recently recognised as a powerful probe of this mechanism, as it is sensitive to the
non-Gaussian features in the seed fluctuations, which in turn are generated by non-linear processes such as the interactions between the fields present in the primordial Universe.

The non-Gaussianity of the CMB, therefore, opens a window on the non-linear physics of the early Universe; the CMB bispectrum is the observable that allows us to look through this window.
However, not all of the observed non-Gaussianity is of primordial origin. Indeed, a bispectrum arises in the CMB even for Gaussian initial conditions due to non-linear dynamics, such as CMB photons scattering off free electrons and their propagation in an inhomogeneous Universe.
This \emph{intrinsic bispectrum} is an interesting signal in its own right as it contains information on such processes. Furthermore, if not correctly estimated and subtracted from the CMB maps, it will provide a bias in the estimate of the primordial non-Gaussianity.

The main purpose of my doctorate has been to quantify the intrinsic bispectrum of the CMB and compute the bias it induces on the primordial signal.
In doing so, I have developed \SONG, a new and efficient code for solving the second-order Einstein-Boltzmann equations and compute primordial and intrinsic bispectra, including polarisation.\footnote{\SONG is open-source and available at \url{https://github.com/coccoinomane/song}.}
While this project might sound eminently technical, it allowed me to gain deep insight on some of the most important aspects of modern cosmology. The purpose of this Ph.D. thesis is to share such insight with the reader, in a plain and accurate way, avoiding the technicalities when possible and making those that cannot be avoided as digestible as possible.

In writing my thesis I employed a pedagogic approach and strived to make it comprehensible to a first-year Ph. D. student with a basic background in physics and statistics. The first four chapters, complemented with the appendices, review the state of the field, while the last chapters detail the original research that I conducted during my Ph. D. at the Institute of Cosmology and Gravitation, University of Portsmouth, UK, which led to the publication of the following paper:

\vspace{0.5cm}
{\smaller
G. W. Pettinari, C. Fidler, R. Crittenden, K. Koyama, and D. Wands. ``The intrinsic bispectrum of the cosmic microwave background''. J. Cosmology Astropart. Phys., 04(2013)003, doi: 10.1088/1475-7516/2013/04/003, \url{http://arxiv.org/abs/1302.0832}, April 2013.
}
\vspace{0.5cm}

Since I obtained my Ph. D. in 2013, my collaborators and I have carried out further research on the topic, extending the work presented in this thesis. In particular, we have found the polarised intrinsic bispectrum to be strongly enhanced with respect to the temperature one; developed a formalism to treat all propagation effects, including lensing, at second order; computed the power spectrum of the second-order B-modes; quantified the CMB spectral distortions in both temperature and polarisation; provided the most precise numerical computation of the intrinsic magnetic field generated around and after recombination. These works and this thesis can be freely accessed as preprints at \url{http://arxiv.org/find/astro-ph/1/au:+Pettinari_G/0/1/0/all/0/1}; their bibliographical references are, respectively:
\vspace{0.5cm}

{\smaller
Pettinari, Fidler, Crittenden, Koyama, Lewis, \& Wands. ``Impact of polarisation on the intrinsic CMB bispectrum''. PRD, 90, 103010, doi: 10.1103/PhysRevD.90.103010, \url{http://arxiv.org/abs/1406.2981}, November 2014.

\vspace{0.5cm}

Fidler, Koyama \& Pettinari. ``A new line-of-sight approach to the non-linear Cosmic Microwave Background.''. JCAP, 04(2015)037, doi: 10.1088/1475-7516/2015/04/037, \url{http://arxiv.org/abs/1409.2461}, July 2014.

\vspace{0.5cm}

Fidler, Pettinari, Crittenden, Koyama \& Wands. ``The intrinsic B-mode polarisation of the Cosmic Microwave Background''. JCAP, 07(2014)011, doi: 10.1088/1475-7516/2014/07/011, \url{http://arxiv.org/abs/1401.3296}, July 2014.

\vspace{0.5cm}

Renaux-Petel, Fidler, Pitrou \& Pettinari. ``Spectral distortions in the cosmic microwave background polarization''. JCAP, 03(2014)033, doi: 10.1088/1475-7516/2014/03/033, \url{http://arxiv.org/abs/1312.4448}, March 2014.

\vspace{0.5cm}

Fidler, Pettinari \& Pitrou. ``A precise numerical estimation of the magnetic field generated around recombination''. \url{http://arxiv.org/abs/1511.07801}, November 2015, to be submitted.
}
\vspace{0.5cm}

I would like to stress that this thesis would not exist without the constant help and encouragement of my Ph. D. supervisor, Prof. Robert Crittenden, and of my collaborator, Dr. Christian Fidler, and, in general, of all the great friends and colleagues that I was lucky enough to meet at the Institute of Cosmology and Gravitation.

\vspace{\baselineskip}
\begin{flushright}\noindent
Rome, December 2015\hfill {\it Guido W. Pettinari}\\
\hfill {\it guido.pettinari@gmail.com}\\
\end{flushright}

\setcounter{tocdepth}{2}
\tableofcontents

\chapter*{List of Abbreviations}

We will adopt the Einstein notation and imply a sum over repeated indices.
Greek letters are space-time indices, $\,\mu=0,1,2,3\,$, latin letters are spatial indices, $\,i=1,2,3\,$, the underlined letters $\underline{a},\underline{b},\underline{c}$ are space-time tetrad indices, while the underlined letters $\underline{i},\underline{j},\underline{k}$ are spatial tetrad indices.
For the metric, we use a $(-,+,+,+)$ signature.

We raise and lower the spatial indices with the Kronecker delta, $\KronUD{i}{j}$. Here are a few examples of this rule:
\begin{enumerate}
  \item $U^i=U_i$ is the spatial part of the four-velocity $U^\mu$; we use no symbol for the spatial part of $U_\mu$, which we shall just denote it as $g_{\mu\,i}\,U^\mu$.
  \item $\DD{\gamma}{i}{j}=\UD{\gamma}{i}{j}=\UU{\gamma}{i}{j}$ is the spatial part of the metric $g_{\mu\nu}$. 
  \item Unless explicitly stated, we shall always use the $(1,1)$-rank versions of the energy-momentum and Einstein tensors. Hence, $\UD{T}{i}{j}=\UU{T}{i}{j}=\DD{T}{i}{j}$ and $\UD{G}{i}{j}=\UU{G}{i}{j}=\DD{G}{i}{j}$ will represent the space-space parts of $\UD{T}{\mu}{\nu}\,$ and $\,\UD{G}{\mu}{\nu}\,$.
  \item $k^i=k_i\,$, $k_1^i=k_{1i}\,$ and $\,k_2^i=k_{2i}\,$ are the Fourier wavemode, and do not have an associated four-tensor.
\end{enumerate}

The cosmological quantities indexed by a `0' are evaluated today, \eg $a_0\equiv a(t_0)\,$, while those with an overbar are evaluated at zero order, \eg $\bar{\rho}(t)\equiv\pert{\rho}{0}(t)$.

The following abbreviations are used in this thesis:
\vspace{0.5cm}
\begin{longtabu} to 0.8\linewidth {X[1,l] X[3,l]}
  $\bullet$ BES   &   Boltzmann-Einstein differential system, \\[0.12cm]
  $\bullet$ CMB   &   Cosmic Microwave Background, \\[0.12cm]
  $\bullet$ CDM   &   Cold Dark Matter, \\[0.12cm]
  $\bullet$ FLRW   &   Friedmann-Lema\^itre-Robertson-Walker, \\[0.12cm]
  $\bullet$ GR   &   General Relativity, \\[0.12cm]
  $\bullet$ ISW   &   Integrated Sachs-Wolfe, \\[0.12cm]
  $\bullet$ LSS   &   Last Scattering Surface, \\[0.12cm]
  $\bullet$ ODE   &   Ordinary Differential equation, \\[0.12cm]
  $\bullet$ PDE   &   Partial Differential equation, \\[0.12cm]
  $\bullet$ SW   &   Sachs-Wolfe \,. \\[0.12cm]
\end{longtabu}
\renewcommand\listfigurename{List of figures}
\listoffigures
\listoftables


\chapter{Introduction}
\label{ch:introduction}

\pagenumbering{gobble}
\pagenumbering{arabic}

\section{Precision cosmology}

During the last three decades, cosmology has undergone a transition from a theory-dominated discipline to a data-driven science.  Currently, numerous Earth and space-based experiments provide observers with a continuous flow of high precision data, allowing us to constrain and rule out many of the models brought forward by theorists. For the first time, we have the tools to study in an accurate and quantitative way the origins and evolution of the Universe. It is unsurprising that our present days are commonly referred to as the era of precision cosmology.

As a result of this process, cosmologists are now converging towards a unified picture of the Universe, similar to when particle physicists built the standard model of particle physics.
The standard model of cosmology depicts the Universe as a mixture of five known particle species (photons, neutrinos, electrons, protons and neutrons), a hypothetical one (cold dark matter) and a mysterious dark energy component that can be interpreted either as a cosmological constant or as a fluid with negative pressure.
The structure that we observe in the Universe (galaxies, clusters, filaments, voids and temperature fluctuations) is thought to have originated from the gravitational enhancement of small initial density perturbations over an otherwise homogeneous and isotropic background.

The standard model of cosmology includes the fundamental observation that the Universe is expanding.
By extrapolating it back in time, today's expansion implies that the Universe was once in very dense and hot state. The limit of infinite temperature and density is called \emph{the Big Bang}, which conventionally marks the beginning of the Universe as we know it.
The existence of this ``primeval fireball'' \cite{peebles:1968a} leads to the prediction that the Universe must be permeated by a relic radiation from the Big Bang, the \emph{cosmic microwave background} (CMB).
The CMB was serendipitously discovered by Penzias and Wilson \cite{penzias:1965a, dicke:1965a} in 1965, thus providing a direct confirmation of the Big Bang scenario.
In the course of the years, the observation of the CMB has provided insight on the Universe that has been crucial to establish the standard model of cosmology.
This was possible thanks to three satellites that measured the CMB temperature map to increasingly high precision:
the NASA Cosmic Background Explorer (COBE) in the $1990$'s \cite{smoot:1999a},
the NASA Wilkinson Microwave Anisotropy Probe (WMAP) in the $2000$'s \cite{bennett:2012a}
and the ESA Planck survey, which has released its first-year results in $2013$ \cite{planck-collaboration:2013a}.
Thanks to these experiments, as well as ground and balloon based observations \cite{crill:2003a,jaffe:2001a}, we are now able to determine the parameters of the standard model of cosmology to percent-level precision.

\section{Cosmic inflation}
The standard model of cosmology, however, leaves open several important theoretical issues.
For example, it cannot explain why the CMB is observed with the same temperature within a part in a $10^5$ in regions of the sky that, in principle, were never in casual contact.
Furthermore, it lacks a mechanism to generate the initial density perturbations that seeded the observed structure of the Universe.
These and other problems are solved by postulating that, at some point in its infancy, the Universe underwent a \emph{cosmic inflation} \cite{guth:1981a, linde:1982a, albrecht:1982a, starobinsky:1980a}, that is, a period of accelerated expansion.
Before this time, our patch of Universe was much smaller than what is predicted by the hot Big Bang model; therefore, regions that are now out of reach were once causally connected and the causality problem is solved.
As for the primordial fluctuations, in the inflationary picture they are generated from microscopic quantum vacuum fluctuations that the accelerated expansion stretches and imprints on superhorizon scales \cite{hawking:1982a, starobinsky:1982a, mukhanov:1981a, bardeen:1983a}.

The simplest model of cosmic inflation involves a hypothetical scalar field slowly rolling down a very flat potential. In this circumstance, the field behaves like a fluid with negative pressure and thus powers an almost exponential cosmic expansion.
This simple picture is very successful as it predicts a nearly scale-invariant power spectrum of perturbations that is actually observed in the CMB \cite{planck-collaboration:2013a} and is compatible with the galaxy data \cite{sanchez:2012a}.
Many different theoretical models of inflation have been put forward that build on this ``vanilla'' model.
Some popular extensions include multiple fields, features in the inflation potential, the presence of a non-canonical kinetic term or non Bunch-Davies vacuum states \cite{martin:2013a, chen:2010a}.
In most cases, it is difficult to distinguish between these models of inflation just from the measurements of the power spectrum.

\section{Non-Gaussianity}
Recently, the three-point function of the primordial perturbation, or \emph{primordial bispectrum}, has aroused the interest of cosmologists for several reasons.
First, it vanishes for a Gaussian field and, therefore, it is the lowest order statistics sensitive to whether a perturbation is Gaussian or non-Gaussian; for this reason, the bispectrum is a measure of \emph{non-Gaussianity}.
Secondly, it is directly related to the angular bispectrum of the cosmic microwave background, which is an observable quantity \cite{planck-collaboration:2013b, komatsu:2001a, komatsu:2010a, bartolo:2010a, yadav:2010a, liguori:2010a}.
Finally, different models of inflation produce specific shapes for the primordial bispectrum, whose amplitudes are usually parametrised by a number denoted $\fnl\,$;
most importantly, the single-field slow-roll inflation produces an effectively Gaussian distribution of primordial density perturbations \cite{maldacena:2003a, acquaviva:2003a}, \ie $\,\fnl\simeq0\,$.
Therefore, the primordial bispectrum as inferred from the CMB has the power of ruling out the simplest models of inflation and to strongly constrain the physics of the early Universe based on the shape and amount of produced non-Gaussianity.

\section{The intrinsic bispectrum of the CMB}
However, we do not expect all of the observed non-Gaussianity to be of primordial origin.
Non-linear evolution will generate some degree of non-Gaussianity even in the absence of a primordial signal, for the simple reason that the product of Gaussian random fields is non-Gaussian.
The propagation of CMB photons in an inhomogeneous Universe and their non-linear collisions with electrons make it possible for Gaussian initial conditions to be non-linearly propagated into a non-Gaussian temperature field.
This results in the emergence of an \emph{intrinsic CMB bispectrum}\index{intrinsic bispectrum}, which is the topic of this thesis.

The primordial bispectrum is hypothetical and its shape and amplitude depend on the largely unknown details of cosmic inflation. The intrinsic CMB bispectrum, on the other hand, is always present and acts as a systematic bias in the measurement of the primordial bispectrum \cite{komatsu:2010a}.
In order to correctly interpret any non-Gaussianity measurement from the CMB bispectrum, and in particular those from the Planck satellite \cite{planck-collaboration:2013b}, it is of crucial importance to quantify this bias, which we label \fnlcon.
In addition, the non-Gaussian signal from non-linear dynamics has an interest of its own, as it might shed light on the details of the gravity theory \cite{gao:2011a}.

The non-linear signal can be quantified theoretically by using second-order perturbation theory; this is the leading order of non-Gaussianity since linear evolution cannot generate non-Gaussian features that are not already present in the initial conditions.
The Einstein and Boltzmann equations at second order have been studied in great detail \cite{bartolo:2006a, bartolo:2007a, pitrou:2007a, pitrou:2009b, beneke:2010a, naruko:2013a} and will be shown below.
They are significantly more complicated than at first order and solving them numerically is a daunting task;
this is testified by the many approximate approaches to the problem that can be found in the literature \cite{boubekeur:2009a, bartolo:2004a, bartolo:2004b, senatore:2009a, khatri:2009a, nitta:2009a, creminelli:2004a, creminelli:2004a, creminelli:2011a, bartolo:2012a, lewis:2012a}, which either neglect some of the physics or focus on a particular bispectrum configuration (we shall comment on these approaches in more detail in \cref{ch:intrinsic}).
Generally, these estimates yield a small non-Gaussianity level, with $\fnlcon\lesssim1$: none of them constitutes a significant bias for Planck, which constrains the local model of non-Gaussianity with an uncertainty of $\,\sigma_{f_{NL}}\sim5\,$.
However, the first full numerical computation of the bias, performed by Pitrou et al. (2010) \cite{pitrou:2010a, pitrou:2011a}, found the much higher value $\fnlcon\sim5\,$, just at the detection threshold for Planck.

The importance of the intrinsic bispectrum for the determination of the primordial non-Gaussianity and the tension between the numerical and analytical results in the literature has motivated us to compute the intrinsic bispectrum of the CMB.
Our purpose is to include all the relevant physical effects at second order in a numerically stable and efficient way.
The result of this effort is \SONG (Second-Order Non-Gaussianity), a numerical code that solves the second-order Einstein-Boltzmann equations for photons, neutrinos, baryons and cold dark matter.
\SONG is written in C, is parallel, and is based on the first-order Boltzmann code \emph{CLASS} \cite{lesgourgues:2011a, blas:2011a}, from which it inherits its modular structure and ease of use.
\SONG is fast enough to perform various convergence tests to check the robustness of the numerical results. 
Utilising this code, we will study the intrinsic non-Gaussianity to quantify the bias in the measurements of primordial non-Gaussianity and evaluate its signal-to-noise ratio.

We have published the results thus obtained in \citet{pettinari:2013a}. While the paper was in preparation, two works appeared that study the intrinsic bispectrum, giving similar results for the bias to the primordial non-Gaussianity templates, but different ones for the signal-to-noise ratio \cite{huang:2012a, su:2012a}. We will discuss in \cref{ch:conclusions} why these references obtained different results.

\section{Summary of the thesis}

The purpose of this thesis is to introduce and compute the intrinsic bispectrum of the cosmic microwave background, and to quantify its observability.
A description of the structure of the thesis follows.

In \cref{ch:homogeneous_universe} we present the standard Model of Cosmology and describe the evolution of the metric and matter species under the assumption of perfect homogeneity and isotropy.
We explain how the cosmic microwave background is originated and discuss the potential of constraining models of cosmic inflation via its bispectrum.

In \cref{ch:perturbation_theory} we use perturbation theory to model the small deviations from homogeneity expected in the early Universe.
The non-linearities in the cosmological perturbations are studied by expanding them up to second order.
We take particular care in separating their stochastical properties from their dynamical evolution by introducing the concept of transfer function.
The main subject of this work, the intrinsic bispectrum, is discussed for the first time.
We also report the Einstein equations up to second order.

In \cref{ch:boltzmann} we introduce the Boltzmann formalism as a general framework to compute the time evolution of the perturbations of the massless and massive species.
To simplify the derivation of the collision term and the interpretation of the energy and momentum of the particles, we work in the local inertial frame via the tetrad formalism.
The angular and positional dependences of the second-order Boltzmann equation are decomposed using plane waves and spherical harmonics, thus resulting in a hierarchy of equations for the Fourier multipoles of the distribution function.
The Boltzmann hierarchies, together with the Einstein equations, form the Boltzmann-Einstein system of differential equations at second order (BES).

In \cref{ch:evolution} we summarise the equations in the BES and illustrate how our code, $\,\SONG\,$, efficiently solves them for the evolution of the metric variables and Fourier multipoles.
We derive and show the initial conditions for the system, which are set deep in the radiation era when the Fourier modes are superhorizon.
The second-order transfer functions are evolved with the differential system until the time of recombination; to obtain their present-day value, we describe and solve the line of sight integral.
The line of sight sources are split into three contributions: the collision sources, the metric sources and the propagation sources.
We also present detailed numerical and analytical tests on the transfer functions computed by \SONG.

In \cref{ch:intrinsic} we compute the intrinsic bispectrum of the cosmic microwave background and quantify its observability and the bias it induces on a measurement of the primordial bispectrum.
We first derive a formula where the intrinsic bispectrum is obtained from a four-dimensional integral over the first and second-order transfer functions.
To quantify its importance we use a Fisher matrix formalism where we consider the intrinsic bispectrum and three primordial ones: local, equilateral and orthogonal.

Finally, in \cref{ch:conclusions} we conclude by summarising our main results.
We also propose other interesting research directions where \SONG will be useful, such as computing the spectrum of the $B$ polarisation of the CMB, studying the impact of modified gravity theories on the intrinsic bispectrum, quantifying the spectral distortions and the generation of magnetic fields at recombination.

\section{Further research}

As mentioned in the preface, since I obtained my Ph. D. in 2013 my collaborators and I have carried out further research on the non-linearities of the CMB, extending the work in this thesis. In particular, we have found the polarised intrinsic bispectrum to be strongly enhanced with respect to the temperature one \cite{pettinari:2014b}; developed a formalism to treat all propagation effects, including lensing, at second order \cite{fidler:2014b}; computed the power spectrum of the second-order B-modes \cite{fidler:2014a}; quantified the intrinsic spectral distortions in the CMB \cite{renaux-petel:2013a}.
These works are all published in peer-reviewed journals, and can be freely accessed as preprints on the arXiv (\url{http://www.arxiv.org}); the reader can refer to the preface for their bibliographical references.

Furthermore, during my Ph.D. I have worked on two projects that are not related to the topic of this thesis.
The first project involved using Active Galactic Nuclei (AGNs) to probe the existence of axion-like particles; in particular, we showed that, while promising, this possibility is unattainable until we understand the nature of AGNs in detail \cite{pettinari:2010a}.
In the second project, we have studied the behaviour of isolated galaxy pairs from a numerical simulation, with the objective of determining whether they contain information about the cosmological expansion \cite{bueno-belloso:2012a}.

\chapterbib


\chapter{The Standard Model of Cosmology}
\label{ch:homogeneous_universe}

\section{Introduction}

The standard model of cosmology encompasses our knowledge of the Universe as a whole. It has matured over the last century, consolidating its theoretical foundations with increasingly accurate observations. The main assumptions on which it rests are:
\begin{itemize}
  \item On sufficiently large scales the Universe is homogeneous and isotropic (the \keyword{cosmological principle}).
  \item The energy content of the Universe is modelled in terms of cosmological fluids with constant equation of state: photons, baryons, neutrinos, cold dark matter and dark energy.
  \item The gravitational interactions between the cosmological fluids are described by Einstein's general relativity (GR).
\end{itemize}
Along with the above theoretical assumptions, the standard model of cosmology includes the fundamental observation that the Universe is expanding.

\subsection{Summary of the chapter}

In this chapter we analyse these features in detail, starting with the cosmological principle in \sref{sec:cosmological_principle}.
The assumptions of isotropy and homogeneity lead to the formulation of the \FLRW metric, which we introduce in \sref{sec:expansion_of_the_universe}.
We derive the dynamic evolution of this metric in \sref{sec:background_evolution} by solving the Einstein equation; in particular, we find that the cosmic expansion is one of the solutions and is favoured by the measured abundances of the various species.
The presence of a cosmic expansion, in turn, indicates that the primordial Universe was in a very hot and dense state where thermal equilibrium between the species was established.
This prediction is spectacularly confirmed by the observation of a cosmic microwave background with a blackbody spectrum, which is the subject of \sref{sec:background_cmb}.
We conclude the chapter by discussing in \sref{sec:inflation} some important problems of the hot Big Bang scenario and one of the possible ways to solve them: the mechanism of cosmic inflation, a phase of accelerated expansion in the early Universe.

Note that in \sref{sec:non_gaussianity} we shall briefly discuss how non-linearities might arise during inflation that generate non-Gaussian signatures. The work described in this thesis is ultimately motivated by the quest to measure said non-Gaussianity.

\section{The Cosmological Principle}
\label{sec:cosmological_principle}
The cosmological principle\index{cosmological principle} (CP) states that on sufficiently large scales the Universe is homogeneous and isotropic. \keyword[homogeneity]{Homogeneous} means that different patches of the Universe have the same average physical properties. In particular, any cosmological fluid has the same energy density, pressure and temperature everywhere. \keyword[isotropy]{Isotropic} means that there are no preferred directions in the Universe. Any observer measuring a cosmological quantity -- \eg the photon flux or a galaxy count -- in two different directions should find the same value.

Homogeneity does not imply isotropy. For example, a Universe filled with a homogeneous magnetic field is homogeneous but not isotropic. On the other hand, isotropy about one location does not guarantee homogeneity. The simplest case is given by an observer at the centre of an isotropic explosion, but there are other examples of inhomogeneous distributions that project isotropically on the sky of one observer \cite{durrer:1997a}. However, isotropy about two locations does guarantee homogeneity and isotropy about all locations (\citet[Page~65-67]{peacock:1999a}).

The cosmological principle is spectacularly violated on small scales. Planets, stars and galaxies should not exist in a perfectly homogeneous Universe. However, when zooming out on scales larger than roughly $ \unit[100 \, h^{-1}]{Mpc} $, where $\unit[1]{Mpc} =  \unit[\sci{3.086}{22}]{m} = \unit[\sci{3.262}{6}]{ly} $ is roughly the average distance between two galaxies, the Universe does become smooth, as we detail in \sref{sec:validity_cosmological_principle}. This allows us to treat the dynamics of the cosmological fluids on the largest scales as if the Universe were perfectly homogeneous and isotropic. In this limit, the physics and the resulting equation are particularly simple, as discussed in \sref{sec:expansion_of_the_universe}.

The cosmological principle also allows us to define a universal time variable, the \keyword{cosmic time}, defined as the time measured by observers at rest with respect to the matter in their vicinity. The homogeneity of the Universe ensures that the clocks of these fundamental observers can be synchronised with respect to the evolution of the universal homogeneous density. We choose the zero of the cosmic time to coincide with the Big Bang, which we shall introduce in \sref{sec:background_evolution}. As a consequence, the cosmic time is interpreted as the age of the Universe.

\subsection{Validity of the Cosmological Principle}
\label{sec:validity_cosmological_principle}
The cosmological principle is crucial in order to make sense of the Universe, as it allows us to give universal significance to our local measurements. Furthermore, as we shall see in \sref{sec:background_evolution}, it leads to an elegant dynamical solution of Einstein's equations. When it was proposed
, however, the cosmological principle was little more than a conjecture. As cosmological observations increased in number and accuracy, it was substantiated by more and more evidence. Nevertheless, the cosmological principle has not been proven unambiguously yet.

The main difficulty lies in the fact that it is impossible to observationally prove the homogeneity of the Universe without first assuming the Copernican principle, according to which we do not occupy a special position in the Universe\footnote{This is also referred to as the \emph{weak cosmological principle} by \citet{ellis:1975a}.}. The reason is that any observation has only access to our past light cone. Even worse, we cannot effectively move in cosmic time or space, so that we can only probe the past light cone of here and now. As a result, our observations mix time and space in such a way that we cannot tell the difference between an evolving homogeneous distribution of matter and an inhomogeneous one with a different time evolution \cite{maartens:2011a}.

If we accept the Copernican principle, however, the existence of isotropy in the observable Universe (that is, isotropy in the past light cone of Earth) would automatically imply the homogeneity of the whole Universe \cite{ellis:1975a, maartens:2011a}. Isotropy, contrary to homogeneity, is well established by many observations. The most relevant ones are the nearly perfect isotropy of the Cosmic Microwave Background \cite{bennett:1996a}, the isotropy of the X-ray background \cite{scharf:2000a} and the isotropies of various source populations, \eg radio galaxies \cite{peebles:1993a}. The isotropy of the CMB also provides a good argument for homogeneity, since its angular distribution is linked to the three-dimensional fluctuations of the gravitational potential during recombination \cite{wu:1999a}.

Not assuming the Copernican principle has two important consequences. First, the observed isotropy could not be used to infer homogeneity, not even in our local Universe. Secondly, observations would need to be interpreted in light of our special position. This is the case in the so-called void models, where the cosmological principle is assumed to be valid but our Galaxy sits close to the center of an under-dense area which is radially inhomogeneous (the void). While some of these models have the benefit of removing the need for a cosmological constant by modifying the redshift-distance relationship (see, \eg, Ref.~\cite{tomita:2000a,nadathur:2011a,moffat:1995a}), they fail to reproduce all the available observations at the same time \cite{caldwell:2008a,clifton:2009a,yoo:2010b,moss:2011a,zhang:2011a,zumalacarregui:2012a,wang:2013a}. For a review of other ways to test the Copernican principle, refer to Ref.~\cite{hamilton:2013a,clarkson:2012a,maartens:2011a}.

A useful check for the homogeneity of the observable Universe consists in counting objects in a galaxy-survey in regions of increasing volume. In a homogeneous Universe, the mean density of galaxies in these regions should approach a constant value at a certain \emph{homogeneity scale}. In order to look for this scale in the data, one needs to assume a cosmological model to convert the measured fluxes of galaxies to distances; hence it is more of a consistency check for homogeneous models rather than a test of homogeneity. \footnote The largest-volume measurement ($V \sim \unit[1 \, h^{-3}]{Gpc^3}$) to date was performed by \citet{scrimgeour:2012a} using the blue galaxies of the WiggleZ survey \cite{drinkwater:2010a}. They found homogeneity for scales larger than $ \unit[70 \, h^{-1}]{Mpc}$, in agreement with what previously obtained by \citet{hogg:2005a} using large red galaxies\footnote{As a comparison consider that the disk of our Galaxy, the Milky Way, which is an average galaxy, measures just around $ \unit[30]{kpc} $.}, and in disagreement with earlier results that suggested a fractal structure of the Universe \cite{pietronero:1987a, syloslabini:2009a}. Interesting discussions about the scale of homogeneity and the fractal Universe can also be found in Ref.~\cite{guzzo:1997a,davis:1997a}. For an observational test of homogeneity that relies only on the angular distances of galaxies, and is therefore less model-dependent, refer to Ref.~\cite{alonso:2013a}.

\section{The expansion of the Universe}
\label{sec:expansion_of_the_universe}

In the 1910's Vesto Slipher had noticed by measuring their light spectra that most of nearby galaxies~--~or \emph{nebulae}, as they were called at the time~--~were quickly receding from us \cite{slipher:1913a, slipher:1915a}.
In 1929, Edwin Hubble \cite{hubble:1929a} independently confirmed that galaxies where receding and found a correlation between their radial velocity and their distance from us. This observation is encoded in \keyword{Hubble's law}, whereby there is a linear relationship between the radial speed with which a galaxy recedes from Earth and its distance to it:
\begin{align}
  \label{eq:hubble_law}
  v \;=\; H_0\,r \;.
\end{align}
The proportionality constant is now called \keyword{Hubble constant}.

If one assumes the cosmological principle, Hubble's law becomes universal: any two galaxies move away from each other with a speed proportional to the distance that separates them. In reality, the cosmological principle alone suffices to enforce the proportionality between distance and radial velocity. Isotropy enforces the radial motion, while homogeneity ensures that the recession velocity is proportional to the distance \cite{harrison:2000a, peacock:1999a}.
However, the cosmological principle alone does not specify the sign of this proportionality, which Hubble found to be positive.

Hubble's discovery was soon linked to previous theoretical papers by Georges Lema\^itre \cite{lemaitre:1927a, lemaitre:1931a} and Alexander Friedmann \cite{friedmann:1922a}. In these pioneering works, the authors found dynamical solutions to Einstein equations where the Universe could expand indefinitely in a homogeneous manner. In this context, Hubble's law is the empirical consequence of a more fundamental concept: space itself is expanding. The apparent recession of galaxies is just one manifestation of the expansion of the Universe, and $H_0$ represents the homogeneous expansion rate\footnote{It is sometimes thought that Hubble discovered the expansion of the Universe in his 1929 paper. This was not the case, as the first connection to Lema\^itre and Friedmann works was made in 1930 by Arthur Eddington and Willem de Sitter. An account by the American Institute of Physics of the fascinating story behind the discovery of the expansion of the Universe can be found at the following URL: \url{http://www.aip.org/history/cosmology/ideas/expanding.htm}.}.
In the expanding Universe picture, the receding galaxies are not thought as projectiles shooting away through space, but as objects at rest in expanding space. Similarly, the recession speed is not the speed of something moving through space, but of space itself; it is not a local phenomenon and this is why it can exceed the speed of light without changing the causal structure of space-time \cite{harrison:2000a}.


The value of $H_0$ cannot be predicted by theoretical means: only observation can pin it down. Since distance measurements are subject to high uncertainty, it is customary to parametrize the Hubble constant by means of the pure number $ h $:
\begin{align}
  \label{eq:hubble_units_of_measure}
  H_0 & \;\equiv\; \unit[100 \:\, h \:]{\frac{km/s}{Mpc}} && \\[0.14cm]
      & \;=\; \frac{h}{\unit[9.77]{Gyr}} & \\[0.14cm]
      & \;=\; \unit[\frac{h}{\sci{4.69}{41}}]{GeV} & \text{( assuming $\hbar=1$ )} & \\[0.14cm]
      & \;=\; \frac{h}{\unit[2998]{Mpc}} & \text{( assuming $c=1$ )} & \;.
\end{align}
In his seminal paper, Hubble estimated $h\sim5\,$.
The most accurate local measurements of $\,h\,$ to date employ Cepheid variables and Type Ia supernovae in low-redshift galaxies, and read
\begin{align}
  \label{eq:hubble_local_measurement}
  &h \;=\; 0.738 \pm 0.024 \;\qquad\text{(\citet{riess:2011a})}\;, \msk
  &h \;=\; 0.743 \pm 0.021 \;\qquad\text{(\citet{freedman:2012a})}\;,
\end{align}
at \onesigma.
The Planck CMB satellite obtained a more precise value \cite{planck-collaboration:2013a}, but it is an indirect estimate as it assumes a cosmological (\LCDM) model:
\begin{align}
  \label{eq:hubble_planck_measurement}
  h \;=\; 0.6780 \pm 0.0077 \;\qquad\text{(Planck+WP+highL+BAO)}\;,
\end{align}
at \onesigma. There is a mild tension between the two measurements, which could be explained by some unknown source of systematic error in the local measurement or by the fact that the \LCDM model assumed in Planck's data analysis is incorrect \cite{planck-collaboration:2013a, verde:2013b}.

On small scales the cosmological principle fails because, over time, gravitational instability creates bound structures such as stars, galaxies and clusters of galaxies. Hence, we expect galaxies to have their own motions decoupled from the Hubble expansion, which are called \keyword{peculiar velocities}. An example of peculiar velocity is the circular motion of the galaxies of a cluster around the common centre of mass. In most cases, the magnitude of the peculiar velocities does not exceed $\unit[10^3]{km/s}$; using the measured values for $H_0$, we expect peculiar velocities to be negligible with respect to the Hubble flow for objects distant more than roughly $\unit[100]{Mpc}$. It is reassuring that such a value is consistent with the homogeneity scale discussed in \sref{sec:cosmological_principle}.

\subsection{The metric}
\label{sec:background_metric}

The dynamics of the expanding Universe are better understood in terms of observers who are at rest with the Hubble expansion, the so-called \keyword{comoving observers}. Comoving observers perceive the Universe as isotropic and see objects receding from them according to Hubble's law. In this section, we shall employ \keyword{comoving coordinates} defined as the coordinate system where all comoving observers have constant spatial coordinates, \ie are static. Any motion in comoving coordinates has the Hubble part subtracted so that the only velocities are the peculiar ones.

In differential geometry the distance $ ds $ between two infinitesimally nearby space-time points $(x^0,$ $x^1,$ $x^2,$ $x^3)$ and $(x^0+dx^0,$ $x^1+dx^1,$ $x^2+dx^2,$ $x^3+dx^3)$ is called the line element and is defined as
\begin{align*}
  ds^2 \,=\, g_{\mu\nu}\left(x\right) \,dx^\mu dx^\nu \; \mbox{ for } \; \mu,\nu = 0, 1, 2, 3 \;.
\end{align*}
Here $g_{\mu\nu}(x)$ is the metric\index{metric}, a (0,2) tensor which determines how distances are computed in the considered space-time manifold. We shall adopt comoving coordinates and set $ x^0 = c\,t $ where $t$ is the cosmic time.

The metric that describes a homogeneous and isotropic expanding space-time is called the Friedmann-Lema\^itre-Robertson-Walker (\FLRW) metric\index{FLRW metric}\index{Friedmann-Lema\^itre-Robertson-Walker metric} \cite{friedmann:1922a, lemaitre:1931a, robertson:1935a, walker:1937a}. In comoving coordinates, it is given by
\begin{align}
  \label{eq:flrw_metric}
  ds^2 \,=\, -(c\,dt)^2 \,+\, a(t)^2 \, \gamma_{ij} \,dx^i\,dx^j \;.
\end{align}
The \keyword{cosmic time} $t\,$, introduced in \sref{sec:cosmological_principle}, is defined so that the Universe has the same density everywhere at each moment in time. The \keyword{scale factor} $a(t)$ parametrises the uniform expansion of the Universe.
We express the spatial part of $ds^2$ so that, in comoving and spherical coordinates $(\rho,\theta,\phi)\,$, it reads
\begin{align}
  \label{eq:metric_flrw_general_spatial}
  \gamma_{ij}\,dx^i\,dx^j \;=\;
  d\rho^2 \,+\, S_k(\rho)^2 \left( d\theta^2 + \sin^2\theta\,d\phi^2 \right) \;.
\end{align}
With this choice, the quantity $\,d\chi^2\equiv\gamma_{ij}\,dx^i\,dx^j\,$ has the meaning of a \keyword{comoving distance} or \keyword{coordinate distance}.
The function $S_k(\rho)$ depends on the \emph{spatial curvature}\index{curvature} of the Universe, which in these models is uniform and is given by $k/a^2$. Even before discussing its form, it should be noted that for radial trajectories ($d\phi=d\theta=0$) the comoving distance coincides with the radial comoving coordinate.

We distinguish three different geometries for the Universe based on the value of the curvature constant $k$:
\begin{align}
  S_k(\rho) \:=\:
  \left\{
  \begin{aligned}
    &\quad \rho        \quad& \text{flat geometry}       \quad  &(k = 0)     &\msk
    &\quad \sin(\rho)  \quad& \text{spherical geometry}  \quad  &(k = +1)    &\msk
    &\quad \sinh(\rho) \quad& \text{hyperbolic geometry} \quad  &(k = -1)    &\;.\msk
  \end{aligned}
  \right.
\end{align}
For $k=0$, the comoving distance is just the usual Euclidean distance: $d\chi^2=\delta_{ij}\,x^ix^j$. The value of the curvature constant $k$ is a free parameter in the \FLRW models and, as the Hubble constant, has to be determined by experiment. Recent results from the WMAP \cite{hinshaw:2012a} and Planck \cite{planck-collaboration:2013a} CMB satellites constrain the spatial curvature to be negligible, thus suggesting that we live in a Universe with a flat geometry. We shall assume $k=0$ for the rest of this work.
This allows us to choose coordinates where $\rho$ and $\chi$ are lengths (measured in Mpc) and the scale factor is a dimensionless quantity such that $a(t_0)=1$ \cite{durrer:2008a}.

Now that we have introduced the concept of scale factor, Hubble's law follows easily. Given an observer at the origin of a spherical coordinate system, we define the physical coordinates of an object as $\vec{r} = a(t)\,\vec{x}$, where $\vec{x} = (x^1, x^2, x^3$) are its comoving coordinates. The distance $r=a(t)\,\chi$ along a radial path is the \keyword{physical distance} and can be thought as the distance that would be measured by stretching a tape measure in a uniformly curved surface \cite{harrison:2000a}. There are two contributions to the velocity $d\vec r/dt$:
\begin{align}
  \frac{d\vec r}{dt} \;=\; \frac{1}{a}\frac{da}{dt}\,\vec r \;+\; a\,\frac{d\vec x}{dt}\;.
\end{align}
We project along the radial direction $\hat{\vec r}$ in order to obtain an expression for the radial velocity $v=d\vec{r}/dt \cdot \hat{\vec r}\,$:
\begin{align}
  v \;=\; \frac{1}{a}\frac{da}{dt}\,r \;+\; a\,\frac{d\vec x}{dt} \cdot \hat{\vec r}\;.
\end{align}
The term $a\,d\vec{x}/dt \cdot \hat{\vec r}$ is the \keyword{peculiar velocity} of the object. For a comoving object ($d\vec x/dt = 0$) we obtain the so-called \keyword{velocity-distance law}:
\begin{align}
  \label{eq:velocity_distance_law}
  v \;=\; \frac{1}{a}\,\frac{da}{dt}\,r \;.
\end{align}
The above equation has the same form of Hubble's law in \eref{eq:hubble_law}. From a direct comparison, we see that the Hubble constant $H_0$ is just the present-day value of the \keyword{Hubble parameter} defined as
\begin{align}
  H \;\equiv\; \frac{1}{a}\;\frac{da}{dt} \;.
\end{align}

\runinhead{Conformal time} The \FLRW metric can be conveniently expressed using the \keyword{conformal time} defined as $\,d\tau = dt/a\,$:
\begin{align}
  \label{eq:conformal_metric_flrw}
  ds^2 \,=\, a(\tau)^2 \;\left\{\;
  -(c\,d\tau)^2 \,+\, \gamma_{ij} \,dx^i\,dx^j \;\right\} 
  \;=\; a(\tau)^2 \: \eta_{\mu\nu} \,dx^\mu\,dx^\nu\;,
\end{align}
where $ \eta_{\mu\nu} $ is the Minkowski metric of special relativity and we have assumed flat space ($k=0$). In the following chapters we shall use $\tau$ instead of $t$ as the evolution variable for the cosmological perturbations, and assume units where $c=1\,$.
It should be noted that, for a radial trajectory, the conformal time is equal to the comoving distance divided by $c$.

\subsection{Light in an expanding Universe}
\label{sec:light_in_an_expanding_Universe}
The cosmological data that we extract from the Universe (temperature and polarisation maps, galaxy surveys, lensing maps, \etc) rely on the observation of light, with the exceptions of neutrinos and, possibly, gravitational radiation. It is therefore crucial to understand how light is affected by the expansion of the Universe.

\subsubsection{Expansion redshift}
\label{sec:expansion_redshift}

All physical lengths are stretched by the expansion of the Universe; the wavelength of a light wave makes no exception. Light emitted by a comoving source at time $t$ with wavelength $\lambda$ will be seen by a comoving observer today with a wavelength $\lambda_0$ given by
\begin{align*}
  \frac{\lambda_0}{\lambda} = \frac{a(t_0)}{a(t)} \;.
\end{align*}
As it travels through the expanding Universe, the light emitted from distant objects experiences an \keyword{expansion redshift}: its spectrum is uniformly shifted to larger wavelength and lower energies by an amount depending solely on the time of emission, regardless of whether the light consists of radio waves or gamma rays.

By adopting the same convention as in spectroscopy, where the fractional wavelength shift $(\lambda_0-\lambda)/\lambda$ is denoted by the letter $z$, we write the \keyword{expansion-redshift law}
\begin{align}
  \label{eq:expansion_redshift_law}
  1 + z(t) = \frac{a(t_0)}{a(t)} \;.
\end{align}
If we assume that the laws governing the emission and absorption of light do not change through cosmic evolution, the expansion redshift of a cosmological source can be inferred from its electromagnetic spectrum. Thanks to spectroscopic galaxy surveys such as 2dF \cite{colless:2001a}, SDSS-II \cite{york:2000a}, WiggleZ \cite{drinkwater:2010a} and BOSS \cite{dawson:2013a}, we have now measured the optical spectra of millions of galaxies and thus determined their redshift. 

In an expanding Universe, the sources with the highest redshift are the ones farthest away from us. Hence, high-redshift objects have to be more luminous than low-redshift ones for us to be able to see them. The highest-redshift galaxy that has been spectroscopically confirmed to date has $z=7.51$ \cite{finkelstein:2013a} \footnote{Note that a galaxy with a spectroscopic redshift of $z=8.6$ had been previously reported in Ref.~\cite{lehnert:2010a}, but it was later found to be a spurious signal in Ref.~\cite{bunker:2013a}.}, and a candidate galaxy with $z=11.9$ \cite{ellis:2013a} has been recently reported. In a \LCDM Universe, the light from these galaxies was emitted about $13$ billion years ago and their distance is now growing at a rate of many times the speed of light.

In the following, we will sometimes use the redshift as a time variable to parametrize the evolution of the Universe. This is correct since $z$ is a monotonically decreasing function of $ a $ which in turn, in an expanding Universe, is a monotonically increasing function of cosmic time. Note also that from \eref{eq:expansion_redshift_law} it follows that today ($a(t_0)=1$) the redshift vanishes: $z(t_0)=0\,$.

\subsubsection{Other redshifts}
\label{sec:other_redshifts}

The expansion redshift should not be confused with the Doppler effect\index{Doppler redshift}. The Doppler effect produces a shift in the observed wavelength of photons because of the relative motion between source and observer. The recession velocity does not give rise to a Doppler shift because it does not describe the motion of objects in space, but the rate at which distances grow in the expanding Universe. Incidentally, this is why recession velocities can be larger than the speed of light. What gives rise to the expansion redshift is the wavelength of photons getting stretched during their trajectory through expanding space. On the other hand, Doppler redshift is generated by the peculiar velocities of the galaxies, which cannot exceed the speed of light.

A third type of redshift, the gravitational redshift\index{gravitational redshift}, arises from the fact that the photons frequencies change as they travel through an inhomogeneous gravitational field.
For example, we expect the light from a cluster of galaxies to be gravitationally redshifted, as the gravitational field at the centre of the cluster is different from that on the surface of Earth.

Expansion redshift, Doppler redshift and gravitational redshift coexist in the spectrum of galaxies and, in general, of all astrophysical sources. When determining the expansion redshift of an object, the non-cosmological Doppler and gravitational redshifts must be subtracted or accounted for in the error budget. The gravitational redshift is usually not too much of a concern as it shifts the spectrum by just $z\sim10^{-3}$ \cite{harrison:2000a}. However, in the local Universe, say for $z<0.01$, the peculiar velocities give rise to a Doppler redshift of the same order of the expansion one. This is a manifestation of the breakdown of the cosmological principle on small scales due to gravitational instability. For more distant objects, peculiar velocities become negligible with respect to recession velocities and one can trust the measured redshift to be due to the expansion of the Universe.

%

\subsection{Comoving distance}
\label{sec:comoving_distance}
In \sref{sec:background_metric} we have introduced the concept of comoving distance\index{comoving distance} $\chi$ as the dimensionless distance between two spatial points on the comoving grid. The great advantage of $\chi$ is that it is constant in time, since its expression only involves comoving coordinates. On the other hand, the physical distance\index{physical distance}, given by $\,r=a(t)\chi\,$, is the tape-measure distance on a grid which is not comoving with the expansion, and hence increases with time.

But how are these theoretical distances related to the measured redshift of an object? Since redshift is intrinsically related to light propagation, we need to study the trajectory of photons from a source to us. This is described by the null geodesics ($ds^2=0$) along a radial path ($d\phi=d\theta=0$)\footnote{It should be noted that, given the choice of the spatial metric in \eref{eq:metric_flrw_general_spatial}, the comoving distance for a radial path is just the radial comoving coordinate.}, which in the case of the \FLRW metric in \eref{eq:flrw_metric} yields
\begin{align}
  d\chi \;=\; \frac{c}{a}\,dt \;.
\end{align}
This result is intuitive: the actual speed of a photon does not vary, but its speed with respect to expanding coordinates is larger when the Universe is small ($a<1$). A photon that was emitted at a time $t_\ems$ and observed at $t_\obs$ will have travelled a comoving distance of
\begin{align}
  \label{eq:comoving_distance_ems_obs}
  \chi\,(t_\ems, t_\obs) \,=\, \int_{t_\ems}^{t_\obs} \, \frac{c}{a(t)} \,dt \;.
\end{align}
Any comoving distance is by construction independent of time. If another photon is emitted soon after the first one (say, at time $t_1+dt_1$), it is obviously observed after the first one (say, at time $t_2+dt_2$), but the comoving distance covered is the same. In formulae, $\chi\,(t_1, t_2) \,=\, \chi\,(t_1 + dt_1, t_2 + dt_2)$. Inserting this identity in \eref{eq:comoving_distance_ems_obs} yields $dt_1/a(t_1) = dt_2/a(t_2)$: the quantity $dt/a(t)$ is conserved along the light cone. This is a formal demonstration of the fact that all time intervals get stretched while propagating through an expanding Universe. Since $d\lambda = c dt\,$, this is true also for all wavelengths.

Using the expansion-redshift law, $\,1+z=a_0/a\,$ and the definition of the expansion rate, $\,aH=da/dt\,$, the comoving distance can be related to the redshift by
\begin{align}
  d\chi \;=\; - \frac{c}{a_0\,H(z)} \, dz \;,
\end{align}
where $a_0\equiv a(t_0)\,$.
Thus, the comoving distance travelled by a photon emitted at a redshift $z$ and received today ($z=0$) is given by
\begin{align}
  \label{eq:distance_redshift_law}
  \chi(z) \;=\; \frac{c}{a_0} \,\int_0^z \frac{dz}{H(z)} \;=\;
  \frac{c}{a_0\,H_0} \,\int_0^z \frac{dz}{E(z)} \;,
\end{align}
where we have defined the dimensionless parameter $E(z) \equiv H(z)/H_0$ \cite{amendola:2010a}. We shall refer to the above formula as the \keyword{distance-redshift law}; it is important because it relates the geometry of the Universe ($\chi$ and $H$) to the measured redshift. By using the velocity-distance relation $\,v=H_0\,r\,$ and the identity $\,r(t,t_0)=a_0\,\chi(t,t_0)\,$, we obtain the \keyword{velocity-redshift law}
\begin{align}
  \label{eq:velocity_redshift_law}
  \frac{v}{c} \;=\; \int_0^z \frac{dz}{E(z)} \;,
\end{align}
which is key to convert a redshift to the recession velocity at the time of emission.

The distance-redshift and velocity-redshift laws tell us that, in order to infer the distances and velocities of an object, we first need to know the expansion history of the Universe $H(z)$ all the way to when the light was emitted. The reason is that our cosmological observations are limited to the region of space-time included in our past light cone. We, as observers, do not have access to a the world map but only to a single world picture taken now and here \cite{harrison:2000a}. The farthest sources in our world picture emitted their light at a time where the expansion rate was significantly different from the current value, $H_0$. Furthermore, the emitted light travelled for a long time in an expanding Universe. Hence, the measured redshift is related to the distance covered by the light by the expansion history between emission time and observation time.

If the object is very close, however, the integral $\int_0^z dz/E(z)$ can be Taylor expanded around $z=0$ \cite{amendola:2010a}:
\begin{align}
  \int_0^z \frac{dz}{E(z)} \;\simeq\; z \;-\; \frac{E'(0)}{2} z^2 \;+\; \frac{1}{6}
  \left[2E'(0)^2 \;-\; E''(0)\right]\,z^3 \;+\; \mathcal{O}(z^4) \;,
\end{align}
where the prime represents a derivative with respect to $z$. By keeping only the first term in the expansion, the distance-redshift and velocity-redshift laws become respectively
\begin{align}
  \label{eq:hubble_law_cz}
  c\,z \,=\, H_0\,r
\end{align}
and
\begin{align}
  \label{eq:fizeau_doppler_law}
  v \,=\, c\,z \;.
\end{align}
In his famous 1929 paper, Hubble interpreted his velocity measurements as peculiar velocities rather than recession velocities. He used the Fizeau-Doppler formula to convert redshifts in velocities, which happens to coincide with the $z \rightarrow 0$ limit of the velocity-redshift law. For this reason, some authors prefer to refer to $cz = H_0\,r$ as the Hubble's law (rather than $v=H_0\,r$) in order to keep clear the distinction between the Doppler redshift and velocity redshift \cite{harrison:2000a}.

\subsection{The Hubble time}
\label{sec:hubble_time}
The \keyword{Hubble time} $t_H$ is defined as the inverse of the Hubble parameter. The current value of the Hubble time is easily obtained from the definition of $H_0$ in \eref{eq:hubble_units_of_measure}:
\begin{align*}
  t_{H_0} \;\equiv\; \frac{1}{H_0} \;=\; \unit[9.77 \, h^{-1}]{Gyr} \;.
\end{align*}
Given constant expansion, \ie $\, d^2a/dt^2 = 0\,$, the Hubble time is the time needed by the Universe to double in size. Equivalently, the solution to:
\begin{align}
  \label{eq:hubble_law_time_double}
  a(t_1) \;+\; \frac{da}{dt} \, \Delta t \;=\; a(t_2) \;,
\end{align}
for $ a(t_2) = 2 a(t_1) $ is $\,\Delta t = H^{-1}(t_1)\,$. If the expansion had been constant after the Big Bang, the Hubble time would be the age of the Universe; to see it, substitute $a(t_1)=0 $ and $a(t_2)=a$ in the above equation.

In a more realistic model where the expansion rate varies, the Hubble time does not correspond anymore to the age of the Universe. It rather sets the time-scale for the expansion of the Universe: in a time comparable to $ H^{-1} $ the expansion parameter increases noticeably. 
In the currently accepted accelerating \LCDM model, $\,t_{H_0}$ is still a good proxy for the current age of the Universe. Using Planck cosmological parameters \cite{planck-collaboration:2013a}, one finds $\,t_0 = \unit[13.817 \pm 0.048]{Gyr}\,$ against $\,t_{H_0} \simeq \unit[14.6]{Gyr}\,$.


\subsection{The Hubble radius}
\label{sec:hubble_radius}

The \keyword{Hubble radius} $L_H$ is defined as the physical distance travelled by light in a Hubble time. From \eref{eq:hubble_units_of_measure}, its current value is given by
\begin{align}
  L_{H_0} \;\equiv\; \frac{c}{H_0} \;=\; \unit[2998 \, h^{-1}]{Mpc} \;.
\end{align}
By virtue of the velocity-distance law ($v=Hr$), objects farther than a Hubble radius recede faster than light%
\footnote{Note that this behaviour does not invalidate special relativity since expansion is uniform everywhere in the Universe and therefore no exchange of information is possible as a result of the super-luminar velocity.}.
Therefore, given a constant expansion, an object located at the centre of a sphere whose radius is equal to the Hubble radius will never be able to interact with objects outside the sphere; a super-luminar motion is necessary for the contrary to be true. In these conditions, the Hubble radius is the maximum extension of the future light cone of any event in the Universe.

However, if the expansion of the Universe slows down, the Hubble sphere swells and an increasing number of regions in the Universe will eventually enter in causal contact. The time-scale needed for this to happen is the Hubble time. On the other hand, if the Universe experiences an accelerated expansion, any object located inside the Hubble sphere now will be out of it after a long enough time; as a result an increasing number of causally disconnected regions will be created. In an accelerating Universe light cannot keep up with the expansion.

Because of this causal interpretation, the Hubble radius is often referred to as \keyword{horizon}. Being defined as
\begin{align*}
  \dfrac{c}{H(t)} \;,
\end{align*}
the horizon is a physical distance, not a comoving one. Its comoving counterpart is obtained by dividing it by the expansion parameter:
\begin{align*}
  \dfrac{c}{a(t)\,H(t)} \;.
\end{align*}
The above quantity, called the \keyword{comoving horizon}, is not to be confused with the \emph{particle horizon}, which we define below and represents the maximum distance a particle could have travelled since the Big Bang until a certain time $ t $.


\subsubsection{Particle horizon and causality}
The distance travelled by a photon from the Big Bang up to a certain time $ t $ is known as the \keyword{particle horizon}. Its expression in comoving coordinates is obtained from \eref{eq:comoving_distance_ems_obs} by setting $t_\ems=0\,$ and $\,t_\obs=t\,$:
\begin{align*}
  \chi(t) \;\equiv\; \int_0^t \,c \, \frac{dt}{a(t)} \;.
\end{align*}
Since the speed of light is the limit velocity, the particle horizon represents the maximum comoving distance any particle could have travelled up to time $t$. Note that the particle horizon is proportional to the conformal time $\tau$ appearing in \eref{eq:conformal_metric_flrw}:
\begin{align}
  \chi(t) \;=\; c\;\tau(t) \;.
\end{align}
In the following we shall use the conformal time and the comoving particle horizon interchangeably.

At any moment $t$ in the evolution of the Universe, the particle horizon $ \chi(t) $ is the maximum extension of the past light cone for all events in the Universe. In particular, for an observer on Earth, the present-day particle horizon sets the size of the observable Universe. Its value depends on the cosmological model adopted; for a \LCDM model, it roughly amounts to $\chi(t_0)\simeq\unit[14,000]{Mpc}\,$. For the same model, $\,c\,t_0\simeq\unit[4,000]{Mpc}\,.$
There is a subtle difference between the particle horizon $\chi(t)$ and the Hubble horizon $c/(aH)$: the former is a measure of the past light cone of an event given the previous expansion history, while the latter sets the extent of its future light cone based on the instantaneous value of $H$. 

\section{The background evolution}
\label{sec:background_evolution}

In order to derive the time evolution of the scale parameter $a(t)$ we need to relate the metric with the energy content of the Universe. This is achieved via the \keyword{Einstein equation}:
\begin{align}
\label{eq:einstein_equations}
  R_{\mu\nu} \;-\; \frac{1}{2} \, g_{\mu\nu} \,R
  \;=\; 8\,\pi\,G \; T_{\mu\nu} \;,
\end{align}
where we have set $c=1$ and
\begin{itemize}
  \item $ R_{\mu\nu}$ is the Ricci tensor, defined as the self-contraction of the Riemann tensor. It can be expressed in terms of the \keyword{Christoffel symbols} or \keyword{affine connection},
  \begin{align}
  \label{eq:christoffel_symbols}
    \christoffel{\mu}{\alpha}{\beta} \;=\;
    \frac{g^{\mu\nu}}{2} \;
    \left[ \;
    \pfrac{g_{\alpha\nu}}{x^\beta} \;+\;
    \pfrac{g_{\beta\nu}}{x^\alpha} \;-\;
    \pfrac{g_{\alpha\beta}}{x^\nu} \;
    \right]
  \end{align}
  as
  \begin{align}
  \label{eq:ricci_tensor}
    R_{\mu\nu} \;=\;
    \pfrac { \christoffel{\alpha}{\mu}{\nu} } { x^\alpha } \;-\;
    \pfrac { \christoffel{\alpha}{\mu}{\alpha} } {x^\nu} \;+\;
    \christoffel{\alpha}{\beta}{\alpha} \, \christoffel{\beta}{\mu}{\nu} \;-\;
    \christoffel{\alpha}{\beta}{\nu} \, \christoffel{\beta}{\mu}{\alpha} \;.
  \end{align}
  \item $ R \;=\; g_{\mu\nu} \, R^{\mu\nu} $ is the Ricci scalar.
  \item $ T_{\mu\nu} $ is the total energy-momentum tensor, source of the gravitational field.
  \item $ G \,$ is Newton's gravitational constant.
\end{itemize}
Inserting the metric for an \FLRW Universe in comoving coordinates (\eref{eq:flrw_metric}), we find that for an isotropic Universe the only non-zero components of the connection, Ricci tensor and Ricci scalar are, respectively,
\begin{align}
  & \christoffel{0}{i}{j} \;=\; \delta_{ij}\,a'\,a
  \qquad\text{and}\qquad
  \christoffel{i}{0}{j} \;=\; \christoffel{i}{j}{0} \;=\; \delta_{ij}\,\dfrac{a'}{a}\;, \msk
  \label{eq:background_ricci_tensor}
  & R_{00} \;=\; -3\;\dfrac{a''}{a}
  \qquad\text{and}\qquad
  R_{ij} \;=\; \delta_{ij} \left(\, 2\,a'^{\,2} \;+\; a\,a'' \,\right)\;, \msk
  \label{eq:background_ricci_scalar}
  & R \;=\; 6\;\left[\; 
  \dfrac{a''}{a} \;+\; \left(\, \dfrac{a'}{a} \,\right)^2 \,
  \,\right] \; ,
\end{align}
where the primes denote differentiation with respect to cosmic time, $a'=da/dt\,$.
The left hand side of the Einstein equation is called the \keyword{Einstein tensor} $\,G_{\mu\nu}\,$ and can be determined using the above relations:
\begin{align}
  \label{eq:background_einstein_tensor}
  & G_{00} \;=\; 3\,\left(\,\frac{a'}{a}\,\right)^2 \;, &&
    G_{ij} \;=\; -\delta_{ij}\;\left(\,a'^{\,2}\,+\,2\,a\,a''\,\right) &&
    G_{i0}\,=\,G_{0i}\,=\,0 \;.
\end{align}

The total energy-momentum tensor is given by the sum of the energy-momentum tensors of the species in the Universe, that is,
\begin{align}
  T_{\mu\nu} \;=\; \sum_a \; T_{a,\mu\nu} \;,
\end{align}
where $\,a=\gamma,b,\nu,c,\Lambda\,$ for photons, baryons, neutrinos, cold dark matter and dark energy, respectively.
The fact that the spatial Einstein tensor is diagonal is a direct consequence of the isotropy of the \FLRW metric. The energy-momentum is forced to be diagonal too, meaning that the cosmological fluids cannot have peculiar velocities or anisotropic stresses. 
Therefore, in the simple \FLRW model a fluid is characterised only by its energy density $\,\rho(t)\,$ and its pressure $\,P(t)\,$.

We shall assume that the fluids that compose the Universe are \emph{barotropic}\index{barotropic fluids}, that is, their pressure is given as an explicit function of their energy density. The relation between $P$ and $\rho$ is called the \keyword{equation of state} of the fluid; we parametrise it via the barotropic parameter $w$ as
\begin{align}
  P \;=\; w(\rho)\;\rho\,.
\end{align}
The energy-momentum tensor of the fluid `$a$' is thus expressed as
\begin{align}
  \label{eq:background_energy_momentum_tensor}
  & T_{a,00} \;=\; \rho_{a} \;, &&
    T_{a,ij} \;=\; \delta_{ij}\;w_a(\rho)\;\rho_{a} \;.
\end{align}
As we shall soon see, knowing the equation of state $\,w(\rho)\,$ of the various species is needed to derive the expansion history of the Universe.
Relativistic species (R), such as the photons, the neutrinos and the massive species while still relativistic, have a constant equation of state: $\,w_\sub{R} = \frac{1}{3}\,$\;.
Non-relativistic species (M), such as the baryons and cold dark matter after decoupling, instead, have no pressure: $\,w_\sub{M} = 0\,$.
Note that, already in a simple mixture of matter and radiation, $w$ ceases to be constant.
In this work we treat dark energy as a cosmological constant, which is equivalent to a negative pressure fluid with constant equation of state: $\,w_\sub{$\Lambda$}=-1\,$.


\subsection{Friedmann equation}
\label{sec:friedmann_equation}

The time-time component of the Einstein equations is called the \keyword{Friedmann equation},
\begin{align}
  \label{eq:friedmann_00}
  H^2 \;=\; \frac{8\,\pi\,G}{3}\,\rho \;-\; \frac{k}{a^2}\;,
\end{align}
where $ H = a'/a $ is the Hubble parameter and $\rho=\sum\rho_a$ is the total energy density of the Universe. We have included the curvature contribution, $k$, to highlight the fact that in a flat universe ($k=0$) the total density always equals the \keyword{critical density} $ \rhoc $, defined as
\begin{align*}
  \rhoc \;\equiv\; \frac{3\,H^2}{8\,\pi\,G} \;.
\end{align*}
The critical density depends on time; its present-day value can be easily computed in terms of the Hubble constant:
\begin{align}
  \label{eq:critical_density_value}
  \rhoc(t_0) \;&=\; \unit[1.878 \; h^2 \; \times \, 10^{-26}\;]{\frac{kg}{m^3}} &&\\[0.14cm]
  &=\;\unit[2.775\;h^{-1} \; \times \; 10^{11}\;]{\frac{M_{\odot}}{\left( h^{-1}\,\text{Mpc} \right)^3}} &&\\[0.14cm]
  &=\;\unit[10.54\;h^2]{\frac{GeV}{m^3}} & \text{( assuming $c=1$ )} & \;.
\end{align}
This is an astonishingly small number: with a density of $\,\unit[1.27]{kg/m^3}\,$, air is around $10^{26}$ times denser than the critical density. However, since $10^{11}$--$10^{12}$ solar masses is close to the mass of a typical galaxy and $\unit[1]{Mpc}$ is the order of magnitude of the typical galaxy separation, the Universe cannot be too distant from the critical density.

The density of the species normalised to the critical density of the Universe is called the \keyword{density parameter}:
\begin{align}
  \Omega_{a}(t) \;\equiv\; \frac{\rho_a(t)}{\rhoc(t)} \;.
\end{align}
Using the information on the equations of state of the various species (\sref{sec:background_continuity_equation}), the Friedmann equation can be recast in terms of the present-day value of the density parameters, $\,\Omega_{a0}\equiv\Omega_{a}(t_0)\,$, as
\begin{align}
  \label{eq:tt_friedmann_alternate_version}
   H^2 \;=\; H_0^2\;\left[\;\frac{\Omega_{\sub{M}0}}{a^3} \;+\; \frac{\Omega_{\sub{R}0}}{a^4} \;
   +\; \frac{\Omega_{k0}}{a^2} \;+\; \Omega_{\Lambda0} \;\right] \;,
\end{align}
where $H_0\equiv H(t_0)$ and
\begin{align}
  &\Omega_{\sub{M}0} \;=\; \frac{\rho_\sub{M}(t_0)}{\rhoc(t_0)} \;,
  &&\Omega_{\sub{R}0} \;=\; \frac{\rho_\sub{R}(t_0)}{\rhoc(t_0)} \;,
  &&\Omega_{k0} \;=\; -\frac{k}{a_0^2\,H_0^2} \;,
  &&\Omega_{\Lambda0} \;=\; \frac{\Lambda}{3\,H_0^2} \;.
\end{align}
(In this thesis, cosmological quantities indexed by a `0' are evaluated today, $X_0\equiv X(t_0)\,$.)

\subsection{Acceleration equation}

In an \FLRW Universe, the spatial components of the Einstein equation reduce to a single expression, the \keyword{acceleration equation}:
\begin{align}
  \label{eq:friedmann_jj}
  \frac{a''}{a} \;=\; -\frac{4\,\pi\,G}{3}\;\left(\;\rho\;+\;3\,P\;\right)\;,
\end{align}
where $\,P=\sum P_a\,$ is the combined pressure of all the species. The acceleration equation holds also in a curved Universe, where $k\neq0$.

The pressure and the density appear in the acceleration equation on equal grounds: they both contribute to increasing the gravitational attraction and thus decelerate the cosmic expansion.
This might seem counter intuitive, as we are used to thinking of pressure as something that powers expansive processes such as explosions.
This is indeed true if a force is supplied by means of a gradient in the pressure field; however, in a homogeneous Universe, $P$ is the same everywhere and no pressure forces are possible.

\subsection{Continuity equation}
\label{sec:background_continuity_equation}

The evolution of the matter species is determined by the conservation of the energy and momentum,
\begin{align}
{T^\mu}_{\nu;\mu} \;=\; 
  \partial_\mu {T^\mu}_\nu \;+\;
  \christoffel{\mu}{\alpha}{\mu} {T^\alpha}_\nu \;-\;
  \christoffel{\alpha}{\nu}{\mu} {T^\mu}_\alpha \;=\; 0 \;.
\end{align}
Due to isotropy, the only meaningful equation is $\nu=0$, the \keyword{continuity equation}:
\begin{align}
  \rho' \;+\; 3\;H\;\left(\,\rho \;+\; P \,\right) \;=\; 0 \;, 
\end{align}
which, in terms of the barotropic parameter, reads
\begin{align}
\label{eq:continuity_equation_flrw}
  \rho' \;+\; 3\;H\;\rho\;\left(\,w\,+\,1\,\right) \;=\; 0 \;.
\end{align}

The continuity equation applies separately to each species as, for the epochs of interest, their particle number is conserved and their energy exchange is negligible.
Then, for a fluid `$a$' with a constant equation of state, $\,P=w\rho\,$, the continuity equation can be solved to yield
\begin{align}
  \label{eq:background_evolution_of_rho_density_1}
  \rho_a \;\propto\; a^{-\,3\,(1\,+\,w_a)} \;.
\end{align}
For radiation ($w=1/3$), cold matter ($w=0$) and the cosmological constant ($w=-1$), the density is thus given by
\begin{align}
  \label{eq:background_evolution_of_rho_density_2}
  &\rho_\sub{R} \;\propto\; a^{-4} \;,
  &&\rho_\sub{M} \;\propto\; a^{-3} \;,
  &&\rho_\Lambda \;=\; \text{constant} \;.
\end{align}
In the more general case of a time-dependent equation of state, $w=w(a)\,$, one has to solve the following integral:
\begin{align}
  \rho \;\propto\; \exp \left(\,-3\,\int_0^a\,\frac{\dd \tilde{a}}{\tilde{a}}\;
  \left[\,1 \,+\, w(\tilde{a})\,\right]\,\right) \;.
\end{align}

\subsection{Expansion history}
\label{sec:expansion_history}

The expansion history of a universe filled by a single species with constant equation of state can be inferred analytically. This is achieved by inserting the general equation of state (\eref{eq:background_evolution_of_rho_density_1}) into the Friedmann equation (\eref{eq:friedmann_00}) and solving for $a(t)\,$. If the curvature $k$ is neglected, we have that \cite{durrer:2008a}
\begin{flalign}
  \label{eq:evolution_of_scale_factor_expansion_history}
  &a \,\propto\, t^{2/(3\,(1+w))} \,\propto\, \tau^{2/(1+3w)} \;,
  && H \,\propto\, t^{-1} \propto\, a^{-3(1+w)/2} \;,
  &&w \,=\, \text{constant} \neq -1 \;,&\\[0.14cm]
  &a \,\propto\, t^{2/3} \,\propto\, \tau^{2} \;,
  && H \,\propto\, t^{-1} \propto\, a^{-3/2} \;,
  &&w \,=\, 0      \quad\text{(cold matter)}\;,\allowbreak&\notag\\[0.14cm]
  &a \,\propto\, t^{1/2} \,\propto\, \tau \;,
  && H \,\propto\, t^{-1} \propto\, a^{-2} \;,
  &&w \,=\, 1/3    \quad\text{(radiation)}\;,&\notag\\[0.14cm]
  &a \,\propto\, e^{H\,t} \,\propto\, 1/|\tau| \;,
  && H \,=\, \text{constant} \;,
  &&w \,=\, -1     \quad\text{(cosmol. constant)}\,.&
  \notag
\end{flalign}
Recall that $t$ is the cosmic time and $\tau$ is the conformal time, $d\tau=dt/a\,$.

In the general case of a mixture of fluids, one has to rely on the full Friedmann equation (\eref{eq:tt_friedmann_alternate_version}):
\begin{align}
  \label{eq:friedmann_equation_scale_factor}
  \frac{1}{a}\frac{da}{dt} \;=\; H_0\;
  \sqrt{\frac{\Omega_{\sub{M}0}}{a^3}\;+\; \frac{\Omega_{\sub{R}0}}{a^4} \;
   +\; \frac{\Omega_{k0}}{a^2} \;+\; \Omega_{\Lambda0}} \;,
\end{align}
which yields a time integral that is easily solved for $a(t)$ once the cosmological parameters are specified.
These have been measured to high accuracy. For the Hubble constant, $\,H_0=100\,h\,\text{km/s/Mpc}\,$, and the density parameter of matter, $\,\Omega_\sub{M}=\Omega_b+\Omega_c\,$, we adopt the best fit values obtained by the Planck experiment \cite{planck-collaboration:2013a},
\begin{align}
  &h = 0.6780 \pm 0.0077 \;,
  &&\Omega_{b0}\,h^2 \;=\; 0.02214 \,\pm\, 0.00024  \;,
  &&\Omega_{c0}\,h^2 \;=\; 0.1187 \,\pm\, 0.0017  \;,
\end{align}
at \onesigma.
The density parameter of the photon fluid is determined by the value of the CMB temperature \cite{fixsen:1996a},
\begin{align}
  \label{eq:cmb_temperature}
  T_0 \;=\; \unit[2.725 \,\pm\, 0.001]{K} \qquad\text{at \twosigma}\;,
\end{align}
which, for a blackbody spectrum, yields
\begin{align}
  \Omega_{\gamma0}\,h^2 \;=\; \sci{2.49}{-5}
  &&\text{and}&&
  \Omega_{\nu0}\,h^2 \;=\; \sci{1.69}{-5} \;,
\end{align}
where we have used the fact that the massless neutrino density is roughly equal to $0.68\,\Omega_\gamma$ because they are fermions rather than bosons and are at a lower temperature.
Finally, we assume a flat Universe ($\Omega_k=0$) so that the density of dark energy can be determined as
\begin{align}
  \Omega_{\Lambda0} \;=\; 1\,-\,\Omega_{\sub{R}0}\,-\,\Omega_{\sub{M}0} \;=\; 0.694 \;.
\end{align}

\begin{figure}[t]
	\centering
		\includegraphics[width=0.75\linewidth]{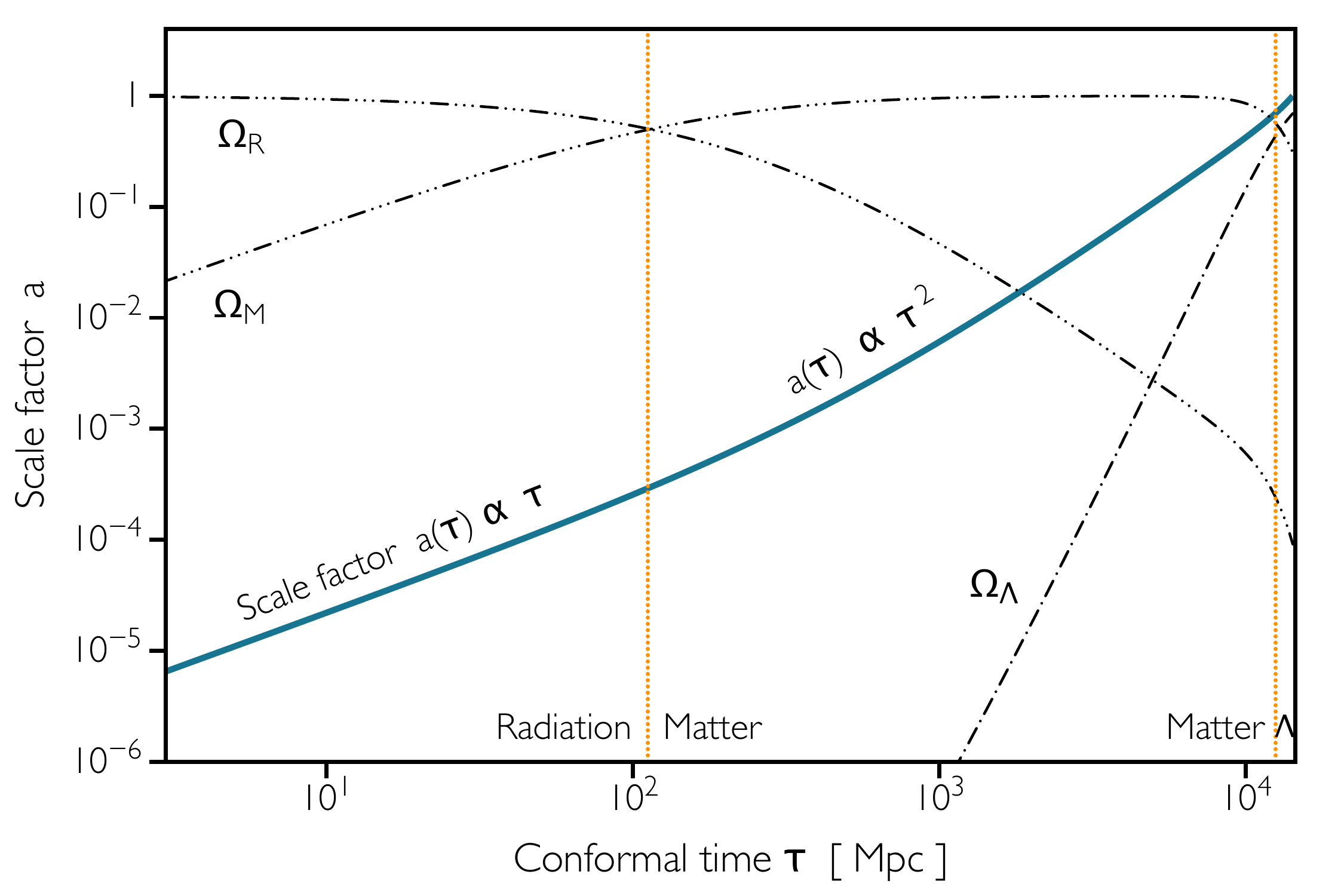}
	\caption[Cosmic history of the Universe]{Cosmic history of the Universe. The blue curve is the scale factor as a function of conformal time, obtained by solving the Friedmann equation in \eref{eq:friedmann_equation_scale_factor}. Today corresponds to $a=1$ and $\tau=\unit[14200]{Mpc}\,$. The three black dot-dashed curves are the density parameters of radiation ($\Omega_\sub{R}$), cold matter ($\Omega_\sub{M}$) and dark energy considered as a cosmological constant fluid ($\Omega_\Lambda$). The intersections between the three $\Omega$'s naturally split the cosmic history in three epochs: the radiation domination era ($a\propto\tau$), the matter domination era ($a\propto\tau^2$) and the dark energy domination era ($a\propto1/\tau$).}
	\label{fig:cosmic_history}
\end{figure}

In \fref{fig:cosmic_history} we show the evolution of the scale factor obtained for the above parameters.
Depending on the species that is the most abundant, we identify three epochs in the cosmic history: the \keyword{radiation dominated era} ($a\propto\tau$), the \keyword{matter domination era} ($a\propto\tau^2$) and the \keyword{dark-energy dominated era} ($a\propto1/\tau$).
The transitions between the three eras take place at
\begin{align}
  &a_\text{eq} \;=\; \frac{\Omega_{\sub{R}0}}{\Omega_{\sub{M}0}} \;=\; \sci{2.96}{-4}
  &&\text{and}&&
  a_\Lambda \;=\; \frac{\Omega_{\sub{M}0}}{\Omega_{\Lambda0}} \;=\; 0.44 \;,
\end{align}
which correspond, respectively, to $\,z_\text{eq}=3380\,$ and $z_\Lambda=1.26\,$.

\runinhead{The Big Bang}
If we inspect the acceleration equation \eref{eq:friedmann_jj},
\begin{align}
  \frac{a''}{a} \;=\; -\frac{4\,\pi\,G}{3}\;\rho\;\left(\;3\,w \;+\; 1\;\right)\;,
\end{align}
we see that in the early Universe when radiation dominates ($w=1/3>0$), the second derivative of $a(t)$ is negative; that is, $a(t)$ is a concave curve. Thus, we expect the scale factor of the Universe to cross the $a=0$ line in a finite amount of time; the moment when this happens is called the \keyword{Big Bang}\footnote{The name was invented during a radio interview by Fred Hoyle, the main supporter of a steady state Universe, as a mockery of the idea of an expanding Universe. Refer to the following URL for the transcript: \url{http://www.joh.cam.ac.uk/library/special_collections/hoyle/exhibition/radio/}.}. 
The Big Bang represents a singularity in the coordinates (the spatial metric vanishes for $a=0$), in the Ricci scalar (\eref{eq:background_ricci_scalar}) and in the density ($\rho_\sub{R}\propto a^{-4}$).

\section{The Cosmic Microwave Background}
\label{sec:background_cmb}

Soon after the Big Bang, the particle density is so high that the species interact at a rate much higher than the expansion rate, with all kinds of particle-antiparticle pairs being created and annihilated. As a result of these continuous collisions, particles of different species are in \keyword{thermal equilibrium}, \ie they can be considered to be part of a single \keyword{cosmic plasma} with a common temperature and average kinetic energy.

Photons in thermal equilibrium obey a \emph{blackbody spectrum}, which is characterised by a simple relation between the energy density $\rho_\gamma$ and the ambient temperature $T$,
\begin{align}
  \label{eq:stefen-boltzmann_law}
  \rho_\gamma \;=\; \alpha\;T^4 \;,
\end{align}
where the proportionality constant is the Stefen-Boltzmann constant times $4/c$, that is, $\,\alpha=\pi^2 k_\text{B}^4/(\hbar^3 c^3)\,$.
Since the energy density of radiation scales with $a^{-4}$, it follows that the temperature of the cosmic plasma scales as $a^{-1}$:
\begin{align}
  \label{eq:photon_temperature_blackbody}
  T \;=\; \frac{\unit[2.725]{K}}{a} \;=\; (z+1)\;\unit[\sci{2.35}{-4}]{eV}  \;,
\end{align}
where we have used the current CMB temperature as normalisation and, in the second equality, we have assumed units where the Boltzmann constant $k_\text{B}=\unit[11605^{-1}\,]{eV/K}$ is equal to one.
To give an idea of the scales involved, we can use the fact that $\,a\propto t^{1/2}\,$ in the radiation dominated era to write
\begin{align}
  T \;\simeq\; \unit[\sci{1.5}{10}]{K}\;\,\sqrt{\frac{\unit[1]{s}}{t}}
  \;\simeq\; \unit[1.3]{MeV}\;\,\sqrt{\frac{\unit[1]{s}}{t}} \;,
\end{align}
Thus, one second after the Big Bang, the average photon has an energy of $\sim\unit[1]{MeV}$ while, after $50,000$ years, its energy has dropped to $\unit[1]{eV}$.

\begin{figure}[t]
 \centering
 \includegraphics[height=3in]{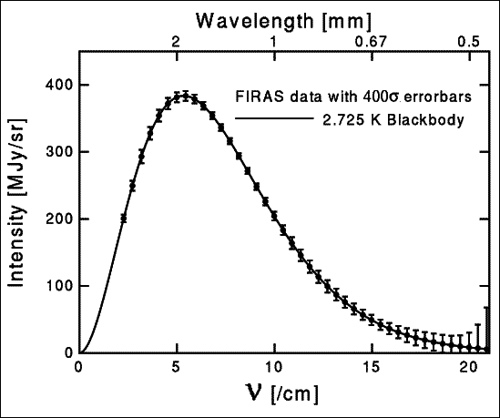}
 \caption[CMB blackbody spectrum (1)]{
   The cosmic microwave background spectrum as measured by FIRAS. The error bars have been multiplied by 400 to make them visible; the line represents the best-fit blackbody spectrum at $T = \unit[2.725]{K}\,$. Source: data from FIRAS \cite{fixsen:1996a}, image courtesy of Edward L. Wright from the website \url{http://www.astro.ucla.edu/~wright/cosmo_01.htm}.
   }
 \label{fig:cmb_spectrum_firas}
\end{figure}

\begin{figure}[t]
 \centering
 \includegraphics[height=3in]{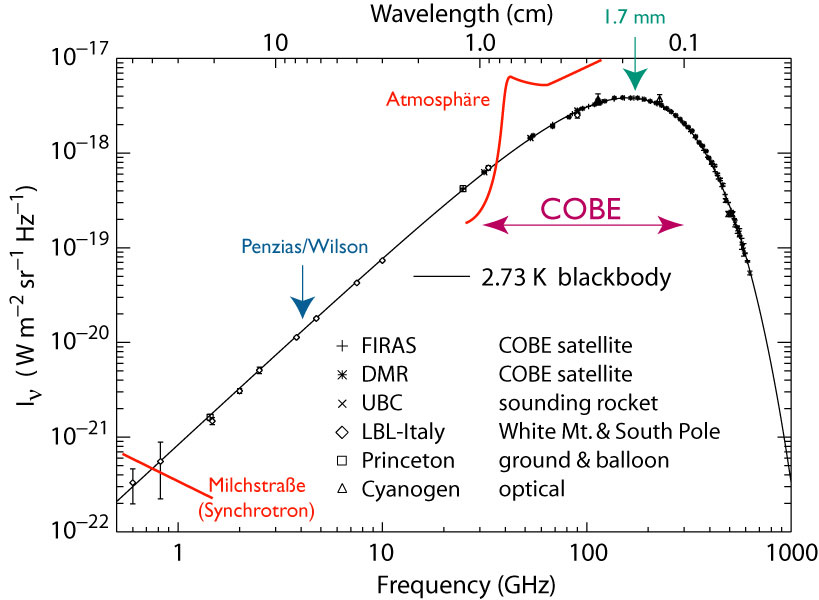}
 \caption[CMB blackbody spectrum (2)]{
 The CMB blackbody spectrum as confirmed by measurements over a broad range of wavelengths. Credit: Fig. 19.1 of Ref.~\cite{caso:1998a}, reproduced with permission of Springer Publishing (\url{http://pdg.lbl.gov/1998/contents_large_sports.html}); coloured additions courtesy of Karl-Heinz Kampert (\url{http://astro.uni-wuppertal.de/~kampert/Cosmology-WS0607.html}).
 }
 \label{fig:cmb_spectrum_many_experiments}
\end{figure}

In an expanding Universe, however, thermal equilibrium does not last forever.
The particles of a given species interact with a rate proportional to their number density, which decays as $a^{-3}$.
The expansion rate $H$, on the other hand, never decays faster than $a^{-3/2}$ (\eref{eq:evolution_of_scale_factor_expansion_history}), meaning that, eventually, it will exceed the interaction rate.
As a result, the thermal equilibrium cannot be maintained anymore and the particle species is said to have \emph{decoupled} from the cosmic plasma.
As we shall see in the next sections, the photons decouple at a redshift of $z\simeq1100$, soon after matter-radiation equality. Then, why do we speak of ``temperature of the photons'', if they are not in thermal equilibrium? 
The answer is simple: the cosmic expansion preserves the blackbody spectrum of the photon fluid even when it is out of thermal equilibrium. Due to its $E/T$ dependence, the distribution function is frozen as it redshifts into a similar distribution with a lower temperature proportional to $1/a$ (we will come back to this point in \sref{sec:photon_distribution_function}).
Thus, after decoupling, the photon fluid possesses an effective temperature rather than a thermodynamical one.

The presence of this blackbody, isotropic background radiation of cosmic origin is a definite prediction of the Big Bang model.
The first measurement that was directly linked \cite{dicke:1965a} to the cosmic background radiation was made serendipitously in 1963 by Penzias and Wilson \cite{penzias:1965a}, who measured an isotropic excess temperature of around $\unit[3.5]{K}\,$.
Since then, many experiments were performed to measure the present-day CMB spectrum over different wavelengths. The most accurate measurement of the CMB spectrum was made by the FIRAS experiment, launched in 1989 on board of the NASA Cosmic Background Explorer (COBE). The spectrum measured by FIRAS \cite{mather:1994a, fixsen:1996a} is blackbody to high accuracy and is shown in \fref{fig:cmb_spectrum_firas}.
The blackbody form of the CMB spectrum has been confirmed by several other experiments for wavelengths outside the millimetre range, as shown in \fref{fig:cmb_spectrum_many_experiments}.
The measured CMB temperature, $\,T_0=\unit[2.725\pm0.001]{K}$ \cite{fixsen:1996a}, implies that the average CMB photon has the following properties:
\begin{align}
 &\text{frequency} \sim \unit[160]{GHz} \;,
 &&\text{wavelength} \sim \unit[2]{mm} \;,
 &&\text{energy} \sim \unit[0.7]{meV} \;.
\end{align}

\subsection{Compton scattering}
\label{sec:background_compton_scattering}

After the temperature of the cosmic plasma has dropped below the electron mass, $\,T\ll\unit[511]{keV}\,$, the only process that maintains the photons in thermal equilibrium are the rapid collisions with the free electrons.
In general, the scattering of a photon by a free charged particle is called \keyword{Compton scattering}. It is an inelastic process, as an incident photon deflected by an angle $\theta$ experiences a wavelength shift $\,\Delta\lambda\equiv\lambda'-\lambda\,$ of
\begin{align}
  \Delta\lambda \;=\; \lambda_c \; \left(\,1\,-\,\cos\theta\,\right) \;,
\end{align}
where $\,\lambda_c\equiv h/(mc)\,$ is the Compton wavelength of the target particle, which is assumed to be at rest. In terms of the photon's energy ($E_\gamma=hc/\lambda$), the formula translates to
\begin{align}
  \frac{\Delta E_\gamma}{E'_\gamma} \;=\; (\cos\theta-1)\;\frac{E_\gamma}{m\,c^2} \;,
\end{align}
which means that the fractional change in the photon's energy is negligible as long as its energy is much smaller than the target's mass. The condition definitely applies to our context, where we consider temperatures of the order of the eV and the target particles are electrons with $m_e c^2=\unit[511]{keV}$.\footnote{Note that, in the context of the cosmological perturbations, even this tiny energy transfer has to be considered, as we shall see in \sref{sec:collision_energy_transfer}.} In this limit, the process is elastic and is called \keyword{Thomson Scattering}.

The total cross-section for the Thomson scattering is given by \cite{dodelson:2003b}
\begin{align}
  \label{eq:thomson_total_cross_section}
  \sigma_T \;&=\; \frac{8 \pi}{3} \;\alpha^2\,\lambda_c^2
  \;=\; \frac{8 \pi}{3}\;\left(\,\frac{\alpha \hbar}{m c}\,\right)^2  &&\\[0.14cm]
  \;&=\; \unit[\sci{6.652}{-29}]{m^{2}} &&\\[0.14cm]
  \;&=\; \unit[\sci{4.328}{-17}]{eV^{-2}} & \text{( assuming $h=c=1$ )} & \;,
\end{align}
where $ \alpha \simeq 1/137 $ is the fine structure constant and in the last equalities we have used the electron mass $m_e c^2=\unit[511]{keV}\,$.
It is important to note that the cross section is inversely proportional to the squared mass of the target particle. Therefore, provided that protons and electrons have the same number density, photon-electron collisions ($m_e c^2=\unit[511]{keV}$) are several million times more likely that photon-proton collisions ($m_p c^2=\unit[938]{GeV}$). For this reason, we shall ignore the latter and focus on the former. 


\subsubsection{Interaction rate and optical depth}
\label{sec:optical_depth}

Here we introduce the interaction rate $\dot\kappa$ and the optical depth $\kappa$ that will be useful in the following chapters to derive and numerically solve the Boltzmann equation.

The cross-section $\sigma$ associated with a scattering process is defined so that
\begin{align}
  \label{eq:cross_section_definition}
  dN \;=\; n\,\sigma\,dx
\end{align}
is the average number of scatterings the incident particle undergoes when covering a distance of $dx$ in a material with a density $n$ of scattering targets. Since $dN/dx$ is the average number of scatterings per unit of length, its inverse is the \keyword{mean free path}:
\begin{align}
  \label{eq:mean_free_path_definition}
  \lambda \;=\; \frac{1}{n \, \sigma} \;,
\end{align}
\ie the average distance a particle covers between two consecutive scatterings. If the velocity $dx/dt$ of the incident particle is known, then it is straightforward to obtain the \keyword{interaction rate} $\,dN/dt\,$, that is the average number of scatterings per unit of time. For a photon,
\begin{align}
  \label{eq:interaction_rate_photon}
  \frac{dN}{dt} \;=\; n \, \sigma \, c \;.
\end{align}
The inverse of the interaction rate is the average time elapsed between two consecutive scatterings; we shall call this quantity \keyword{mean free time}. For a photon it is given by:
\begin{align}
  \label{eq:mean_free_time_photon}
  t_\gamma \;=\; \frac{1}{n \, \sigma \, c} \;.
\end{align}

In the context of the cosmic microwave background, the \keyword{optical depth} or \keyword{optical depth}, $\,\kappa\,$, is the average number of Thomson scatterings a photon undergoes from the time $t$ up to now,
\begin{align}
  \label{eq:optical_depth_definition}
  \kappa(t) \;=\; \int_t^{t_0}\,dt'\;n_e\,\sigma_T\,c  \;.
\end{align}
The optical depth is a monotonically decreasing function of time; its time derivative is just the interaction rate with a negative sign
\begin{align}
  \frac{d \kappa}{dt} \;=\;  - n_e\,\sigma_T\,c \;.
\end{align}
In terms of conformal time, $\,d\tau=dt/a\,$, the interaction rate reads
\begin{align}
  \label{eq:kappa_dot}
  \dot\kappa \;=\; \frac{d\kappa}{d\tau} \;=\; -a\,n_e\,\sigma_T\,c \;.
\end{align}

\subsection{Recombination and decoupling}
\label{sec:recombination_and_decoupling}

The frequent Thomson scatterings between the photons and the electrons before recombination keep the two fluids in thermal equilibrium. Together with the protons, which are tightly coupled with the electrons via Coulomb scattering, the three species form a unique fluid with a common temperature.

The photons are maintained in thermal equilibrium as long as their interaction rate with the electrons, $\,n_e\,\sigma_T\,c\,$, exceeds the cosmic expansion rate, $H\,$.
If we assume that the electrons remain free throughout cosmic evolution, such decoupling happens only at a redshift of $z\sim40\,$ \cite{dodelson:2003b}.
The electrons, however, do not stay free as it is energetically favourable for them to combine with the free protons to form hydrogen atoms via the reaction
\begin{align}
  &e^- \,+\,p \;\quad\longrightarrow\;\quad \text{H} \,+\, \gamma\,(\unit[13.6]{eV}) \;.
\end{align}
In the early Universe, the energy and the density of photons are so high that the hydrogen atoms thus formed are rapidly disrupted via the inverse reaction;
thus, most of the electrons are free and the abundance of neutral hydrogen is very low.
As the Universe expands and cools, however, more and more atoms are able to form and endure in a process that is called \keyword{recombination}.

During recombination, the number density of free electrons quickly drops and so does the rate of photon scatterings, $\,|d\kappa/dt|=n_e\sigma_T c\,$.
When the interaction rate is surpassed by the expansion rate, the photon fluid goes out of equilibrium and decouples from the electron fluid.
As a result, the photons can stream freely in a now transparent Universe.
This process is called \keyword{decoupling}.
As we shall see below, decoupling happens during recombination.

Recombination is a complicated process that involves non-equilibrium physics and is usually treated using the Boltzmann formalism.
In principle, to obtain the ionisation history of the Universe requires solving a system with $300+$ differential equations, one per energy level of the hydrogen atom \cite{seager:1999a}.
In practice, however, one can model the hydrogen atom as having effectively three energy levels: ground state, first excited state and continuum \cite{peebles:1968a} (see also \sref{sec:perturbed_recombination}).
Numerical codes such as \emph{RECFAST} \cite{seager:1999a} start from this 3-level approximation to compute the ionisation history of the Universe in less than a second with sub-percent accuracy over a wide range of redshifts.
The code \emph{HyRec} \cite{ali-haimoud:2011a} implements an even more accurate numerical treatment of recombination where four energy levels are considered that is mathematically equivalent to the multi-level approach \cite{ali-haimoud:2010a}.

However, it is still possible to make general statements about recombination and decoupling without resorting to a numerical computation, and we shall do so in the following two subsections.
One of the major simplifications that we shall adopt is to assume that all the protons are in hydrogen nuclei, thus ignoring the $\sim25\%$ contribution in mass that is expected from the helium nuclei. Since about $1$ proton out of every $8$ is in a Helium nucleus, this results in an error of roughly $10\%\,$.


\subsubsection{Recombination}
\label{sec:recombination}

The quantity of interest is the \keyword{free electron fraction} or \keyword{ionisation fraction},
\begin{align}
  \label{eq:free_electron_fraction_definition}
  x_e \;\equiv\; \frac{n_e}{n_e + n_H} \;,
\end{align}
where $ n_e $ and $ n_H $ are respectively the number densities of free electrons and neutral hydrogen atoms; note that, since the Universe is globally neutral, $n_e=n_p$, the number density of free protons. If we neglect the small number of electrons and protons in Helium nuclei, the denominator is equal to the number density of baryons: $n_e + n_H\simeq n_b\,$.



Before recombination begins, the reaction $\inlinereaction{e}{p}[\longleftrightarrow]{\text{H}}{\gamma}$ is in equilibrium and we use the \keyword{Saha ionisation equation} \cite{dodelson:2003b, durrer:2008a} to describe it:
\begin{align}
  \frac{x_e^2}{1-x_e} \;=\; \frac{1}{n_e+n_H} \;
  \left(\,\frac{m_e\,T}{2\,\pi}\,\right)^{3/2} \; e^{-\epsilon/T} \;.
\end{align}
If we approximate $n_e + n_H\simeq n_b$ and multiply and divide the right hand side by the blackbody density of the photons, $\,n_\gamma=2/\pi^2\,T^3\,\zeta(3)\,$, where $\zeta(3)\simeq1.2021\,$, we obtain
\begin{align}
  \label{eq:saha_equation_recombination}
  \frac{x_e^2}{1-x_e} \;\simeq\; 0.265\;\,\frac{n_\gamma}{n_b}\;
  \left(\,\frac{m_e}{T}\,\right)^{3/2} \; e^{-\epsilon/T} \;.
\end{align}
The $\,n_\gamma/n_b\,$ factor is the photon to baryon ratio, which is constrained by observations \cite{hinshaw:2012a} to be equal to $\sim\sci{1.64}{9}\,$, while $\epsilon=\unit[13.6]{eV}$ is the hydrogen ionisation energy.

\begin{figure}[t]
  \centering
    \includegraphics[width=0.7\linewidth]{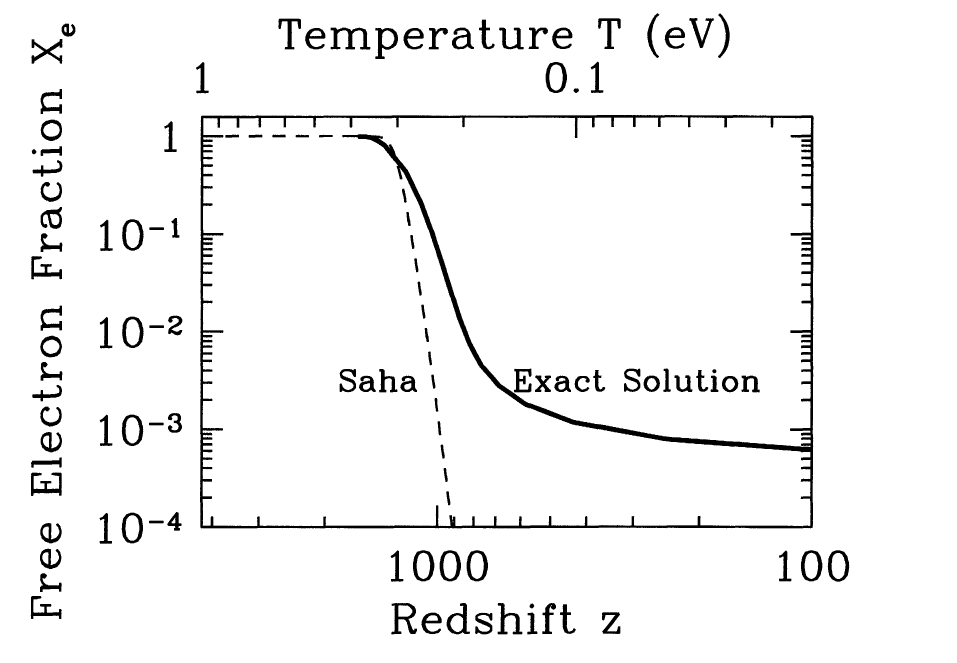}
  \caption[Recombination history]{
  Ionisation history of recombination. The free electron fraction is plotted against redshift and temperature. Recombination starts when $x_e$ begins to drop and is a quick process. The Saha approximation (\eref{eq:saha_equation_recombination}) correctly describes the beginning of recombination, but fails when the average energy of the photons becomes too small to maintain the $\,e+p\leftrightarrow H+\gamma\,$ reaction in equilibrium. Note that the exact solution does not drop to zero but, due to the reaction ``freezing'' when $\sigma_T\,x_e\,n_b\,c\ll H\,$, it asymptotes to $x_e\simeq10^{-3}\,$. Source: \citet[Page~72]{dodelson:2003b}, reproduced with permission from Elsevier Books.
  }
  \label{fig:recombination_evolution}
\end{figure}

The function $\,x_e(z)\,$ from the Saha equation is shown in \fref{fig:recombination_evolution}. Due to the presence of the exponential term, we see that recombination is a sudden process. If we conventionally set the recombination temperature $T_\text{rec}$ as the temperature when $x_e(T_\text{rec})=0.5\,$, the Saha equation yields
\begin{align}
  T_\text{rec} \;=\; \unit[0.32]{eV} \;=\; \unit[3700\,]{K} \;,
  &&\text{and}&&
  z_\text{rec} \;=\; 1360 \;.
\end{align}
Because of the steep slope of the $x_e$ curve, these values are not particularly sensitive to the choice of $\,x_e(T_\text{rec})\,$. 
It should be noted that $T_\text{rec}$ is considerably smaller than the energy needed to ionise an hydrogen atom.
The reason is that the large value of $n_\gamma/n_b$ pushes $x_e$ to unity and significantly delays recombination; the photons are so abundant that, even at sub-eV energies, there are still enough of them in the high-energy tail of the Planck distribution to keep the Universe ionised \cite{durrer:2008a}.

The Saha equation is meant to be accurate only when recombination happens in quasi-equilibrium.
In \fref{fig:recombination_evolution}, we show the Saha solution together with the ``exact'' ionisation history as obtained from solving the Boltzmann equation. As expected, the Saha approximation is accurate in determining the redshift when recombination starts but it fails at lower redshifts when the system goes out of equilibrium. 
It should be noted that the $x_e$ curve flattens at low redshift, as if recombination at some point had become ineffective in binding electrons and protons.
This is indeed what happens after the recombination rate drops below the expansion rate, so that recombination ``freezes'' and the ionisation fraction remains constant.


\subsubsection{Decoupling}
\label{sec:decoupling}

Two particle species decouple from each other when their interaction rate drops below the cosmic expansion rate.
Roughly speaking, if a photon scatters an electron less than once in an expansion time, equilibrium between the two species cannot be maintained.
As we mentioned above, all the species are doomed to decouple at some point due to the expansion rate decreasing slower than any interaction rate.
For the photons, the process of recombination anticipates this moment by suddenly removing most of the free electrons from the Universe.

We estimate the redshift of photon decoupling by equating the rate of photon scatterings with the cosmic expansion rate:
\begin{align}
  \label{eq:decoupling_condition}
  n_e(z_\text{dec})\,\sigma_T\,c \;=\; H(z_\text{dec}).
\end{align}
Provided that we neglect the helium nuclei, we can express the fraction of free electrons as
\begin{align*}
  n_e \;=\; x_e\,n_b \;=\; x_e\;\frac{\Omega_{b0}\;\rhoc}{m_p}\;\,(1+z)^3 \;.
\end{align*}
where we have used $\,n_b=\rho_{b0}/m_p\,a^{-3}\,$. The Hubble parameter is given by the Friedmann equation \eref{eq:tt_friedmann_alternate_version},
\begin{align}
  H^2 \;=\; H_0^2 \;\, (1+z)^{3}\;\,\Omega_{\sub{M}0}\;\left(\,1 \,+\, \frac{1+z}{1+z_\text{eq}}\,\right) \;,
\end{align}
where we have neglected the cosmological constant and the curvature because they were insignificant at the high redshifts considered.
By enforcing the condition in \eref{eq:decoupling_condition} we obtain
\begin{align}
  \label{eq:zdec_messy_equation}
  x_e \; (1+z_\text{dec})^{3/2}
  \left(\,1\,+\,\frac{1+z_\text{dec}}{1+z_\text{eq}}\,\right)^{-1/2} \;=\;
  \left[\;\frac{m_p\,H_0\,\Omega_\sub{M0}^{1/2}}{c\,\rhoc\,\sigma_T\,\Omega_{b0}}\;\right] \;.
\end{align}
Inserting the cosmological parameters considered in \sref{sec:expansion_history}, the term in the right hand side evaluates to $236$ and $z_\text{eq}\simeq3380\,$. The ionisation fraction $x_e$ needs to be computed numerically (Saha's equation is of no use when $x_e$ is small) and we do so by using \emph{RECFAST} \cite{seager:1999a}.
This results in the values $\,z_\text{dec}\simeq 900\,$ and $x_e(z_\text{dec})\simeq10^{-2}\,$, which imply that photon decoupling takes place during recombination (recombination ends when the ionisation fraction reaches the freeze-out value of $x_e\simeq10^{-3}$, see \fref{fig:recombination_evolution}).
It is interesting to note that if recombination did not happen the photons would have decoupled only at $z\simeq40\,$; this can be seen by setting $x_e=1$ in the above equation.

In \sref{sec:line_of_sight} (and in \SONG) we shall use a more sophisticated method to determine the time of photon decoupling, making use of the \keyword{visibility function}, the probability that a photon last scattered at a given redshift. In particular, we shall see that the visibility function peaks at $\,z_\text{dec}\simeq1100\,$, a redshift slightly higher than what we have inferred by enforcing $n_e\,\sigma_T\,c=H\,$.
For a standard \LCDM model, a redshift of $\,z_\text{dec}\simeq1100\,$ correponds to
\begin{align}
  \label{eq:decoupling_}
  &\chi\,(z_\text{dec})\;\simeq\;\unit[280]{Mpc}\;,
  &&t\,(z_\text{dec})\;\simeq\;\unit[380,000]{yr}\;,
  &&T\,(z_\text{dec})\;\simeq\;\unit[0.26]{eV}\;.
\end{align}
The three-dimensional spatial surface identified by the time of decoupling is called the \keyword{last scattering surface} (LSS). Note that the comoving particle horizon at the LSS, $\,\chi(z_\text{dec})\;\simeq\;\unit[280]{Mpc}\,$, is roughly $80$ times smaller than the one today, $\,\chi_0\simeq\unit[14200]{Mpc}\,$.

We conclude this section by noting that the electrons remain coupled to the photons even after recombination ends and the photons go out of thermal equilibrium. That is, the photons decouple from the electrons but not viceversa.
This happens because the mean free path of an electron is much shorter than that of a photon, for the simple reason that there are many more photons than electrons.
Equivalently, the interaction rate of the free electrons ($\sigma_T\,n_\gamma\,c$) is much larger than that of the photons ($\sigma_T\,x_e\,n_b\,c$) because $n_\gamma/n_b\gg1\,$.
Therefore, the temperature of the electrons does not decay as $1/a^2$, as it would be expected from a thermal fluid of massive particles, but follows that of the CMB until low redshifts.

\section{Cosmic inflation}
\label{sec:inflation}


The standard hot Big Bang model introduced in the previous sections succesfully accounts for the observed expansion of the Universe (\sref{sec:expansion_history}), for the blackbody spectrum of the cosmic microwave background (\sref{sec:background_cmb}) and for the abundances of the light nuclei created via nuclesynthesis (see, for example Ref.~\cite{dodelson:2003b} and \cite{durrer:2008a}).
The model, however, is unable to answer several important observational and theoretical questions that we list below.

\begin{itemize}
  \item \textbf{The Big Bang singularity}\quad The most obvious issue is the presence of a a singularity in the finite past, the Big Bang (\sref{sec:background_evolution}), when the curvature and the density of the Universe are divergent.
  \item \textbf{The Horizon problem}\quad Any sign of correlations between regions of the Universe separated by a distance larger than the particle horizon cannot be explained by the standard model (\sref{sec:hubble_radius}).
This is, however, what we observe: the cosmic microwave background has the same temperature with a precision of a part over $10^{5}$ regardless of the direction of observation.
The particle horizon at decoupling was $80$ times smaller than the current value (\sref{sec:decoupling}), meaning that we would expect to observe fluctuations of order unity in the temperature of the CMB sky on angular scales of about $1\deg\,$.
The fact that we do not observe such fluctuations poses a causality problem that is referred to as the \keyword{horizon problem}: how can regions of the Universe be so similar if they did not have enough time to interact?
  \item \textbf{The Flatness problem}\quad The Friedmann and acceleration equations (\eref{eq:tt_friedmann_alternate_version} and \ref{eq:friedmann_jj}) can be combined to obtain an evolution equation for the total density parameter $\,\Omega(t)\equiv\rho/\rhoc=1-k/(a^2H^2)$:
\begin{align}
  \label{eq:flatness_equation}
  \frac{d}{dt}\bigl[\,\Omega(t)-1\,\bigr]\;=\;\bigl[\,\Omega(t)-1\,\bigr]\;
  \Omega(t)\,\left(\,1\,+\,3w\,\right) \;.
\end{align}
This equation shows that, for a Universe with an equation of state of $w>-1/3\,$, such as in a mixture of matter and radiation, the solution  $\Omega(t)=1$ is dynamically unstable; in fact, the sign of the derivative is positive for $\Omega(t)>1$ and negative for $\Omega(t)<1\,$ so that $\Omega(t)$ will always evolve away from unity.
This means that, for the Universe to be close to the critical density today as observations suggest, it had to be much more so in the past.
For example, for a current value of $\,0.1<\Omega_0<2\,$, it can be shown \cite{durrer:2008a} that $|\Omega-1|\leq10^{-15}$ at nucleosynthesis ($z\simeq10^9$) and $|\Omega-1|\leq10^{-60}$ at the Planck time ($t_P=\sqrt{\hbar G/c^5}\simeq\unit[\sci{5.4}{-44}]{s}$).
The smallness of these values poses a fine-tuning issue that is called the \keyword{flatness problem}:
how can the Universe be still so close to the critical density?
  \item \textbf{The structure problem}\quad 
  We observe tiny anisotropies in the CMB with an amplitude of $\Delta T/T\approx10^{5}$ and, more evidently, the observed Universe is highly inhomogeneous with a strongly clustered distribution of galaxies on small scales. By which mechanism was this structure formed?
\end{itemize}

These shortcomings of the hot Big Bang model are all connected to the initial conditions of the Universe. In this section we shall see that, apart from the Big Bang singularity, they can be solved by postulating the existence of a phase of accelerated expansion in the early Universe, the so-called \keyword{cosmic inflation}.
We first describe in \sref{sec:inflation_accelerated_expansion} how inflation solves the aforementioned cosmological problems.
Then, in \sref{sec:inflation_single_field_models} we show that the inflationary expansion can be achieved if the early Universe was dominated by a slowly-evolving scalar field, the so-called inflaton.
In section \sref{sec:primordial_fluctuations} we briefly discuss how inflation generates the density fluctuations that have seeded the observed structure on large scales.
In particular, we shall focus on the possibility that these primordial fluctuations are non-Gaussian, thus opening a window on interesting new physics.
(Note that to do so we use the concepts of cosmological perturbations and $n$-point functions, which are described only in the next chapter.)

In this section we shall only mention the fundamental properties of inflation. A detailed description of the topic can be found in several textbooks. For example, Chapter 6 of \citet{dodelson:2003b} provides a pedagogical introduction to inflation while \citet{liddle:2000a} treat inflation from a more advanced point of view; we refer the reader to these references for the omissions of this section.
Technical reviews focussed on the generation of non-Gaussianity during inflation can be found in Ref.~\cite{bartolo:2004c, chen:2010a}.


\subsection{The accelerated expansion}
\label{sec:inflation_accelerated_expansion}

\begin{figure}[p]
	\centering
		\includegraphics[width=0.8\linewidth]{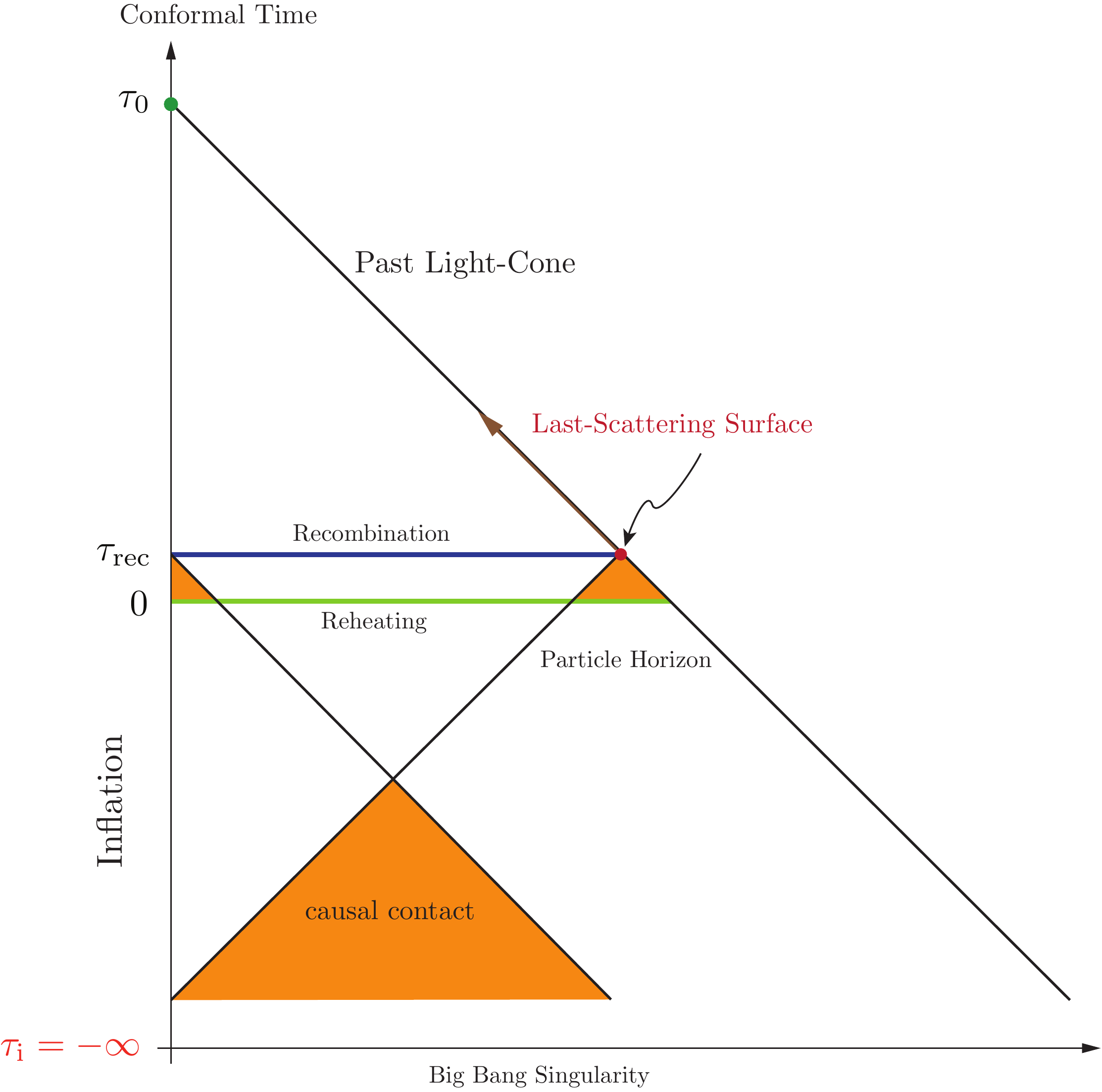}
	\caption[Conformal diagram of inflation]{Conformal diagram of inflation. The $y$-axis is conformal time, while the $x$-axis is distance. Our vantage point is today ($\tau_0$), on the $x=0$ vertical line.
The standard Big Bang model predicts that the dynamical evolution of the Universe started at $\tau=0$ (green horizontal line). In this picture, the past light cones of two distant CMB patches (small orange triangles) do not intersect, because the particle horizon at the time where the CMB is formed (horizontal line at $\tau_\text{rec}$) is much smaller than $\tau_0$. Therefore, we expect order-unity differences in the CMB temperature on large scales. However, we observe the CMB today to be almost perfectly isotropic on all scales; this is the horizon problem.
In the inflationary scenario, the horizon problem is solved by postulating the existence of a period where the two CMB patches were in casual contact (big orange triangle). This is achieved by extending the time axis below $\tau=0$ in order to allow the past-light cones of the two CMB patches to intersect.
A period of accelerated expansion, \emph{cosmic inflation}, is needed in order to bridge the gap between the ``small Universe'' where the casual contact was established, and the large value of today's particle horizon.
In this context, $\tau=0$ is not a singularity but an apparent Big Bang, as it marks the end of inflation and the decay of the inflaton (\sref{sec:inflation_single_field_models}) into a thermal mix of elementary particles. The actual Big Bang singularity sits at $\tau\rightarrow-\infty$.
Source: courtesy of Daniel Baumann, from Fig.~9 of Baumann (2009) \cite{baumann:2009a}.}
	\label{fig:inflation}
\end{figure}

The mechanism of cosmic inflation \cite{guth:1981a, linde:1982a, albrecht:1982a, starobinsky:1980a} consists of postulating the existence of a period in which the Universe was much smaller than what
one would infer based on the standard Big Bang model.
In this period, the same regions of the Universe that we see today as separate and independent, were actually in causal contact.
In order to link this ``small universe'' with the size of the universe today, one needs to postulate a phase in between where the Universe has expanded much quicker than the normal rate; hence the name cosmic inflation.
In \fref{fig:inflation} we explain this process in terms of a conformal diagram of cosmic inflation.

Cosmic inflation solves the horizon problem by connecting regions that, in a standard Big Bang model, would be causally disconnected.
For this to happen, the comoving Hubble radius, which we defined in \sref{sec:hubble_radius} to be $\,c/(aH)\,$, at the beginning of inflation had to be larger than the largest scale observable today, that is the current comoving Hubble radius.
Since after inflation the horizon grows with time (\sref{sec:expansion_history}), it follows that during inflation it has to decrease; the expansion during inflation must therefore satisfy
\begin{align}
  \frac{d}{dt}\,\left[\,\frac{1}{aH}\,\right] \;<\; 0
  \quad\Rightarrow\quad
  \frac{d^2a}{dt^2} \;>\; 0 \;,
\end{align}
that is, the expansion had to be \emph{accelerated}.
It is important to remark that it is not the accelerated expansion that solves the horizon problem: the causal connection (\ie the Universe becoming uniform) is established \emph{before inflation} and what inflation does is to put those regions out of reach again, because this is how we see them today.

The accelerated expansion, however, does solve the flatness problem, because it washes out any curvature, stretching the geometry of the Universe so much that it becomes spatially flat \cite{hawley:2005a}.
More quantitatively, we see from the acceleration equation (\eref{eq:friedmann_jj}),
\begin{align}
  \frac{a''}{a} \;=\; -\frac{4\,\pi\,G}{3}\;\left(\;\rho\;+\;3\,P\;\right)\;,
\end{align}
that the Universe undergoes an accelerated expansion only if $\,\rho+3\,P<0\,$ or, in terms of the barotropic parameter, if $w<-\frac{1}{3}\,$.
If we inspect \eref{eq:flatness_equation}, we realise that this is the same condition needed to make $\,\Omega(t)=1\,$ an attractor solution; that is, if cosmic inflation lasted long enough, the flatness problem would be solved without the need to fine tune the initial curvature.
In fact, we can ask the question: how many times must the Universe double in size during inflation to justify the fact that today's Universe is so close to the critical density?
The answer comes from the Friedmann equation for a constant equation of state (\eref{eq:tt_friedmann_alternate_version}):
\begin{align}
  |\Omega(t)-1| \;=\; \frac{3\,|k|}{8\,\pi\,G\,a^2\,\rho}
  \;\propto\; a^{1\,+\,3\,w} \;.
\end{align}
If we assume that during the inflationary phase $w=-1\,$, then $|\Omega(t)-1|$ decreases like $a^{-2}$\,; to bring $|\Omega(t)-1|$ to today's value of order unity from $\sim10^{-60}$ at the Planck time would require that
\begin{align}
  N \;\equiv\; \ln\left(\,\frac{a_\text{end}}{a_\text{ini}}\,\right)
  \;\simeq\; 30\;\ln(10) \;\simeq\; 70 \;,
  \label{eq:duration of }
\end{align}
where $N$ is called the number of e-foldings and $\,a_\text{ini}\,$ and $\,a_\text{end}\,$ mark the beginning and the end of inflation, respectively.

Cosmic inflation provides a solution to the structure problem that is rooted in quantum mechanics; we postpone this discussion until \sref{sec:primordial_fluctuations}.


\subsection{Single field model}
\label{sec:inflation_single_field_models}

Inflation is a mechanism rather than a theory of the early Universe, a phase of accelerated expansion before which the comoving horizon was larger than the largest scale observable today.
We have seen that to realise the accelerated expansion it is necessary for the matter content of the Universe to have an equation of state of $w<-\frac{1}{3}\,$, which corresponds to a negative pressure, $\,\rho+3\,P<0\,$.
Neither cold matter ($w=0$) nor radiation ($w=\frac{1}{3}$) are suitable candidates as they have positive pressure; the cosmological constant ($w=-1$) can produce an accelerated expansion but is completely negligible in the early Universe, so it cannot be responsible for inflation.

Let us see how the presence of a \emph{scalar field}, which we call the \keyword{inflaton} $\phi\,$, can trigger the mechanism of cosmic inflation.
The scalar field Lagrangian is given by
\begin{align}
  \mathcal{L}_\phi \;=\; -\,\frac{1}{2}\,\partial_\mu\,\phi\,\partial^\mu\,\phi
  \;-\; V(\phi) \;,
\end{align}
where $V(\phi)$ is the potential for the field, which we assume to be positive. In principle $\mathcal{L}$ should include terms to account for the interactions with the other species, but we postulate that they are negligible during inflation.
The pressure and the energy density of the inflaton field can be inferred from its energy-momentum tensor:
\begin{align}
  T_{\mu\nu} \;=\; \partial_\mu\,\phi\,\partial_\nu\,\phi \;-
  \;\frac{1}{2}\,g_{\mu\nu}\,\partial_\alpha\,\phi\:\partial^\alpha\,\phi \;-
  \;g_{\mu\nu}\,V(\phi) \;.
\end{align}
Here we assume that the Universe is homoegenous, so that $\,g_{\mu\nu}\,$ is the conformal \FLRW metric in \eref{eq:conformal_metric_flrw} and the spatial gradients of $\phi$ vanish. It follows that
\begin{align}
  &\rho_\phi \;=\; -\UD{T}{0}{0} \;=\;
  \frac{1}{2}\,\phi'^{\,2} \;+\; V(\phi)
  &&\text{and}&&
  P_\phi \;=\; \frac{1}{3}\,\UD{T}{i}{i} \;=\;
  \frac{1}{2}\,\phi'^{\,2} \;-\; V(\phi) \;.
\end{align}
where $\phi'=d\phi/dt\,$.
The expression for the energy density is reminiscent of that of a particle moving in a potential $V$ with velocity $\phi'\,$ and kinetic energy $\,\frac{1}{2}\,\phi'^{\,2}\,$.
In this picture, a field with negative pressure is one with more potential energy than kinetic.
In the limit where the inflaton field is constant ($\phi'=0$), its kinetic energy vanishes and we have a constant energy density: $\,\rho_\phi=V(\phi)=\text{constant}\,$.
If we assume that the energy density and pressure of the Universe are dominated by the inflaton's contribution, 
the expansion rate of the Universe is determined by $\,\rho_\phi\,$ via the Friedmann equation \eref{eq:friedmann_00}:
\begin{align}
  \label{eq:inflation_H_constant}
  H \;=\; \frac{1}{a}\,\frac{da}{dt} \;=\; \sqrt{\frac{8\,\pi\,G\,\rho_\phi}{3}} \;=\; \text{constant} \;,
\end{align}
It follows that a Universe whose dynamical evolution is determined by a \emph{constant} scalar field expands at an exponential rate: $\,a\propto e^{H\,t}\,$, where $H\propto\sqrt{\rho_\phi}\,$ constant. Inflation is therefore realised.

The Friedmann equation (\eref{eq:friedmann_00}) during inflation reads
\begin{align}
  H^2 \;=\; \frac{1}{3\,m_\sub{P}^2}\,
  \left(\,\frac{1}{2}\,\phi'^{\,2}\,+\,V(\phi)\,\right) \;,
\end{align}
where we have introduced the Planck mass $\,m_\sub{P}\equiv(8\pi G)^{-1/2}\simeq\unit[\sci{2.4}{18}]{GeV}\,$.
The Friedmann and acceleration (\eref{eq:friedmann_jj}) equations can be combined to yield the background evolution of the inflaton,
\begin{align}
  \label{eq:inflation_inflaton_background_evolution}
  \phi'' \;+\; 3\,H\,\phi' \;+\; V_{,\phi} \;=\; 0 \;,
\end{align}
where the primes denote derivatives with respect to cosmic time $t$ and $V_{,\phi}=\partial V/\partial\phi\,$.

\subsubsection{The slow-roll condition}

\begin{figure}[t]
	\centering
		\includegraphics[width=0.55\linewidth]{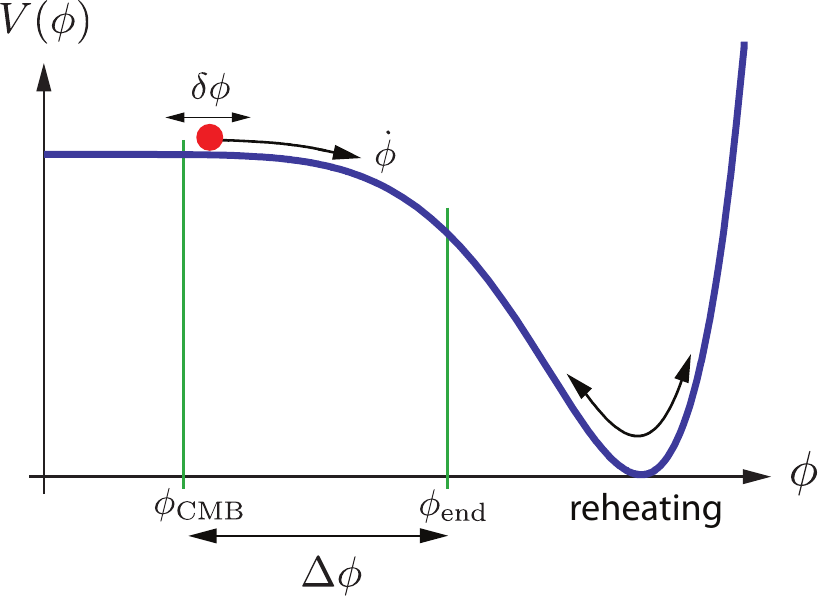}
	\caption[Inflationary potential]{Example of a slow-roll inflationary potential. As long as the inflaton's kinetic energy, $\,\frac{1}{2}\,\phi'^{\,2}\,$, is negligible with respect to its potential energy, $\,V(\phi)\,$, the Universe expands in an accelerated fashion; this limit corresponds to the constant part of the potential. When $\,\frac{1}{2}\,\phi'^{\,2}\simeq V(\phi)\,$, the acceleration can no longer be sustained and inflation ends. When the inflaton reaches the minimum of the potential, reheating occurs and the energy density of the inflaton is converted into a thermal mix of elementary particles. Source: courtesy of Daniel Baumann, from Fig.~10 of Baumann (2009) \cite{baumann:2009a}.}
	\label{fig:slow_roll}
\end{figure}

We have just proved that a scalar field can drive inflation as long as it does not evolve significantly, $\,\phi'^{\,2}\ll V(\phi)\,$.
The issue now is to determine the potential $V(\phi)$ that keeps $\phi$ nearly constant for the number of e-foldings necessary to solve the horizon and flatness problems.
Most models of inflation satisfy the \keyword{slow-roll condition} \cite{linde:1982a, albrecht:1982a}, whereby the inflaton stays nearly constant by slowly rolling down a potential that is almost flat.
We show an example of a slow-roll potential in \fref{fig:slow_roll}.
Because inflation cannot last forever, the potential needs to have a minimum; as time goes on, the inflaton approaches this minimum and, due to the increased slope of the potential, it starts to evolve faster.
Inflation comes to an end when the kinetic energy $\,\frac{1}{2}\,\phi'^{\,2}\,$ grows to be of the order of the potential $V(\phi)\,$.
When the inflaton eventually reaches the minimum of the potential, the coupling with the other fields becomes significant so that it decays into a thermal mix of elementary particles \cite{durrer:2008a}, leading to a radiation dominated universe in a process called \keyword{reheating}.
In practice, we can think of the reheating process after inflation as the moment when the hot Big Bang occurs, in which matter and radiation as we know them start to be created.

Many different potentials can be devised that satisfy the slow-roll condition.
It is customary to parametrise them with two variables that vanish in the limit where $\phi$ is constant.
The first slow-roll parameter $\eta$ quantifies the variation in the Hubble factor, and is related to the first derivative of the inflaton potential. It is defined as
\begin{align}
  \label{eq:inflation_epsilon}
  \epsilon \;\equiv\; \frac{d}{dt}\,\left(\frac{1}{H}\right)
  \;=\; -\frac{H'}{H^2} \;\approx\; \frac{m_P^2}{2}\,\left(\frac{V_{,\phi}}{V}\right)^2\;.
\end{align}
Whenever the inflaton field is constant, $\,\phi'=0\,$, then also $H\propto\sqrt{\rho_\phi}$ is constant (\eref{eq:inflation_H_constant}) meaning that the $\epsilon$ parameter vanishes.
In fact, the slow-roll condition requires $\epsilon\ll1$, an assumption that implies an approximate time-translation invariance of the background.
On the other hand, in the radiation dominated era $\epsilon=2\,$; in fact, one can define the inflationary epoch as $\epsilon<1\,$.
The second slow-roll parameter, $\,\eta\,$, is directly related to the second derivative of the potential\footnote{In defining the slow-roll parameters, we are using the notation of the review by \citet{bartolo:2004c}. \citet{chen:2010a}, on the other hand, denotes the quantity in \eref{eq:inflation_eta} as $\eta_\sub{V}$ and uses the symbol $\eta$ for a third slow-roll parameter:
\begin{equation}
  \eta \;\equiv\; -2\,\eta_\sub{V} \;+\; 4\,\epsilon
  \;=\; \frac{\epsilon'}{\epsilon\,H} \;.
\end{equation}},
\begin{align}
  \label{eq:inflation_eta}
  \eta \;\equiv\; m_P^2\;\left(\frac{V_{,\phi\phi}}{V}\right) \;.
\end{align}
Again, in the case of a constant field or potential this parameter vanishes.
As we shall see below, the most important predictions of inflation can be recast in terms of the slow-roll parameters $\epsilon$ and $\eta\,$.

\subsection{Primordial fluctuations}
\label{sec:primordial_fluctuations}

Cosmic inflation was originally proposed to solve the horizon and flatness problems \cite{guth:1981a, linde:1982a, albrecht:1982a, starobinsky:1980a}, but it was soon realised that it also provided a mechanism to generate primordial density fluctuations \cite{hawking:1982a, starobinsky:1982a, mukhanov:1981a, bardeen:1983a}.
The idea is that the structure that we observe today, such as the CMB anisotropies and the galaxy distribution, formed starting from tiny quantum fluctuations set during inflation and later enhanced throughout cosmic history via gravitational instability.
These primordial fluctuations were generated as microscopic quantum vacuum fluctuations in the inflaton field that, during inflation, were stretched and imprinted on superhorizon scales by the accelerated expansion.
These density fluctuations reentered the horizon after inflation ended and served as initial conditions for the anisotropy and the growth of structure in the Universe.

In what follows, we briefly describe the main features of the primordial fluctuations generated during inflation. 
To do so, we need to use some concepts that will be formally defined only in the next chapter, like the idea that the primordial fluctuations generated during inflation are stochastic in nature and, therefore, their magnitude is determined in terms of their variance (in real space) or their power spectrum (in Fourier space).
We will also use of the concepts of scalar and tensor (\sref{sec:perturbations_svt_decomposition}) perturbations (\sref{sec:stochastic_perturbations}), power spectrum (\sref{sec:two_point_function}) and bispectrum (\sref{sec:three_point_function}).

\subsubsection{Scalar fluctuations}

The primordial fluctuations generated during slow-roll inflation are expected to have nearly the same variance on all spatial scales.
The reason is that the slow-roll condition $\epsilon=-H'/H^2\ll1\,$ results into an approximate time-translation invariance of the background.
Therefore, the primordial fluctuations are produced with approximately the same background expansion rate regardless of the scale considered.
This \keyword{scale invariance} is usually quantified in terms of the \keyword{scalar spectral index}\index{spectral index}, $\,n_s\,$, defined to be the slope of the \keyword{dimensionless power spectrum} of the primordial curvature perturbation,
\begin{align}
  \mathcal{P}_\mathcal{R} \;\propto\; k^{\,n_s-1} \;.
\end{align}
The condition of scale invariance translates to $n_s=1\,$.
However, the presence of structure in the inflaton potential affects the expansion rate and, therefore, it generates deviations from scale invariance. In a slow-roll inflationary model where the potential is nearly flat, these deviations are small \cite{bartolo:2004c, chen:2010a}:
\begin{align}
  n_s \;=\; 1 \;-\; 6\,\epsilon \;+\; 2\,\eta \;.
\end{align}
Because the slow-roll parameters $\epsilon$ and $\eta$ describe, respectively, the first and second derivative of the inflaton potential $V(\phi)\,$, measuring $n_s$ is equivalent to constraining the shape of $V(\phi)\,$.
The cosmic microwave background is strongly affected by the tilt of the primordial fluctuations and, as a result, it can be used to constrain $n_s$ \cite{planck-collaboration:2013a}:
\begin{align}
  n_s \;=\; 0.9603 \,\pm\, 0.0073 \qquad\text{at \onesigma} \;.
\end{align}
This measurement is in agreement with the slow-roll inflationary models and suggests that the two slow-roll parameters have a value of $\O(10^{-2})\,$.

Another important observable of inflation is the amplitude $A_s$ of the primordial fluctuations, which is defined as
\begin{align}
  \mathcal{P}_\mathcal{R}(k) \;=\; A_s\;\left(\,\frac{k}{k_0}\,\right)^{\,n_s-1},
\end{align}
where $k_0$ is the pivot scale.
In the slow-roll limit, the amplitude $A_s$ is connected to the ratio between the inflaton potential and the slow-roll parameter $\epsilon\,$ \cite{planck-collaboration:2013d}:
\begin{align}
  A_s \;=\; \frac{V}{24\,\pi^2\,m_\sub{P}^4\,\epsilon} \;.
\end{align}
By measuring the amplitude of the CMB angular spectrum, the Planck team \cite{planck-collaboration:2013d} found the value $\ln(10^{10}\,A_s) = 3.089^{+0.024}_{-0.027}$ at \onesigma for a pivot scale of $\,k_0 = \unit[0.05]{Mpc^{-1}}\,$, which translates to a constraint on the energy scale of inflation, $V^{1/4}$, and on $\epsilon$:
\begin{align}
  \frac{V^{1/4}}{\epsilon^{1/4}} \;=\; 0.027\,m_\sub{P} \;=\; \unit[\sci{6.6}{16}]{GeV} \;.
\end{align}

\begin{figure}[t]
	\centering
		\includegraphics[width=0.95\linewidth]{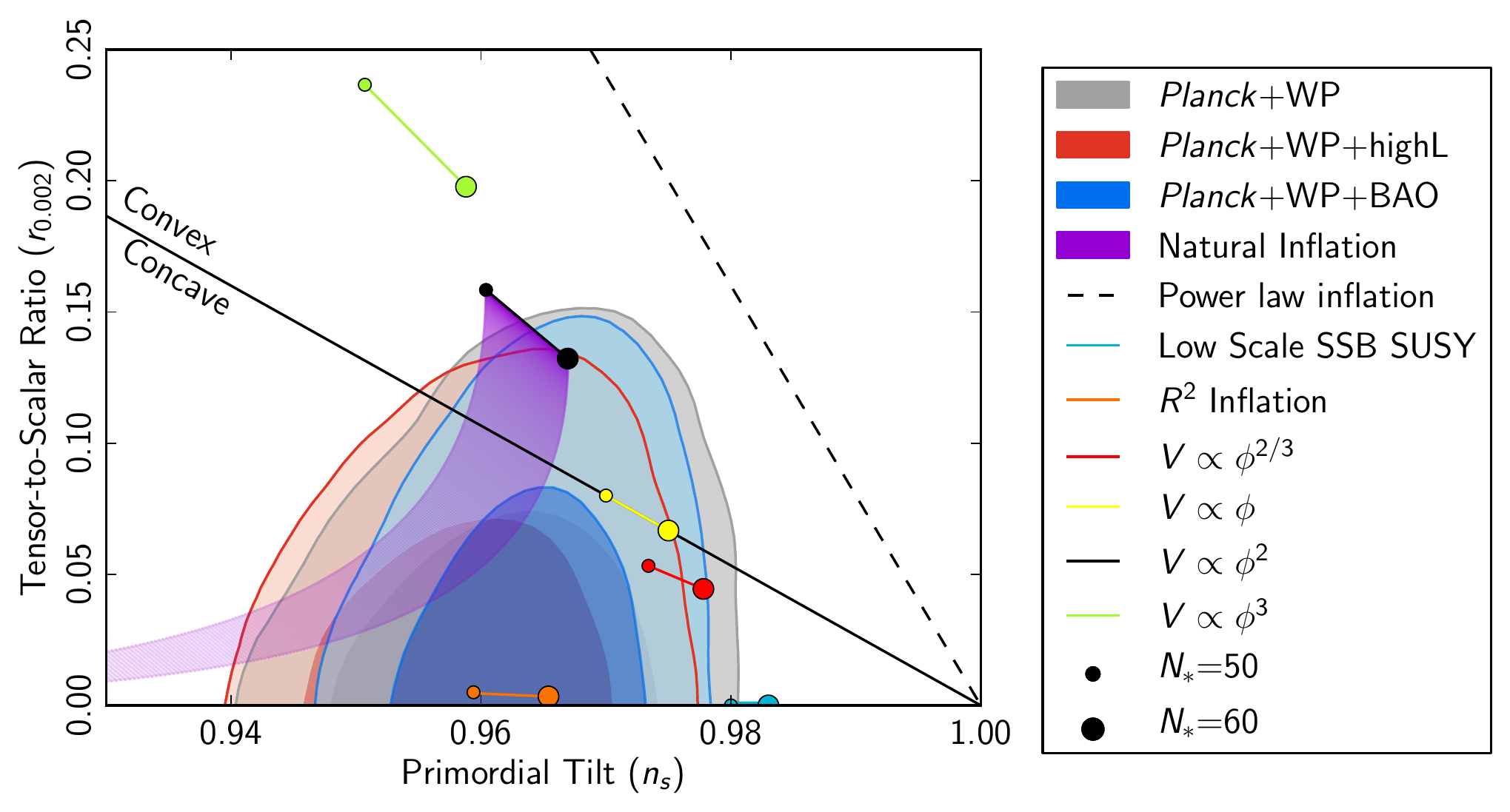}
	\caption[Planck constraints on inflation]{
  Constraints on the spectral tilt and the tensor-to-scalar ratio $r$ from Planck \cite{planck-collaboration:2013d}. The ellipses represent the $68\%$ and $95\%$ confidence limits on $n_s$ and $r\,$ for various combinations of datasets (WP is WMAP polarisation, BAO is baryon acoustic oscillation, highL is high-resolution CMB data). The theoretical predictions of several inflationary models are also shown.
  Credit: Fig.~1 on page 10 of Ref.~\cite{planck-collaboration:2013d} by the Planck collaboration, A\&A, reproduced with permission \textcopyright\xspace ESO.}
	\label{fig:planck_inflation}
\end{figure}


\subsubsection{Gravitational waves}

Another prediction from inflation is the presence of a background of primordial gravitational waves.
These are generated with the same mechanism as the scalar fluctuations and are thus also expected to be nearly scale invariant.
The power spectrum of tensor fluctuations,
\begin{align}
  \mathcal{P}_{\,t}(k) \;=\; A_t\;\left(\,\frac{k}{k_0}\,\right)^{\,n_t},
\end{align}
defines the tensor amplitude $A_t$ and the \keyword{tensor spectral index} $\,n_t\,$, which vanishes for a scale-invariant spectrum. For a slowly rolling scalar field, they are given by \cite{dodelson:2003b, planck-collaboration:2013d}
\begin{align}
  &A_t \;=\; \frac{2\,V}{3\,\pi^2\,m_\sub{P}^4} \;,
  &&\text{and}&& n_t = -2\,\epsilon \,.
\end{align}
In the slow-roll limit, a \keyword{consistency relation} links the spectral index $n_t$ to the amplitudes of the scalar and tensor power spectra:
\begin{align}
  \label{eq:consistency_relation}
  r \;\equiv\; \frac{\mathcal{P}_{\,t}}{\mathcal{P}_\mathcal{R}} \;=\; -8\,n_t \;,
\end{align}
where we have defined the \keyword{tensor-to-scalar ratio} $r\,$.
Since $A_s$ has already been experimentally determined, measuring the value of $r$ would automatically yield the amplitude of the tensor perturbations $A_t$ and, through the consistency relation, the tilt $n_t$ of the tensor spectrum.
Furthermore, a determination of $r$ would imply also an indirect detection of the gravitational waves.
So far, only upper limits for the tensor-to-scalar exist; in \fref{fig:planck_inflation} we show the joint measurement of $r$ and $n_s$ produced by the Planck experiment \cite{planck-collaboration:2013d}.

\subsection{Non-Gaussianity}
\label{sec:non_gaussianity}

The inflation observables that we have introduced in the previous subsection, the spectral index $\,n_s\,$ and the tensor-to-scalar ratio $\,r\,$, are defined with respect to the power spectrum of the primordial curvature perturbation, $\,\mathcal{P}_\mathcal{R}\,$.
The power spectrum, however, is just one of the infinite series of $n$-point functions that characterise the primordial field (\sref{sec:stochastic_perturbations}).
In the case of a Gaussian random field, these moments can be expressed as products of $\mathcal{P}_\mathcal{R}\,$; for an arbitrary field, this is not the case: the higher-order moments contain extra information that eludes the power spectrum and that, as we shall soon see, is precious to understand the non-linear physics at work in the early Universe.
We shall refer to this extra information as \keyword{non-Gaussianity}, simply because it is absent for Gaussian perturbations.

In this thesis, we focus on the three-point function of the primordial curvature perturbation, or \keyword{primordial bispectrum}.
The full formalism to characterise the bispectrum and its observability in the cosmic microwave background will be introduced in \cref{ch:intrinsic}.
The purpose of this subsection is to explain our motivations for studying the bispectrum; therefore, for now, we shall keep the technical details to a minimum.


The primordial bispectrum is important for two reasons.
First, it is the lowest order statistic sensitive to whether a perturbation is Gaussian or non-Gaussian. This follows from the fact that the three-point function of a Gaussian random field with zero mean vanishes.
Secondly, it is directly related to the angular bispectrum of the cosmic microwave background, which is an observable quantity \cite{komatsu:2001a, komatsu:2010a, yadav:2010a}.
Therefore, the primordial bispectrum as inferred from the CMB has the power of discriminating models of inflation based on the amount of non-Gaussianity they produce.

The standard slow-roll inflation models that we have described above, where the accelerated expansion is driven by a non-interacting scalar field, produce a bispectrum of the order of the slow-roll parameters \cite{maldacena:2003a, acquaviva:2003a}; for all practical purposes, this non-Gaussianity can be considered negligible. This is intuitive as the bispectrum is inherently related to the non-linearities in the propagation of the field. In the ``vanilla'' models, the inflaton propagates freely along a very flat potential ($\epsilon,|\eta|\ll1$), so that any self-interaction term of the inflaton potential and the gravitational coupling must be very small; consequently, the non-linearities are also suppressed \cite{bartolo:2004c}.

Measuring a significant bispectrum would therefore rule out the simplest models of inflation.
It should be stressed that these models are otherwise highly successful in reproducing the required duration of inflation and the observed shape of the power spectrum.
The non-Gaussianity measurement is thus complementary to the usual inflation observables, $n_s$ and $r$, and it provides extra information on the physics of the early Universe that is useful to break degeneracies between models that would otherwise be observationally equivalent.


The constraining power of the primordial bispectrum and its observability prompted particle physicists and cosmologists to join forces and investigate many well-motivated extensions to the inflationary vanilla model.
The multiple-field models, for example, postulate that two or more fields are present during inflation.
These models are appealing also because, from the point of view of particle physics, it is natural to have several other fields that contribute to the inflationary dynamics.
If the fields interact, the Lagrangian will include non-linear contributions that ultimately lead to deviations from pure Gaussian statistics \cite{seery:2005a, byrnes:2010a, bartolo:2004c}.
This is not, however, the only mechanism to create non-Gaussianity in a multi-field model.
In the \keyword{curvaton scenario} \cite{linde:1997a, enqvist:2002a,lyth:2002a,moroi:2001a,moroi:2002a}, for example, the inflaton field drives the accelerated expansion as in a single field model, while a subdominant second field, the curvaton, is responsible for generating the curvature perturbations.
In this case, the non-Gaussianity is produced by the non-linear evolution of the curvature perturbation on superhorizon scales. 

Other extensions to the vanilla model include features in the inflaton potential, the presence of a non-canonical kinetic term, non-linearities in the initial vacuum state or modifications to the theory of gravity \cite{chen:2010a}. 
These features generally translate to non-Gaussian signatures in the primordial curvature perturbation and, thus, in specific shapes of the bispectrum. For a review on these models and their observability, refer to the reviews in Ref.~\cite{komatsu:2010a, bartolo:2010a, yadav:2010a, liguori:2010a}.

In summary, the non-Gaussianity of the cosmological perturbations opens a window on the non-linear physics of the early Universe; the CMB bispectrum is the observable that allows us to look through this window.
The subject of this thesis is the connection between the primordial non-Gaussianity and the CMB bispectrum.
In the following chapters, we shall answer the questions: how is the measured CMB bispectrum affected by the non-linear evolution that happens \emph{after} inflation? Would this effect significantly bias a measurement of the primordial signal?

The answers can be found in \cref{ch:intrinsic}.

\chapterbib


\chapter{Perturbation theory}
\label{ch:perturbation_theory}

\section{Introduction}
\label{sec:perturbations_intro}
According to the hot Big Bang cosmology introduced in the previous chapter, all observations are expected to be perfectly homogeneous and isotropic about our location. This prediction is in clear disagreement with the observed distribution of galaxies in the sky, which shows strong clustering properties on scales smaller than \unit[100]{Mpc} (\sref{sec:cosmological_principle}), and with the measured temperature of the cosmic microwave background, which is characterised by tiny direction-dependent fluctuations \cite{smoot:1999a,bennett:2012a,planck-collaboration:2013a}.
The ultimate origin of this structure was explained in \sref{sec:primordial_fluctuations} in terms of the primordial fluctuations generated in the early Universe via cosmic inflation.
In this chapter, we introduce a formalism that is useful to study their subsequent evolution.

The \keyword[perturbation theory]{theory of cosmological perturbations} has been extremely successful in describing the clustering of galaxies and the angular distribution of the CMB temperature. The key aspect of perturbation theory is to consider the Universe as being described by a homogeneous background with small position-dependent perturbations that are assumed not to affect the background itself. The background is modelled as a hot Big Bang Universe with an \FLRW metric, as discussed in \cref{ch:homogeneous_universe}, while the perturbations evolve according to a form of the Einstein and Boltzmann equations obtained by expanding them around the homogeneous solution. The advantage of this approach is that the perturbed equations have a recursive structure that can be truncated at the desired level of accuracy.  


The temperature map of the cosmic microwave background is particularly well suited to be treated with a perturbative approach, because it is almost perfectly smooth, with deviations from isotropy of a part in $10^5$ \cite{smoot:1999a,bennett:2012a,planck-collaboration:2013a}. The reason for this behaviour is that photons, being relativistic particles, tend to stream freely rather than cluster, thus preserving the amplitude of the small initial fluctuations that were set in the early Universe. The only time where photons clustered was before recombination, when they strongly interacted with baryons through Thomson scattering; this is the reason why the observed fluctuations in the CMB peak on the angular scale, $\sim 1\deg$, corresponding to the size of the sound horizon at recombination (see \sref{sec:background_cmb}).

Since the CMB anisotropies are small, their basic properties are well described by the first order in perturbation theory, where the Boltzmann and Einstein equations are linearised. There are, however, many aspects of the CMB that cannot be predicted by linear theory. One of them, which is the main topic of this thesis, in the generation of non-Gaussian features in the CMB due to the propagation of photons through an inhomogeneous Universe. Other notable effects are the generation of vorticity and anisotropic stresses at recombination, which ultimately leads to the generation of magnetic fields and $B$-mode polarisation in the CMB, and the momentum transfer between photon and electrons due to Compton scattering, which gives raise to distortions in the frequency spectrum of the CMB. These non-linearities of the CMB can be treated in the framework of the standard relativistic perturbation theory by going to second order in the cosmological perturbations, a technique that we shall review in this chapter.

Contrary to the CMB, the density of the cold matter grows in time due to gravitational collapse, to the point that, eventually, the assumption of small perturbations on a homogeneous background breaks down. This is the so-called \keyword{non-linear regime}, which happens at late times and on scales that are well inside the horizon. The non-linear regime is better described by ad-hoc perturbative techniques that are generally more involved than the standard perturbation theory; for an extensive review, refer to Ref.~\cite{bernardeau:2002a}. However, because the CMB photons were emitted at a time ($t\sim400,000$ years) where the non-linear effects in the matter distribution were subdominant, in this thesis we only treat the standard relativistic perturbative approach.

\subsection{Summary of the chapter}
In \sref{sec:perturbations_intro} we explain why perturbation theory is needed to describe our inhomogeneous Universe, and provide a literature review of the field.
In \sref{sec:perturbations_formalism} we introduce the mathematical definition of perturbations and show the general properties of first and second-order equations.
In \sref{sec:perturbations_metric} we illustrate what a gauge is and we pick one, the Newtonian gauge, to build our perturbed metric.
In \sref{sec:stochastic_perturbations} we focus on the statistical properties of the cosmological perturbations and explain why they are described in terms of stochastic fields that satisfy statistical homogeneity and isotropy.
In \sref{sec:transfer_functions} we explain why going to Fourier space is a good idea, and we introduce the concept of a transfer function as a way of separating the deterministic part of a perturbation from its stochastic one.
In \sref{sec:einstein_equations} we introduce the fluid variables and show the Einstein equations up to second order in the cosmological perturbations.
The main cosmological observables, the power spectrum and the bispectrum, are treated in \sref{sec:spectra_and_bispectra}, where we also introduce the concepts of primordial, linear and intrinsic bispectra.

\subsection{Literature review}

\citet{lifshitz:1946a} and, later, \citet{lifshitz:1963a}, first developed the relativistic linear theory of the cosmological perturbations in a \FLRW Universe, and used it to derive the evolution of the density perturbations in the synchronous gauge. A more general gauge-invariant treatment that did not rely on a choice of the coordinates, was introduced by \citet{bardeen:1980a} and later generalised by \citet{kodama:1984a} (see also \citet{gerlach:1979a}). The subject of cosmological perturbations on a \FLRW background is treated in great detail in the following reviews and books: \citet{peebles:1980a}, \citet{kodama:1984a}, \citet{mukhanov:1992a}, \citet{durrer:1994a}, \citet{ma:1995a}, \citet{bertschinger:1996a}, \citet{tsagas:2008a}, \citet{malik:2008a} and \citet{malik:2009a}, with the last review treating also the second-order perturbations. A more general approach dealing with general space-times is given in \citet{stewart:1974a}. For a pedagogical introduction to the topic, refer to the review by \citet{knobel:2012a}.

Among the earliest works that applied the linear perturbation theory to the cosmic microwave background were \citet{sachs:1967a}, who showed how density perturbations generate fluctuations in the angular distribution of the CMB photons, and \citet{peebles:1970a}, who first integrated the collision equation of the photon distribution function and introduced the tight coupling approximation. \citet{kaiser:1983a}, \citet{bond:1984a} and \citet{polnarev:1985a} were among the first authors to study the linear polarisation induced by Thomson scattering in the CMB, thus finding an alteration of the CMB anisotropy pattern of the $10\%$ level. A systematic study of the two-point statistics of the CMB in real and harmonic space can be found in \citet{bond:1987a}. Further analytic insight on the relation between the matter components (baryons and cold dark matter) and the CMB photons at recombination was given, \eg, in \citet{hu:1996a}. The correlation between the dark matter structure and the CMB anisotropies was discussed in \citet{crittenden:1996a} and later measured in, \eg, \citet{giannantonio:2012a}, to yield a confirmation of the presence of dark energy. Reviews on the theory of CMB perturbations can be found in \citet{ma:1995a}, \citet{hu:1997a}, \citet{durrer:2001a}, \citet{hu:2002a}, \citet{challinor:2004a}, \citet{challinor:2009a}, \citet{lesgourgues:2013a}, and in the book by \citet{dodelson:2003b}.

The first author to study the relativistic cosmological perturbations beyond linear order was \citet{tomita:1967a} who, extending Lifshitz's theory, computed the growth of the second-order density perturbations in synchronous gauge. A general way to relate higher-order perturbations in different gauges was given by \citet{bruni:1997a} and \citet{sonego:1998a}, and was later used by \citet{matarrese:1998a} to study the relativistic perturbations in an Einstein-de Sitter Universe in both the synchronous and Newtonian gauges. More recently, the second-order equations and their gauge invariance were discussed by \citet{bartolo:2004a}, \citet{pitrou:2009a}, \citet{beneke:2010a}, \citet{nakamura:2011a} and \citet{naruko:2013a}. 

When relaxing the approximation of linear perturbations, a number of effects arise that alter the anisotropies in the cosmic microwave background \cite{pyne:1996a, mollerach:1997a, maartens:1999a}, such as the Rees-Sciama effect \cite{rees:1968a}, the time-delay effect \cite{hu:2001a}, the gravitational lensing of CMB photons \cite{lewis:2012a, hanson:2009a, smith:2011a, serra:2008a, lewis:2011a, lewis:2006a}, the emergence of $B$-mode polarisation from the vector and tensor modes in the metric \cite{mollerach:2004a} and in the baryon-photon scattering \cite{beneke:2011a}, and, in general, a number of new quadratic contributions to the electron-photon scattering during recombination and reionisation \cite{hu:1994a, dodelson:1995a, bartolo:2006a, senatore:2009b, pitrou:2009a, beneke:2010a}. Most of these effects can be estimated with a second-order Boltzmann approach, which is what our code, \SONG, does and is the topic of the next chapters.

\section{General formalism}
\label{sec:perturbations_formalism}

A cosmological field $X(\tx)$ is perturbatively expanded around its background value $\pert{X}{0}(t)$ according to
\begin{align}
  X(\tx) \,=\, \pert{X}{0}(t) \,+\, \sum\limits_{i=i}^\infty\,\epsilon^{i}\,\pert{X}{i}(\tx) \;,
\end{align}
where $\epsilon$ is the expansion parameter\index{expansion parameter} and \pert{X}{n} is the $n$-th order perturbation of $X$. We identify the background value \pert{X}{0}, often indicated also as $\overline{X}$, as the value that $X$ would have if the Universe were perfectly homogeneous; this is why it depends on cosmic time alone. The other terms in the expansion form the perturbed part of $X$, which is by definition inhomogeneous and thus depends on both time and position. The first-order term, $\pert{X}{1}(\tx)$, is usually called the linear term. 

When $\epsilon$ is smaller than unity, the sum can be truncated at a certain order $n$ to yield $X$ up to the $n$-th order:
\begin{align}
  X(\tx) \,\simeq\, \pert{X}{0}(t) + \epsilon\,\pert{X}{1}(\tx) + \ldots + \epsilon^n\,\pert{X}{n}(\tx) \;.
\end{align}

For the sake of readability, we absorb the expansion parameter $\epsilon$ in the perturbed variables by setting $\epsilon^n\pert{X}{n}\,\rightarrow\,\pert{X}{n}$. For the same reason, we shall often omit to specify the space-time dependence of the perturbations.

\runinhead{First-order perturbations} The observed isotropy of the CMB suggests that in the early Universe ($z>1000$) the perturbations had an amplitude $10^{5}$ times smaller than the background. It is then an excellent approximation to truncate the sum at linear order
\begin{align}
  X(\tx) \,\simeq\, \pert{X}{0}(t) + \pert{X}{1}(\tx) \;.
\end{align}
At later times, the CMB stays linear because, as we pointed out in the introduction to the chapter, the photon perturbations do not grow with time. We are then justified in using the linearised equations to describe most of the CMB physics all the way to today.

\runinhead{Second-order perturbations} There are, however, important effects in the CMB that cannot be predicted by linear perturbation theory. In particular, by employing a first-order approach, one would ignore all the complexity in the non-linear structure of the Einstein and Boltzmann equations. Unless the primordial perturbations are non-Gaussian to start with, doing so ultimately yields to a vanishing 3-point function for the CMB. Hence, in order to study the generation of non-Gaussianity, we shall expand all variables up to second order according to
\begin{align}
  X(\tx) \,\simeq\, \pert{X}{0}(t) + \pert{X}{1}(\tx) + \pert{X}{2}(\tx) \;.
\end{align}



\subsection{Perturbing functions}
\label{sec:perturbing_functions}

The most common exercise in perturbation theory is to expand a perturbed variable inside a function or an equation. A simple but relevant case is the product of two perturbations
\begin{align*}
  X\,Y = \left(\pert{X}{0}+\pert{X}{1}+\pert{X}{2}+\dotsb\right)\,
  \left(\pert{Y}{0}+\pert{Y}{1}+\pert{Y}{2}+\dotsb\right) \;,
\end{align*}
that is easily split into orders:
\begin{align}
  \nonumber\pert{(XY)}{0} &= \pert{X}{0} \pert{Y}{0} \\
  \nonumber\pert{(XY)}{1} &= \pert{X}{0} \pert{Y}{1} + \pert{X}{1} \pert{Y}{0}\\
  \pert{(XY)}{2} &= \pert{X}{0} \pert{Y}{2} + \pert{X}{2} \pert{Y}{0} + \pert{X}{1} \pert{Y}{1} \;,
\end{align}
and so on. The above expansion shows that perturbation theory is ``verbose'' in the sense that it produces long equations; even stopping at second order, a simple product yields 6 terms. However, many perturbations have a vanishing background value. This is the case of all 3-vectors, including velocity, because if they had a background value they would violate the requirement of homogeneity and isotropy. When $\pert{X}{0} = \pert{Y}{0} = 0$, the product $XY$ simplifies to 
\begin{align}
  \nonumber\pert{(XY)}{0} &= 0 \\
  \nonumber\pert{(XY)}{1} &= 0 \\
  \pert{(XY)}{2} &= \pert{X}{1} \pert{Y}{1} \;.
\end{align}

A generic function of the perturbed variable $X$ can be Taylor expanded around $\pert{X}{0}\equiv\overline{X}\,$ as
\begin{align}
  f(X) \;\simeq\; f(\overline{X}) \,+\, \diffp*{f}{X}{\overline{X}}\!\!(X-\overline{X}) \,+\, \frac{1}{2}\,\diffp*{f}{{X^2}}{\overline{X}}\!\!(X-\overline{X})^2\;.
\end{align}
If we also expand $X \simeq \pert{X}{0} + \pert{X}{1} + \pert{X}{2}$ and split $f(X)$ into orders, we obtain
\begin{align}
  \label{eq:perturbations_function_of_x}
  \pert{f(X)}{0} &= f(\overline{X}) \phantom{\diffp{f}{x}} \nmsk
  \pert{f(X)}{1} &= \diffp*{f}{X}{\overline{X}}\!\!\pert{X}{1} \nmsk
  \pert{f(X)}{2} &= \diffp*{f}{X}{\overline{X}}\!\!\pert{X}{2} \,+\, \frac{1}{2}\,\diffp*{f}{{X^2}}{\overline{X}}\!\!\pert{X}{1}\pert{X}{1} \;.
\end{align}
Two useful examples are $(1+x)^\alpha$ and $e^{\,x}$, with $x=(X-\overline{X})/\overline{X}$, which are expanded up to second order as
\begin{align}
  \label{eq:perturbations_fourier_1_plus_x}
  (1+x)^\alpha \;\simeq\;
  1 +
  \alpha\,\pert{x}{1} +
  \alpha\,\pert{x}{2} + \frac{\alpha(\alpha-1)}{2}\,\pert{x}{1}\,\pert{x}{1} \;
\end{align}
and
\begin{align}
  \label{eq:perturbations_fourier_exp_x}
  e^{\,x} \;\simeq\;
  1 +
  \pert{x}{1} +
  \pert{x}{2} + \frac{1}{2}\,\pert{x}{1}\,\pert{x}{1} \;.
\end{align}
In particular, we have that
\begin{align}
  &\sqrt{1+2\,x} \;\simeq\; 1\,+\,\pert{x}{1}\,+\,\pert{x}{2}
  \,-\,\frac{1}{2}\,\pert{x}{1}\,\pert{x}{1} \;,\nmsk
  &\frac{1}{\sqrt{1+2\,x}} \;\simeq\; 1\,-\,\pert{x}{1}\,-\,\pert{x}{2}
  \,+\,\frac{3}{2}\,\pert{x}{1}\,\pert{x}{1} \;.
  \label{eq:perturbations_expansions}
\end{align}


\subsection{Perturbing equations}
\label{sec:perturbing_equations}

The main advantage of perturbation theory is that the perturbed equations can be solved order by order. An equation is split into a background part, a first-order part, a second-order part and so on. The equation for the $n$-th order is solved using the solutions for the preceding orders, from the $(n-1)$-th order all the way to the $0$-th order, or background, solution. The solution for the $(n+1)$-th order is not needed because it is negligible with respect to the $n$-th order one.

The last line of \eref{eq:perturbations_function_of_x} implies that a second-order equation can be always split in a \keyword{purely second-order part}, which is linear in the second-order perturbations, and in a \keyword{quadratic part}, involving the product of first-order perturbations. The purely second-order part, as can be seen from the second line of \eref{eq:perturbations_function_of_x}, has the same structure of the linearised equation.

In this and in the next chapter, we shall expand the Boltzmann and Einstein equations up to the second order in the cosmological perturbations. This will result in a system of coupled ordinary differential equations (ODEs) where the time evolution of the second-order variables is the unknown. The quadratic part of each equation, whose evolution is known from the solution of the first-order system, acts as a time-dependent source term for the second-order structure. If these \keyword{quadratic sources} are neglected, the second-order system is equivalent to the first-order one. This is an important property of perturbation theory that generalises to any order: a perturbed system of equations at the $n$-th order, as intimidating as it may look, has the same structure as the linear system with the addition of extra sources that are known from solving the previous orders.


\runinhead{Conventions} In the following, we shall ofter refer to the equations at second perturbative order simply as ``second-order equations''. These should not to be confused with the second-order differential equations, which instead we shall always call with their full name\footnote{The ambiguity is minimal also because we shall almost always solve first-order differential equations. The only second-order differential equation we shall deal with is the one for the tensor modes of the metric, $\gamma_{[\pm2]}$.}. Furthermore, we shall often omit showing the perturbative order in our expressions. There is no ambiguity in doing so because we never go beyond second order; a quadratic term will always be made of two first-order perturbations while a term which is alone is necessarily a purely second-order variable.

\section{The perturbed metric}
\label{sec:perturbations_metric}

We parametrise the metric as
\begin{align}
  \label{eq:perturbations_the_metric}
  \dd s^2 \;=\; a^2(\tau) \, \left\{ - (1+2\Psi) \dd \tau^2
    + 2\,\omega_i\,\dd x^i \dd \tau
  + \,\left[\,(1-2\Phi) \delta_{ij} + 2\,\gamma_{ij} \,\right]\, \dd x^i \dd x^j \right\} \;,
\end{align}
where the variables $\Psi$, $\Phi$, $\omega^i$ and $\gamma_{ij}$ are perturbations with vanishing background value. Since $\gamma_{ij}$ is by construction traceless and symmetric, the perturbed variables contain $10$ independent components ($1+1+3+5$, respectively) as expected from a symmetric space-time tensor. By expanding the above metric according to $g_{\mu\nu} \,\simeq\, \pert{g_{\mu\nu}}{0} \,+\, \pert{g_{\mu\nu}}{1} \,+\, \pert{g_{\mu\nu}}{2}$, we see that its background value is given by the homogeneous flat \FLRW metric in \eref{eq:conformal_metric_flrw}. Note that we are assuming a vanishing spatial curvature of the Universe at the background level, $k=0$, as suggested by the observations of the cosmic microwave background and of other geometrical probes \cite{hinshaw:2012a, planck-collaboration:2013a, planck-collaboration:2013c}; for a discussion of perturbations on a curved background, refer to \eg \citet{hu:1998a,zaldarriaga:1998b, lewis:2000a}.

The first and second-order parts of the metric each have $10$ independent components whose time-evolution is given by the second-order Einstein equations. In \sref{sec:perturbations_svt_decomposition} we shall split these components in scalar, vector and tensor parts that evolve independently by virtue of the decomposition theorem. In \sref{sec:perturbations_gauge} we shall show how the $10$ components can be cut down to only $6$ degrees of freedom by picking a specific gauge; in this work we choose to use the Newtonian gauge.

\subsection{Scalar-Vector-Tensor decomposition} 
\label{sec:perturbations_svt_decomposition}

Under a spatial coordinate transformation $x^{\,i} \rightarrow \tilde{x}^{\,i} = \tilde{x}^{\,i}\,(x^1,x^2,x^3) $, the components of a space-time tensor $T$ transform as a 3-scalar ($T_{00}$), a 3-vector ($T_{0i}$) and a 3-tensor ($T_{ij}$). This follows directly from the tensor transformation rule,
\begin{align}
  \tilde{T}_{\mu\nu}
  \;=\; \frac{\partial\,{x}^\alpha}{\partial \tilde{x}^\mu}\,\,
  \frac{\partial\,{x}^\beta}{\partial \tilde{x}^\nu}\,\,T_{\alpha\beta} \;,
\end{align}
after noting that, for a spatial transformation, $\partial{x}^i/\partial \tilde{x}^0=\partial{x}^0/\partial \tilde{x}^i=0$.
\annotate{This can be restated as: the parts of the 4D tensor transform as a scalar, a vector and a tensor on spatial hypersurfaces \cite{malik:2009a}.}

The split is not complete, though, as the $3$ components of $T_{i0}$ and the $6$ independent components of $T_{ij}$ still are a mixture of scalar, vector and tensor degrees of freedom. These can be extracted in a systematic way by using the projection vectors $\xi^i_{[m]}$ and matrices $\chimatrix{2}{m}{ij}$, which we detail in Appendix \ref{app:sphere_projection}. The contraction 
\begin{align}
  \chimatrix{2}{m}{ij}\,\DD{T}{i}{j}
\end{align}
yields an $m$-dependent object that represents the scalar ($m=0$), vector ($m=\pm1$) and tensor ($m=\pm2$) components of \DD{T}{i}{j}. The remaining scalar component of $T_{ij}$ is in the trace,
\begin{align}
  \frac{\KronUD{i}{j}\,T_{ij}}{3} \;.
\end{align}
Similarly, the vector \DD{T}{i}{0} can be contracted with the vectors \xivector{m}{i},
\begin{align}
  \xivector{m}{i}\,\DD{T}{i}{0}
\end{align}
to yield one scalar component ($m=0$) and two vector ones ($m=\pm1$). To sum up, any symmetric space-time tensor $T$ can be decomposed into 4 scalar, 4 vector and 2 tensor components according to the following scheme:

\begin{table}[h]
  \taburulecolor{Blue}
  \arrayrulewidth=0.4mm
  \extrarowsep=2mm
  \begin{tabu} to \textwidth {X[2,l,m]| X[1,l,m] X[1,l,m] X[1,l,m] X[-1,l,m]}
    $m=0$ (scalar)    & \chimatrix{2}{0}{ij}\,\DD{T}{i}{j}    & \xivector{0}{i}\,\DD{T}{i}{0} & \KronUD{i}{j}\,\DD{T}{i}{j}/3 & \DD{T}{0}{0} \\
    \hline
    $m=\pm1$ (vector) & \chimatrix{2}{\pm1}{ij}\,\DD{T}{i}{j} & \xivector{\pm1}{i}\,\DD{T}{i}{0} &  & \\
    \hline
    $m=\pm2$ (tensor) & \chimatrix{2}{\pm2}{ij}\,\DD{T}{i}{j} &  &  & 
  \end{tabu}
\end{table}

\noindent This separation is called the \keyword{scalar-vector-tensor (SVT) decomposition}\index{SVT decomposition}.  In the following, we shall use $m=0$, $m=\pm1$ and $m=\pm2$ as shorthands for scalar, vector and tensor degrees of freedom, respectively. We shall collectively refer to them as \keyword{azimuthal modes} as they are ultimately connected to the $m$ index in the spherical harmonic $Y_{lm}$. For further details, refer to Appendix \ref{app:sphere_projection}.

The metric is decomposed in its SVT components in the same way. After defining
\begin{align}
  \gamma_{[m]} \,\equiv\, \chimatrix{2}{m}{ij}\,\DD{g}{i}{j}
\end{align}
and
\begin{align}
  \omega_{[m]} \,\equiv\, \xivector{m}{i}\,\DD{g}{i}{0} \;,
\end{align}
it is straightforward to see that $\Phi$, $\Psi$, $\gamma_{[0]}$ and $\omega_{[0]}$ are the scalar components of the metric, $\gamma_{[\pm1]}$ and $\omega_{[\pm1]}$ are the vector ones and $\gamma_{[\pm2]}$ are the tensor ones.


\subsection{The decomposition theorem}
\label{sec:decomposition_theorem}

In the following chapters, we shall decompose the Einstein and Boltzmann equations into azimuthal modes by contracting them with the projection vectors $\xi^i_{[m]}$ and matrices $\chimatrix{2}{m}{ij}\,$.
The main advantage of doing so is that, at first order, the resulting differential system will be decoupled in its scalar ($m=0$), vector ($m=\pm1$) and tensor ($m=\pm2$) components.
For example, the Einstein equations that dictate the evolution of the scalar modes will not contain either the vector or the tensor degrees of freedom. Similarly, the evolution of $\gamma_{[\pm1]}$ will be completely decoupled from $\gamma_{[\pm2]}$, and, since $\gamma_{[\pm2]}$ is the only tensor degree of freedom, its evolution will not involve any other metric perturbation.
This separation in the evolution of different $m$-modes is called the \keyword{decomposition theorem}, and is widely used at first order (see, for example, Appendix B of Ref.~\cite{kodama:1984a}, Sec.~4.2 of Ref.~\cite{bertschinger:1996a} and Sec.~3.2 of Ref.~\cite{knobel:2012a}) as it considerably simplifies the treatment of the vector and tensor perturbations. In particular, from the numerical point of view, the decomposition theorem allows to solve three simple differential systems, one for each of the considered modes, rather than a single one where the modes are coupled in a complicated way.

At second and higher order, the decomposition theorem does not hold anymore, because the various azimuthal modes mix and source each other. This SVT mixing is a direct consequence of the non-linear structure of the quadratic sources, as we shall show explicitly in \sref{sec:perturbations_energy_momentum_tensor} for the energy-momentum tensor and in \sref{sec:spherical_projection_of_functions} for the Boltzmann equation.
Nonetheless, it is still possible to solve the second-order Boltzmann-Einstein system separately for each $m$-mode. In fact, the linear structure of the second-order system coincides with that of the first-order one (\sref{sec:perturbing_equations}), and it is therefore decoupled in $m$. The internal structure of the quadratic sources still couples different $m$-modes but, since the sources are known from the solution of the first-order system, they can be precomputed without interfering with the evolution of the second-order system, which can thus be solved separately for each $m$.




\subsection{Gauge choice}
\label{sec:perturbations_gauge}

The split of the metric into background and perturbed parts implies the presence of two separate manifolds, namely the background and perturbed spacetimes.
To compare the two metrics and perform the usual tensorial operations such as addition and subtraction, it is therefore required to define a correspondence between the points of the two aforementioned manifolds.
A gauge transformation is exactly that: an infinitesimal, invertible diffeomorphism that relates the points in the background manifold with those in the perturbed one\footnote{For details on the definition of a gauge transformation (and on its active and passive interpretations), refer to Refs.~\cite{malik:2008a, mukhanov:1992a, bruni:1997a}. See Refs.~\cite{bruni:1997a, malik:2008a, malik:2009a, nakamura:2011a} for details on gauge transformations in a second-order context. See also Sec.~3.1.1 and 3.4 of Ref~\cite{knobel:2012a} for a pedagogical approach to gauge transformations. Finally, we refer to Refs.~\cite{naruko:2013a, pitrou:2009a} for a discussion of the gauge invariance of the second-order Boltzmann equation.}.
Because the theory of general relativity is diffeomorphism invariant, there is no preferred gauge; the perturbations themselves, however, are gauge dependent.



While all the gauges are theoretically equivalent, one gauge choice might be better suited than another depending on the problem at hand. Historically, many different gauges have been used to study the cosmological perturbations; a list can be found in Sec.~7 of \citet{malik:2009a}.
In this work and in \SONG, we choose the \emph{Newtonian} or \keyword{Poisson gauge}\index{Newtonian gauge} \cite{bertschinger:1996a} whereby both the $\,g_{0i}\,$ and $\,g_{ij}\,$ perturbations are transverse or, in terms of the metric variables in \eref{eq:perturbations_the_metric},
\begin{align}
  \partial^i\,\omega_i \;=\; 0 \qquad\text{and}\qquad
  \partial^j\gamma_{ij} \;=\; 0 \;.
\end{align}
We shall see that, in Fourier space and for \k configurations along the polar axis, this choice is equivalent to setting $\,\omega_{[0]}=0\,$ and $\,\gamma_{[0]}=\gamma_{[\pm1]}=0\,$.
It follows that in the Poisson gauge there are two scalar potentials ($\Phi$ and $\Psi$), one transverse vector potential ($\omega$), and one transverse-traceless tensor potential ($\gamma$), for a total of 6 degrees of freedom.

Another popular gauge choice is the \keyword{synchronous gauge} \cite{bertschinger:1996a}, whereby the perturbations are confined to the spatial part of the metric:
\begin{align}
  \Psi \;=\; 0 \qquad\text{and}\qquad
  \omega_i\;=\;0 \;.  
\end{align}
The synchronous gauge, however, leads to a more complicated angular dependence in the Boltzmann equation at second order, which contains terms that are cubic and quartic in the photon's direction, $n^{(i)}\,$ (see Eq.~3.29 of Ref.~\cite{naruko:2013a}). The multipole expansion of these terms is much more complicated than that of the equivalent ones in Newtonian gauge, which are at most quadratic in $n^{(i)}\,$ (see Appendix~\ref{app:sphere_projection}).
Nonetheless, it is our intention to implement the synchronous gauge in \SONG at a later stage, for two reasons. First, verifying that the observables such as the bispectrum do not depend on the gauge would be an important check of the implementation of the differential system and of the line of sight sources (see \cref{ch:evolution}).
Secondly, we could further test our transfer functions by making use of the gauge transformation between the Newtonian and synchronous gauges up to second order, which can be found in \citet{bruni:1997a}.


\runinhead{The exponential metric}
Another way to express the metric in Newtonian gauge is using exponentials, as it is done in, \eg, Ref.~\cite{bartolo:2006a, senatore:2009b, maldacena:2003a}:
\begin{align}
  \dd s^2 \;=\; a^2(\tau) \, \left[
  -e^{2\Psi_e}\,\dd\tau^2\,
  +\,2\,\omega_i\,\dd x^i \dd\tau\,
  +\,\left(\,e^{-2\Phi_e}\,\delta_{ij} \,+\, 2\,\gamma_{ij}\,\right)\,\dd x^i \dd x^j \right] \;,
  \label{eq:bmr_metric}
\end{align}
where the suffix `$e$' serves the purpose to distinguish the potentials thus defined from the ones in the usual metric in \eref{eq:perturbations_the_metric}. The resulting equations are slightly simpler due to the properties of the exponential, especially for the Liouville term in the Boltzmann equation. After expanding the two metrics up to the second order and equating them ($1+2\Psi=e^{2\Psi_e}$ and $1-2\Phi=e^{-\Phi_e}$), it is clear that the $\Psi$ and $\Phi$ potentials in the two representations differ only at the second-order level:
\begin{align}
  \Psi \;=\; \Psi_e\,(1\,+\,\Psi_e) \quad\quad\text{and}\quad\quad \Phi \;=\; \Phi_e\,(1\,-\,\Phi_e) \;.
  \label{eq:bmr_ours_potential}
\end{align}
In particular, the following equalities hold that are useful for computations that involve the tetrad (\sref{sec:tetrad_formalism}):
\begin{align}
  & \sqrt{1+2\Psi} \;=\; e^{\Psi_e} \;\;,
  && \frac{1}{\sqrt{1+2\Psi}} \;=\; e^{-\Psi_e} \;,\nmsk
  & \sqrt{1-2\Phi} \;=\; e^{-\Phi_e} \;\;,
  && \frac{1}{\sqrt{1-2\Phi}} \;=\; e^{\Phi_e} \;.
  \label{eq:bmr_ours_potential_sqrt}
\end{align}
In the computations that follow we always use the metric in \eref{eq:perturbations_the_metric}; we refer to the ``exponential'' metric only to compare our results with the ones in the literature.

\runinhead{Relation with the literature}
In \cref{ch:boltzmann}, we will often refer to the results found in the second-order literature. Here we provide the rules to convert from our metric variables to those adopted by the following authors:
\begin{itemize}
  \item Beneke \& Fidler \cite{beneke:2010a,beneke:2011a}:
  \begin{align}
    A^\sub{BF} \,=\, \Psi\;, \quad\; D^\sub{BF} \,=\, -\Phi\;,
    \quad\; B^\sub{BF}_i \,=\, -\omega_i\;, \quad\; E^\sub{BF}_{ij} \,=\, \gamma_{ij} \;;
    \label{eq:beneke_fidler_metric}
  \end{align}
  \item Pitrou et al. \cite{pitrou:2009a,pitrou:2010a}:
  \begin{align}
    \Phi^\sub{P} \,=\, \Psi\;, \quad\; \Psi^\sub{P} \,=\, \Phi\;,
    \quad\; B^\sub{P}_i \,=\, \omega_i\;, \quad\; H^\sub{P}_{ij} \,=\, \gamma_{ij} \;;
  \end{align}
  \item Senatore et al. \cite{senatore:2009b,senatore:2009a} (see also \eref{eq:bmr_ours_potential}):
  \begin{align}
    \Psi^\sub{S} \,=\, \Psi\,(1-\Psi)\;, \quad\; \Phi^\sub{S} \,=\, \Phi\,(1-\Phi)\;,
    \quad\; \omega^\sub{S}_i \,=\, \omega_i\;, \quad\; \chi^\sub{S}_{ij} \,=\, 2\,\gamma_{ij} \;;
  \end{align}
  \item Bartolo, Matarrese \& Riotto \cite{bartolo:2006a, bartolo:2007a, nitta:2009a} (see also \eref{eq:bmr_ours_potential}):
  \begin{align}
    \Phi^\sub{B} \,=\, \Psi\,(1-\Psi)\;, \quad\; \Psi^\sub{B} \,=\, \Phi\,(1-\Phi)\;,
    \quad\; \omega^\sub{B}_i \,=\, \omega_i\;, \quad\; \chi^\sub{B}_{ij} \,=\, 2\,\gamma_{ij} \;.
  \end{align}
\end{itemize}

\section{Statistical description of the perturbations}
\label{sec:stochastic_perturbations}

According to the mechanism of cosmic inflation, the structure that we observe in the CMB and in the galaxy distribution is due to quantum-mechanical fluctuations that were set soon after the Big Bang. Due to the stochastic nature of quantum processes, the Universe should be considered as just one of the potential outcomes of a statistical ensemble of realisations that could have arisen from inflation. Since all stochastic processes have a variance, any two realisations differ and, if we were to live in a realisation different from ours, we would observe a different sky. This intrinsic discrepancy between what is accessible by observations and the underlying description of the perturbations is called \keyword{cosmic variance}.


Because of their stochastic nature, we shall treat the cosmological perturbations as \indexword{random fields}. In the next subsection, we shall detail the properties of random fields and characterise them in terms of their connected correlation functions. In \sref{sec:fair_sample_hypothesis} we shall introduce the concepts of statistical homogeneity and isotropy, and briefly discuss how to relate the abstract idea of an ensemble of realisations to the observable Universe. In \sref{sec:gaussian_random_fields}, we shall discuss the Gaussian random fields, which are particularly important in the study of the cosmological perturbations, and give details on their two-point correlation function.

\subsection{Random fields}
\label{sec:random_fields}

A random field, $\R(\vecx)$, is a set of random variables, one for each points in space, characterised by a probability functional, $\probab[\hat\R(\vecx)]$, which specifies the probability for the occurrence of a particular realisation of the field. A realisation of the field, $\hat\R(\vecx)$, is a deterministic\footnote{Here and in the following, we shall use the adjective ``deterministic'' to mean non-stochastic, non-random.} function of position, $\vecx$, that represents one of the possible outcomes of the random field \cite{porciani:2009a}.

The main difference between a random field, $\R(\vecx)$, and a set of random variables, $r_i$, is that the former is continuous. The PDF of the field is therefore expressed as a functional of one realisation, $\probab[\hat\R(\vecx)]$, rather than a function of the discrete set of random variables, $\probab(r_1,\dotsc,r_n)$. Accordingly, the expectation value of any functional, $\mathcal{F}(\R[\vecx])$, is obtained by a functional convolution with the PDF:
\begin{align}
  \avg{\mathcal{F}[\R(\vecx)]} \;=\; \int\ddRhat\;\probab[\hat\R(\vecx)]\;\mathcal{F}[\hat\R(\vecx)] \;,
  \label{eq:random_field_expectation}
\end{align}
where $\int\ddRhat$ stands for the product of the integrals at each space point \vecx \cite{kleinert2:2001a,zinn-justin:2010a},
\begin{align}
  \int\ddRhat \,\equiv\, \int\prod\limits_\vecx\dd\hat\R(\vecx) \;.
\end{align}
As an example, consider the functional $\mathcal{F}[\R]=\R(\vecxp)$, which is the value of the field at a given position \vecxp. The expectation value of $\R(\vecxp)$ is given by the value of the field in \vecxp averaged over the infinite ensemble of possible realisations of the field. This way of averaging is impossible to do in practice, since observations can only probe the single realisation we live in; in \sref{sec:fair_sample_hypothesis} we shall see that we can still relate these abstract averages with the observed quantities by assuming the fair sample hypothesis.

The cosmological perturbations are usually described by either 2D or 3D random fields. The temperature of the CMB, for example, is modelled as a two-dimensional random field, $T(\n)$, because all CMB photons were emitted from the last scattering surface, whose distance does not depend significantly on the direction of observation. On the other hand, the density of the cold dark matter component can be observationally traced by measuring the redshift of galaxies at various distances, and thus is described by a three-dimensional random field, $\rho(\vecx)$. In this section we shall not specify a dimension, so that the obtained results shall be general. It is also important to remember that all the cosmological perturbations also have a time dependence, \eg $T=T(\n,\tau)$ and $\rho=\rho(\vecx,\tau)$, that we shall often omit for clarity.

It is convenient to define the cosmological perturbations as zero-mean quantities. For example, rather than dealing with mass densities, $\rho(\vecx,\tau)$, we define the \emph{fractional \indexword{overdensity}} or \emph{\indexword{density contrast}} field as
\begin{align}
  \delta(\vecx,\tau) \;\equiv\; \frac{\rho(\vecx,\tau)-\bar{\rho}(\tau)}{\bar{\rho}(\tau)} \;,
\end{align}
where $\bar{\rho}=\avg{\rho(\vecx,\tau)}$, so that $\avg{\delta(\vecx,\tau)}=0$, and the field of \emph{temperature fluctuations} as
\begin{align}
  \Theta(\vecx,\tau) \;\equiv\; \frac{T(\vecx,\tau)-\bar{T}(\tau)}{\bar{T}(\tau)} \;,
  \label{eq:temperature_fluctuations_def_tbar}
\end{align}
where $\bar{T}=\avg{T(\vecx,\tau)}$, so that $\avg{\Theta(\vecx,\tau)}=0$. In writing the definitions above, we have implicitly set the average value of the fields not to depend on position, $\avg{\rho(\vecx,\tau)}=\bar\rho(\tau)$ and $\avg{T(\vecx,\tau)}=\bar T(\tau)$. As we shall see in \sref{sec:fair_sample_hypothesis}, this is justified by the requirement of statistical homogeneity.
\annotate{Note that you cannot write $\avg{T}=\pert{T}{0}$ because how can you say that $\avg{\pert{T}{1}}=0$? Backreaction problem...}

\subsubsection{The $n$-point functions and the partition functional}
A simple way to characterise a random field is through its $n$-point functions, that is the expectation values of the product of $n$ perturbations in different positions and at the same time $\tau$,
\begin{align}
  \label{eq:n-point_functions}
  \avg{\R(\vec{x_1})\,\dotsc\,\R(\vec{x_n})}
  \;\equiv\; \int \ddR\;\probab[\R(\vecx)]\;\R(\xone)\,\dotsc\,\R(\vec{x_n}) \;.
\end{align}
For a completely uncorrelated random field, the probability is given by $\probab[\R(\vecx)] = \prod\limits_\vecx \probab(\R(\vecx))$ and the $n$-point functions reduce to products of one-point functions.

The $n$-point functions in \eref{eq:n-point_functions} can be defined in terms of the partition functional,
\begin{align}
  \mathcal{Z}[f(\vecx)]
  \;&=\; \avg{\exp\left[\int\dd\vecxp\;\R(\vecxp)\,f(\vecxp)\right]} \nmsk
  \;&=\; \int \ddR\;\probab[\R(\vecx)]\;\exp\left[\int\dd\vecxp\;\R(\vecxp)\,f(\vecxp)\right] \;,
\end{align}
where $f(\vecx)$ is a realisation.
The partition functional is the generalisation to the continuum of the characteristic function of a discrete set of random variables $\vec r$,
\begin{align}
  C_{\vec r}\,(\vec{b}) \;=\; \avg{e^{\,\scalarP{b}{r}}} \;=\; \int \dd\vec r\;\probab(\vec{r})\;e^{\,\scalarP{b}{r}} \;.
  \label{eq:characteristic_function}
\end{align}
The realisation $\R(\vecx)$ plays the role of the vector $r_i$ and the location $\vec x$ the role of the index $i$ \cite{zinn-justin:2010a}.
\annotate{From comment to eq 2.42 in \citet{zinn-justin:2010a}.}
Taking the derivatives of the characteristic function with respect to the components of $\vec b$ directly yields the moments of the distribution,
\begin{align}
  \avg{r_{k_1}\dotsc r_{k_n}} \;=\; \diffp{}{{\,b_{k_1}}}\,\dotsb\,\diffp{}{{\,b_{k_n}}}\;
  C_{\vec r}(\vec{b})\bigg|_{\vec b=0} \;.
\end{align}
Similarly, functional differentiation can be used to obtain the $n$-point functions from the partition functional,
\begin{align}
  \avg{\R(\xone)\dots\R(\vec{x_n})} \;=\;
  \frac{\delta}{\delta f(\xone)}\,\dotsb\,\frac{\delta}{\delta f(\vec{x_n})}\;\mathcal{Z}[f(\vecx)]\bigg|_{f=0} \;,
\end{align}
where we have used the property of the functional differentiation,
\begin{align}
  \frac{\delta}{\delta f(\vec{x_i})}\,\int\dd\vecx\;\R(\vecx)\,f(\vecx) \;=\; \R(\vec{x_i}) \;.
\end{align}
Thus, the $n$-point functions are just the MacLaurin coefficients of the partition functional.

\subsubsection{The connected functions}
The $n$-point functions, also known as the \indexword{disconnected correlation functions}, are not the only way to characterise a random field. It is sometimes convenient to use the \indexword{connected correlation functions}, which are defined as the Maclaurin coefficients of the logarithm of the partition functional,\footnote{Note that we use commas to separate the variables in $\,\con{\R(\xone),\dotsc,\R(\vec{x_n})}\,$ to make it clear that the connected functions are not obtained as the average of a product of random fields.}
\begin{align}
  \con{\R(\xone),\dotsc,\R(\vec{x_n})} \;\equiv\;
  \frac{\delta}{\delta f(\xone)}\,\dotsb\,\frac{\delta}{\delta f(\vec{x_n})}\;
  \ln\mathcal{Z}[f(\vecx)]\bigg|_{f=0} \;.
  \label{eq:connected_functions}
\end{align}
The connected functions are the generalisation of the cumulants of a discrete set of variables, just as the $n$-point functions are the generalisations of the non-central moments.

The main advantage of the connected functions is that they vanish if any of their arguments are independent. To prove this, let us assume that the space where $\vecx$ lives can be divided into two sets, $\mathcal{X}_1$ and $\mathcal{X}_2$, where the random field is causally disconnected. We can then think of the random field as being described by two disjoint probability distribution functionals, one for the points in $\mathcal{X}_1$ and another for those in $\mathcal{X}_2$,
\begin{align}
  \probab[\R(\vecx)] \;=\; \probab[\R(\vecx)]_{\mathcal{X}_1}\,\times\,\probab[\R(\vecx)]_{\mathcal{X}_2} \;.
\end{align}
The probability measure is separable, too,
\begin{align}
  \int\ddR \;=\;
  \int\prod\limits_{\vecx\in\mathcal{X}_1}\dd\R(\vecx) \,\times\,
  \int\prod\limits_{\vecx\in\mathcal{X}_2}\dd\R(\vecx) \;,
\end{align}
which, after using \eref{eq:n-point_functions}, implies that the $n$-point functions, $\,\avg{\R(\xone)\dotsc\R(\vec{x_n})}\,$, break down according to whether the points belong to $\mathcal{X}_1$ or $\mathcal{X}_2$. For example, if $\xone, \xtre \in \mathcal{X}_1$ and $\xtwo \in \mathcal{X}_2$, we obtain
\begin{align}
  &\avg{\R(\xone)\,\R(\xtwo)} \,=\, \avg{\R(\xone)}\avg{\R(\xtwo)} \nmsk
  &\avg{\R(\xone)\,\R(\xtwo)\,\R(\xtre)} \,=\, \avg{\R(\xone)\R(\xtre)}\avg{\R(\xtwo)} \;.
\end{align}

The connected functions are a different story. Because the scalar product behaves linearly,
\begin{align}
  \int\dd\vecx\;\R(\vecx)\,f(\vecx)
  \,=\, \int_{\mathcal{X}_1}\dd\vecx\;\R(\vecx)\,f(\vecx)
  \,+\, \int_{\mathcal{X}_2}\dd\vecx\;\R(\vecx)\,f(\vecx) \;,
\end{align}
we have that the partition function of $\R(\vecx)$ is given by the product
\begin{align}
  \mathcal{Z}[f(\vecx)] \,=\, \mathcal{Z}[f(\vecx)]_{\mathcal{X}_1}\,\mathcal{Z}[f(\vecx)]_{\mathcal{X}_2} \;.
\end{align}
The generating function for the connected correlation functions is the logarithm of $\mathcal{Z}$:
\begin{align}
  \ln\mathcal{Z}[f(\vecx)] \,=\, \ln\mathcal{Z}[f(\vecx)]_{\mathcal{X}_1}
  \;+\; \ln\mathcal{Z}[f(\vecx)]_{\mathcal{X}_2} \;.
\end{align}
By virtue of the definition of connected correlation functions in \eref{eq:connected_functions}, we have that
\begin{align}
  \con{\R(\xone),\,\dotsc\,,\R(\vec{x_n})}\,
  &=\,\frac{\delta}{\delta f(\xone)}\,\dotsb\,\frac{\delta}{\delta f(\vec{x_n})}
    \;\ln\mathcal{Z}[f(\vecx)]_{\mathcal{X}_1}\bigg|_{f=0} \nmsk
  &+\,\frac{\delta}{\delta f(\xone)}\,\dotsb\,\frac{\delta}{\delta f(\vec{x_n})}
    \;\ln\mathcal{Z}[f(\vecx)]_{\mathcal{X}_2}\bigg|_{f=0} \;.
\end{align}
This means that the connected functions vanish unless all of the points are either in $\mathcal{X}_1$ or in $\mathcal{X}_2$, simply because
\begin{align}
  \frac{\delta}{\delta f(\vec{x})}\,
  \int_{\mathcal{X}}\dd\vec{x}^\prime\;\R(\vec{x}^\prime)\,f(\vec{x}^\prime) \;=\; 0 \;
\end{align}
if $\vec{x}$ does not belong to $\mathcal{X}$.

We have proven that the connected correlation functions, $\,\avgc{\R(\xone),\dotsc,\R(\vec{x_n})}\,$, vanish if at least two points belong to casually disconnected regions (hence the adjective ``connected''). As a consequence, each independent region has its own set of connected correlations functions that, under the assumption of statistical homogeneity, coincide with those of any other region. One could say that each disconnected region behaves as a realisation within the realisation. This statement is particularly important for the cosmological study of the Universe, where a structure of disconnected regions arises naturally due to the finite speed of light; we shall treat the consequences of this statement in \sref{sec:fair_sample_hypothesis}.


\subsubsection{Wick's theorem}
Like the moments and the cumulants of a distribution, the disconnected and connected correlation functions of a random field are related by simple polynomial expressions. The coefficients of the polynomials can be determined by the repeated application of the chain rule to the logarithmic function in \eref{eq:connected_functions}; there is however a simpler version of the formula in terms of set partitions \cite{speed:1983a, rota:2000a}, which we report here:
\begin{align}
  \avg{\R_1\dotsb\R_n} \;=\; \sum\limits_\pi\,\prod\limits_{b\in\pi}\,\con{b}\;.
  \label{eq:connected_functions_set_partitions}
\end{align}
The sum goes over all the possible partitions $\pi$ of the set $\{\R_1,\dotsc,\R_n\}$, while the product goes over each block $b$ of the considered partition, and $\R_i$ stands for $\R(\vec{x_i})$. For example, the set $\{\R_1,\R_2\}$ has only two partitions: the one-block partition $\{\{\R_1,\R_2\}\}$ and the two-block partition $\{\{\R_1\},\{\R_2\}\}$; hence, the average of $\R_1\R_2$ includes two terms involving, respectively, one and two unconnected functions:
\begin{align}
  \avg{\R_1\R_2} \;=\; \con{\R_1,\R_2} \,+\, \con{\R_1}\con{\R_2} \;.
\end{align}
Since $\avg{\R_1}=\con{\R_1}\,$, the above formula tells us that $\con{\R_1,\R_2}$ is just the covariance of the field between $\R(\xone)$ and $\R(\xtwo)$.

The combinatorics formula in \eref{eq:connected_functions_set_partitions} is usually referred to as Wick's theorem and is widely used in particle physics to compute Feynman diagrams. Here, we use it to find the first four $n$-point functions for a zero-mean random field:
\begin{align}
  &\avg{\R_1} = \con{\R_1} = 0\nmsk
  &\avg{\R_1\R_2} = \con{\R_1,\R_2}  \nmsk
  &\avg{\R_1\R_2\R_3} = \con{\R_1,\R_2,\R_3} \nmsk
  &
  \begin{aligned}
    \avg{\R_1\R_2\R_3\R_4} &= \con{\R_1,\R_2,\R_3,\R_4} + \con{\R_1,\R_2}\con{\R_3,\R_4} \msk
    &+ \con{\R_1,\R_3}\con{\R_2,\R_4} + \con{\R_1,\R_4}\con{\R_2,\R_3} \;.
  \end{aligned}
  \label{eq:wick_example}
\end{align}
In this work we shall mostly deal with the two and three-point functions (that is, spectra and bispectra), which, as can be seen by the above expression, coincide with their corresponding connected functions. Sometimes, we will need to evaluate the four-point function of a Gaussian random field; in that case, all the connected functions apart from the covariance vanish, and we are left with
\begin{align}
  \avg{\R_1\R_2\R_3\R_4} = \avg{\R_1\R_2}\avg{\R_3\R_4} + \avg{\R_1\R_3}\avg{\R_2\R_4}
  + \avg{\R_1\R_4}\avg{\R_2\R_3} \;.
  \label{eq:four_point_function_wick}
\end{align}

\subsection{Statistical homogeneity and isotropy}
\label{sec:fair_sample_hypothesis}

The stochastic nature of the cosmological perturbations poses the problem of connecting the observations to the underlying theory. Theoretical investigation is only able to compute quantities averaged over the ensemble of possible realisations of the Universe, such as the $n$-functions in \eref{eq:n-point_functions}; it cannot predict the details of our peculiar realisation which is just the final outcome of a random process that took place during inflation. On the other hand, cosmological observations probe just a portion of the single realisation we live in; a measurement is always an average over a finite volume of some observable quantity. For example, cosmologists count the number of galaxies as a function of direction and redshift and then compute their correlation functions as an average over the probed volume. Similarly, the temperature of the CMB is averaged over all directions to obtain the angular power spectrum.

Observation can be still used to constrain the theory if the statistical properties of the Universe do not vary from region to region. Then, sampling different regions in our realisation is equivalent to sampling different realisations. Therefore, we can compensate the fact that we observe only one realisation of the Universe by observing as much Universe as we can.
In principle, if we could access arbitrary large regions of the Universe we would be able to probe the statistics of the primordial density fluctuations on any scale.  In practice, this is obviously not possible because the finite size of our past light cone still limits the maximum volume we can probe to $\sim(\unit[14]{Gpc})^3$.

We shall therefore demand that the random fields describing the cosmological perturbations are statistically homogeneous and isotropic.
\index{statistical homogeneity}A random field is statistically homogeneous if the joint probability distribution for any finite set of points is invariant under a spatial translation, that is
\begin{align}
  \probab\left(\R(\xone),\dotsc,\R(\vec{x_n})\right) \;=\;
  \probab\left(\R(\xone+\vecx),\dotsc,\R(\vec{x_n}+\vecx)\right) \;
\end{align}
for any $n$. This property, also called stationarity, is directly transferred to the $n$-point functions of the field; for instance, the homogeneity condition implies that $\avg{\R(\vecx)}$ is spatially independent and that $\avg{\R(\xone)\R(\xtwo)}$ is a function only of the relative separation, $\vecr\equiv\xtwo-\xone$.
\index{statistical isotropy}Statistical isotropy instead means invariance of the finite joint probability under a global rotation of its arguments. Thus, in a statistically isotropic and homogeneous Universe, $\avg{\R(\xone)\R(\xtwo)}$ depends solely on the distance, $r=\abs{\vecr}$, between \xtwo and \xone.

The statistical homogeneity and isotropy are far less stringent requirements than the cosmological principle, whereby all realisations must be perfectly homogeneous. The statistical version of the cosmological principle still allows for distant regions in the Universe to look different from each other, just because of the variance which is intrinsic in the stochastic nature of the perturbations. However, the variance itself should not depend on the location, and taking averages of different patches of the Universe should yield similar results.

\runinhead{Fair sample and ergodicity hypotheses}
The requirement of statistical homogeneity and isotropy is closely related to the \keyword{fair sample hypothesis}, whereby well separated regions of the Universe can be thought as being independent realisations of the underlying distribution; thus, spatial averages over many of such regions are equal to expectations over the ensemble \cite{peebles:1980a}. The fair sample hypothesis, which implies the statistical homogeneity and isotropy, provides an operational way to perform a volume average that is directly related to the ensemble average: first perform a volume average over a representative patch of the Universe, and secondly an average over many independent patches within your past light cone \cite{coles:2003a}. Another related hypothesis is that of \emph{ergodicity}\index{ergodic hypothesis}, whereby volume averages over the full extent of a realisation are equal to the expectations over the ensemble. Ergodicity is of less practical importance than the fair sample hypothesis because it requires averaging over an infinite volume; its advantage, however, is that it is automatically satisfied for all the homogeneous Gaussian fields with a continuous power spectrum \cite{adler:1981a}.

\subsection{Gaussian Random Fields}
\label{sec:gaussian_random_fields}

In the simplest scenario of cosmic inflation, the primordial perturbations are Gaussianly distributed and can therefore be described by \keyword{Gaussian random fields}. The probability distribution functional for one of such fields, \G, is given by
\begin{align}
  \probab[\G] \;=\; (\det K)^{1/2} \,
  \exp\left(-\frac{1}{2} \int\dd\xone\,\dd\xtwo\;\G(\xone)\;K(\xone,\xtwo)\;\G(\xtwo)\right) \;,
  \label{eq:pdf_gaussian_random_field}
\end{align}
where $K(\xone,\xtwo)$ is a symmetric, invertible operator. An important property of Gaussian random fields is that they are completely characterised by their two-point connected function, which we denote as $\xi(\xone,\xtwo)$ and is given by the functional inverse of $K$:
\begin{align}
  \xi(\xone,\xtwo) \;=\; K^{-1}(\xone,\xtwo) \;.
\end{align}
All the other connected functions vanish. This property greatly simplifies the task of deriving the $n$-point functions of the field, which can be expressed in terms of sums of products of $\xi$ by virtue of the Wick's theorem in \eref{eq:connected_functions_set_partitions}.

The fact that a Gaussian field is completely characterised by its two-point connected function is easily proven when considering a finite set of points, rather than a full realisation. The probability of measuring the finite number of values $\{g_1,\,\dotsc\,g_n\}$ in the space points $\{\xdep{n}\}$ is given by a multivariate Gaussian distribution:
\begin{align}
  \probab(g_1,\,\dotsc,\,g_n) = \frac{1}{\sqrt{(2\pi)^n\det K}}\,\exp \left( -\frac{1}{2}\,g_i\,K^{-1}_{ij}\,g_j \right),
  \label{eq:pdf_gaussian_random_field_discrete}
\end{align}
where $ K_{ij} = \avg{g_i\,g_j} $ is the (symmetric) covariance matrix. The above expression is just the discrete version of \eref{eq:pdf_gaussian_random_field}. Because the solution of the Gaussian integral with a linear term is analytically known, the characteristic function of $\probab$ is simply given by
\begin{align}
  C_\vec{g}(\vec{b}) \;=\; \exp\left( \frac{1}{2}\,b_i\,K_{ij}\,b_j \right) \;.
\end{align}
The joint cumulants, $\kappa$, of the random variables $\vec{g}$ can be obtained by differentiating $\ln C$ with respect to $\vec{b}$:
\begin{align}
  \kappa\,(x_1,\dotsc,x_n) \;=\; \diffp{}{{b_1}}\dotsb\diffp{}{{b_n}}\,\ln C_\vec{g}(\vec{b})\,\Bigg|_{\vec b=0}\;.
\end{align}
Since $\ln C$ is quadratic in $\vec b$, it is clear that the only non-vanishing cumulants of a set of Gaussian variables are
\begin{align}
  \kappa\,(x_i,x_j) \;=\; M_{ij} \;,
\end{align}
a statement that, after taking the limit $n\rightarrow\infty$, applies also to a Gaussian random field and its two-point connected correlation functions.

\section{Transfer functions}
\label{sec:transfer_functions}

The evolution of the cosmological perturbations is dictated by the Einstein and Boltzmann equations, which, as we shall see in the following chapters, form a system of coupled partial differential equations (PDEs). The differential system can be turned into a hierarchy of ordinary differential equations (ODEs), which are easier to treat numerically, by projecting the positional dependence, $\vec{x}$, into a basis of plane waves with wavevector $\vec{k}$. We shall introduce the formalism necessary to do so in \sref{sec:fourier_formalism}.

As we pointed out in \sref{sec:stochastic_perturbations}, the cosmological perturbations are stochastic three-dimensional fields. Rather than evolving a single realisation of such fields, we are interested into predicting their expectation values such as power spectra and bispectra. In \sref{ssec:transfer_functions} we show how to do so by introducing the concept of the transfer function.

\subsection{Fourier formalism}
\label{sec:fourier_formalism}

We shall solve the Einstein-Boltzmann differential system in Fourier space. This is achieved by applying to both sides of the equations the Fourier operator,
\begin{align}
  \mathcal{F_{\,\veck}}\left[\,f\,\right] \;=\; \int\dd\vec{x}\:e^{-i \vec{k}\cdot\vec{x}}\:f(\vec{x}) \;,
  \label{eq:fourier_operator}
\end{align}
where $f(\vec{x})$ is a generic function of the position. Note that, being linear, the Fourier operator acts separately on all the addends of its argument. The function $\mathcal{F_{\,\veck}}\left[\,f\,\right]$ is called the \keyword{Fourier transform} of $f(\vec{x})$ and we shall denote it simply as $f(\vec{k})$. (Note that, although we adopt the same symbol to denote them, the functions $f(\vec{x})$ and $f(\vec{k})$ generally have a different functional dependence.) The inverse Fourier transformation is given by
\begin{align}
  f(\vecx) \;=\; \int\frac{\dd\vec{k}}{(2\pi)^3}\:e^{i \vec{k}\cdot\vec{x}}\:{f}(\vec{k}) \;,
  \label{eq:fourier_expansion}
\end{align}
from which it follows that the Fourier transform of a real valued function obeys ${f}(-\veck) =  {f}^*(\veck)$.
It is important to note that both $\vec{x}$ and $\vec{k}$ are comoving quantities, that is, they are unaffected by the expansion of the Universe.

In Fourier space, partial derivatives transform to products,
\begin{align}
  \label{eq:fourier_partial_derivative}
  \mathcal{F_{\,\veck}}\left[\,\pfrac{f}{x^i}\,\right] \;=\; i\,k^i\,f(\vec{k}) \;,
\end{align}
as a direct consequence of the properties of the exponential function with respect to differentiation. The Laplacian operator $\nabla^2 = \partial^i\partial_i$ also has a simple Fourier transform,
\begin{align}
  \label{eq:fourier_laplacian}
  \mathcal{F_{\,\veck}}\left[\,\nabla^2 f\,\right] \;=\; -\,\abs{\vec{k}}^2\,f(\vec{k}) \;,
\end{align}
where $\abs{\vec{k}}^2 \equiv k^i\,k_i $. Therefore, going to Fourier space has the desirable property of turning our system of PDEs into an easier-to-treat system of ODEs by eliminating the partial derivatives with respect to the position.

The components of the wavevector $\vec{k}=\left(k^1,k^2,k^3\right)$ enter the Fourier-space equations as external parameters. In principle, to obtain the time evolution of the perturbations, one has to solve $N^3$ independent differential systems, where $N$ is the number of sampling points in each $\vec{k}$-direction.
In practice, however, the statistical isotropy of the cosmological perturbations allows us to choose a coordinate system for each wavevector $\vec{k}$ where the zenith is aligned with $\vec{k}$ itself. As a result, the solution for a given wavevector $\vec{k}$ will depend only on its magnitude, $k \equiv \abs{\vec{k}}$, and on conformal time, $\tau$.

As an example, consider the time-time component of Einstein equations in Newtonian gauge, also known as energy-constraint equation. In real space and at first perturbative order, it reads
\begin{align}
  \label{eq:perturbations_timetime_real}
  \dot{\Phi} \;+\; \Hc\,\Psi \;-\; \frac{1}{3\,\Hc}\,\nabla^2\Phi
  \;+\; \frac{a^2}{2\,\Hc}\,\sum\,\UD{T}{0}{0} \,=\, 0\;,
\end{align}
where a dot denotes differentiation with respect to conformal time, $\tau$, the sum is over all the matter species, and $\Phi$, $\Psi$, $\UD{T}{0}{0}$ are first-order quantities with a $(\taux)$ dependence. In Fourier space and with the zenith aligned with $\vec{k}$, the time-time equation reads
\begin{align}
  \label{eq:perturbations_timetime_fourier}
  \dot{\Phi} \,+\, \Hc\,\Psi \;+\; \frac{k^2}{3\,\Hc}\,\Phi
  \;+\; \frac{a^2}{2\,\Hc}\,\sum \,\UD{T}{0}{0} \,=\, 0 \;,
\end{align}
where, now, all perturbed variables have a $(\tau,k)$ dependence. Even though they look almost identical, the Fourier-space equation is much easier to solve than the real-space one as it does not involve partial derivatives. However, it contains a parameter, $k$, that has to be sampled in a range and with a frequency suitable to capture the physics of perturbations on all scales. We shall discuss the best sampling strategies for the wavemode $k$ in \sref{sec:sampling_strategies}.

\subsubsection{Sub and super-horizon scales}

The value of a random field in Fourier space, $X(k)$, quantifies the correlation between pairs of points separated by a distance of $r=2\pi/k$. This follows directly from the harmonic behaviour of the exponential in the Fourier transform, and it is sometimes known as the Wiener-Khinchin theorem\index{Wiener-Khinchin theorem}.

In the case of cosmological perturbations, the correlation length $2\pi/k$ defines a comoving scale \index{comoving scale} with an important causal meaning.
A given wavemode is said to be inside \index{sub-horizon modes} or outside \index{super-horizon modes} the horizon if its comoving scale, $2\pi/k$, is respectively smaller or larger than the particle horizon, $c\tau$. Modes inside the horizon, or sub-horizon, have $\tau k>2\pi/c$, while modes outside the horizon, or super-horizon, have $\tau k<2\pi/c$. Since the particle horizon, which we have defined in \sref{sec:comoving_distance}, is the maximum length a particle can travel since the Big Bang, no causal physics can take place on super-horizon scales; hence, we expect the observable correlators to evolve only on sub-horizon scales.

\subsection{Mode coupling}
\label{sec:perturbations_mode_coupling}

As we have seen in \sref{sec:perturbing_functions}, a second-order equation always includes a quadratic source term  consisting of products of first-order perturbations. The Fourier transform of a generic quadratic term, $A(\vecx)B(\vecx)$, yields a convolution integral:
\begin{align}
  \label{eq:fourier_convolution}
  \mathcal{F_{\,\veck}}[\,A(\vecx)\,B(\vecx)\,]\;
  &=\; \int\,\frac{\dd\kone}{(2\pi)^3}\; A(\kone)\:B(\veck-\kone)\\[0.25cm]
  \label{eq:fourier_convolution_dirac}
  &=\; \int\,\frac{\dd\kone\dd\ktwo}{(2\pi)^3}\; A(\kone)\:B(\ktwo)\:\diracMinus \;,
\end{align}
where \diracMinus is a Dirac delta and forces the three wavevectors \k, \kone, \ktwo to form a triangle.
The second form of expressing the convolution is particularly useful for reasons that will be clear after we introduce the transfer functions in \sref{ssec:transfer_functions}. For the sake of readability, we shall adopt the shorthand notations $A_1 \equiv A(\kone)$, $B_2 \equiv B(\ktwo)$ and denote the convolution integral as \cite{pitrou:2010a}
\begin{align}
  \K{f} \;\equiv\; \int\,\frac{\dd\kone\dd\ktwo}{(2\pi)^3}\; f(\kone,\ktwo)\:\diracMinus \;.
\end{align}

Due to the presence of these non-local terms, the evolution of the mode \veck of a second-order perturbation is determined by all other modes, which in \eref{eq:fourier_convolution} are represented by \kone. Equivalently, the behaviour of perturbations on a given scale is influenced by all other scales. This important property is typical of non-linear system and is referred to as \keyword{mode coupling}. At linear order, where there are no quadratic sources, all modes evolve independently.

Let us see with an example what the quadratic sources look like in Fourier space. In real space, the quadratic sources $S$ of the time-time equation are given by (see \sref{sec:einstein_equations})
\begin{align}
  \label{eq:perturbations_timetime_quadratic_real}
  \notag
  S\,(\taux) \;
  &=\; 4\,\Hc\,\Psi\,\Psi \;+\; 4\,\Psi\,\dot\Phi \;-\; 4\,\Phi\,\dot\Phi \\[0.2cm]
  &+\; \frac{1}{3\,\Hc}
  \left( 8\,\Phi\,\nabla^2\Phi \:+\: 3\,\partial_i\Phi\,\partial^i\Phi \:+\: 3\,\dot\Phi\,\dot\Phi \right) \;,
\end{align}
where $\partial_i$ is a shorthand for $\partial/\partial x^i$ and all perturbations have the same $(\taux)$ dependence. The full second-order time-time equation is obtained by adding $S$ to the left hand side of \eref{eq:perturbations_timetime_real}. In Fourier space, we have that $S(\tauk)\,=\,\K{S\,(\kone,\ktwo)}$ where the \keyword{convolution kernel} is given by
\begin{align}
  S\,(\kone,\ktwo) \;&=\; 4\,\Hc\,\Psi_1\,\Psi_2
  \;+\; 4\,\Psi_1\,\dot\Phi_2 \;-\; 4\,\Phi_1\,\dot\Phi_2 \nmsk
  &+ \frac{1}{3\,\Hc}
  \left[ - \left(\,8\,k_2^2 \:+\:
  3\:\scalarP{k_1}{k_2}\,\right)\,\Phi_1\,\Phi_2 \;+\, \dot\Phi_1\,\dot\Phi_2 \,\right] \;.
  \label{eq:perturbations_timetime_quadratic_fourier}
\end{align}
To obtain the above equation, we have just transformed the Laplacian term, $8\,\Phi\,\nabla^2\Phi$, into $-8\,\Phi_1\,k_2^2\,\Phi_2$ according to \eref{eq:fourier_laplacian}, and the gradient product, $3\,\partial_i\Phi\,\partial^i\Phi$, into $-3\,\scalarP{k_1}{k_2}\,\Phi_1\,\Phi_2$ according to \eref{eq:fourier_partial_derivative}.

\runinhead{Symmetrisation}
The convolution wavevectors \kone and \ktwo are dummy variables, thus there is no unique way to express the quadratic source terms. In the above example, we could have written the $\Phi_1\,\Phi_2$ coefficient as $8\,k_1^2+3\,\scalarP{k_1}{k_2}$ or as $4\,k_1^2 + 4\,k_2^2 + 3\:\scalarP{k_1}{k_2}$. In \SONG, we shall solve the second-order equations by symmetrising the quadratic sources with respect to the exchange of \kone and \ktwo, because doing so cuts the computation time by half (see \cref{ch:intrinsic}). In this thesis, instead, we shall report the lowest possible number of terms, except for the quadratic terms in the same variable (\eg $\Phi_1\,\Phi_2$), which we shall symmetrise.

\subsection{Transfer functions}
\label{ssec:transfer_functions}

One of the purposes of \SONG is to predict the current value of the cosmological perturbations by numerically evolving them from an initial state, according to a given cosmological model. The perturbations, however, are three-dimensional stochastic fields of which the observable Universe, that is our sky, is just a realisation (\sref{sec:stochastic_perturbations}). Since all stochastic processes have a variance, any two realisations differ; thus, the physical insight lies in the expectation values of the field rather than in the stochastic fluctuations of a single realisation.

In order to separate the stochastic part of the perturbations from their deterministic evolution, we introduce the concept of \keyword{transfer function}. The transfer function of a given cosmological field is an operator that maps a realisation of the field in the early Universe to its state today. The stochastic process is relegated to the initial realisation, which is drawn from the probability distribution of whatever physics took place in the early Universe. The transfer function, instead, is completely deterministic as it describes the subsequent physical processes, which are dictated by the Einstein and Boltzmann equations.

We shall express a perturbation field $X$ in terms of its \emph{linear} and \emph{second order} transfer functions, $\pert{\mathcal{T}}{1}$ and $\pert{\mathcal{T}}{2}$ respectively, as
\begin{align}
  X(\tauk) \;=\;\, &\T_X^{(1)} (\tauk) \; \Phi(\tauini,\veck) \nmsk
  &+\,\K{\T_X^{(2)} (\tau,\kone,\ktwo,\k) \: \Phi(\tauini,\kone)\:\Phi(\tauini,\ktwo)} \;,
  \label{eq:transfer_function_definition}
\end{align}
where $\Phi(\tauini,\veck)$ is the curvature potential at the initial time \tauini, a stochastic quantity. As we shall see in \cref{ch:evolution}, the initial time should be chosen to be deep in the radiation era, where the evolution of the perturbations is known analytically. Note that, in principle, the full perturbation $X$ is given by an infinite sum of terms, each involving a higher-order transfer function and an extra primordial potential; we truncate the sum at $\pert{\T}{2}$ because all of the other terms are at least third order.

The choice of $\Phi$ as the reference field is arbitrary and choosing another perturbation results in a simple rescaling of the transfer functions; in fact, many authors prefer to choose the curvature perturbation \R instead. Note that, contrary to $\Phi$ and $X$, the  linear and non-linear transfer functions are not perturbed quantities and are of order unity. Nonetheless, we denote them with a perturbative order with a small abuse of notation.

\subsubsection{Linear transfer functions}
\label{ssec:transfer_function_first}

If follows from \eref{eq:transfer_function_definition} that the evolution of the first-order part of a perturbation is completely determined by its linear transfer function:
\begin{align}
  \label{eq:transfer_function_first}
  \pert{X}{1}(\tauk) \;\equiv\; \pert{\T}{1}_X (\tauk) \; \pert{\Phi}{1}(\tauini,\veck) \;.
\end{align}
If we take a first-order equation and express all the perturbations in terms of their linear transfer functions, we can factor out the primordial stochastic field, $\Phi(\tauini,\k)$, because it does not have a time dependence. This leads to a fully deterministic equation. For example, the time-time equation (\eref{eq:perturbations_timetime_fourier}) becomes
\begin{align}
  \pert{\dot{\T}}{1}_\Phi \,-\, \Hc\,\pert{\T}{1}_\Psi \;+\; \frac{k^2}{3\,\Hc}\,\pert{\T}{1}_\Phi
  \;+\; \frac{a^2}{2\,\Hc}\,\sum \,\pert{\T}{1}_{\UD{T}{0}{0}} \,=\, 0 \;,
\end{align}
which is an ordinary differential equation that can be solved to yield the time evolution of $\pert{\T}{1}_\Phi$. In general, numerical solutions for the linear transfer functions of the cosmological perturbations can be computed in the matter of seconds for a number of different cosmological model, by using any of the publicly available Boltzmann codes \cite{seljak:1996a, lewis:2000a, doran:2005a, lesgourgues:2011a, huang:2012a}.

The time-time equation example shows that, in order to derive the time evolution of the transfer functions, it is not needed to know the details of the primordial field, $\Phi(\tauini,\k)$. Note that this is possible because we have defined the transfer functions in Fourier space. Had we defined them in real space as $\pert{X}{1}(\taux)=\pert{\T}{1}_X(\taux)\,\pert{\Phi}{1}(\tauini,\vecx)$, the partial derivatives in the evolution equations would have made it impossible to factor out the primordial potential. As a result, the same equations in Fourier space would have had convolution integrals over $\pert{\T}{1}_X\,\pert{\Phi}{1}(\tauini)$ even at first order.

\subsubsection{Second-order transfer functions}
\label{ssec:transfer_function_second}

The second-order part of a perturbation is determined by both the linear and the second-order transfer functions:
\begin{align}
	\pert{X}{2}(\tauk) \;&=\;
  \pert{\T}{1}_X(\tauk)\,\pert{\Phi}{2}(\tauini,\k) \nmsk
  &+\;\K{\pert{\T}{2}_X(\tau,\kone,\ktwo,\k) \:
  \pert{\Phi}{1}(\tauini,\kone)\:\pert{\Phi}{1}(\tauini,\ktwo)} \;.
	\label{eq:transfer_function_second}
\end{align}
Similarly to the linear case in \eref{eq:transfer_function_first}, the evolution of the second-order transfer functions is deterministic and is independent of the primordial potential.
We can see that this is the case by inserting the above expression in the second-order time-time equation (given by \eref{eq:perturbations_timetime_fourier} and \eref{eq:perturbations_timetime_quadratic_fourier}). The first part of the resulting expression involves only the linear transfer functions,
\begin{align*}
  &\left(\pert{\dot{\T}}{1}_\Phi \,-\, \Hc\,\pert{\T}{1}_\Psi
  \;+\; \frac{k^2}{3\,\Hc}\,\pert{\T}{1}_\Phi
  \;+\; \frac{a^2}{2\,\Hc}\,\sum \,\pert{\T}{1}_{\UD{T}{0}{0}}
  \right)\,\pert{\Phi}{2}(\tauini,\k) \;,
\end{align*}
and it vanishes identically because it corresponds the first-order time-time equation.
The second part is a convolution over the \kone and \ktwo wavemodes:
\begin{align*}
  &\begin{aligned}
    \mathcal{K}\,\biggl\{
    \biggl(\,
    \pert{\dot{\T}}{2}_\Phi
    \,-\, \Hc\,\pert{\T}{2}_\Psi \;+\; \frac{k^2}{3\,\Hc}\,\pert{\T}{2}_\Phi
    \;+\; \frac{a^2}{2\,\Hc}\,\sum \,\pert{\T}{2}_{\UD{T}{0}{0}}
    \biggr)
    \Phi_1(\tauini)\:\Phi_2(\tauini)
    \,+\, S\,(\kone,\ktwo)
    \,\biggr\} \,=\, 0 \;,
  \end{aligned}
\end{align*}
where the quadratic source term $S\,(\kone,\ktwo)$ is given in \eref{eq:perturbations_timetime_quadratic_fourier}:
\begin{align}
  S\,(\kone,\ktwo) \;&=\; 4\,\Hc\,\Psi_1\,\Psi_2 \;+\; 4\,\Psi_1\,\dot\Phi_2 \notag
  \;-\; 4\,\Phi_1\,\dot\Phi_2 \\[0.2cm]
  &+ \frac{1}{3\,\Hc}
  \left[ - \left(\,8\,k_2^2 \:+\: 3\:\scalarP{k_1}{k_2}\,\right)\,\Phi_1\,\Phi_2
   \;+\, \dot\Phi_1\,\dot\Phi_2 \,\right] \;.
\end{align}
The important point here is that the whole expression is inside a convolution integral over \kone and \ktwo. If we drop the integral, we are left with
\begin{align}
  \left(\,   \notag
  \pert{\dot{\T}}{2}_\Phi \,-\, \Hc\,\pert{\T}{2}_\Psi
  \;+\; \frac{k^2}{3\,\Hc}\,\pert{\T}{2}_\Phi
  \;+\; \frac{a^2}{2\,\Hc}\,\sum \,\pert{\T}{2}_{\UD{T}{0}{0}}
  \right) \; \Phi_1(\tauini)\:\Phi_2(\tauini)
  \,+\, S\,(\kone,\ktwo)
  \,=\, 0 \;,
\end{align}
which is an expression where \kone and \ktwo appear now as external parameters, at the same level of \k.
Let us now divide this expression by $\Phi_1(\tauini)\:\Phi_2(\tauini)$. Then, the $\Phi$ potentials in the purely second-order part are simply factored out, while for the quadratic sources we have
\begin{align}
  \frac{S\,(\kone,\ktwo)}{\Phi_1(\tauini)\:\Phi_2(\tauini)} \;,
\end{align}
which reduces to products of linear transfer functions, like in
\begin{align}
  4\;\;\frac{\Psi_1(\tau)}{\Phi_1(\tauini)} \; \frac{\dot\Phi_2(\tau)}{\Phi_2(\tauini)}
  \;=\; 4\;\pert{\T}{1}_\Psi(\kone) \; \pert{\dot\T}{1}_\Phi(\ktwo) \;.
\end{align}
As in the first-order case, the time-time equation now contains only deterministic transfer functions and can be solved numerically to yield the evolution of $\pert{T}{2}_\Phi(\k,\kone,\ktwo)$. The same applies for all other equations at second order. Computing numerically the second-order transfer functions is indeed one of the main features of our code \SONG, which shall be described in \cref{ch:evolution}.

A final remark is in order. The second-order transfer functions $\pert{\T}{2}_X(\k,\kone,\ktwo)$ are mathematical objects introduced to parametrise the evolution of the second-order perturbations. Being defined inside a convolution integral where \kone and \ktwo are dummy variables, they are neither unique nor observable. The observable quantities, such as spectra and bispectra, will depend on the actual perturbations that result from convolving the transfer functions with the initial conditions by means of \eref{eq:transfer_function_second}.


\section{The Einstein equations}
\label{sec:einstein_equations}


We consider the following form of the Einstein equation:
\begin{align}
  \UD{G}{\mu}{\nu} \;=\; \UD{R}{\mu}{\nu} \,-\,
  \frac{1}{2}\,\KronUD{\mu}{\nu} \,R \;=\; \kappa\,\UD{T}{\mu}{\nu} \;,
  \label{eq:updown_einstein_equation}
\end{align}
where $\kappa=8\pi G/c^4$. We prefer to work with the up-down version of Einstein equations because in this configuration the energy-momentum tensor has a simpler form, for reasons that will be clear after introducing the tetrad formalism in \sref{sec:tetrad_formalism}.

We project the Einstein equation in its scalar, vector and tensor components by using the projection vectors, $\xi_{[m]}$, and matrices, $\chi_{[m]}$, according to the SVT decomposition detailed in \sref{sec:perturbations_svt_decomposition}. We shall refer to the projected equations as follows:
\begin{align}
  & \text{Time-time}            &&\UD{G}{0}{0}\;=\;\kappa\:\UD{T}{0}{0}      \nmsk
  & \text{Trace}                &&\delta^{ij}\:G_{ij}\;=\;\kappa\:\delta^{ij}\:T_{ij}    \nmsk
  & \text{Space-time}           &&i\,\xivector{m}{i}\:G_{i0}\;=\;\kappa\:i\,\xivector{m}{i}\:T_{i0} \nmsk
  & \text{Space-space}          &&\chimatrix{2}{m}{ij}\:G_{ij}\;=\;\kappa\:\chimatrix{2}{m}{ij}\:T_{ij}
  \label{eq:einstein_equations_schematic_SVT}
\end{align}
The time-time and trace equations each describe 1 scalar degree of freedom, the space-time one describes 3 DOFs (1 scalar, 2 vector) and the space-space equation describes 5 DOFs (1 scalar, 2 vector, 2 tensors), for a total of 10 degrees of freedom.
The spatial indices refer to the up-down version of the Einstein equation (\eref{eq:updown_einstein_equation}).
They are raised and lowered with the Euclidean metric $\delta_{ij}$ and its inverse $\delta^{ij}$ so that, for instance, $T_{ij}$ is the spatial part of $\UD{T}{\mu}{\nu}$ and not that of $T_{\mu\nu}$.

\subsection{The metric}

As discussed in \sref{sec:perturbations_metric}, we shall adopt the Newtonian gauge and neglect the first-order parts of the vector and tensor degrees of freedom. The resulting metric up to second order reads
\begin{align}
  &\pert{g}{2}_{00} \;=\; - a^2\,(1+2\PsiFO+2\PsiSO) \nmsk
  &\pert{g}{2}_{0i} \;=\; \pert{g}{2}_{i0} \;=\; a^2\,\omegaSO_i \nmsk
  &\pert{g}{2}_{ij} \;=\; a^2\,(1-2\PhiFO-2\PhiSO)\,\delta_{ij} \;+\; 2\,a^2\,\gammaSO_{ij} \;.
  \label{eq:the_expanded_ng_metric}
\end{align}
The spatial perturbation $\gamma_{ij}$ is traceless by definition while the Newtonian gauge conditions enforce that both the vector and spatial perturbations are transverse: $\partial^i\omega_i = 0$ and $\partial^j\gamma_{ij}=0$. The number of independent degrees of freedom in $g_{\mu\nu}$ is therefore $6$. As pointed out in \sref{sec:perturbations_svt_decomposition}, we further decompose the metric in scalar, vector and tensor degrees of freedom by introducing the variables
\begin{align}
  \omegaSOm{m} \;=\; \xivector{m}{i}\,\omegaSO_i
\end{align}
and
\begin{align}
  \gammaSOm{m} \;=\; \chimatrix{2}{m}{ij}\,\gammaSO_{\,ij} \;.
\end{align}
In Fourier space, after aligning the zenith with the \k wavemode, the gauge conditions read $\omegaSOm{0}=0$ and $\gammaSOm{0}=\gammaSOm{\pm1}=0$, which means that in Newtonian gauge the vector modes are only in the $g_{i0}$ part of the metric.

\subsection{The energy-momentum tensor}
\label{sec:perturbations_energy_momentum_tensor}

The energy-momentum tensor for a given species is rigorously defined as the momentum integral over the one-particle distribution function, $f$,
\begin{align}
  \UD{T}{\mu}{\nu}(\taux) \;=\; \frac{1}{\sqrt{-g}}\;\int\dd\vecp\;\frac{\U{p}{\mu}\,\D{p}{\nu}}{\U{p}{0}}
  \;f(\tau,\vecx,\vecp) \;,
\end{align}
\annotate{The proper definition includes a $1/(2\pi)^3$ factor and the degree of freedom factor $g$. They do not make a difference in our treatment because they are constant factor that do not affect the fluctuations.}
where $p^\nu$ is the four-momentum of one particle of the considered species. A useful way to parametrise the energy momentum tensor is by means of a fluid,
\begin{align}
  \UD{T}{\mu}{\nu} \;=\; (\rho+P)\,\U{U}{\mu}\,\D{U}{\nu} \;+\;
  \KronUD{\mu}{\nu}\,P \;+\; \UD{\Sigma}{\mu}{\nu} \;,
  \label{eq:emt_perfect_fluid}
\end{align}
where $\rho$ is the \indexword{energy density}, $P$ is the \indexword{pressure}, $U^\mu$ is the \indexword{four-velocity}, $\Sigma^\mu_{\,\nu}\,$ is the \indexword{anisotropic stress tensor}\index{shear tensor}, a symmetric and traceless tensor, and we have assumed $c=1$. These variables are defined in the energy frame of the species; we shall refer to them collectively as the \keyword{fluid variables}.
\annotate{Careful here. From now on you will be implicitly setting $\Sigma^{00}=\Sigma^{i0}=0$, which is a condition that only applies for locally flat coordinates in the fluid rest frame (Bertschinger, pag.~44). Clearly not our case... investigate. Update: Ma \& Berti, Eq.~25 on pag.~11 arXiv, they do the same as use, and note ``where we have allowed an anisotropic shear perturbation ${\Sigma^i}_j$ in ${T^i}_j$._0''}
\annotate{Pitrou 2008 calls the four-velocity as the timelike tangent vector to the fluid world lines.}
\annotate{Bertschinger, pag.~44, says that $T_{00}$, ${T^i}_i$, etc, are the energy density, pressure, etc, measured in a locally flat coordinate system (local inertial frame?), so that $\rho$ and $P$ are the energy density and pressure measured in locally flat coordinates by an observer comoving with the fluid (because in that case $T_{00}=\rho$ and $T^{ij}=P\delta^{ij}$). Ma \& Berti, pag.~11 arXiv, say that the pressure $P$ and the energy density $\rho$ are defined to be the pressure and energy density measured by a comoving observer at rest with the fluid.}


The fluid description is particularly apt to treat the baryons and the cold dark matter because, being massive particles, they can be approximated as dust ($P=0$ and $\Sigma^\mu_{\,\nu}=0$) for all relevant cosmological epochs. However, it captures only part of the energetics of the photons and the neutrinos, which are relativistic particles and need to be described by the full distribution function. In \cref{ch:boltzmann} we shall introduce a more general framework where we treat both relativistic and non-relativistic particles by expanding the distribution function into a hierarchy of multipole moments; the components of the energy-momentum tensor will be just the lowest moments of such expansion. Nonetheless, we shall refer to the fluid limit often because it is still a valuable tool to relate the abstract multipole moments to the familiar energy density, velocity, pressure and shear.

\subsubsection{The 4-velocity}
Before expanding the energy-momentum tensor up to second order, let us study the behaviour of the four-velocity of the fluid, $U^\mu$. At the background level, the cosmological principle forbids the existence of any preferred direction in the fluids' motions \cite{malik:2009a}. Thus, we have that, in comoving coordinates,
\begin{align}
  U^\mu_{(0)} \,=\, \left(\;U_{(0)}^0,\: 0,\: 0,\: 0\;\right) \;.
\end{align}
(Note that, for the same reason, the shear tensor $\Sigma^i_j$ vanishes at the background level).
The time component, $\,U^0\,$, can be obtained from the other ones, at any order, by noting that the four-velocity, $\,U^\mu = dx^\mu/ds\,$, satisfies the normalisation condition
\begin{align}
  g_{\mu\nu}\,U^\mu\,U^\nu \;=\; -1 \;,
\end{align}
which, up to second order, yields
\begin{align}
  &g_{00}\,U^0\,U^0 \;+\;\delta_{ij}\,U^i\,U^j \;=\; -1 &&\Rightarrow&&
  U^0 \;=\; \frac{1}{\sqrt{-g_{00}}}\;\sqrt{1\,+\,U^iU_i} \;,
\end{align}
where we have considered the vector and tensor modes to be at least second order. In Newtonian gauge, where $\,g_{00}=-a^2(1+2\Psi)\,$, and if we choose the positive root of $\,\sqrt{-g_{00}}\,$, the above relation reads
\begin{align}
  U^0 \;=\; \frac{1}{a\,\sqrt{1+2\,\Psi}}\;\sqrt{1\,+\,U^iU_i} \;,
\end{align}
which up to second order is equivalent to
\begin{align}
  U^0 \;=\; \frac{1}{a}\;\left(\,1\,-\,\Psi\,+\,\frac{3}{2}\,\Psi\,\Psi\,+\,\frac{U^iU_i}{2}\,\right) \;.  
\end{align}
Note that, had we not imposed $\,\omega^{(1)}_i=0\,$, the expression would have included a $\,\omega_i\,U^i\,$ term.

\runinhead{Helmholtz decomposition}
In the following, we parametrise the spatial part of the four-velocity of a fluid as
\begin{align}
  U^i \;\equiv\; \frac{V^i}{a} \;,
\end{align}
and further decompose $V^i$ into a scalar field $v$ and a divergence-less vector field $v^i$,
\begin{align}
  V^i \;=\; \partial^i\,v \;+\; v^i \;,
\end{align}
in what is called the \indexword{Helmholtz decomposition}. The two parts are, respectively, curl-free and divergence-free, and are known as the longitudinal and solenoidal parts of the vector field \cite{malik:2009a}. In  Fourier space, after aligning the zenith with the \k vector, the decomposed velocity field reads
\begin{align}
  V_i \;=\; (v_1,\,v_2,\,ikv) \;.
  \label{eq:V_v1_v2_kv}
\end{align}
Note that, for an \indexword{irrotational fluid} (that is, a fluid whose velocity is curl-free), $v_i$ vanishes and the velocity field is completely described by its longitudinal part.

\runinhead{Spherical decomposition}
Like for any other three-vector, we decompose the fluid velocity $V^i$ into its spherical components as
\begin{align}
  V_{[m]} \;=\; \xivector{m}{i}\,V_i \;.
\end{align}
The Helmholtz and spherical decompositions of a three-vector are closely related. By using the expression for $V^i$ in Fourier space from \eref{eq:V_v1_v2_kv} and the definition of the $\xi$ vectors from Appendix~\ref{app:sphere_projection}, we obtain
\begin{align}
  &V_{[0]} \;=\; i\,k\,v \;,\nmsk
  &V_{[\pm1]} \;=\; \frac{1}{\sqrt{2}} \, (\mp v_1\,+\,i\,v_2) \;.
\end{align}
Thus, the longitudinal and solenoidal parts of the Helmholtz decomposition correspond, respectively, to the scalar ($m=0$) and vector ($m=\pm1$) parts of the vector field.

\subsubsection{Perturbative expansion of $\UD{T}{\mu}{\nu}$}
We now have all the ingredients to expand the fluid energy momentum tensor up to second order
\begin{align}  
  &\UD{T}{0}{0} \;=\; - \rho \;-\; (\overline{\rho}\,+\,\overline{P})\,V^i\,V_i \;,\nmsk
  &\UD{T}{i}{0} \;=\; -(\rho\,+\,P) \, (1\,+\,\Psi)\,V^i  \;, \nmsk
  &\UD{T}{0}{i} \;=\; (\rho\,+\,P) \, (1\,+\,\Psi\,+\,2\,\Phi)\,(V^i\,+\,\omega^i) \;, \nmsk
  &\UD{T}{i}{j} \;=\; \KronUD{i}{j}\,P\;+\;\UD{\Sigma}{i}{j}\;+\;(\overline{\rho}\,+\,\overline{P})\:V^i\,V^j \;,
  \label{eq:fluid_emt_expanded}
\end{align}
Since the only quantities with a background value are $\rho$ and $P$, the energy-momentum tensor up to first order is free from metric perturbations:
\begin{align}  
  &\UD{T}{0}{0} \;=\; - \rho \;,&&
  \UD{T}{i}{0} \;=\; -(\overline{\rho}\,+\,\overline{P}) \, V^i  \;, \nmsk
  &\UD{T}{0}{i} \;=\; - \UD{T}{i}{0} \;,&&
  \UD{T}{i}{j} \;=\; \KronUD{i}{j}\,P\;+\;\UD{\Sigma}{i}{j} \;.
  \label{eq:fluid_emt_expanded_first_order}
\end{align}

We can obtain the spherical components of the energy-momentum tensor by applying the SVT decomposition described in \sref{sec:perturbations_svt_decomposition}:
\begin{align}  
  &\UD{T}{0}{0} \;=\; - \rho \;-\; (\overline{\rho}\,+\,\overline{P})\,V^i\,V_i \;,\nmsk
  &\UD{T}{i}{i} \;=\; 3\,P \;+\; (\overline{\rho}\,+\,\overline{P})\,V^i\,V_i \;,\nmsk
  &\xivector{m}{i}\,\DD{T}{i}{0} \;=\; -(\rho\,+\,P) \, (1\,+\,\Psi)\,V_{[m]}  \;, \nmsk
  &\chimatrix{2}{m}{ij}\,\DD{T}{i}{j} \;=\; \Sigma_{[m]}\;+\;(\overline{\rho}\,+\,\overline{P})\:\tensorP{m}{V}{V} \;,
  \label{eq:fluid_emt_spherical_expanded}
\end{align}
where we have introduced the shorthands $\,\tensorP{m}{V}{V} = \chimatrix{2}{m}{ij}\,V_i\,V_j\,$ and $\,V_{[m]} = \xivector{m}{i}\,V_i$.
We remark that all the quadratic sources in the above expression mix different azimuthal modes, thus violating the decomposition theorem, as expected from the discussion in \sref{sec:decomposition_theorem}.
For example, the vector part of the third line, $\,\xivector{\pm1}{i}\,\DD{T}{i}{0}\,$, includes the term $\,\Psi\,V_{[\pm1]}\,$ which involves the scalar potential $\Psi$.
Similarly, the scalar $\,\UD{T}{0}{0}\,$ in the first line contains the quadratic term $\,V^iV_i=\sum\limits_{m=-1}^1\,V_{[m]}\,V^*_{[m]}\,$ (see \sref{sec:the_projection_vectors_xi}), which is in itself a scalar but has contributions from the vector part $V_{[\pm1]}$ of the velocity.


\subsection{The Einstein equations at second order}
\label{ssec:the_einstein_equation_at_second_order}

We derive the Einstein equations up to second order in Newtonian gauge by first inserting the perturbed metric in \eref{eq:the_expanded_ng_metric} into the Einstein equation in \eref{eq:updown_einstein_equation}. We then decompose the resulting expression into its scalar, vector and tensor parts according to \eref{eq:einstein_equations_schematic_SVT}, and project it to Fourier space using the Fourier operator in \eref{eq:fourier_operator}. It is crucial at this point to align the zenith to the \k wavevector, so that $k_x=k_y=0$ or, in spherical coordinates, $k_{[\pm1]}=0$; only in this way the mixing between the different azimuthal modes is forbidden explicitly.

Below, we show the Einstein equations in Fourier space as obtained with the procedure described above. For the real space equations,
refer to, \eg, Appendix A of \citet{pitrou:2010a}.
Also note that, due to the gauge conditions, only six out of the ten independent Einstein equations are independent.

\subsubsection{Purely second-order structure}

The purely second-order Einstein equations read
\begin{itemize}
  \item Time-time, or energy constraint, equation:
\begin{align}
  6\,\Hc^2\,\II\Psi \;+\; 6\,\Hc\,\II\Phid \;+\; 2\,k^2\,\II\Phi \;+\; \I\QTT
  \;=\; a^2\kappa\,\II T^0_{\,0} \;.
  \label{eq:einstein_pure_timetime}
\end{align}
  \item Trace equation:
\begin{align}
  6\,\II\Phidd \;+\; \II\Psi\,(6\,&\Hc^2\,+\,12\,\dot\Hc)
  \;+\; 6\,\Hc\,(\II\Psid+2\,\II\Phid)
  \;+\;2\,k^2\,(\II\Phi-\II\Psi) \;+\; \I\QTR
  \;=\; a^2\kappa\,\II T^i_{\,i} \;.
  \label{eq:einstein_pure_trace}
\end{align}
  \item Space-time equations for $m=0$ and $m=\pm1$,
\begin{align}
  &-2\,k\,(\II\Phid\,+\,\Hc\II\Psi) \;+\; \I\QST{_{[0]}}
  \;=\; a^2\kappa\; (i\,\xivector{0}{i}\,\II T_{i0}) \;, \nmsk
  &\frac{i}{2}\,\II\omega_{[\pm1]}\,(4\,\Hc^2\,-\,4\,\dot\Hc\,+\,k^2)
  \;+\; \I\QST{_{[\pm1]}}
  \;=\; a^2\kappa\;(\,i\,\xivector{\pm1}{i}\,\II T_{i0}\,) \;.
  \label{eq:einstein_pure_spacetime}
\end{align}
  \item Space-space, or anisotropic stresses, equations for $m=0$, $m=\pm1$ and $m=\pm2$,
\begin{align}
  &-\frac{2\,k^2}{3}\,(\II\Phi\,-\,\II\Psi) \;+\; \I\QSS{_{[0]}}
  \;=\; a^2\kappa\;(\,\chimatrix{2}{0}{ij}\,\II T_{ij}\,) \;, \nmsk
  &-\frac{i\,k}{\sqrt{3}}\,(\II{\dot\omega}_{[\pm1]}
  \,+\,2\,\Hc\,\II\omega_{[\pm1]})
  \;+\; \I\QSS{_{[\pm1]}}
  \;=\; a^2\kappa\;(\,\chimatrix{2}{\pm1}{ij}\,\II T_{ij}\,) \;, \nmsk
  &\II{\ddot\gamma}_{[\pm2]} \;+\; 2\,\Hc\,\II{\dot\gamma}_{[\pm2]}
  \;+\; k^2\,\II\gamma_{[\pm2]}
  \;+\; \I\QSS{_{[\pm2]}}
  \;=\; a^2\kappa\;(\,\chimatrix{2}{\pm2}{ij}\,\II T_{ij}\,) \;.
  \label{eq:einstein_pure_spacespace}
\end{align}
\end{itemize}
The dots denote differentiation with respect to the conformal time, $\tau$, and $\kappa=8\pi G$. The symbols $Q$ stands for the the quadratic part of the Einstein tensor, which we shall show below. The right hand side of each equation contains the spherical decomposition of the energy-momentum tensor. This is given by a sum of the energy-momentum tensors of the single species (photons, neutrinos, baryons and cold dark matter). Its form in the fluid limit can be read from \eref{eq:fluid_emt_spherical_expanded}; however, in \SONG, it is computed using the Fourier multipoles $\Delta_\lm(\vec{k})$ defined in \cref{ch:evolution} rather than the fluid variables.
Note that the four scalar equations can be directly compared with Eq. (23a) to (23d) in \citet{ma:1995a}.

\subsubsection{Quadratic sources}

We denoted the quadratic sources for the Einstein tensor with the letter $Q$:
\begin{align}
  &\I\QTT \;=\; a^2\,{G^0_{\,0}}^{(1)(1)}\;,
  &&\I\QTR \;=\; a^2\,{G^i_{\,i}}^{(1)(1)}\;, \nmsk
  &\I\QST{_{[m]}} \;=\; i\,\xivector{m}{i}\;a^2\,{G_{i0}}^{(1)(1)}\;,
  &&\I\QSS{_{[m]}} \;=\; \chimatrix{2}{m}{ij}\;a^2\,{G_{ij}}^{(1)(1)}\;. \notag
\end{align}
Their explicit form is given by
\begin{align}
  & 
  \begin{aligned}
  \I\QTT \;=\;
  -12\,\Hc^2\,\I\Psi_1\,\I\Psi_2
  \;+\;(3\,\kokt+4\,k_1^2+4\,k_2^2)\,\I\Phi_1\,\I\Phi_2
  \;+\;12\,\Hc\,\I\Phid_2 \, (\I\Phi_1 - \I\Psi_1)
  \;-\;3\,\I\Phid_1\,\I\Phid_2 \notag
  \end{aligned}
  \displaybreak[0]
  \lmsk& 
  \begin{aligned}
  \I\QTR \;=\;
  &-12\,\I\Psi_1\,\I\Psi_2\,(\Hc^2\,+\,2\,\dot\Hc)
  \;+\; (k^2\,+\,k_1^2\,+\,k_2^2)\,\I\Psi_1\,\I\Psi_2\, \nmsk
  &\;+\;(3\,\kokt+4\,k_1^2+4\,k_2^2)\,\I\Phi_1\,\I\Phi_2
  \;+\; (2\,\kokt\,-\,4\,k_2^2)\,\I\Phi_1\,\I\Psi_2 \nmsk
  &\;+\;12\,(\I\Phidd_2\,+\,2\,\Hc\,\I\Phid_2)\:(\I\Phi_1 - \I\Psi_1)
  \;-\; 6\,\I\Psid_2\,(4\,\Hc\,\I\Psi_1\,+\,\I\Phid_1)
  \;+\;3\,\I\Phid_1\,\I\Phid_2
  \end{aligned}
  \displaybreak[0]
  \lmsk& 
  \begin{aligned}
  \I\QST{_{[m]}} \;=\;
  2\,\koneM{m}\,\Bigl[\,
  2\,\Hc\,\I\Psi_1\,(\I\Psi_2-\I\Phi_2)
  \,-\,2\,\I\Phi_1\,\I\Phid_2
  \,-\,4\,\I\Phid_1\,\I\Phi_2
  \,+\,\I\Psi_1\,\I\Phid_2\,\Bigr] \notag
  \end{aligned}
  \displaybreak[0]
  \lmsk& 
  \begin{aligned}
  \I\QSS{_{[m]}} \;
  &=\;\tensorP{m}{k_1}{k_2}\,\Bigl[2\,\I\Phi_1\,\I\Psi_2
  \,-\,3\,\I\Phi_1\,\I\Phi_2 \,-\, \I\Psi_1\,\I\Psi_2\Bigr]
  \;+\;\tensorP{m}{k_1}{k_1}\,\Bigl[\,2\,\I\Psi_1\,\I\Phi_2
  \,-\, 4\,\I\Phi_1\,\I\Phi_2\,-\,2\,\I\Psi_1\,\I\Psi_2
  \Bigr] \;.
  \end{aligned}
  \label{eq:einstein_quadratic_sources}
\end{align}
The subscripts indicate the dependence on the convolution wavemodes, \eg $\Phi_1=\Phi(\kone)$ and $\Phi_2=\Phi(\ktwo)$. Because \kone and \ktwo are dummy variables that will be eventually integrated out (\sref{sec:perturbations_mode_coupling}), there is no unique way to write down the quadratic sources. In writing the above expression, we have favoured brevity and we have written the quadratic sources using as few terms as possible. In \SONG, for the purpose of optimisation, we shall symmetrise the sources with respect to the exchange of \kone and \ktwo (\cref{ch:evolution}).

%
%

\subsubsection{Modified gravity theories}\index{modified gravity}
In this work, we shall always assume that the theory of general relativity (GR) holds. There are, however, other viable theories of gravitation than GR. In fact, while GR is well tested for scales smaller than the size of the solar system \cite{bertotti:2003a,kapner:2007a},
there is still room for different formulations of gravity on larger scales \cite{turyshev:2009a}. The possibility is particularly interesting because the least understood components of the Universe, that is cold dark matter and dark energy, are known to be relevant on large scales.

It has been proposed that the observed flatness of the galaxy rotation curves on kiloparsec scales might be due to a modification of Newton's law \cite{milgrom:1983a,milgrom:1983b,milgrom:1983c,bekenstein:2004a} rather than to the presence of dark matter. Similarly, it was shown that including more structure in the Lagrangian of GR results in a richer phenomenology that can ultimately yield to cosmic acceleration, without the need of a cosmological constant (see, \eg, Ref.~\cite{caldwell:2009a, starobinsky:2007a, capozziello:2003a}). In this work, and in \SONG, we assume that the gravitational interaction is well described by the standard Einstein field equations; as we shall discuss in \sref{sec:future_prospects}, using a different theory of gravity might have interesting effects on the CMB bispectrum and is left for future work.

\annotate{Bertotti use the frequency shift of radio photons to and from the Cassini spacecraft as they passed near the Sun to find that GR is correct with $10^-5$ accuracy. Kapenr use a Cavendish-like experiment (torsion of suspended balls caused by heavier balls around them) to find that the inverse square law holds down to scales of $\lambda=56\mu m$.}

\annotate{From Caldwell \& Kamionkowski, 2009:
In scalar-tensor theories, the Einstein-Hilbert action
$S_{\mathrm{EH}} =(16\pi G)^{-1} \int d^4x \sqrt{-g} R$ for
gravity is replaced by an action (see, e.g.,
Ref.~\cite{Carroll:2004st}),
\begin{equation}
    S=\int d^4x \sqrt{-g} \left[ b(\lambda) R -\frac{1}{2}
    h(\lambda) g^{\mu\nu} (\partial_\mu
    \lambda)(\partial_\nu\lambda) -U(\lambda) + {\cal
    L}_M(g_{\mu\nu},\psi_i) \right],
\label{eqn:STaction}
\end{equation}
where $\lambda(\vec x,t)$ is the eponymous scalar field; ${\cal
L}(g_{\mu\nu},\psi_i)$, the matter Lagrangian, is a function of
the metric and matter fields $\psi_i$; and $b(\lambda)$,
$h(\lambda)$, and $U(\lambda)$ are functions that determine the
form of the scalar-tensor theory.}

\section{Spectra \& bispectra}
\label{sec:spectra_and_bispectra}



Given the stochastic nature of the cosmological perturbations, both predictions and observables must be expressed in terms of the probability distribution function (PDF) of the perturbed fields. In \sref{sec:stochastic_perturbations} we have introduced the $n$-point functions as a simple way to characterise the PDF. In this section, we focus on the two and three-point functions, which, in the case of the temperature of the cosmic microwave background, have been observed to high precision by the WMAP \cite{bennett:2012a} and Planck \cite{planck-collaboration:2013a, planck-collaboration:2013b} satellites.

\subsection{The two-point function}
\label{sec:two_point_function}

\subsubsection{The power spectrum}
Given a random field $\R$, we denote its two-point function, or autocorrelation, with the symbol $\xi(\vecr)$:
\begin{align}
  \xi(\vecr) \,\equiv\, \avg{\R(\vecx)\,\R(\vecx+\vecr)} \;.
\end{align} 
In principle, the autocorrelation depends on both the point, $\vecx$, and the separation, $\vecr$. However, enforcing statistical homogeneity (\sref{sec:fair_sample_hypothesis}) removes the $\vecx$ dependence. As a consequence, the expectation value for the product of two Fourier modes,
\begin{align*}
  \avg{\R(\kone)\,\R(\ktwo)} \,=\, \int \dd\vecx\,\dd\vecy \,
  e^{-i\,(\,\scalarP{k_1\,}{\,x}\,+\,\scalarP{k_2\,}{\,y}\,)} \, \avg{\R(\vecx)\,\R(\vecy)} \;.
\end{align*}
collapses to a Dirac delta after the change of variable $\vecy=\vecx+\vecr$:
\begin{align}
  \avg{\R(\kone)\,\R(\ktwo)} \,=\, (2\pi)^3 \,\delta (\kone + \ktwo) \, P(\k_1) \;,
  \label{eq:wiener_khintchine_theorem}
\end{align}
where we have defined the \keyword{power spectrum}, $P(\k)$, as the Fourier transform of the two-point function:
\begin{align}
\label{eq:pk_definition}
  P(\k) \,\equiv\, \int \dd\vecr \, \xi(\vecr) \, e^{-i \scalarP{k\,}{\,r}} \;.
\end{align}
Therefore, the two-point function of a homogeneous field in Fourier space vanishes unless the two considered wavevectors are equal and opposite. In other words, the homogeneity enforces that the covariance matrix in Fourier space is diagonal. This useful result is known as the \keyword{Wiener-Khintchine theorem}.

The two-point function is readily obtained by taking the inverse Fourier transform of $ P(\vec k) $:
\begin{align*}
  \xi(\vec r) = \int \frac{\dd\k}{(2\pi)^3} \, P(\vec k) \, e^{i \scalarP{k\,}{\,r}} \;.
\end{align*}
One can also enforce statistical isotropy, $\xi(\vecr)=\xi(r)$, to reduce the integration to one dimension:
\begin{align}
	&\xi(\vecr) \,=\, \xi(r) \,=\, \frac{1}{2 \pi^2} \int \dd k \, k^2 \, \frac{\sin(kr)}{kr} \, P(k) \;, \nmsk
	&P(\k) \,=\, P(k) \,=\, 4 \pi \int \dd r \, r^2 \, \frac{\sin(kr)}{kr} \, \xi(r) \;. 
\end{align}
In the limit where $\vecr \rightarrow 0$, the two-point function reduces to the variance of the field:
\begin{align}
  \sigma^2 \,=\, \avg{\R(\vecx)^2} \,=\, \int \frac{\dd\k}{(2\pi)^3} \, P(\vec k) \;.
  \label{eq:sigma2_pk}
\end{align}
It follows that the product $\dd\k\,P(\k)/(2\pi)^3$ is the contribution to the variance of the field coming from the volume element $\dd\k$; that is, the power spectrum quantifies the power in the fluctuations per unit-volume of $\k$-space. Any non-trivial random field has a non-vanishing power spectrum which, if measured, provides important information on its PDF. In the case of Gaussian random fields, the power spectrum, being the Fourier transform of the two-point correlation function, uniquely determines the PDF of the field.

\subsubsection{Perturbative expansion}

After adopting the transfer function representation in \eref{eq:transfer_function_definition},
\begin{align*}
  X(\tauk) \;=\;\, &\T_X^{(1)} (\tauk) \; \Phi(\tauini,\veck) \nmsk
  &+\,\K{\T_X^{(2)} (\tau,\kone,\ktwo,\k) \: \Phi(\tauini,\kone)\:\Phi(\tauini,\ktwo)} \;,
\end{align*}
the leading term of the two-point function of a cosmological perturbation $X$ is given by
\begin{align}
  \avg {X(\kone)\,X(\ktwo)} \,\simeq\,
  \pert{\T}{1}_X(\kone)\,\pert{\T}{1}_X(\ktwo)\,\avg{\Phi(\kone)\,\Phi(\ktwo)} \;,
  \label{eq:pk_leading}
\end{align}
and is second-order in the primordial perturbation. (Note that we have dropped the time dependence, as the potential $\Phi$ is always evaluated at the initial time $\tauini$ and the transfer functions at the arbitrary time $\tau$.)

The next-to-leading order contribution is a product of the linear transfer function with the second-order one,
\begin{align}
  \K{\pert{\T}{1}(\kone)\,\pert{\T}{2}(\ktwo,\konep,\ktwop)\,
  \avg{\Phi(\kone)\,\Phi(\konep)\,\Phi(\ktwop)}} \; + \; \text{1 permutation},
  \label{eq:twopoint_corrections}
\end{align}
where \konep and \ktwop are convolution variables and the permutation consists of the same term with \kone and \ktwo switched. This contribution is penalised with respect to \eref{eq:pk_leading} by the presence of an extra power of the primordial potential, which is of order $10^{-5}$. The penalisation can be compensated either by a strong initial non-Gaussianity, manifesting itself in a large value of the three-point function, or by a growth of the perturbation with time, which would correspond to a large value of the second-order transfer function. The former case has been excluded observationally, as we shall detail in \sref{sec:three_point_function}; the latter, while being certainly possible for cold dark matter, cannot happen to photon perturbations, because they do not grow with time. Thus, we can safely use the linear term in \eref{eq:pk_leading} to approximate the two-point function in Fourier space:
\begin{align}
  \avg {X(\kone)\,X(\ktwo)}
  \;\simeq\; \pert{\T}{1}_X(\kone)\,\pert{\T}{1}_X(\ktwo)\,\avg{\Phi(\kone)\,\Phi(\ktwo)} \;.
\end{align}

By enforcing the statistical homogeneity of the cosmological perturbations (\eref{eq:wiener_khintchine_theorem}), we obtain a relation between the \keyword{primordial power spectrum}, $P_\Phi$, and that of the considered perturbation, $P_X$:
\begin{align*}
  P_X(\veck) \;\simeq\; \pert{\T}{1}_X(\k)\,\pert{\T}{1}_X(-\k)\;P_\Phi(\k) \;,
\end{align*}
Because of statistical isotropy, we also have that $P_\Phi(\veck)=P_\Phi(k)$ and  $\pert{\T}{1}_X(\k)=\pert{\T}{1}_X(-\k)=\pert{\T}{1}_X(k)$. Hence, we obtain
\begin{align}
  P_X(\tau,k) \;\simeq\; \pert{\T}{1}_X(\tau,k)^2\;P_\Phi(\tauini,k) \;,
  \label{eq:transfer_variance}
\end{align}
where we have reintroduced the time dependence. 
Therefore, measuring the power spectrum of a cosmological perturbation today, gives valuable information on the product between the primordial power spectrum, whose shape and amplitude are dictated by the physical processes at work in the early Universe, and the first-order transfer function, which depends on the way the perturbations evolved from the initial conditions all the way to today. As we pointed out before, this is true only if the higher-order corrections such as that in \eref{eq:twopoint_corrections} are negligible.

The power spectrum of the photon temperature field has been measured to great precision by the WMAP \cite{bennett:2012a} and Planck \cite{planck-collaboration:2013a} experiments. The simplified description of \eref{eq:transfer_variance}, where we only consider the leading contribution to the fluctuations, complemented by the simple \LCDM model, fits the angular power spectrum of the CMB with impressive precision. Such agreement is an important confirmation of the fact that photon perturbations do not grow and that, therefore, the higher-order corrections like the one in \eref{eq:twopoint_corrections} can be neglected. However, it should be noted that these corrections can still play a role at the power spectrum level if one aims to a precision below the percent level.

\subsection{The three-point function}
\label{sec:three_point_function}

\subsubsection{The bispectrum}
We denote the three-point function of a cosmological perturbation $X$ as
\begin{align*}
  \xi(\vecs,\vect) \,=\, \avg{\R{(\vecx)}\,\R{(\vecx+\vecs)}\,\R{(\vecx+\vect)}} \;.
\end{align*}
The statistical homogeneity ensures that $\xi(\vecs,\vect)$ does not depend on the point where it is evaluated, $\vecx$, but only on the separations, $\vecs$ and $\vect$. The statistical isotropy takes out three more degrees of freedom from $\xi(\vecs,\vect)$ by forcing it to depend only on the three combinations of \vecs and \vect that are rotationally invariant: their magnitudes, $s$ and $t$, and their scalar product.

If we take the expectation value of three perturbations in Fourier space,
\begin{align*}
  \avg{\R(\kone)\,\R(\ktwo)\,\R(\ktre)} \,=\, \int \dd\vecx\,\dd\vecy\,\dd\vecz \,
  e^{-i\,(\,\scalarP{k_1\,}{\,x}\,+\,\scalarP{k_2\,}{\,y}\,+\,\scalarP{k_3\,}{\,z}\,)}\,
  \avg{\R(\vecx)\,\R(\vecy)\,\R(\vecz)} \;,
\end{align*}
and  introduce the variables $\vecs=\vecy-\vecx$ and $\vect=\vecz-\vecx$ eliminating \vecy and \vecz,
\begin{align*}
  \avg{\R(\kone)\,\R(\ktwo)\,\R(\ktre)} \,=\,
  \int \dd\vecx\, e^{-i\,\vecx\,\!\cdot\,\!(\kone+\ktwo+\ktre)}\,
  \int \dd\vecs\,\dd\vect \,
  \,e^{-i\,(\,\scalarP{k_2\,}{\,s}\,+\,\scalarP{k_3\,}{\,t}\,)}\,\,
  \xi(\vecs,\vect) \;,
\end{align*}
we see that the statistical homogeneity makes it possible to substitute the $\vecx$ integral with a Dirac delta function:
\begin{align}
  \avg{\R(\kone)\,\R(\ktwo)\,\R(\ktre)} \,=\, (2\pi)^3 \,\delta (\kone+\ktwo+\ktre) \, B(\ktwo,\ktre) \;,
  \label{eq:bispectrum_definition}
\end{align}
where we have defined the \keyword{bispectrum} $B(\ktwo,\ktre)$ as the Fourier transform of the three-point function:
\begin{align}
  B(\ktwo,\ktre) \,\equiv\, \int\dd\vecs\,\dd\vect\;
  e^{-i\,(\,\scalarP{k_2\,}{\,s}\,+\,\scalarP{k_3\,}{\,t}\,)} \,
  \xi(\vecs,\vect) \;.
\end{align}
For a zero-mean Gaussian random field the three-point function, $\xi(\vecs,\vect)$, vanishes (see \sref{sec:gaussian_random_fields}) and so does the bispectrum. The bispectrum is therefore the lowest-order statistic which is sensitive to the non-Gaussianity of the field. In particular, measuring a non-vanishing bispectrum for a cosmological perturbation would prove that the perturbation has undergone some non-Gaussian (or, equivalently, non-linear) process at some point in the evolution of the Universe.

In an isotropic Universe, the bispectrum can only depend on the magnitudes of the wavevectors, $k_1$ and $k_2$, and on the angle between them.
Because of the presence of the Dirac delta function, $\delta (\kone+\ktwo+\ktre)$, the wavevector $\ktre$ can be used to parametrise the bispectrum, too; in fact, in the literature it is customary to express the bispectrum using the magnitudes of the $\k$-vectors:
\begin{align}
  \avg{\R(\kone)\,\R(\ktwo)\,\R(\ktre)} \,=\, (2\pi)^3 \,\delta (\kone+\ktwo+\ktre) \, B(k_1,k_2,k_3) \;.
  \label{eq:bispectrum_definition_isotropic}
\end{align}

Assuming the statistical isotropy and homogeneity of the Universe brings down the number of independent degrees of freedom in the three-point function from 9 to 3; the bispectrum is just a convenient way of expressing these 3 DOFs in Fourier space. The freedom in choosing how to parametrise the bispectrum might lead to ambiguities in the notation. We shall avoid them by denoting the bispectrum with its full dependence on the wavevectors, $B(\kone,\ktwo,\ktre)$.

\runinhead{Higher-order spectra}
In general, the $n$-point connected function of a homogeneous field can be always expressed in Fourier space in terms of its \keyword{polyspectrum}, $S(\ktwo,\dotsc,\kn)$:
\begin{align}
  \avg{\R(\kone)\,\dotsb\,\R(\kn)} \,=\, (2\pi)^3 \,\delta (\kone+\dotsb+\kn) \, S(\ktwo,\dotsc,\kn) \;.
  \label{eq:polyspectra_definition}
\end{align}
The polyspectrum is defined as the Fourier transform of the $n$-point correlation function:
\begin{align}
  S(\ktwo,\dotsc,\kn) \;\equiv\; \int\dd\vec{r_2}\dotsb\dd\vec{r_n}\;
  e^{-i\,(\,\scalarP{k_2\,}{\,r_2}\,+\dotsb+\,\scalarP{k_n\,}{\,r_n}\,)} \,
  \xi(\vec{r_2},\dotsc,\vec{r_n}) \;.
\end{align}
Because of homogeneity, the polyspectrum only depends on $n-1$ out of the $n$ wavevectors in the $n$-point function.
Note that for Gaussian random fields all odd-$n$ polyspectra vanish, because they are defined out of the connected correlation functions (\sref{sec:gaussian_random_fields}).

\subsubsection{Perturbative expansion}
We expand the three-point function of a cosmological perturbation $X$ in terms of its transfer functions via \eref{eq:transfer_function_definition}:
\begin{align*}
  X(\tauk) \;=\;\, &\T_X^{(1)} (\tauk) \; \Phi(\tauini,\veck) \nmsk
  &+\,\K{\T_X^{(2)} (\tau,\kone,\ktwo,\k) \: \Phi(\tauini,\kone)\:\Phi(\tauini,\ktwo)} \;,
\end{align*}
The resulting expression involves several terms, the leading order ones being of order $\O(\Phi^3)$ and $\O(\Phi^4)$.
The $\O(\Phi^3)$ part is
\begin{align}
  \pert{\T}{1}_X(\kone)\,\pert{\T}{1}_X(\ktwo)\:\pert{\T}{1}_X(\ktre)\;
  \avg{\Phi(\kone)\,\Phi(\ktwo)\,\Phi(\ktre)} \;,
  \label{eq:leading_order_3pf}
\end{align}
and, according to \eref{eq:bispectrum_definition}, corresponds to a bispectrum that is proportional to the bispectrum of the primordial potential:
\begin{align}
  B_X^\lin(\kone,\ktwo,\ktre) \,=\, 
  \pert{\T}{1}_X(\kone)\,\pert{\T}{1}_X(\ktwo)\:\pert{\T}{1}_X(\ktre)\;
  B_\Phi(\kone,\ktwo,\ktre) \;.
  \label{eq:linear_bispectrum_k1k2k3}
\end{align}
We shall call this contribution the \keyword{linearly propagated bispectrum}, because it involves only linear transfer functions\footnote{Note that some authors refer to $B^\lin$ as the \keyword{primary bispectrum}.}. The above relation implies that, at leading order in the perturbations, any non-Gaussianity present in the initial conditions is linearly transferred throughout the evolution of the Universe. In a linear Universe, any non-Gaussian feature observed in the sky today can be traced back to some process that took place in the early Universe. In particular, if the early Universe was Gaussian, as predicted by the simplest models of inflation \cite{maldacena:2003a}, all the observables including the CMB sky and the distribution of galaxies would be normally distributed. Equivalently, at linear order there is no mechanism to produce non-Gaussianities that were not already in the initial conditions. 

The next-to-leading order contribution to the bispectrum is of order $\O(\Phi^4)$ and it involves the second-order transfer function of the considered perturbation:
\begin{align}
  \pert{\T}{1}_X(\kone)\,\pert{\T}{1}_X(\ktwo)\;
  &\int\frac{\dd\konep\,\dd\ktwop}{(2\pi)^3}\,\dirac{\konep+\ktwop-\ktre} \:
  \pert{\T}{2}_X(\konep,\ktwop,\ktre) \; \nmsk
  &\times\,\avg{\Phi(\kone)\,\Phi(\ktwo)\,\Phi(\konep)\,\Phi(\ktwop)}
  \;\;+\;\; \text{2 permutations} \;,
  \label{eq:next_to_leading_order_3pf}
\end{align}
where the permutations consist of two extra terms where $\pert{\T}{2}_X$ is assigned \kone and \ktwo, respectively. Using the Wick's theorem for a zero-mean field (\eref{eq:wick_example}), we expand the four-point function as
\begin{align}
  \avg{\Phi_1\,\Phi_2\,\Phi'_1\,\Phi'_2} &=
  \con{\Phi_1,\Phi_2,\Phi'_1,\Phi'_2} + \avg{\Phi_1\,\Phi_2}\avg{\Phi'_1\,\Phi'_2} \nmsk
  &+ \avg{\Phi_1\,\Phi'_1}\avg{\Phi_2\,\Phi'_2} + \avg{\Phi_1\,\Phi'_2}\avg{\Phi_2\,\Phi'_1} \;.
  \label{eq:wick_with_four_point}
\end{align}
The three products involving the two-point function can be expressed in terms of the power spectrum via \eref{eq:wiener_khintchine_theorem}; the resulting Dirac delta functions combine with the one in \eref{eq:next_to_leading_order_3pf}. We neglect the combination that arises from $\avg{\Phi_1\,\Phi_2}\avg{\Phi'_1\,\Phi'_2}$ because, being proportional to \dirac{\ktre}, would imply evaluating a perturbation with infinite wavelength. The other two terms collapse in the usual Dirac delta, $\dirac{\kone+\ktwo+\ktre}$, which can be extracted to yield the following contribution to the total bispectrum:
\begin{align}
  B_X^\intr(\kone,\ktwo,\ktre) \,=\, 2\;
  \pert{\T}{1}_X(\kone)\;
  \pert{\T}{1}_X(\ktwo)\;
  \pert{\T}{2}_X(-\kone,-\ktwo,\,\ktre)\,
  P_\Phi(-\kone) \, P_\Phi(-\ktwo) 
  \;+\; \text{2 perm.} \;,
  \label{eq:intrinsic_bispectrum_k1k2k3}
\end{align}
where the factor 2 comes from the fact that we choose the second-order transfer function to be symmetric with respect to the exchange of \kone with \ktwo.

We shall denote the contribution to the bispectrum in \eref{eq:intrinsic_bispectrum_k1k2k3} as the \keyword{intrinsic bispectrum}. The intrinsic bispectrum is always present no matter what the initial conditions are: the very existence of the perturbations ensure that the power spectrum of $\Phi$ is non-vanishing, while the non-linearity of the gravitational interactions always sources the second-order transfer function. This is in stark contrast with the linearly propagated bispectrum in \eref{eq:linear_bispectrum_k1k2k3}, which, instead, strongly depends on the statistics of the primordial field, to the point that it vanishes when $\Phi$ is Gaussian.

The connected four-point function in \eref{eq:wick_with_four_point} can be expressed in terms of the primordial trispectrum, $S_\Phi(\kone,\ktwo,\konep,\ktwop)$, according to \eref{eq:polyspectra_definition}:
\begin{align*}
  \con{\Phi(\kone),\,\Phi(\ktwo),\,\Phi(\konep),\,\Phi(\ktwop)} \,=\,
  (2\pi)^3 \,\delta (\kone+\ktwo+\konep+\ktwop) \: S_\Phi(\kone,\ktwo,\konep,\ktwop) \;,
\end{align*} 
which, inserted into \eref{eq:next_to_leading_order_3pf}, yields the contribution from the primordial trispectrum to the observed bispectrum:
\begin{align}
  B_X^\text{trisp}(\kone,\ktwo,\ktre) \,=\,
  \pert{\T}{1}_X(\kone)\,\pert{\T}{1}_X(\ktwo)\;
  \K{\pert{\T}{2}_X(\konep,\ktwop,\ktre)\,
  S_\Phi(\kone,\ktwo,\konep,\ktwop)} 
  \;+\; \text{2 perm.} \;,
  \label{eq:next_to_leading_order_bispectrum_trisp}
\end{align}
where \konep and \ktwop are convolution variables. Note that in a statistically isotropic and homogeneous Universe, the trispectrum can only depend on 6 scalars parameters.

The bispectrum of the cosmological perturbation $X$, today, is given by the sum of $B^\lin$ (\eref{eq:linear_bispectrum_k1k2k3}), $B^\intr$ (\eref{eq:intrinsic_bispectrum_k1k2k3}) and $B^\text{trisp}$ (\eref{eq:next_to_leading_order_bispectrum_trisp}):
\begin{align}
  \label{eq:total_bispectrum_k1k2k3}
  B_X(\kone,\ktwo,\ktre) \,&\simeq\,
  \pert{\T}{1}_X(\kone)\,\pert{\T}{1}_X(\ktwo)\:\pert{\T}{1}_X(\ktre)\;
  B_\Phi(\kone,\ktwo,\ktre) \msk
  &+\,2\,\pert{\T}{1}_X(\kone)\,\pert{\T}{1}_X(\ktwo)\;
  \pert{\T}{2}_X(\kone,\ktwo,\ktre)\,
  P_\Phi(\kone) \, P_\Phi(\ktwo)  \,+\, \; \text{2 perm.} \nmsk
  &+\,\pert{\T}{1}_X(\kone)\,\pert{\T}{1}_X(\ktwo)\;
  \K{\pert{\T}{2}_X(\konep,\ktwop,\ktre)\,
  S_\Phi(\kone,\ktwo,\konep,\ktwop)} \,+\, \; \text{2 perm.} \; \notag
\end{align}
where the permutations refer only to those terms including the second-order transfer function, $\pert{\T}{2}$. The above relation neglects the infinite series of terms of order $\O(\Phi^5)$ or higher which involve the higher-order transfer functions. These terms are naturally suppressed due to the smallness of the primordial potential. Since photon perturbations do not grow, their transfer function stays small, too; as a result, they are negligible and the CMB bispectrum is well approximated by \eref{eq:total_bispectrum_k1k2k3}. When considering cold dark matter, however, the smallness of the potential is compensated by the quick growth of the high-order transfer functions on subhorizon scales, so that the relation in \eref{eq:total_bispectrum_k1k2k3} ceases to be accurate on small scales and at late times.

\subsubsection{The importance of the intrinsic bispectrum of the CMB}

We shall now focus on the bispectrum of the CMB temperature perturbation, $\Theta$. 
The first question to ask is: which of the three contributions to the CMB bispectrum in \eref{eq:total_bispectrum_k1k2k3} is dominant? If we assume that all the transfer functions are of order unity, which is a reasonable assumption for the photon perturbations during all epochs, the relative size of the various terms is determined by the statistics of the primordial field, $\Phi$. Since the amplitude $A_s$ of the primordial power spectrum is known from the CMB ($A_s \simeq \sci{2.5}{-9}$ \cite{planck-collaboration:2013a, hinshaw:2012a, smoot:1992a}), it makes sense to express the primordial bispectrum\index{primordial bispectrum} in terms of $P(k)\,$. In the simple local template \cite{komatsu:2001a, gangui:1994a, verde:2000a}, the bispectrum is parametrised by a single amplitude, \fnl:
\begin{align}
  B_\Phi(\kone,\ktwo,\ktre) \,=\, 2\,\fnl\,P_\Phi(\kone)\,P_\Phi(\ktwo) \;+\; \text{2 permutations} \;.
  \label{eq:perturbations_local_shape}
\end{align}
The local shape is just one of the several physically motivated shapes that are commonly used in the literature to parametrise the primordial bispectrum. In \cref{ch:intrinsic}, we shall introduce the other shapes and relate them to actual models of cosmic inflation; for the time being we shall assume the local shape only to provide order-of-magnitude estimates of the various bispectra.

\runinhead{Estimate of the linearly propagated bispectrum} It is not simple to make an estimate of \fnl based on physical insight, because its exact value depends on the largely unknown details of cosmic inflation. For a mildly non-Gaussian random field, we would expect the three-point function to be of order $\Phi_\text{rms}^3 = P(k)^{3/2}$, which, given that $A_s^{\nicefrac{1}{2}}\sim\sci{5}{-5}$ corresponds to a value of $\fnl \sim 10^4$; most models of inflation, however, tend to favour lower values. The Planck collaboration \cite{planck-collaboration:2013b} has recently produced the most stringent constraints to date on the non-Gaussianity of the cosmic microwave background by measuring its bispectrum. Their result highlights that the CMB is almost perfectly Gaussian, with an estimate of $\fnl = 2.7 \pm 5.8$ for the local shape. If we take into account the definition of \fnl in \eref{eq:perturbations_local_shape}, this constraint translates to an upper limit for the absolute value of $B_\Phi(\kone, \ktwo, \ktre)$ of roughly $20\times P(k)^2$ at 95\% CL.

\runinhead{Estimate of the intrinsic bispectrum} Chapters \ref{ch:evolution} and \ref{ch:intrinsic} will be devoted to the numerical computation of the intrinsic bispectrum of the CMB. This is a formidable task that requires solving the Boltzmann-Einstein system of differential equations and estimating several multi-dimensional oscillating integrals. An order of magnitude estimate, however, is already possible at this stage. Armed with the knowledge that the photon perturbations do not grow with time, and that their transfer functions start with an amplitude close to unity, we can see from \eref{eq:intrinsic_bispectrum_k1k2k3} that the intrinsic bispectrum should be roughly of the same order of magnitude as $2\times P(k)^2$. 

\runinhead{Estimate of the trispectrum term}
The primordial trispectrum is usually parametrised using two amplitudes, \taunl and \gnl. The former, \taunl, is not independent from \fnl and, for most models of inflation, is proportional to $\fnl^2$. The latter, \gnl, is the independent degree of freedom that represents the amplitude of the actual intrinsic cubic non-linearities in the primordial potential \cite{planck-collaboration:2013b}. The two amplitudes appear as proportionality constants between the primordial trispectrum and terms of order $A_s^3$ involving the product of three power spectra. Thus, for the trispectrum contribution in \eref{eq:total_bispectrum_k1k2k3} to be of the same order as the linear and intrinsic ones, either \taunl or \gnl needs to be of order $A_s^{-1} \sim \sci{4}{8}$. Both the upper limits from the Planck team, $\abs{\taunl} < 2800$ at 95\% CL, and from \citet{smidt:2010a}, $\abs{\gnl} < \sci{8}{5}$ at 95\% CL, fall short of that value. Therefore, in the following we shall always neglect the trispectrum contribution to the observed bispectrum.\\

In summary, the observed bispectrum of the CMB is well approximated by two contributions of potentially comparable size: the linearly propagated bispectrum, which is directly related to the physics of the early Universe and vanishes for Gaussian initial conditions (\eref{eq:linear_bispectrum_k1k2k3}), and the intrinsic bispectrum, whose amplitude and shape are fixed by the non-linear physics of gravity and radiation transfer (\eref{eq:intrinsic_bispectrum_k1k2k3}). The linear bispectrum carries information on the early Universe that is directly linked to the parameters of the many models of cosmic inflation, as we shall see in \cref{ch:intrinsic}. However, Planck has posed strong constraints on the linear bispectrum which suggest that, if it exists, then it must be of comparable size or smaller than the intrinsic one. In order to extract the primordial information from the CMB bispectrum, it is therefore needed to precisely compute the shape and amplitude of the intrinsic signal, which in this context acts as a source of systematic. In fact, this is one of the main reason that motivated us into developing \SONG and computing the intrinsic bispectrum.

\chapterbib


\chapter{The Boltzmann equation}
\label{ch:boltzmann}

\section{Introduction}

The Einstein equations
(Eq.~\ref{eq:einstein_pure_timetime} to \ref{eq:einstein_quadratic_sources})
need to be completed by a model of matter that specifies the form and evolution of the energy-momentum tensor in terms of the matter and metric variables. The fluid model
that we have introduced in \sref{sec:perturbations_energy_momentum_tensor}
provides a good description for the massive species of the Universe (cold dark matter and baryons) but is not adequate to represent the detailed evolution of the relativistic species (photons and neutrinos).
An alternative and more general model of matter is provided by the \keyword{kinetic theory} \emph{of gases in general relativity}, which is the main subject of this chapter.

The kinetic theory postulates that dilute\annotate{In a dilute gas the number of occupied states (sum of the occupation number) is much smaller than the number of available states.}
matter is formed by a discrete system of particles whose overall dynamics can be interpreted as a stochastic process. The physically relevant and macroscopic properties of the system, such as the energy density or pressure, are described by smooth expectation values \cite{ehlers:1971a}.
The main ingredient of the theory is the \keyword{phase-space density} or \keyword{one-particle distribution function}, $f(\tau,\vec{x},\vec{p})$, defined so that, for an observer sitting at the space-time point $(\taux)$ and adopting a local inertial frame,
\begin{equation*}
  \dd N \;=\; f \left(\tau,\vec{x},\,\vec{p}\right) \dd\vec{x}\;\dd\vec{p}
\end{equation*}
is the average number of particles in the volume element $\dd\vec{x}\;\dd\vec{p}$ at the position $(\vecx,\vec{p})$ in phase space. This definition highlights the statistical nature of the kinetic treatment: rather than focussing on the behaviour of the single particles, the system is characterised by a probability distribution in phase space. All possible measurements of numbers, energies, and directions of travel of a flux of particles can be described as an integral over the distribution function.

In the kinetic theory picture, the interactions between the particles in the system can be divided in \emph{long range forces} and \emph{short range forces} according to the following scheme. The long range forces are described by a mean field generated collectively by the particles through macroscopic field equations. Gravity belongs to this category, as the gravitational field is sourced by the particles through the Einstein field equations, with the particles, in turn, following geodesic trajectories under the action of the field. (Another example of long range force is the Lorentz force generated through the Maxwell equations, but we will not consider it.) The short range forces, instead, are treated in terms of point-collisions whose probability of occurrence is governed by cross-sections taken from a special-relativistic scattering theory \cite{ehlers:1974a}. This is the case for the interaction between the photons and the baryons prior to recombination and after reionisation, which is governed by the Compton scattering cross-section. Another assumption of the theory is that, between collisions, the particles move like test particles in the mean field.

The two types of interaction determine the form of the distribution function of a system of particles through the \keyword{Boltzmann equation}:
\begin{align}
  \diff{f}{\lambda} \;=\; C[f] \;.
  \label{eq:raw_boltzmann_equation}
\end{align}
The \keyword{Liouville term}, $\dd f/\dd\lambda$, represents the change of $f$ as measured by an observer that follows the flow of the particles. Said flow is caused by the action of the long range forces and, since we consider only the gravitational interaction, it consists of geodesic trajectories parametrised by the affine parameter $\lambda$. The short range forces, on the other hand, are encoded in the \keyword{collision term}, $C[f]$, that is the average rate at which the particles' momenta change due to collisions.

In the absence of collisions, the Boltzmann equation is called the \keyword{Liouville equation}\index{collisionless Boltzmann equation},
\begin{align}
  \diff{f}{\lambda} \;=\; 0 \;,
  \label{eq:collisionless_boltzmann_equation}
\end{align}
which implies that the distribution function is conserved along geodesic trajectories.
Stated differently, observers that drift along with the particles that surround them do not perceive a change in the local density. However, if the particles start interacting through collisions, even the geodesic observers will witness a change in their momenta and, therefore, in the local phase-space distribution. Note that, $f$ being an average quantity, the Liouville equation applies also in the presence of collisions that are in detailed balance, \ie as long as the direct collisions are equilibrated by the inverse ones. This is the case for fluids in thermal equilibrium, such as the photons and the baryons before recombination.
\annotate{Detailed balance: At equilibrium, each elementary process should be equilibrated by its reverse process.}


\subsection{Summary of the chapter}
The raw Boltzmann equation in \eref{eq:raw_boltzmann_equation} is of little practical use. In this chapter we shall turn it into an evolution equation for the temperature and polarisation anisotropies of the CMB by
\begin{enumerate}
  \item expressing it in terms of the metric and matter variables, up to second order in the cosmological perturbations, and by
  \item projecting its positional (\vecx), angular (\n) and momentum ($p$) dependences so that it turns into a system of ordinary differential equations that is numerically tractable.
\end{enumerate}

To do so, we first introduce in \sref{sec:local_inertial_frame} the local inertial frame as a convenient tool to derive the collision term and to express the energetics of the system. In \sref{sec:distribution_function} we show how to expand the CMB distribution function around its equilibrium form, the blackbody spectrum; we shall also treat the issue of defining a temperature at second order. In \sref{sec:liouville_term} we derive the Liouville term, that is the part of Boltzmann equation that encodes the effect of the geodesic motion of the particles on the distribution function. In \sref{sec:collision_term} we shall compute the collision term for the Compton scattering at recombination that, complemented with the Liouville term, will allow us to obtain the evolution equation for the temperature and polarisation anisotropies of the CMB.

\subsection{Literature review}
For a detailed review of kinetic theory and of its many uses in cosmology and astrophysics, refer to the works by Ehlers \cite{ehlers:1971a, ehlers:1974a} and Lindquist \cite{lindquist:1966a}, and to the book by Bernstein \cite{bernstein:1988a}. An early application of the theory to predict the first-order CMB fluctuations can be found in \citet{peebles:1970a}. 

The collision term at second order in the cosmological perturbations was obtained independently by \citet{dodelson:1995a} and \citet{hu:1994a} in a systematic way, in the context of cosmic reionisation, assuming azimuthal symmetry of the perturbations. This assumption does not hold in general at second order, where vorticity naturally arises even for scalar initial conditions (\sref{sec:perturbations_svt_decomposition}). \citet{bartolo:2006a} computed the collision term in the general case and complemented it with the Liouville term in Newtonian gauge.\footnote{Note that some mistakes in their equations were reported and corrected by \citet{pitrou:2009a} and \citet{senatore:2009b}.} \citet{senatore:2009b} provided a way to compute the evolution of the perturbed electron density, thus completing the derivation of the second-order collision term for the CMB temperature fluctuations. 

\citet{pitrou:2009a} and \citet{beneke:2010a} paved the way to a precise numerical integration of the system by independently including the effect of polarisation in the second-order Boltzmann equation. More recently, \citet{naruko:2013a} did the same but without fixing a particular gauge; they also studied in detail the generation of spectral distortions in the CMB temperature and polarisation. Note that the authors of Ref.~\cite{senatore:2009b, pitrou:2009a, beneke:2010a, naruko:2013a} performed their computations in the local inertial frame by employing a tetrad approach.



\section{The local inertial frame}
\label{sec:local_inertial_frame}

The collision term in the Boltzmann equation, $C[f]$, is determined by the cross-section of the Compton scattering, a local quantity that is known in the flat Minkowskian space of special relativity. Rather than deriving the collision term in a curved space-time, it is preferable to adopt a frame where $C[f]$ assumes the simple Minkowskian form. This is achieved by employing a set of orthonormal \keyword{tetrads} whereby the components of the metric are equal to those of the flat Minkowski metric. In this so-called \keyword{local inertial frame}, we can use the Compton scattering cross-section computed in flat space and thus derive a collision term that is free from metric fluctuations; in fact, all the metric fluctuations will be confined to the Liouville term \cite{senatore:2009b, naruko:2013a}.

Another advantage of computing the Boltzmann equation in the local inertial frame is that it allows to separate the energy, momentum and direction of a particle in a covariant manner. (For example, in the local inertial frame the mass shell relation assumes the special relativity form, $E^2=p^2+m^2$.) At linear order, this property can be used to simplify the Boltzmann equation without making the tetrads machinery explicit \cite{ma:1995a, dodelson:2003b}; at second order, however, this is no longer the case.

In the next subsection, we briefly introduce the tetrad formalism following the approach in Chapter~1 of \citet{chandrasekhar:1992a} and Appendix~J of \citet{carroll:2004a}. In \sref{sec:tetrads_NG} we show the explicit form of the tetrad in Newtonian gauge up to second order, while in the rest of the section we give formulae for the four-momentum (\sref{sec:tetrad_four_momentum}) and the energy-momentum tensor (\sref{sec:tetrad_emt}) that relate their components in the tetrad and coordinate frames.



\subsection{Tetrad formalism}
\label{sec:tetrad_formalism}

The tangent space of a space-time point is spanned by a basis of four contravariant vectors which are collectively called the \keyword{tetrad}. The choice of the tetrad is arbitrary and it defines the \keyword{reference frame} in that point. Because all vectors and tensors, most notably the four-momentum and the energy-momentum tensor, live in the tangent space, their components depend on the chosen tetrad.

Being geometrical objects, the tetrads exist regardless of the coordinate system. Once we pick one, however, it is natural to define a \keyword{coordinate tetrad} as the directional derivatives with respect to the coordinates, $\partial/\partial x^\mu$. Following the notation used in Chapter~1 of \citet{chandrasekhar:1992a}, we express a general tetrad in terms of the coordinate ones as
\begin{align}
  \tD{\vece}{a} \;=\; \e{a}{\mu}\;\pfrac{}{x^\mu} \; \quad\quad (a=0,1,2,3) \;,
  \label{eq:tetrad_using_coordinate}
\end{align}
where the tetrad indices are underlined to distinguish them from the usual coordinate ones.
To make the distinction clearer, we shall also use the Latin letters $a,b,c$ to denote the tetrad indices instead of the Greek ones ($a=0,1,2,3$).
We can also define an \emph{inverse tetrad} that spans the dual tangent space:
\begin{align}
  \tU{\vece}{a} \;=\; \eInv{a}{\mu}\;\dd \vecx^\mu \;,
  \label{eq:inverse_tetrad_using_coordinate}
\end{align}
with the inverse coefficient matrix, $\eInv{a}{\mu}$, given by
\begin{align}
  \e{a}{\mu}\,\eInv{b}{\mu}\,=\,\tUD{\delta}{b}{a} \quad\quad\text{and}\quad\quad
  \e{a}{\mu}\,\eInv{a}{\nu}\,=\,{\delta^{\,\mu}}_{\nu} \;.
  \label{eq:inverse_tetrads_coefficient_matrix}
\end{align}
The existence of the inverse tetrad allows us to express the coordinate bases in terms of the tetrad ones by contracting Eq.~\ref{eq:tetrad_using_coordinate} and \ref{eq:inverse_tetrad_using_coordinate} with $\eInv{a}{\nu}$ and $\e{a}{\nu}$, respectively:
\begin{align}
  \pfrac{}{x^\nu} \;=\; \eInv{a}{\nu}\,\tD{\vece}{a} \quad\quad\text{and}\quad\quad
  \dd\vecx^\nu \;=\; \e{a}{\nu}\,\tU{\vece}{a} \;.
  \label{eq:coordinate_using_tetrad}
\end{align}

Any vector $\vec{V}$, 1-form $\vec\omega$ or tensor $\vec{T}$ can be represented using either the coordinate basis or the tetrad basis:
\begin{align}
  &\vec{V} \;=\; V^\mu\,\pfrac{}{x^\mu} \;=\; \tU{V}{a}\,\tD{\vece}{a} \;, \nmsk
  &\vec{\omega} \;=\; \omega_\mu\,\dd \vecx^\mu \;=\; \tD{\omega}{a}\,\tU{\vece}{a} \;, \nmsk
  &\vec{T} \;=\; {T^\mu}_\nu\,\left(\pfrac{}{x^\mu}\,\otimes\,\dd \vecx^\nu\right) \;=\;
  \tUD{T}{a}{b}\,\left(\tD{\vece}{a}\,\otimes\,\tU{\vece}{b}\right) \;.
\end{align}
After expanding the tetrad in the above identities according to Eq.~\ref{eq:tetrad_using_coordinate} and \ref{eq:inverse_tetrad_using_coordinate}, we see that the components in the two frames are related by
\begin{align}
  &V^\mu\,=\,\tU{V}{a}\,\e{a}{\mu} 
  &\text{and}\quad\quad\quad
  &\tU{V}{a}\,=\,V^\mu\,\eInv{a}{\mu} \;,\nmsk
  &\omega_\mu\,=\,\tD{\omega}{a}\,\eInv{a}{\mu} 
  &\text{and}\quad\quad\quad
  &\tD{\omega}{a}\,=\,\omega_\mu\,\e{a}{\mu} \;,\nmsk
  &{T^\mu}_\nu\,=\,\tUD{T}{a}{b}\,\e{a}{\mu}\,\eInv{b}{\nu}
  &\text{and}\quad\quad\quad
  &\tUD{T}{a}{b}\,=\,{T^\mu}_\nu\,\eInv{a}{\mu}\,\e{b}{\nu} \;,  
\end{align}
for vectors, 1-forms and tensors, respectively. Therefore, a covariant (contravariant) tetrad index can be turned into a covariant (contravariant) coordinate index by contraction with the (inverse) tetrad coefficient matrix. This also implies that the contraction between two tensors yields the same result regardless of whether it is carried over their tetrad or coordinate indices. For example,
\begin{align}
  V^\mu\,\omega_\mu \;=\; \tU{V}{a}\,\tD{\omega}{a} \qquad\text{and}\qquad
  {T^\mu}_\nu\,\omega_\mu\,V^\nu\,=\, \tUD{T}{a}{b}\,\tD{\omega}{a}\,\tU{V}{b} \;.
\end{align}

The metric in tetrad indices, $\tDD{M}{a}{b}$, is obtained by contracting the coordinate metric, $g_{\mu\nu}$, with two tetrads:
\begin{align}
  g_{\mu\nu}\,\e{a}{\mu}\,\e{b}{\nu} \;=\; \tDD{M}{a}{b} \;.
\end{align}
Unsurprisingly, the inverse relation involves the contraction with two inverse tetrads:
\begin{align}
  g_{\mu\nu} \;=\; \tDD{M}{a}{b}\,\eInv{a}{\mu}\,\eInv{b}{\nu} \;.
\end{align}
The metric $\tDD{M}{a}{b}$ and its inverse $\tUU{M}{a}{b}$ can be used to lower and raise the tetrad indices, respectively. For a vector $\vec{V}$, this can be proven by expanding $V^\nu=\e{a}{\nu}\,\tU{V}{a}$ and $V_\nu=\eInv{a}{\nu}\,\tD{V}{a}$ in the identity $g_{\mu\nu}V^\nu=V_\mu$ and by later contracting the result with $\e{b}{\mu}$. 
In general, it is easy to prove all the following relations:
\begin{align}
  &\tD{V}{a} \,=\, \tDD{M}{a}{b}\,\tU{V}{b} \;,
  &&\tU{\omega}{a} \,=\, \tUU{M}{a}{b}\,\tD{\omega}{b} \; &&\text{and}
  &&\tUD{T}{a}{b} \,=\, \tDD{M}{b}{c}\,\tUU{T}{a}{c} \;.
\end{align}

As we pointed out in the introduction to the section, it is convenient to express the Boltzmann equation in terms of a tetrad that is orthonormal:
\begin{align}
  g_{\mu\nu}\,\e{a}{\mu}\,\e{b}{\nu} \;=\; \tDD{\eta}{a}{b} \;,
  \label{eq:orthonormal_tetrads}
\end{align}
where $\tDD{\eta}{a}{b}$ are the components of Minkowski's metric and are constant. (This is equivalent to setting $\tDD{M}{a}{b}=\tDD{\eta}{a}{b}$ in the above equations.) The resulting frame is called the \keyword{local inertial frame}.

The orthonormality condition determines only 10 out of the 16 components of the tetrad matrix, $\e{a}{\mu}$. The remaining 6 degrees of freedom correspond to a Lorentz boost and to a rotation of the tetrad base with respect to the coordinate axes (see Appendix~J in \citet{carroll:2004a} and the note 15 in \citet{senatore:2009b}). We choose the tetrad so that they are at rest with a comoving observer, \ie an observer with constant spatial coordinates. This is achieved by setting $\tD{e}{0} \;\propto\; \partial/\partial\tau$, where $\tau$ is the time coordinate\footnote{Note that \citet{senatore:2009b} (Sec.~4.1) and \citet{beneke:2010a} (Sec.~I) use the same convention, while \citet{pitrou:2009a} (Sec.~4.2.2) and \citet{naruko:2013a} (Sec.~2.1), instead, choose the tetrad to be orthogonal to constant time hypersurfaces, that is $\tU{e}{0}\,\propto\,\dd\tau$. See Sec.~5.3.1 of \citet{pitrou:2009a} for further details.},
which, by virtue of \eref{eq:tetrad_using_coordinate}, is equivalent to have
\begin{align}
  \e{0}{i} \;=\;0 \;.
  \label{eq:tetrad_choice_velocity}
\end{align}
We fix the other three degrees of freedom by setting 
\begin{align}
  \e{i}{j} \;=\; \e{j}{i} \;,
\end{align}
which corresponds to asking that there is no rotation between the background and the perturbed tetrads. 
The two constraints that we have just discussed correspond to choosing one out of the infinitely many local inertial frames; for simplicity, from now on we shall use the term ``local inertial frame'' to denote this particular choice. We shall also refer to an observer with vanishing spatial velocity in the inertial frame as an inertial observer.

\subsection{Tetrads in Newtonian gauge}
\label{sec:tetrads_NG}

The tetrad components for the local inertial frame in Newtonian gauge are obtained by applying the orthonormality condition,
\begin{align}
  &g_{\mu\nu}\,\e{a}{\mu}\,\e{b}{\nu} \;=\; \tDD{\eta}{a}{b} \;,
\end{align}
to the expanded metric in \eref{eq:the_expanded_ng_metric} and by fixing the velocity and orientation of the local frame with respect to the coordinate axes,
\begin{align}
  &\e{0}{i} \;=\;0 \quad\quad\text{and}\quad\quad \e{i}{j} \;=\; \e{j}{i} \;.
\end{align}
The components of the inverse tetrad can be obtained from the direct ones as $\eInv{a}{\mu}\,=\,g_{\mu\nu}\,\tUU{\eta}{a}{b}\e{b}{\nu}$.
By doing so, we obtain the following expression up to second-order accuracy\footnote{The expression coincides with the one in Eq.~4.4 of Ref.~\cite{senatore:2009b} once we convert our potentials to the ``exponential'' ones using \eref{eq:bmr_ours_potential_sqrt}, but differs from the one in Ref.~\cite{naruko:2013a} due to the different choice of tetrads.}:
\begin{align}
  \label{eq:tetrad_explicit}
  & a\;\e{0}{0} \;=\; \frac{1}{\sqrt{1+2\Psi}}\;,
  &&\Nicefrac{\eInv{0}{0}}{a} \;=\; \sqrt{1+2\Psi} \;, \nmsk
  & a\;\e{0}{i} \;=\; 0\;,
  &&\Nicefrac{\eInv{0}{i}}{a} \;=\; -\omega_i \;,\nmsk
  & a\;\e{i}{0} \;=\; \omega_i \;,
  &&\Nicefrac{\eInv{i}{0}}{a} \;=\; 0 \;, \nmsk
  & a\;\e{i}{j} \;=\; \frac{{\delta^j}_{i}}{\sqrt{1-2\Phi}} \,-\, {\gamma^{j}}_i \;,
  &&\Nicefrac{\eInv{i}{j}}{a} \;=\; {\delta^i}_{j}\,\sqrt{1-2\Phi} \,+\, {\gamma^{i}}_j \;,
\end{align}
which is straightforwardly expanded into perturbative orders by enforcing \eref{eq:perturbations_expansions}. To obtain the same expression in terms of the ``exponential'' potentials of \eref{eq:bmr_metric}, one has to substitute the square root factors with exponentials according to \eref{eq:bmr_ours_potential_sqrt}. Note that, had we not neglected the vector and tensor modes at first order, the tetrad components would have included extra quadratic contributions (for example, see Eq.~2.6 and 2.7 of \citet{naruko:2013a}).

\subsection{The four-momentum}
\label{sec:tetrad_four_momentum}

We parametrise the four-momentum of a particle in the local inertial frame as
\begin{align}
  \tU{p}{a} \;=\; \left( E,\, p\,\tU{n}{i} \right) \;,
  \label{eq:E_p_n_split_of_momentum}
\end{align}
where we have introduced the energy, $E$, the momentum, $p$, and the direction of propagation, $\tU{n}{i}$, of the particle. The momentum is defined as $p=\sqrt{\tD{p}{i}\tU{p}{i}}$, which implies that $\tD{n}{i}\tU{n}{i}=1$. The energy and the momentum are related by the mass-shell relation:
\begin{align}
  \DD{g}{\mu}{\nu}\,p^\mu\,p^\nu \,=\,
  \tDD{\eta}{a}{b}\,\tU{p}{a}\,\tU{p}{b} \,=\, -m^2 \;,
\end{align}
which, given the diagonal form of \tDD{\eta}{a}{b}, implies that
\begin{align}
  E^2 \,=\, p^2 \,+\, m^2 \;,
  \label{eq:mass_shell}
\end{align}
where $m$ is the rest mass of the considered particle. For this reason, the tetrad momentum $\tU{p}{a}$ is also called the \keyword{proper momentum} \cite{bertschinger:1996a, ma:1995a}. In general, being able to split energy, momentum and direction in a covariant way is one of the advantages of using orthonormal tetrads. We also define the velocity in the local inertial frame as
\begin{align}
  \tU{v}{i} \,\equiv\, \frac{\tU{p}{i}}{\tU{p}{0}} \,=\, \frac{p}{E}\,\tU{n}{i} \:.
\end{align}
For massless particles such as photons, $p=E$ and the velocity is just $\tU{v}{i}=\tU{n}{i}$.

In order to study the trajectory of a particle as seen in the local inertial frame, we need a dictionary to translate the tetrad four-momentum into the coordinate one. This is provided by the relation
\begin{align}
  p^\mu \;=\; \e{a}{\mu}\,\tU{p}{a} \;,
\end{align}
which, up to second-order accuracy, results in
\begin{align}
  & p^0 \;=\;\frac{E}{a\,\sqrt{1+2\Psi}}
  \,\left(1 \,+\, \frac{p}{E}\:\omega_i\,\tU{n}{i}\right)
  \;,\nmsk
  & p^i \;=\;\frac{p\,\tU{n}{i}}{a\,\sqrt{1-2\Phi}}\,
  \left({\delta^i}_j \,-\,{\gamma^i}_j\right) \;,
  \label{eq:four_momentum_tetrad_coordinate_dictionary}
\end{align}
or, in terms of the exponential potentials $\Psi_e$ and $\Phi_e$,
\begin{align}
  & p^0 \;=\;\frac{E}{a}\;e^{-\Psi_e}\;
  \left(1 \,+\, \frac{p}{E}\:\omega_i\,\tU{n}{i}\right)
  \;,\nmsk
  & p^i \;=\;\frac{p\,\tU{n}{i}}{a}\;e^{\Phi_e}\;
  \left({\delta^i}_j \,-\,{\gamma^i}_j\right) \;.
  \label{eq:four_momentum_tetrad_coordinate_dictionary_exponentials}
\end{align}
(Note that we have used the fact that $\omega$ and $\gamma$ are second-order quantities to pull out of the parentheses the scalar potentials.) By explicitly expanding the perturbations up to second order, we obtain
\begin{align}
  & p^0 \;=\; \frac{E}{a}\, \left(1\,-\,\pert{\Psi}{1}\,-\,\pert{\Psi}{2}\,+\,\frac{3}{2}\,
  \pert{\Psi}{1}\,\pert{\Psi}{1} \,+\, \frac{p}{E}\:\omegaSO_i\,\tU{n}{i}\right) \;,\nmsk
  & p^i \;=\; \frac{p\,\tU{n}{i}}{a}\,\left[\,{\delta^i}_j
  \left(1\,+\,\pert{\Phi}{1}\,+\,\pert{\Phi}{2}\,
  +\,\frac{3}{2}\,\pert{\Phi}{1}\,\pert{\Phi}{1}\right)
  \,-\,{\gamma^{(2)i}}_j\,\right] \;.
\end{align}
It should be noted that an observer who stands still in the local frame ($\tU{p}{i}/\tU{p}{0}=0$) is comoving with the coordinates ($\dd x^i/\dd\tau=p^i/p^0=0$); this is a direct consequence of having chosen the tetrads such as $\tD{e}{0} \propto \partial/\partial \tau$ in \sref{sec:tetrad_formalism}. Had we chosen, for example, $\tU{e}{0} \propto \dd\tau$, we would have had $\dd x^i/\dd\tau \propto \omega^i$ when $\tU{p}{i}/\tU{p}{0}=0$, instead.

In the following, we shall use the variables of the local inertial frame, $p$ and $\tU{n}{i}$, to reparametrise the momentum dependence in the distribution function. With an abuse of notation, we denote the functional dependence in the new variables with the same letter, $f$:
\begin{align}
  f(\tau,x^i,p,\tU{n}{i}) \,=\, f(\tau,x^i,p^i(\tau,x^i,p,\tU{n}{i})) \;.
\end{align}
Moreover, for the sake of readability we shall, drop the underlining of the tetrad index for the direction of propagation of a particle in the local inertial frame: $n^i=\tU{n}{i}$.

\subsection{The energy momentum tensor}
\label{sec:tetrad_emt}

We compute the evolution of the matter species (photons, neutrinos, baryons and cold dark matter) by solving the Boltzmann equation in the local inertial frame; the matter perturbations thus obtained source the Einstein equation via the energy momentum tensor, $\,\tUD{T}{a}{b}\,$,
\begin{align}
  \UD{G}{\mu}{\nu} \;=\; \kappa\;\UD{T}{\mu}{\nu} \;=\;
  \kappa\;\e{a}{\mu}\,\eInv{b}{\nu}\:\tUD{T}{a}{b} \;.
\end{align}
In this subection we address three important questions, that is
\begin{enumerate}
  \item what is the explicit transformation that relates the energy-momentum tensor in the local inertial frame (which is what we obtain by evolving the Boltzmann equation) to that in the coordinate frame (which is the one that appears in the Einstein equation);
  \item how to relate the moments of the distribution function, $f_\lm\,$, to the energy-momentum tensor, and
  \item what is the relation between such multipoles and the fluid variables (energy density, pressure, velocity and shear) that we have introduced in \sref{sec:perturbations_energy_momentum_tensor}.
\end{enumerate}

\subsubsection{From $\,\tUD{T}{a}{b}\,$ to $\,\UD{T}{\mu}{\nu}\,$}

The energy-momentum tensor in the coordinate frame is related to $\tUD{T}{a}{b}$ by
\begin{align}
  {T^\mu}_\nu \;=\; \e{a}{\mu}\,\eInv{b}{\nu}\:\tUD{T}{a}{b} \;.
  \label{eq:emt_tetrad_coordinate_relation}
\end{align}
The explicit form of ${T^\mu}_\nu$ in terms of inertial-frame variables is obtained by inserting in the above expression the tetrad components of \eref{eq:tetrad_explicit}. The result up to second order is remarkably simple:
\begin{align}
  & \UD{T}{0}{0} \;=\; \tUD{T}{0}{0} \;, \nmsk
  & \UD{T}{i}{0} \;=\; \tUD{T}{i}{0} \; (1\,+\,\Psi\,+\,\Phi) \;, \nmsk
  & \UD{T}{0}{i} \;=\; \tUD{T}{0}{i} \; (1\,-\,\Psi\,-\,\Phi)
  \,-\, \tUD{T}{0}{0}\;(w\,+1)\;\omega_i\, \;, \nmsk
  & \UD{T}{i}{j} \;=\; \tUD{T}{i}{j} \;.
  \label{eq:emt_tetrad_coordinate_relation_explicit}
\end{align}
where we have used the fact that $\tUD{T}{i}{0}$ and $\tUD{T}{0}{i}$ vanish in the isotropic background, and we have introduced the \indexword{barotropic parameter} $w$, a background quantity defined by
\begin{align}
  \tUD{{\pert{T}{0}}}{i}{j} \,=\, -\,\KronUD{i}{j}\;w\;\tUD{{\pert{T}{0}}}{0}{0} \;,
\end{align}
which in terms of the fluid variables (\sref{sec:perturbations_energy_momentum_tensor}) simply reads $\,w=\frac{\bar{P}}{\bar{\rho}}\,$.
The relation between the energy-momentum tensor in the inertial and coordinate frames is particularly simple for two reasons. First, the tetrad components are simple to start with, because we are neglecting the first-order part of the vector and tensor modes in the metric. Secondly, and more subtly, the formula for the up-down version of the energy-momentum tensor, \eref{eq:emt_tetrad_coordinate_relation}, contains the product of a tetrad with its inverse, which results in a cancellation when both of $T$'s indices are either temporal or spatial. Had we instead used the up-up version, $\,T^{\mu\nu}=\e{a}{\mu}\,\e{b}{\nu}\,\tUU{T}{a}{b}\,$, we would have obtained a more complicated relation whereby $\,T^{00} \neq \tUU{T}{0}{0}\,$ and $\,T^{ij} \neq \tUU{T}{i}{j}\,$.

It should be noted that, to first order accuracy, the components of the energy-momentum tensor are the same in the coordinate and tetrad frames. This is a confirmation of what we anticipated in the introduction to the section: at first order introducing the tetrads is not necessary to derive the correct equations. At second order, however, there are corrections to $\UD{T}{i}{0}$ and $\UD{T}{0}{i}$ that cannot be neglected.

\subsubsection{Multipole decomposition of $\,\tUD{T}{a}{b}\,$}

In the local inertial frame, the volume element of momentum space has the standard Lorentz invariant measure (see Sec~3.6 of \citet{ehlers:1971a} or Appendix~A.1 of \citet{senatore:2009b}).
Thus, the energy momentum tensor is simply given by
\begin{align}
  \tUU{T}{a}{b} \;=\; \int\dd \vecp\;\:\frac{\tU{p}{a}\;\tU{p}{b}}{E}\;f \;,
\end{align}
where $\dd\vecp/E$ is the (invariant) measure in the inertial frame ($\dd\vecp=\dd\tU{p}{1}\dd\tU{p}{2}\dd\tU{p}{3}$) and $f$ is the one-particle distribution function. If we separate the magnitude of the momentum from its direction as in \eref{eq:E_p_n_split_of_momentum}, the components of the energy-momentum tensor read
\begin{align}
  &\tUD{T}{0}{0} \,=\, -\int\dd p\;p^2\;E\;\int\dd\Omega\;f \;, \nmsk
  &\tUD{T}{i}{0} \,=\, -\tUD{T}{0}{i} \,=\,
  -\int\dd p\;p^2\;p\,\int\dd\Omega\;\tU{n}{i}\;f \;, \nmsk
  &\tUD{T}{i}{j} \,=\, \int\dd p\;p^2\;\frac{p^2}{E}\,\int\dd\Omega\;\tU{n}{i}\,\tD{n}{j}\;f \;,
  \label{eq:emt_tetrad_frame}
\end{align}
where we have lowered one of the indices of $\,\tUU{T}{a}{b}\,$ by contracting it with $\,\tDD{\eta}{a}{b}\,$.

We decompose the energy-momentum tensor in its spherical components using the projection vectors $\xi$ and the projection matrices $\chi$ according to the scheme shown in \eref{eq:einstein_equations_schematic_SVT}. In particular, we use the relations
\begin{align}
  \xivector{m}{i}\,n_i \;=\; \sqrt{\frac{4\,\pi}{3}}\;Y_{1m}^* \; \qquad\text{and}\qquad
  \chimatrix{2}{m}{ij}\,n_i\,n_j \;=\; \frac{2}{3}\,\sqrt{\frac{4\,\pi}{5}}\,Y_{2m}^* \;,
\end{align}
from \sref{sec:projection_of_tensors}, and the expansion in spherical harmonics of $f\,$,
\begin{align*}
  f(\n) \;=\; \sum\limits_{\L=0}^\infty\,\sum\limits_{m=-\L}^{\L}\,(-i)^\L\,\sqrt{\frac{4\pi}{2\L+1}}
  \,f_{\lm}\,Y_\lm(\n) \;,
\end{align*}
which, with respect to the usual expansion, includes extra $\ell$-dependent factors in order to simplify the Boltzmann equation (see also comment after \eref{eq:alm_vs_thetalm_1}).
By inserting the first three multipoles of $f$,
\begin{align}
  & f_{00}\;=\;\int\frac{\dd\Omega}{4\,\pi}\;f \;,
  &&f_{1m}\;=\;i\;\sqrt{\frac{3}{4\,\pi}}\;\int\dd\Omega\;f\;Y^*_{1m} \;,
  &&f_{2m}\;=\;-\sqrt{\frac{5}{4\,\pi}}\;\int\dd\Omega\;f\;Y^*_{2m} \;,
\end{align}
in \eref{eq:emt_tetrad_frame}, we find
\begin{align}
  &\tUD{T}{0}{0}  \;=\; -4\,\pi\,\int\dd p\,p^2\;E\;f_{00} \;,
  &&\tUD{T}{i}{i}  \;=\; 4\,\pi\,\int\dd p\;p^2\;\frac{p^2}{E}\;f_{00} \;, \nmsk
  &i\,\xivector{m}{i}\:\tDD{T}{i}{0}  \;=\;  -\frac{4\,\pi}{3}\,\int\dd p\;p^2\;p\;f_{1m} \;,
  &&\chimatrix{2}{m}{ij}\:\tDD{T}{i}{j} \;=\;  -\,\frac{4\,\pi}{5}\,\frac{2}{3}\,\int\dd p\;p^2\;\frac{p^2}{E}\;f_{2m} \;.
  \label{eq:energy_momentum_tensor_multipoles_f}
\end{align}
The energy-momentum tensor is therefore completely determined by the first three angular multipoles of the distribution function: the monopole $f_{00}\,$, the dipole $f_{1m}$ and the quadrupole $f_{2m}\,$. (Note that, with our conventions, $\,\tDD{T}{i}{j}\,$ are the spatial components of $\,\tUD{T}{a}{b}\,$, and not of $\,\tDD{T}{a}{b}\,$.)

\runinhead{Relativistic case}
If we consider a relativistic fluid ($p=E$), we can express the energy-momentum tensor in terms of the brightness $\Delta$ (defined in \eref{eq:brightness_definition}),
\begin{align}
  &\tUD{T}{0}{0}  \;=\; -\rhoz\;(1+\Delta_{00}) \;,
  &&\tUD{T}{i}{i}  \;=\; \rhoz\;(1+\Delta_{00}) \;, \nmsk
  &i\,\xivector{m}{i}\:\tDD{T}{i}{0}  \;=\;  -\frac{1}{3}\,\rhoz\;\Delta_{1m} \;,
  &&\chimatrix{2}{m}{ij}\:\tDD{T}{i}{j} \;=\;  -\,\frac{2}{15}\,\rhoz\;\Delta_{2m} \;,
  \label{eq:energy_momentum_tensor_multipoles_delta}
\end{align}
where
\begin{align}
  \rhoz \;=\; \int\dd\vecp\,E\,\pert{f}{0} \;=\; 4\,\pi\,\int\dd p\,p^2\,E\,\pert{f}{0} \;.
\end{align}
(Note that $\tUD{T}{a}{a}=0$, as expected from a fluid of relativistic particles.)
Because the brightness multipoles are the quantities that we actually evolve in \SONG, the above equation, complemented with the tetrad transformation in \eref{eq:emt_tetrad_coordinate_relation_explicit}, allows us to build the right hand side of the Einstein equation. In particular, it should be stressed that the second-order space-time equation will contain an extra quadratic term in $\Psi+\Phi$,
\begin{align}
  i\,\xivector{m}{i}\:\UD{T}{i}{0} \;=\; -\frac{1}{3}\,\rhoz\;\Delta_{1m}\;(1\;+\;\Psi\;+\;\Phi) \;,
\end{align}
which comes from the tetrad transformation.

\runinhead{General case} 
In order to describe an arbitrary fluid, be it relativistic or non relativistic, we introduce the \keyword{beta-moments},
\begin{align}
  1\,+\,{}_n\Delta\,(\taux,\n) \;\equiv\; \frac{1}{\int\dd p\,p^3\,\bar{f}(\tau,p)}
  \int\dd p\,p^3\;\left(\,\frac{p}{E}\,\right)^{n-1}\,f\,(\taux,p,\n) \;,
  \label{eq:n_brightness_definition}
\end{align}
so that the energy-momentum tensor in \eref{eq:energy_momentum_tensor_multipoles_f} can be recast as
\begin{align}
  &\tUD{T}{0}{0}  \;=\; -\rhoz\;(1+\bmult{\Delta}{0}{0}{0}) \;,
  &&\tUD{T}{i}{i}  \;=\; \rhoz\;(1+\bmult{\Delta}{2}{0}{0}) \;, \nmsk
  &i\,\xivector{m}{i}\:\tDD{T}{i}{0}  \;=\;  -\frac{1}{3}\,\rhoz\;\bmult{\Delta}{1}{1}{m} \;,
  &&\chimatrix{2}{m}{ij}\:\tDD{T}{i}{j} \;=\;  -\,\frac{2}{15}\,\rhoz\;\bmult{\Delta}{2}{2}{m} \;.
  \label{eq:energy_momentum_tensor_multipoles_betamoments}
\end{align}
The $\beta_n$ operator defines an expansion in the powers of the dimensionless velocity of the particle, $\,\beta=p/E\,$, hence the name. For relativistic or massless species ($p/E=1$) the beta-moments reduce to the brighness moments, that is $\,\bmult{\Delta}{n}{\L}{m}=\Delta_{lm}\,$. For non-relativistic species ($p\ll E$) the higher order beta-moments are suppressed so that only the lowest multipoles count, as in the fluid limit.
Therefore, the beta-moments allow us to treat massive and massless particles within the same framework; we shall use this property in writing the Boltzmann equation for the baryon and CDM fluids in \sref{sec:the_evolved_equations}.
As a final note, we remark that our beta-moments are equivalent to the momentum-integrated multipoles defined in \citet{lewis:2002b} (see also Ref.~\cite{ellis:1983a}).

\subsubsection{Fluid limit}
To relate the fluid variables to the moments of the distribution function, we enforce the following equality,
\begin{align}
  \tUD{T}{a}{b} \;
  &=\; \int\dd \vecp\;\:\frac{\tU{p}{a}\;\tD{p}{b}}{E}\;f \nmsk
  &=\; (\rho+P)\;\tU{U}{a}\;\tD{U}{b} \;+\; \KronUD{a}{b}\,P \;+\; \tUD{\Sigma}{a}{b} \;,
\end{align}
where the first line is the energy-momentum tensor in the local inertial frame, expressed in terms of the beta-moments via \eref{eq:energy_momentum_tensor_multipoles_betamoments}, and the second line is the fluid representation, which is expanded up to second order according to\footnote{The expansion is obtained by following the procedure in \sref{sec:perturbations_energy_momentum_tensor}, with the difference that now we are adopting the local intertial frame and, therefore, the metric is Minkowskian. In particular, we have defined $\tU{v}{i}\equiv\tU{U}{i}/a$ and we have used $U^0=(1+U^iU_i)/a$.}
\begin{align}  
  &\tUD{T}{0}{0} \;=\; - \rho \;-\; (\rhoz\,+\,\Pz)\,\tU{v}{i}\,\tD{v}{i} \;,
  &&\tUD{T}{i}{i} \;=\; 3\,P \;+\; (\rhoz\,+\,\Pz)\,\tU{v}{i}\,\tD{v}{i} \;, \nmsk
  &\tUD{T}{i}{0} \;=\; -(\rho\,+\,P)\,\tU{v}{i}  \;,
  &&\tUD{T}{i}{j} \;=\; \KronUD{i}{j}\,P\;+\;\tUD{\Sigma}{i}{j}
  \;+\;(\rhoz\,+\,\Pz)\:\tU{v}{i}\,\tD{v}{j} \;.
\end{align}
The correspondence between the moments of the distribution function and the fluid variables, up to second order, is therefore given by
\begin{align}
  \label{eq:fluid_limit}
  &\rhoz\;(1+\bmult{\Delta}{0}{0}{0}) \;=\; \rho \;+\; (\rhoz\,+\,\Pz)\,\tU{v}{i}\,\tD{v}{i} \;,
  \quad\;&&\rhoz\;(1+\bmult{\Delta}{2}{0}{0}) \;=\;
  3\,P \;+\; (\rhoz\,+\,\Pz)\,\tU{v}{i}\,\tD{v}{i} \;,\msk
  &\rhoz\;\bmult{\Delta}{1}{1}{m} \;=\; 3\;(\rho\,+\,P)\;i\,v_{[m]}  \;,
  &&\rhoz\;\bmult{\Delta}{2}{2}{m} \;=\; -\frac{15}{2}\;
  \bigl[\,\Sigma_{[m]}\;+\;(\rhoz\,+\,\Pz)\:\tensorP{m}{v}{v}\,\bigr] \;,
  \notag
\end{align}
where we have introduced the shorthand $\,\tensorP{m}{v}{v} = \chimatrix{2}{m}{ij}\,\tD{v}{i}\,\tD{v}{j}\,$.
It is clear that, at second order, the moments of $f$ do not correspond to the fluid variables. The reason is that $\rho$ and $P$ represent the energy density and the pressure for an inertial observer at rest with the fluid, while our inertial observer is at rest with the coordinates (let us recall that, in \eref{eq:tetrad_choice_velocity}, we have chosen the tetrad to correspond to observers with constant spatial coordinates, \ie $\tD{e}{0}\,\propto\,\partial/\partial\tau$). In fact, the quadratic terms in the above equation represent the Lorentz boost that brings our observer at rest with the fluid. These terms matter only at second order, so that, up to first order, the moments of the distribution function do correspond to the fluid variables,
\begin{align}
  &\rhoz\;(1+\bmult{\Delta}{0}{0}{0}) \;=\; \rho \;, 
  &&\rhoz\;(1+\bmult{\Delta}{2}{0}{0}) \;=\; 3\,P \;, \nmsk
  &\rhoz\;\bmult{\Delta}{1}{1}{m} \;=\; 3\;(\rhoz\,+\,\Pz)\;i\,v_{[m]}  \;,
  &&\rhoz\;\bmult{\Delta}{2}{2}{m} \;=\; -\frac{15}{2}\;\Sigma_{[m]} \;.
  \label{eq:fluid_limit_first_order}
\end{align}
At the background level, we have that $\,(1+\bmult{\Delta}{2}{0}{0})/3=\Pz/\rhoz\equiv w\,$.

It is convenient to express the dictionary between the moments and the fluid variables in terms of the density contrast, $\,\delta=(\rho-\rhoz)/\rhoz\,$, the barotropic parameter, $\,w(\rho)=P/\rho\,$,  and the sound of speed, $\,c_s^2=\partial P/\partial \rho\,$,
\begin{align}
  &\bmult{\Delta}{0}{0}{0} \;=\; \delta \;+\; (w+1)\,\tU{v}{i}\,\tD{v}{i} \;, \nmsk
  &1+\bmult{\Delta}{2}{0}{0} \;=\;
  3\,\left[\,\,w\,+\,\delta\,c_s^2\,
  +\,\frac{\rhoz}{2}\:\pfrac{c_s^2}{\rho}\:\delta^2 \;\right]\;
  +\; (w+1)\,\tU{v}{i}\,\tD{v}{i} \;,
  \displaybreak[0]\nmsk
  &\bmult{\Delta}{1}{1}{m} \;=\; 3\;i\,v_{[m]}\,
    \left[\,(w+1)\,+\,\delta\,(c_s^2+1)\,\right] \;, \nmsk
  &\bmult{\Delta}{2}{2}{m} \;=\; -\frac{15}{2}\;
  \left[\;\frac{\Sigma_{[m]}}{\rhoz}\;+\;(w+1)\:\tensorP{m}{v}{v}\;\right] \;,
  \label{eq:fluid_limit_w_cs}
\end{align}
where we have used the following relation for the adiabatic pressure,
\begin{align}
  P(\rho) \;=\; \rhoz\;\left[\;\,w\,+\,\delta\,c_s^2\,
  +\,\frac{\rhoz}{2}\:\pfrac{c_s^2}{\rho}\:\delta^2
  \;\right]\biggr|_{\rho=\rhoz} \;\,,
\end{align}
obtained by Taylor expanding around $\,\rho=\bar{\rho}\,$ the relation $\,P=w\,\rho\,$ up to second order.
In the following we shall treat only fluids with a constant equation of state, such as the photons ($w=c_s^2=1/3$) or the cold dark matter ($w=c_s^2=0$); in that case, the above expression reduces to
\begin{align}
  \label{eq:fluid_limit_w}  
  &\bmult{\Delta}{0}{0}{0} \;=\; \delta \;+\; (w+1)\,\tU{v}{i}\,\tD{v}{i} \;,
  \quad&&1+\bmult{\Delta}{2}{0}{0} \;=\; 3\,w\,(1+\delta)\;  
  +\; (w+1)\,\tU{v}{i}\,\tD{v}{i} \;, \msk
  &\bmult{\Delta}{1}{1}{m} \;=\; 3\;(w+1)\;i\,v_{[m]}\,(1+\delta) \;,
  &&\bmult{\Delta}{2}{2}{m} \;=\; -\frac{15}{2}\;
  \left[\;\frac{\Sigma_{[m]}}{\rhoz}\;+\;(w+1)\:\tensorP{m}{v}{v}\;\right] \;.
  \notag
\end{align}

To connect with the existing literature, we take into consideration $\,\theta\,$ and $\,\sigma\,$, the first-order fluid variables defined in \citet{ma:1995a},
\begin{align}
  &(\rhoz\,+\,\Pz)\:\theta \;\equiv\; i\,k^j\,\UD{T}{0}{j}  &&\text{and}
  &&(\rhoz\,+\,\Pz)\:\sigma \;\equiv\; -\left(\,\hat{k}_i\,\hat{k}_j\,-\,\frac{\delta_{ij}}{3}\,\right)\:\UU{\Sigma}{i}{j} \;.
\end{align}
Being a first-order definition, we can identify $\,\UD{T}{0}{j}=\tUD{T}{0}{j}=-\tUD{T}{j}{0}\,$ and $\,\UD{\Sigma}{i}{j}=\tUD{\Sigma}{i}{j}\,$ by using \eref{eq:emt_tetrad_coordinate_relation_explicit}. After we align \k with the zenith, it follows that $\,\xivector{0}{j}=k^j/k\,$ and $\,\chimatrix{2}{0}{ij}\,=\,\hat{k}^i\hat{k}^j-\delta^{ij}/3\,$. Thus, using \eref{eq:fluid_limit_first_order} yields
\begin{align}
  &\theta\;=\;\frac{k}{3}\;\frac{\bmult{\Delta}{1}{1}{0}}{w+1}\; &&\text{and}&&
  \sigma \;=\; \frac{2}{15}\;\frac{\bmult{\Delta}{2}{2}{0}}{w+1} \;,
\end{align}
where we have used $\,\Pz/\rhoz=w\,$.

\section{The distribution function}
\label{sec:distribution_function}

In this section we use the concept of thermal equilibrium to specify a simple form for the distribution functions of the photon (\sref{sec:photon_distribution_function}) and electron (\sref{sec:electron_distribution_function}) fluids; this ansatz will considerably simplify the computation of the collision term in \sref{sec:collision_term}.
In \sref{sec:photon_distribution_function}, we also discuss the ambiguity of defining the CMB temperature at second order due to the presence of spectral distortions.

\subsection{The photon distribution function}
\label{sec:photon_distribution_function}

Before the epoch of recombination, the CMB photons frequently interact with the free electrons via Compton scattering due to the high density of the early Universe. As a result, they are in a state of thermal equilibrium which is well described by the  \keyword{Bose-Einstein distribution function} with vanishing chemical potential, or \keyword{blackbody spectrum}:
\begin{align}
  f_\sub{BB}(\tau,p) \;=\; \left[\;\exp\left(\frac{p}{T(\tau)}\right)-1\;\right]^{-1} \;,
  \label{eq:bose_einstein}
\end{align}
where $p$ is the photon momentum in the local inertial frame and $T$ is the CMB temperature.
This simple picture is complicated by two circumstances.
First, in an inhomogeneous Universe, different observers would measure a different distribution function according to their position and to the direction they look at; this can be accommodated by including a positional and directional dependence in the temperature: $T=T(\tau,\vecx,n^i)$.
Secondly, as the Universe expands and cools down, the Compton scattering rate decreases and the photons eventually cease to be in thermal equilibrium. Thus, one has to allow for deviations from the blackbody spectrum, or \keyword{spectral distortions}, which amounts to $f$ having a momentum dependence more complicated than the one in \eref{eq:bose_einstein}.

According to the above considerations, we assume for the photon distribution function the following ansatz:
\begin{align}
  \label{eq:distribution_function_ansatz}
	f(\tau,\vec{x},p,n^i) = \left[\exp\left(\frac{p}
	{\overline{T}(\tau) \,[\,1+\Theta(\tau,\vec{x},p,n^i)\,]}\right)-1\right]^{-1}\;,
\end{align}
where $\overline{T}$ is the background temperature and we have introduced the \keyword{temperature fluctuation}, $\Theta=(T-\overline{T})/\,\overline{T}$\,.
After Taylor expanding $f$ about $\Theta=0\,$,
\begin{align}
  f \;=\; f\,\Bigr|_{\Theta=0} \;+\;
  \pfrac{f}{\Theta}\,\biggr|_{\Theta=0} \;\Theta \;+\;
  \frac{1}{2}\,\ppfrac{f}{\Theta}\,\biggr|_{\Theta=0} \;\Theta^2  \;,
\end{align}
and setting $\,\Theta=\pert{\Theta}{1}+\pert{\Theta}{2}\,$, we find the relation between the temperature fluctuation and the distribution function up to second order:
\begin{align}
  \label{eq:f_of_theta_up_to_second_order}
  f \;=\; \bar{f} \;-\; p\;\pfrac{\bar{f}}{p}\;\Theta \;+\;
  \left(\;\frac{p^2}{2}\,\ppfrac{\bar{f}}{p}
  \;+\;\pfrac{\bar{f}}{p}\;\right)\;\Theta^2 \;,
\end{align}
where $\bar{f}\equiv\pert{f}{0}\,$.

By choosing the form in \eref{eq:distribution_function_ansatz} for $f$, we have implicitly assumed that, at the background level, the blackbody shape of the spectrum is preserved throughout the cosmic evolution,
\begin{align}
	\bar{f}(\tau,p) \;=\; \left[\;\exp\left(\frac{p}
	{\overline{T}(\tau)}\right)-1\;\right]^{-1} \;.
  \label{eq:bose_einstein_background}
\end{align}
This occurs for two reasons. First, as we shall see in \sref{sec:collision_energy_transfer}, during recombination the energy transfer between the photons and the electrons is so small that the background collision term is negligible and cannot induce spectral distortions.
Secondly, after recombination, when the collisions are unimportant, both the energy of the photon and the temperature decay as $1/a\,$, leaving $\,p/T\,$ unchanged during the cosmic expansion. Thus, the blackbody spectrum of the background CMB, which was established before recombination by the frequent Compton collisions, is not altered and survives all the way to today\footnote{It should be noted that the cosmic expansion not altering the CMB spectrum is not a coincidence; in fact, the spectral distortions cannot be induced by the geodesic motion encoded in the Liouville operator, for the simple reason that a photon follows the same geodesic trajectory regardless of its energy. Therefore, we expect the spectral distortions to arise only at the level of the collision term.}.
As a matter of fact, in section \sref{sec:collision_contributions} we shall see that the negligible energy transfer between photons and electrons preserves the blackbody shape also at the first-order level. It follows that the spectral distortions are confined to the higher-order fluctuations; up to second order, this corresponds to setting
\begin{align}
  \Theta \;=\; \pert{\Theta}{1}(\tau,\vec{x},n^i)
  \,+\, \pert{\Theta}{2}(\tau,\vec{x},p,n^i) \;.
  \label{eq:theta_second_order_depends_on_the_momentum}
\end{align}



\subsubsection{Temperature definition}

The presence of spectral distortions makes it impossible to unambiguously define a temperature for the CMB. 
This is clear by looking at the moments of the distribution function,
\begin{align}
  M_m \;\equiv\; \int\dd p\;p^2\:E^{\,m}\,f \;.
\end{align}
For the blackbody spectrum in \eref{eq:bose_einstein}, all the moments can be expressed in terms of powers of the temperature \footnote{This can be proven by integrating $M_m[f_\sub{BB}]=4\pi\int\dd p\,p^{2+m}\,f_\sub{BB}$ by parts and using the fact that $\partial f_\sub{BB}/\partial p = -T/p\;\partial f_\sub{BB}/\partial T\,$.}
\begin{align}
  \frac{M_m}{M^{(0)}_{m}} \;=\; \left(\,\frac{T}{\overline{T}}\,\right)^{3+m} \;,
\end{align}
where we have normalised the moments and the temperature with respect to their background values.
In particular, the number density $n\equiv M_0$ and the brightness $\mathcal{I}\equiv M_1$ satisfy
\begin{align}
  \frac{n_\sub{BB}}{\overline{n}} \;=\; \left(\,\frac{T}{\overline{T}}\,\right)^3
  \quad\quad\text{and}\quad\quad
  \frac{\mathcal{I_\sub{BB}}}{\overline{\mathcal{I}}} \;=\; \left(\,\frac{T}{\overline{T}}\,\right)^4  \;.
\end{align}

On the contrary, the moments of an arbitrary spectrum $f$ are in general independent and cannot be expressed in terms of a single temperature function. If we parametrise them as
\begin{align}
  \frac{M_m}{M^{(0)}_{m}} \;\equiv\; \left(\,\frac{T_m}{\overline{T}}\,\right)^{3+m} \;,
  \label{eq:different_temperatures}
\end{align}
we see that $\,T_m\,$ is the temperature of a blackbody spectrum whose $m$-th moment is equal to that of $f$. For a blackbody spectrum, all these effective temperatures are equal; it follows that the existence of a scatter in the $T_m$'s indicates the presence of spectral distortions.

One could pick one of the effective temperatures $T_m$ to represent the CMB temperature, but this is clearly an arbitrary choice. In \citet{pitrou:2010b}, however, it was shown that the CMB bispectrum is insensitive to the specific moment of the distribution function that is chosen to define the temperature. 
We therefore follow what is commonly done in the literature \cite{nitta:2009a, pitrou:2010a, beneke:2011a} and define the temperature $T$ via the first moment of the distribution, the brightness,
\begin{align}
  \left(\,\frac{T}{\overline{T}}\,\right)^4 \;\equiv\;
  \frac{\mathcal{I}}{\overline{\mathcal{I}}} \;.
  \label{eq:bolometric_temperature_T_I}
\end{align}
which is the temperature of the blackbody spectrum with the same energy density as the CMB, and is referred to as the \keyword{bolometric temperature}.\footnote{Note that \citet{pitrou:2010b} proposed another definition of temperature, the \keyword{occupation number temperature}, $\,T_\sub{\#}\,$, which is the temperature associated to the blackbody spectrum with the same number density as the CMB,
\begin{equation}
  \left(\,\frac{T_\sub{\#}}{\overline{T}}\,\right)^3 \;\equiv\; \frac{n}{\overline{n}} \;.
\end{equation}
For a more detailed discussion on temperature moments and on their relation to what is measured by CMB experiment, refer to \citet{pitrou:2014a}.}

\subsubsection{The brightness fluctuation $\Delta$}

We introduce the \keyword{brightness fluctuation}, $\,\Delta\,$, as
\begin{align}
  \mathcal{I} \;\equiv\; \overline{\mathcal{I}}\,(1\,+\,\Delta) \;.
\end{align}
Because $\mathcal{I}=\int\dd p\,p^3\,f$, the brightness fluctuation is explicitly given by
\begin{align}
  1\,+\,\Delta\,(\taux,\n) \;\equiv\; \frac{1}{\int\dd p\,p^3\,\overline{f}(\tau,p)}
  \;\int\dd p\,p^3\,f\,(\taux,p,\n) \;.
  \label{eq:brightness_definition}
\end{align}
In general we define the brightness operator, $\beta$, as
\begin{align}
 \beta\,[\,F\,] \;\equiv\; \frac{1}{\int\,\dd p\,p^3\,\overline{F}} \;
 \int\,\dd p\,p^3\,F \;.
 \label{eq:boltzmann_brightness_operator}
\end{align}
so that $\beta\,[\,f\,]=1+\Delta$. The evolution of the brightness fluctuation is dictated by the brightness-projected Boltzmann equation,
\begin{align}
  \beta\,\left[\,\diff{f}{\tau}\,-\,\frac{1}{p^0}\,C[f]\,\right] \;=\; 0 \;,
  \label{eq:brighness_equation}
\end{align}
which we shall call the \keyword{brightness equation}.

The bolometric temperature fluctuation $\Theta$, defined as $T=\overline{T}\,(1+\Theta)\,$, is related to $\Delta$ via \eref{eq:bolometric_temperature_T_I},
  \begin{align}
  \left(\,1\,+\,\Theta\,\right)^4 \;=\; 1\,+\,\Delta \;.
  \label{eq:bolometric_temperature_DELTA_THETA}
\end{align}
Up to first order, the relation translates to $\,\Delta=4\Theta\,$ while, up to second order, it reads
\begin{align}
  & \Delta \;=\; 4\,\Theta \,+\, 6\,\Theta\,\Theta \;,\msk
  & \Theta \;=\; \frac{1}{4}\,\Delta \,-\, \frac{3}{32}\,\Delta\,\Delta \;.
  \label{eq:delta_theta_relation_brightness}
\end{align}
To compute the anisotropies of the CMB, we need to first solve the brightness equation up to second-order for $\pert{\Delta}{2}$, and then relate it to the bolometric temperature through the above equation.

\citet{huang:2013a} have recently proposed a different parametrisation for the brightness using the \DeltaT variable,
\begin{equation}
  \mathcal{I} \;\equiv\; \overline{\mathcal{I}}\;e^\DeltaT \;,
\end{equation}
which differs from $\Delta$ at the second-order level,
\begin{align}
  \DeltaT \;=\; \Delta\,-\,\frac{1}{2}\,\Delta\,\Delta \;=\; 4\,\Theta \,-\, 2\,\Theta\,\Theta \;.
  \label{eq:deltatilde_theta_relation_brightness}
\end{align}
In principle, there is no difference in using one or the other expansion but, as we shall see in \sref{sec:redshift_term_deltatilde}, a specific term in the left hand side of Boltzmann equation is simpler to integrate when using the \DeltaT variable.

We conclude this subsection showing some relations that will be useful to compute the brightness-projected Liouville and collision terms:
\begin{align}
  &\beta\,\left[\,p\,\pfrac{f}{p}\,\right] \,=\, -4\;(1+\Delta) \;, 
  &\beta\,\left[\,p^2\,\ppfrac{f}{p}\,\right] \,=\, 20\;(1+\Delta) \;, \nmsk
  &\beta\,\left[\,\pfrac{f}{\tau}\,\right] \,=\, \pfrac{\Delta}{\tau} \:-\: 4\,\Hc\,(1+\Delta) \;.
  \label{eq:boltzmann_brightness_integrals}
\end{align}
We have obtained them by repeated application of integration by parts and, for the last one, by enforcing the zero order Boltzmann equation, $\dot{\overline{f}}=\Hc\,p\,\partial \overline{f}/\partial p\,$. Note that the relations can also be inferred by those for the more general $\beta$-moments (see \sref{sec:the_evolved_equations}).

\subsubsection{Projected distribution function}

To characterise the spatial and directional dependence of the brightness fluctuation $\Delta$, we project it on plane waves using the Fourier-space operator $\mathcal{F}$ (\eref{eq:fourier_operator}) and on spherical harmonics using the multipole-space operator $L_\lm$ (\eref{eq:L_operator}):
\begin{align}
  \Delta_{\lm}(\tau,\k) \;&\equiv\;
  \left(\mathcal{F_{\,\veck}}\circ L_\lm\circ \beta\right)[\,f\,] \nmsk
  &=\; i^\L\,\sqrt{\frac{2\L+1}{4\pi}}\,\int\dd\vecx\;\dd\Omega\;
  e^{-i \vec{k}\cdot\vec{x}}\;Y^*_\lm(\n)\;\Delta(\taux,\n) \;.
  \label{eq:brightness_fourier_multipoles_projection}
\end{align}
The evolution equations for $\Delta_{\lm}(\tau,\k)$ are given by the \keyword{projected Boltzmann equation}:
\begin{align}
  \left(\mathcal{F_{\,\veck}}\circ L_\lm\circ
  \beta\right)\,\left[\,\diff{f}{\tau}\,-\,\frac{1}{p^0}\,C[f]\,\right] \;=\; 0 \;.
  \label{eq:projected_boltzmann_equation}
\end{align}
Being linear, the three operators act on the Boltzmann equation on a term-by-term basis, so that the formulae we have provided in \sref{sec:fourier_formalism}, \sref{sec:spherical_projection_of_functions} and in  \eref{eq:boltzmann_brightness_integrals} are sufficient to obtain the evolution equation for $\Delta_{\lm}(\tau,\k)$.

The advantage of following this approach is that the Boltzmann equation, originally a partial differential equation in time, position, momentum and direction, turns into a system of ordinary differential equations for the time evolution of $\Delta_{\lm}(\tau,\k)$ which is numerically tractable.


\subsection{The electron distribution function}
\label{sec:electron_distribution_function}

During all epochs of interest, the Coulomb collision rate between free electrons and protons is much larger than the expansion rate of the Universe \cite{dodelson:2003b}, meaning that they are kept in thermal equilibrium. Furthermore, until the end of recombination, the electrons share the same temperature with the photons as they frequently interact through Compton scattering. Around recombination, this common temperature is much smaller than the electron mass so that electrons and protons can be treated as non-relativistic particles. Therefore, both fluids must be described by the \keyword{Maxwell-Boltzmann}\index{distribution function of baryons} distribution function, which for the electrons reads
\begin{align}
  g\,(\taux,\vecq) \;=\; n_e(\taux)\,\left(\,\frac{2\pi}{m_e\,T_e(\tau)}\,\right)^{3/2}\,
  \exp\left\{\,\frac{-\left[\,\vecq-m_e\,\vec{v_e}(\taux)\,\right]^2}{2\,m_e\,T_e(\tau)}\,\right\} \;,
  \label{eq:maxwell_boltzmann}
\end{align}
where $T_e(\tau)$, $\vec{v_e}(\taux)$ and $n_e(\taux)$ denote respectively the electron temperature, the bulk velocity of the electron fluid and the number density of free electrons,
\begin{align} 
  n_e(\taux) \,=\, \int\,\frac{\dd\vecq}{2\pi^3}\:g\,(\taux,\vecq) \;.
\end{align}
Note that the distribution function is normalised so that $\avg{g}=\int\dd\vecq/(2\pi^3)g=n_e$.
The total momentum of an electron, $\vecq$, has two contributions: the bulk velocity of the electron fluid, $\vecq_B=m_e\,\vecv_e$, which coincides with that of the proton fluid due to the tight coupling between the two fluids induced by Coulomb scattering, and the thermal motion, $\vecq_T=\vecq-\vecq_B$, which appears in the numerator of the exponential in \eref{eq:maxwell_boltzmann}.

We report the moments of $g$ that will be useful in the derivation of the collision term:
\begin{align}
 & \avg{g} \;\equiv\; \int\frac{\dd\vecq}{(2\pi)^3}\,g \;=\; n_e\;,\nmsk
 & \avg{g\,q^i} \;\equiv\;\int\frac{\dd\vecq}{(2\pi)^3}\,q^i\,g \;=\; n_e\,m_e\,v_e^i\;,\nmsk
 & \avg{g\,q^i\,q^j} \;\equiv\; \int\frac{\dd\vecq}{(2\pi)^3}\,q^i\,q^j\,g \;=\;
 \delta^{ij}\,n_e\,m_e\,T_e\,+\,n_e\,m_e^2\,v_e^i\,v_e^j \;.
 \label{eq:electron_moments}
\end{align}
To derive the second equality, one has to perform the variable substitution $\vecq_T=\vecq-m_e\vecv_e$ and realise that the integral
\begin{align}
  \int\dd\vecq\:q_T^i\:
  \exp\left\{\,-\frac{q_T^2}{2\,m_e\,T_e}\,\right\} \;
\end{align}
vanishes. (Note that this is a direct consequence of the assumed isotropy of the thermal motion of particles.)

Let us establish some relations between the magnitudes of the various momenta and velocities, an exercise that will prove itself useful in computing the collision term in \sref{sec:collision_term}. If follows from the Maxwell-Boltzmann distribution that the average thermal momentum of an electron is of order $q_T\simeq\sqrt{m_e\,T_e}$. Due to Compton scattering, the temperature of the electron fluid is nearly identical to that of the photons until the end of recombination: $T_e \simeq T_\gamma = T$. Therefore, on average, the momentum of a photon, $p=T$, is much smaller than that of an electron:
\begin{align}
  \frac{p}{q_T} \;\simeq\; \sqrt{\frac{T}{m_e}} \;=\; \O(10^{-3}) \;,
  \label{eq:collision_p_smaller_than_q}
\end{align}
where we have used $T\simeq \unit[1]{eV}$ during recombination and $m_e\simeq\unit[511]{keV}$. The average thermal momentum of an electron, however, is still much smaller than its mass,
\begin{align}
  v_T\,\equiv\,\frac{q_T}{m_e} \;\simeq\; \sqrt{\frac{T}{m_e}} \;=\; \O(10^{-3}) \;.
  \label{eq:collision_q_smaller_than_mass}
\end{align}
Because $v_T=q_T/m_e$ is the average thermal velocity, the free electrons are non-relativistic (hence the Maxwell-Boltzmann distribution). It is important to note that the bulk velocity of the electrons, being of the same order as the metric perturbations,
\begin{align}
  v_e \;=\; \O(10^{-5}) \;,
\end{align}
it is on average much smaller than the thermal component.

\section{The Liouville term}
\label{sec:liouville_term}

The Liouville term appears in the left hand side of the Boltzmann equation:
\begin{align}
  \diff{f}{\lambda} \;=\; C[f] \;,
\end{align}
and describes the evolution of the considered species in the absence of interactions. This is in turn determined by the geodesic motion of the species particles, which propagate in a perturbed metric.
The geodesic flow is parametrised by the affine parameter $\lambda$. Using
\begin{align}
  \U{p}{\mu} \;\equiv\; \diff{x^\mu}{\lambda} \;,
\end{align}
\annotate{A parameter $\lambda$ is affine to a curve if the tangent vector $\dd x^\alpha/\dd\lambda$ has constant magnitude.}
where $x^\mu(\lambda)$ is a geodesic curve, we write the Boltzmann equation as
\begin{align}
  \diff{f}{\tau} \;=\; \coll[f] \;,
  \label{eq:boltzmann_equation_schematic_time}
\end{align}
where $\dd\tau=\dd t/a\,$ is the conformal time and we have defined $\coll[f]\equiv C[f]/p^0$. With a small abuse of terminology, we shall sometimes refer to the left and right hand sides of \eref{eq:boltzmann_equation_schematic_time} as the Liouville and collision terms, respectively.

As we have mentioned in the previous section, we shall solve the Boltzmann equation in the local inertial frame, where the four-momentum of a particle is split into its magnitude, $p$, and its direction, $n^i$ (\eref{eq:E_p_n_split_of_momentum}). Being a scalar, the distribution function has the same value in the coordinate and inertial frames,
\begin{align}
    f(\tau,x^i,p,n^i) \;=\; f(\tau,x^i,p^i(\tau,x^i,p,n^i)) \;,
\end{align}
and, therefore, we can expand the Liouville term in terms of the partial derivatives of $f$ with respect to $p$ and $n^i$:
\begin{align}
  \pfrac{f}{\tau}\;+\;\pfrac{f}{x^i}\,\diff{x^i}{\tau}\;+\;\pfrac{f}{p}\,
  \diff{p}{\tau}\;+\;\pfrac{f}{n^i}\,\diff{n^i}{\tau} \;=\; \coll[f] \;.
  \label{eq:boltzmann_partial_derivatives}
\end{align}
As we shall see, each of the terms in the Liouville term affects the CMB anisotropies in a different way. The first two terms encodes free streaming, that is the propagation of perturbations from the small to the large multipoles. At higher order this term also includes \keyword{gravitational time delay} effects. The third term, at background level, causes the redshifting of photons, and at higher-order includes the well-known Sachs-Wolfe (SW), integrated Sachs-Wolfe (ISW) and Rees-Sciama (RS) effects. The fourth term vanishes to first order and describes the small-scale effect of gravitational lensing on the CMB. We shall refer to these terms as the \emph{free-streaming}\index{free-streaming term}, \emph{redshift}\index{redshift term} and \emph{lensing}\index{lensing term} terms, respectively.

We now express the three parts of the Liouville term in terms of the metric and matter variables, and integrate out the momentum dependence of the resulting expressions.


\subsection{The free streaming term}
\label{sec:free_streaming term}

The free-streaming term,
\begin{align}
  \FST \;\equiv\; \pfrac{f}{\tau}\;+\;\pfrac{f}{x^i}\,\diff{x^i}{\tau} \;,
\end{align}
contains the coordinate velocity, which we can express in terms of $p$ and $n^i$ up to second order
using \eref{eq:four_momentum_tetrad_coordinate_dictionary}:
\begin{align}
  \diff{x^i}{\tau} \;=\; \frac{\U{p}{i}}{\U{p}{0}} \;=\;
  \frac{p\,n^j}{E}\,\sqrt{\frac{1+2\Psi}{1-2\Phi}}\;
  \left[\, \KronUD{i}{j}\,(1-\frac{p}{E}\,\D{\omega}{i}\,n^i) \,-\,\UD{\gamma}{i}{j}\,\right] \;.
\end{align}
The second-order part of the particle's velocity is not needed, because it multiplies a quantity, $\partial f/\partial x^i$, that is at least first-order due to the fact that the background distribution function is position-independent (\eref{eq:bose_einstein_background}). Thus, the free-streaming term, up to second order, reads
\begin{align}
  \FST \;=\; \dot f\;+\;n^i\,\partial_i\,f\;\frac{p}{E}\:(\,1\,+\,\Psi\,+\,\Phi\,) \;,
  \label{eq:free_streaming_f}
\end{align}
where the dot denotes a partial derivative with respect to conformal time and $\partial_i=\partial/\partial x^i$.


\subsubsection{Momentum integrated \FST}
For the photons ($p=E$) and in terms of the brightness fluctuation $\Delta$, the free streaming term reads
\begin{align}
  \beta\,[\,\FST\,] \;=\; \dot\Delta \;-\; 4\,\Hc\,(1+\Delta)
  \;+\; n^i\;\partial_i\:\Delta\,(1\,+\,\Phi\,+\,\Psi) \;,
  \label{eq:free_streaming_brightness}
\end{align}
where we have used \eref{eq:boltzmann_brightness_integrals} to compute the time derivative.
The term multiplied by \Hc comes from taking the time derivative of the background distribution function in the denominator of \eref{eq:boltzmann_brightness_operator}, and represents the universal redshift due to the expansion. It will cancel out with the equal but opposite term in the redshift term (\eref{eq:redshift_term_brightness}), thus leaving no effect on the temperature perturbation.

\subsection{The redshift term}
\label{sec:redshift_term}

The redshift term,
\begin{align}
  \RT \;\equiv\; \pfrac{f}{p}\,\diff{p}{\tau} \;,
\end{align}
encodes the change of the phase-space density caused by the energy variations of the particles as they travel in a curved Universe. To obtain an expression for $\dd p/\dd\tau$ valid up to second order, we use the geodesic equation:
\begin{align}
  \diff{\U{p}{0}}{\tau} \;=\; -\,\Gamma^{0}_{\alpha\beta} \,
  \frac{\U{p}{\alpha}\,\U{p}{\beta}}{\U{p}{0}} \;.
  \label{eq:geodesic_unexpanded_0}
\end{align} 
The computation is lengthy and is more easily carried using the exponential potentials $\Psi_e$ and $\Phi_e$ in \eref{eq:bmr_metric}. Using the expression for $\U{p}{0}$ in terms of the proper momentum (\eref{eq:four_momentum_tetrad_coordinate_dictionary}), the left hand side of \eref{eq:geodesic_unexpanded_0} reads
\begin{align}
  \diff{\U{p}{0}}{\tau} \,
  &=\, \diff{}{\tau}\,\left[\,\frac{E}{a}\,e^{-\Psi_e}\,(1+\frac{p}{E}\,\D{\omega}{i}\,n^i) \,\right] \nmsk
  &=\,
  \frac{1}{a}\diff{p}{\tau}\,\left(\D{\omega}{i}\,n^i + \frac{p}{E}\,e^{-\Psi_e}\right) \;
  -\frac{E}{a}\,e^{-\Psi_e}\,(\dot\Psi_e\,+\,\Hc) \;
  -\frac{p}{a}\,e^{\Phi_e}\,\partial_i\Psi_e\,n^i \nmsk
  &\quad\quad-\,\frac{p}{a}\,(\,\Hc\,\D{\omega}{i}\,n^i - \D{\dot\omega}{i}\,n^i -
  \frac{p}{E}\,\partial_i\,\D{\omega}{j}\,n^i\,n^j \,) \;,
  \label{eq:dpdtau_lhs}
\end{align}
where a dot denotes a partial derivative with respect to time, $\partial/\partial \tau$, and we have used the following identities:
\begin{align}
  & \diff{}{\tau}\left(\frac{E}{a}\right) \,=\, \frac{p}{a\,E}\,\diff{p}{\tau}\,-\,\frac{E}{a}\,\Hc  \;,\nmsk
  & \diff{\Psi_e}{\tau} \,=\, \pfrac{\Psi_e}{\tau}\,+\,\pfrac{\Psi_e}{x^i}\,\diff{x^i}{\tau} \,=\,
  \dot\Psi_e \,+\, \frac{p}{E}\,\partial_i\Psi_e\,n^i\,e^{\Psi_e+\Phi_e} \;,\nmsk
  & \diff{}{\tau}\left(\frac{p}{E}\,\D{\omega_i}\,n^i\right) \,=\,
  \frac{1}{E}\,\diff{p}{\tau}\,\left( 1 - \frac{p^2}{E^2} \right)\,\D{\omega_i}\,n^i
  \,+\frac{p}{E}\,\D{\dot\omega}{i}\,n^i
  \,+\, \frac{p^2}{E^2}\,\partial_i\,\D{\omega_j}\,n^i\,n^j \;.
\end{align}
The right hand side of \eref{eq:geodesic_unexpanded_0} is expanded using the components of the Levi-Civita connection at second order
and, again, the dictionary in \eref{eq:four_momentum_tetrad_coordinate_dictionary}:
\begin{align}
  -\,\Gamma^{0}_{\alpha\beta}\,\frac{\U{p}{\alpha}\,\U{p}{\beta}}{\U{p}{0}} \;
  &=\, \frac{p^2}{a\,E}\,e^{-\Psi_e}\,(\dot\Psi_e-\Hc)\,-\,\frac{E}{a}\,e^{-\Psi_e}\,(\dot\Psi_e\,+\,\Hc)\,
  -2\,\frac{p}{a}\,e^{\Phi_e}\,\partial_i\Psi_e\,n^i \nmsk
  &\quad\quad+\,\Hc\,\frac{p}{a}\,\left(\frac{p^2}{E^2}-3\right)\,\D{\omega_i}\,n^i\,
  +\,\frac{p^2}{a\,E}\,\left(\,\partial_i\,\D{\omega}{j} - \DD{\dot\gamma}{i}{j}\,\right)\,n^i\,n^j \;.
  \label{eq:dpdtau_rhs}
\end{align}

We then equate Eq.~\ref{eq:dpdtau_lhs} and \ref{eq:dpdtau_rhs} and multiply both sides of the resulting expression by $aE/p^2\,e^{\Psi_e}$ in order to isolate the fractional rate of change in the particle momentum, $d\ln p/d\tau$. As a result, several terms cancel; in particular, after enforcing the zeroth-order version of the equation, $d \ln p/d\tau=-\Hc$, all the terms involving $\D{\omega_i}\,n^i$ can be grouped into a single one,
\begin{align}
  \Hc\,\left(\frac{p}{E}-\frac{E}{p}\right)\,\omega_i\,n^i\,
  =\,-\Hc\,\frac{m^2}{E\,p}\,\omega_i\,n^i \;.
\end{align}
Thus, we obtain the so-called \keyword{redshift formula} up to second order:\footnote{Our expression for $\dd p/\dd\tau$ matches the one given in Eq.~4.14 by \citet{senatore:2009b}
but is different form the one in Eq.~3.14 of \citet{bartolo:2006a}. The reason for this discrepancy is explained in the footnote 11 of the former paper.}
\begin{align}
  \frac{1}{p}\,\diff{p}{\tau} \;=\; &-\Hc\,+\,\dot\Phi_e\,
  -\,\frac{E}{p}\,n^i\,\partial_i\Psi_e\,(1\,+\,\Psi_e\,+\,\Phi_e) \nmsk
  &\quad\quad
  -\,\Hc\,\frac{m^2}{E\,p}\:\omega_i\,n^i\,
  -\,\frac{E}{p}\,\dot\omega_i\,n^i\,
  -\,\DD{\dot\gamma}{i}{j}\,n^i\,n^j \;,
  \label{eq:redshift_equation_bmr}
\end{align}
where the dots denote partial differentiation with respect to the conformal time, $\partial/\partial\tau$, and $\partial_i=\partial/\partial x^i$. The redshift formula can be recast in terms of the usual potentials, $\Psi$ and $\Phi$, using the relations in \eref{eq:bmr_ours_potential},
\begin{align}
  & \dot\Phi_e \,=\; \dot\Phi \,+\, 2\,\Phi\,\dot\Phi \;,\nmsk
  & \partial_i\Psi_e \,=\; \partial_i\Psi \,-\, 2\,\Psi\,\partial_i\Psi \;,
\end{align}
at the cost of introducing two extra quadratic terms:
\begin{align}
  \frac{1}{p}\,\diff{p}{\tau} \;=\; &-\Hc\,+\,\dot\Phi\,
  -\,\frac{E}{p}\,n^i\,\partial_i\Psi\,\sqrt{\frac{1+2\Psi}{1-2\Phi}}
  \,+\,2\,\left(\,\Phi\,\dot\Phi \,+\,\frac{E}{p}\,\Psi\,n^i\,\partial_i\Psi\,\right) \nmsk 
  &\quad\quad
  -\,\Hc\,\frac{m^2}{E\,p}\:\omega_i\,n^i\,
  -\,\frac{E}{p}\,\dot\omega_i\,n^i\,
  -\,\DD{\dot\gamma}{i}{j}\,n^i\,n^j \;.
\end{align}
Up to second order, this is equivalent to
\begin{align}
  \frac{1}{p}\,\diff{p}{\tau} \;=\; &-\Hc\,+\,\dot\Phi\,
  (1+2\,\Phi)\,-\,\frac{E}{p}\,n^i\,\partial_i\Psi\,(1+\Phi-\Psi) \nmsk
  &\quad\quad 
  -\,\Hc\,\frac{m^2}{E\,p}\:\omega_i\,n^i\,
  -\,\frac{E}{p}\,\dot\omega_i\,n^i\,
  -\,\DD{\dot\gamma}{i}{j}\,n^i\,n^j \;.
  \label{eq:redshift_equation}
\end{align}


\subsubsection{Momentum integrated \RT}
For the photons ($p=E$) and in terms of the brightness fluctuation $\Delta$, the redshift term up to second order reads
\begin{align}
  &\beta\,[\,\RT\,] \;=\; 4\,\Hc\,(1+\Delta)\;-\;4\,\left(1\,+\,\Delta\right)\,
  \left(\dot\Phi\,-\,n^i\,\partial_i\Psi\right) \nmsk
  &\quad\quad-4\,\left[\,2\,\Phi\,\dot\Phi\,-\,(\Phi-\Psi)\,n^i\,\partial_i\Psi
  \,-\,n^i\,\dot\omega_i\,-\,n^i\,n^j\,\dot\gamma_{ij}\,\right] \;,
  \label{eq:redshift_term_brightness}
\end{align}
where we have used the relation $\beta\,[\,p\,\partial f/\partial p\,]=-4\,(1+\Delta)$ from \eref{eq:boltzmann_brightness_integrals}. Note that the first term, which encodes the uniform redshift of the spectrum, cancels with the equal but opposite one in \eref{eq:free_streaming_brightness}.

\subsection{The lensing term}
\label{sec:lensing_term}
The lensing term,
\begin{align}
  \LT \;\equiv\; \pfrac{f}{n^i}\,\diff{n^i}{\tau} \;,
\end{align}
describes the change in the direction of propagation of the particles induced by the matter distribution; for photons, this is known as the gravitational lensing. Because the background distribution function (\eref{eq:bose_einstein_background}) does not depend on the particle's direction, the term $\partial f/\partial n^i$ is at least first order and, therefore, we only need to compute $\dd n^i/\dd\tau$ up to first order. Using the geodesic equation,
\begin{align}
  \diff{\U{p}{i}}{\tau} \;=\; -\,\Gamma^{i}_{\alpha\beta} \,
  \frac{\U{p}{\alpha}\,\U{p}{\beta}}{\U{p}{0}} \;,
  \label{eq:geodesic_unexpanded_i}
\end{align}
it can be shown that, up to first order \cite{bartolo:2006a, senatore:2009b},
\begin{align}
  \diff{n^i}{\tau} \;=\; -(\UU{\delta}{ij}\,-\,n^i\,n^j)\,
  \left(\frac{E}{p}\,\partial_i\Psi\,+\,\frac{p}{E}\,\partial_i\Phi \right) \;.
  \label{eq:lensing_equation}
\end{align}
The operator in the first parentheses, $\,\UU{\delta}{ij}\,-\,n^i\,n^j\,$, extracts from a vector the part that is transverse to $n^i$, the direction of propagation of the particle. Therefore, the bending of the particle's trajectory is determined only by the transverse gradients of the scalar potentials. Since $p/E$ is the velocity of the particle in the local inertial frame, the coefficients of the potentials have a precise physical meaning: relativistic particles ($p/E=1$) are deflected twice as much with respect to the non-relativistic ones ($p/E\ll1$).

\subsubsection{Momentum integrated \LT}
For the photons ($p=E$) and in terms of the brightness fluctuation $\Delta$, the lensing term up to second order reads
\begin{align}
  \beta\,[\,\LT\,] \;=\; -\left(\,\UU{\delta}{ij}\,-\,n^i\,n^j\,\right)
  \,\pfrac{\Delta}{n^i} \, \left(\,\partial_i\Psi+\partial_i\Phi\right) \;.
  \label{eq:lensing_term_brightness}
\end{align}


\subsection{The momentum-integrated Liouville term}
\label{sec:momentum_integrated_liouville}

The momentum-integrated Liouville term is given by
\begin{align}
  \beta\,\left[\,\diff{f}{\tau}\,\right] \;=\;
  \beta\,\left[\,\FST\,\right] \;+\; \beta\,\left[\,\RT\,\right] \;+\; \beta\,\left[\,\LT\,\right] \;.
\end{align}
Inserting the expressions in \eref{eq:free_streaming_brightness}, \eref{eq:redshift_term_brightness} and \eref{eq:lensing_term_brightness}, we obtain up to second order\footnote{Note that, with respect to what we have written in \cite{pettinari:2013a}, we have corrected a typo in the sign of $\dot\omega_i$.}
\begin{multline}
  \beta\,\left[\,\diff{f}{\tau}\,\right] \;=\;
  \dot\Delta \;+\; n^i\;\partial_i\:\Delta\,(1\,+\,\Phi\,+\,\Psi) \;\msk
  -\;4\,\left(1\,+\,\Delta\right)\,
  \left(\dot\Phi\,-\,n^i\,\partial_i\Psi\right) 
  \;-4\;\left[\,2\,\Phi\,\dot\Phi\,-\,(\Phi-\Psi)\,n^i\,\partial_i\Psi
  \,-\,n^i\,\dot\omega_i\,-\,n^i\,n^j\,\dot\gamma_{ij}\,\right] \msk
  \;-\;\left(\,\UU{\delta}{ij}\,-\,n^i\,n^j\,\right) 
  \,\pfrac{\Delta}{n^i} \, \left(\,\partial_i\Psi+\partial_i\Phi\right) \;.
  \label{eq:liouville_brightness}
\end{multline}
Up to first order, all the quadratic terms and the non-scalar perturbations can be neglected; what is left are two contributions from the free streaming term and two from the redshift term,
\begin{align}
  \beta\,\left[\,\diff{f}{\tau}\,\right] \;=\; 
  \dot\Delta \;+\; n^i\;\partial_i\:\Delta \;-\; 4\,\left(\dot\Phi\,-\,n^i\,\partial_i\Psi\right) \;.
  \label{eq:liouville_brightness_first}
\end{align}

At the background level, the brightness fluctuation vanishes by definition (\eref{eq:brightness_definition}) and so does the Liouville term. Therefore, we use the redshift formula (\eref{eq:free_streaming_f}),
\begin{align}
  \frac{1}{p}\,\diff{p}{\tau} \;=\; -\Hc
\end{align}
to obtain
\begin{align}
  \diff{\overline{f}}{\tau} \;=\; \dot{\overline{f}} \;-\; \Hc\,p\,\pfrac{\overline{f}}{p} \;.
\end{align}
As we shall see in \sref{sec:collision_energy_transfer}, during and after recombination, the zero-order collision term vanishes due to the negligible energy transfer between photons and electrons.
Thus, the evolution equation for $\dot{\overline{f}}$ simply reads
\begin{align}
  \dot{\overline{f}} \;=\; \Hc\,p\,\pfrac{\overline{f}}{p} \;.
\end{align}
Using the relation $\partial f_\sub{BB}/\partial p = -T/p\;\partial f_\sub{BB}/\partial T\,$, we find that the background temperature scales as the inverse of the scale factor,
\begin{align}
  \overline{T} \;\propto\; \frac{1}{a} \;,
\end{align}
as expected from the thermodynamical argument of \sref{sec:background_cmb}.

\annotate{Before recombination, the collision term vanishes at any order because the direct and inverse reaction rates are equal. It vanishes after recombination (excluding reionisation) because there are no collisions. At zero order, it vanishes also during recombination because the energy transfer between photons and electrons is much smaller than the temperature at recombination.}

\section{The Collision term}
\label{sec:collision_term}

In order to obtain a time evolution equation for the distribution function $f$, one needs to specify the form of the collision term in the Boltzmann equation. The collision term for a particle species described by $f$,
\begin{align}
 C[f] \;=\; C[f](\taux,\vecp) \;,
\end{align}
is the average rate of collisions happening in the neighbourhood of $(\taux)$ that result in the creation or annihilation of a particle with momentum $\vecp$. If more than one interaction can create or annihilate that type of particle, then its collision term will consist of a sum over the various contributions. 

In this section we derive the collision term for the Compton scattering between a photon and a free electron to second order in the cosmological perturbations. The period of interest is the recombination ($z\simeq1100$), when the photons progressively go out of thermal equilibrium as the electrons combine with the protons to form neutral hydrogen. Due to the low thermal energy of photons during recombination, $E_\gamma\simeq\unit[0.25]{eV}$, with respect to the electrons rest mass, $m_e\simeq\unit[511]{keV}$, one can assume that, at first order, the scattering processes are well described by the low-energy limit of the Klein-Nishina formula for the Compton scattering, that is the Thomson cross-section. We shall see that at second order one has to also consider corrections of the order of the energy transfer.
It should be noted that the photons also interact with protons; however, the proton collisions are penalised with respect to the electron ones by a factor $(m_p/m_e)^2 \simeq 1836^2$ by virtue of the mass-dependence in the Thomson scattering cross section.

The ionisation and expansion histories of the Universe play a crucial role in determining the collision term.
Before recombination ($z\lesssim1100$), all the electrons are free and the Universe is very dense. As a result, the Compton collisions between photons and electrons are so frequent that the two fluids are in thermal equilibrium, the direct collisions balancing, on average, the inverse ones.
After recombination, there are no more free electrons for the photons to scatter with, meaning that collisions cannot take place. As a result, the photons free stream in a transparent Universe. At $z\sim10$, however, the Universe undergoes a second phase transition as a result of the light from the first galaxies ionising the hydrogen in the intergalactic medium. This process is known as \keyword{reionisation}; there is now evidence from quasars that the Universe was completely ionised at $z\sim6$ \cite{becker:2001a, fan:2002a}. Reionisation is not physically different from recombination, and can be modelled within the same kinetic treatment \cite{hu:1994a, dodelson:1995a}. The main difference lies in the fact that reionisation happens when the density of the Universe is a million times smaller than at recombination, thus reducing the collision rate and making the Universe effectively transparent to radiation \cite[Sec.~3.3]{dodelson:2003b}. For this reason, in this work we do not treat reionisation.\footnote{It should be noted, however, that reionisation does play a role at second-order as it generates spectral distortions in the CMB \cite{pitrou:2010b}; we have investigated the effect of reionisation spectral distortions on the CMB spectrum both in temperature and polarisation with \SONG in Ref.~\cite{renaux-petel:2013a}.}
\annotate{Super quick introduction to 21cm and reionisation here: \url{http://www.astron.nl/moon/pdf/Koopmans_Bremen_EoR_DA.pdf}}

We shall derive the collision term up to second order following the approach of \citet{dodelson:1995a}, where only the temperature perturbations are considered. For a complete treatment including polarisation, refer to \citet{pitrou:2009a} and \citet{beneke:2010a}, and to the references therein. Note that, in \SONG, we have included the full collision term including the $E$ and $B$-modes of polarisation.

\subsection{General form of the collision term}
\label{sec:collision_general_form}

We consider the reversible reaction
\begin{align}
  \gamma(\vecp) \,+\, e(\vecq) \quad\;\longleftrightarrow\;\quad \gamma(\vecp') \,+\, e(\vecq') \;,
  \label{eq:compton_scattering}
\end{align}
representing the Compton scattering of a photon with momentum $\vecp$ off a free electron with momentum $\vecq$, that results into a photon with momentum $\vecp'$ and a free electron with momentum $\vecq'$. We assume that the electrons are thermally distributed about some bulk velocity $v_e$, as in \eref{eq:maxwell_boltzmann}. At this stage, we do not specify the form of the distribution function of the photons, $f$.

The collision term is the rate of change of the number of photons with momentum \vecp, and is therefore given by the differential cross-section for the scattering, $|M|^2$, weighted by the occupation number and integrated over all the possible momentum configurations that sum up to \vecp:
\begin{align}
  C[f](\vecp) \;=\;&
  \int\frac{\dd\vecq}{(2 \pi)^3\,2\,E_q}\, \int\frac{\dd\vecp'}{(2 \pi)^3\,2\,E_{p'}}\,
  \int\frac{\dd\vecq'}{(2 \pi)^3\,2\,E_{q'}} \;|M|^2 \nmsk
  &\times\: (2\pi)^4\,\delta(\vecp+\vecq-\vecp'-\vecq')\,\delta(E_p+E_q-E_{p'}-E_{q'})   \nmsk
  &\times\: \Bigl\{\:f_{p'}\,g_{q'}\,[1+f_{p}]\,[1-g_{q}] \,-\, f_p\,g_q\,[1+f_{p'}]\,[1-g_{q'}]\:\Bigr\} \;,
  \label{eq:collision_term_master}
\end{align}
where we have adopted the shorthand notation $E_p=E(\vecp)$, $f_p=f(\vecp)$, $g_{q'}=g(\vecq')$ and similarly for the other momenta. Because we have assumed the interaction to be reversible, the balance between the direct and inverse collisions is dictated by the relative abundances of the reagents and products of the reaction. As a result, the production and annihilation rates of $\gamma(\vecp)$ are respectively proportional to $f_{p'}\,g_{q'}$ and $f_p\,g_q$; we shall call the two terms in curly brackets the \keyword{gain term} and the \keyword{loss term}, respectively. The $1+f$ and $1-g$ factors encode the Bose enhancement and the Pauli suppression, \ie the fact that the reaction is favoured (disfavoured) if photons (electrons) with the same final state already exist; in the following, we shall approximate $1-g\simeq1$ because of the smallness of the electron density, $n_e$. The two Dirac delta functions enforce energy and momentum conservation in the local inertial frame. We are assuming that the mass-shell relation is valid, so that $E_p=p^2+m^2$, and similarly for the other momenta. As a matter of fact, to obtain \eref{eq:collision_term_master} we have already performed the integration over the energies of the particles by enforcing
\begin{align}
  \int\limits_0^\infty\,\dd E\,\delta(E^2-p^2-m^2) \;=\; \int\limits_0^\infty\,
  \dd E\;\frac{\delta\left(E-\sqrt{p^2+m^2}\right)}{2\,E} \;,
\end{align}
which explains the presence of the $2\,E$ factors.

\subsection{Energy transfer as an expansion parameter}
\label{sec:collision_energy_transfer}

We perform the first integration over $\vecq'$ by enforcing $\vecq'=\vecp-\vecp'+\vecq$ via the three-dimensional Dirac delta function:
\begin{align}
  C[f](\vecp) \;=\;& \frac{1}{8\,\pi}\,
  \int\dd p' \,p'\,\frac{\dd\Omega(\vec{\n'})}{4\pi}\:\int\frac{\dd\vecq}{(2 \pi)^3}\,
  \;\frac{|M|^2}{E_q\,E_{p-p'+q}} \nmsk
  &\times\: \delta(p-p'+E_q-E_{p-p'+q})   \nmsk
  &\times\: \Bigl\{\:f_{p'}\,g_{p-p'+q}\,[1+f_{p}] \,-\, f_p\,g_q\,[1+f_{p'}]\:\Bigr\} \;,
  \label{eq:collision_term_master_noqprime}  
\end{align}
where we have split the $\vecp'$ integration into its radial and angular parts, and we have enforced $E(p')=p'$ and $E(p)=p$. The next step is to realise that the energy transferred in the scattering, $\,p-p'=E_{q'}-E_q\,$, is much smaller than the energy scale at recombination, which is given by the ambient temperature $T$. The energy transfer is given by the difference in the kinetic energy of the electron,
\begin{align}
  E(\vecq)\,-\,E(\vecq') \;&=\; E(\vecq)\,-\,E(\vecp-\vecp'+\vecq) \;=\;
  \frac{q^2}{2\,m_e} \,-\, \frac{(\vecp-\vecp'+\vecq)^2}{2\,m_e} \nmsk
  &\simeq\; \frac{\vecq\cdot(\vecp'-\vecp)}{m_e} \;,
  \label{eq:collision_energy_transfer}
\end{align}
where, after expanding the scalar product in the last term of the first line, we have neglected the term $(\vecp-\vecp')^2$ because it is much smaller than $\vecq\cdot(\vecp-\vecp')$ by virtue of \eref{eq:collision_p_smaller_than_q}.
Since for thermal photons $|\vecp'-\vecp|=\O(T)$, it follows that the energy transfer over the temperature is of the same order as the electron velocity, $q/m_e$, which, as we have proven in \eref{eq:collision_q_smaller_than_mass}, is very small\footnote{It is interesting to note that, even if the energy transfer is very small, $p-p'=E_{q'}-E_q=\O(T\,q/m_e)$, it is still possible for a photon to scatter with a large angle, $|\vecp'-\vecp|=\O(T)$, so that $\frac{p'-p}{|\vecp'-\vecp|}=\O(q/m_e)$.} (order $10^{-3}$). Therefore, we can expand all the parts in the collision term -- energies, squared matrix element, delta functions and distribution functions -- using the energy transfer as an expansion parameter \cite{dodelson:1995a}.

The distribution function of the electrons is expanded up to second order in the energy transfer as\footnote{At zero order in the energy transfer, neither the momentum nor the direction of propagation of an electron is changed by the scattering ($\vecq'=\vecq$) because the electrons have a large mass compared to the energy of the incident photon. This is reflected in  \eref{eq:collisions_g_expanded} by the fact that, at zero order, $g(\vecq')=g(\vecq)$. This is not the case for the scattering photon, whose direction can change even if the momentum stays constant (see previous footnote).}
\begin{align}
  g\,(\vecp-\vecp'+\vecq) \;=\;&g\,(\vecq)\,\biggl\{\:1\,
  -\,\frac{(\vecp-\vecp')\cdot(\vecq-m\vecv)}{m_e\,T_e}
  -\,\frac{(\vecp-\vecp')^2}{2\,m_e\,T_e} \nmsk
  &\quad+\frac{1}{2}\,\left[\,\,\frac{(\vecp-\vecp')\cdot(\vecq-m\vecv)}{m_e\,T_e}\,\right]^2 + \dotsb \biggr\}\;.
  \label{eq:collisions_g_expanded}
\end{align}
similarly for the Dirac delta function,
\begin{align}
  \delta\,\Bigl(\,p&-p'+E(\vecq)-E(\vecp-\vecp'+\vecq)\,\Bigr) \;=\;
  \delta\,(p-p') \,+\, \frac{\vecq\cdot(\vecp-\vecp')}{m_e}\,\pfrac{\,\delta\,(p-p')}{p'} \nmsk
  &+\,\frac{(\vecp-\vecp')^2}{2\,m_e}\,\,\pfrac{\,\delta\,(p-p')}{p'} \,
  +\frac{1}{2}\,\left[\, \frac{\vecq\cdot(\vecp-\vecp')}{m_e} \,\right]^2 \,
  \ppfrac{\,\delta\,(p-p')}{p'} \;,
  \label{eq:collisions_delta_expanded}
\end{align}
where the momentum derivatives of $\delta$ makes sense only when integrated by parts.
On the other hand, we expand the photon distribution function up to second order in the cosmological perturbations:
\begin{align}
  f(\vecp) \;=\; \fz\,+\,\pert{f}{1}(\vecp)\,+\,\pert{f}{2}(\vecp) \;,
  \label{eq:collisions_f_expanded}
\end{align}
where $\fz=\pert{f}{0}(p)$ is the background blackbody distribution.
We perform the two types of perturbative expansion at the same time\footnote{It should be noted that the perturbative expansion in the energy transfer is different from the one in the metric variables. For more details on this topic, refer to the discussion in Sec.~7.2 of \citet{pitrou:2009a}.} and neglect all the terms that are higher than second order, including the mixed terms such as $\pert{f}{2}\,q/m_e$. 

The leading order in both expansions corresponds to a homogeneous Universe ($f(\vecp)=\bar{f}(p)$) where photons and electrons scatter elastically ($p'=p$). Equivalently,
\begin{align}
  \label{eq:collision_term_vanishes_leading_order}
  g(\vecq')=g(\vecq) \quad\quad\text{and}\quad\quad f(\vecp')=f(\vecp) \;.
\end{align}
It follows that the gain and loss terms in \eref{eq:collision_term_master_noqprime} are equal and opposite, so that the whole collision term vanishes at the leading order.
This has two important consequences. First, because spectral distortions can only be induced by collisions, we have proven that the zero-order CMB spectrum retains its blackbody shape even after the photons cease to be in thermal equilibrium.
Secondly, the other parts of the integrand function in \eref{eq:collision_term_master_noqprime} need to be expanded only up to first order in the energy transfer. In particular, the two energies in the denominator can be simply replaced by $m_e^2$ and the Compton matrix element is expanded as \cite{dodelson:1995a}
\begin{align}
  |M|^2 \;=\; 6\,\pi\,\sigma_T\,m_e^2\,\left[\,
  (1+\cos^2\theta) \,-\, 2\,\cos\theta\,(1-\cos\theta)\,\,
  \frac{\vecq\cdot(\vec{n}+\vec{n}')}{m_e}\,\right] \;,
  \label{eq:collisions_M_expanded}
\end{align}
where $\cos\theta=\vec{n}\cdot\vec{n}'$ and $\sigma_T$ is the Thomson cross section (\eref{eq:thomson_total_cross_section}). The first term in brackets is the angular dependence of Thomson scattering, while the second one is the first-order correction coming from the Klein-Nishina formula \cite{klein:1929a}.
\annotate{The Klein-Nishina formula is given by
\begin{align}
  \frac{\alpha^2}{2}\,(\frac{\hbar}{m_e\,c^2})^2\,P^2\,\left[P+\frac{1}{P}-1+\cos(\theta)^2\right]
\end{align}
where $\theta$ is the scattering angle and
\begin{align}
  P(E_\gamma,\theta) \;=\; \frac{E_\gamma'}{E_\gamma} \;=\;
  \left[1+\frac{E_\gamma}{m_e\,c^2}(1-\cos\theta)\right]
\end{align}
is the ration between the incident and scattered energy of the photon.}

\subsection{Contributions to the collision term}
\label{sec:collision_contributions}

The next step consists of inserting the perturbed expressions for $g$ (\eref{eq:collisions_g_expanded}), $\delta$ (\eref{eq:collisions_delta_expanded}), $f$ (\eref{eq:collisions_f_expanded}) and $|M|^2$ (\eref{eq:collisions_M_expanded}) in the collision term (\eref{eq:collision_term_master_noqprime}) and to keep only the terms up to second order. As a result, the integrand function has a simple $\vecq$ dependence that can be integrated out using the moments of the Maxwell distribution function in \eref{eq:electron_moments}. Following the approach of \citet{dodelson:1995a}, we write the resulting expression as the sum of a first-order contribution and 4 second-order ones:
\begin{align}
  \label{eq:collision_6_contributions}
  \coll[f](\vecp) \;=\; \frac{1}{\U{p}{0}}\;C[f](\vecp) \;=\;
  &-\;\frac{3}{4\,p}\;\dot\kappa\;\frac{n_e}{\overline{n}_e}\; \,
  \int\dd p'\,p'\,\frac{\dd\Omega(\vec{\n'})}{4\,\pi}\;\biggl[\;(1+\Psi)\;\pert{c}{1}(\ppp)\;\msk
  &+\;\pert{c}{2}(\ppp)\;
  +\;\pert{c}{2}_{f v}(\ppp)\;+\;
  \pert{c}{2}_{vv}(\ppp)\;+\;\pert{c}{2}_K(\ppp) \,\biggr] \;, \notag
\end{align}
where $\overline{n}_e=\pert{n_e}{0}$ is the background numer density of free electrons and we have introduced the \keyword{Thomson scattering rate},
\begin{align}
  \dot\kappa \;=\; -\,\overline{n}_e\,\sigma_T\,a \;,
\end{align}
whose meaning is explained in \sref{sec:background_compton_scattering}.
With respect to what is reported in Ref.~\cite{dodelson:1995a}, we have explicitly included the $1/p^0$ factor from \eref{eq:boltzmann_equation_schematic_time}, which is expanded to first order as 
\begin{align}
  \frac{1}{\U{p}{0}} \;=\; \frac{a}{p}\;(1+\Psi) \;.
\end{align}
The factor $(1+\Psi)$ is important as it encodes the change in the photon energy from the coordinate frame to the local inertial one. Note, however, that it is not part of the collision term, which cannot contain metric perturbations in the local inertial frame. 

A list with the form of each contribution follows.\footnote{The below equations slightly differ from the ones in \citet{dodelson:1995a} in that we have merged the purely second-order terms into $c^{(2)}$ and we have implemented the corrections that were pointed out in Appendix~C of \citet{senatore:2009b}. For an alternative splitting strategy, refer to Eq.~6 of \citet{hu:1994a}, where the photon distribution function is left unperturbed and the integrand function is expressed in terms of 7 contributions.}

\begin{itemize}

  \item The part linear in the metric perturbations consists of a damping term, also called the anisotropy suppression term, and a Doppler term:
  \begin{align}
  \pert{c}{1}&(\ppp) \,=\,\left(1+\cos^2\theta\right)\,\Bigg[\,
    \delta(p-p')\,\left(\fps1-\f1\right) \nmsk
    &+\,\Bigl(\,\fpz\,-\,\fz\,\Bigr)\,\pert{\vecv}{1}\cdot(\vecp-\vecp')\;\pfrac{\delta(p-p')}{p'} \Bigg]\;.
  \end{align}
  Once integrated in $p'$, the first term can be expressed as $\pert{f}{1}=-p\,\partial\bar{f}/\partial p\,\pert{\Theta}{1}\,$ by using \eref{eq:f_of_theta_up_to_second_order}; the second one, due to the presence of the derivative of the delta function, is proportional to $\,p\,\partial \bar{f}/\partial p\,$. Therefore, the momentum dependence of the linear collision term is encoded in an overall factor $\,p\,\partial\bar{f}/p\,$. Similarly, the Liouville term, once it is expressed in terms of $\dd\left(\pert{\Theta}{1}\right)/\dd\tau\,$, has exactly the same dependence.\annotate{The Liouville term has the $\,p\,\partial \bar{f}/\partial p\,$ dependence also at second order, see Eq. 3.24 of \cite{bartolo:2006a}.}
  This means that $\,p\,\partial\bar{f}/p\,$ can be eliminated from both sides of the Boltzmann equation, thus resulting in a momentum-independent $\pert{\Theta}{1}\,$: the linear CMB is free from spectral distortions and is therefore well described by a blackbody distribution.
  In general, all the terms in the collision term that are proportional to $\,p\,\partial\bar{f}/p\,$ result in a momentum independent temperature perturbation and, thus, in a blackbody distribution.
  
  \item The purely second-order part has the same structure of the first-order one,
  \begin{align}
  \pert{c}{2}&(\ppp) \,=\,\left(1+\cos^2\theta\right)\,\Bigg[\,
    \delta(p-p')\,\left(\fps2-\f2\right) \nmsk
    &+\,\Bigl(\,\fpz-\fz\,\Bigr)\,\pert{\vecv}{2}\cdot(\vecp-\vecp')\;\pfrac{\delta(p-p')}{p'} \Bigg]\;,
  \end{align}
  and, therefore, it does not induce spectral distortions.
  
  \item A quadratic part that mixes the photon perturbation with the electron velocity:
  \begin{align}
    \pert{c}{2}_{f v}&(\ppp) \,=\,
    \left(\fps1 - \f1\right) \,
    \Bigg[\,\left(1+\cos^2\theta\right)\,\pert{\vecv}{1}\cdot(\vecp-\vecp')\;\pfrac{\delta(p-p')}{p'}\nmsk
    &-\,2\,\cos\theta\,(1-\cos\theta)\,\delta(p-p')\,
    \pert{\vecv}{1}\cdot(\n + \n')\,\Bigg] \,.
  \end{align}
  The first term in brackets, after integration over $p'$, has the form $p\,\partial \pert{f}{1}/\partial p\,$. If we substitute $\pert{f}{1}=-p\,\partial\bar{f}/\partial p\,\pert{\Theta}{1}$, we see that, even if $\pert{\Theta}{1}$ does not depend on $p$, this term generates an explicit momentum dependence in the equation for $\pert{\Theta}{2}$ which is not of the ``blackbody'' form $\,p\,\partial\bar{f}/p\,$; that is, $\,\pert{c}{2}_{f v}\,$ does generate a spectral distortion.

  \item A part quadratic in the electron velocity:
  \begin{align}
    \pert{c}{2}_{vv}&(\ppp) \,=\, \Bigl(\,\fpz-\fz\,\Bigr)\,\pert{\vecv}{1}\cdot(\vecp-\vecp')\,
    \Bigg[\,\left(1+\cos^2\theta\right)\,\frac{\pert{\vecv}{1}\cdot(\vecp-\vecp')}{2}\,
    \ppfrac{\delta(p-p')}{p'} \nmsk
    & \,-\, 2\,\cos\theta\,(1-\cos\theta)\,
    \pert{\vecv}{1}\cdot(\n + \n')\,\;\pfrac{\delta(p-p')}{p'}\,\Bigg] \;.
  \end{align}
  The second derivative of the delta function generates $\,p^2\,\partial^2\bar{f}/p^2\,$ contributions that ultimately spoil the blackbody shape of the distribution.
  
  \item The so-called \emph{Kompaneets}\index{Kompaneets term} part,
  \begin{align}
    \pert{c}{2}_K&(\ppp) \,=\, \left(1+\cos^2\theta\right)\,\frac{(\vecp-\vecp')^2}{2\,m_e}\,
  	\Bigg[\,\Bigl(\,\fpz-\fz\,\Bigr)\,T_e\,\ppfrac{\delta(p-p')}{p'} \nmsk
    &-\Bigl(\,\fpz\,+\,\fz\,+\,2\,\fpz\,\fz\,\Bigr)\,\pfrac{\delta(p-p')}{p'}\,\Bigg] 
    \;+\;\frac{2\,(p-p')\,\cos\theta\,(1-\cos^2\theta)}{m_e}\,\nmsk
  	&\times\;\Bigg[\,\delta(p-p')\,\fpz\,\Bigl(\,1+\fz\,\Bigr)\,
    -\,T_e\,\Bigl(\,\fpz-\fz\,\Bigr)\,\pfrac{\delta(p-p')}{p'}\,\Bigg] \;,
  \end{align}
  induces spectral distortions via the terms quadratic in the distribution function and those including the second derivative of the delta function.
  The Kompaneets part is the only one with neither photon nor electron perturbations, as it is already second order in the energy transfer. It vanishes in the limit where the photon and electron temperatures coincide and we neglect it \cite[Sec.~7.4]{pitrou:2009a}.
\end{itemize}

It should be noted that we have not expanded $n_e$ yet. The density of free electrons is defined as the product between the density of all electrons and the ionisation fraction: $n_e\,=\,N_e\,x_e$. 
Because the collision term vanishes at leading order (\eref{eq:collision_term_vanishes_leading_order}), $\,n_e$ needs to be expanded only up to first order:
\begin{align}
  \frac{n_e}{\overline{n}_e} \;=\; 1\,
  +\,\frac{N_e^{(1)}}{\overline{N}_e}\,
  +\,\frac{x_e^{(1)}}{\overline{x}_e}\;.
  \label{eq:free_electron_density_perturbation}
\end{align}
The second term in parentheses is the density contrast of the electrons, which is equal to the protons' because of the tight coupling between the two fluids induced by Coulomb scattering; we denote such common value as the baryons density contrast, $\delta_b$. The third term is determined by perturbing the recombination process up to first order, and is the subject of \sref{sec:perturbed_recombination}.
After perturbing $n_e$ according to \eref{eq:free_electron_density_perturbation}, the collision term reads
\begin{align}
  \coll[f](\vecp) \;=\; &-\;\frac{3}{4\,p}\;\dot\kappa\; \,
  \int\dd p'\,p'\,\frac{\dd\Omega(\vec{\n'})}{4\,\pi}\;\biggl[\;\left(1+\pert{\Psi}{1}  
  +\;\pert{\delta}{1}_b+\frac{x_e^{(1)}}{\overline{x}_e}\right)\;\pert{c}{1}(\ppp)\notag\msk
  &+\;\pert{c}{2}(\ppp)\;
  +\;\pert{c}{2}_{f v}(\ppp)\;+\;
  \pert{c}{2}_{vv}(\ppp)\;+\;\pert{c}{2}_K(\ppp) \,\biggr] \;.
  \label{eq:collision_final}
\end{align}

All the contributions to the collision term listed above are in the form of an integral over the momentum of the scattered photon, $\vecp'$, that can be solved analytically. To do so, one needs to expand the quantities that depend on the direction of $\vecp'$ in terms of spherical harmonics, so that the $\dd\Omega(\vecp')$ integral can be solved by using the orthogonality properties of the $Y_\lm$'s (\sref{sec:spherical_harmonics}). The remaining integrals on the magnitude of the scattered momentum, $\dd p'$, is computed by enforcing the properties of the Dirac Delta function, after integration by parts. The detailed steps are explained in \citet{bartolo:2006a}; the correct formula of the second-order collision term for the CMB temperature as a function of $\vecp$ is reported in Eq.~C.1 of \citet{senatore:2009b}.\footnote{The expression obtained in Ref.~\cite{bartolo:2006a} is not correct because it assumes that the first-order distribution function only has scalar components, \ie $f^{(1)}_\lm(\kone)\propto\delta_{m0}$. This is the case only if the polar axis is chosen to coincide with the wavemode $\kone$. In a second-order expression, however, the first-order quantities are evaluated in the convolution wavevectors, \kone and \ktwo; since the polar axis was already chosen to be aligned with $\veck$, one cannot assume $f^{(1)}_\lm(\kone)\propto\delta_{m0}$; as explained in Appendix~\ref{app:perturbations_geometry}, the angular dependence of $f^{(1)}(\kone)$ is given by $f^{(1)}_\lm(\kone)\propto \tilde{f}^{(1)}_{\L0}(k_1)\,Y_\lm(\veck)$.}

\subsection{Polarisation}
\label{sec:polarisation}

So far, we have neglected the fact that Compton scattering also induces a change in the polarisation state of the photon. For example, the cross section includes terms like
\begin{align}
  |M|^2 \;\supset\; |\:\vec\epsilon\cdot\vec\epsilon'\:|^2 \;,
\end{align}
where $\,\vec\epsilon\,$ and $\,\vec\epsilon'\,$ are the incident and scattered polarisation directions of the photon, respectively.\annotate{CF: this $ee^2$ shape is only to first order. at second order there are also $ee$, $ep$ and $eq$ contributions.}
In the early Universe, the frequent interactions force the photons and the baryons to be tightly coupled in a highly isotropic fluid, the only non-negligible anisotropy being the Doppler dipole from the electrons' bulk flow; as a result, the CMB cannot develop a net polarisation.
During recombination, however, the interaction rate slows down so that the inhomogeneities in the photon fluid can convert to anisotropies. In particular, the quadrupolar variation in the incident flux of the photons, as seen by the electrons, makes it possible for the CMB to acquire a net linear polarisation through Compton scattering.
Thus, the polarisation of the CMB is due to those photons that scattered after a quadrupole anisotropy was generated. However, by the time a significant quadrupole develops, the Universe is already optically thin, that is, the scatterings are already very rare. As a result, only about $10\%$ of the CMB photon anisotropies are polarised \cite{kaiser:1983a, bond:1984a, hu:1997a, challinor:2009a}.

To describe the polarised radiation in the Boltzmann formalism, one has to introduce a Hermitian tensor-valued distribution function, $\,f_{\mu\nu}(\taux,\vecp)\,$, such that
\begin{align}
  \epsilon^\mu\;\epsilon^{*\nu}\;f_{\mu\nu}\,(\taux,\vecp)
\end{align}
is the number density of photons at ($\vecx$,\,$\vecp$) in phase space with polarisation state $\epsilon$ (see \cite{pitrou:2009b,beneke:2010a} and references therein). 
The \indexword{polarised distribution function} can be decomposed on the so-called \indexword{helicity basis} of the spherical coordinate system,
\begin{align}
  f^{\mu\nu} \;=\; \sum\limits_{ab}\;f_{ab}\;\,
  \hat{\vec\epsilon}^{*\mu}_a\;\,\hat{\vec\epsilon}^{\nu}_b \;,
\end{align}
given by the two vectors
\begin{align}
  \hat{\vec\epsilon}_+ \;=\; -\frac{1}{\sqrt{2}}\,(\,\vec e_\theta\,+\,i\,\vec e_\phi\,)
  \qquad\text{and}\qquad
  \hat{\vec\epsilon}_- \;=\; -\frac{1}{\sqrt{2}}\,(\,\vec e_\theta\,-\,i\,\vec e_\phi\,) \;,
\end{align}
where $\,\vec e_\theta=\partial_\theta\n\,$ and $\,\vec e_\phi=\partial_\phi\n/\sin\theta\,$ are the two orthonormal vectors that span the plane orthogonal to the direction of propagation of the photon, $\,\n\,$.
The $a$ and $b$ indices are called helicity indices and can assume the values $\,ab=++,\,--,\,-+,\allowbreak\,+-\,$.

The four physical degrees of freedom of $\,f_{ab}\,$ can also be expressed in terms
of the Stokes parameters,
\begin{align}
  f_{ab} \;=\; 
  \left(\begin{array}{cc} f_{++} & f_{+-} \\ 
  f_{-+} & f_{--}\end{array}\right) \;=\; 
  \left(\begin{array}{cc} f_I-f_V & f_Q-if_U \\ 
  f_Q+if_U & f_I+f_V\end{array}\right) \;,
  \label{eq:Stokes}
\end{align}
where $f_I$ is the intensity, $f_V$ the circular polarisation, $f_Q$ and $f_U$ the two components of linear polarisation. The intensity is related to the photon temperature; what we have been referring to as $f$ in the previous sections is, in the formalism of polarised radiation, $f_I$. The linear polarisation of the CMB is usually described in terms of its curl-free and gradient-free components, the $E$ and $B$ polarisation modes \cite{kamionkowski:1997a, seljak:1997a, hu:1997b}, which are obtained from the $Q$ and $U$ parameters as
\begin{align}
  f_{E,\lm} \,\pm\, i\,f_{B,\lm} \;=\; i^\L\,\sqrt{\frac{2\,\L+1}{4\,\pi}}\,
  \int\dd\Omega\;Y_\lm^{\mp2*}(\n)\,\left[\,f_Q(\n)\,\pm\,i\,f_U(\n)\,\right] \;,
\end{align}
where $Y_\lm^s(\n)$ is the spin-weighted spherical harmonic with spin $s$. In the following, we shall refer to the $E$ and $B$ polarisation modes of the photon fluid as $E$-modes and $B$-modes, respectively.
\annotate{From Beneke \& Fidler 2011, the polarisation basis is take to consist of the two circular polarisation vectors
\begin{align}
  \vec{\epsilon}_{\pm} \,=\, -\frac{1}{\sqrt{2}}\,(\vec{e}_\theta\,\pm\,i\,\vec{e}_\phi) \;.
\end{align}}
The circular polarisation, $f_V$, is not sourced by the Compton scattering or by any mechanism in the standard cosmological paradigm; we shall therefore ignore it.

The evolution of polarised light is described by a tensor-valued Boltzmann equation for $\,f_{\mu\nu}\,$, which can be recast as a system of differential equations for $\,f_{I,\lm}\,$, $\,f_{B,lm}\,$ and $\,f_{E,\lm}\,$. We shall report them in the next section, following \citet{beneke:2010a}.

\section{The final form of the Boltzmann equation}
\label{sec:final_Boltzmann_equation}


In the unpolarised case, the brightness equation is obtained by equating the Liouville term in \eref{eq:liouville_brightness} with the collision term in \eref{eq:collision_6_contributions}, after integrating out the momentum dependence of the latter using the $\beta$ operator in \eref{eq:boltzmann_brightness_operator}. The resulting expression is a partial differential equation in $\Delta(\taux,\n)$, which can be turned into a system of differential equations by projecting it into Fourier and multipole space,
\begin{align}
  \left(\mathcal{F_{\,\veck}}\circ L_\lm\circ
  \beta\right)\,\left[\,\diff{f}{\tau}\,-\,\coll[f]\,\right] \;=\; 0 \;,
\end{align}
where the three projection operators are defined, respectively, in Eq.~\ref{eq:boltzmann_brightness_operator}, \eref{eq:L_operator} and \eref{eq:fourier_operator}. 

To include polarisation, one has to follow the approach outlined in \sref{sec:polarisation}. For the details, we refer to \citet{pitrou:2009a} (\PONE, hereafter) and \citet{beneke:2010a} (\BFONE, hereafter), who independently derived the Boltzmann equation in the polarised case, up to second order and in Newtonian gauge. The two groups used different methods to derive the collision term: \PONE first computed it in the rest frame of the electron, and then performed a Lorentz boost to the coordinate frame, while \BFONE followed an approach more similar to the one we have outlined in the previous section, which consists in describing the electrons with a Maxwell-Boltzmann distribution from the beginning. Another difference is that \PONE used projected symmetric trace-free tensors to perform the angular projections, while \BFONE used spin-weighted spherical harmonics. Nonetheless, their results match up to a few minor discrepancies, as pointed out in Sec.~5 of \BFONE.

Here we report the brightness equation for the three types of photon perturbations ($\INs$, $\EMs$ and $\BMs$) by applying the $\beta$ operator to Eqs.~143 to 146 of \BFONE. Following their notation, we employ the coupling coefficients $C,D$ and $R,K$ as shorthands for the multipole decompositions of $\,n^i\,f\,$ and of $\,(\delta_{ij}-n_in_j)\,\partial f/\partial n^j\,$, respectively; we give their explicit form in Eqs.~\ref{eq:coupling_coefficients_explicit_CD} and \ref{eq:coupling_coefficients_explicit_RK}.
In writing the equations, we adopt the following conventions:
\begin{itemize}
  \item We denote the brightness multipoles with the symbols $\INs$, $\EMs$ and $\BMs$, so that
  \begin{align}
    & \IN{\L}{m}(\k) \;=\; \beta\,[\,f_{I,\lm}\,] \;,\quad&&
    & \EM{\L}{m}(\k) \;=\; \beta\,[\,f_{E,\lm}\,] \;,\quad&&
    & \BM{\L}{m}(\k) \;=\; \beta\,[\,f_{B,\lm}\,] \;.
  \end{align}
  \item We drop the perturbation suffix.
  \item We drop the explicit $\veck$ dependence in the purely second-order terms.
  \item We write the equations in terms of $\tilde\omega_i\equiv i\,\omega_i$ and $u_e^i\equiv i\,v_e^i$ in order to absorb all the imaginary factors. The variables $\tilde\omega_i$ and $u_e^i$ are the ones that are actually numerically evolved in \SONG.
\end{itemize}
In multipole space, the metric variables in \BFONE are related to ours by
\begin{align}
  A^\sub{BF}\,\rightarrow\,\Psi\;, \quad\; D^\sub{BF}\,\rightarrow\,-\Phi\;,\quad\; 
  B^\sub{BF}_{[m]}\rightarrow\,-\tilde\omega_{[m]}\,, \quad\; \alpha_m\,E^\sub{BF}_{[m]}\,\rightarrow\,-\gamma_{[m]} \;,
\end{align}
as follows from the correspondences given in \eref{eq:beneke_fidler_metric} and in footnotes~\ref{ftn:spherical_vector} to \ref{ftn:spherical_tensor} of \sref{sec:projection_of_tensors}.
Furthermore, due to the different definition of the spherical components (see footnote~\ref{ftn:spherical_vector} in \sref{sec:projection_of_tensors}), we have that $i\,k^\sub{BF}_{[m]}=-k_{[m]}$ and $v^\sub{BF}_{e[m]}=u_{e[m]}$.

We recall that the equations that follow were obtained in conformal Newtonian gauge,
\begin{align}
  \dd s^2 = a^2(\tau) \, \left\{ - (1+2\Psi) \dd \tau^2
    + 2 \omega_i \dd x^i \dd \tau
  + \,\left[\,(1-2\Phi) \delta_{ij} + 2\,\gamma_{ij} \,\right]\, \dd x^i \dd x^j \right\} \;,
\end{align}
for phase-space densities defined in an inertial frame locally at rest and aligned with the coordinate axes, under the assumption that the first-order vector and tensor perturbations in the metric vanish ($\pert{\omega_i}{1}=\pert{\gamma_{ij}}{1}=0$). For the expansion in spherical harmonics, we have chosen the zenith to be aligned with the \k wavemode.

\subsection{Purely second-order structure}

The linear structure of the Boltzmann equation follows. We group the quadratic parts of the Liouville and collision terms for the species $X$ using the symbols  $L_\lm\left[\,Q^L_X\,\right]$ and $L_\lm\left[\,Q^\mathfrak{C}_X\,\right]$, respectively.

\begin{itemize}
  \item Photon temperature:
  \begin{align}
    \label{eq:boltzmann_pure_intensity}
    \dot\INs^\L_m \;
    &+\;k\,\left(\,\IN{\L+1}{m}\,C^{+,\L}_{m\,m}\;-\;\IN{\L-1}{m}\,C^{-,\L}_{m\,m}\,\right) \;
     -\;\delta_{\ell0}\,4\,\Phid\; \;\msk
    &\quad\;-\;4\,\delta_{\ell1}\,\left(\,\delta_{m0}\,k\,\Psi\,
      -\,\delta_{m1}\,\dot{\tilde\omega}_{[1]}\,\right) \;
     -\;\delta_{\ell2}\,\delta_{m2}\,4\,\dot\gamma_{[m]}
     +\;L_\lm\left[\,\QLI\,\right] \nmsk
    &=\;\dot\kappa\,\left(\,-\IN{\L}{m} \;
     +\;\delta_{\ell0}\,\IN{0}{0} \;
     +\;\delta_{\ell1}\,4\,u_{[m]} \;
     +\;\delta_{\ell2}\,\Pi_m
    \,\right)\;
    +\;L_\lm\left[\,\QCI\,\right]  \tag*{}\;,
  \end{align}
  where we have defined
  \begin{align}
    \label{eq:Pi_polarisation_definition}
    \Pi_m \;=\; \frac{1}{10}\,\left(\,\IN{2}{m}\,-\,\sqrt{6}\,\EM{2}{m}\,\right) \;.
  \end{align}
  \item Photon $E$-mode polarisation:
  \begin{align}
    \label{eq:boltzmann_pure_emodes}
    \dot\EMs^{\,l}_{\,m} \;
    &+\;k\,\left(\,\EM{\L+1}{m}\,D^{+,\L}_{m\,m}\;-\;\EM{\L-1}{m}\,D^{-,\L}_{m\,m}\;
    +\;\BM{\L}{m}\,D^{0,\L}_{m\,m} \right) \;
    +\;L_\lm\left[\,\QLE\,\right]\; \msk
    &=\;\dot\kappa\,\left(\,-\EM{\L}{m} \;-\; \delta_{\ell2}\;\sqrt{6}\;\Pi_m \,\right) \;
    +\;L_\lm\left[\,\QCE\,\right] \;.\notag
  \end{align}
  \item Photon $B$-mode polarisation:
  \begin{align}
    \label{eq:boltzmann_pure_bmodes}
    \dot\BMs^{\,l}_{\,m} \;
    &+\;k\,\left(\,\BM{\L+1}{m}\,D^{+,\L}_{m\,m}\;-\;\BM{\L-1}{m}\,D^{-,\L}_{m\,m}\;
    -\;\EM{\L}{m}\,D^{0,\L}_{m\,m} \right) \;
    +\;L_\lm\left[\,\QLB\,\right]\; \msk
    &=\;-\,\dot\kappa\;\BM{\L}{m} \;+\;L_\lm\left[\,\QCB\,\right] \;.\notag
  \end{align}  
\end{itemize}

It is important to remark that the linear structure of the Boltzmann equation does not mix the azimuthal modes, that is, all the above expressions have the same mode, $m$, on both sides. As we have already noted in \sref{sec:perturbations_metric} and \sref{sec:spherical_projection_of_functions}, this is a consequence of having chosen the polar axis of the spherical coordinate system to coincide with the wavemode \k.

The $E$ polarisation and the temperature are directly coupled through the quadrupole of the collision term. This means that, today, we expect at least a fraction of the CMB photon anisotropies to be polarised \cite{kaiser:1983a, bond:1984a}, a circumstance that was experimentally verified \cite{kovac:2002a, bennett:1996a}. Before recombination, however, polarisation is quenched by the high scattering rate, as we shall explicitly show in \sref{sec:initial_conditions_matter} when discussing the tight-coupling approximation.

On the other hand, at first order the $B$ polarisation couples only indirectly to the temperature, through the $E$ polarisation. The coupling appears in the free-streaming part of the Liouville term, that is the first line of Eq.~\ref{eq:boltzmann_pure_emodes} and \ref{eq:boltzmann_pure_bmodes}, which means that the mixing between the $E$ and $B$-modes, at linear order, is a propagation effect rather than a scattering one. The coupling is active only for the non-scalar modes, as the coupling coefficient, $D^{0,\L}_{m\,m}$, vanishes for $m=0$. As a result, at linear order and in the standard cosmological scenario, the presence of $B$-mode polarisation today has to be linked to the presence of non-scalar perturbations in the initial conditions. In principle, because the vector modes decay with time \cite{hawking:1966a}, measuring the $B$-modes would be a smoking gun for the presence of gravitational waves in the early Universe \cite{kamionkowski:1997a, seljak:1997a, hu:1997b}. In practice, as we shall soon see, there are other sources of $B$-mode polarisation from second-order effects that need to be considered.


\subsection{Quadratic sources}

The quadratic sources of the Boltzmann equation are a convolution integral over two dummy wavemodes, \kone and \ktwo (\sref{sec:perturbations_mode_coupling}). Here, for brevity, we report the kernels of the convolution, so that, for example, when we write
\begin{align}
   L_\lm\left[\,\QLI\,\right] \;=\; \text{kernel}(\kone,\ktwo)
\end{align}
we mean
\begin{align}
   L_\lm\left[\,\QLI\,\right] \;=\; \int\,\frac{\dd\kone\dd\ktwo}{(2\pi)^3}\;
   \;\text{kernel}(\kone,\ktwo)\;\,\diracMinus \;.
\end{align}
We also omit writing the explicit $\kone$ and $\ktwo$ dependence of the first-order perturbations and assume that the first term in a product is assigned $\kone$ and the second $\ktwo$, \eg $4\,\dot\Phi\,\IN{\L}{m}=4\,\dot\Phi(\kone)\,\IN{\L}{m}(\ktwo)\,$.

The mode coupling mixes not only the wavemodes but also the azimuthal modes, as explained in \sref{sec:spherical_projection_of_functions}; in what follows, we introduce the indices $\,m_1\,$ and $\,m_2=m-m_1\,$, and implicitly assume a sum over $m_1=-1,0,+1$. Expressions for the coupling coefficients $C,D$ and $R,K$ can be found in Eqs.~\ref{eq:coupling_coefficients_explicit_CD} and \ref{eq:coupling_coefficients_explicit_RK}. 

Note that, in principle, the quadratic sources for the $E$ and $B$ polarisation should also include terms involving $\BM{\L}{m}$ at first order. However, we shall ignore them because the first-order $B$-modes vanish unless the initial conditions contain non-scalar modes, a circumstance that we do not explore in this work. For the full expression including the first-order $B$-modes, refer to Eqs.~144 and 145 of \BFONE.

\begin{itemize}
  \item Photon temperature:
  \begin{flalign}
    \label{eq:boltzmann_quad_liouville_intensity}
    L_\lm\left[\,\QLI\,\right] \;&=\;
    \sum\limits_\pm \pm\,(\Psi\,+\,\Phi)\;k_2^{[m_2]}\;\IN{\L\pm1}{m_1}\,C^{\pm,\L}_{m_1\,m} &\msk
    &+\;4\,\Bigl[\,
    -\dot\Phi\;\IN{\L}{m} \;
    +\;\sum\limits_\pm \pm\,k_1^{[m_2]}\;\Psi\;\IN{\L\pm1}{m_1}\,C^{\pm,\L}_{m_1\,m} \;
    -\;\delta_{\ell0}\;2\,\dot\Phi\,\Phi\;
    -\;\delta_{\ell1}\,k_1^{[m]}\,\Psi\,\left(\Phi\,-\,\Psi\right)\;\Bigr]\; &\nmsk
    &+\;\sum\limits_\pm \pm\,k_1^{[m_2]}\;(\Psi\,+\,\Phi)\;\IN{\L\pm1}{m_1}\,R^{\pm,l}_{m_1\,m}
    \notag\;. &
  \end{flalign}
  \begin{flalign}
    L_\lm\left[\,\QCI\,\right] \;&=\;
    \left(\;\Psi\,+\,\delta_b\,+\,\frac{x_e^{(1)}}{\bar{x}_e}\;\right)\;
    \coll_\lm\,[\,\INs\,] \;
    +\;\dot\kappa\;u_e^{[m_2]}\;\Biggl\{
    \;\sum\limits_\pm \mp\,\;\IN{\L\pm1}{m_1}\,C^{\pm,\L}_{m_1\,m} &\nmsk
    &+\;\delta_{\ell0}\;\left(\,2\,\IN{1}{m_1}\,-\,4\,u_e^{[m_1]}\right)\;C^{+,0}_{m_1\,m} \;
    +\;\delta_{\ell1}\;3\;\IN{0}{m_1}\;C^{-,1}_{m_1\,m} \;&\nmsk
    &+\;\delta_{\ell2}\;\left(\,7\,u_e^{[m_1]}\,-\,\frac{1}{2}\,\IN{1}{m_1}\right)\;C^{-,2}_{m_1\,m} \;
    +\;\delta_{\ell3}\;5\;\Pi_{m_1}\;C^{-,3}_{m_1\,m} \;\Biggr\} 
    \label{eq:boltzmann_quad_collision_intensity} \;, &
  \end{flalign}
  where $\Pi_m$ is given in \eref{eq:Pi_polarisation_definition} and $\coll_\lm$ is the first-order collision term for the intensity,
  \begin{flalign}
    \coll_\lm\,[\,\INs\,] \;=\; \dot\kappa\,\left(\,-\IN{\L}{m} \;
    +\;\delta_{\ell0}\,\IN{0}{0} \;+\;\delta_{\ell1}\,4\,u_{[m]} \;+\;\delta_{\ell2}\,\Pi_m \,\right) \;.
  \end{flalign}
  Note that the collision term, contrary to the Liouville one, does not include gradient terms (\ie an explicit \k, \kone or \ktwo dependence) because collisions are local in space.
  \item Photon $E$-mode polarisation:
  \begin{flalign}
    L_\lm\left[\,\QLE\,\right] \;&=\;
    \sum\limits_\pm \pm\,(\Psi\,+\,\Phi)\;k_2^{[m_2]}\;\EM{\L\pm1}{m_1}\,D^{\pm,\L}_{m_1\,m} &\nmsk
    &+\;4\,\Bigl[\,
    -\dot\Phi\;\EM{\L}{m} \;
    +\;\sum\limits_\pm \pm\,k_1^{[m_2]}\;\Psi\;\EM{\L\pm1}{m_1}\,D^{\pm,\L}_{m_1\,m}\;\Bigr]\; &\nmsk
    &+\;\sum\limits_\pm \pm\,k_1^{[m_2]}\;(\Psi\,+\,\Phi)\;\EM{\L\pm1}{m_1}\,K^{\pm,l}_{m_1\,m}
    \label{eq:boltzmann_quad_liouville_emodes} \;. &
  \end{flalign}
  \begin{flalign}
    L_\lm\left[\,\QCE\,\right] \;&=\;
    \left(\;\Psi\,+\,\delta_b\,+\,\frac{x_e^{(1)}}{\bar{x}_e}\;\right)\;
    \coll_\lm\,[\,\EMs\,] \;
    +\;\dot\kappa\;u_e^{[m_2]}\;\Biggl\{
    \;\sum\limits_\pm \mp\,\;\EM{\L\pm1}{m_1}\,D^{\pm,\L}_{m_1\,m} &\nmsk
    &+\;\delta_{\ell2}\;\frac{\sqrt{6}}{2}\;\left(\,\IN{1}{m_1}\,-\,2\,u_e^{[m_1]}\right)\;C^{-,2}_{m_1\,m} \;
     +\;\delta_{\ell3}\;5\,\sqrt{6}\;\Pi_{m_1}\;C^{-,3}_{m_1\,m} \;\Biggr\}
    \label{eq:boltzmann_quad_collision_emodes} \;, &
  \end{flalign}
  where $\Pi_m$ is given in \eref{eq:Pi_polarisation_definition} and $\coll_\lm$ is the first-order collision term for the $E$ polarisation,
  \begin{flalign}
    \coll_\lm\,[\,\EMs\,] \;=\; 
    \dot\kappa\,\left(\,-\EM{\L}{m} \;-\; \delta_{\ell2}\;\sqrt{6}\;\Pi_m \,\right) \;.
  \end{flalign}
  \item Photon $B$-mode polarisation:
  \begin{flalign}
    L_\lm\left[\,\QLB\,\right] \;=\;&
    -\;\left(\,\Psi\,+\,\Phi\,\right)\,k_2^{[m_2]}\,\EM{\L}{m_1}\,D^{0,\L}_{m_1\,m} &\nmsk
    &-\;4\;k_1^{[m_2]}\;\Psi\;\EM{\L}{m_1}\,D^{0,\L}_{m_1\,m} &\nmsk
    &-\;k_1^{[m_2]}\,\left(\,\Psi\,+\,\Phi\,\right)\,\EM{\L}{m_1}\,K^{0,\L}_{m_1\,m} 
    \label{eq:boltzmann_quad_liouville_bmodes} \;.
  \end{flalign}  
  \begin{flalign}
    L_\lm\left[\,\QCB\,\right] \;&=\;
    \dot\kappa\;u_e^{[m_2]}\;\Biggl\{ \;
    \EM{\L}{m_1}\;D^{0,\L}_{m_1\,m} \;
    -\;\delta_{\ell2}\;2\,\sqrt{6}\;\Pi_{m_1}\;D^{0,2}_{m_1\,m} \; \Biggr\}
    \label{eq:boltzmann_quad_collision_bmodes} \;. &
  \end{flalign}
\end{itemize}

The full second-order Boltzmann equation shows that the $B$ polarisation is generated even in the absence of vector and tensor modes. In particular, the $B$-modes are sourced by the propagation of photons through an inhomogeneous Universe, via \eref{eq:boltzmann_quad_liouville_bmodes}, and by the collisions with the electrons, via \eref{eq:boltzmann_quad_collision_bmodes}.
The former is a well known mechanism \cite{zaldarriaga:1998c,lewis:2006a} that converts E into $B$ polarisation, in analogy with the linear streaming term in \eref{eq:boltzmann_pure_bmodes}; it is dominated by the conversion due to the weak gravitational lensing \cite{hu:2001a}.
The latter mechanism includes the conversion of non-scalar $E$-modes into $B$-modes through collisions, via $\,\dot\kappa\,u_e^{[m_2]}\,\EM{\L}{m_1}\,D^{0,\L}_{m_1\,m}\,$, and the generation of the $B$-modes directly from the temperature quadrupole \cite{beneke:2010a} due to the term 
\begin{align}
  -\;\delta_{\ell2}\;\dot\kappa\;\frac{\sqrt{6}}{5}\;u_e^{[m_2]}\;
  \left(\,\IN{2}{m_1}\,-\,\sqrt{6}\,\EM{2}{m_1}\right)\;D^{0,2}_{m_1\,m} \;
\end{align}
of \eref{eq:boltzmann_quad_collision_bmodes}.
We remark that these collisional sources for the B-modes are purely kinematic in nature, as they do not exist in the electron's rest frame, that is, they are proportional to the electron velocity $u_e^{[m_2]}\,$.
Their efficiency in generating the $B$ polarisation was found to be negligible with respect to the weak lensing contribution by \citet{beneke:2011a}. For a comprehensive description and computation of the $B$-modes generated at second order, refer to \citet{fidler:2014a}.

In writing the quadratic Liouville term for $\INs$, $\EMs$ and $\BMs$, we have confined the free streaming ($\pfrac{f}{x^i}\frac{\dd x^i}{\dd\tau}$), redshift ($\pfrac{f}{p}\frac{\dd p}{\dd\tau}$) and lensing ($\pfrac{f}{n^i}\frac{\dd n^i}{\dd\tau}$) contributions to the first, second and third lines, respectively. The \L-dependence of the three effects is determined by their coupling coefficients, which we have reported in \eref{eq:coupling_coefficients_explicit_CD} and \ref{eq:coupling_coefficients_explicit_RK}. The free-streaming and redshift terms are proportional to $C^{\pm,\L}$, for the intensity, and to $D^{\pm,\L}$, for the $E$-modes; both coefficients are of order unity for large $\L$'s. On the other hand, the gravitational lensing is determined by $R^{\pm,\L}$, for the intensity, and by $K^{\pm,\L}$, for the $E$-modes; since they both grow proportionally to $\L$, we expect that, for temperature and $E$ polarisation, the gravitational lensing dominates over the other second-order propagation effects on small angular scales. For the $B$ polarisation, however, the three effects are of comparable importance as they all involve the coefficients $D^{0,\L}$ and $K^{0,\L}$, which are of order $1/\L$ for large $\L$.
Thus, in principle, the time-delay and the redshift effects are expected to be as efficient as weak lensing in converting the $E$-modes into $B$ polarisation. In practice, however, it was shown that the generation of $B$-modes through the time-delay effect is suppressed for geometrical reasons \cite{hu:2001a, creminelli:2004a}.



\subsection{A compact form of the Boltzmann equation}
\label{sec:compact_form_boltzmann}

We now introduce a compact notation for the Boltzmann equation that will be useful in the next chapter, when we shall introduce the line of sight formalism.
Following \citet{beneke:2011a}, we introduce a single composite index, $n$, to express the harmonic dependence, $(\ell,m)$, and the kind of photon perturbation (temperature, $E$ polarisation or $B$ polarisation). The Boltzmann equation at second order then reads\footnote{Note that this notation is the same that we have adopted in \citet{pettinari:2013a} and in \citet{fidler:2014a}.}
\begin{equation}
	\label{eq:compact_boltzmann}
	\dot{\Delta}_{n} \;+\; k\,\Sigma_{nn'}\,\Delta_{n'} \;+\; \M_{n}  \;+\;
  \mathcal{Q}^L_{n} \;=\; \coll_n \;,
\end{equation}
where a sum over the composite index $n'$ in implicit, and:
\begin{itemize}
  \item $\Sigma_{nn'}$ is the free streaming matrix that arises from the decomposition of $\,n^i\partial_i\Delta\,$ into spherical harmonics. Its form can be read from Eqs.~\ref{eq:boltzmann_pure_intensity} and \ref{eq:boltzmann_pure_bmodes}:
\begin{align}
  &\Sigma_{nn'}\,\Delta_{n'} \qquad\xrightarrow{\quad\INs\quad}\qquad 
  \IN{\L+1}{m}\,C^{+,\L}_{m\,m}\;-\;\IN{\L-1}{m}\,C^{-,\L}_{m\,m}\;, \nmsk
  &\Sigma_{nn'}\,\Delta_{n'} \qquad\xrightarrow{\quad\EMs\quad}\qquad
  \EM{\L+1}{m}\,D^{+,\L}_{m\,m}\;-\;\EM{\L-1}{m}\,D^{-,\L}_{m\,m}\;
  +\;\BM{\L}{m}\,D^{0,\L}_{m\,m} \;, \nmsk
  &\Sigma_{nn'}\,\Delta_{n'} \qquad\xrightarrow{\quad\BMs\quad}\qquad
  \BM{\L+1}{m}\,D^{+,\L}_{m\,m}\;-\;\BM{\L-1}{m}\,D^{-,\L}_{m\,m}\;
  -\;\EM{\L}{m}\,D^{0,\L}_{m\,m} \;.
  \label{eq:compact_free-streaming_sigma_matrix}
\end{align}
Note that free streaming mixes the $E$ and $B$-modes in an efficient way. We shall see in the next chapter (Eqs~\ref{eq:los_integral_emodes} and \ref{eq:los_integral_bmodes}) that, as a result of this coupling, the two types of polarisation directly source each other in the line of sight integral.
  \item $\M_{n}$ groups all the terms, pure and quadratic, that consist exclusively of metric perturbations. Because the polarisation multipoles do not couple directly to the metric perturbations\annotate{They are generated only through collisions, that are metric free, or through redshift-lensing-time-delay, that couple with a $\Delta$}, we have that $\M_n$ exists only for the temperature perturbations. By inspecting \eref{eq:boltzmann_partial_derivatives}, we identify \M with the only part of the Boltzmann equation that does not involve the perturbed distribution function, that is
  \begin{align}
    \M \;=\; {\pfrac{f}{p}}^{(0)}{\diff{p}{t}}^{(2)} \;.
  \end{align}
The explicit form of $\M_n$ can be read from Eqs.~\ref{eq:boltzmann_pure_intensity} and \eref{eq:boltzmann_quad_liouville_intensity}:
  \begin{align}
    \label{eq:definition_M_metric}
    &\M_{\INs,\lm} \;=\; 
      -\;\delta_{\ell0}\;4\,\left[\;\Phid\,+\,2\,\dot\Phi\,\Phi\;\right] \msk
      &\qquad-\;\delta_{\ell1}\;4\,\left[\;\delta_{m0}\,k\,\Psi\,
      +\,k_1^{[m]}\,\Psi\,\left(\Phi\,-\,\Psi\right)
      \,-\,\dot{\tilde\omega}_{[m]}\;\right] \;
      -\;\delta_{\ell2}\;4\,\dot\gamma_{[m]}\;, \nmsk
    &\M_{\EMs,\lm} \;=\; \M_{\BMs,\lm} \;=\; 0 \;.
  \end{align}
  \item $\Q{L}_{\,n}$ groups the quadratic terms in the left hand side of the Boltzmann equation that do include the perturbed photon distribution function; each of them is the product of a metric perturbation ($\Psi$ or $\Phi$ or their derivatives) with a photon perturbation ($\INs$, $\EMs$ or $\BMs$).
  Its explicit form can be obtained as
  \begin{align}
    \Q{L}_{\,n} \;=\; Q^L_{\,n} \;-\; \M_n \;,
  \end{align}
  where $\,Q^L_n\,$, depending on the index $n$, is either $\,L_\lm\left[\,\QLI\,\right]\,$, $\,L_\lm\left[\,\QLE\,\right]\,$ or $\,L_\lm\left[\,\QLB\,\right]\,$, which are reported in Eqs.~\ref{eq:boltzmann_quad_liouville_intensity}, \ref{eq:boltzmann_quad_liouville_emodes} and \ref{eq:boltzmann_quad_liouville_bmodes}, respectively.
\end{itemize}

As for the collision term, we split its second-order part in two contributions, so that it reads
\begin{equation}
	\label{eq:compact_collision_term}
	\mathfrak{C}_n \;=\; \dot\kappa\,\left(\;-\Delta_{n} \;+\;
  \Gamma_{n}\,\Delta_{n'} \;+\; \mathcal{Q}^{\,\mathfrak{C}}_{\,n}\,\,\right) \;,
\end{equation}
where $\,\mathcal{Q}^{\,\mathfrak{C}}_{\,n}\,$ is the quadratic contribution. We have introduced the split in view of building the line of sight sources in the next chapter, which, by construction, do not include the $\,-\dot\kappa\,\Delta_{n}$ term. The explicit form of the term with the $\Gamma$ matrix is immediately obtained by inspecting Eqs.~\ref{eq:boltzmann_pure_intensity} to \ref{eq:boltzmann_pure_bmodes},
\begin{align}
  &\Gamma_{nn'}\,\Delta_{n'} \qquad\xrightarrow{\quad\INs\quad}\qquad \delta_{\ell0}\,\IN{0}{0} \;
   +\;\delta_{\ell1}\,4\,u_{[m]} \;
   +\;\delta_{\ell2}\,\left(\,\IN{2}{m}\,-\,\sqrt{6}\,\EM{2}{m}\,\right)/10 \;,\nmsk
  &\Gamma_{nn'}\,\Delta_{n'} \qquad\xrightarrow{\quad\EMs\quad}\qquad -\delta_{\ell2}\;\sqrt{6}\;
  \left(\,\IN{2}{m}\,-\,\sqrt{6}\,\EM{2}{m}\,\right)/10 \;, \nmsk
  &\Gamma_{nn'}\,\Delta_{n'} \qquad\xrightarrow{\quad\BMs\quad}\qquad 0 \;.
  \label{eq:compact_collision_term_gammamatrix}
\end{align}
As for $\,\mathcal{Q}^{\,\mathfrak{C}}_{\,n}\,$, depending on the index $n$, it is either $\,L_\lm\left[\,\QCI\,\right]\,$, $\,L_\lm\left[\,\QCE\,\right]\,$ or $\,L_\lm\left[\,\QCB\,\right]\,$, which are reported in Eqs.~\ref{eq:boltzmann_quad_collision_intensity}, \ref{eq:boltzmann_quad_collision_emodes} and \ref{eq:boltzmann_quad_collision_bmodes}, respectively.


\chapterbib


\chapter{Evolution of the second-order perturbations}
\label{ch:evolution}

\section{Introduction}

The main results of the last two chapters are the Einstein and Boltzmann equations up to second order in the cosmological perturbations. The Boltzmann equation dictates the evolution of the matter fields (photons, neutrinos, baryons and cold dark matter) in an inhomogeneous Universe, while the Einstein equation describes how the curvature is affected by the distribution of matter, energy and momentum. 
By studying the structure of the equations we have seen that, at second order, several non-linear effects arise that:
\begin{itemize}
  \item couple different scales, ultimately generating an intrinsic bispectrum in the cosmic microwave background (\sref{sec:spectra_and_bispectra}) even for Gaussian initial conditions;
  \item couple the scalar, vector and tensor modes, resulting in the presence of vector and tensor modes even for purely scalar initial conditions (\sref{sec:perturbations_metric});
  \item generate $B$ polarisation both from the $E$ polarisation and from the temperature fluctuations (\sref{sec:final_Boltzmann_equation});
  \item perturb the blackbody shape of the photon spectrum (\sref{sec:photon_distribution_function}).
\end{itemize}
In order to accurately quantify these effects, the first step is to numerically solve the Boltzmann-Einstein system of coupled ODEs (BES, hereafter) at second order in the cosmological perturbations, which is the topic of this chapter. 
Even though the purpose of this Ph.~D. thesis is to compute the intrinsic bispectrum of the CMB, the results of this chapter are general and can be used to explore the other effects mentioned above.

\subsection{Summary of the chapter}

In \sref{sec:diff_system} we explain how \SONG solves the Boltzmann equation for photons, massless neutrinos, baryons and cold dark matter, including the effect of perturbed recombination.
This is a complex task that involves solving the inherent stiffness of the differential system and devising efficient sampling techniques for the time and wavemode grids.

To numerically solve the equations, we choose the initial conditions that correspond to the fastest growing mode of the density perturbations, which, in Newtonian gauge, is constant \cite{ma:1995a}. Therefore, one has to carefully match the initial conditions with the analyical solution of the differential system in the early Universe, in order to avoid exciting the decaying mode. We discuss these issues in \sref{sec:initial_conditions}.


In principle, once suitable initial conditions are specified deep in the radiation dominated era, the second-order system can be solved all the way to today. In practice, however, the CMB anisotropies cannot be computed in this way because of the size of the differential system; in fact, after the time of recombination more and more multipoles are excited and it soon becomes impractical to follow their evolution.
Instead, we use the line of sight (LOS) formalism to directly compute the today's transfer functions in a numerically efficient way. The key ingredient of the formalism is the line of sight source function, which encodes the physical effects that alter the CMB anisotropy pattern. We shall identify three contributions to the LOS source function: collision, metric and propagation sources. To build them, we still need to evolve the BES, but only until shortly after recombination. We introduce the line of sight formalism and its implementation in \SONG in \sref{sec:line_of_sight}.

Finally, \sref{sec:evolution_checks_of_robustness} we compare the numerical results of \SONG against some analytical limits known in the literature.



\section{The code, \SONG}
\label{sec:song}

\SONG is a numerical code to compute the effect of the non-linear dynamics on the CMB observables.
The reason for writing \SONG was \emph{not} to provide a more accurate version of the already existing first-order Boltzmann codes. Rather, \SONG is a tool that, given a cosmological model, provides predictions for ``new'' observables or probes that do not exist at first order, such as
\begin{itemize}
  \item the intrinsic bispectrum of the CMB,
  \item the angular power spectrum of the spectral distortions,
  \item the power spectrum of the magnetic fields generated at recombination, and
  \item the angular power spectrum of the $B$-mode polarisation.
\end{itemize}
So far, \SONG only computes the intrinsic bispectrum. It is our intention to include the other effects in the near future. This task is achievable with a comparatively smaller effort, because all these observables can be built starting from the second-order transfer functions; as we shall describe in the rest of the chapter, \SONG already implements the complex framework needed to compute the second-order transfer functions up to today.

\SONG is able to compute the polarised intrinsic bispectrum of the CMB to $5\%$ precision in about $4$ CPU-hours, which is roughly equivalent to $4$ minutes on a $60$-core machine or one hour on a standard laptop with four cores; a $10\%$ run takes about a quarter of this time, thus making it possible to compute the intrinsic bispectrum in $15$ minutes on a standard laptop. Once they are implemented, the other observables will take considerably less time, because they do not involve the computation of the non-separable bispectrum integral. These numbers have to be compared with the two weeks taken by CMBquick \cite{pitrou:2010a, pitrou:2011a} and the few days needed by CosmoLib2nd \cite{huang:2013a} for a full bispectrum run. (Note that these are rough estimates based on private communications with the authors of the aforementioned codes.)

The structure of \SONG is based on that of \emph{CLASS}, a recently released first-order Boltzmann code \cite{lesgourgues:2011a, blas:2011a}. In particular, \SONG inherits the philosophy of \emph{CLASS}, that is to provide an easy-to-use interface that builds on a modular and flexible internal structure. Special care is taken to avoid the use of hard-coded numerical values, or ``magic numbers''; the physical and numerical parameters are controlled through two separate input files by the user, who needs to set only those parameters of their interest, the others taking default values. In writing \SONG we have followed the principle of encapsulation, so that a programmer who wants to modify or add a feature to \SONG has to ``hack'' the code only in a few localised portions of the source files. When in doubt, said programmer can resort to the internal documentation, that comprises more than $10,000$ lines of comments.

We conclude this subection with a summary of the most relevant properties of \SONG:
\begin{itemize}
  \item \SONG is written in C using only freely distributed libraries.
  \item It inherits from \emph{CLASS} \cite{blas:2011a} a modular and flexible structure (work is in progress to implement a Python interface, also adapted from the one used by \emph{CLASS}).
  \item It employs an ad hoc differential evolver designed for stiff systems to solve the BES.
  \item It is OpenMP parallelised.
  \item Its source code is extensively documented with more than $10,000$ lines of comments.
  \item It uses novel algorithms for Bessel convolution, bispectrum integration and 3D interpolation.
  \item It implements the concept of beta-moments, whereby the non-realitivistic and relativistic species are treated in a unified way in terms of the moments of the distribution function.
\end{itemize}

\SONG is open-source and is available since August 2015 on the website \url{https://github.com/coccoinomane/song}.


\section{The differential system} 
\label{sec:diff_system}

The numerical integration of the Boltzmann-Einstein system at second order presents several challenges. The most obvious one comes from the sheer size of the system. Having projected the equations to Fourier and multipole space, we have introduced five external parameters in our equations: the three wavemode magnitudes, ($k_1,k_2,k_3$), and the two harmonic indices, ($\L,m$); this parameter space has to be sampled for each of the four considered species (photons, neutrinos, baryons and cold dark matter) and for the metric. In \sref{sec:the_evolved_equations}, we introduce several simplifying assumptions such as truncating the photon hierarchies to $\L_\max\sim\O(10)$ or considering baryons and CDM as perfect fluids whereby $\L_\max=1$.
Similarly, in \sref{sec:sampling_strategies}, we devise a strategy to sample the Fourier space and the time evolution grid in an optimised way, so that the regions where the transfer functions are expected to vary slowly are sampled less finely than the rest.
Even after adopting these optimisations, the system remains sizeable; in a typical run of \SONG, we evolve a system of $\sim100$ differential equations for $\sim10^6$ independent ($k_1,k_2,k_3$) triplets.
Another difficulty arises from the stiffness of the Boltzmann equation in the tight coupling regime. In \sref{sec:stiffness}, we shall explain why this is the case and show that it is a purely numerical issue which can be solved by adopting an implicit differential solver; for this purpose, we use \emph{ndf15}, the solver from the first-order Boltzmann code \emph{CLASS} \cite{blas:2011a}.
Finally, in \sref{sec:perturbed_recombination} we outline \SONG's implementation of inhomogeneous recombination, a linear effect that changes the position of the last scattering surface.

\subsection{The evolved equations} 
\label{sec:the_evolved_equations}

In this subsection, we review the differential system that is solved by \SONG and explore some of the numerical approximations employed in doing that.
A first important property is that the system is coupled in \L but decoupled in $m$, so that each $m$-mode is described by a separate differential system; in other words, the scalar ($m=0$), vector ($m=\pm1$) and tensor ($m=\pm2$) equations are decoupled from each other\footnote{It is important to note that this property is not a consequence of the decomposition theorem, which holds only at first order, but of the fact that the second-order system shares the same linear structure with the first-order one. Mode details can be found in \sref{sec:decomposition_theorem}.}.
Furthermore, we only need to evolve the $m\geq0$ modes as we consider real-valued transfer functions whereby
\begin{align}
  T_{\L-m} \;=\; (-1)^m\;T_{\lm} \;.
\end{align}
The second-order transfer functions are sourced by terms quadratic in the first-order ones, so that we first need to solve the BES at the background and linear level. For this purpose we employ \emph{CLASS}, a recently released linear Boltzmann code \cite{lesgourgues:2011a, blas:2011a}.
The linear transfer functions thus obtained are computed only in the direction of the polar axis, so that they need to be ``rotated'' according to \eref{eq:rotation} before being inserted in the quadratic sources,

\subsubsection{Einstein sector} 

In principle, the metric in \eref{eq:perturbations_the_metric} has ten degrees of freedom. After imposing the Newtonian gauge conditions ($\omega_{[0]}=\gamma_{[0]}=\gamma_{[\pm1]}=0$), and using the fact that $\omega_{[-1]}=-\omega_{[1]}$ and $\gamma_{[-2]}=\gamma_{[2]}$, we see that only four of them are independent: $\,\Psi\,$, $\,\Phi\,$, $\,\omega_{[1]}\,$ and $\,\gamma_{[2]}\,$. This means that, in order to obtain the time evolution of the metric, we only need four out of the ten Einstein equations; the remaining ones can be used to check the consistency of the numerical results and the initial conditions, as we shall do in \sref{sec:evolution_checks_of_robustness}. A list follows of the four Einstein equations that we employ in \SONG; the quadratic sources for each equation can be read from \eref{eq:einstein_quadratic_sources}.
\begin{itemize}
  \item We evolve the curvature potential $\Phi$ using the time-time equation (\eref{eq:einstein_pure_timetime}),
  \begin{align}
    \dot{\II\Phi} \;=\; -\Hc\,\II\Psi \;
    -\;\frac{k^2}{3\,\Hc}\,\II\Phi \;
    -\;\frac{1}{6\,\Hc}\,\kappa\,a^2\,\sum\,\rhoz\;\bmult{\II\Delta}{0}{0}{0} \;
    -\;\frac{\I\QTT}{6\,\Hc} \;.
    \label{eq:evolution_equation_phi_timetime}
  \end{align}
  Alternatively, \SONG supports evolving $\Phi$ with the space-time equation or the trace equation; the latter option is claimed to be numerically stabler by \citet{huang:2012b}.
  \item We determine the Newtonian potential $\Psi$ using the constraint from the scalar part of the space-space, or anisotropic stress, equation (\eref{eq:einstein_pure_spacespace}),
  \begin{align}
    \II\Psi \;=\; \II\Phi \;
    -\; \frac{1}{5\,k^2} \; \kappa\,a^2\,\sum\,\rhoz\;\bmult{\II\Delta}{2}{2}{0} \;
    +\; \frac{3}{2\,k^2} \; \I\QSS{_{[0]}} \;.
  \label{eq:evolution_equation_psi}
  \end{align}
  It should be noted that, unlike the first-order case, at second order the quadrupole includes a contribution from the non-relativistic fluids (baryons and cold dark matter), in the form of terms quadratic in their velocity. This is due to the fact that the quadrupoles do not correspond to the shear, as is clear from the discussion below \eref{eq:fluid_limit}.
  \item We evolve the vector potential $\tilde\omega_{[1]}\equiv i\,\omega_{[1]}$ using the vector part of the space-space equation (\eref{eq:einstein_pure_spacespace}),
  \begin{align}
    i\,\dot{\II\omega} \;=\;
    -\;2\,\Hc\,i\,\II\omega \;
    +\;\frac{2\,\sqrt{3}}{15\,k}\,\kappa\,a^2\,\sum\,\rhoz\;\bmult{\II\Delta}{2}{2}{1} \;
    +\;\frac{\sqrt{3}}{k}\,\I\QSS{_{[1]}} \;.
  \label{eq:evolution_equation_omega}
  \end{align}
  \item We evolve the tensor potential $\gamma_{[2]}$ using the only tensorial equation, that is the $m=2$ part of the space-space equation (\eref{eq:einstein_pure_spacespace}),
  \begin{align}
    \ddot{\II\gamma}_{[2]} \;=\;
    -\;2\,\Hc\,\dot{\II\gamma}_{[2]} \;
    -\;k^2\,{\II\gamma}_{[2]} \;
    -\;\frac{2}{15}\,\kappa\,a^2\,\sum\,\rhoz\;\bmult{\II\Delta}{2}{2}{2} \;
    -\;\I\QSS{_{[2]}} \;.
  \label{eq:evolution_equation_gamma}    
  \end{align}
\end{itemize}
The sum symbol refers to the sum over the different species, so that the $\Delta$'s appearing in the above equation are understood as
\begin{align}
  \sum\,\rhoz\,\bmult{\Delta}{n}{\L}{m} \;\:=\;\:
  \rhoz_\gamma\,\,\IN{\L}{m} \;+\;
  \rhoz_\nu\,\,\NE{\L}{m} \;+\;
  \rhoz_b\,\BA{n}{\L}{m} \;+\;
  \rhoz_c\,\CD{n}{\L}{m} \;,
\end{align}
where the terms in the right hand side correspond to the background density and moments of the photon, neutrino, baryon and cold dark matter distribution functions, respectively.
Note that we have denoted the moments of the baryon and cold dark matter fluids as $\BA{n}{\L}{m}$ and $\CD{n}{\L}{m}\,$, respectively.
We recall that the $\,\bmult{\Delta}{n}{\L}{m}\,$ variables are the moments of the distribution function, and are related to the energy-momentum tensor and to the fluid variables according to \eref{eq:energy_momentum_tensor_multipoles_betamoments} and \ref{eq:fluid_limit_w}, respectively.

One could choose a different set of equations to determine the four degrees of freedom in the metric. For example, one could evolve the curvature potential $\Phi$ by using its second time-derivative from the trace Einstein equation in \eref{eq:einstein_pure_trace}, as shown in \citet{huang:2012b}, or the first time-derivative from the longitudinal equation in \eref{eq:einstein_pure_spacetime}, as it is now done in \emph{CLASS}.
Not all solutions, however, are numerically stable. In fact, $\Phi$ was initially determined in \SONG by using the constraint equation obtained by combining the time-time and space-time equations,
\begin{align}
  \Phi \;=\;
  -\;\frac{\kappa\,a^2\,\rhoz}{2\,k^3}\;\left(\, \Hc\,\Delta_{10} \;+\; k\,\Delta_{00} \,\right) \;
  -\frac{1}{2\,k^2}\,{\QTT}_{[0]}\;-\;\frac{3\,\Hc}{2\,k^3}\,{\QST}_{[0]} \;.
  \label{eq:unstable_phi}
\end{align}
This equation turns up to be numerically unstable at first and second order because, at early times, the two terms in parentheses cancel each other, that is $\pert{\Delta_{10}}{1} \simeq -\pert{\Delta_{00}}{1}\,k/\Hc$. The loss of significant digits due the cancellation is then enhanced by the $1/k^3$ factor, which can be as large as $10^{18} \text{Mpc}^3$ on large scales\footnote{The \emph{CLASS} code initially used \eref{eq:unstable_phi} to evolve $\Phi$; this was changed in v1.4 after we communicated with the authors about the numerical instability. \emph{CLASS} now uses the space-time equation.}.


\subsubsection{Relativistic sector} 

We evolve the multipoles for the photon temperature and polarisation using the Boltzmann equation in harmonic and Fourier space; its linear structure is reported in \eref{eq:boltzmann_pure_intensity} to \ref{eq:boltzmann_pure_bmodes}, while its quadratic sources can be found in \eref{eq:boltzmann_quad_liouville_intensity} to \ref{eq:boltzmann_quad_collision_bmodes}. The neutrino multipoles, being collisionless and assumed to be massless, obey the same equations but without a collision term.
The linear structure of the Boltzmann equation is such that adjacent multipoles are coupled to each other, thus defining an infinite hierarchy of equations where the evolution of, say, $\IN{\L}{m}$ is determined by $\IN{\L-1}{m}$ and $\IN{\L+1}{m}\,$. The azimuthal modes, on the other hand, do not couple thanks to the decomposition theorem.

Before recombination, all moments vanish apart from the monopole, the dipole and, at second order, the quadrupole. As the time of decoupling approaches and the mean free path of the photons increases, the $\L$ coupling in the BES has the effect of propagating the anisotropies from these small multipoles to the large ones. In physical terms, we can say that the inhomogeneities begin to generate anisotropies. The efficiency of this transmission of power is proportional to $k$, due to the gradient term in Liouville equation. As a result, the time of excitation $\tau$ of the multipole $\L$ obeys the following approximate relation,
\begin{align}
  \L \;\sim\; k\;(\,\tau\;-\;\taurec\,) \;.
  \label{eq:free_streaming_transmission_of_power}
\end{align}
(Note that the neutrinos obey a similar relation where $\,\L=k\tau\,$ because, being collisionless, they always stream freely.) These arguments apply equally to the first and second-order differential systems, as both share the same structure of equations.

To solve the BES numerically, one has to truncate the $\L$-hierarchy at some multipole $\,L_\text{cut}\,$. The simplest approach consists in setting all the multipoles with $\L>\Lcut$ to zero. Doing so, however, disrupts the symmetry of the system by preventing the higher moments with $\L>\Lcut$ to feed back into the lower ones, thus generating numerical noise. Following the argument that led to \eref{eq:free_streaming_transmission_of_power}, we expect this disruption to affect the lower moments in a time which is inversely proportional to $k$; namely,
\begin{align}
  \Lcut \;\sim\; \L \;+\; k\;\frac{(\,\tau\;-\;\taurec\,)}{2} \;.
  \label{eq:free_streaming_truncation_disruption}
\end{align}
The reason for the factor $1/2$ is that the anisotropies have to propagate first from $\L$ to $\Lcut$, where the disruption is created, and then back to $\L$.

In the line of sight approach (\sref{sec:line_of_sight}), we sample the multipoles up to the quadrupole ($\L=2$) until the decay of the visibility function, which corresponds to $\tau-\tau_\rec\simeq\unit[120]{Mpc}$ for a standard \LCDM cosmology.  If we consider that the smallest scale probed usually corresponds to $\,k=\unit[0.2]{Mpc^{-1}}\,$, we see from \eref{eq:free_streaming_truncation_disruption} that to accomplish this goal we have to evolve at least $\,14\,$ multipoles in the Boltzmann hierarchy. While this is certainly a viable option, there are more efficient truncation schemes than a simple cutoff of the hierarchy.
The most widely used truncation scheme is the one described in Ref.~\cite{ma:1995a}, which uses the fact that, in the absence of scattering, the first-order multipoles behave like spherical Bessel functions, $\,\IN{\L}{m}\propto\,j_\L(k\tau)\,$. Then, the recurrence properties of the Bessel functions can be used to express the last element in the $\L$-hierarchy without reference to the higher-order ones \cite{seljak:1996a}. In \SONG we adopt this truncation scheme for the four relativistic hierarchies, applying the general \keyword{closure relations} provided in Appendix~D by \citet{pitrou:2010a},
\begin{align}
  &\dot\INs^{\L}_m \;=\; k\;\left[\;
  \sqrt{\frac{\L+|m|}{\L-|m|}}\;\frac{2\L+1}{2\L-1}\;\IN{\L-1}{m}
  \;-\; \frac{\L+1+|m|}{k\,\tau}\;\IN{\L}{m} \;\right] \;,\nmsk
  &\dot\EMs^{\L}_m \;=\; k\;\left[\;
  \sqrt{1\,-\,\frac{m^2}{\L^2}}\;\sqrt{\frac{\L+2}{\L-2}}\;
  \frac{2\L+1}{2\L-1}\;\EM{\L-1}{m} \;-\; \frac{\L+3}{k\,\tau}\;\EM{\L}{m}
  \;-\;\frac{m}{\L}\;\BM{\L}{m} \;\right] \;,\nmsk
  &\dot\BMs^{\L}_m \;=\; k\;\left[\;
  \sqrt{1\,-\,\frac{m^2}{\L^2}}\;\sqrt{\frac{\L+2}{\L-2}}\;
  \frac{2\L+1}{2\L-1}\;\BM{\L-1}{m} \;-\; \frac{\L+3}{k\,\tau}\;\BM{\L}{m}
  \;+\;\frac{m}{\L}\;\EM{\L}{m} \;\right] \;.
  \label{eq:closure_relations}
\end{align}
For the neutrinos, we use the same relations as for the photons. At second order, the presence of the quadratic sources undoes the spherical Bessel solution; nonetheless, the above closure relations represent an improvement over the simple cutoff scheme, and allow us to obtain a percent convergence in the spectrum and in the bispectrum already for $\,\Lcut=8\,$.




\subsubsection{Cold matter sector} 

In \SONG we treat the baryons and the cold dark matter as pressureless perfect fluids, which are described only by their energy density and velocity. We are justified in doing so because the baryon fluid is non-relativistic, since the masses of the electron ($m_ec^2=\unit[511]{keV}$) and of the proton ($m_pc^2=\unit[938]{MeV}$) are much larger than the background temperature for all considered times. As for dark matter, it has to be non-relativistic, or cold, in order to explain the formation of structure in the observable Universe \cite{dodelson:2003b}.

The usual approach at second order is to evolve the energy density and the velocity of the massive species using the continuity and Euler equations \cite{pitrou:2010a, beneke:2011a}. In \SONG, we prefer to adopt a unified treatment where all the species are described by the Boltzmann equation in terms of the moments of the distribution function.
In order to do so, in \sref{sec:tetrad_emt} we have introduced the beta-moments, an expansion of the one-particle distribution function in terms of the powers the particle's velocity,
\begin{align}
  1\,+\,{}_n\Delta\,(\taux,\n) \;\equiv\; \frac{1}{\int\dd p\,p^3\,\bar{f}(\tau,p)}
  \int\dd p\,p^3\;\left(\,\frac{p}{E}\,\right)^{n-1}\,f\,(\taux,p,\n) \;.
\end{align}
The beta-moments are directly related to the energy-momentum tensor,
\begin{align}
  &\tUD{T}{0}{0}  \;=\; -\rhoz\;(1+\bmult{\Delta}{0}{0}{0}) \;,
  &&\tUD{T}{i}{i}  \;=\; \rhoz\;(1+\bmult{\Delta}{2}{0}{0}) \;, \nmsk
  &i\,\xivector{m}{i}\:\tDD{T}{i}{0}  \;=\;  -\frac{1}{3}\,\rhoz\;\bmult{\Delta}{1}{1}{m} \;,
  &&\chimatrix{2}{m}{ij}\:\tDD{T}{i}{j} \;=\;  -\,\frac{2}{15}\,\rhoz\;\bmult{\Delta}{2}{2}{m} \;.
\end{align}
The equivalent expression for the fluid variables (\eref{eq:energy_momentum_tensor_multipoles_delta}) includes extra quadratic terms in the fluid's velocity, which need to be accounted for when computing the right hand side of Einstein equations; by evolving directly the beta-moments, we can avoid performing this step. 
The relation of the beta-moments with the fluid variables can be read from \eref{eq:fluid_limit_w}.

The main advantage of the beta-moments is that they can be used to describe any particle regardless of its mass. For the photons and the massless neutrinos ($p/E=1$) they reduce to the usual brighness moments $\,\bmult{\Delta}{n}{\L}{m}=\Delta_{lm}\,$, while for the baryons and the cold dark matter ($p\ll E$), only the lowest order beta-moments survive, and we recover the fluid limit.
In general, one can project the Boltzmann equation into a hierarchy of ODEs for three indices, ($n,l,m$), using 
\begin{align}
  \left(\mathcal{F_{\,\k}}\circ L_\lm\circ
  \beta_n\right)\,\left[\,\diff{f}{\tau}\,-\,\frac{1}{p^0}\,C[f]\,\right] \;=\; 0 \;,
  \label{eq:projected_boltzmann_equation_betamoments}
\end{align}
where $\beta_n\,$ is the operator that projects a function into its $n$-th beta-moment,
\begin{align}
  \beta_n[\,F\,] \;\equiv\; \frac{1}{\int\,dp\,p^2\,E\,\overline{F}}\;
  \int\,dp\,p^3\,\left(\frac{p}{E}\right)^{n-1}\,F(p) \;.
  \label{eq:betamoments_operator}
\end{align}
The standard brightness equation for the photons is just the special case of \eref{eq:projected_boltzmann_equation_betamoments} where $n=1$. To project the Boltzmann equation into its beta-moments, the following relations are needed,
\begin{align}
  &\betaop{\left(\frac{p}{E}\right)^m\,f}{n} \,=\, {\,}_{n+m}\Delta\;,
  &&\betaop{E\,\pfrac{f}{p}}{n} \,=\, -(n+2)\,{\,}_{n-1}\Delta \:+\: (n-2)\,{\,}_{n+1}\Delta\;, \nmsk
  &\betaop{\pfrac{f}{\tau}}{n} \,=\, \pfrac{{\,}_n\Delta}{\tau} \:-\: 3\,\Hc\,{\,}_n\Delta \:(1 + w)\;,
  &&\betaop{p\,\pfrac{f}{p}}{n} \,=\, -(n+3)\,{\,}_n\Delta \:+\: (n-1)\,{\,}_{n+2}\Delta\;.
  \label{eq:boltzmann_betamoments_integrals}
\end{align}
The expressions are obtained by performing simple integration by parts and by using the on-shell relation $E(p)=\sqrt{p^2 + m^2}$.  By setting $\,{}_n\Delta=\Delta$ and substituting $n=1$ in the coefficients, one recovers the usual relations for the photon brightness (see \eref{eq:boltzmann_brightness_integrals}).

We denote the beta-moments of the baryon and cold dark matter fluids as $\BA{n}{\L}{m}$ and $\CD{n}{\L}{m}\,$, respectively.
Since we treat them as perfect-fluids, the only moments that survive are the $n=0$ and $n=1$ ones. Their evolution is governed by the following equations:
\begin{align}
  \label{eq:boltzmann_pure_baryons}
  &{}_0\dot{b}^0_0 \;=\; -\; \Hc\;\BA{2}{0}{0}\;
  -\;\frac{k}{3}\;\BA{1}{1}{0} \;+\; 3\,\dot\Phi \
  \;-\; \left(\,L_{00}\circ\beta_0\,\right)\,[\QLb]
  \;-\; r\,\coll[\,\INs\,]_{00} \;, \msk
  &\begin{aligned}
    {}_1\dot{b}^1_m \;=\; &-\Hc\,\BA{1}{1}{m} \;+\; k\,\left(\;
    C^{-,1}_{m\,m}\;\BA{2}{0}{0} \;-\; C^{+,1}_{m\,m}\;\BA{2}{2}{m}
    \;\right) \;-\; \left(\,L_{1m}\circ\beta_1\,\right)\,[\QLb] \msk
    \;&+3\;\delta_{m0}\,k\,\Psi
    \;-\;3\,\delta_{m1}\,\left(\,\dot{\tilde{\omega}}_{[1]}
    \,+\;\Hc\,\tilde{\omega}_{[1]}\,\right)
    \;-\;r\,\coll[\,\INs\,]_{1m} \;,
  \end{aligned}
  \notag
\end{align}
where $\,r=\rhoz_\gamma/\rhoz_b\,$ and $\,\coll[\,\INs\,]_\lm\,$ is the collision term for the photons, which coincides with the right hand side of \eref{eq:boltzmann_pure_intensity}.
Let us stress that the collision term for the baryons has a very simple form; were we evolving the energy density and the velocity instead of the monopole and the dipole, the equations would have included extra quadratic terms in the fluid's velocity. The cold dark matter moments obey identical equations, but with the collision term set to zero.

It should be noted that $\,\BA{2}{0}{0}\,$ and $\,\BA{2}{2}{m}\,$ enter the evolution equations for the monopole and the dipole. In principle, to obtain their value we would need to evolve the $n=2$ moment of the Boltzmann equation.  However, using \eref{eq:fluid_limit}, we see that, at second order, they are respectively related to the pressure and to the anisotropic stress,
\begin{align}
  \rhoz\,\BA{2}{0}{0} \;=\; 3\,P \;+\; (\rhoz\,+\,\Pz)\,\tU{v}{i}\,\tD{v}{i}
  \qquad\text{and}\qquad
  \rhoz\,\BA{2}{2}{m} \;=\; -\frac{15}{2}\;
  \bigl[\,\Sigma_{[m]}\;+\;(\rhoz\,+\,\Pz)\:\tensorP{m}{v}{v}\,\bigr] \;.
\end{align}
As the anisotropic stress vanishes for a perfect fluid like the baryons, we can simply set
\begin{align}
  \BA{2}{0}{0} \;=\; \tU{v}{i}\,\tD{v}{i}
  \qquad\text{and}\qquad
  \BA{2}{2}{m} \;=\; -\frac{15}{2}\;\tensorP{m}{v}{v} \;,
\end{align}
where the quadratic velocity term are known from the solution of the first-order differential system.

As a final note, we remark that using the beta-moments to treat a perfect fluid is more a matter of preference rather than necessity.
However, when it comes to species that are neither relativistic nor cold, like massive neutrinos and other non-cold relics, the beta-moments are an efficient way to solve the Boltzmann equation.
Indeed, the first-order code \emph{CAMB} \cite{lewis:2000a} implements the massive neutrinos using a momentum-integrated Boltzmann hierarchy which is equivalent to the beta-moments \cite{lewis:2002b}.
The usual way to treat massive neutrinos in a first-order Boltzmann code consists in evolving the perturbation of the distribution function as a partial differential equation, on a momentum grid \cite{lesgourgues:2011b}.
Using a velocity expansion, instead, the problem would be that of solving a hierarchy of ODEs in the beta-moments ($n$), in complete analogy with the Fourier projection in wavemodes ($\k$) and the harmonic one in spherical harmonics ($\lm$).



\subsection{Sampling strategies} 
\label{sec:sampling_strategies}

In this subsection we discuss the strategy adopted in \SONG to sample the Fourier and time grids. In doing so, we use some of the optimisation introduced in first-order Boltzmann codes such as \emph{CLASS} \cite{lesgourgues:2011a} and \emph{CAMB} \cite{lewis:2000a}. This is possible because the second-order differential system, apart from the obvious differences of having the non-scalar modes and three Fourier modes instead of one, is similar to the first-order one. Furthermore, the physical scales involved -- age of the Universe, distance to recombination, sound horizon at recombination, epoch of matter-radiation equality -- are all background quantities.

Below, we shall introduce a few numerical parameters and choose reference values for them; although such choices might seem arbitrary at this stage, we shall back them up with extensive convergence tests in the next chapter, in \sref{sec:convergence_tests}.

\subsubsection{Sampling of $k_1$ and $k_2$} 

Due to mode coupling (\sref{sec:perturbations_mode_coupling}), the second-order system has to be solved on a three-dimensional grid in Fourier space. In \SONG, we parametrise the $k$-space using the magnitudes of the three comoving wavevectors, $k_1$, $k_2$ and $k_3$, and take $k_3$ as the one satisfying the triangular condition,
\begin{align}
  |k_1-k_2| \;\leq\; k_3 \;\leq\; k_1+k_2 \;.
\end{align}
Therefore, our transfer functions depend on four parameters, \eg, $\pert{\Psi}{2}(k_1, k_2, k_3, \tau)$. We recall that the actual second-order perturbations are obtained as a convolution of the transfer functions with two primordial potentials, $\Phi(\kone)\,\Phi(\ktwo)$ (\eref{eq:transfer_function_definition}); every observable quantity, including the bispectrum, depends on such integrals rather than on the transfer functions themselves, which are just mathematical objects.

In \SONG, we fix a lower and an upper limit for all the wavemodes, regardless of whether they are $k_1$, $k_2$ or $k_3\,$; we denote such limits as $\,\kmin$ and $\kmax\,$. Their value is determined by two numerical parameters, $\,K_\text{min}\,$ and $\,K_\text{max}\,$, as
\begin{align}
  &\kmin \;=\; \frac{1}{\tau_\sub{0}}\;K_\text{min} \qquad\text{and}\qquad
  \kmax \;=\; \frac{\lmax}{\tau_\sub{0}}\;K_\text{max} \;,
\end{align}
where $\lmax$ is the maximum angular multipole that we want to probe. The choice of the parametrisation follows from the fact that a comoving scale $k$ at recombination is projected onto our sky, today, at an angular scale of $\,\L\simeq k(\tau_\sub{0}-\tau_\text{rec})\simeq k\tau_\sub{0}\,$.
We find that choosing $\,K_\text{min}\leq0.5\,$ and $\,K_\text{max}\geq1.5\,$ gives a percent level convergence in the bispectrum of the cosmic microwave background for $\,\lmin=2\,$ and $\,\lmax>1000\,$.

In a typical run of \SONG, we employ the resolution of the Planck experiment, $\,\lmax\simeq2000\,$, which, based on the above arguments, corresponds to sampling the Fourier space in the range between $\,\kmin\simeq\unit[10^{-5}]{Mpc}\,$ and $\,\kmax\simeq\unit[0.2]{Mpc}\,$ (assuming a standard \LCDM model where $\tauz\simeq\unit[14,000]{Mpc}$). 
This amounts to 4 orders of magnitude in Fourier space that, for high precision runs, can extend to 5 or 6. Given that we are dealing with a 3D space, it is clear that the sampling strategy should be optimised as much as possible to avoid wasting precious computational time.

\begin{figure}[t]
	\centering
		\includegraphics[width=0.7\linewidth]{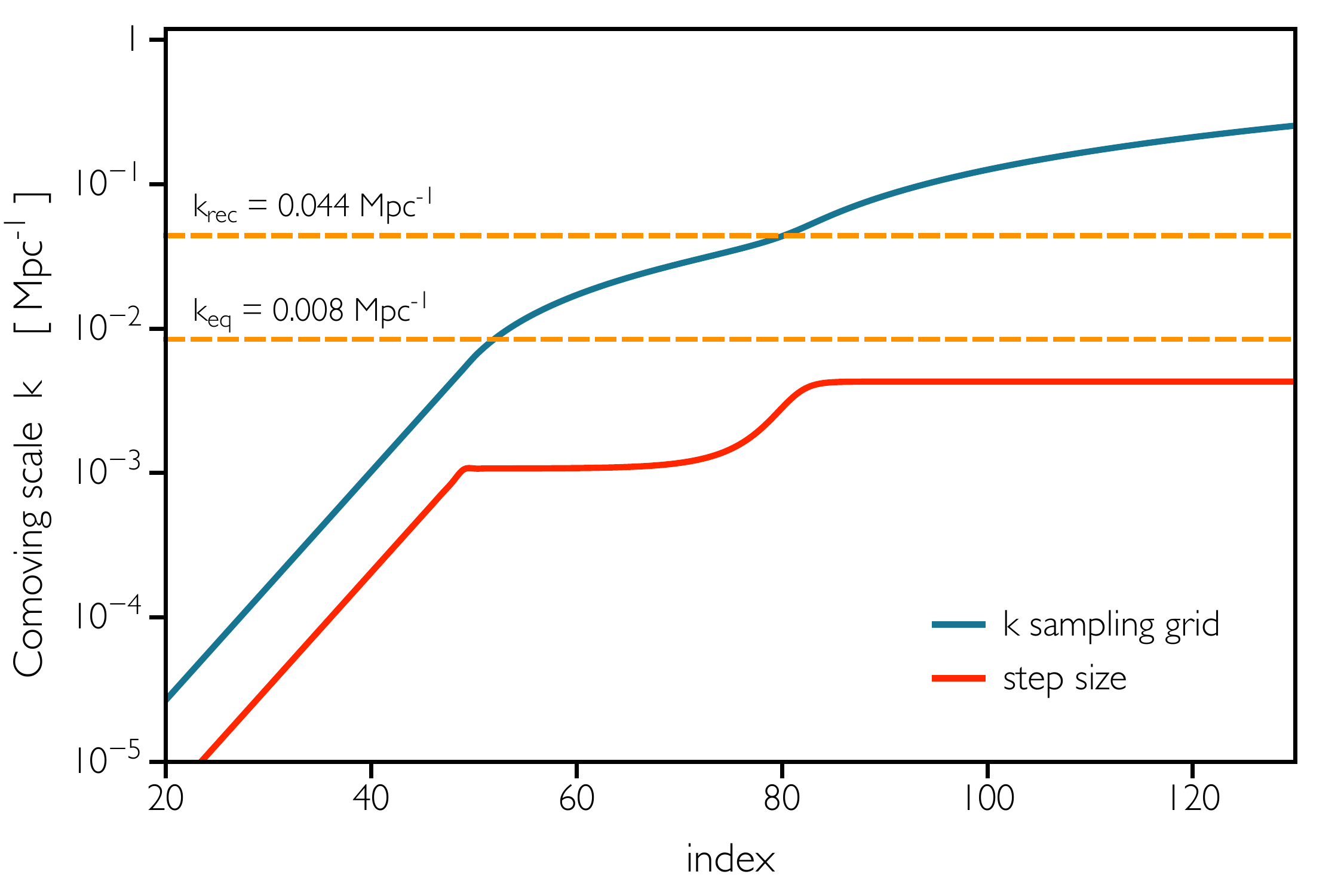}
	\caption[Sampling of the transfer functions in $k$]{
  Example of the comoving Fourier grid that is used to sample $k_1$ and $k_2$ in \SONG. In order to sample equally well the four orders of magnitude spanned, we define three regimes: a logarithmic regime up to $k=\unit[10^{-3}]{Mpc^{-1}}$, a first linear regime up to $\krec=\unit[0.044]{Mpc^{-1}}$, and a second linear one all the way to $\kmax$.
	}
	\label{fig:k_sampling}
\end{figure}

Using a linear $k$-sampling obviously neglects the large-scale details of the system, unless the step is chosen to be of the same order as $\kmin\,$, a prohibitive choice from the computational point of view. On the other hand, a 
logarithmic sampling would fail to capture the oscillations in $k$ experienced by the transfer functions on scales 
that are smaller than the sound horizon at recombination, $\,\krec\,$.
The approach of the \emph{CLASS} code is to use two linearly sampled intervals with different steps: a fine one 
from $\kmin$ to $\krec$ and a coarse one from $\krec$ to $\kmax$. (Note that, for a standard \LCDM cosmology, 
$\,\krec=\sqrt{3}\,(2\pi)/\taurec\simeq\unit[0.04]{Mpc}\,$.) To smooth the transition between the two linear 
regimes, an arctangent function with variable width is used. The two steps are parametrised in units of $\,\krec$ with the parameters 
$\,K^\text{super}_\text{lin}\,$ and $\,K^\text{sub}_\text{lin}\,$. 

In \SONG, we slightly modify the strategy used by \emph{CLASS} by including a logarithmic sampling, $\,K_\text{log}\,$, which is used starting from $\,\kmin\,$ and is kept as long as the step is smaller than both $\,\krec\,K^\text{super}_\text{lin}\,$ and $\,\krec\,K^\text{sub}_\text{lin}\,$. After that, \emph{CLASS}' strategy is used all the way to $\,\kmax\,$. Schematically, this corresponds to having the logarithmic step
\begin{align}
  k_{n+1}\;=\;k_{n}\;K_\text{log} \quad\;\text{until}\quad\;
  k_{n+1}\,-\,k_{n} \;<\; \min\,\left(\,\krec\,K^\text{super}_\text{lin},
  \;\,\krec\,K^\text{sub}_\text{lin}\,\right) \;.
\end{align}
The inclusion of a logarithmic regime makes it possible to obtain a convergence in the bispectrum using fewer $k$-values. In \fref{fig:k_sampling} we show the $k$-grid thus obtained for our standard set of parameters, $K_\text{min}=0.1\,$, $\,K_\text{max}=2\,$, $\,\,K_\text{log}=1.2\,$, $K^\text{super}_\text{lin}=0.025$ and $K^\text{sub}_\text{lin}=0.1$, which, for a \LCDM universe, gives rise to about $N_k=130$ values.


\subsubsection{Sampling for $k_3$} 

We draw the magnitudes of the wavemodes $k_1$ and $k_2$ from the $k$-grid that we have obtained following the procedure outlined above (hereafter, we shall refer to such grid as \k). An important optimisation that can be made at this stage is to symmetrise the quadratic sources of the BES with respect to the exchange of $k_1$ and $k_2$; by doing so, we are allowed to solve the system only for those ($k_1,k_2$) couples whereby $k_1\,\geq\,k_2$. This results in a two-dimensional grid with $N(N+1)/2$ nodes, where $N$ is the number of points in \k. 

For each couple ($k_1,k_2$), we need to create a second grid for $k_3$ that satisfies the triangular condition, \ie $k_3 \in [\,|k_1-k_2|,\,k_1+k_2]$.
In order to minimise the number of parameters in the code, we sample $k_3$ using the points in \k, taking care of including only those $k$-values that fall into the triangular regime for the considered ($k_1$, $k_2$). One of the consequences of this choice is that the $k_3$ wavemode will never take values below $\kmin$ or above $\kmax$, even if they were allowed by the triangular condition.\footnote{There are obviously other ways to sample the triangular wavemode, $k_3$. In fact, in CMBquick \cite{pitrou:2010a} a different technique is used where, for each $k_1$ and $k_2$, the $k_3$ grid is chosen so that the angle between $\kone$ and $\ktwo$ is linearly sampled for a fixed number of time (16 in the latest version of CMBQuick).}

If either $k_1$ or $k_2$ is very small, it is likely that none of the values in \k satisfies the triangular condition; when this happens, we just sample $k_3$ linearly between $\,|k_1-k_2|\,$ and $\,k_1+k_2\,$ using a fixed number of points. We find that the bispectrum is insensitive to this number; this is expected, because the size of these regions in the 3D Fourier space is very small, and therefore they contribute only marginally to any observable.

For the standard set of \SONG parameters, \k counts around $140$ elements, while the total number of nodes in the $(k_1,k_2,k_3)$ mesh amounts to about $150,000$. This means that, using the above strategy, the average size of a $k_3$ grid is of 8 elements.


\subsubsection{Time sampling of the line of sight sources} 

Rather than evolving the photon multipoles all the way to today, we sample them only up to a certain time to build their line of sight sources, as discussed in detail in \sref{sec:line_of_sight}.
Therefore, we need to devise a time sampling of the transfer functions that captures all of their relevant features. The time steps of the differential solver can be used for this purpose as, by definition, they closely follow the variations in the transfer functions; in fact, this is how we store the time evolution of the background quantities. However, this method is computationally inefficient, as the differential solver always performs more steps, typically $\O(1000)$, than what is strictly needed to sample the transfer functions, typically $\O(200)$.  When it comes to second order, where we evolve about $100$ transfer functions for more than $10^5$ wavemodes, this option is impractical, from both points of view of memory usage and computational speed.

To optimise the time sampling of the transfer functions, we adopt the same strategy of \emph{CLASS}. We start sampling the transfer functions when the Universe starts to become transparent to the CMB photons, that is when the Compton interaction rate has slowed down enough to be comparable with the expansion rate.
The exact time is determined  by the parameter $\,T_\text{start}\,$, defined as
\begin{align}
  \frac{\tau_\sub{\Hc}(\tau_\text{start})}{\tau_c(\tau_\text{start})} 
  \;=\; \frac{\dot\kappa(\tau_\text{start})}{\Hc(\tau_\text{start})} \;=\; T_\text{start} \;.
\end{align}
We find a percent convergence in the spectrum and in the bispectrum for values of $\,T_\text{start}\leq0.01\,$, which in conformal time correspond to $\tau_\text{start}\leq\unit[230]{Mpc}$ for a \LCDM model where the peak of recombination is at $\taurec\simeq\unit[280]{Mpc}$. It is important to note that $\tau_\text{start}$ is \emph{not} the time at which we start evolving the sytem, $\,\tau_\text{ini}\,$, which is much smaller and of order $\O(\unit[0.1]{Mpc})$.

\begin{figure}[t]
	\centering
		\includegraphics[width=0.7\linewidth]{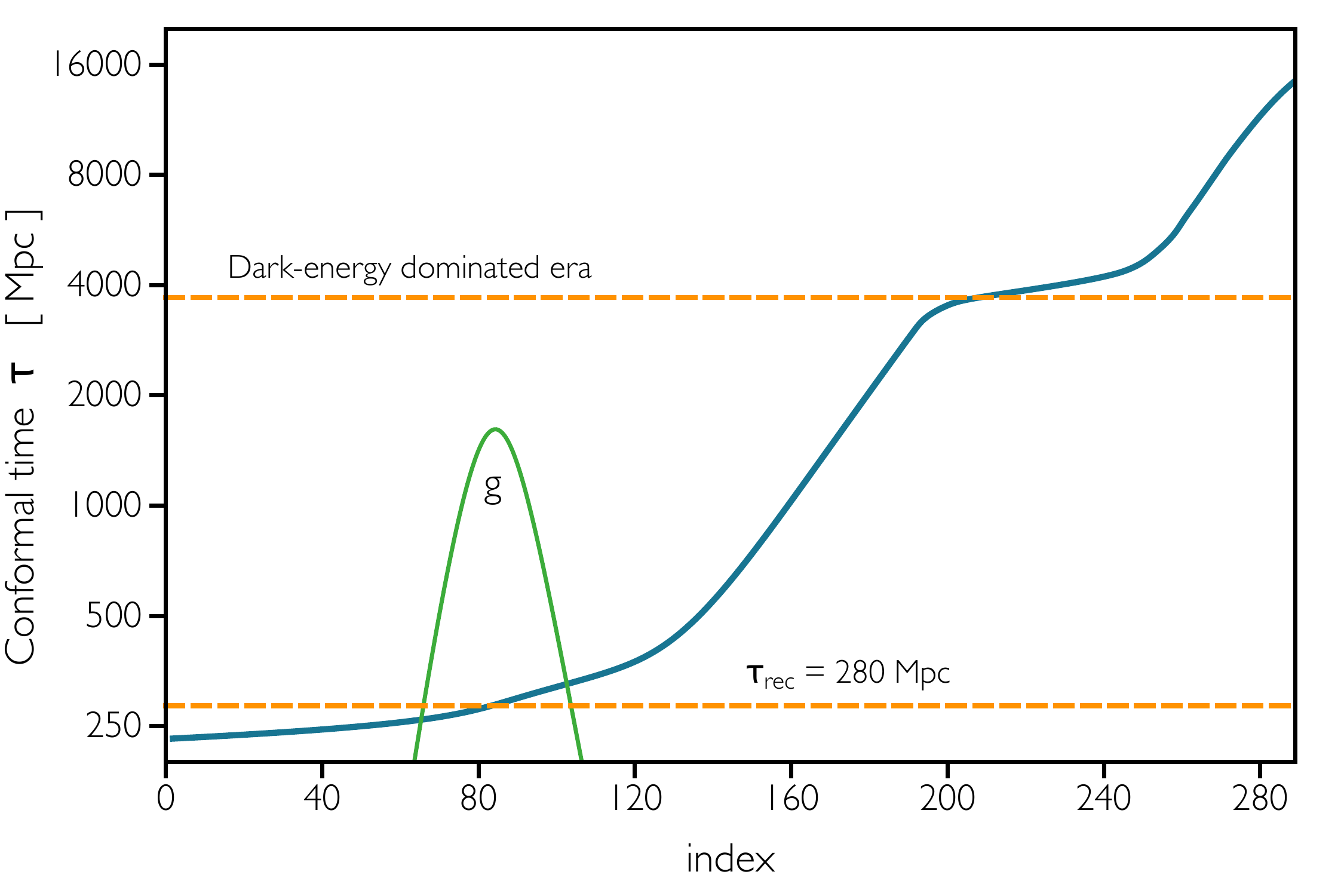}
	\caption[Sampling of the second-order transfer functions in time]{
  Example of the conformal time grid that is used to sample the line of sight sources in \SONG. The sampling is devised so that the regions close to the two phase transitions -- the one from an opaque to a transparent Universe and that from a matter dominated to a dark-energy dominated era -- are sampled more finely than the others. In green, we show the shape of the visibility function around recombination.
	}
	\label{fig:tau_sampling}
\end{figure}

We then define at each time two timescales: the time variation of the visibility function and that of the cosmic expansion, that is
\begin{align}
  \Delta\tau_\text{rec} \;=\; \frac{g}{\dot{g}} \;\quad\text{and}\quad\;
  \Delta\tau_\text{exp} \;=\; \frac{1}{\sqrt{\abs{2\,\frac{\ddot{a}}{a}{\;-\;\Hc^2}}}} \;,
\end{align}
respectively. (Note that the second timescale is the usual Hubble time with a correction to include extra points during a phase of accelerated expansion, such as the one induced by late time dark energy.)
The sampling points that follow $\tau_\text{start}$ are determined by the lowest of the two timescales,
\begin{align}
  \tau_{n+1} \;=\; \tau_n \;+\; T_\text{step}\,\Delta\tau \;\;\quad\text{with}\quad\;\;
  \Delta_\tau \;=\; \left[\;\frac{1}{\Delta\tau_\text{rec}}\;
  +\;\frac{1}{\Delta\tau_\text{exp}}\;\right]^{-1} \;.
\end{align}
Smaller values of the numerical parameter $\,T_\text{step}\,$ correspond to finer time samplings; a percent-level convergence in the spectrum and in the bispectrum is found by setting $T_\text{step}\leq0.2$

In \fref{fig:tau_sampling} we show the time sampling of the second-order line of sight sources which, adopting the typical parameters $\,T_\text{start}=0.008\,$ and $\,T_\text{step}=0.2\,$, consists of $\,N_\tau=290\,$ points between $\,\tau_\text{start}=\unit[230]{Mpc}\,$ and $\,\tauz=\unit[14,300]{Mpc}\,$, for a standard \LCDM model where $\,\taurec=\unit[280]{Mpc}\,$.



\subsection{The differential solver} 
\label{sec:stiffness}

\subsubsection{Stiffness in the differential system}

One of the major difficulties in deriving the evolution of the photon anisotropies is that the Boltzmann equation is numerically stiff.
Stiffness in a differential equation of the form $y'=f(t,y)$ arises when its exact solution, $y(t)$, contains a term that decays exponentially to zero, but whose derivative is much larger and of opposite sign with respect to the term itself. A simple example of stiff system is given by
\begin{align}
  y' \;=\; -c\:y\;, \quad\quad t\,>\,0\;,\quad\quad y(0)\,=\,1 \;,
  \label{eq:stiff_exponential_example}
\end{align}
where $\,c\,$ is a large and positive constant; the exact solution is the exponentially decaying function $y=e^{\,-c\,t}$. If we numerically solve the equation using the simple Euler's method with a step size of $h$, we obtain for the $n$-th iteration 
\begin{align}
  y_{n+1} \;=\; y_n\,+\,h\,y'_n \;=\; (1\;-\; h\,c)\;y_n \;,
\end{align}
which yields the solution $\,y_{n}=(1-h\,c)^n\,$. The numerical solution correctly converges to zero for $\,n\rightarrow \infty\,$ only if the step is chosen so that $\,h<2/c\,$, otherwise it is a diverging and exponentially growing succession that alternately undershoots and overshoots the exact solution.

By looking at the evolution equation for the photons, \eref{eq:boltzmann_pure_intensity}, we see that the scattering rate, $\,\dot\kappa=a\,n_e\,\sigma_T\,$, plays the same role that $\,c\,$ had in the previous example, making the system potentially stiff. 
To follow the evolution of the differential system with an explicit integration method, such as Euler or Runge-Kutta, the time step $h$ needs to be smaller than $\,1/\dot\kappa\equiv\tau_c\,$, the mean time between two collisions.
This is clearly not an issue after recombination, where the collisions are absent ($\dot\kappa=0$). In that case, the evolution of the system is determined on super-horizon scales by the conformal Hubble time, $\,1/\Hc\equiv\tau_\sub{\Hc}\,$, and on sub-horizon scales by $\,1/k\equiv\tau_k\,$; both are typically of order \unit[1]{Mpc} or larger, meaning that the system can be evolved until today, $\tauend\simeq\unit[14000]{Mpc}$, in roughly $\,\O(3000)\,$ steps, the exact number depending on the considered wavemode.
However, before recombination the interaction time $\,\tau_c\,$, which is proportional to $\propto a^{-2}$, is much smaller than both $\,\tau_\sub{\Hc}\,$ ($\propto a^{-1}$) and $\,\tau_k\,$ ($\propto a^0$), and the time step needs to be similarly small. In a typical run of \SONG, we set the initial conditions at $\tauini=\unit[0.5]{Mpc}$ when the interaction time, $\,\tau_c\simeq\unit[10^{-6}]{Mpc}\,$, is at least 5 orders of magnitudes smaller than $\,\tau_k\,$ or $\,\tau_\sub{\Hc}\,$. To evolve the system with a step size of $\,\tau_c\,$ up to the end of recombination, $\tauend\simeq\unit[400]{Mpc}$, requires about $\sci{4}{8}$ time steps.
This approach is not practical as we need to solve the system for more than $10^5$ different configurations of the wavemodes; furthermore, it is unsatisfactory to use so many time steps to sample a function that we know to be smooth.

Stiff systems are more easily treated using an implicit integration method, that is, a method where information from the next step, in the form of $\,y'_{n+1}\,$, is used to estimate $\,y_{n+1}\,$. The simplest implicit method is the backward Euler's method, whereby $\,y_{n+1}=y_n+h\,y'_{n+1}\,$. Going back to the example of \eref{eq:stiff_exponential_example}, this is equivalent to using
\begin{align}
  y_{\,n+1} \;=\; y_n\;+\; h\,y'_{n+1} \;=\; y_n \;-\; h\,c\,y_{\,n+1} \;,
\end{align}
whose solution,
\begin{align}
  y_{\,n+1} \;=\; \frac{y_n}{1+h\,c} \;\;\Rightarrow\;\; y_n\;=\;\left(\frac{1}{1+h\,c}\right)^n \;,
\end{align}
correctly decays to zero as $n$ increases, for any step size and without oscillations, thus solving the stiffness of the system. The drawback of using an implicit method is that $y_{n+1}$ can be obtained only after solving an implicit algebraic equation. In the general case of a system of coupled differential equations, one has to solve a system of algebraic equations in the vector-valued $\vecy_{n+1}$ at each time step.

\subsubsection{An implicit evolver}

To evolve the Boltzmann-Einstein system of coupled ODEs in \SONG, we use \emph{ndf15} \cite{blas:2011a}, the ODE solver of the first-order Boltzmann code \emph{CLASS} \cite{lesgourgues:2011a}. The principle of \emph{ndf15} is similar to that of the simple backward Euler's method that we have discussed above, in that it is an implicit method built to overcome the stiffness of the system. It uses, however, the more elaborated numerical differentiation formulae in Ref.~\cite{shampine:1997a} which are built to ensure a faster convergence using fewer time steps. 

The implicit formulae for $\vecy_{n+1}$ form a linear system of algebraic equations which is solved numerically, at each step, by using Newton's method. In principle, this requires the computation of the Jacobian of the system at each time step, which, for a typical run where $N\sim100$ cosmological perturbations are evolved, is an $N\times N$ matrix. This part is optimised in two ways. First, each step reuses the previous Jacobian unless the convergence of Newton's method is too slow.\footnote{The Jacobian is computed only for the purpose of accelerating the convergence of Newton's method; it is not used in building the differentiation formulae. Therefore, reusing it does not imply a loss of precision, but just a slightly slower convergence.} Secondly, a sparse matrix method is used to optimise the storage and access of the Jacobian matrix, using the fact that most of the Jacobian's entries are zeros due to the system being only partially coupled. (As an example, consider the fact that the neutrino hierarchy is coupled only to the metric, and that the polarisation and intensity hierarchies are mutually coupled only through the $\L=2$ and $\L=3$ moments.)

By using \emph{ndf15} and the optimisation techiques outlined above, we manage to evolve the Boltzmann-Einstein system of coupled ODEs up to the end of recombination for a given ($k_1,k_2,k_3$) triplet in $\O(1000)$ time steps and with $\O(50)$ Jacobian computations, where we have considered a scalar ($m=0$) system consisting of roughly $100$ equations, with a requested tolerance of $10^{-4}$. In a complete run, we solve the same system for about $10^5$ independent ($k_1,k_2,k_3$) configurations in about $1$ hour on a quad-core machine.

Another approach to solve the stiffness problem is the so-called \indexword{tight-coupling approximation} \cite{blas:2011a, peebles:1970a, ma:1995a}, where the photon hierarchy is expanded in powers of the interaction time, $t_c=1/\dot\kappa$, to obtain equations that are numerically well behaved. The resulting differential system is drastically reduced in size as the anisotropies with $\L>2$ are tight-coupling suppressed. While we do use the tight-coupling approximation to find the initial conditions of the photon fluid in \sref{sec:initial_conditions}, we have not implemented it yet in \SONG to solve the differential system; we plan to do so in the near future as it is likely to reduce the computation time considerably.

We conclude this subsection by noting that the above considerations are valid at any order in perturbation theory. In particular, the stiffness is always present as it pertains to the linear structure of Boltzmann equation; this is the reason why the differential solver from \emph{CLASS} is well suited for the task at hand. Note, however, that at second order the quadratic sources depend on two wavemodes, $\kone$ and $\ktwo$, meaning that the timescale $\tau_k$ is given by $1/\max(k_1,k_2,k_3)$ rather than by $1/k$.


\subsection{Perturbed recombination} 
\label{sec:perturbed_recombination}

The existence of the density perturbations make the recombination process inhomogeneous, in the sense that different regions of the Universe have different ionisation histories according to the local density of free electrons; this effect is known as \keyword{perturbed recombination} and slightly alters the time of decoupling and the visibility function.
The perturbed recombination is encoded by the presence in the collision term (\eref{eq:collision_final}) of the term
\begin{align}
  \Bigl(\;\pert{\delta}{1}_b \;+\; \pert{\delta}{1}_x\;\Bigr)\;\;\pert{\coll}{1} \;,
  \label{eq:collision_perturbed_recombination}
\end{align}
where $\,\pert{\delta}{1}_x \;\equiv\; x_e^{\,(1)}/\bar{x}_e\,$ is the perturbation in the fraction of the free electrons. Note that, since the collision term vanishes in a homogeneous Universe, the CMB is affected by the perturbed recombination only at the second order level.

At the background level, the recombination process is usually treated by using the 3-level atom approximation \cite{peebles:1968a}, whereby the hydrogen is considered as an atom with effectively 3 energy levels: ground state, first excited state and continuum.
As a result, the ionisation history is determined by a single differential equation for the free electron density,
\begin{align}
  \dot{\bar{x}}_e \;=\; a\,\bar{\mathcal{Q}}\,\bar{x}_e \;,
\end{align}
where the collision term is a complicated function of four parameters, $\,\mathcal{Q}\,(\,x_e,\,n_b,\,T,\,H)\,$; its expression can be obtained as $\,\mathcal{Q}=Q/n_e\,$ from Eq.~2.10 of \citet{senatore:2009b}.
In \SONG, we compute the background ionisation history by implementing the code RECFAST \cite{seager:1999a, wong:2008a}, which is indeed based on a slightly modified version of the three-level approximation.
\annotate{From Seager et al.~1999: our results can be reproduced by artificially speeding up recombination [at z<800] in the standard calculation, simply by multiplying the recombination and ionisation coefficient by a “fudge factor” F.}

\begin{figure}[t]
	\centering
		\includegraphics[width=1.03\linewidth]{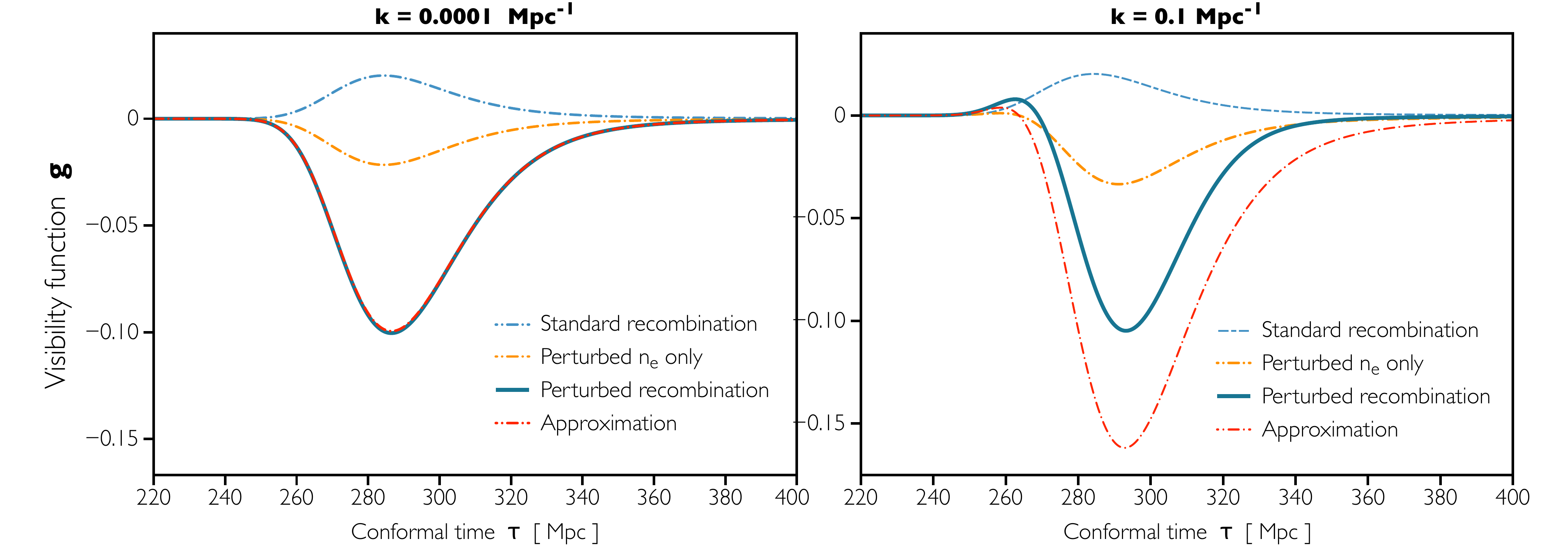}
	\caption[Perturbed recombination]{
  Effect of perturbing recombination on the visibility function for a super-horizon (left panel) and a sub-horizon (right panel) mode. 
  From top to bottom: the usual unperturbed visibility function, $\,\bar{g}=\dot\kappa e^{\,-\kappa}\,$; the visibility function including the effect of the perturbed electron density, $\,g=\bar{g}\,\delta_b\,$; the same with the addition of the perturbed ionisation fraction $\,g=\bar{g}\,(\delta_b+\delta_x)\,$; the superhorizon approximation, $\,g=\bar{g}\,\delta_b\,[\,1 - \dot{x}_e/(3\,x_e\,\Hc)]\,$, obtained by considering  $\delta_e$ as a time delay of the homogeneous solution (see Sec.~3.3 of Ref.~\cite{senatore:2009b} for details). For the super-horizon mode, the approximation is very precise and the two curves are indistinguishable. In general, we see that including the perturbation to $x_e$ enhances the visibility function.
  It is important to note that, at second order, the perturbed visibility function enters only that part of the collision term shown in \eref{eq:collision_perturbed_recombination}; the rest of $\coll[f]$ is multiplied by the standard unperturbed $g$.
	}
	\label{fig:perturbed_recombination}
\end{figure}

The physics of the perturbed recombination has been treated by several authors \cite{novosyadlyj:2006a, lewis:2007a, lewis:2007b, senatore:2009b}. In particular, \citet{senatore:2009b} have rigorously proved that the perturbed ionisation fraction, $\,\delta_x\,$, is still well described by the recombination equation for the 3-level atom, as long as it is expressed in terms of the perturbed variables. The resulting equation for $\delta_x\,$ is given by
\begin{align}
  \dot\delta_x \;=\; a\,\left[\;\Psi\,\bar{\mathcal{Q}}\;+\;\pert{\mathcal{Q}}{1}\,\right] \;,
  \label{eq:perturbed_recombination_delta_x_dot}
\end{align}
where the perturbed source function $\pert{\mathcal{Q}}{1}$ is obtained by expanding the arguments of $\,\mathcal{Q}\,(\,x_e,\,n_b,\,T,\,H)\,$ up to first order,
\begin{align}
  \pert{\mathcal{Q}}{1} \;=\;
  \pfrac{\mathcal{Q}}{x_e}\;\bar{x}_e\,\pert{\delta}{1}_x \;+
  \pfrac{\mathcal{Q}}{n_b}\;\bar{n}_b\,\pert{\delta}{1}_b \;+
  \pfrac{\mathcal{Q}}{T}\;\bar{T}\,\pert{\Theta}{1} \;+
  \pfrac{\mathcal{Q}}{x_e}\;H\,\pert{\delta}{1}_\sub{H} \;.
  \label{eq:perturbed_recombination_Q_derivatives}
\end{align}
The temperature perturbation can be expressed in terms of the energy perturbations of photons as $\,\pert{\Theta}{1}=\delta_g/4\,$, while, at first order, $H$ assumes the meaning of the local divergence of the baryons,
\begin{align}
  \pert{\delta}{1}_\sub{H} \;=\; -\Psi \;-\; \frac{\dot\delta_b}{3\,\Hc} \;\;.
\end{align}

We have implemented the perturbed recombination in \SONG using \eref{eq:collision_perturbed_recombination}, \ref{eq:perturbed_recombination_delta_x_dot} and \eref{eq:perturbed_recombination_Q_derivatives}. We have considered the photon and electron temperatures to coincide and we have not included the effect of Helium recombination; it was shown in Ref.~\cite{senatore:2009b} that both are very good approximation for the computation of the CMB anisotropies. As we shall see in \cref{ch:intrinsic}, we find that the perturbed recombination does not affect the intrinsic bispectrum of the CMB at a significant level.
In \fref{fig:perturbed_recombination} we show our numerical results for the perturbed recombination, which are in perfect agreement with those obtained by \citet{senatore:2009b}. In particular, we confirm that $\,\delta_x\,$ is 2-5 times larger than $\,\delta_b\,$ around recombination, depending on the considered $k$-mode.



\section{The initial conditions} 
\label{sec:initial_conditions}

In this section we derive the initial conditions of the second-order transfer functions for the differential system. Because the transfer functions are decoupled from the details of the primordial potential (\sref{sec:transfer_functions}), such as the amplitude of the primordial spectrum or the non-Gaussianity, we do not choose a specific model of the early Universe yet.
At this stage, we only assume that the primordial perturbations are \emph{adiabatic}\index{adiabatic initial conditions}, meaning that the relative abundances of the different species (photons, neutrinos, baryons and cold dark matter) are spatially constant.
Another approach would be to keep the total energy density spatially constant (thus leaving the curvature unperturbed) but to allow the relative abundances to vary, in what are called \indexword{isocurvature initial conditions} \cite{bucher:2000a}. The simplest models of single-field inflation generate adiabatic initial conditions, while the isocurvature modes naturally arises in the context of multifield inflation \cite{linde:1997a, enqvist:2002a,lyth:2002a,moroi:2001a,moroi:2002a}. However, CMB observations constrain the fractional contribution to the primordial power spectrum from the isocurvature modes to be below a few percent \cite{valiviita:2012a, planck-collaboration:2013d}, thus motivating our choice of adiabatic initial conditions.

We set the initial conditions deep in the radiation dominated era when all the evolved Fourier modes are super-horizon, so that we can expand the system in terms of $\,k\tau\ll1\,$ and neglect all the terms that are of order $(k\tau)^2$ or larger. 
In the typical \SONG run, we start evolving the system at $a\simeq10^{-6}$; back then, the baryon and the cold dark matter fluids make a negligible contribution to the total energy density, so that $\rhoz_\sub{tot}=\rhoz_\gamma+\rhoz_\nu$ and $\Hc=1/\tau$ (\sref{sec:expansion_history}).
Note that, in the Newtonian gauge, a constant mode and a decaying mode exist for the density perturbations \cite{ma:1995a}. We shall assume that, when we set our initial conditions, the decaying mode is already negligible, so that the energy density of the various species and the two scalar potentials are time independent.

Under the assumptions of adiabatic and super-horizon perturbations in the radiation dominated era, it is possible to compute the evolution of the transfer functions analytically, up to second order, by solving the Boltzmann-Einstein system. In doing so below, we recover the results obtained by \citet{pitrou:2010a}, and derive a new formula for the adiabatic velocity perturbations, \eref{eq:adiabatic_velocity_initial_condition}. We set the initial conditions in this way only for the scalar modes ($m=0$), and assume vanishing initial conditions for the non-scalar ones ($m\neq0$). This is equivalent to assuming that no vector nor tensor modes were produced in the primordial Universe and, since the non-scalar modes do not have a monopole, to ignoring the terms that grow like $k\tau$ or faster, which is reasonable as long as we set our initial conditions early enough.


\subsection{Initial conditions for the matter perturbations}
\label{sec:initial_conditions_matter}

For purely \indexword{adiabatic initial conditions}, all the fluids in the early Universe (photons, neutrinos, baryons and cold dark matter) share a common velocity field,
\begin{align}
  v_{\gamma[m]}\;=\;v_{\nu[m]}\;=\;v_{b[m]}\;=\;v_{c[m]} \;,
  \label{eq:adiabaticity_velocity}
\end{align}
and their density perturbations are locked together so that the ratios between $\,\rho_\gamma^{\nicefrac{1}{4}}\,$, $\,\rho_\nu^{\nicefrac{1}{4}}\,$, $\,\rho_b^{\nicefrac{1}{3}}\,$ and $\,\rho_c^{\nicefrac{1}{3}}\,$ remain spatially constant.
In particular, we have that the energy density of any relativistic species, $\,\rho_\sub{R}\,$, is related to that of a non-relativistic one, $\,\rho_\sub{M}\,$, by
\begin{align}
  \frac{\rho_\sub{R}^{\,\nicefrac{1}{4}}}{\rhoz_\sub{R}^{\,\nicefrac{1}{4}}} \;=\;
  \frac{\rho_\sub{M}^{\,\nicefrac{1}{3}}}{\rhoz_\sub{M}^{\,\nicefrac{1}{3}}} \,
\end{align}
which, after introducing the density contrast $\delta=(\rho-\rhoz)/\rhoz$, reads
\begin{align}
  (1\,+\,\delta_\sub{R})^{\nicefrac{1}{4}} \;=\; (1\,+\,\delta_\sub{M})^{\nicefrac{1}{3}}
  \quad\Rightarrow\quad
  \frac{\delta_\sub{R}}{4}\,-\,\frac{3}{16}\,\delta_\sub{R}^2 \;=\; \frac{\delta_\sub{M}}{3}\,-\,\frac{2}{9}\,\delta_\sub{M}^2
\end{align}
where in the second line we have expanded the expression up to second order using \eref{eq:perturbations_fourier_1_plus_x}.
It follows that, at first order, the two energy densities are related by a $3/4$ factor,
\begin{align}
  \frac{\pert{\delta_\sub{R}}{1}}{4} \;=\;  \frac{\pert{\delta_\sub{M}}{1}}{3} \;.
\end{align}
Thus, the expression for $\delta_\sub{R}$ up to second order is
\begin{align}
  \frac{\delta_\sub{M}}{3} \;=\; \frac{\delta_\sub{R}}{4}\;-\;\frac{1}{16}\,\delta_\sub{R}^2 \;,
  \label{eq:adiabaticity_density}
\end{align}
where $\delta_\sub{R}$ refers to either $\delta_\gamma$ of $\delta_\nu\,$, and $\delta_\sub{M}$ to either $\delta_b$ or $\delta_c\,$.

Thanks to the adiabaticity relations \eref{eq:adiabaticity_velocity}  and \ref{eq:adiabaticity_density}, we only need to find the initial conditions for the common adiabatic velocity, which we denote $v$, and for the density perturbation of one of the fluids. To do so, in the next two subsections, we use the space-time and time-time Einstein equations, respectively.
In the last two subsections, we shall also compute the initial conditions for the photon and neutrino quadrupoles. Whereas they are in principle negligible, because of order $(k\tau)^2\,$, they need to be considered in order to compute the initial values of the metric potentials due to a cancellation in the anisotropic stress equation, as we shall see in \sref{sec:initial_conditions_metric}.

\subsubsection{Dipoles}
The space-time Einstein equation in the Newtonian gauge (\eref{eq:einstein_pure_spacetime}) reads
\begin{align}
  \dot\Phi \;=\;
  -\Hc\,\Psi\;
  +\; \frac{1}{6\,k}\,\kappa\,a^2\,\sum\,\rhoz\,\bmult{\Delta}{1}{1}{0}\;
  +\; \frac{\QST{_{[0]}}}{2\,k} \;.
  \label{eq:evolution_equation_phi_longitudinal}
\end{align}
At early times, if we only consider the constant mode of the initial conditions, we can set $\dot\Psi=0$, while the expression for the quadratic contribution in $\QST$ is found in \eref{eq:einstein_quadratic_sources}.
The term containing the dipole can be expanded as
\begin{align}
  \kappa\,a^2\,\sum\,\rhoz\;\bmult{\Delta}{1}{1}{m} \;&=\;
  \kappa\,a^2\,\left[\vphantom{\Omega_\gamma}\;\rhoz_\gamma\;\IN{1}{m}\,+\,\rhoz_\nu\;\NE{1}{m}\,
  +\,\rhoz_b\,\BA{1}{1}{m}\,+\,\rhoz_c\,\CD{1}{1}{m}\;\right] \nmsk
  &=\;3\,\Hc^2\,\left[\;\Omega_\gamma\;\IN{1}{m}\,+\,\Omega_\nu\;\NE{1}{m}\,
  +\,\Omega_b\,\BA{1}{1}{m}\,+\,\Omega_c\,\CD{1}{1}{m}\;\right] \;,
\end{align}
where in the second line we have extracted $\rhoz_\sub{tot}$ and used the Friedmann equation to write $\,\kappa\,a^2\,\rhoz_\sub{tot}=3\,\Hc^2\,$. Note that we are allowed to simplify the above expression by setting $\Omega_c=\Omega_b=0$, but we refrain from doing so in order to get a slightly more accurate result. We now enforce the relation between the dipole of a given species and its velocity, \eref{eq:fluid_limit_w},
\begin{align}
  &\IN{1}{m} \;=\; \NE{1}{m} \;=\; 4\,i\,\left(\,v_{[m]}\,
  +\,\delta_\sub{R}\,v_{[m]}\,\right) \;, \nmsk
  &\BA{1}{1}{m} \;=\; \CD{1}{1}{m} \;=\; 3\,i\,\left(\,v_{[m]}\,
  +\,\delta_\sub{M}\,v_{[m]}\,\right) \;,
  \label{eq:dipole_initial_condition}
\end{align}
to express the velocities of all the species in terms of the common adiabatic velocity, $v$,
\begin{align}
  \kappa\,a^2\,\sum\,\rhoz\,\bmult{\Delta}{1}{1}{m} \;=\;
  3\,\Hc^2\;i\,v_{[m]}\;\left[\,3\,\Omega_\sub{M}\,
  (1+\delta_\sub{M})\,+\,4\,\Omega_r\,(1+\delta_\sub{R})\,\right] \;,
\end{align}
where we have used the adiabaticity to set $\,\delta_\gamma=\delta_\nu\equiv\delta_\sub{R}\,$ and $\,\delta_b=\delta_c\equiv\delta_\sub{M}\,$, and we have collected $\,\Omega_\gamma+\Omega_\nu=\Omega_\sub{R}\,$ and $\,\Omega_b+\Omega_c=\Omega_\sub{M}\,$.
We can now insert the expression back in the time-space equation to obtain a formula for the velocity shared by all the fluids in the early Universe, up to second order:
\begin{align}
  u_{[0]} \;=\; \left\{\;
  2\,\frac{k}{\Hc}\,\left[\,\Psi\,-\,\frac{\QST{_{[0]}}}{2\,k\,\Hc}\,\right]\;
  -\;u_{[0]}\,\left(\,3\,\Omega_\sub{M}\,\delta_\sub{M}\,+\,4\,\Omega_\sub{R}\,\delta_\sub{R}\,\right)
  \;\right\} \frac{1}{3\,\Omega_\sub{M}\,+\,4\,\Omega_\sub{R}} \;,
  \label{eq:adiabatic_velocity_initial_condition}
\end{align}
where we have introduced $u=i\,v$.
In \SONG, however, we evolve the dipoles of the distribution function rather than the velocities. The initial conditions for the former are obtained from $\,u_{[0]}\,$ by using the correspondence in \eref{eq:dipole_initial_condition}.

All the elements appearing in the adiabatic velocity $\,u_{[0]}\,$ are known from the solution of the first-order differential system, except $\Psi$, which is constant. In particular, the first-order adiabatic velocity is given by\footnote{The expression matches with Eq.~98 of \citet{ma:1995a}, that is $\,\theta=(k^2\,\tau)\,\Psi\,$, once we realise that, at first order, $\,\theta=i\,k_j\,v^j=i\,k\,v_{[0]}\,$.}
\begin{align}
  u_{[0]} \;=\; 2\;\frac{k}{\Hc}\;\frac{\Psi}{3\,\Omega_\sub{M}\,+\,4\,\Omega_\sub{R}} \;
  \simeq\; \frac{1}{2}\,k\,\tau\,\Psi \;,
\end{align}
where we have set $\,\Omega_\sub{R}=1\,$, $\,\Omega_\sub{M}=0\,$ and $\,\Hc=\tau^{-1}\,$. The term in the quadratic source is also proportional to $k\tau$, as can be verified by inspecting \eref{eq:einstein_quadratic_sources}, while we know that, for the constant mode, the $\delta$'s are constant. Thus, at early times, both the first and second-order adiabatic velocity are proportional to $k\tau$. An interesting consequence of this dependence is that any term quadratic in the velocity can be safely ignored in the early Universe. As an example, consider the relation between the monopole and the density perturbation (\eref{eq:fluid_limit_w}),
\begin{align}
  &\IN{0}{0} \;=\; \NE{0}{0} \;=\; \delta_\sub{R} \;-\; \frac{4}{3}\;\tU{u}{i}\;\tD{u}{i} \;, \nmsk
  &\BA{0}{0}{0} \;=\; \CD{0}{0}{0} \;=\; \delta_\sub{M} \;-\; \tU{u}{i}\;\tD{u}{i} \;.
\end{align}
Since the adiabatic velocity goes as $k\tau$, we can ignore the terms quadratic in the velocity; what is left is the density perturbation of the two relativistic fluids, which, for adiabatic initial conditions, coincide. Therefore, up to first order in $k\tauini$ and up to second order in the cosmological perturbations, the monopoles correspond to the energy densities: $\,\IN{0}{0}=\NE{0}{0}=\delta_\sub{R}\,$ and $\,\BA{0}{0}{0}=\CD{0}{0}{0}=\delta_\sub{M}\,$. Similarly, in the early Universe, the quadrupole corresponds to the shear.

\subsubsection{Monopoles}
The time-time Einstein equation (\eref{eq:einstein_pure_timetime}) reads
\begin{align}
  \dot\Phi \;=\; -\Hc\,\Psi \;
  -\;\frac{k^2}{3\,\Hc}\,\Phi \;
  -\;\frac{1}{6\,\Hc}\,\kappa\,a^2\,\sum\,\rhoz\;\bmult{\Delta}{0}{0}{0} \;
  -\;\frac{\QTT}{6\,\Hc} \;.
\end{align}
On super-horizon scales, we can ignore $\dot\Phi$, because we focus on the constant mode, and the term in $\Phi$, because it is suppressed by a factor $(k\tau)^2$ with respect to $-\Hc\,\Psi$. For the same reasons, the only term in the quadratic source (see \eref{eq:einstein_quadratic_sources}) which is non negligible with respect to $\,-\Hc\,\Psi\,$ is $\,2\,\Hc\,\Psi\,\Psi\,$. Thus,
\begin{align}
  \Hc\,\Psi \;=\; 2\,\Hc\,\Psi\,\Psi\;
  -\;\frac{1}{6\,\Hc}\,\kappa\,a^2\,\sum\,\rhoz\;\bmult{\Delta}{0}{0}{0} \;.
\end{align}
If we neglect the baryon and cold dark matter contributions, the matter term can be recast as
\begin{align}
  \kappa\,a^2\,\sum\,\rhoz\;\bmult{\Delta}{0}{0}{0} \;\simeq\;
  3\,\Hc^2\,(\,\Omega_\gamma\,\IN{0}{0}\,+\,\Omega_\nu\,\NE{0}{0}\,) \;,
\end{align}
where we have used the Friedmann equation, $\,\kappa\,a^2\,\rhoz_\sub{tot}=3\,\Hc^2\,$.
Because of adiabaticity, the two monopoles coincide, and we can write
\begin{align}
  \kappa\,a^2\,\sum\,\rhoz\;\bmult{\Delta}{0}{0}{0} \;=\;
  3\,\Hc^2\,\IN{0}{0}\,(\,\Omega_\gamma\,+\,\Omega_\nu\,) \;\simeq\; 3\,\Hc^2\,\IN{0}{0}\;,
\end{align}
which, inserted in the time-time Einstein equation, leads to
\begin{align}
  \IN{0}{0} \;=\; \NE{0}{0} \;=\; -2\,\Psi \;+\; 4\,\Psi\,\Psi \;.
  \label{eq:monopole_radiation_initial_condition}
\end{align}
This expression is valid up to second order and is used in \SONG to set the initial conditions for the monopoles of the relativistic species. For the non-relativistic species, we use the adiabaticity condition in \eref{eq:adiabaticity_density} which, up to first order in $k\tau$, reads
\begin{align}
  \BA{0}{0}{0} \;=\; \CD{0}{0}{0} \;=\; \frac{3}{4}\,
  \left(\,\IN{0}{0}\;-\;\frac{1}{4}\,\IN{0}{0}\,\IN{0}{0}\,\right) \;.
  \label{eq:monopole_massive_initial_condition}
\end{align}

\subsubsection{Photon quadrupole}
To derive the initial conditions for the photon perturbations, we enforce the tight-coupling approximation at zero order (TCA0, hereafter). The TCA0 approximation consists in assuming that the interaction rate between the photons and the baryons is infinite. This is a good approximation of the physics in the pre-recombination epoch, when the extremely high density of photons and free electrons renders the Universe opaque to radiation. At the level of the Boltzmann equation, the TCA0 is equivalent to neglecting all the terms that do not appear multiplied by $\dot\kappa$, which implies that the collision term as a whole must be equated to zero.

At first order, the collision terms for the temperature and $E$ polarisation read
\begin{align}
  &\coll_\lm\,[\,\INs\,] \;=\; \;\dot\kappa\,\left(\,-\IN{\L}{m} \;
  +\;\delta_{\ell0}\,\IN{0}{0} \;+\;\delta_{\ell1}\,4\,u_{e[m]} \;+\;\delta_{\ell2}\,\Pi_m \,\right) \;, \nmsk
  &\coll_\lm\,[\,\EMs\,] \;=\; 
  \dot\kappa\,\left(\,-\EM{\L}{m} \;-\; \delta_{\ell2}\;\sqrt{6}\;\Pi_m \,\right) \;,
\end{align}
where $\,\Pi_m = (\IN{2}{m}-\sqrt{6}\,\EM{2}{m})/10\,$. Using the TCA0 approximation, we set $\,\coll_\lm\,[\,\INs\,]=0\,$. For the dipole, this implies $\,\IN{1}{m} \,=\, 4\,u_{e[m]}\,$, which, using the correspondence between moments and fluid variables in \eref{eq:fluid_limit_w}, simply tells us that the baryon and photon fluids have the same velocity, $\,u_{\gamma[m]}=u_{e[m]}\,$, a statement that is true at all orders for tightly coupled fluids (and consistent with the adiabaticity condition).
If we also set $\,\coll_\lm\,[\,\EMs\,]\,$ to vanish, we obtain for $\L=2$ an algebraic system that admits only the solutions $\,\IN{2}{m}=\EM{2}{m}=0\,$. Similarly, for $\L>2$, the TCA0 relation reduces to the identities $\,\IN{\L}{m}=0\,$ and $\,\EM{\L}{m}=0\,$.

Thus, at first order, the tight-coupling between the photons and the electrons forces all the anisotropies except from the dipole to vanish; this result confirms the physical intuition that in a fluid where the mean free path of the particles is infinitely short, there is no way for the inhomogeneities to turn into anisotropies.

\runinhead{Second-order dipole}
The second order expression for the dipole in the TCA0 approximation is given by $\,\coll_{1m}\,[\,\INs\,]=0\,$, with $\coll$ taken from \eref{eq:boltzmann_quad_collision_intensity}:
\begin{align}
  -\IN{1}{m} \;+\; 4\,u_{e[m]}\;
  +\;(\,\Psi+\delta_e\,)\,(\,-\IN{1}{m}\,+\,4\,u_{e[m]}\,)\;
  +\;4\,u_{e[m]}\,\IN{0}{0}\;=\;0\;,
\end{align}
where we have set the first-order multipoles with $\L\geq2$ to zero and used $C^{-,1}_{0,0}=1$. The third term in the expression vanishes after enforcing the first-order TCA0 relation, $\,\IN{1}{m} = 4\,u_{e[m]}\,$; we are thus left with
\begin{align}
  \IN{1}{m} \;=\; 4\,u_{[m]} \,+\, \IN{1}{m}\,\IN{0}{0} \;=\;
  4\,(\,u_{e[m]} \,+\, u_{\gamma[m]}\,\delta_\gamma) \;.
\end{align}
Again, if use the moments-fluid correspondence in \eref{eq:fluid_limit_w},
\begin{align}
  \IN{1}{m} \;=\; 4\,(u_{\gamma[m]}\,+\,u_{\gamma[m]}\,\delta_\gamma) \;,
\end{align}
we see that the expression enforces $\,u_{\gamma[m]}=u_{e[m]}\,$, that is, the velocities of the baryon and photon fluid during tight coupling coincide also at second order, as expected.

\runinhead{Second-order quadrupole}
The expression for the second-order quadrupole at zero order in the tight coupling approximation is given by $\,\coll_{2m}\,[\,\INs\,]\,=\,0\,$:
\begin{align*}
  -\IN{2}{m}\;+\;\frac{1}{10}\,\IN{2}{m}\;-\;\frac{\sqrt{6}}{10}\,\EM{2}{m}\;
  +\;u_{e[m_2]}\,\IN{1}{m_1}\,C^{-,2}_{m_1m}\;
  +\;u_{e[m_2]}\,(\,7\,u_{e[m_1]}\,-\,\frac{1}{2}\,\IN{1}{m_1}\,)\,C^{-,2}_{m_1m} \;=\; 0 \;.
\end{align*}
By enforcing the first-order relation $u_{e[m]}=\IN{1}{m}/4$, the sum collapses to
\begin{align}
  \IN{2}{m} \;=\; \frac{5}{8}\;C^{-,2}_{m_1m}\;\IN{1}{m_2}\;\IN{1}{m_1}
  \;-\; \frac{\sqrt{6}}{9}\,\EM{2}{m} \;.
\end{align}
If we insert the above expression into the TCA0 equation for the $E$ polarisation, $\,\coll_{2m}\,[\,\EMs\,]\,=\,0\,$, where $\coll[\EMs]$ is taken from \eref{eq:boltzmann_quad_collision_emodes}, we obtain the identity $\,\EM{2}{m}=0\,$. By inspecting the structure of \eref{eq:boltzmann_quad_collision_bmodes}, it is straightforward to verify that this is the case also for the $B$ polarisation, that is, $\,\BM{2}{m}=0\,$. Thus, at second order, the photon quadrupole during tight coupling is given by
\begin{align}
  \IN{2}{m} \;&=\; \frac{5}{8}\;C^{-,2}_{m_1m}\;\IN{1}{m_2}\;\IN{1}{m_1}\nmsk
  \;&=\; -\,10\;\tensorP{m}{v}{v}\;,
  \label{eq:photon_quadrupole_initial_condition_tca0}
\end{align}
where the last equality stems from a geometrical identity involving the tensor product $\,\tensorP{m}{v}{v} \equiv \chimatrix{2}{m}{ij}\,v_\g\,v_\g\,$ and the coupling coefficients $C$ defined in \eref{eq:coupling_coefficients_explicit_CD}. (Let us recall that a sum over $m_2=-1,0,1$ is implicit and that $m_1=m-m_2$.)

We verify below (in \fref{fig:tight_coupling_quadrupole} on page \pageref{fig:tight_coupling_quadrupole}) that \SONG indeed reproduces the quadrupole limit in \eref{eq:photon_quadrupole_initial_condition_tca0}. It should be noted that the presence of a quadrupole is still compatible with the absence of anisotropic stresses. In fact, the last relation of \eref{eq:fluid_limit_w} can be used to show that the shear, $\,\Sigma_{[m]}\,$, does vanish in the tight coupling regime;
the velocity squared terms in \eref{eq:photon_quadrupole_initial_condition_tca0} encode the Lorentz boost needed to bring our observer to the rest frame of the photon fluid.

Finally, we note that during the tight-coupling regime all the photon moments with $\L>2$ vanish at second order, because they are sourced by first-order multipoles with $\L\geq2$.

\subsubsection{Neutrino quadrupole}
The evolution of the neutrino quadrupole in the radiation dominated era can be inferred from the first moments of the Boltzmann equation,
\begin{align}
  & \dot\NEs^{\,1}_{\,0} \;=\; k\,\left(\,\NE{0}{0} \;-\; \frac{2}{5}\,\NE{2}{0}\,\right)
  \;+\; 4\,k\,\Psi \;-\; L_{10}[\QLN] \;, \nmsk
  & \dot\NEs^{\,2}_{\,0} \;=\; k\,\left(\,\frac{2}{3}\,\NE{1}{0} \;-\; \frac{3}{7}\,\NE{3}{0}\,\right)
  \;-\; L_{20}[\QLN] \;,
  \label{eq:evolution_neutrino_moments}
\end{align}
where the quadratic sources $\,L_\lm[\QLN]\,$ are equal to those of the photons in \eref{eq:boltzmann_quad_liouville_intensity} with $\INs$ substituted with $\NEs$.

The dipole equation can be recast into
\begin{align}
  \dot\NEs^{\,1}_{\,0} \;=\; 2\,k\,(\,\Psi \,+\, 2\,\Psi^2\,) \;-\; L_{10}[\QLN] \;,
\end{align}
after neglecting the quadrupole term ($\,\NE{2}{0}/\NE{0}{0}=\O(k\tau)^2\ll1\,$) and using the monopole initial condition in \eref{eq:monopole_radiation_initial_condition}, that is $\,\NE{0}{0}=-2\,\Psi+4\,\Psi^2\,$. 
The quadratic source can be schematically written as
\begin{align}
  L_{10}[\QLN] \;=\; (\,\text{metric}\,)^2 \;\;+\;\; k\;\NE{0}{0}\,\times\;\text{metric}
  \;\;+\;\; k\;\NE{2}{m}\,\times\,\text{metric} \;.
\end{align}
If we drop the terms in the first-order quadrupole ($\,\NE{2}{0}\ll\NE{0}{0}\,$) and use $\,\NE{0}{0}=-2\,\Psi\,$, we see that the quadratic source at early times is constant. Since all the terms in the right hand side of $\,\dot\NEs^{\,1}_{\,0}\,$ are constant, the dipole equation can be solved analytically to yield
\begin{align}
  \NEs^{\,1}_{\,0} \;=\; 2\,k\,\tau\,(\Psi \,+\, 2\,\Psi^2) \,-\, \tau\,L_{10}[\QLN] \;.
  \label{eq:neutrino_dipole_initial_condition_boltzmann}
\end{align}
It can be verified that the above expression for the neutrino dipole is compatible with the one in \eref{eq:dipole_initial_condition}, which was obtained by solving the longitudinal Einstein equation.

If we insert the solution for the neutrino dipole into the quadrupole equation in \eref{eq:evolution_neutrino_moments} and neglect the octupole term (\,$\NE{3}{0}/\NE{1}{0}=\O(k\tau)^2\ll1\,$), we obtain
\begin{align}
  \dot\NEs^{\,2}_{\,0} \;=\; \frac{4}{3}\,k^2\,\tau\,\left(\,\Psi\,+\,2\,\Psi^2\,\right) \;
  -\;\frac{2}{3}\,k\,\tau\,L_{10}[\QLN] \;-\; L_{20}[\QLN] \;.
  \label{eq:evolution_neutrino_quadrupole}
\end{align}
By inspecting \eref{eq:einstein_quadratic_sources}, we see that the second quadratic term can be schematically written as
\begin{align}
  L_{20}[\QLN] \;=\; k\;\NE{1}{0}\,\times\,\text{metric} \;\;+\;\; k\;\NE{3}{m}\,\times\,\text{metric} \;,
\end{align}
meaning that $\,L_{20}[\QLN]\,\propto\,\tau\,$. Therefore, the right hand side of $\,\dot\NEs^{\,2}_{\,0}\,$ contains only terms proportional to $\tau$ that can be integrated to yield a $\tau^2$ proportionality for $\NE{2}{0}\,$,
\begin{align}
  \NE{2}{0} \;=\; \frac{2}{3}\,(k\,\tau)^2\,\Psi \;+\; \mathcal{Q}_{{\cal N}_2} \;,
  \label{eq:neutrino_quadrupole_initial_condition}
\end{align}
where we have grouped all the quadratic sources in 
\begin{align}
  \mathcal{Q}_{{\cal N}_2} \;\equiv\; (k\,\tau)^2\,\left(\;\frac{4}{3}\,\Psi^2\,
  -\,\frac{1}{3}\,\frac{L_{10}[\QLN]}{k}\,-\,\frac{1}{2}\,\frac{L_{20}[\QLN]}{k^2\,\tau}\;\right) \;.
  \label{eq:QN2_definition}
\end{align}

\subsection{Initial non-Gaussianity}
\label{sec:initial_non_gaussianity}


In the previous section, we have enforced the Einstein and Boltzmann equations to express up to second order the initial conditions of the matter perturbations in terms of the metric potentials $\Psi$ and $\Phi$.
The latter, however, cannot be determined without first knowing the amount and type of primordial non-Gaussianity produced in the early Universe.
We need therefore to choose a model of inflation and to relate the non-Gaussianity produced by such model to the gravitational potentials at the time where the initial conditions for the non-linear transfer functions are set.
To do so, we employ the gauge-invariant curvature perturbation $\,\zeta\,$, the same variable used in Maldacena (2003) \cite{maldacena:2003a}, which up to second order is given by \cite{malik:2004a, vernizzi:2005a}\footnote{In order to facilitate the comparison with the literature, we express $\zeta$ in terms of the perturbation \R used in Pitrou et al.~(2010) \cite{pitrou:2010a}. The two variables are unperturbatively related by $e^{2\zeta}=1-2\R$, which translates to $R=-\zeta-\zeta^2$ up to second order.
We also note that \eref{eq:R_curvature_perturbation_definition} is the same as Eq.~3.6b of Ref.~\cite{pitrou:2010a}, with $\Phi\leftrightarrow\Psi$ and a multiplicative factor $\nicefrac{1}{2}$ in the quadratic part, to account for the fact that we use the perturbative expansion $X\approx\pert{X}{1}+\pert{X}{2}$ instead of $X\approx\pert{X}{1}+\frac{1}{2}\pert{X}{2}$.}
\begin{align}
  \zeta = -\;\R \;-\; \R^2 \;,
  \label{eq:Z_curvature_perturbation_definition}
\end{align}
with
\begin{align}
  \label{eq:R_curvature_perturbation_definition}
  \R \;=\; &\Phi \;+\; \frac{2}{3\,\Hc\,(w+1)} \;
  \left[\;\dot\Phi\,+\,\Hc\,\Psi\,-\,4\,\Hc\,\Psi^2\,
  -\,\frac{\dot\Phi^2}{\Hc}\,-\,4\,\left(\Psi-\Phi\right)\,\dot\Phi\;\right]\msk
  &+\;(1+3\,c_s^2\,)\;\left[\;\frac{\delta}{3\,(w+1)}\;\right]^2\;
  +\;\frac{4}{3\,(w+1)}\,\delta\;\Phi \;,\notag
\end{align}
where the density contrast $\delta=(\rho-\bar\rho)/\bar\rho$, the barotropic parameter $w$ and the adiabatic sound of speed $c_s^2\,$ refer to the total fluid.
The expression for $\zeta$ simplifies considerably in the radiation dominated era ($w=c_s^2=\frac{1}{3}$) and on super-horizon scales ($\dot\Psi=\dot\Phi=0$):
\begin{align}
  \label{eq:Z_comoving_curvature_perturbation_radiation_dominated}
  \zeta \;=\; -\;\Phi \;-\; \frac{1}{2}\,\Psi \;+\; \frac{1}{2}\,\Psi^2
  \;-\; \Phi^2 \;.
\end{align}

The advantage of using $\zeta$ is that, for adiabatic perturbations, it is conserved on super-horizon scales regardless of the perturbative order \cite{lyth:2005a, malik:2004a, vernizzi:2005a, lyth:2003a}.
Being conserved, $\zeta$ provides a convenient way to relate the primordial curvature fluctuations created during the inflationary period to the gravitational potentials at the time where we set our initial conditions.
Therefore, once the post-inflationary transfer function of $\zeta$, $\,\pert{T}{2}_\zeta(\kone,\ktwo,\ktre)\,$, is specified, the relation in \eref{eq:Z_comoving_curvature_perturbation_radiation_dominated} can be used together with the Einstein equations to infer the initial values of $\,\pert{T}{2}_\Phi(\kone,\ktwo,\ktre)\,$ and $\,\pert{T}{2}_\Psi(\kone,\ktwo,\ktre)\,$, which are the numerically-evolved quantities in \SONG.

Because the topic of this thesis is the intrinsic bispectrum, which is independent of the initial non-Gaussianity \cite{pitrou:2010a}, in what follows we shall assume Gaussian initial conditions.
Following the discussion in \sref{sec:transfer_functions} and \sref{sec:three_point_function}, this requirement translates into the absence of mode coupling in the $\zeta$ random field and, ultimately, in a vanishing initial transfer function:
\begin{align}
  \pert{T}{2}_\zeta(\kone,\ktwo,\ktre) \;=\; 0 \;.
\end{align}
The above condition is indeed used to compute the intrinsic bispectrum in \SONG and to derive the results presented in the next chapter.
For the rest of this section, however, we shall keep the form of  $\,\zeta\,$ unspecified, so that the initial conditions derived below can be used for an arbitrary model of inflation.

\subsection{Initial conditions for the metric perturbations}
\label{sec:initial_conditions_metric}


As discussed in \sref{sec:initial_non_gaussianity}, we parametrise the initial conditions for the scalar potentials, $\,\Phi\,$ and $\,\Psi\,$, in terms of the curvature perturbation, $\,\zeta\,$.
The initial values of the two potentials are determined by the algebraic system consisting of the equation defining $\zeta\,$, \eref{eq:Z_comoving_curvature_perturbation_radiation_dominated}, and of the anisotropic stress equation for $m=0$, \eref{eq:evolution_equation_psi},
\begin{align}
  &\Phi \;=\; -\;\zeta \;-\; \frac{1}{2}\,\Psi \;+\; \frac{1}{2}\,\Psi^2
  \;-\; \Phi^2 \;,
  \label{eq:phi_as_a_function_of_R}\msk
  &\Psi \;=\; \Phi \;
  -\; \frac{1}{5\,k^2} \; \kappa\,a^2\,\sum\;\rhoz\;\bmult{\Delta}{2}{2}{0} \;+\; \Q{A} \;,
\end{align}
where we have introduced the shorthand
\begin{align}
  \Q{A} \;\equiv\; \frac{3}{2\,k^2} \; \QSS{_{[0]}} \;,
  \label{eq:qa_definition}
\end{align}
and $\,\QSS{_{[0]}}\,$ is given in \eref{eq:einstein_quadratic_sources}. The only species that are relevant in the radiation dominated era are the photons and the neutrinos, so we can ignore the contributions to the quadrupole from the baryon and the cold dark matter fluids. (Note that they do contribute to the quadrupole in later epochs, even if their anisotropic stresses vanish, via a quadratic contribution in their velocity, see \eref{eq:fluid_limit_w}.) Therefore, the anisotropic stress equation can be written as
\begin{align}
  \Psi \;=\; \Phi \;
  -\; \frac{3}{5}\,\left(\,\frac{\Hc}{k}\,\right)^2\,
  \left(\;\Omega_\gamma\,\IN{2}{0}\,+\,\Omega_\nu\,\NE{2}{0}\;\right) \;+\; \Q{A} \;,
  \label{eq:evolution_anisotropic_stress_1}
\end{align}
where the extra $\,\Hc^2\,$ factor comes from enforcing the Friedmann equation. To close the system, we need the initial values of the quadrupoles of the photon and neutrino fluids. In principle, both quantities are of order $(k\tau)^2$, and thus negligible. However, they appear in the above equation multiplied by a factor $(\Hc/k)^2\simeq(k\tau)^{-2}$, meaning that their contribution to the $\Psi$ potential is of order unity, and should therefore be considered.
If we insert the expression for the neutrino quadrupole (\eref{eq:neutrino_quadrupole_initial_condition}) into the anisotropic stress equation, we obtain
\begin{align}
  \Psi \;&=\; \Phi \;-\; \frac{6}{15}\,\Omega_\nu\,\Psi \;-\;
  \frac{3}{5}\,\frac{1}{(k\,\tau)^2}\;\left[\;
  \Omega_\gamma\,\IN{2}{0} \;+\; \Omega_\nu\,\mathcal{Q}_{{\cal N}_2} 
  \;\right] \;+\; \Q{A} \;,
  \label{eq:psi_initial_condition_intermediate}
\end{align}
which, after substituting $\Phi$ using \eref{eq:phi_as_a_function_of_R}, becomes an algebraic equation for $\Psi$ that can be easily solved to yield the initial condition for the Newtonian potential up to second order,
\begin{align}
  \left[\;1\,+\,\frac{4}{15}\,\Omega_\nu\;\right]\;\Psi \;=\;
  \frac{2}{3}\;\left[\;-\;\zeta\;+\;\frac{1}{2}\,\Psi^2\;-\;\Phi^2\;+\;\Q{B}\;\right] \;,
  \label{eq:psi_initial_condition}
\end{align}
where we have grouped the quadratic sources in
\begin{align}
  \Q{B} \;=\; \Q{A} \;-\; \frac{3}{5}\,\frac{1}{(k\,\tau)^2}\;\left[\;\Omega_\gamma\,\IN{2}{0}\;
  +\;\Omega_\nu\,\mathcal{Q}_{\mathcal{N}_2}\;\right] \;.
  \label{eq:psi_initial_condition_quadratic}
\end{align}
Let us reiterate our notation. All the terms in $\Q{B}$ are quadratic: $\,\Q{A}$ is the quadratic part of the anisotropic stress equation, as defined in \eref{eq:qa_definition}; $\,\IN{2}{0}$ is the photon quadrupole, whose form is dictated by the tight coupling condition and grows as $(k\tau)^2$, as shown in \eref{eq:photon_quadrupole_initial_condition_tca0}; $\,\mathcal{Q}_{\mathcal{N}_2}$ is the quadratic part of the neutrino quadrupole, as defined in \eref{eq:QN2_definition}, and also grows as $(k\tau)^2$.  The density parameters are defined as $\Omega_\gamma=\rhoz_\gamma/\rhoz_\sub{tot}$ and $\Omega_\nu=\rhoz_\nu/\rhoz_\sub{tot}$ and are related by $\Omega_\gamma=1-\Omega_\nu\,$ in the radiation dominated era, when $\rhoz_\sub{tot}\simeq\rhoz_\gamma+\rhoz_\nu$.
The initial value of $\Phi$, up to second order, can be found by going back to \eref{eq:psi_initial_condition_intermediate},
\begin{align}
  \Phi \;=\; \left[\;1\,+\,\frac{2}{5}\,\Omega_\nu\;\right]\;\Psi \;-\; \Q{B} \;.
  \label{eq:phi_initial_condition}  
\end{align}
Note that, at first order, our initial conditions read
\begin{align}
  &\Psi \;=\; -\frac{10}{15\,+\,4\,\Omega_\nu}\;\zeta \; &&\text{and}
  &&\Phi \;=\; \left[\;1\,+\,\frac{2}{5}\,\Omega_\nu\;\right]\;\Psi \;,
\end{align}
and are in agreement with those found in the literature. In particular, from the comparison with Eq.~98 of \citet{ma:1995a}, we find that $\zeta=-2\,C$, where $C$ is the variable used in that reference to denote the amplitude of the fastest-growing mode.

To sum up, the numerical initial conditions in \SONG are set using: \eref{eq:monopole_radiation_initial_condition} and \ref{eq:monopole_massive_initial_condition} for the monopoles, \eref{eq:dipole_initial_condition} for the dipoles, \eref{eq:photon_quadrupole_initial_condition_tca0} and \ref{eq:neutrino_quadrupole_initial_condition} for the quadrupoles, \eref{eq:psi_initial_condition} and \ref{eq:phi_initial_condition} for the scalar potentials. All the other perturbations, including the non-scalar ones, are evolved starting from vanishing values.



\section{The line of sight sources} 
\label{sec:line_of_sight}

\SONG efficiently implements the Boltzmann-Einstein system of differential equations in the Newtonian gauge (\sref{sec:diff_system}) with correct initial conditions set deep into the radiation era (\sref{sec:initial_conditions}) and passes all the numerical tests that we could devise  (\sref{sec:evolution_checks_of_robustness}).
Therefore, in principle, we could compute the transfer functions for any perturbation at any time after the initial conditions are set. In particular, we could obtain the value of the photon moments today in order to build the CMB observables at first and second order, such as the angular power spectra and bispectra of the CMB temperature and polarisation.

In practice, however, one has to first face a major numerical issue. The current CMB experiments have angular resolutions of $\,\lmax=\O(1000)\,$, meaning that in order to fully use the data to constrain the theoretical predictions, the latter need to be computed with a similar resolution. Because the Boltzmann equation in multipole space forms a hierarchy which is coupled in $\L\,$, we cannot solve it for a number of $\L$ values and later interpolate the results; to obtain a resolution of $\,\lmax=2000\,$, one needs to evolve at least $\,\lmax=2001\,$ coupled differential equations for each of the considered wavemodes. Furthermore, one has to consider the issue of numerical reflection in the Boltzmann hierarchy, discussed in \sref{sec:the_evolved_equations}. Thus, the number of evolved equations in the photon hierarchy needs to be larger than $\,\Lcut=\lmax+k_\text{max}\,\tauz/2\simeq3500\,$, even using the clever Bessel truncation scheme.
This was indeed the standard procedure adopted by the cosmological community before $1996$ (see, \eg, \cite{crittenden:1993a, ma:1995a}). As an example of the required computational effort, the first-order COSMICS code \cite{bertschinger:1995a} took about 90 hours on the 16 processors of the Cray C90 supercomputer to compute the $C_l$ spectrum up to $\,\lmax=3000\,$.

In 1996, a new method to compute the anisotropies of the cosmic microwave background was proposed by \citet{seljak:1996a} that neatly separates the geometrical evolution of the multipoles from the physical effects that source them.
In this \keyword{line of sight approach}, the multipoles at $\tauz$ are obtained as a convolution integral along the past light cone of the photon, hence the name, that involves a source function, smooth in $k$ and time, and a spherical Bessel function, oscillatory in both.
By applying the line of sight (hereafter, LOS) approach, the current value of the first-order transfer functions up to $\lmax=2000$ can be numerically computed in a matter of seconds, without the sacrifice of precision; it is no surprise that all the recent first-order Boltzmann codes, including \emph{CLASS} and \emph{CAMB}, implement the LOS formalism.

Although it was developed with the purpose of solving the first-order BES, the LOS formalism can be adapted to obtain the transfer functions at any order \cite{nitta:2009a, pitrou:2010a, beneke:2011a, huang:2013a}, as we shall describe in \sref{sec:los_line_of_sight_formalism}, where we also discuss \SONG's implementation. The main result will be the expression for the line of sight integral in multipole space for the intensity, $E$-modes and $B$-modes, reported in \eref{eq:los_integral_intensity}, \ref{eq:los_integral_emodes} and \ref{eq:los_integral_bmodes}, respectively.
In \sref{sec:los_source_function}, we discuss the form of the LOS sources and identify three kinds of contributions: the scattering sources, the metric sources and the propagation sources. The propagation sources include the time-delay, redshift and lensing effects which are numerically challenging to integrate; one of them, however, can be computed via a clever change of variables introduced in \sref{sec:redshift_term_deltatilde}. We conclude the section in \sref{sec:integration_by_parts} with a brief note on integration by parts, a technique that is commonly used at first order but whose interpretation at second order is still not clear.

\subsection{The line of sight formalism} 
\label{sec:los_line_of_sight_formalism}

We first introduce the LOS formalism for the intensity perturbation, which will lead to \eref{eq:los_integral_intensity}, and later extend it to include the $E$ and $B$ polarisation, in \eref{eq:los_integral_emodes} to \ref{eq:los_integral_bmodes}.

\subsubsection{The LOS integral}
The brightness equation for the photon intensity can be written, before multipole decomposition, as
\begin{align}
  \dot\Delta \;+\; \bigl(\;
  i\,\scalarP{k}{n} \;+\; \dot\kappa\;\bigr)\,\Delta \;=\; S \;,
  \label{eq:line_of_sight_starting_point}
\end{align}
where $\Delta$ is the brightness moment of the one-particle distribution function (\eref{eq:brightness_definition}), $\,\n$ is the photon's direction and $\,\dot\kappa=a\,n_e\sigma_\sub{T}\,$ is the Thomson scattering rate.
The source function, $\,S\,$, groups all the other terms of the Boltzmann equation; both the source function and the brightness are functions of $(\tau,\k,\n)$. The left hand side of the above expression can be written as
\begin{align}
  \diff{}{\tau}\;\left[\;e^{\,i\,\scalarp{k}{n}\,\tau\,+\,\kappa(\tau)}\;\Delta\;\right]
  \;e^{\,-i\,\scalarp{k}{n}\,\tau\,-\,\kappa(\tau)}\;\;,
\end{align}
which leads to an integral solution for the Boltzmann equation:
\begin{align}
  \int\limits_{\tauini}^{\tauz}\;\dd\,\left[\;e^{\,i\,\scalarp{k}{n}\,
  +\,\kappa(\tau)}\;\Delta(\tau)\;\right]
  \;=\; \int\limits_{\tauini}^{\tauz}\;\dd\tau\;e^{\,i\,\scalarp{k}{n}\,\tau\,+\,\kappa(\tau)}\;\S(\tau) \;,
  \label{eq:integral_solution}
\end{align}
where we have introduced an arbitrary lower limit for the integral, $\,\tauini\,$.
After expanding the left hand side,
\begin{align}
  e^{\,i\,\scalarp{k}{n}\,\tauz\,+\,\kappa(\tauz)}\;\Delta(\tauz) \;=\;
  e^{\,i\,\scalarp{k}{n}\,\tauini\,+\,\kappa(\tauini)}\;\Delta(\tauini) \;+\;
  \int\limits_{\tauini}^{\tauz}\;\dd\tau\;e^{\,i\,\scalarp{k}{n}\,\tau\,+\,\kappa(\tau)}\;\S(\tau) \;,
\end{align}
we can get an expression for the brightness perturbation today,
\begin{align}
  \Delta(\tauz) \;=\;
  e^{\,i\,\scalarp{k}{n}\,(\tauini-\tauz)\,-\,\kappa(\tauini,\tauz)}\;\Delta(\tauini) \;+\;
  \int\limits_{\tauini}^{\tauz}\;\dd\tau\;
  e^{\,i\,\scalarp{k}{n}\,(\tau-\tauz)\,-\,\kappa(\tau,\tauz)}\;\S(\tau) \;.
\end{align}
Because $\dot\kappa$ is the number of scatterings in the unit of time, the \keyword{optical depth} $\kappa$,
\begin{align}
  \kappa\,(\tauini,\tauz) \;\equiv\;
  \int\limits_{\tauini}^{\tauz}\;\dd\tau\;\dot\kappa \;=\;
  \kappa(\tauz)\;-\;\kappa(\tauini) \;,
\end{align}
is the average number of scatterings experienced by a photon between the initial time $\tauini$ and today. If we set the initial time of integration before the time of recombination, this number becomes extremely large, so that the term in $\Delta(\tauini)$ is completely negligible. We are thus left with the so-called \keyword{line of sight integral}:
\begin{align}
  \Delta(\tauz,\k,\n) \;=\; \int\limits_{\tauini}^{\tauz}\;\dd\tau\;
  e^{\,i\,\scalarp{k}{n}\,(\tau-\tauz)\,}\;e^{-\,\kappa(\tau,\tauz)\,}\;\S(\tau,\k,\n) \;,
  \label{eq:los_integral}
\end{align}
where we have reestablished both the Fourier and directional dependences.
The line of sight integral is an \emph{exact} representation of the photon distribution function, in the sense that no approximations where made in its derivation from the Boltzmann equation; furthermore, it has the desirable property of separating the geometrical and dynamical contributions to the anisotropies \cite{seljak:1996a}.
Note also that the precise value of the initial time of integration, $\,\tauini\,$, is not important as long as it is set before the beginning of recombination; in fact, any earlier contribution is suppressed by the $\,e^{-\,\kappa(\tau,\tauz)\,}\,$ term.

The LOS representation makes evident an important property of the CMB anisotropies. The factor $\,e^{-\,\kappa(\tau,\tauz)\,}\,$ acts as a step function that penalises the contributions to $\,\Delta(\tauz)\,$ from before the time of recombination, when $\kappa(\tau,\tauz)$ was huge. Thus, only the last scattering undergone by a photon is important.
In the limit of instantaneous recombination, the LOS integral reduces to 
\begin{align}
  \Delta(\tauz,\k,\n) \;=\; \int\limits_{\taurec}^{\tauz}\;\dd\tau\;
  e^{\,i\,\scalarp{k}{n}\,(\tau-\tauz)\,}\;\S(\tau,\k,\n) \;,
\end{align}
and, if we make the assumption that after recombination the photons stream freely, measuring the CMB gives us information on the source function $\S$ at the time of recombination. This is the reason why the CMB is often referred to as an instantaneous picture of the Universe at the redshit $z_\text{rec}\simeq1100$. Note, however, that the photons \emph{do not} stream freely after recombination, as both scattering (\eg reionisation, Sunyaev-Zeldovich effect) and gravitational effects (\eg time delay and gravitational lensing, see \sref{sec:liouville_term}) slightly alter the anisotropy and spectral patterns of the CMB.

\subsubsection{Multipole decomposition}

To solve the LOS integral numerically, we first need to find its multipole representation. For intensity ($\,\Delta\rightarrow\INs$ and $\S\rightarrow \S^{\,\INs}\,$), we have that
\begin{align}
  \IN{\L}{m}(\tauz,\k) \;=\; \int\limits_{\tauini}^{\tauz}\;\dd\tau\;
  e^{-\kappa}\;
  L_\lm\,\left[\;e^{\,i\,\scalarp{k}{n}\,(\tau-\tauz)\,}\;S^{\,\INs}(\tau,\k,\n)\;\right] \;,
\end{align}
where the spherical projection operator, $\,L_\lm\,$ is defined in \eref{eq:L_operator}.
The spherical harmonic decomposition of a plane wave is given by the Rayleigh formula \cite{mehrem:2011a},
\begin{align}
  \label{eq:rayleigh_expansion}
  e^{\,i\,\scalarp{k}{r}} \;&=\; \sum\limits_{{\lone}=0}^\infty
  \;i^{\lone}\,(2\,{\lone}+1)\;j_{\lone}(k\,r)\;P_{\lone}(\hat{\k}\cdot\hat{\vecr}) \msk
  &=\; \sum\limits_{{\lone}=0}^\infty\;\sum\limits_{m_1=-{\lone}}^{{\lone}}
  \;i^{\lone}\,(4\,\pi)\;j_{\lone}(k\,r)\;Y_\lmone(\hat{\k})\,Y^*_\lmone(\hat{\vecr}) \;, \notag
\end{align}
where $\vecr = (\tau-\tauz)\,\n\,$, $\,j_{\lone}\,$ is the spherical Bessel function of order ${\lone}$, and in the second line we have used the addition theorem (\eref{eq:addition_theorem}) to express the Legendre polynomials $P_{\L}\,$ in terms of two $\,Y_\lm$'s.
If we choose the polar axis of the spherical coordinate system to be aligned with $\k$, we have that
\begin{align}
  Y_{\lm}(\hat{k}) \;=\; Y_\lm(\theta=0,\phi=0) \;=\;
  \delta_{m0}\;\sqrt{\frac{2\,{\lone}+1}{4\,\pi}} \;,
\end{align}
and the plane wave expansion reduces to
\begin{align}
  e^{\,i\,\scalarp{k}{r}} \;=\; \sum\limits_{{\lone}=0}^\infty
  \;i^{\lone}\,\sqrt{4\,\pi\,(2\,{\lone}+1)}\;j_{\lone}(k\,r)\;\,Y_{\lone0}(\hat{\vecr}) \;.
  \label{eq:plane_wave_expansion}
\end{align}
This is, again, a manifestation of the decomposition theorem: when $\k$ is aligned with the zenith, the coupling between the azimuthal modes vanish.
The source function also depends on the direction of propagation, $\,\n\,$, so we expand it in spherical harmonics,
\begin{align}
  S^{\,\INs}(\tau,\k,\n) \;=\; \sum\limits_{L M}\;(-i)^L\;\sqrt{\frac{4\,\pi}{2\,L+1}}\;
  S^{\,\INs}_{L M}(\tau,\k)\;Y_{L M}(\n) \;.
\end{align}
Thus,
\begin{align}
  &L_\lm\,\left[\;e^{\,i\,\scalarp{k}{n}\,(\tau-\tauz)\,}\;S^{\,\INs}(\tau,\k,\n)\;\right] \;=\;
  i^\L\;\sqrt{\frac{2\,\L+1}{4\,\pi}}\;\sum\limits_{\lone L M}\;
  i^{\lone-L}\;4\,\pi\;\sqrt{\frac{2\,\L+1}{2\,L+1}} \nmsk
  &\qquad\qquad\qquad\qquad j_\lone\,(k\,(\tau-\tauz))\;S^{\,\INs}_{L M}(\tau,\k)\;
  \int\dd\Omega(\n)\;Y^*_\lm(\n)\;Y_{\lone0}(\n)\;Y_{L M}(\n) \;.
\end{align}
The final step consists in substituting the expression for the Gaunt integral in the last line,
\begin{align}
  &\int\dd\Omega(\n)\;\,Y^*_\lm(\n)\;Y_{\lone0}(\n)\;Y_{L M}(\n) \;=\; \nmsk &\qquad\
  (-1)^m\;\sqrt{\frac{(2\,\L+1)(2\,\lone+1)(2\,L+1)}{4\,\pi}}\;
  \threej{\L}{\lone}{L}{0}{0}{0}\;
  \threej{\L}{\lone}{L}{-m}{0}{m} \;,
\end{align}
set $M=m\,$ enforcing the Wigner 3$j$ symmetry, and to use the relation
\begin{align}
  \,j_\lone\,(k\,(\tau-\tauz))=(-1)^\lone\,j_\lone\,(k\,(\tauz-\tau))\, \;.
\end{align}
Then, we can express the photon multipoles as a convolution between a geometrical projection function and the source function:
\begin{align}
  \IN{\L}{m}(\tauz,\k) \;=\; \int\limits_{\tauini}^{\tauz}\;\dd\tau\;
  e^{-\kappa}\;\,\sum\limits_{L=0}^{\Lmax}\;\;
  J_{L\L m}\,(k\,r)\;\;S^{\,\INs}_{Lm}(\tau,\k) \;,
  \label{eq:los_integral_intensity}
\end{align}
where we have set $r\equiv\tauz-\tau$ and $\Lmax$ is, in principle, infinity. We have introduced the \emph{line of sight projection function}\index{projection function}\index{J$_{L\lm}$}\index{projection function}\index{line of sight projection function} as \footnote{Note that our projection functions are related to those defined in \citet{hu:1997b} by
\begin{equation}
  J_{L\L m}(x) \;=\; i^{\L}\;\sqrt{4\,\pi\,(2\,\L+1)}\;\,j_\L^{(Lm)} \;.
\end{equation}}
\begin{align}
  J_{L\L m}(x) \;\equiv\; (-1)^m\;(2\,\L+1)
  \sum\limits_{\lone=|\L-L|}^{\L+L}\;i^{\,\L-\lone-L}\;(2\,\lone+1)\;
  \threej{\L}{\lone}{L}{0}{0}{0}
  \threej{\L}{\lone}{L}{-m}{0}{m}\;\,
  \,j_\lone\,(x) \;.
  \label{eq:J_projection_function}
\end{align}

The projection function $\,J_{L\L m}(x)\,$ encodes the excitation of higher multipoles through streaming. It oscillates in both conformal time and comoving scale, and it is real valued as the Gaunt structure forces $\L+\lone+L$ to be even and, therefore, $i^{\,\L+\lone+L}$ to be real. For the monopole, $\L=0$ and $m=0$, $J_{L\lm}$ reduces to a simple spherical Bessel function,
\begin{align}
  J_{L00}(x) \;=\; (-1)^L\;j_L(x) \;,
\end{align}
but, in general, for a given $L$, it is a linear combination of $2\,L+1$ spherical Bessel functions with  coefficients of similar magnitudes.

\subsubsection{Numerical advantages of the LOS formalism}

There are several reasons why solving the line of sight integral in \eref{eq:los_integral_intensity} is more advantageous than obtaining $\,\IN{\L}{m}(\tauz,\k)\,$ by directly solving the differential system:
\begin{enumerate}

  \item In \sref{sec:los_source_function} we shall see that, for all the terms in the source function apart from the quadratic propagation sources, the sum in the LOS integral can be truncated at $\Lmax<10\,$. The LOS integral can be therefore computed efficiently for any value of $\L$ and $m$ using only a reduced number of precomputed sources. In particular, one can build an $\L$-grid that goes up to $\L=O(1000)$ without having to sample every single $\L$-value, as it would be the case if solving the coupled differential system.
 We shall see in the next chapter that, as far as the intrinsic bispectrum is concerned, a grid of $\,N_\L\simeq100\,$ points up to $\,\lmax=2000\,$ yields a $1\%$-level convergence.

  \item The features of the projection function are transferred to $\,\IN{\L}{m}(\tauz,\k)\,$, which is therefore a highly oscillating function in $k\,$. In particular, any feature of the source function at the time of recombination will generate oscillations in $\,\IN{\L}{m}(\tauz,\k)\,$ of wavelength $\,1/(\tauz-\taurec)\simeq1/\tauz\simeq\unit[10^{-4}]{Mpc^{-1}}\,$.
  On the other hand, the source function is a slowly varying function of $\k\,$ and therefore only requires the cruder $k$-sampling that we have discussed in \sref{sec:sampling_strategies}.

  \item The projection function is a purely geometrical object that does not depend on any cosmological parameter. In \SONG, it is computed and stored in a table and later interpolated for quick access.
\end{enumerate}


\subsubsection{Polarisation} 
The LOS integral (\eref{eq:los_integral_intensity}) was derived in multipole space assuming that $\Delta$ was an intensity perturbation, $\,\Delta=\INs\,$. As described in \BFTWO, the result can be generalised to the polarised case by simply substituting $\,\Delta\,$ for $\,\Delta_{ab}\,$, where $ab=++,+-,-+,--\,$ are the helicity indices (\sref{sec:polarisation}), and by introducing a spin factor in the 3$j$ symbol,
\begin{align}
  \threej{\L}{\lone}{L}{0}{0}{0} \qquad\longrightarrow\qquad
  \threej{\L}{\lone}{L}{-s}{0}{s},
\end{align}
where $\,s=2\,$ for $\,ab=+-\,$, $\,s=-2\,$ for $\,ab=-+\,$ and $\,s=0\,$ for $\,ab=++\,$ or $--$. The multipoles for the $E$ and $B$ polarisation are obtained by enforcing the transformations
\begin{align}
  &\EM{\L}{m} \;=\; \frac{1}{2}\;\left(\;
  \Delta_{+-,\lm}\;+\;\Delta_{-+,\lm}\;\right) \;, \nmsk
  &\BM{\L}{m} \;=\; \frac{i}{2}\;\left(\;
  \Delta_{+-,\lm}\;-\;\Delta_{-+,\lm}\;\right) \;.
\end{align}
The spin integer $s$ introduces a sign swap that, after inserting \eref{eq:los_integral_intensity} in the above expression, ultimately leads to a mixing between the $E$ and $B$-modes,
\begin{align}
  \label{eq:los_integral_emodes}
  &\EM{\L}{m}(\tauz,\k) \;=\; \int\limits_{\tauini}^{\tauz}\;\dd\tau\; 
  e^{-\kappa}\;\,\sum\limits_{L=2}^{\Lmax}\;\;\biggl[\;
  J^\sub{\EMs\!\!\EMs}_{L\L m}\,(kr)\;\;S^{\,\EMs}_{Lm}(\tau,\k) \;+\;
  J^\sub{\EMs\!\!\BMs}_{L\L m}\,(kr)\;\;S^{\,\BMs}_{Lm}(\tau,\k)
  \;\biggr]\;,\msk
  \label{eq:los_integral_bmodes}
  &\BM{\L}{m}(\tauz,\k) \;=\; \int\limits_{\tauini}^{\tauz}\;\dd\tau\; 
  e^{-\kappa}\;\,\sum\limits_{L=2}^{\Lmax}\;\;\biggl[\;
  J^\sub{\BMs\!\!\BMs}_{L\L m}\,(kr)\;\;S^{\,\BMs}_{Lm}(\tau,\k) \;+\;
  J^\sub{\BMs\!\!\EMs}_{L\L m}\,(kr)\;\;S^{\,\EMs}_{Lm}(\tau,\k)
  \;\biggr]\;,
\end{align}
and to slightly different projection functions (see Eq. B.12 of \BFTWO),
\begin{align}
  \label{eq:J_EE_projection_function}
  J^\sub{\EMs\!\!\EMs}_{L\L m}\,(x) \;=\; J^\sub{\BMs\!\!\BMs}_{L\L m}\,(x) \;&=\;
  (-1)^m\;(2\,\L+1)\sum\limits_{\lone=|\L-L|}^{\L+L}\;\text{even}\,(\L-\lone-L)\;\msk
  &\times \;i^{\,\L-\lone-L}\;(2\,\lone+1)\;
  \threej{\L}{\lone}{L}{-2}{0}{2}
  \threej{\L}{\lone}{L}{m}{0}{-m}\;\,
  \,j_\lone\,(x) \;, \nmsk
  \displaybreak[0]
  \label{eq:J_EB_projection_function}
  J^\sub{\EMs\!\!\BMs}_{L\L m}\,(x) \;=\; -J^\sub{\BMs\!\!\EMs}_{L\L m}\,(x) \;&=\;
  (-1)^m\;(2\,\L+1)\sum\limits_{\lone=|\L-L|}^{\L+L}\;\text{odd}\,(\L-\lone-L)\;\msk
  &\times \;i^{\,\L-\lone-L-1}\;(2\,\lone+1)\;
  \threej{\L}{\lone}{L}{-2}{0}{2}
  \threej{\L}{\lone}{L}{m}{0}{-m}\;
  \,j_\lone\,(x) \;. \notag
\end{align}
(Note the different exponent of the $i$ factor in the $\EMs\BMs$ case.) The functions ``odd'' and ``even'' are equal to one if their argument is, respectively, odd or even, and vanish otherwise.



\subsection{The source function} 
\label{sec:los_source_function}

In the general case of polarised radiation, the line of sight integral can be written using the composite index notation (\sref{sec:compact_form_boltzmann}) as
\begin{align}
	\label{eq:los_integral_compact}
	\Delta_{n}(\tauz,\k) \;=\;
  \int\limits_{\tauini}^{\tauz}\;\dd\tau\;
  e^{-\kappa(\tau)}\; J_{\,nn'}(kr) \;\;\S_{\,n'}(\tau,\k) \;,
\end{align}
where $\,r=\tauz-\tau\,$. A sum over the composite index $n'$ is implicit and it includes both the perturbation indices ($\INs, \BMs, \EMs$) and the $L$ one. This compact expression encloses the three formulae for the intensity, $E$-modes and $B$-modes that we have derived, respectively, in \eref{eq:los_integral_intensity}, \ref{eq:los_integral_emodes} and \ref{eq:los_integral_bmodes}.  The source function $\,\S_{n}\,$ is defined by the multipole decomposition of \eref{eq:line_of_sight_starting_point}, which reads
\begin{align}
	\dot{\Delta}_{n} \;+\; k\;\Sigma_{nn'}\,\Delta_{n'} \;+\; \dot{\kappa}\,\Delta_{n} 
	\;=\; \S_{\,n} \;.
\end{align}
where $\Sigma_{nn'}$ is the free streaming matrix that arises from the decomposition of $\,n^i\partial_i\Delta\,$ (\sref{sec:compact_form_boltzmann}).
By equating the above expression with the compact Boltzmann equation in \eref{eq:compact_boltzmann} and the collision term in \eref{eq:compact_collision_term}, we see that the source function is given by
\begin{align}
	\label{eq:line_of_sight_sources_compact}
	\S_{\,n} \;=\; 
  \dot{\kappa}\;\left(\;\Gamma_{nn'}\,\Delta_{n'} \;+\; 
  \mathcal{Q}^{\,\mathfrak{C}}_{\,n}\;\right) \;
  -\;\M_{n} \;-\; \mathcal{Q}^L_{\,n} \;.
\end{align}
We shall refer to the three addends in the right hand side as the \keyword{collision sources}\index{line of sight - collision}, the \keyword{metric sources}\index{line of sight - metric} and the \keyword{propagation sources}\index{line of sight - propagation}, respectively.

\subsubsection{Collision sources}

\begin{figure}[t]
	\centering
		\includegraphics[width=0.7\linewidth]{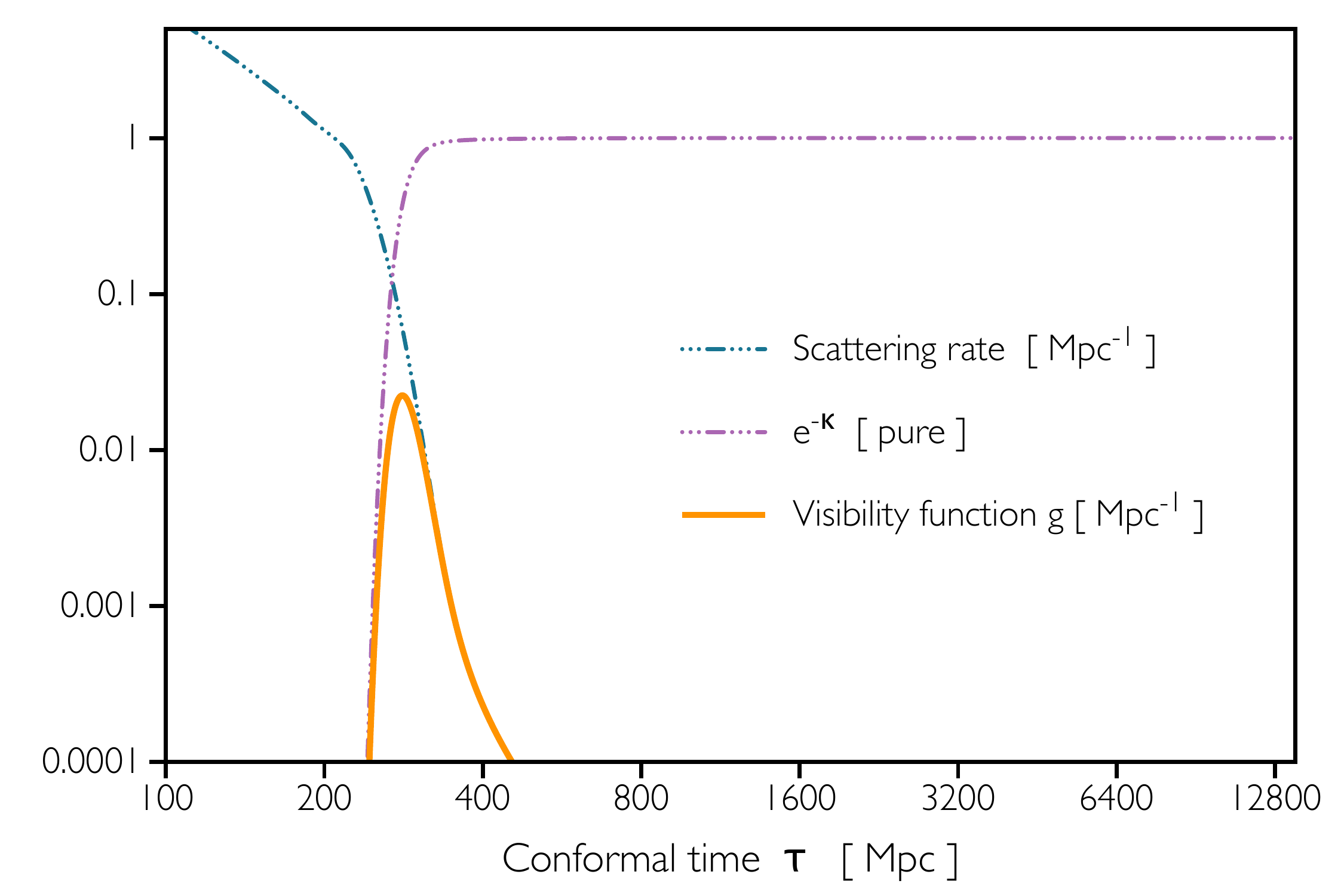}
	\caption[Visibility function]{
  The visibility function, $g$, is obtained as the product between the scattering rate $\dot\kappa$ and $e^{-\kappa}$. The former vanishes after recombination, the latter before. As a result, $g$ is sharply peaked at the time of recombination. In fact, the redshift of recombination is defined in \SONG as the time when the visibility function peaks; for a standard \LCDM model, it corresponds to $\tau_\text{dec}\simeq\unit[280]{Mpc}\,$ or $z_\text{dec}\simeq1100\,$. The redshift of decoupling, instead, is defined using the $T_\text{end}$ parameter in \eref{eq:tauend_decoupling_time}.
	}
	\label{fig:visibility}
\end{figure}

The contribution to the photon anisotropies from the collision sources is
\begin{align}
  \Delta_{n}(\tauz) \;\supset\;
  \int\limits_{\tauini}^{\tauz}\;\dd\tau\;
  \; J_{\,nn'}(kr) \;\;g(\tau)
  \left(\;\Gamma_{n'n''}\,\Delta_{n''} \;+\; 
  \mathcal{Q}^{\,\mathfrak{C}}_{\,n'}\;\right) \;,
\end{align}
where we have introduced the \keyword{visibility function} as
\begin{align}
  g(\tau) \;\equiv\; \dot\kappa\;e^{-\kappa} \;.
\end{align}
The visibility function is the probability that a photon scatters off an electron for the last time around the time $\tau$, and is therefore strongly peaked at the time of recombination; this feature of the visibility function can be appreciated in \fref{fig:visibility}. 

The presence of $\Delta_n$ in the linear structure of the collision sources makes it impossible to compute the line of sight integral without a prior knowledge of the solution of the Boltzmann equation. This apparent paradox holds regardless of the perturbative order, and can be solved after inspecting the form of $\,\Gamma_{nn'}\,\Delta_{n'}\,$, as reported in \eref{eq:compact_collision_term_gammamatrix}:
\begin{align}
 &\Gamma_{nn'}\,\Delta_{n'} \qquad\xrightarrow{\quad\INs\quad}\qquad \delta_{L0}\,\IN{0}{0} \;
   +\;\delta_{L1}\,4\,u_{[m]} \;
   +\;\delta_{L2}\,\left(\,\IN{2}{m}\,-\,\sqrt{6}\,\EM{2}{m}\,\right)/10 \nmsk
 &\Gamma_{nn'}\,\Delta_{n'} \qquad\xrightarrow{\quad\EMs\quad}\qquad -\delta_{L2}\;\sqrt{6}\;
 \left(\,\IN{2}{m}\,-\,\sqrt{6}\,\EM{2}{m}\,\right)/10 \;, \nmsk
 &\Gamma_{nn'}\,\Delta_{n'} \qquad\xrightarrow{\quad\BMs\quad}\qquad 0 \;.
 \label{eq:gamma_coupling}
\end{align}
Because of the geometry of Thomson scattering, the only multipoles that contribute to $\,\Gamma_{nn'}\,\Delta_{n'}\,$ are the monopole, the dipole and the quadrupole.
Armed with this knowledge, we can simply truncate the sum over the purely second-order scattering sources in the LOS integral to $\Lmax=2$, without loss of precision. For the $E$-modes, this amounts to considering only the quadrupole, while the $B$-modes do not have any purely second-order source.

The value of the multipoles up to $\,\Lmax=2\,$ is computed by directly solving the BES system at second order, as described in \sref{sec:diff_system}. The evolution of the photon hierarchies can be stopped at the time of decoupling, just after recombination (\sref{sec:decoupling}), as any other contribution to the LOS integral would be suppressed by the visibility function, which is strongly peaked there. This is indeed what we do in \SONG, where the final time of integration for the scattering sources is determined as the time where the visibility function drops below a certain value $T_\text{end}$ relative to its height at the peak,
\begin{align}
  \frac{g(\tauend)}{g(\taurec)} \;=\; T_\text{end} \;.
  \label{eq:tauend_decoupling_time}
\end{align}
It is important to tune the parameter $T_\text{end}$ as much as possible because \Lcut, the number of equations to follow in the photon hierarchy, has to increase proportionally to $\,\tauend\,$ due to numerical reflection (\sref{sec:the_evolved_equations}).
After running convergence tests, we find that a $1\%$-level convergence in the bispectrum is obtained for $\,\Lcut\gtrsim 8\,$ and $\,T_\text{end} \gtrsim 100\,$; the latter choice corresponds to evolve the differential system up to $\tauend\gtrsim\unit[500]{Mpc}$ for a \LCDM model where the peak of recombination is at $\taurec\simeq\unit[280]{Mpc}$, as can be inferred from \fref{fig:visibility}.

The quadratic collision sources $\,\mathcal{Q}^{\,\mathfrak{C}}_{\,n}\,$ can be built from the solution of the BES at first order. They are multiplied by the visibility function, so that they contribute to the observed anisotropies only at the time of recombination. Contrary to the purely second-order collision sources, the quadratic ones exist also for $L>2\,$; for example, the intensity sources $\,\mathcal{Q}^{\,\mathfrak{C}}_{\,\INs}\,$ (\eref{eq:boltzmann_quad_collision_intensity}) includes the following terms that are present at any angular scale:
\begin{align*}
  \left(\;\Psi\,+\,\delta_b\,+\,\frac{x_e^{(1)}}{\bar{x}_e}\;\right)\;
  \coll_{Lm}\,[\,\INs\,] \;+\;\dot\kappa\;u_e^{[m_2]}\;
  \sum\limits_\pm \mp\,\;\IN{L\pm1}{m_1}\,C^{\pm,L}_{m_1\,m} \;.
\end{align*}
The $L>2\,$ contributions, however, are subdominant with respect to those with $L\leq2$, as they always involve first-order multipoles above the dipole, which are tight-coupling suppressed during recombination.
In \SONG, we set the maximum number of multipoles to include in $\,\mathcal{Q}^{\,\mathfrak{C}}_{\,n}\,$ using the $\,\Lmax\,$ parameter, whose convergence will be discussed in \sref{sec:convergence_tests}.

\subsubsection{Metric sources}
The explicit form of the metric sources $\M_n$ can be read off from \eref{eq:boltzmann_pure_intensity} and \eref{eq:boltzmann_quad_liouville_intensity}:
\begin{align}
  &\M_n \;=\; 
    -\;\delta_{L0}\;4\,\left[\;\Phid\,+\,2\,\dot\Phi\,\Phi\;\right] \msk
    &\qquad-\;\delta_{L1}\;4\,\left[\;\delta_{m0}\,k\,\Psi\,
    +\,k_1^{[m]}\,\Psi\,\left(\Phi\,-\,\Psi\right)
    \,-\,i\,\dot{\omega}_{[m]}\;\right] \;
    -\;\delta_{L2}\;4\,\dot\gamma_{[m]}\;. \notag
\end{align}
These sources exist only for the photon intensity, as polarisation is not sourced by the metric. They are qualitatively different from those in the collision term as they do not involve moments higher than the quadrupole, as expected for the metric modes, and they are active throughout cosmic evolution all the way to today. 

We have already seen that solving the relativistic hierarchies after recombination is computationally inefficient; however, it is not needed do so in order to compute the metric sources up to today. In fact, after the epoch of matter-radiation equality, the relativistic species (photons and neutrinos) become subdominant in the total energy density with respect to the cold ones (baryons and cold dark matter). Their effect on the curvature of the Universe is therefore negligible, and the Einstein equations can be safely evolved without considering the
four relativistic Bolzmann hierarchies in their entirety. Under this assumption, the BES reduces to just $10$ equations (four for the metric variables and six for the cold species) that are well behaved numerically; in particular, the problem of numerical reflection in the relativistic hierarchies is removed. We can therefore obtain the value of the second-order metric sources by evolving this reduced system of ODEs all the way to today. As for the quadratic part of the metric sources, we build them from the first-order solutions of the system evolved in \emph{CLASS}.

From the numerical point of view, we activate this no-radiation approximation (NRA) only after the time $\,\tau_\sub{NRA}\,$ when the ratio between the energy density of matter and that of radiation has exceeded the numerical parameter $T_\sub{NRA}\,$:
\begin{align}
  \frac{\rhoz_\sub{M}(\tau_\sub{NRA})}{\rhoz_\sub{R}(\tau_\sub{NRA})}
  \;=\; \frac{a(\tau_\sub{NRA})}{a_\sub{eq}} \;=\; T_\sub{NRA} \;,
\end{align}
where $a_\text{eq}$ is the scale factor at equality. We find that the transfer functions of the second-order metric variables are not affected by the NRA as long as $\,T_\sub{NRA}\,>100\,$, which corresponds to a redshift of $z\,(T_\sub{NRA})<32\,$ for a standard \LCDM model.
Note that the smaller is the scale considered, the earlier can the NRA be turned on, as what matters in the Einstein equation is the product between $\rhoz$ and the density perturbation $\delta$, and $\delta$ grows much faster for matter than for radiation on subhorizon scales.

\subsubsection{Propagation sources}
The propagation sources contain all the terms in the Boltzmann equation that are products of a metric potential with a photon perturbation (\sref{sec:compact_form_boltzmann}). These are only present in the Liouville term, and can be read for the intensity, $E$-modes and $B$-modes from \eref{eq:boltzmann_quad_liouville_intensity}, \ref{eq:boltzmann_quad_liouville_emodes} and \ref{eq:boltzmann_quad_liouville_bmodes}, respectively. 
In real space and before multipole decomposition, the propagation sources for the photon intensity can be read from \eref{eq:liouville_brightness},
\begin{align}
	\label{eq:propagation_sources_intensity}
  \mathcal{Q}^L_{\,n} \;&=\; 
  n^i\;\partial_i\:\Delta\,(\Phi\,+\,\Psi) \;\msk
  &-\;4\,\Delta\,
  \left(\dot\Phi\,-\,n^i\,\partial_i\Psi\right) \nmsk
  &-\;\left(\,\UU{\delta}{ij}\,-\,n^i\,n^j\,\right) 
  \,\pfrac{\Delta}{n^i} \, \left(\,\partial_i\Psi+\partial_i\Phi\right) \;,\notag
\end{align}
where the first line, second and third lines are the contributions from the free-streaming ($\pfrac{f}{x^i}^{(1)}\frac{\dd x^i}{\dd\tau}^{(1)}$), redshift ($\pfrac{f}{p}^{(1)}\frac{\dd p}{\dd\tau}^{(1)}$) and lensing ($\pfrac{f}{n^i}^{(1)}\frac{\dd n^i}{\dd\tau}^{(1)}$) terms, respectively.

The contribution $\,\mathcal{Q}^L_{\,n}\,$ is purely quadratic in first-order terms and, in principle, can be computed without the need to solve the differential system at second order. However, it comprises a sum over first-order multipoles which are important over all angular scales and times. To compute the $\Delta$'s in the standard line of sight approach up to today would require evolving thousands of equations in the first-order system with an extremely fine sampling in the wavemode $k$, and later solving the LOS integral with $\Lmax=\O(1000)\,$. This is clearly impractical, and special techniques need to be introduced in order to treat the propagation sources, as we shall do for the redshift contribution in the next section.


\subsection{Treating the redshift contribution} 
\label{sec:redshift_term_deltatilde}


As shown by Huang and Vernizzi \cite{huang:2012a}, the redshift contribution to $\,\mathcal{Q}^L_{\,n}\,$ in \eref{eq:propagation_sources_intensity},
\begin{align}
  -\;4\,\Delta\,\left(\,\dot\Phi\,-\,n^i\,\partial_i\Psi\,\right)\;,
\end{align}
can be absorbed by using the new variable
\begin{align}
 	\DeltaT \,\equiv\, \ln \,(1 + \Delta) \,,
	\label{eq:delta_tilde_transformation}
\end{align}
which is expanded up to second order as
\begin{align}
  \DeltaT \;=\; \Delta\,-\,\frac{1}{2}\,\Delta\,\Delta \;.
	\label{eq:delta_tilde_transformation_up_to_secondorder}
\end{align}
The time derivative of $\,\DeltaT\,$ up to second order is then given by
\begin{align}
	\dot{\DeltaT} \;=\; \dot{\Delta}\; - \;\Delta\dot{\Delta} \;=\;
  &-\,n_i\,\partial^i\,\DeltaT \;-\; \M 
  \;+\;\coll\,(1-\Delta) \nmsk
  &\;-\; \mathcal{Q}^L \;-\; 4 \, \Delta\,(\dot{\Phi} \,-\, n^i\partial_i \Psi)  \;,
  \label{eq:delta_tilde_absorption}
\end{align}
where we have used the first-order Boltzmann equation
\begin{align}
	\dot{\Delta} \;=\; -\,n_i\,\partial^i\,\Delta \;+\; 4\,(\dot{\Phi}\;
  - \;n^i\partial_i \Psi) \;+\; \coll \;,
\end{align}
to replace the quadratic term $\,\Delta\dot{\Delta}\,$ and the second-order one \eref{eq:compact_boltzmann} to replace $\,\dot{\Delta}\,$.
The new contribution $\,-4\,\Delta\,(\dot{\Phi}-n^i\partial_i \Psi)\,$ exactly cancels the redshift term in $\,\mathcal{Q}^L\,$, so that the second line of \eref{eq:delta_tilde_absorption} reduces to only the time-delay and lensing contributions. In addition, the collision term $\,\coll\,$ is replaced by $\,\coll\,(1-\Delta)\,$.

As can be seen, the transformation is effective because the second order source we are eliminating is the first order $\Delta$ times part of the first order source.  The price is to make the scattering term more complex by introducing an extra quadratic source of the form $-\coll\,\Delta\,$, which is tractable with the standard line of sight approach. Thus, in \SONG we evolve the BES for the usual intensity brightness $\Delta\,$, but build the line of sight sources for the transformed brightness $\DeltaT$. These are equal to those for $\Delta$ but for the extra $-\coll\,\Delta\,$ term and the lack of the redshift term. We treat photon polarisation in a similar way, using the generalised $\DeltaT$ transformation that we have developed in \citet{fidler:2014a}.

It should be noted that the effect of the $\DeltaT$ transformation is not that of moving the time-integrated redshift term to the last scattering surface. Like $\,\Delta\,$, also $\,\DeltaT\,$ is non-linearly related to the observed temperature anisotropies. This leads to an additional quadratic contribution to the temperature bispectrum arising from the first-order evolution, as we shall show in \sref{sec:temperature_bispectrum}.

Unfortunately, the $\DeltaT$ transformation still leaves other problematic terms in $\,\mathcal{Q}^L\,$, the lensing and time-delay terms (first and third lines of \eref{eq:propagation_sources_intensity}). These do not relate to the first-order sources, and cannot be removed by a similar change of variables.  We will not include them in the line of sight integration in this thesis, and leave them for future work.  Note, however, that we do include all terms in $\,\mathcal{Q}^L\,$ when solving the differential system given in \sref{sec:final_Boltzmann_equation} up to recombination.

\subsection{A note on integration by parts} 
\label{sec:integration_by_parts}

It is often a good technique to use integration by parts in order to separate recombination effects from time-integrated effects. By doing so, \eref{eq:los_integral_compact} becomes
\begin{eqnarray}
	\int\limits_{\tauini}^{\tauz}\;\dd\tau\;e^{-\kappa}\,\S_{\,n'} 
  \;J_{nn'}(kr) \;=\;
	\int\limits_{\tauini}^{\tauz}\;\dd\tau\;e^{-\kappa}\;
  \left(\;\frac{\dot{\S}_{\,n'}}{k} \;-\;
  \frac{\dot{\kappa}\,\S_{\,n'}}{k}\;\right)\;j_{nn'}(kr) \;,
\end{eqnarray}
where we have chosen to integrate $J_{nn'}$ and $j_{nn'}$, its antiderivative, can still be expressed in terms of spherical Bessel functions. This is usually done at first order, where the source is equal to the gradient of the potential $\,\S_{\,n'} = k_{n'}\,\Phi \,$ and gives rise to the usual SW ($\dot\kappa \Phi$) and ISW ($\dot\Phi$) split. This separation is useful because $\,\dot\Phi\,$ is much smaller than $\,k\Phi\,$ as the potential is slowly changing. The second-order metric terms $\M$ can be treated in the same way.

However, the quadratic sources $\mathcal{Q}^L$ are problematic as they contain the first-order photon fluctuations, which oscillate with frequency $k$ so that $\,k^{-1} \dot{\mathcal{Q}^L}_n \sim \mathcal{Q}^L_n\,$. Integration by parts then generates two terms: one with $\dot\kappa$, which is clearly located on the last scattering surface, and a second one which is comparable to the original integral. That second term itself can be decomposed by using integration by parts, and will yield a non-negligible LSS contribution. Therefore, the technique fails to single out a unique LSS contribution.

When we exclude sources such as lensing, we exclude them in their entirety rather than imposing an arbitrary split. In this way, our results can be complemented by the known non-perturbative approaches, see Ref.~\cite{lewis:2012a, hanson:2009a, smith:2011a, serra:2008a, lewis:2011a, lewis:2006a, su:2014a} for lensing.



\begin{figure}[t]
	\centering
		\includegraphics[width=0.93\linewidth]{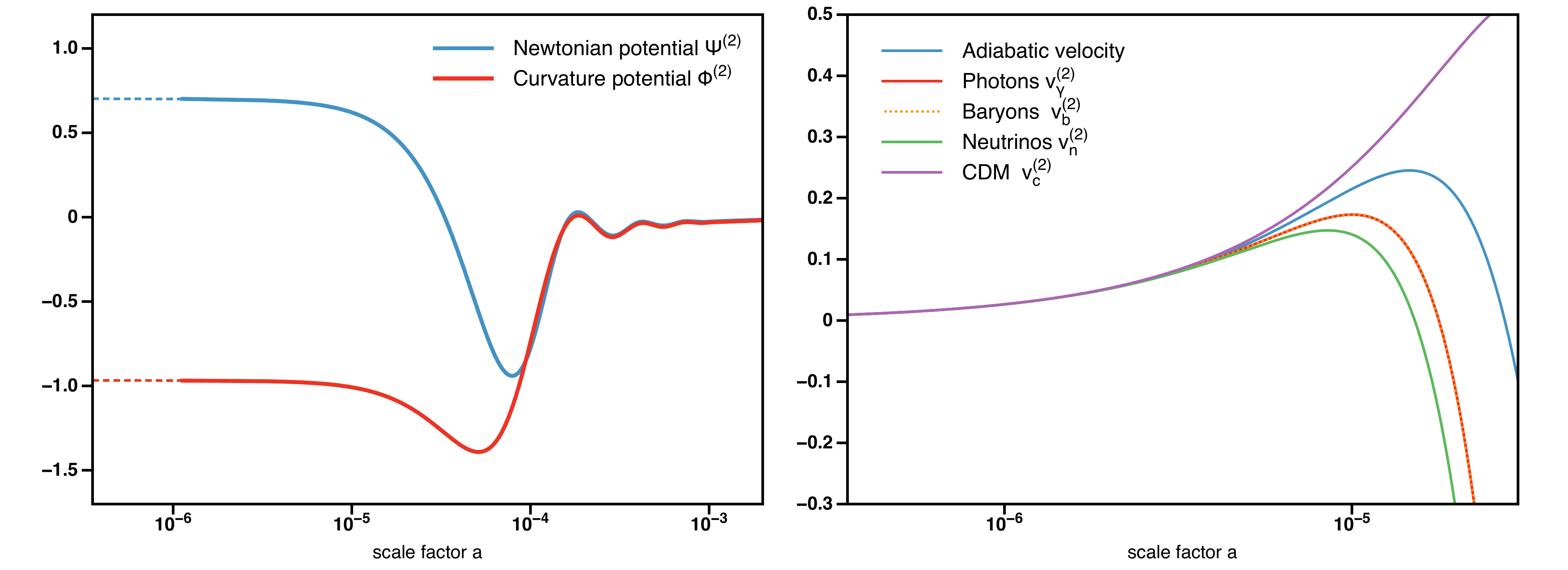}
	\caption[Initial conditions at second order]{
  Initial conditions computed by \SONG in Newtonian gauge. In the left panel, the Newtonian and curvature potentials pick up the constant mode as expected; this does not change when the initial time of evolution is varied (dashed curves). In the right panel, the velocities of the species converge at early times to a common value, as expected for adiabatic initial condtions.  (Wavemodes: $k_1=\unit[0.2]{Mpc^{-1}}\,$, $k_2=\unit[0.1]{Mpc^{-1}}\,$, $k_3=\unit[0.15]{Mpc^{-1}}$.)
	}
	\label{fig:initial_conditions}
\end{figure}

\section{Robustness of \SONG's transfer functions} 
\label{sec:evolution_checks_of_robustness}

We have tested \SONG against a number of analytical limits and consistency checks. In this section we show the most relevant ones.

\subsection{Initial conditions}

\begin{figure}[t]
	\centering
		\includegraphics[width=0.55\linewidth]{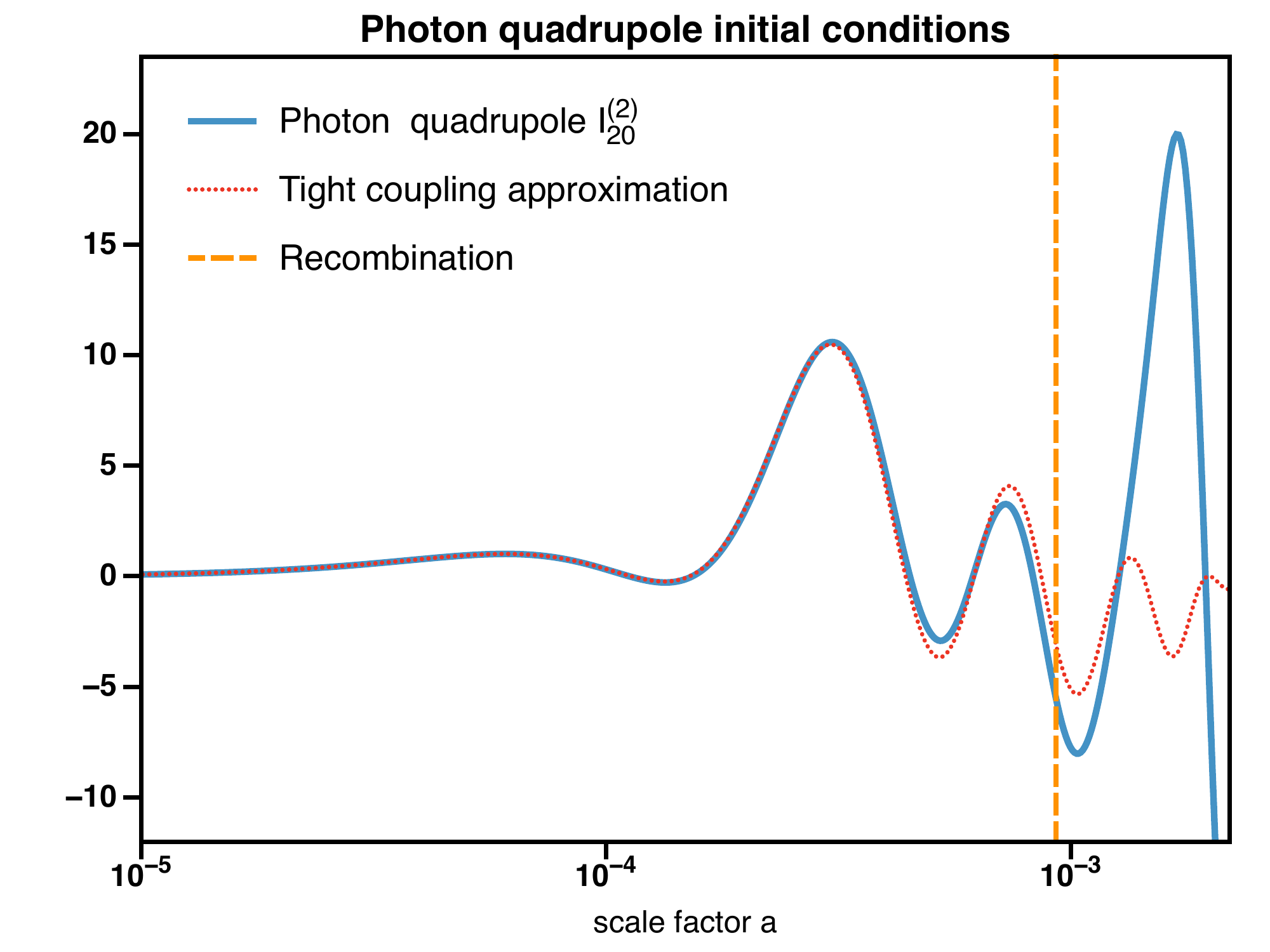}
	\caption[Tight-coupling quadrupole at second order]{
  Transfer function of the photon quadrupole before recombination. The numerical curve matches the tight coupling approximation obtained in \eref{eq:photon_quadrupole_initial_condition_tca0}. (Wavemodes: $k_1=\unit[0.087]{Mpc^{-1}}\,$, $k_2=\unit[0.069]{Mpc^{-1}}\,$, $k_3=\unit[0.081]{Mpc^{-1}}$.)
	}
	\label{fig:tight_coupling_quadrupole}
\end{figure}

We provide \SONG with the initial conditions that we have derived in \sref{sec:initial_conditions}. 
In the left panel of \fref{fig:initial_conditions} we show the transfer functions of the scalar potentials $\Psi$ and $\Phi$ thus obtained. At early times, they are time independent, meaning that \SONG picks the constant mode of the Newtonian gauge immediately.
This is an important test of the consistency of the differential system, as even a small displacement of the initial conditions spoils the flatness of the potentials.
The adiabaticity of the initial conditions is tested in the right panel of the same figure. Also in this case, \SONG's transfer functions respect the analytical expectations, whereby the cosmological fluids all share a common velocity in the early Universe (\eref{eq:adiabatic_velocity_initial_condition}).

\begin{figure}[t]
	\centering
		\includegraphics[width=0.95\linewidth]{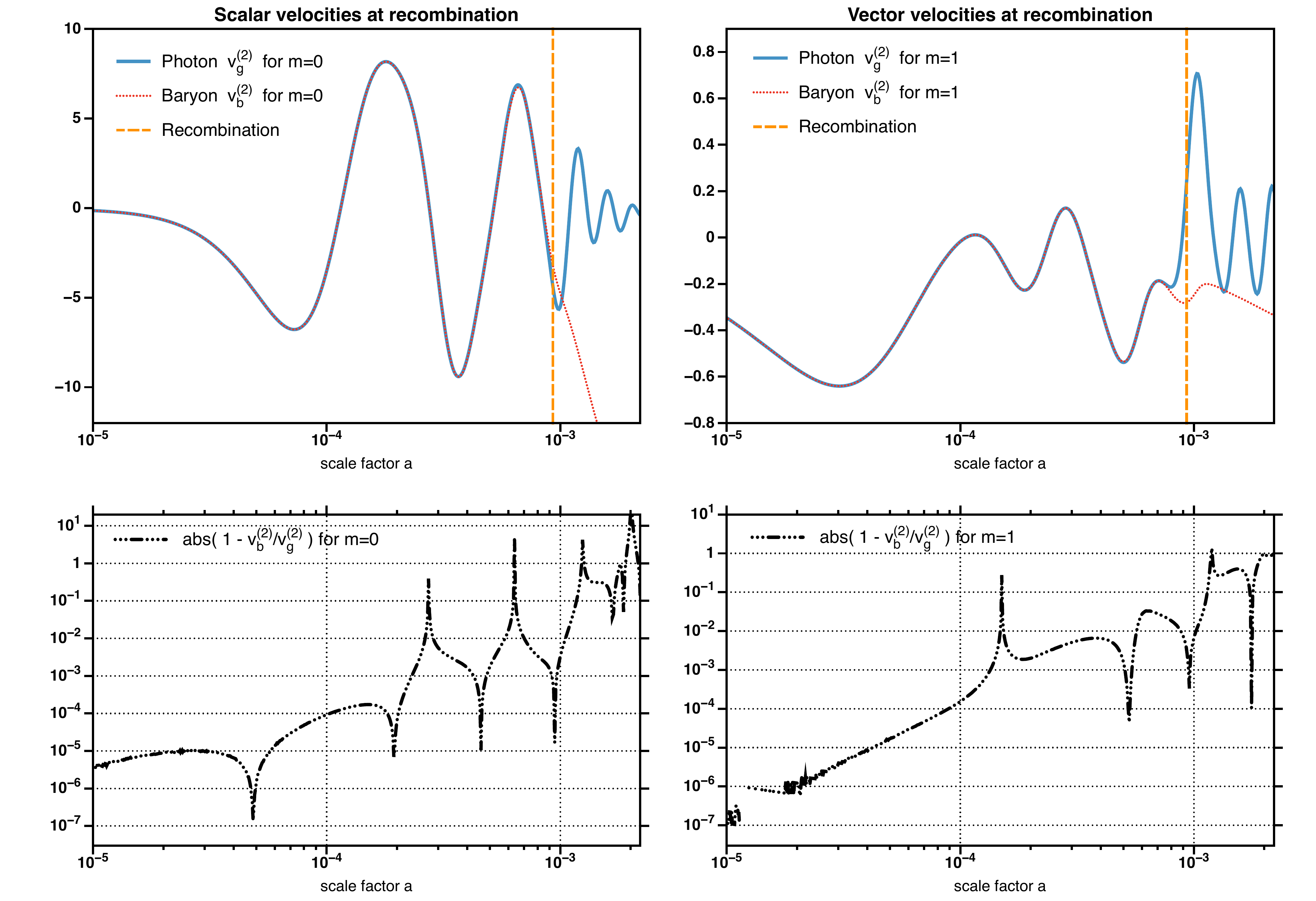}
	\caption[Tight-coupling velocity at second order]{
  Transfer functions of the photon and baryon velocities before recombination, as computed in \SONG for a sub-horizon mode at recombination. Both the scalar (left panels) and vector (right panels) velocities coincide until recombination due to tight coupling. Note that the scalar baryon velocity after recombination grows in amplitude as $\Hc\,a^2$, as predicted by the sub-horizon approximation in Eq.~40 of \citet{bernardeau:2002a}. The spikes in the lower panels correspond to the zero crossing. (Wavemodes: $k_1=\unit[0.2]{Mpc^{-1}}\,$, $k_2=\unit[0.1]{Mpc^{-1}}\,$, $k_3=\unit[0.133]{Mpc^{-1}}$.)
	}
	\label{fig:tight_coupling}
\end{figure}

As we have proven in \sref{sec:initial_conditions_matter}, the velocities of the photon and baryon fluids coincide before the epoch of recombination due to Compton scattering.
We show how \SONG reproduces this limit in \fref{fig:tight_coupling}. The precision of the match is a good test of the implementation of the initial conditions and of the full second-order collision term. In \fref{fig:tight_coupling_quadrupole} we also show the agreement between the numerical quadrupole and the approximate one that we have derived in \sref{sec:initial_conditions_matter}.

\subsection{Constraint equations}
\SONG employs only a subset of the Einstein equations to compute the evolution of the four metric perturbations of Newtonian gauge ($\Psi$, $\Phi$, $\omega_{[1]}$ and $\gamma_{[2]}$). The redundant equations are useful to check the numerical consistency of the differential system and of the initial conditions.

\begin{figure}[p]
	\centering
		\includegraphics[width=1.03\linewidth]{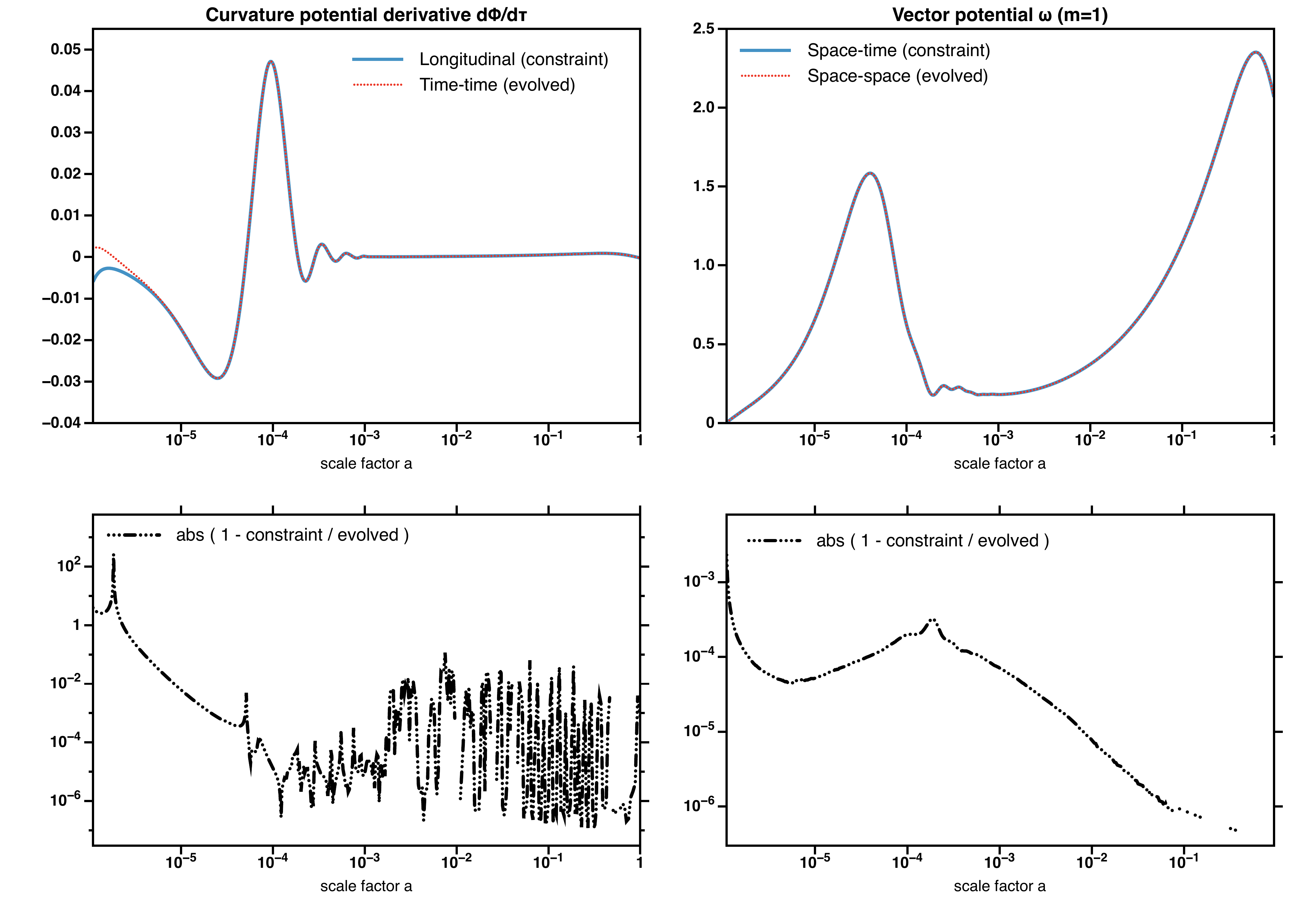}
	\caption[Comparison between the evolution and constraint Einstein equations]{
  Comparison between the evolution and constraint Einstein equations. In the left panels, we show $\dot\Phi$ from the time-time and longitudinal equations. The two curves do not match at early times (hinting some minor issue with the initial conditions); furthermore, the time-time curve develops a small numerical noise after $a=10^{-3}$. In the right panels, we show the evolved $\omega_{[1]}$ against the constraint one. In this case, the match is very good. The tiny difference between the two curves at the initial conditions ($a=10^{-6}$) is due to the fact that we have assumed the starting value of $\omega_{[1]}=0$  to be zero (\sref{sec:initial_conditions_metric}). The consistency between the two curves at late times suggests that such approximation is appropriate. (Wavemodes: $k_1=\unit[0.2]{Mpc^{-1}}\,$, $k_2=\unit[0.1]{Mpc^{-1}}\,$, $k_3=\unit[0.15]{Mpc^{-1}}$.)
	}
	\label{fig:constraints}
\end{figure}

In the left panel of \fref{fig:constraints}, we compare the derivative of the curvature potential $\dot\Phi$ as obtained from the time-time equation (red curve, \eref{eq:einstein_pure_timetime}) and from the longitudinal equation, that is the $m=0$ part of the space-space Einstein equation in \eref{eq:einstein_pure_spacetime} (blue curve). The time-time equation is used to evolve $\Phi$, while the longitudinal one is just a constraint.
We can see that the two curves start slightly displaced but then rapidly converge. After recombination, however, some numerical noise is introduced in the time-time curve $\dot\Phi$ that prevents the match to improve below the $1\%$ level. We have made separate runs of \SONG using either of the equations to evolve $\Phi$ and found no significative difference in the final bispectrum; nonetheless, we plan to discover the origin of this small numerical instability and solve it.

In the right panel of the figure, we compare the vector mode of the metric, $\omega_{[1]}$, as obtained by direct evolution via \eref{eq:evolution_equation_omega} (red curve) and from the $m=1$ constraint equation in \eref{eq:einstein_pure_spacetime} (blue curve). In this case, the match is precise and improves over time.

\subsection{Einstein-de Sitter limit}

The evolution of the density contrast of cold dark matter, $\,\delta_c$, can be analytically computed on sub-horizon scales in the Einstein-de Sitter limit, whereby $\,\Omega_\sub{M}\,=1\,$ \cite{bernardeau:2002a, goroff:1986a, jain:1994a, makino:1992a}:
\begin{align}
  \pert{\delta}{2}(\k) \;=\;
  \K{\,F_2(\kone,\ktwo)\;\pert{\delta}{1}_c(k_1)\;\pert{\delta}{1}_c(k_2)\,} \;,
  \label{eq:deltac_analytical}
\end{align}
where the convolution kernel $F_2$ is given by
\begin{align}
  F_2(\kone,\ktwo) \;=\; \frac{5}{7}\;+\;\frac{1}{2}\;\frac{\scalarP{k_1}{k_2}}{k_1\,k_2}\;
  \left(\,\frac{k_1}{k_2}\,+\,\frac{k_2}{k_1}\,\right)\;+\;\frac{2}{7}\;
  \left(\frac{\scalarP{k_1}{k_2}}{k_1\,k_2}\right)^2 \;.
\end{align}
The Newtonian potential is related to $\delta_c$ by the time-time equation, so that
\begin{align}
  \pert{\Psi}{2}(\k) \;=\; -\frac{2}{3\,\Hc^2\,k^2}\,\;\K{\,k_1^2\,k_2^2\;
  F_2(\kone,\ktwo)\;\pert{\delta}{1}_c(k_1)\;\pert{\delta}{1}_c(k_2)\,}  \;.  
  \label{eq:psi_analytical}
\end{align}
Similarly, the form of the vector and tensor modes of the metric can be analytically computed in the EdS limit, for any scale, to yield \cite{boubekeur:2009a, matarrese:1998a}
\begin{align}
  \label{eq:omega_analytical}
  & i\,\pert{\omega}{2}_{[m]} \;=\; \frac{4}{3\,\Hc\,k^2}\;\;\K{\,(\,k_1^2\;k_{2[m]}\,+\,k_2^2\;k_{1[m]}\,)\;
  \pert{\Psi}{1}(k_1)\,\pert{\Psi}{1}(k_2)\,} \;, \msk
  \label{eq:gamma_analytical}
  & \pert{\gamma}{2}_{[m]} \;=\; -10\,\left(\;\frac{1}{3}\,-\,\frac{j_1(k\tau)}{k\tau}\,\right)\;
  \;\K{\,\frac{\tensorP{m}{k_1}{k_2}}{k^2}\;
  \pert{\Psi}{1}(k_1)\,\pert{\Psi}{1}(k_2)\,} \;  
\end{align}
where $\tensorP{m}{k_1}{k_2}=\chimatrix{2}{m}{ij}\,k_1^i\,k_2^j\,$.

\begin{figure}[p]
	\centering
		\includegraphics[width=1.0\linewidth]{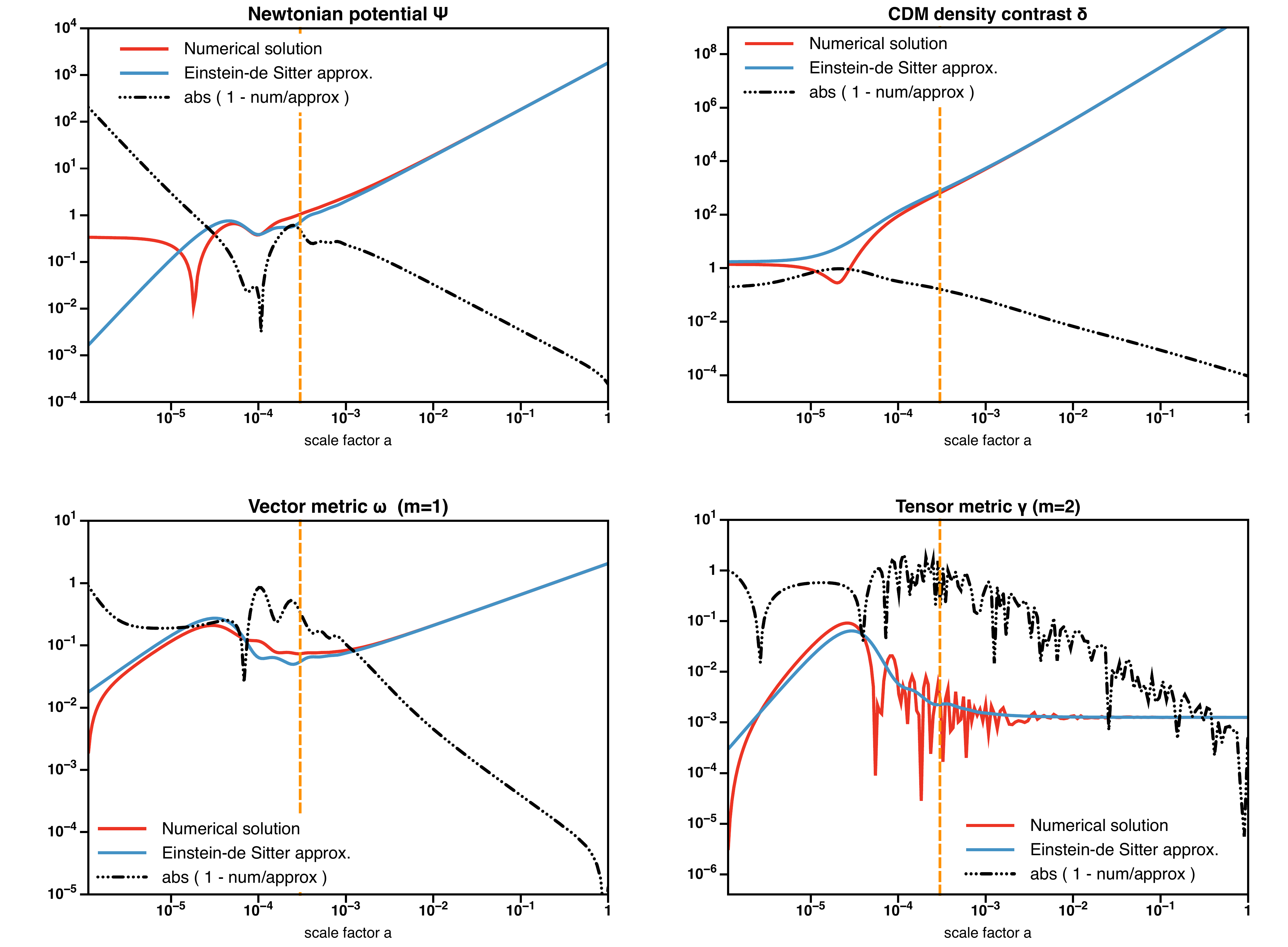}
	\caption[Transfer functions at second order in the EdS limit]{
  Transfer functions in the Einstein-de Sitter limit. In clockwise order, we show the analytical curves (blue) vs \SONG's numerical results (red) for the transfer functions of the Newtonian potential (\eref{eq:psi_analytical}), density contrast (\eref{eq:deltac_analytical}), vector (\eref{eq:omega_analytical}) and tensor (\eref{eq:gamma_analytical}) metric modes. We have considered a universe without dark energy. As expected, a match between the analytical and numerical curves is obtained after the epoch of matter-radiation equality, which is indicated by the vertical dashed line. The numerical noise in the lower-right panel is due to a poor choice of sampling for the $x$-axis, insufficient to follow the frequent oscillations in time of $\gamma_{[2]}\,$; this does not affect \SONG's results as they are obtained using a finer sampling than in the figure. Note that the density contrast of CDM grows as $\,a^2\,$ during matter domination, as predicted by the sub-horizon approximation in Eq.~40 of \citet{bernardeau:2002a}. (Wavemodes: $k_1=\unit[0.26]{Mpc^{-1}}\,$, $k_2=\unit[0.14]{Mpc^{-1}}\,$, $k_3=\unit[0.32]{Mpc^{-1}}$.)
	}
	\label{fig:cdm_approximations}
\end{figure}

In \fref{fig:cdm_approximations} we show that \SONG's numerically-computed transfer functions match the aforementioned analytical results to high precision. The match improves as the ratio between the the matter and radiation densities increases with time, as expected. This is an important test of \SONG's implementation of the Einstein equation and of the description of cold dark matter.

\clearpage

\subsection{Squeezed limit}

\begin{figure}[t]
	\centering
		\includegraphics[width=1.03\linewidth]{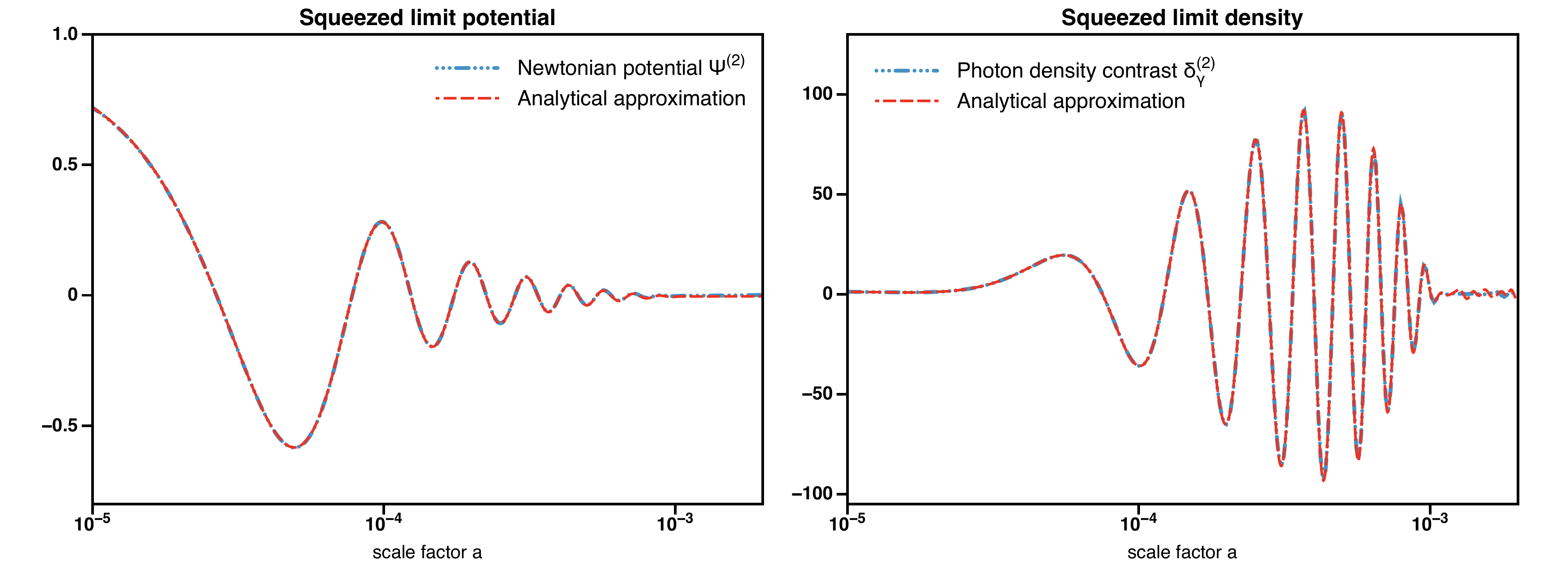}
	\caption[Squeezed limit of the second-order transfer functions]{
  Squeezed limit transfer functions for the Newtonian potential (left panel) and the density contrast of the photon fluid (right panel). In both cases \SONG's result (blue curves) matches the analytic approximation in \eref{eq:squeezed_transfer_analytical} (red curves) to sub-percent accuracy. We adopt a configuration where the long wavemode is $3,000$ times smaller than the short wavemode ($k_1=\unit[0.0001]{Mpc^{-1}}\,$, $k_2=\unit[0.3]{Mpc^{-1}}\,$, $\cos\theta=-0.5$). These plots are similar to those in Fig.~2 of Ref.~\cite{creminelli:2011a}.
	}
	\label{fig:squeezed_limit_transfers}
\end{figure}

When one of the two convolution wavemodes of a second-order perturbation is much smaller than the other, say $\,k_1\ll k_2\,$, its effect can be understood as a time-dependent modulation of the first-order perturbation in $k_2$. Then, in this so-called squeezed limit, the transfer functions for the Newtonian potential and the photon density contrast read \cite{creminelli:2011a, bartolo:2012a}
\begin{align}
  & \pert{\Psi}{2}(k_1,k_2,k) \;=\; f(\tau)\,\pfrac{\,\pert{\Psi}{1}(k_2)}{\,\ln\tau}\;
  -\; \pfrac{\,\pert{\Psi}{1}(k_2)}{\,\ln k_2} \;, \msk
  & \pert{\delta_\gamma}{2}(k_1,k_2,k) \;=\; -4\,f(\tau)\,\tau\,\Hc\,\delta_\gamma(k_2) \;
  +\;f(\tau)\,\pfrac{\,\pert{\delta_\gamma}{1}(k_2)}{\,\ln\tau}\;
  -\; \pfrac{\,\pert{\delta_\gamma}{1}(k_2)}{\,\ln k_2} \;, \notag
  \label{eq:squeezed_transfer_analytical}
\end{align}
where the modulating function $f(\tau)$ is defined as
\begin{align}
  f(\tau) \;=\; -\frac{20+15\,\alpha\tau+3\,\alpha^2\tau^2}{15\,(2+\alpha\tau)^2}
  \quad\text{with}\quad \alpha\;=\;\frac{1}{\sqrt{8}} \;.
\end{align}
In \fref{fig:squeezed_limit_transfers} we show that \SONG indeed matches this analytical limit.

\subsection{Green functions}
Finally, we have compared the results obtained with \SONG with those of an updated version of the code used in Ref.~\cite{beneke:2011a}, which is based on Green's functions rather than transfer functions. Green's functions provide an orthogonal method of reducing the stochastic Boltzmann equations to algebraic differential equations, that can be solved efficiently. The Green's function $G_{nm}(k,\tau_1,\tau_2)$ depends on two times and describes the impact of a mode $m$ at time $\tau_2$ on the mode $n$ at time $\tau_1$. The differential equations for the Green's functions are especially simple as they are independent of the quadratic source terms. It is also not necessary to introduce the additional wavevectors $k_1$ and $k_2$. However, the Green's functions do depend on an additional time, $\tau_2$, and have one additional composite index $m$. For runs with average precision, the methods have a comparable speed, but, when refining the numerical parameters, we find that the transfer function approach scales better. Comparing the results between these different approaches, we obtain a sub-percent level agreement.


\chapterbib


\chapter{The intrinsic bispectrum of the CMB}
\label{ch:intrinsic}

\section{Introduction}

The formalism that we have developed in the previous chapters makes it possible to efficiently compute the first and second-order transfer functions of the cosmic microwave background all the way to today.
The transfer functions can be then used to build observables such as the bispectrum of the temperature anisotropies.
As we have discussed in \sref{sec:spectra_and_bispectra}, it is possible to identify two major cosmological contributions to the CMB bispectrum: the \keyword{linearly propagated bispectrum}, sensitive to the the non-Gaussianity of primordial origin, and the \keyword{intrinsic bispectrum}, arising from the subsequent non-linear evolution of the cosmological perturbations.
In this chapter, we describe how the intrinsic and linear bispectra are computed in \SONG and we constrain their observability and the bias that the former induces on a measurement of the latter.

%

The linearly propagated bispectrum is hypothetical and, depending on the considered model of the early Universe, it assumes specific shapes that have been extensively investigated in the literature.
In models such as the curvaton one \cite{linde:1997a, enqvist:2002a,lyth:2002a,moroi:2001a,moroi:2002a}, where non-Gaussianity arises due to the non-linear evolution of the primordial curvature perturbation on super-horizon scales, the bispectrum peaks at squeezed configurations where one of the momenta is much smaller than the other two momenta. This is called the local type non-Gaussianity \cite{komatsu:2001a, gangui:1994a, verde:2000a} as the non-linearity appears locally in real space. On the other hand, the non-linearity of quantum fluctuations on sub-horizon scales during inflation generally produces a bispectrum that peaks for more equilateral configurations \cite{alishahiha:2004a, silverstein:2004a}. Theoretical templates for the bispectra have been developed to optimally measure these two distinct types of non-Gaussianity. In addition, an orthogonal template with minimal overlap was developed to measure the bispectrum that cannot be captured by the local and equilateral templates \cite{senatore:2010a}. These three templates have been applied to CMB anisotropies measured by WMAP, giving constraints $-3<\fnl^{\local}<77$, $-221<\fnl^{\equil}<323$, $-445 < \fnl^{\orth} < -45$ at $95\%$ confidence level \cite{bennett:2012a}.
The Planck satellite \cite{planck-collaboration:2013b} dramatically improved these constraints but still found values of \fnl compatible with a Gaussian Universe,
\begin{align}
  \label{eq:planck_fnl_measurement}
  \fnl^{\local} \;=\; 2.7\,\pm\,5.8 \;, \quad\quad
  \fnl^{\equil} \;=\; -42\,\pm\,75 \;, \quad\quad
  \fnl^{\orth} \;=\; -25\,\pm\,39 \;.
\end{align}
at $68\%$ confidence level.


The intrinsic bispectrum, on the other hand, is always present, as it is generated by the well known gravitational and collisional effects that we have treated in \cref{ch:boltzmann}; computing its shape and amplitude numerically is the major effort of this chapter.
Due to the difficulty of this task, many approximate approaches to the problem can be found in the literature that either neglect some of the physics or focus on a particular bispectrum configuration.
On super-horizon scales at recombination, where only gravitational effects are important, it is well established that $\fnlcon \sim -1/6$ for the local model \cite{boubekeur:2009a, bartolo:2004a, bartolo:2004b}. On small angular scales, one has to consider the interactions taking place between photons and baryons before the time of decoupling. The contribution to \fnlcon arising from the fluctuations in the free-electron density has been shown to be of order unity \cite{senatore:2009a, khatri:2009a}, and likewise for the contribution from the other quadratic sources in the Boltzmann equation \cite{nitta:2009a}. An alternative approach consists of focussing on the squeezed limit, where the local template peaks. The recombination bispectrum in this limit can be obtained by a coordinate rescaling \cite{creminelli:2004a} and yields a contamination to the local signal again of order unity \cite{creminelli:2004a, creminelli:2011a, bartolo:2012a, lewis:2012a}.

\subsection{Summary of the chapter}
In \sref{sec:source_to_bispectrum}, we derive the formula needed to compute the intrinsic bispectrum, which is now fully implemented in \SONG. In the same section we also explain how to compute the linearly propagated bispectrum and give the shape of the local, equilateral and orthogonal templates of primordial non-Gaussianity.

To quantify the observability of the various bispectra and their correlations, in \sref{sec:bispectrum_to_fnl} we shall adopt a Fisher matrix approach.
We will be particularly interested in the observability of the intrinsic bispectrum, quantified by its signal-to-noise ratio, and in the bias that its presence induces in the measurements of the primordial non-Gaussianity.

The main results of this thesis are illustrated in \sref{sec:results}, where we find that the amplitude of the intrinsic bispectrum is beyond the sensitivity of the Planck CMB survey, with a signal-to-noise ratio of $\sim1/3$ and biases smaller than the error bars.

In \sref{sec:bispectrum_numerical_checks} we conclude the chapter with a number of numerical and analytical checks on \SONG's results.
These include extensive convergence tests on the most important numerical parameters in \SONG and a successful comparison which the well-known analytical limit for the squeezed configurations of the bispectrum.


\subsection{Cosmological parameters}
Throughout the chapter we employ a \LCDM model with WMAP9 parameters \cite{hinshaw:2012a}, whereby $h = 0.697$, $\Omega_b = 0.0461$, $\Omega_{\text{cdm}} = 0.236$, $\Omega_\Lambda = 0.718$, $A_s = \sci{2.43}{-9}$, $n_s = 0.965$, $\tau_\text{reio} = 0.08$, $N_\text{eff} = 3.04$. In this model, the age of the Universe is $\unit[13.75]{Gyr}$, the conformal age $c\,\tau_0 = \unit[14297]{Mpc}$ and recombination happens at $z=1088$, corresponding to a conformal time of $c\,\tau_\text{rec} = \unit[284]{Mpc}$.
We recall that we use purely scalar adiabatic initial conditions (\sref{sec:initial_conditions}). For the power spectrum of the primordial perturbations, we assume the following form:
\begin{align}
  P_\Phi(k) \;=\; \frac{2\,\pi^2}{k^3} \; A_s \; \left(\,\frac{k}{k_0}\,\right)^{n_s-1}\;.
\end{align}
where the pivot scale is taken to be $k_0=\unit[0.002]{Mpc^{-1}}\,$, following the WMAP team \cite{hinshaw:2012a}.

\section{From the sources to the bispectrum} 
\label{sec:source_to_bispectrum}

In this section we derive the formulae used in \SONG to compute the bispectrum of the cosmic microwave background.
The starting point is the definition of the angular bispectrum for the brightness perturbation,
\begin{align}
  \avg{\Delta^3} \;\equiv\;
  \avgbig{\Delta_\lmone(\tauz,\vecx_0)\,\Delta_\lmtwo(\tauz,\vecx_0)\,\Delta_\lmtre(\tauz,\vecx_0)} \;,
\end{align}
which we evaluate here ($\vecx_0$) and now ($\tauz$) in order to relate it to the observations.
In Fourier space, the angular bispectrum reads
\begin{align}
  \avg{\Delta^3} \;=\;
  \int\frac{\dd\kone\,\dd\ktwo\,\dd\ktre}{(2\,\pi)^9}\;e^{\,i\,\vecx_0\,(\kone+\ktwo+\ktre)}\;
  \avgbig{\Delta_\lmone(\tauz,\kone)\,\Delta_\lmtwo(\tauz,\ktwo)\,\Delta_\lmtre(\tauz,\ktre)}\;.
\end{align}
In a statistically homogeneous Universe the real-space bispectrum cannot depend on the position. This is reflect by the presence in the Fourier-space bispectrum of the Dirac delta function $\delta(\kone+\ktwo+\ktre)\,$, as shown in \sref{sec:three_point_function}. Therefore, the exponential can be set to unity:
\begin{align}
  \avg{\Delta^3} \;=\;
  \int\frac{\dd\kone\,\dd\ktwo\,\dd\ktre}{(2\,\pi)^9}\;
  \;\avgbig{\Delta_\lmone(\tauz,\kone)\,\Delta_\lmtwo(\tauz,\ktwo)\,\Delta_\lmtre(\tauz,\ktre)}\;.
\end{align}
The brightness perturbation can be expressed in terms of its transfer function using \eref{eq:transfer_function_definition},
\begin{align}
  \label{eq:transfer_function_definition_bispectrum}
  &\Delta_\lm(\tauk) \;=\;\, \pert{\T_\lm}{1}(\tauk)\;\Phi(\tauini,\veck) \msk
  &\qquad+\;\,\int\,\frac{\dd\konep\,\dd\ktwop}{(2\pi)^3}\;\,\delta(\konep+\ktwop-\k)\;\,
  \pert{\T_\lm}{2}(\tau,\konep,\ktwop,\k) \;\, \Phi(\tauini,\konep)\;\Phi(\tauini,\ktwop) \;. \notag
\end{align}
As we have explained in \sref{sec:three_point_function}, this results into three contributions to the bispectrum: the \keyword{linearly propagated bispectrum} (\eref{eq:linear_bispectrum_k1k2k3})\footnote{Note that from now on we shall omit writing the time dependence. This does not create ambiguity as the transfer functions $\T$ are always evaluated today, $\tauz$, and the potentials $\Phi$ at the initial time $\tauini$.}
\begin{flalign}
  \label{eq:linear_angular_bispectrum_k1k2k3}
  &\avg{\Delta^3}_\lin \;=\; \int\frac{\dd\kone\,\dd\ktwo\,\dd\ktre}{(2\,\pi)^6}\;\,
  \delta\,(\kone+\ktwo+\ktre)\;
  \pert{\T_\lmone}{1}(\kone)\,
  \pert{\T_\lmtwo}{1}(\ktwo)\,
  \pert{\T_\lmtre}{1}(\ktre)\;
  B_\Phi(\kone,\,\ktwo,\,\ktre) \;,  &
\end{flalign}
where the primordial bispectrum $B_\Phi$ is defined as
\begin{align}
  \avg{\Phi(\kone)\,\Phi(\ktwo)\,\Phi(\ktre)} \;=\;
  (2\pi)^3\;\delta\,(\kone+\ktwo+\ktre)\;B_\Phi(\kone,\ktwo,\ktre)\;,
\end{align}
which vanishes for Gaussian initial conditions; the \keyword{intrinsic bispectrum} (\eref{eq:intrinsic_bispectrum_k1k2k3})
\begin{flalign}
  \label{eq:intrinsic_angular_bispectrum_k1k2k3}
  &\avg{\Delta^3}_\intr \;=\; 
  \int\,\frac{\dd\kone\,\dd\ktwo\,\dd\ktre}{(2\,\pi)^6}\;\,
  \delta(\kone+\ktwo+\ktre)\msk
  &\qquad\qquad\times\Bigl[\;2\;\pert{\T}{1}_\lmone(\kone)\;
  \pert{\T}{1}_\lmtwo(\ktwo)\;
  \pert{\T}{2}_\lmtre(-\kone,-\ktwo,\,\ktre)\;
  P_\Phi(-\kone) \, P_\Phi(-\ktwo) 
  \;+\; \text{2 perm.}\;\Bigr] \;,\notag &
\end{flalign}
which exists no matter what the initial conditions are; and the trispectrum contribution (\eref{eq:next_to_leading_order_bispectrum_trisp})
\begin{flalign}
  &\avg{\Delta^3}_\trisp \;=\; 
  \int\,\frac{\dd\kone\,\dd\ktwo\,\dd\ktre}{(2\,\pi)^6}\;\,
  \delta(\kone+\ktwo+\ktre)\;\msk
  &\qquad\qquad\times\Bigl[\;\pert{\T}{1}_\lmone(\kone)\;
  \pert{\T}{1}_\lmtwo(\ktwo)\;
  \K{\pert{\T_\lmtre}{2}(\konep,\ktwop,\ktre)\;
  S_\Phi(\kone,\ktwo,\konep,\ktwop)} 
  \;+\; \text{2 perm.}\;\Bigr] \;,\notag
\end{flalign}
which involves the trispectrum of the primordial potential.

The three contributions to the CMB bispectrum -- linear, intrinsic and trispectrum -- add linearly.
Understanding their relative importance is crucial for interpreting the observed bispectrum as it allows us to separate the effect of the primordial non-Gaussianity, encoded in $\,B_\Phi\,$ and $\,S_\Phi\,$, from the post-inflationary evolution of the signal, given by $\,\pert{\T}{2}_\lm\,$; indeed, this was one of our main motivations in developing \SONG.
According to the order-of-magnitude estimate provided in \sref{sec:three_point_function}, the latest observations from the Planck satellite \cite{planck-collaboration:2013b} suggest that the linear bispectrum has an amplitude similar to or smaller than what is expected from the intrinsic one.
The trispectrum contribution, on the other hand, is constrained to be negligible \cite{planck-collaboration:2013b, smidt:2010a}.
We remark that these considerations apply only to the forms of the primordial bispectrum (local, equilateral and orthogonal) and trispectrum ($\taunl$ and $\gnl$ models) that we take into account.
It is possible that a yet-to-be constrained model of inflation generates a larger non-Gaussianity than the intrinsic bispectrum for a specific $(\lone,\ltwo,\ltre)$ limit.
However, the purpose of this thesis is to quantify the amplitude and shape of the intrinsic bispectrum, which is independent from the details of inflation and is a guaranteed contribution to the total CMB bispectrum.

In what follows, we obtain a numerically viable formula for the intrinsic bispectrum (\sref{sec:intrinsic_bispectrum_derivation}) and explain how it is implemented in \SONG (\sref{sec:intrinsic_bispectrum_numerical}).
We also briefly describe the templates that are usually employed to parametrise the primordial non-Gaussianity (\sref{sec:templates}).


\subsection{The intrinsic bispectrum formula} 
\label{sec:intrinsic_bispectrum_derivation}

We shall now derive in four steps the formula that is used by \SONG to compute the intrinsic bispectrum of the cosmic microwave background.
We shall express it as a sum over the azimuthal modes,
\begin{align}
  B^\intr_\ldep \;=\; \sum\limits_m\;B_\ldep^{\{m\}}\;,
\end{align}
where the scalar ($m=0$) contribution resembles the well known expression for the primordial bispectrum \cite{komatsu:2001a, fergusson:2007a}. 
We remark that the bispectrum formula, which is reported in its final form in \eref{eq:intrinsic_angular_bispectrum_final_formula}, was first derived by Christian Fidler and is going to be included in a paper in preparation.

\subsubsection{Enforce statistical isotropy}
In \SONG, we compute the second-order transfer functions assuming that the zenith, that is the polar axis of the spherical coordinate system, is aligned with the $\ktre$ direction; this choice makes it possible to solve the differential systems for the different $m$-modes separately.
The transfer functions thus computed can be inserted in the formula for the intrinsic bispectrum, \eref{eq:intrinsic_angular_bispectrum_k1k2k3}, only after rotating the coordinate system to align the zenith with the $\ktre$ vector; the statistical isotropy of the Universe ensures that the angular bispectrum is invariant under such rotation.
To do so, we contract each of the transfer functions in \eref{eq:intrinsic_angular_bispectrum_k1k2k3} with the Wigner rotation matrices \cite{hu:1997b, komatsu:2002a}
\begin{align}
  \label{eq:wigner_rotation_matrix}
  \mathcal{D}^{\,(\L)}_{m',m}(\phi,\theta,\psi) \;=\;
  \sqrt{\frac{4\,\pi}{2\,\L+1}}\;Y_\lm^{-m'}\,(\theta,\phi)\,e^{\,i\,m'\,\psi} \;,
\end{align}
where $Y_\lm^{s}\,(\theta,\phi)$ is the spin-weighted spherical harmonic of spin $s$ and $(\phi,\theta,\psi)$ are the Euler angles that map the zenith in the unrotated coordinate system to $\ktre$. The last rotation about $z$ is clearly not needed, so that we can set $\psi=0$; the $\theta$ and $\phi$ angles are the polar and azimuthal angles of $\ktre$ in the unrotated coordinate system. Then, the rotation amounts to performing the following substitutions:
\begin{align}
  \label{eq:rotation_bispectrum}
  \pert{\T}{1}_\lmone(\kone) \quad&\longrightarrow\quad
  \sqrt{\frac{4\,\pi}{2\,\lone+1}}\;
  Y_\lmone^{-m_1'}(\vec{\hat{k}_3})\;\,\pert{\T}{1}_{\lone m_1'}(\konep) \;,
  \displaybreak[0]\msk
  \pert{\T}{1}_\lmtwo(\ktwo) \quad&\longrightarrow\quad
  \sqrt{\frac{4\,\pi}{2\,\ltwo+1}}\;
  Y_\lmtwo^{-m_2'}(\vec{\hat{k}_3})\;\,\pert{\T}{1}_{\ltwo m_2'}(\ktwop) \;,
  \displaybreak[0]\nmsk
  \pert{\T}{2}_\lmtre(-\kone,-\ktwo,\,\ktre) \quad&\longrightarrow\quad
  \sqrt{\frac{4\,\pi}{2\,\ltre+1}}\;
  Y_\lmtre^{-m}(\vec{\hat{k}_3})\;\,\pert{\T}{2}_{\ltre m}(-\kone',-\ktwo',\,k_3) \;, \notag
\end{align}
where $\konep$, $\ktwop$ and $\ktrep$ are the rotated axes in Fourier space and sums over the $m_1'$, $m_2'$ and $m$ indices are implicit; we have used a different notation for the $m$ index for reasons that will be clear soon.
It should be noted that, after the rotation, the second order $\T$ depends only on the magnitude of the third wavevector, $\,k_3={k_3}'\,$, and not anymore on its direction.

Applying the rotation to the intrinsic bispectrum, \eref{eq:intrinsic_angular_bispectrum_k1k2k3}, results in
\begin{align}
  \label{eq:intrinsic_angular_bispectrum_intermediate}
  &\avg{\Delta^3}_\intr \;=\; 
  \sqrt{\frac{(4\,\pi)^3}{(2\,\lone+1)(2\,\ltwo+1)(2\,\ltre+1)}}\;
  \int\,\frac{\dd\kone\,\dd\ktwo}{(2\,\pi)^6}\;\,
  \int\,\dd k_3\;k_3^2\;\,\delta(\kone+\ktwo+\ktre)\msk
  &\qquad\qquad\times\vphantom{\int}\;2\;\pert{\T}{1}_{\L_1m_1'}(\kone)\;
  \pert{\T}{1}_{\L_2m_2'}(\ktwo)\;
  \pert{\T}{2}_{\L_3m}(-\kone,-\ktwo,\,k_3)\;
  P_\Phi(-\kone) \, P_\Phi(-\ktwo) \nmsk
  &\qquad\qquad\times\int\dd\Omega(\vec{\hat{k}_3})\;Y_\lmone^{-m_1'}(\vec{\hat{k}_3})\;
  Y_\lmtwo^{-m_2'}(\vec{\hat{k}_3})\;Y_\lmtre^{-m}(\vec{\hat{k}_3})\;
  \;+\; \text{2 perm.}\;\;,\notag
\end{align}
where we have split the $\ktre$ integral in its radial and angular parts and we have dropped the prime indices for the wavemodes\footnote{Note that we have also assumed that the Dirac delta function does not depend on $\vec{\hat{k}_3}\,$; we shall prove this point later in the comment to \eref{eq:dirac_delta_bispectrum_expansion}.}. The latter ($\dd\Omega$) can be immediately solved using the Gaunt relation for the spin weighted spherical harmonics \cite[Appendix~A.1]{shiraishi:2011a} to yield
\begin{align}
  &\avg{\Delta^3}_\intr \;=\;
  4\,\pi\; \threej{\lone}{\ltwo}{\ltre}{m_1}{m_2}{m_3}\;\threej{\lone}{\ltwo}{\ltre}{m_1'}{m_2'}{m}
  \int\,\frac{\dd\kone\,\dd\ktwo}{(2\,\pi)^6}\;\,
  \int\,\dd k_3\;k_3^2\;\,\delta(\kone+\ktwo+\ktre)\nmsk
  &\qquad\times\vphantom{\int}\;2\;\pert{\T}{1}_{\L_1m_1'}(\kone)\;
  \pert{\T}{1}_{\L_2m_2'}(\ktwo)\;
  \pert{\T}{2}_{\L_3m}(-\kone,-\ktwo,\,k_3)\;
  P_\Phi(-\kone) \, P_\Phi(-\ktwo)
  \;+\; \text{2 perm.} \;,
  \label{eq:intrinsic_angular_bispectrum_after_rotation}
\end{align}
where a sum over the $m'$ indices is implicit.
Thus, after enforcing the statistical isotropy of the Universe, the $m$-dependence of the bispectrum assumes the simple form of a 3$j$ symbol. The information content of $\,\avg{\Delta^3}_\intr\,$ can be therefore compressed in the \keyword{angle-averaged bispectrum} $\,B_\ldep\,$ defined as\footnote{The adjective ``angle-averaged'' comes from the fact that, using \eref{eq:3j_symbols_squared}, $\,B_\ldep\,$ can be written as
\begin{equation}
  B_\ldep[\,\Delta\,] \;=\; \sum\limits_{m_1m_2m_3}\;
  \threej{\lone}{\ltwo}{\ltre}{m_1}{m_2}{m_3}\;
  \avgbig{\Delta_\lmone\,\Delta_\lmtwo\,\Delta_\lmtre} \;.
\end{equation}}
\begin{align}
  \avgbig{\Delta_\lmone\,\Delta_\lmtwo\,\Delta_\lmtre} \;=\;
  \threej{\lone}{\ltwo}{\ltre}{m_1}{m_2}{m_3}\;
  B_\ldep[\,\Delta\,] \;.
\end{align}
The angle-averaged bispectrum depends only on the three angular scales $\,\lone\,$, $\,\ltwo\,$ and $\,\ltre\,$ but, as it clear from \eref{eq:intrinsic_angular_bispectrum_after_rotation}, it contains a sum over the different azimuthal modes; this is an analogy with the angular power spectrum $C_\L$, which includes contributions from the scalar, vector and tensor modes.

The linear transfer functions computed by \SONG, $\,\pert{\widetilde{\T}}{1}_{\L\,0}(k)\,$, do not depend on the direction of the wavemode as they are obtained assuming that the zenith is aligned with $\k\,$; on the other hand, those appearing in the bispectrum formula, $\,\T_\lm(\k)\,$, are for an arbitrary coordinate system and include the full $\k$ dependence. The multipoles in the two coordinate systems are related by the rotation matrices,
\begin{align}
  \pert{\T}{1}_\lmone(\kone) \;=\; \sqrt{\frac{4\,\pi}{2\,\lone+1}}\;
  Y_\lmone(\vec{\hat{k}_1})\;\,\pert{\widetilde{\T}}{1}_{\lone\,0}(k_1) \;, \msk
  \pert{\T}{1}_\lmtwo(\ktwo) \;=\; \sqrt{\frac{4\,\pi}{2\,\ltwo+1}}\;
  Y_\lmtwo(\vec{\hat{k}_2})\;\,\pert{\widetilde{\T}}{1}_{\ltwo\,0}(k_2) \;,
\end{align}
where a sum over $m'$ is absent because we are assuming that at linear order the non-scalar modes are negligible.
It is important to note that this is \emph{not} a rotation of the axes but a simple substitution; in fact, had we performed a rotation to align the zenith with $\kone$ or $\ktwo$, we would have undone what was gained with the rotation in \eref{eq:rotation_bispectrum}. If we insert the above identities in \eref{eq:intrinsic_angular_bispectrum_after_rotation}, we obtain an expression for the angle-averaged bispectrum,
\begin{align}
  \label{eq:intrinsic_angular_bispectrum_after_parity_inversion}
  &B^\intr_\ldep[\,\Delta\,] \;=\;
  (-1)^{\lone+\ltwo}\;\sqrt{\frac{(4\,\pi)^4}{(2\,\lone+1)(2\,\ltwo+1)}} \;
  \threej{\lone}{\ltwo}{\ltre}{m_1'}{m_2'}{m} \msk
  &\quad\;
  \times\int\,\frac{\dd\,k_1\,\dd\,k_2\,\dd\,k_3}{(2\,\pi)^6}\;
  (k_1\,k_2\,k_3)^2\; \;\pert{\widetilde{\T}}{1}_{\lone\,0}(k_1)\;
  \pert{\widetilde{\T}}{1}_{\ltwo\,0}(k_2)\; P_\Phi(k_1) \, P_\Phi(k_2)\nmsk
  &\quad\;
  \times\int\dd\Omega(\vec{\hat{k}_1})\;\dd\Omega(\vec{\hat{k}_2})\;\;
  \delta(\kone+\ktwo-\ktre)
  \;Y_{\lone m_1'}(\vec{\hat{k}_1})\;Y_{\ltwo m_2'}(\vec{\hat{k}_2})\;
  2\;\pert{\T}{2}_{\L_3m}(\kone,\ktwo,k_3)\;
  \;+\; \text{2 perm.} \;, \notag
\end{align}
where
we have enforced again the statistical isotropy to set $\,P_\Phi(\kone)=P_\Phi(k_1)\,$ and $\,P_\Phi(\ktwo)=P_\Phi(k_2)\,$, and we have performed the parity inversions $\,\kone\rightarrow-\kone\,$ and $\,\ktwo\rightarrow-\ktwo\,\,$; the $\,(-1)^{\lone+\ltwo}\,$ factor comes from the relations
\begin{align*}
  Y_{\lone m_1'}(-\vec{\hat{k}_1})\;=\;(-1)^\lone\;Y_{\lone m_1'}(\vec{\hat{k}_1})
  \qquad\text{and}\qquad
  Y_{\ltwo m_2'}(-\vec{\hat{k}_2})\;=\;(-1)^\ltwo\;Y_{\ltwo m_2'}(\vec{\hat{k}_2})\;.
\end{align*}

\subsubsection{Isolate the azimuthal dependence of \pert{\T}{2}}

The second-order transfer function in the bispectrum formula, $\,\pert{\T}{2}_{\L_3m}(\kone,\ktwo,k_3)\,$, is characterised by 7 degrees of freedom: $k_1, \theta_1, \phi_1,\allowbreak k_2, \theta_2, \phi_2, k_3\,$, where $\theta$ and $\phi$ are the polar and azimuthal angles, respectively.
Due to the presence of the Dirac delta function, however, the integral has support only for those configurations where $\kone+\ktwo-\ktre=0\,$. The relation allows us to express 3 of the 7 coordinates as a function of the remaining 4, which we choose to be $k_1, \phi_1, k_2, k_3\,$. In particular, we remark that $\theta_1$ is obtained as
\begin{align}
  \label{eq:cosine_k1k2k3}
  \cos\theta_1 \;=\; \frac{k_3^2\,+\,k_1^2\,-\,k_2^2}{2\,k_3\,k_1} \;.
\end{align}
(For further details, refer to Appendix~\ref{app:perturbations_geometry}.)
In the bispectrum formula (\eref{eq:intrinsic_angular_bispectrum_after_parity_inversion}) we can thus substitute
\begin{align}
  \pert{\T}{2}_{\L_3m}(\kone,\ktwo,k_3) \;\;\longrightarrow\;\;
  \pert{\T}{2}_{\L_3m}(k_1,\phi_1,k_2,k_3) \;.
\end{align}
This is still not enough because, as discussed in Appendix~\ref{app:perturbations_geometry}, in \SONG we compute the transfer functions assuming that $\phi_1=0$ and $\phi_2=\pi\,$ or, equivalently, $k_{1y}=k_{2y}=0\,$. Therefore, \SONG's transfer functions, which we denote as $\,\pert{\widetilde{\T}}{2}(k_1,k_2,k_3)\,$, are related to those in the bispectrum integral by a rotation about the zenith,
\begin{align}
  \label{eq:azimuthal_rotation}
  \pert{\T}{2}_{\L_3m}(k_1,\phi_1,k_2,k_3) \;=\;
  e^{\,im\phi_1}\;\pert{\widetilde{\T}}{2}_{\L_3m}(k_1,k_2,k_3) \;.
\end{align}

The term $\,e^{\,im\phi_1}\,$ has to be included in the bispectrum integral and thus complicates the $\,\dd\Omega(\vec{\hat{k}_1})\,$ integration considerably. One strategy is to expand $\,e^{\,im\phi_1}\,$ into spherical harmonics and use the orthogonality relations to integrate it out, thus introducing an extra pair of multipole indices. This can be avoided if we note that $\,e^{\,im\phi_1}\,$ can be expressed in terms of the spherical harmonic $Y_{|m|m}$, which is given in \sref{sec:spherical_harmonics} as
\begin{align}
  Y_{|m|m}(\vec{\hat{k}_1}) \;=\;
  e^{\,im\phi_1}\;\sqrt{\frac{2|m|+1}{4\,\pi}}\;
  \frac{\sqrt{(2|m|)!}}{2^{|m|}\,|m|!}\;\sin^{|m|}\theta_1\;
  \times\;
  \left\{\;\begin{aligned}
  &(-1)^m\;
  &&\quad\text{for $m\geq0$}\msk
  &\phantom{(}+1
  &&\quad\text{for $m<0$} \;.
  \end{aligned}\right.
\end{align}
Then, we can write 
\begin{align}
  e^{\,im\phi_1}\;\pert{\widetilde{\T}}{2}_{\L_3m}(k_1,k_2,k_3) \;=\;
  (-1)^m\;\sqrt{\frac{4\,\pi}{2|m|+1}}\;\;\pert{\overline{\T}}{2}_{\L_3m}(k_1,k_2,k_3)\;\;
  Y_{|m|m}(\vec{\hat{k}_1})\;,
\end{align}
where we have defined the \keyword{rescaled transfer function} as
\begin{align}
  \label{eq:tbar_definition}
  \pert{\overline{\T}}{2}_{\L_3m}(k_1,k_2,k_3) \;\equiv\;
  \pert{\widetilde{\T}}{2}_{\L_3m}(k_1,k_2,k_3)\;
  \frac{1}{\sin^{|m|}\theta_1}\;
  \frac{2^{|m|}\,|m|!}{\sqrt{(2{|m|})!}}\;\times\;
  \left\{\;\begin{aligned}
  &\phantom{(}+1
  &&\quad\text{for $m\geq0$}\msk
  &(-1)^m\;
  &&\quad\text{for $m<0$} \;.
  \end{aligned}\right.
\end{align}
The crucial point here is that the rescaled transfer function does \emph{not} depend on the azimuthal angle $\phi_1$ but only on the magnitude of the three wavemodes. Furthermore, it is immediately obtained by multiplying \SONG's transfer function, $\pert{\widetilde{\T}}{2}\,$, by a simple factor. The azimuthal dependence is confined to $Y_{|m|m}(\theta_1,\phi_1)$, which, as we shall soon see, will be integrated out using the orthogonality properties of the spherical harmonics. 

In summary, we have found that, by using the properties of the Dirac delta function, we can substitute the second-order transfer function in the bispectrum formula (\eref{eq:intrinsic_angular_bispectrum_after_parity_inversion}) with
\begin{align}
  \label{eq:tbar_substitutions}
  \pert{\T}{2}_{\L_3m}(\kone,\ktwo,k_3) \;\;\longrightarrow\;\;
  (-1)^m\;\sqrt{\frac{4\,\pi}{2m+1}}\;\;\pert{\overline{\T}}{2}_{\L_3m}(k_1,k_2,k_3)\;\;
  Y_{|m|m}(\vec{\hat{k}_1}) \;,
\end{align}
where $\overline{\T}$ is defined in \eref{eq:tbar_definition} and is numerically computed in \SONG.
This is a substantial advancement because the angular part of the transfer function is now completely separated from the radial one, without the need of performing additional multipole expansions.

\subsubsection{Integrate out the angular dependence}
At this stage, two strategies are possible.
One can integrate out the $\ktwo$ dependence using the Dirac delta function and then solve numerically the resulting 4D integral in
\begin{align}
  \int \dd\,k_1\;\dd\,k_3\;\dd\,\theta_1\;\dd\,\phi_1 \;,
\end{align}
which involves the highly oscillating spherical harmonics and transfer functions. Instead, we choose to expand the delta function in spherical harmonics and then solve the angular integration analytically; as we shall see, the final result is still a 4D integral, but its computation is numerically advantageous since it presents two smooth directions.

The integral form of the Dirac delta function,
\begin{align}
  (2\,\pi)^3\;\delta\,(\kone+\ktwo-\ktre) \;=\; 
  \int\dd\vecx\;e^{\,i\,\vecx\cdot(\kone+\ktwo-\ktre)} \;,
\end{align}
includes three plane waves that can expanded via the Rayleigh formula (\eref{eq:rayleigh_expansion}). If we do so, we obtain an expression for the delta function that involves spherical harmonics and Bessel functions \cite{mehrem:2011a}:
\begin{align}
  \label{eq:dirac_delta_bispectrum_expansion}
  &\delta\,(\kone+\ktwo-\ktre) \;=\;
  8\;i^{\,L_1+L_2+L_3}\;\sqrt{\frac{(2\,L_1+1)(2\,L_2+1)(2\,L_3+1)}{4\,\pi}}\;
  \threej{L_1}{L_2}{L_3}{0}{0}{0}\;\threej{L_1}{L_2}{L_3}{M_1}{M_2}{M_3} \;\nmsk
  &\qquad\times\;
  Y_{L_1M_1}(\vec{\hat{k}_1})\;Y_{L_2M_2}(\vec{\hat{k}_2})\;(-1)^{L_3}\,
  Y_{L_3M_3}(\vec{\hat{k}_3})\;
  \int\dd r\;r^2\;j_{L_1}(rk_1)\;j_{L_2}(rk_2)\;j_{L_3}(rk_3)\;
\end{align}
where a sum over the $L$ and $M$ indices is intended and the $(-1)^{L_3}$ factor comes from the parity inversion of $Y_{L_3M_3}(-\vec{\hat{k}_3})\,$.
The presence of $\,Y_{L_3M_3}(\vec{\hat{k}_3})\,$ is suspicious, because we have already integrated out the angular dependence of $\ktre\,$. However, since $\,\ktre$ is aligned with the zenith, we see that the dependence on $\ktre$ is only apparent:
\begin{align}
  Y_{L_3M_3}(\vec{\hat{k}_3}) \;=\; Y_{L_3M_3}(\theta=0,\phi) \;=\;
  \delta_{M_30}\;\sqrt{\frac{2\,L_3+1}{4\,\pi}} \;.
\end{align}
This is indeed the reason why we were allowed to take $\,\delta\,(\kone+\ktwo+\ktre)\,$ out of the $\dd\Omega$ integral in \eref{eq:intrinsic_angular_bispectrum_intermediate}.
If we insert the delta function expansion (\eref{eq:dirac_delta_bispectrum_expansion}) and the rescaled transfer function (\eref{eq:tbar_substitutions}) in the bispectrum integral (\eref{eq:intrinsic_angular_bispectrum_after_parity_inversion}), we obtain
\begin{align}
  \label{eq:intrinsic_angular_bispectrum_after_dirac_expansion}
  &B^\intr_\ldep[\,\Delta\,] \;=\;
  8\;i^{\,L_1+L_2+L_3}\;(-1)^{\lone+\ltwo+L_3+m}\;
  \sqrt{\frac{(4\,\pi)^3\,(2\,L_1+1)(2\,L_2+1)(2\,L_3+1)^2}{(2\,\lone+1)(2\,\ltwo+1)(2m+1)}}\;
  \msk&\quad\times\;
  \threej{\lone}{\ltwo}{\ltre}{m_1'}{m_2'}{m}
  \threej{L_1}{L_2}{L_3}{0}{0}{0}\;\threej{L_1}{L_2}{L_3}{M_1}{M_2}{0} \;
  \int\,\frac{\dd\,k_1\,\dd\,k_2\,\dd\,k_3\,\dd\,r}{(2\,\pi)^6}\;
  (k_1\,k_2\,k_3\,r)^2\;
  \displaybreak[0]\nmsk&\quad\times\;\vphantom{\int}
  \pert{\widetilde{\T}}{1}_{\lone\,0}(k_1)\;
  \pert{\widetilde{\T}}{1}_{\ltwo\,0}(k_2)\;
  2\;\pert{\overline{\T}}{2}_{\L_3m}(k_1,k_2,k_3)\;
  P_\Phi(k_1)\,P_\Phi(k_2)\;
  j_{L_1}(rk_1)\;j_{L_2}(rk_2)\;j_{L_3}(rk_3)\;
  \nmsk&\quad\times\;
  \int\dd\Omega(\vec{\hat{k}_1})\;
  Y_{\lone m_1'}(\vec{\hat{k}_1})\;
  Y_{L_1M_1}(\vec{\hat{k}_1})\;
  Y_{|m|m}(\vec{\hat{k}_1})\;
  \int\dd\Omega(\vec{\hat{k}_2})\;
  Y_{\ltwo m_2'}(\vec{\hat{k}_2})\;
  Y_{L_2M_2}(\vec{\hat{k}_2})\;
  \;+\; \text{2 perm.}\;, \notag
\end{align}
We recall that the $L$ and $M$ indices come from the delta function expansion while the $m'$ and $m$ indices come from the axes rotation; all 8 indices are summed. It should also be noted that the $i^{\,L_1+L_2+L_3}$ factor is always real because the second 3$j$ symbol vanishes when $\,L_1+L_2+L_3\,$ is odd.
The two angular integrals in the last line can be solved analytically using the Gaunt equality:
\begin{align}
  &\begin{aligned}
  &\int\dd\Omega(\vec{\hat{k}_1})\;
  Y_{\lone m_1'}(\vec{\hat{k}_1})\;
  Y_{L_1M_1}(\vec{\hat{k}_1})\;
  Y_{|m|m}(\vec{\hat{k}_1})
  \msk&\qquad\qquad\qquad
  =\;\sqrt{\frac{(2\,\lone+1)(2\,L_1+1)(2m+1)}{4\,\pi}}
  \;\GAUNT{\lone}{L_1}{|m|}{m_1'}{M_1}{m} \;,
  \end{aligned}
  \displaybreak[0]\llmsk
  &\int\dd\Omega(\vec{\hat{k}_2})\;
  Y_{\ltwo m_2'}(\vec{\hat{k}_2})\;
  Y_{L_2M_2}(\vec{\hat{k}_2}) \;=\;
  \sqrt{(2\,\ltwo+1)(2\,L_2+1)}\;
  \GAUNT{\ltwo}{L_2}{0}{m_2'}{M_2}{0} \;. \notag
\end{align}
We could express the last integral simply as $\,\delta_{\ltwo L_2}\,\delta_{m_2' M_2}\,$, but by doing so we would not be able to spot the following identity:
\begin{align}
  \label{eq:horrible_identity}
  &\sum\limits_{m_1'm_2'M_1M_2} \;
  \threej{\lone}{\ltwo}{\ltre}{m_1'}{m_2'}{m}\;
  \threej{L_1}{L_2}{L_3}{M_1}{M_2}{0}\;
  \threej{\lone}{L_1}{|m|}{m_1'}{M_1}{m}\;
  \threej{\ltwo}{L_2}{0}{m_2'}{M_2}{0}\nmsk
  &\qquad\qquad\qquad\qquad
  \;=\;(-1)^{\lone+\ltwo+L_3+m}\;\frac{\delta_{\ltwo L_2}}{\sqrt{2\,\ltwo+1}}\;
  \threej{\ltre}{L_3}{|m|}{m}{0}{-m}\;\sixj{\lone}{\ltre}{\ltwo}{L_3}{L_1}{|m|}\;,
\end{align}
where the term in curly brackets is Wigner's 6j symbol. 
To derive the identity, one has to introduce an extra factor in the sum,
\begin{align}
  \sqrt{2m+1}\;\threej{|m|}{0}{|m|}{\widetilde{m}}{0}{-\widetilde{m}} \;=\;
  \delta_{m\widetilde{m}} \;.
\end{align}
Then, the whole sum over the 5 azimuthal indices ($m_1',\,m_2',\,\allowbreak M_1,\,M_2,\,\widetilde{m}$) collapses to the product between a 3j and a 9j symbol (see Eq.~34.6.1 of Ref.~\cite{dlmf_website}). The latter contains a vanishing entry and therefore collapses to a 6j symbol, thus yielding the result in \eref{eq:horrible_identity}. Note that we have verified every step of this derivation using the Mathematica software \cite{wolfram:1991a}.

Expanding the Dirac delta function in spherical harmonics has allowed us to solve the angular integrations and all the azimuthal sums but $m$ analytically. As we shall show in the next subsection, what is left is a 4D integral that can be tackled numerically.

\subsubsection{Final formula}
After inserting the geometrical identity (\eref{eq:horrible_identity}) in the bispectrum integral (\eref{eq:intrinsic_angular_bispectrum_after_dirac_expansion}), we obtain the final formula for the angle-averaged intrinsic bispectrum:
\begin{align}
  &B^\intr_\ldep[\,\Delta\,] \;=\;
  \sum\limits_{m=-\infty}^\infty\;\;
  \sum\limits_{L_3=|\ltre-|m||}^{\ltre+|m|}\;\;
  \sum\limits_{L_1=|\lone-|m||}^{\lone+|m|}\;\;
  8\;i^{\,L_1+\ltwo+L_3}\;
  4\,\pi\,(2\,L_1+1)(2\,\ltwo+1)(2\,L_3+1)\;
  \nmsk&\quad\times\;
  \threej{L_1}{\ltwo}{L_3}{0}{0}{0}\;\threej{\L_1}{L_1}{|m|}{0}{0}{0}\;
  \threej{\ltre}{L_3}{|m|}{m}{0}{-m}\;\sixj{\lone}{\ltre}{\ltwo}{L_3}{L_1}{|m|}\;
  \int\,\frac{\dd\,k_1\,\dd\,k_2\,\dd\,k_3\,\dd\,r}{(2\,\pi)^6}\;
  (k_1\,k_2\,k_3\,r)^2\;
  \displaybreak[0]\nmsk&\quad\times\;\vphantom{\int}
  \pert{\widetilde{\T}}{1}_{\lone\,0}(k_1)\;
  \pert{\widetilde{\T}}{1}_{\ltwo\,0}(k_2)\;
  2\;\pert{\overline{\T}}{2}_{\L_3m}(k_1,k_2,k_3)\;
  P_\Phi(k_1)\,P_\Phi(k_2)\;
  j_{L_1}(rk_1)\;j_{\ltwo}(rk_2)\;j_{L_3}(rk_3)
  \;+\; \text{2 perm.}
  \label{eq:intrinsic_angular_bispectrum_final_formula}
\end{align}
We recall that $\,B^\intr_\ldep[\,\Delta\,]\,$ is the bispectrum of the brightness perturbation, and that the transfer functions in the integral are accordingly defined with respect to $\Delta$ (\eref{eq:transfer_function_definition_bispectrum}).
In \sref{sec:temperature_bispectrum}, we shall see that the observed intrinsic bispectrum, $\,B^\intr_\ldep[\,\Theta\,]\,$, is obtained from the above by the simple relation
\begin{align}
  B^\intr_\ldep[\,\Theta\,] \;&=\; \hat{B}^\intr_\ldep[\,\Delta\,] \;
  -\; 3\;h_\ldep\;\left(\;C_\lone C_\ltwo\;+\;C_\ltwo C_\ltre\;+\;C_\ltre C_\lone\;\right) \;,
\end{align}
where $\hat{B}^\intr$ is a simple rescaling of $B^\intr_\ldep\,$ (\eref{eq:bhat_definition}) and $\,h_\ldep\,$ is the purely geometrical factor defined in \eref{eq:gaunt_factor_I}:
\begin{align*}
  h_{\lone\ltwo\ltre} \,=\, \sqrt{\frac{(2\lone+1)(2\ltwo+1)(2\ltre+1)}{4\pi}}
  \,\threej{\lone}{\ltwo}{\ltre}{0}{0}{0} \;.
\end{align*}

We invite the reader not to be intimidated by the long expression in \eref{eq:intrinsic_angular_bispectrum_final_formula}. In fact, the formula is a substantial improvement over the starting point of our computation (\eref{eq:intrinsic_angular_bispectrum_k1k2k3}) because all the involved quantities are in a form that can be numerically evaluated. The first-order transfer functions, $\,\pert{\widetilde{\T}}{1}\,$, can be produced in the matter of seconds by any linear Boltzmann code, while the second-order one, $\,\pert{\overline{\T}}{2}\,$ is a direct product of \SONG. The 3$j$ symbols and the spherical Bessel functions are purely geometrical factors that can be precomputed and stored in tables using publicly available libraries such as \emph{SLATEC} \cite{vandevender:1982a} or \emph{GSL} \cite{galassi:2009a}.

\runinhead{Squeezed limit}
The squeezed limit of the bispectrum consists in considering only those configurations where one of the $\L$'s is much smaller than the other two; thus, the squeezed bispectrum encodes the correlations between large and small angular scales.
The projection functions in the line of sight integral enforce that the Fourier modes contributing to such configurations are also squeezed, that is, one of the three wavemodes has to be much smaller than the other two.
Since we align $k_3$ to the polar axis, the triangular condition implies that, for squeezed configurations, at least one between $k_1$ or $k_2$ is also aligned with the polar axis;
it follows that the quadratic sources of the Einstein and Boltzmann equations always contain at least one first-order perturbation with a polar angle $\theta\simeq0$.
In the absence of first-order vector and tensor modes, any linear perturbation is proportional to $Y_{\lm}(\vec{\hat{k}})$ (\eref{eq:rotation}) which, in turn, is proportional to $\sin^m\theta$.
Therefore, the quadratic sources are suppressed for squeezed configurations unless $m=0$.
Because we assume that the $m\neq0$ modes are only sourced by the quadratic sources (\ie we assume the absence of primordial vector and tensor modes), it follows that the $m\neq0$ transfer functions vanish in the squeezed limit and so do the $m\neq0$ contributions to the intrinsic bispectrum.

Thus, the dominant contribution to the intrinsic bispectrum in the squeezed limit comes from the scalar modes, that is, by setting $m=0$ in \eref{eq:intrinsic_angular_bispectrum_final_formula}:
\begin{align}
  \label{eq:intrinsic_bispectrum_m0_scalar}
  &\hat{B}^\intr_\ldep[\,\Delta\,]\,\Bigr|_{m=0} \;=\;
  h_\ldep \;\left(\,\frac{2}{\pi}\,\right)^3\;
  \int\dd\,r\;r^2\;\,
  \int\dd\,k_1\;k_1^2\;\,P_\Phi(k_1)\;\pert{\widetilde{\T}}{1}_{\lone}(k_1)\;\,j_{\lone}(rk_1)\;
  \msk&\quad\times\;
  \int\dd\,k_2\;k_2^2\;\,P_\Phi(k_2)\;\pert{\widetilde{\T}}{1}_{\ltwo}(k_2)\;\,j_{\ltwo}(rk_2)\;
  \int\dd\,k_3\;k_3^2\;\,2\;\pert{\overline{\T}}{2}_{\ltre}(k_1,k_2,k_3)
  \;\,j_{\ltre}(rk_3)
  \;+\; \text{2 perm.} \;,\notag
\end{align}
where we have introduced the notation $\T_\L \equiv \T_{\L\,0}/(2\,\L+1)\,$ and used the identities
\begin{align}
  \threej{\L}{L}{0}{0}{0}{0} \;=\;
  \delta_{\L L}\;\frac{(-1)^\L}{\sqrt{2\L+1}}
  &&\text{and}&&
  \sixj{\lone}{\ltre}{\ltwo}{\ltre}{\lone}{0} \;=\;
  \frac{(-1)^{\lone+\ltwo+\ltre}}{\sqrt{(2\lone+1)(2\ltre+1)}} \;.
\end{align}
The $m=0$ formula is accurate to study the overlap between the intrinsic bispectrum and the local template, the latter being strongly peaked on squeezed configuration. This is what we have done in Ref.~\cite{pettinari:2013a}, as we shall detail in \sref{sec:results}.

\subsection{Numerical estimation} 
\label{sec:intrinsic_bispectrum_numerical}

We express the bispectrum formula schematically as
\begin{align}
  \label{eq:intrinsic_angular_bispectrum_schematic_formula}
  B^\intr_\ldep[\,\Delta\,] \;=\;
  \sum\limits_{m=-\infty}^\infty\;\;
  \sum\limits_{L_3=|\ltre-|m||}^{\ltre+|m|}\;\;
  \sum\limits_{L_1=|\lone-|m||}^{\lone+|m|}\;\;
  \Gamma\,_\ldep^{mL_1L_3}\;\times\;I\,_\ldep^{mL_1L_3}\;
  \;+\; \text{2 perm.} \;,
\end{align}
where $\Gamma$ groups the terms in \eref{eq:intrinsic_angular_bispectrum_final_formula} before the integral sign, and $I$ the rest.
The computation of $\,B^\intr_\ldep\,$ is then split in two parts: estimating the 4D integral, $\,I\,$, and performing the three summations over the geometrical factors, $\,\Gamma\,$.

The two permutations in the formula refer to the exchange of $\kone$, $\ktwo$ and $\ktre\,$ (see comment to \eref{eq:next_to_leading_order_3pf}). By looking back at \eref{eq:intrinsic_angular_bispectrum_k1k2k3}, we see that they are equivalent to permutations in $(\lmone)$, $(\lmtwo)$ and $(\lmtre)$. Therefore, they can be accounted for in the last step of the computation as
\begin{align}
  B^\intr_\ldep[\,\Delta\,] \;=\;
  B^\text{asymm}_{\lone\ltwo\ltre}[\,\Delta\,] \;+\;
  B^\text{asymm}_{\ltre\lone\ltwo}[\,\Delta\,] \;+\;
  B^\text{asymm}_{\ltwo\ltre\lone}[\,\Delta\,] \;,
\end{align}
where $\,B^\text{asymm}\,$ is the first term in the right hand side of \eref{eq:intrinsic_angular_bispectrum_schematic_formula}. Note that by doing so, we also ensure that the intrinsic bispectrum is symmetric.

\subsubsection{Integral estimation} 

The integral in the intrinsic bispectrum reads
\begin{align}
  \label{eq:bispectrum_integral_only}
  &I\,_\ldep^{mL_1L_3} \;=\; \frac{1}{(2\,\pi)^6}
  \int\dd\,r\;r^2\;\,
  \int\dd\,k_1\;k_1^2\;P_\Phi(k_1)\;\,\pert{\widetilde{\T}}{1}_{\lone\,0}(k_1)\;\,j_{L_1}(rk_1)\;
  \msk&\quad\times\;
  \int\dd\,k_2\;k_2^2\;P_\Phi(k_2)\;\,\pert{\widetilde{\T}}{1}_{\ltwo\,0}(k_2)\;\,j_{\ltwo}(rk_2)\;
  \int\dd\,k_3\;k_3^2\;\,2\;\pert{\overline{\T}}{2}_{\L_3m}(k_1,k_2,k_3)\;\,j_{L_3}(rk_3)\;. \notag
\end{align}
A similar integral has been efficiently treated in \citeauthor{fergusson:2007a} \cite{fergusson:2007a, fergusson:2009a}, where the role of the second-order transfer function was played by the separable primordial bispectrum $B_\Phi(k_1,k_2,k_2)\,$. Our case is more complicated as $\,\pert{\overline{\T}}{2}_{\L_3m}(k_1,k_2,k_3)\,$ is not separable; however, we can still numerically solve the integral in an efficient way by exploiting other useful properties of $\,\pert{\overline{\T}}{2}_{\L_3m}(k_1,k_2,k_3)\,$.

\runinhead{Sampling in $k_1$ and $k_2$}
The non-linear transfer function $\,\pert{\overline{\T}}{2}_{\L_3m}(k_1,k_2,k_3)\,$ is rapidly oscillating in $k_3$ but it is smooth in the $k_1$ and $k_2$ directions. This is clear by looking at the line of sight integral (\eref{eq:los_integral_compact}), which is used to compute $\pert{\widetilde{\T}}{2}\,$:
\begin{align}
	\pert{\widetilde{\T}}{2}_n(k_1,k_2,k_3) \;=\;
  \int\limits_{\tauini}^{\tauz}\;\dd\tau\;
  e^{-\kappa(\tau)}\; J_{\,nn'}(k_3\,(\tauz-\tau))\;\;\S_{\,n'}(k_1,k_2,k_3) \;.
\end{align}
Any feature in the source at the time of recombination, $\,\taurec\,$, generates oscillations of frequency $\tauz-\taurec$ in the $k_3$ direction of $\pert{\widetilde{\T}}{2}\,$, through the projection function $J\,$. The $k_1$ and $k_2$ directions of $\pert{\widetilde{\T}}{2}\,$, on the other hand, inherit the oscillation frequency of $S\,$, which, at the time of recombination, is dictated to be of order $\taurec/\sqrt{3}$ by the tight coupling between the photon and baryon fluids. Because $\,\tauz\simeq80\,\taurec\,$ for a standard \LCDM Universe, $\,\pert{\widetilde{\T}}{2}\,$ oscillates in the $k_1$ and $k_2$ directions with a frequency $\sim80$ times slower than that of $k_3\,$. The same argument applies to $\,\pert{\overline{\T}}{2}\,$, which is related to $\,\pert{\widetilde{\T}}{2}\,$ by the smooth rescaling shown in \eref{eq:tbar_definition}.
The smoothness of the $k_1$ and $k_2$ directions substantially reduces the execution time, as the $k_3$ integral can be solved and tabulated on the small $(k_1,k_2)$ grid discussed in \sref{sec:sampling_strategies}.
It should be noted that, had we directly integrated out the delta function in \eref{eq:intrinsic_angular_bispectrum_after_parity_inversion} instead of expanding it in spherical harmonics, we could not have used this property; in fact, in that case, the dependence of the transfer function on the wavemodes would have been mixed, thus spoiling its smoothness in $k_1$ and $k_2$.

\runinhead{Sampling in $r$}
The projection function in the line of sight formula, above, is effectively a spherical Bessel function (see comment to \eref{eq:J_projection_function}); similarly, in the bispectrum formula, for $\ltre\gg m$ we can approximate $j_{L_3}\simeq j_\ltre\,$. Thus, the $k_3$ integral in \eref{eq:bispectrum_integral_only} is roughly given by
\begin{align}
  \int_0^\infty\dd\,k_3\;k_3^2\;\,j_\ltre(k_3r)\;j_\ltre(k_3(\tauz-\taurec))\,
  S_{\ltre m}(k_1,k_2,k_3) \;,
\end{align}
where we have also assumed that all the sources are localised on the last scattering surface.
The source function is smooth in $k_3\,$, meaning that it acts as a modulation of the two oscillating functions in the integrand. In the limit of a flat source, we can use the closure relation of the spherical Bessel functions \cite{mehrem:2011a} to find
\begin{align}
  \int_0^\infty\dd\,k_3\;k_3^2\;\,j_\ltre(k_3r)\;j_\ltre(k_3(\tauz-\taurec))\,
  \;\propto\; \delta(r-(\tauz-\taurec)) \;.
\end{align}
Thus, we expect the integrand of the bispectrum integral to be peaked around $r\simeq\tauz-\taurec\,$. The same argument applies to the $k_1$ and $k_2$ integrals, so that any contribution to the bispectrum from regions where $r$ is far from $\tauz-\taurec$ is threefold suppressed.
The argument breaks down when we consider the propagation sources (\eref{eq:propagation_sources_intensity}), which are not localised on the last scattering surface and can therefore couple with the late-time effects encoded in the linear transfer functions. This is the case of the gravitational lensing, that couples with the integrated Sachs-Wolfe effect to give a squeezed bispectrum \cite{lewis:2012a, hanson:2009a, smith:2011a, serra:2008a, lewis:2011a, lewis:2006a} that has been actually measured by the Planck satellite \cite{planck-collaboration:2013b}.
In this work, however, we do not consider lensing. By including only the scattering and metric sources, we obtain a sub-percent level convergence in the bispectrum with an $r$-grid of $\,\O(100)$ points around $\tauz-\taurec$ (\sref{sec:convergence_tests}).

\runinhead{Order of the integrations}
Armed with the knowledge that the $r$, $k_1$ and $k_2$ directions are smooth, we estimate the bispectrum integral in a straightforward way. Below, we describe the order of integration that we adopt; we also assume that $(m,L_3,L_1)\,$ is fixed.
\begin{enumerate}
  \item We first compute the $\dd\,k_3$ integral,
  \begin{align}
    I_\ltre(r,k_1,k_2) \;=\; 
    2\;\,\int\dd\,k_3\;k_3^2\;\;\pert{\overline{\T}}{2}_{\L_3m}(k_1,k_2,k_3)\;\,j_{L_3}(rk_3) \;,
  \end{align}
  and store the result as a table in $r$, $k_1$, $k_2$ and $\ltre\,$. For an average precision run where each of these parameters is sampled in $\O(100)$ points, this corresponds to solving the integral for about $10^8$ times for each $\,(m,L_3,L_1)\,$ configuration that is considered.
Note that we only need to compute $\,I_\ltre(r,k_1,k_2)\,$ for the $k_1\geq k_2$ configurations, as the behaviour of the rescaled transfer function (\eref{eq:tbar_definition}) with respect to the exchange $k_1\leftrightarrow k_2$ ensures that
  \begin{align}
    \frac{I_\ltre(r,k_1,k_2)}{I_\ltre(r,k_2,k_1)}
    \;=\; (-1)^m\;\left(\,\frac{\sin\theta_2}{\sin\theta_1}\right)^{|m|}
    \;=\; (-1)^m\,\left(\,\frac{k_1}{k_2}\right)^{|m|} \;,
  \end{align}
  where we have used the relation $\,k_1\sin\theta_1=k_2\sin\theta_2\,$ (\eref{eq:sin_theta}). The $(-1)^m$ factor comes from exchanging $k_1\leftrightarrow k_2$ in the unrescaled transfer functions $\,\pert{\widetilde{\T}}{2}\,$ (\eref{eq:alternating_sign_exchange_k1_k2}).
  \item Then, we use the results of the previous integration to compute the $\dd\,k_2$ integral,
  \begin{align}
    I_{\ltwo\ltre}(r,k_1) \;=\; 
    \int\dd\,k_2\;k_2^2\;\;P_\Phi(k_2)\;\,\pert{\widetilde{\T}}{1}_{\ltwo\,0}(k_2)\;\,j_{\ltwo}(rk_2)\;
    I_\ltre(r,k_1,k_2)\;,
  \end{align}
  and store the result as a table in $r$, $k_1$, $\ltwo$ and $\ltre\,$. The presence of the power spectrum does not require an ad-hoc treatment as it is usually a smooth function of $k_2\,$. Because  $\,\pert{\widetilde{\T}}{1}_{\ltwo\,0}(k_2)\,$ oscillates rapidly in $k_2$ but $\,I_\ltre(r,k_1,k_2)\,$ does not, we interpolate the latter in $k_2\,$. 
  \item The $\dd\,k_1$ integral,
  \begin{align}
    I_{\lone\ltwo\ltre}(r) \;=\; 
    \int\dd\,k_1\;k_1^2\;\;P_\Phi(k_1)\;\,\pert{\widetilde{\T}}{1}_{\lone\,0}(k_1)\;\,j_{L_1}(rk_1)\;
    I_{\ltwo\ltre}(r,k_1) \;,
  \end{align}
  is equivalent to that in $\dd\,k_2\,$, so that it also requires the interpolation of $\,I_{\ltwo\ltre}(r,k_1)\,$ in $k_1\,$. The result is stored in a table in $\lone$, $\ltwo$ and $\ltre$.
  \item The last integral in $\dd\,r\,$,
  \begin{align}
    I_\ldep \;=\; \frac{1}{(2\,\pi)^6}
    \int\dd\,r\;r^2\;\,I_{\lone\ltwo\ltre}(r) \;.
  \end{align}
   is the simplest one as it does not involve oscillations and only depends on $\lone$, $\ltwo$ and $\ltre\,$.
\end{enumerate}
We remark that the three integrals in $k$ are similar as they always involve the convolution of a rapidly oscillating function with a spherical Bessel function; in fact, in \SONG they are all solved using the same integration routine via a simple trapezoidal rule.


\subsubsection{Angular summations} 

In the bispectrum formula of \eref{eq:intrinsic_angular_bispectrum_schematic_formula},
\begin{align}
  B^\intr_\ldep[\,\Delta\,] \;=\;
  \sum\limits_{m=-\infty}^\infty\;\;
  \sum\limits_{L_3=|\ltre-|m||}^{\ltre+|m|}\;\;
  \sum\limits_{L_1=|\lone-|m||}^{\lone+|m|}\;\;
  \Gamma\,_\ldep^{mL_1L_3}\;\times\;I\,_\ldep^{mL_1L_3}\;
  \;+\; \text{2 perm.} \;,
\end{align}
the sum over the azimuthal modes is in principle infinite and needs to be truncated at some $m_\text{max}\,$.
At its present state, \SONG implements the intrinsic bispectrum for any value of $m$, but we have not yet performed a full convergence test to assess the optimal value of $m_\text{max}\,$. However, we expect the largest contribution to the intensity bispectrum to come from the $m\leq2$ modes, because the other modes correspond to multipoles that are tight-coupling suppressed during recombination.

For $m\leq2$, the summations over $L_1$ and $L_3$ contain a small number of addends. The number is further reduced if one considers that, for the photon intensity, only even values of $\lone+L_1+m$ and $\ltre+L_3+m$ are allowed. Thus, for $m=0$, there is only one contribution to the bispectrum while for $m=1$ and $m=2$ there are 4 and 9, respectively.
This is indeed a welcome simplification, since the bispectrum integral in \eref{eq:bispectrum_integral_only} needs to be solved for each combination of $m$, $L_3$ and $L_1\,$.

Another major simplification in the computation of $\,B^\intr_\ldep\,$ comes from the fact that the $m<0$ elements of the sum can be inferred from the $m>0$ ones. In fact, from \eref{eq:tbar_definition} it follows that, for the intensity, the rescaled transfer function is invariant under a sign-flip of $m$,
\begin{align}
  \pert{\overline{\T}}{2}_{\L_3-m} \;=\;
  \pert{\overline{\T}}{2}_{\L_3m} \;,
\end{align}
as the $(-1)^m$ factor in the definition of $\pert{\overline{\T}}{2}_{\ltre m}$ cancels with that coming from $\,\pert{\widetilde{\T}}{2}_{\ltre -m}=(-1)^m\,\pert{\widetilde{\T}}{2}_{\ltre m}\,$.
Since the only term apart from $\,\pert{\widetilde{\T}}{2}_{\ltre m}\,$ that depends on the sign of $m$ in the bispectrum formula \eref{eq:intrinsic_angular_bispectrum_final_formula} is
\begin{align}
  \threej{\ltre}{L_3}{|m|}{m}{0}{-m} \;,
\end{align}
we infer that, for a given $|m|$, the contribution to the bispectrum is proportional to
\begin{align}
  \threej{\ltre}{L_3}{|m|}{m}{0}{-m} \;+\; \threej{\ltre}{L_3}{|m|}{-m}{0}{m} \;=\;
  \left[\;1\;+\;(-1)^{\,\ltre+L_3+|m|}\;\right]\;\threej{\ltre}{L_3}{|m|}{m}{0}{-m} \;,
\end{align}
which forces the intensity bispectrum to vanish for odd values of $\,\ltre+L_3+|m|\,$ and yields a factor 2 otherwise. That is, the negative azimuthal modes contribute to the intrinsic bispectrum as much as their positive counterparts.
We also note that, for the intensity, the angle-averaged bispectrum $\,B^\intr_\ldep[\,\Delta\,]\,$ vanishes when $\lone+\ltwo+\ltre$ is odd. This follows directly from the fact that the sums $\,L_1+\ltwo+L_3\,$, $\,\lone+L_1+m\,$ and $\,\ltre+L_3+m\,$ must all be even.

\runinhead{$B$-modes} The above considerations have to be slightly adjusted when treating bispectra involving $B$ polarisation. In fact, the $B$-mode transfer functions satisfy
\begin{align}
  \pert{\widetilde{\T}}{2}_{\ltre -m} \;=\; (-1)^{m+1}\;\pert{\widetilde{\T}}{2}_{\ltre m}
  \quad\Rightarrow\quad
  \pert{\overline{\T}}{2}_{\ltre -m} \;=\; -\pert{\overline{\T}}{2}_{\ltre m} \;.
\end{align}
This implies that, when considering an odd number of B-modes (\eg $\avg{BTT}$ or $\avg{BEE}\,$), the intrinsic bispectrum in \eref{eq:intrinsic_angular_bispectrum_final_formula} is proportional to
\begin{align}
  \threej{\ltre}{L_3}{|m|}{m}{0}{-m} \;-\; \threej{\ltre}{L_3}{|m|}{-m}{0}{m} \;=\;
  \left[\;1\;-\;(-1)^{\,\ltre+L_3+|m|}\;\right]\;\threej{\ltre}{L_3}{|m|}{m}{0}{-m} \;,
\end{align}
and therefore vanishes when $\,\ltre+L_3+|m|\,$ is even. If we consider that $\,L_1+\ltwo+L_3\,$ and $\,\lone+L_1+m\,$ still have to be even due to the 3$j$ symmetries, if follows that a bispectrum with an odd number of $B$-modes possesses \emph{odd parity}, that is, it vanishes when $\,\lone+\ltwo+\ltre\,$ is even. On the other hand, a bispectrum with an even number of $B$-modes possesses \emph{even parity} and vanishes when $\,\lone+\ltwo+\ltre\,$ is odd. This latter case includes the bispectra involving exclusively intensity or $E$-modes, such as $\avg{TTT}$, $\avg{EEE}$ and $\avg{TEE}\,$.




\subsection{Linearly propagated bispectrum} 
\label{sec:templates}

The linearly propagated bispectrum, $\,\avg{\Delta^3}_\lin\,$, describes how the primordial non-Gaussianity of the CMB evolves throughout cosmic history. It is therefore crucial to accurately compute $\,\avg{\Delta^3}_\lin\,$ to relate the current CMB observations to the non-Gaussian properties of the early Universe.

The linear bispectrum has a simple form,
\begin{align}
  &\avg{\Delta^3}_\lin \;=\; \int\frac{\dd\kone\,\dd\ktwo\,\dd\ktre}{(2\,\pi)^6}\;\,
  \delta\,(\kone+\ktwo+\ktre)\;
  \pert{\T_\lmone}{1}(\kone)\,
  \pert{\T_\lmtwo}{1}(\ktwo)\,
  \pert{\T_\lmtre}{1}(\ktre)\;
  B_\Phi(\kone,\,\ktwo,\,\ktre) \;,  & \notag
\end{align}
where the primordial bispectrum $B_\Phi$ is defined by (\sref{sec:three_point_function})
\begin{align}
  \avg{\Phi(\kone)\,\Phi(\ktwo)\,\Phi(\ktre)} \;=\;
  (2\pi)^3\;\delta\,(\kone+\ktwo+\ktre)\;B_\Phi(\kone,\ktwo,\ktre)\;.
\end{align}
The numerical computation of $\,\avg{\Delta^3}_\lin\,$ requires a simplified treatment with respect to the intrinsic bispectrum, because of the absence of the complicated second-order transfer function.
Schematically, the steps involved are:
\begin{enumerate}
  \item Substitute the three linear transfer functions with
  \begin{align}
    \pert{\T}{1}_\lm(\k) \;=\; \sqrt{\frac{4\,\pi}{2\,\L+1}}\;
    Y_\lm(\vec{\hat{k}})\;\,\pert{\widetilde{\T}}{1}_{\L\,0}(k)
  \end{align}
  to express the integrand in terms of the transfer functions in the coordinate system where the zenith is aligned with $\k\,$, which are those actually computed by a Boltzmann code.
  \item Expand the Dirac delta function in spherical harmonics according to \eref{eq:dirac_delta_bispectrum_expansion}; this introduces 6 sums in ($L_1M_1$), ($L_2M_2$) and ($L_3M_3$) and the Gaunt coefficient $\,\gaunt{L_1}{L_2}{L_3}{M_1}{M_2}{M_3}\,$.
    \item Enforce the statistical isotropy of the Universe to set the primordial bispectrum to depend only on the magnitudes of the wavevectors: $\,B_\Phi(\kone,\ktwo,\ktre)=\allowbreak B_\Phi(k_1,k_2,k_3)\,$ (\sref{sec:three_point_function}).
  \item Solve the simple angular integrals in $\,\dd\Omega(\vec{\hat{k}_1})\,$, $\,\dd\Omega(\vec{\hat{k}_2})\,$ and $\,\dd\Omega(\vec{\hat{k}_3})$ exploiting the orthogonality property of the spherical harmonics; the resulting Kronecker deltas can be used to enforce $L=\L$ and $M=m$ and thus solve the summations introduced by the delta function expansion.
\end{enumerate}
As a result, one is left with the following formula for the linear bispectrum:
\begin{align}
  \label{eq:final_linear_angular_bispectrum_k1k2k3}
  \avg{\Delta^3}_\lin \;&=\; 
  \gaunt{\lone}{\ltwo}{\ltre}{m_1}{m_2}{m_3} \;
  \left(\frac{2}{\pi}\right)^3
  \;i^{\,\lone+\ltwo+\ltre}\;
  \sqrt{\frac{(2\,\lone+1)(2\,\ltwo+1)(2\,\ltre+1)}{(4\pi)^3}}\;
  \\&\times\;
  \int\dd\,r\;r^2\;\,
  \int\dd\,k_1\;k_1^2\;\,
  \frac{\pert{\widetilde{\T}}{1}_{\lone\,0}(k_1)}{2\,\lone+1}\;\,j_{\lone}(rk_1)\;
  \int\dd\,k_2\;k_2^2\;\,
  \frac{\pert{\widetilde{\T}}{1}_{\ltwo\,0}(k_2)}{2\,\ltwo+1}\;\,j_{\ltwo}(rk_2)\;\notag
  \\&\times\;
  \int\dd\,k_3\;k_3^2\;\,
  \frac{\pert{\widetilde{\T}}{1}_{\ltre\,0}(k_3)}{2\,\ltre+1}\;\,j_{\ltre}(rk_3)\;
  B_\Phi(k_1,k_2,k_3)\;. \notag
\end{align}

At first order, the temperature bispectrum is related to the brightness one by
\begin{align}
  4^3\;\avgbig{a_\lmone\,a_\lmtwo\,a_\lmtre}_\lin \;=\;
  \avg{\Delta^3}_{\lin} \;
  i^{-\lone-\ltwo-\ltre} \;
  \sqrt{\frac{(4\pi)^3}{(2\,\lone+1)(2\,\ltwo+1)(2\,\ltre+1)}}\;
\end{align}
The factor $4^3$ comes from the fact that, at the linear level, $\,\Delta=4\,\Theta\,$ (\eref{eq:delta_theta_relation_brightness}), while the remaining coefficients are due to the different convention for the $Y_\lm$ expansions of $\Delta$ and $\Theta\,$ (\eref{eq:alm_vs_thetalm_2}).
Furthermore, due to the absence of non-scalar modes, it is customary to express the transfer functions in terms of their Legendre coefficients rather than the spherical multipoles; the two are related by a $2\L+1$ factor:
\begin{align}
  \pert{\widetilde{\T}}{1}_{\L}(k) \;=\; \frac{\pert{\widetilde{\T}}{1}_{\L\,0}(k)}{2\,\L+1} \;.
\end{align}
With these notational changes, our formula for the linearly propagated bispectrum reads
\begin{align}
  \label{eq:final_linear_angular_bispectrum_k1k2k3_literature}
  &4^3\;\avgbig{a_\lmone\,a_\lmtwo\,a_\lmtre}_\lin \;=\;
  \gaunt{\lone}{\ltwo}{\ltre}{m_1}{m_2}{m_3} \;\left(\,\frac{2}{\pi}\,\right)^3\;
  \int\dd\,r\;r^2\;\,
  \int\dd\,k_1\;k_1^2\;\,\pert{\widetilde{\T}}{1}_{\lone}(k_1)\;\,j_{\lone}(rk_1)\;
  \msk&\quad\times\;
  \int\dd\,k_2\;k_2^2\;\,\pert{\widetilde{\T}}{1}_{\ltwo}(k_2)\;\,j_{\ltwo}(rk_2)\;
  \int\dd\,k_3\;k_3^2\;\,\pert{\widetilde{\T}}{1}_{\ltre}(k_3)\;\,j_{\ltre}(rk_3)\;
  B_\Phi(k_1,k_2,k_3)\;, \notag
\end{align}
which is the usual form found in the literature \cite{komatsu:2001a, fergusson:2007a}.

It should be noted that the formula for the linearly propagated bispectrum, above, resembles that for the scalar intrinsic bispectrum, shown in \eref{eq:intrinsic_bispectrum_m0_scalar}. In fact, the two formulae are equivalent if we substitute
\begin{align}
  \pert{\widetilde{\T}}{1}_{\ltre}(k_3)\;B_\Phi(k_1,k_2,k_3) \quad\rightarrow\quad
  2\;\pert{\overline{\T}}{2}_{\ltre}(k_1,k_2,k_3)\;P_\Phi(k_1)\;P_\Phi(k_2) \;.
\end{align}
This result was expected since the same transformation relates \eref{eq:linear_angular_bispectrum_k1k2k3} and \ref{eq:intrinsic_angular_bispectrum_k1k2k3}.

\subsubsection{The primordial templates}
Many models of the early Universe exist that give definite predictions for the shape and amplitude of the primordial bispectrum $B_\Phi(k_1,k_2,k_3)\,$. In principle, they can be falsified or constrained by comparing the measured CMB bispectrum with the predicted one, via \eref{eq:final_linear_angular_bispectrum_k1k2k3_literature}.
To facilitate the comparison between theory and observations, three theoretical templates have been put forward that capture most of the physics in the models of the early Universe:
\begin{itemize}
  \item The local shape \cite{komatsu:2001a, gangui:1994a, verde:2000a},
  \begin{align}
    B^\local_\Phi(k_1,k_2,k_3) \;=\;
    2\;\fnl^\local\;\bigl[\;
    P_\Phi(k_1)\,P_\Phi(k_2) \;+\; P_\Phi(k_2)\,P_\Phi(k_3) \;+\; P_\Phi(k_3)\,P_\Phi(k_1) \;\bigr] \;,
    \label{eq:local_template}
  \end{align}
  is produced in a wide class of multi-field models, including the curvaton one \cite{linde:1997a, enqvist:2002a,lyth:2002a,moroi:2001a,moroi:2002a}. It peaks at the so-called ``squeezed'' triangles where one of the sides is much smaller than the other two.
  \item The equilateral shape \cite{creminelli:2006a},
  \begin{align}
    \label{eq:equilateral_template}
    &B^\equil_\Phi(k_1,k_2,k_3) \;=\; 6\;\fnl^\equil
    \msk&\qquad\times\Bigl\{\;
    -\, P_\Phi(k_1)\,P_\Phi(k_2) \,-\, P_\Phi(k_1)\,P_\Phi(k_3) \,-\, P_\Phi(k_2)\,P_\Phi(k_3)\;
    \allowbreak\nmsk&\qquad
    -\; 2\;\bigl[\;P_\Phi(k_1)\,P_\Phi(k_2)\,P_\Phi(k_3)\;\bigr]^{\nicefrac{2}{3}} \,+\, \text{5 perm.} \;
    \nmsk&\qquad
    +\; \bigl[\;P_\Phi(k_1)\,P_\Phi(k_2)^2P_\Phi(k_3)^3\;\bigr]^{\nicefrac{1}{3}} \,+\, \text{5 perm.}
    \;\Bigr\} \;, \notag
  \end{align}
   arises in single-field models with non-standard kinetic terms such as DBI inflation \cite{alishahiha:2004a, silverstein:2004a} or, in general, in models where the Lagrangian involves higher-order derivative operators. As the name suggests, it peaks when the three wavemodes have similar values. The local and equilateral shapes are almost orthogonal.
  \item The orthogonal shape \cite{senatore:2010a},
  \begin{align}
    \label{eq:orthogonal_template}
    &B^\orth_\Phi(k_1,k_2,k_3) \;=\; 6\;\fnl^\orth
    \msk&\qquad\times\Bigl\{\;
    -\, 3\,P_\Phi(k_1)\,P_\Phi(k_2) \,-\, 3\,P_\Phi(k_1)\,P_\Phi(k_3) \,-\, 3\,P_\Phi(k_2)\,P_\Phi(k_3)\;
    \allowbreak\nmsk&\qquad
    -\; 8\;\bigl[\;P_\Phi(k_1)\,P_\Phi(k_2)\,P_\Phi(k_3)\;\bigr]^{\nicefrac{2}{3}} \,+\, \text{5 perm.} \;
    \nmsk&\qquad
    +\; 3\;\bigl[\;P_\Phi(k_1)\,P_\Phi(k_2)^2P_\Phi(k_3)^3\;\bigr]^{\nicefrac{1}{3}} \,+\, \text{5 perm.}
    \;\Bigr\} \;, \notag
  \end{align}
  was constructed to be as orthogonal as possible to the local and orthogonal shapes; a few models of inflation are known to produce this shape, one of them being the DBI Galileon inflation \cite{renaux-petel:2011a}.
\end{itemize}
The three shapes of non-Gaussianity have the advantage of being separable in $k_1$, $k_2$ and $k_3\,$, thus allowing the CMB bispectrum to be quickly estimated via \eref{eq:final_linear_angular_bispectrum_k1k2k3_literature} by solving four one-dimensional integrals.

In \SONG, we have implemented the computation of the three primordial templates in the ``bispectrum.c'' module. The module computes the linearly propagated bispectrum of the CMB once the primordial bispectrum function $\,B_\Phi(k_1,k_2,k_3)\,$ is provided.
The non-separable shapes are implemented following the same procedure used for the intrinsic bispectrum, described in \sref{sec:bispectrum_numerical_checks}.
We have used the bispectrum module to produce the Fisher matrices of Ref.~\cite{koyama:2013a}, where we have considered the two non-separable shapes from the DBI Galileon model of inflation; the results we have obtained match with those of the WMAP team \cite{bennett:2012a}, thus confirming our computation.

\section{From the bispectrum to \fnl} 
\label{sec:bispectrum_to_fnl}


The primordial and intrinsic contributions coexist in the observed CMB bispectrum.
To disentangle them and quantify their amplitude requires a detailed knowledge of the expected signals and of their correlation for a given CMB survey.
In this section, we introduce a Fisher matrix approach whereby the elements of the matrix are scalar products between the considered bispectra (local, equilateral, orthogonal, intrinsic) that quantify their overlap on the sky.
In particular, the diagonal elements will represent the potential of the considered CMB survey to measure the single bispectra, while the off-diagonal ones quantify how the presence of the other bispectra might bias such measurement.

Before introducing the Fisher matrix approach, however, we define the observed temperature bispectrum and relate it to the theoretical one for the brightness, which we have derived in \eref{eq:intrinsic_angular_bispectrum_final_formula}.

\subsection{The temperature bispectrum} 
\label{sec:temperature_bispectrum}

In \sref{sec:photon_distribution_function}, we have shown that it is not possible to unambiguously define the temperature in a perturbed Universe, because the perturbations provoke an unbalanced transfer of momentum between photons and baryons that breaks the blackbody spectrum of the photon distribution function.
As a result, one can choose between a number of ``effective'' temperatures, each corresponding to a different moment of the distribution function (\eref{eq:different_temperatures}); while this choice is in general arbitrary, it was shown that the CMB bispectrum is insensitive to it \cite{pitrou:2010b}.

In \SONG, we adopt the commonly used \keyword{bolometric temperature} $T$ \cite{pitrou:2010b}, that is the temperature of the blackbody spectrum with the same energy density as the CMB.
It is related to the brightness perturbation $\Delta$ by
\begin{align}
   \left(\,\frac{T}{\overline{T}}\,\right)^4 \;=\; \frac{\mathcal{I}}{\overline{\mathcal{I}}}
   \qquad\Longrightarrow\qquad \left(\,1\,+\,\Theta\,\right)^4 \;=\; 1\,+\,\Delta \;,
\end{align}
which, up to second order, reads
\begin{align}
  \label{eq:deltas_bolometric_relations_second_order}
  &\Delta \;=\; 4\,\Theta \;+\; 6\,\Theta\,\Theta \;\qquad\text{and}\qquad\;
  \DeltaT\;=\; 4\,\Theta \,-\, 2\,\Theta\,\Theta \;,
\end{align}
where $\,\DeltaT=\Delta-\Delta^2/2\,$ is the variable introduced in \sref{sec:redshift_term_deltatilde} to treat the redshift contribution.

We define the temperature angle-averaged bispectrum as 
\begin{align}
  \label{eq:temperature_bispectrum_definition_alm_theta}
  B_\ldep[\,\Theta\,] \;&\equiv\;
  \avgbig{a_\lmone\,a_\lmtwo\,a_\lmtre} \;
  \threej{\lone}{\ltwo}{\ltre}{m_1}{m_2}{m_3} \;,
\end{align}
where the $a_\lm$'s are the multipoles of the observed CMB temperature map:
\begin{align}
  \Theta(\n) \;=\; \sum\limits_{\lm}\;a_{\lm}\,Y_\lm(\n) \;,
\end{align}
which are conventionally related to the $\Theta_\lm$'s by \eref{eq:alm_vs_thetalm_2}:
\begin{align}
  \label{eq:alm_vs_thetalm_2_in_bispectrum_chapter}
  a_\lm  \;=\; i^{-\L}\,\sqrt{\frac{4\pi}{2\L+1}}\;\Theta_{\lm}\;.
\end{align}
Using the identities in \ref{eq:deltas_bolometric_relations_second_order} we can relate the temperature bispectrum
to the analogous bispectra constructed using the brightness moments $\Delta\,$ and $\DeltaT\,$:
\begin{align}
  \label{eq:bispectrum_theta_to_delta}
  B^\intr_\ldep[\,\Theta\,] \;
  &=\;\hat{B}^\intr_\ldep[\,\Delta\,] \;
  -\; 3\;h_\ldep\;\left(\;C_\lone C_\ltwo\;+\;C_\ltwo C_\ltre\;+\;C_\ltre C_\lone\;\right) \lmsk
  &=\;\hat{B}^\intr_\ldep[\,\DeltaT\,]\;
  +\; h_\ldep\;\left(\;C_\lone C_\ltwo\;+\;C_\ltwo C_\ltre\;+\;C_\ltre C_\lone\;\right) \;,\notag
\end{align}
where $\,h_\ldep\,$ is the purely geometrical factor defined in \eref{eq:gaunt_factor_I}.
The angular power spectrum of temperature fluctuations, $C_\L\,$, is obtained from linear perturbation theory as $\,\avg{a_\lm\,a_\lmp}=(-1)^m\,C_\L\,\delta_{\L\L'}\,\delta_{m-m'}\,$. The rescaled bispectrum $\hat{B}$ is defined as
\begin{align}
  \label{eq:bhat_definition}
  \hat{B}_\ldep\,[\,\Delta\,] \;=\;
  \frac{1}{4^3}\;
  \avgbig{\Delta_\lmone\,\Delta_\lmtwo\,\Delta_\lmtre}\;
  i^{-\lone-\ltwo-\ltre} \;
  \sqrt{\frac{(4\pi)^3}{(2\,\lone+1)(2\,\ltwo+1)(2\,\ltre+1)}}\;,
\end{align}
in order to counter the $4$ coefficients in \eref{eq:deltas_bolometric_relations_second_order} and the $\L$ factors in the definition of the $a_\lm$'s with respect the $\Theta_\lm$'s (\eref{eq:alm_vs_thetalm_2_in_bispectrum_chapter}).
Note that to derive the identities in \eref{eq:bispectrum_theta_to_delta} we have inserted \eref{eq:deltas_bolometric_relations_second_order} into the temperature bispectrum $\avg{a_\lmone\,a_\lmtwo\,a_\lmtre}\,$ and used Wick's theorem to obtain the terms quadratic in the $C_\L$'s.

In principle, the temperature bispectrum can be obtained by either computing $B_\ldep\allowbreak[\,\Delta\,]\,$ or $\,B_\ldep[\,\DeltaT\,]\,$. In practice, as we have explained in \sref{sec:redshift_term_deltatilde}, using the latter is advantageous because the $\,\DeltaT\,$ variable includes by construction the numerically challenging redshift contribution.
Thus, in \SONG we first compute the bispectrum formula in \eref{eq:intrinsic_angular_bispectrum_final_formula}, using the transfer functions for $\,\DeltaT\,$, and then build the temperature bispectrum with the relation in the second line of \eref{eq:bispectrum_theta_to_delta}.


\subsection{The estimator} 
\label{sec:the_estimator}

We quantify the importance of the intrinsic bispectrum by using a Fisher matrix approach. The Fisher matrix element between two temperature bispectra $B^{\,i}$ and $B^{\,j}$ is given by \cite{komatsu:2001a, smith:2011a}
\begin{align}
	\label{eq:estimator}
	\F{\,i}{j} \;=\; f_\text{sky} \sum\limits_{2\leq\lone\leq\ltwo\leq\ltre}^{\lmax} \,
  \frac{B^{\,i}_\ldep\,B^{\,j}_\ldep}
  {\widetilde{C}_\lone \widetilde{C}_\ltwo \widetilde{C}_\ltre\;\Delta_\ldep} \;,
\end{align}
where $\,\widetilde{C}_\L\,$ is the observed spectrum, \ie the signal plus noise, \lmax and $f_\text{sky}$ are, respectively, the maximum angular resolution and fraction of covered sky attainable with the considered CMB survey, and $\Delta_\ldep$ is equal to $1,2,6$ for triangles with no, two or three equal sides.
The bispectrum appearing in the estimator is the angle-averaged one, defined as
\begin{align}
  \avg{a_\lmone\,a_\lmtwo\,a_\lmtre} \;=\;
  \threej{\lone}{\ltwo}{\ltre}{m_1}{m_2}{m_3}\;B_{\ldep} \;.
\end{align}
For the intrinsic bispectrum, this corresponds to the one in \eref{eq:bispectrum_theta_to_delta}.

The observability of a given bispectrum $B$ is quantified by its signal-to-noise: $\,\SN = \sqrt{\F{B}{B}}\,$. If the signal-to-noise is smaller than unity, the considered survey will not be able to distinguish $B$ from the intrinsic variance of the temperature field, which is given by the $C_\L$ product in the denominator of \eref{eq:estimator}.
The amplitude of the primordial templates is parametrised by the \fnl parameter, so that
\begin{align}
  \label{eq:sigma_fnl_fisher}
  \sigma^\sub{B}_{\fnl} \;=\; (\SN)^{-1} \;=\; \frac{1}{\sqrt{\F{B}{B}}}
\end{align}
is the minimum value of $\,\fnl^\sub{B}\,$ that is needed for the survey to be able to detect the bispectrum $B\,$.

Several effects contribute to the bispectrum of the cosmic microwave background and one wants to be able to distinguish them.
For example, a measurement of the primordial signal is subject to a number of contaminants from Galactic emissions (synchrotron, free-free, thermal dust, CO molecular lines), extra-Galactic point sources and cosmological effects such as the ISW-lensing bispectrum \cite{planck-collaboration:2013b}.
\annotate{Foregrounds from Planck XXIV: [foregrounds include] diffuse Galactic emissions (synchrotron, free-free, thermal dust, Anomalous Microwave Emission and CO molecular lines), emission from compact objects (thermal SZ effect, kinetic SZ effect, radio sources, infrared sources, correlated far-infrared background and ultra-compact H ii regions).}
A contaminant $C$ generally induces a bias on the \fnl measurement of a primordial template $T$; if the bispectrum generated by the contaminant is theoretically known, its bias can be quantified using the Fisher matrix as
\begin{align}
  \label{eq:estimator_bias}
  \fnl^\sub{C} \;=\; \frac{\F{C}{T}}{\F{T}{T}} \;.
\end{align}
The bias $\fnl^\sub{C}$ is the amplitude of primordial non-Gaussianity that would be (wrongly) inferred by applying the estimator to the bispectrum produced by the contaminant $C\,$.
We shall use this formula in \sref{sec:results} to quantify the contamination to the primordial signal caused by the intrinsic bispectrum.  

The computation of the estimator, the noise model and the interpolation of the bispectra are implemented in \SONG in a separate module called ``fisher.c''. No assumptions are made in the module on the input bispectra, which can be of any type, \eg template, intrinsic or analytical bispectra. It is, in this respect, a general and flexible tool to produce Fisher matrices and \fnl estimates for any number of bispectra. Furthermore, the experiment parameters (resolution, number of frequency channels, their beam and noise) can be specified via \SONG's input file in a straightforward way.

\subsubsection{Noise model} 
In what follows, we shall assume a Planck-like experiment with homogeneous noise, where the observed CMB spectrum is given by 
\begin{align}
  \widetilde{C}_\L \;\equiv\; C_\L \;+\; N_\L \;.
\end{align}
The noise power spectrum, $\,N_\L\,$, is a combination of the noise from each frequency channel $c\,$:
\begin{align}
  N_\L \;=\; \left[\;\sum\limits_c\,N_{\L,c}^{-1}\;\right]^{-1} \;.
\end{align}
We assume that the noise in the channel $c$ is due to the \keyword{instrument beam}\index{beam for Planck}, taken to be Gaussian and parametrised by $\theta_{\sub{FWHM,c}}\,$, and to the limited sensitivity of the experiment, represented by the variance $\,\sigma^2_c\,$ per pixel of size $\theta_{\sub{FWHM},c}\,$ \cite{pogosian:2005a}:
\begin{align}
  N_{\L,c} \;=\; \left(\,\frac{\sigma_c\;\theta_{\sub{FWHM},c}}{\overline{T}}\,\right)^2
  \exp\left[\;\frac{\L\,(\L+1)\,\theta_{\sub{FWHM},c}^2}{8\ln2}\;\right] \;.
\end{align}

\begin{table}[t]
  \caption[Beam and noise parameters for Planck]{Beam and noise parameters for the frequency channels of Planck where the CMB signal dominates over the foregrounds. The values are taken from the Planck Explanatory Supplement, which can be found at the following URL: \url{http://wiki.cosmos.esa.int/planckpla/index.php/Main_Page}.}
  \label{tab:planck_noise_and_beam}
  \taburulecolor{Blue}
  \arrayrulewidth=0.4mm
  \extrarowsep=1mm
  \begin{tabu} to 0.9\linewidth {X[c] X[c] X[c] X[c] X[c]}
    $\nu$ & $\theta_\sub{FWHM}$ & $\sigma$ & $f_\text{sky}$ & $\lmax$ \\
    \hline
    \unit{100\,}{GHz} & 9.66 & \unit{10.77\,}{$\mu$K} & \multirow{3}{*}{100\%} & \multirow{3}{*}{2500} \\
    \unit{143\,}{GHz} & 7.27 & \unit{6.40\,}{$\mu$K}  & & \\
    \unit{217\,}{GHz} & 5.01 & \unit{12.48\,}{$\mu$K} & &
  \end{tabu}
\end{table}

In our analysis we include the $100$, $143$ and $\unit[217]{GHz}$ frequency channels measured by the HFI instrument on board of Planck, where the CMB signal dominates over the foregrounds. As for the noise and beam parameters, we use those provided by the Planck team, which we report in \tref{tab:planck_noise_and_beam}.
By doing so, we find the following Fisher matrix for the local, equilateral and orthogonal shapes (\sref{sec:templates}):
\begin{align}
  \label{eq:fisher_matrix_templates_planck}
  F \;=\;
  \begin{pmatrix}
    398   &  6.95  &    -28.7    \\
    6.95  &  2.59    &  -0.200  \\
    -28.7  &  -0.200  &   10.1  \\
  \end{pmatrix} \times 10^{-4} \;,
\end{align}
where to compute the transfer functions we have used the best-fit cosmological parameters from Planck (dataset Planck+WP+highL+BAO) \cite{planck-collaboration:2013a}.
The diagonal elements can be converted to uncertainties on the \fnl parameters via \eref{eq:sigma_fnl_fisher},
\begin{align}
  \sigma^\local_{\fnl} \;=\; 5.01 \;, \quad\quad
  \sigma^\equil_{\fnl} \;=\; 62.1 \;, \quad\quad
  \sigma^\orth_{\fnl} \;=\;  31.5 \;,
\end{align}
that are in line with the errors of the Planck experiment \cite{planck-collaboration:2013b} quoted in  \eref{eq:planck_fnl_measurement}.\footnote{More precisely, our uncertainties are about $15\!-\!20\%$ smaller than Planck's. The reason is that the error budget in Planck's analysis includes uncertainties from more subtle effects such as incomplete foreground removal. By setting $f_\text{sky}=0.74\,$ in our Fisher matrix estimator, we obtain a percent-level match.}



\subsubsection{Interpolation strategy} 

The Fisher matrix in \eref{eq:estimator} is given by a sum over all the independent bispectra configurations up to $\lmax\,$. For a typical run where $\lmax=2000\,$, this corresponds to computing the intrinsic bispectrum for almost a billion configurations, a task that would take weeks even on a supercomputer.
The transfer functions, however, are determined by the acoustic oscillations at the time of recombination and thus oscillate with a period of $\,\L=\O(100)$; the bispectrum, which is the correlation of three transfer functions, inherits this property. Therefore, the features of the intrinsic bispectrum can be captured using an \L-sampling with a step of $\L=\O(10)\,$.
In \SONG, we build a grid in \L which starts as logarithmic and, when the logarithmic step exceeds a fixed linear step, continues linearly up to $\,\lmax\,$. In this way, we ensure that the low-\L regions are sampled more finely than the large-\L ones.
Using this inhomogeneous sampling, we build a bidimensional grid in $\lone$ and $\ltwo$ and then choose for each node an $\ltre$-sampling that satisfies the triangular condition, in analogy to what is done for the wavemodes sampling (\sref{sec:sampling_strategies}).

To compute the Fisher matrix, we resort to interpolation.
The main difficulty in interpolating the bispectrum is that it is not defined on a cubic grid. In fact, the triangular condition,
\begin{align}
  |\ell_i-\ell_j| \;\leq\; \ell_k \;\leq\; \ell_i+\ell_j \qquad\text{with $i,j,k=1,2,3$} \;,
\end{align}
results in a mesh for ($\lone,\ltwo,\ltre$) that has the shape of a ``tetrapyd'', the union of two triangular pyramids through the base (see Fig.~2 of Ref.~\cite{fergusson:2012a}).
A simple trilinear method can be used to interpolate the bispectrum, but it is inaccurate near the edges of the tetrapyd as it inherently assumes that the domain is cubic. The problem can be circumvented by deforming the allowed region to a cube via a geometrical transformation and then using trilinear interpolation \cite{fergusson:2009a}. While viable, this approach would force us to discard the points that do not fall in the transformed grid, thus requiring a finer $\L$-sampling.

Rather than relying on a cubic method, we devise a general interpolation technique that is valid on any mesh. 
We first define a correlation length $L$ and divide the tetrapyd domain in boxes of side $L\,$. To compute the interpolation in an arbitrary point $\vec{\L}=\allowbreak(\lone,\ltwo,\ltre)\,$, we consider the values of all the nodes in the box where $\vec\L$ falls and in the adjacent ones. To each node, we assign a weight that is inversely proportional to its distance from $\vec\L\,$.
The problem with this approach is that, the mesh being inhomogeneous, there might be a group of close nodes in one direction that influences the interpolated value in $\vec\L$ much more than a closer point in the opposite direction.
In order to prevent this, we weight down the nodes that have a high local density within a certain distance from them. This mesh interpolation technique relies on two free parameters:
\begin{enumerate}
  \item The correlation length $L$, which sets the size of the local region influencing the interpolation.
  It should correspond roughly to the largest distance of two neighbouring points.
  \item The grouping length, that is the distance below which many close nodes are considered as a single one. It is used to avoid the interpolation being determined by a bunch of close nodes in one direction.
  The grouping length should roughly correspond to the shortest distance between two points.
\end{enumerate}

We have found the optimal values for the logarithmic step, the linear step, the correlation length and the grouping length through extensive convergence tests.
As a result, \SONG can now compute the signal-to-noise of the intrinsic bispectrum at the percent-level accuracy using only $60$ points per $\L$-direction up to $\lmax=2000\,$ (\sref{sec:convergence_tests}).
The mesh interpolation technique is used with success also to compute the Fisher matrix for the separable bispectra such as the local, equilateral and orthogonal templates; as an example, we can compute the signal-to-noise of the equilateral model for a given cosmology with $\sim 1\%$ accuracy in the matter of seconds on a quad-core machine.




\section{Results}
\label{sec:results}

We present results for the intrinsic bispectrum considering three different combinations of line of sight sources. The first considered bispectrum ($B^R$) includes only sources located on the surface of last scattering, that is the $|\dot\kappa|$ sources in \eref{eq:line_of_sight_sources_compact} plus the second-order Sachs-Wolfe effect, $\,4\,|\dot{\kappa}|\,\Psi\,$, which only contributes to the monopole.
The second ($B^{R+Z}$) also includes the redshift term of $\mathcal{Q}^L\,$, that is $\,4\,(n^i\partial_i\,\Psi -\dot{\Phi})\,\Delta\,$. This is computed using $\DeltaT\,$, as discussed in \sref{sec:redshift_term}, and it is the same bispectrum presented in Huang and Vernizzi (2012) \cite{huang:2012a}.
Finally, $B^{R+Z+M}$ consists of the above sources plus all the terms in $\mathcal{M}$ (\eref{eq:definition_M_metric}). One of such terms gives rise to the second-order integrated Sachs-Wolfe effect, or Rees-Sciama effect \cite{rees:1968a, boubekeur:2009a, mollerach:1995a, munshi:1995a}, which is given by $4\,(\dot\Psi+\dot\Phi)\,$. 
The latter bispectrum contains all terms in the Boltzmann equation but the time-delay and lensing contributions (first and third line of \eref{eq:propagation_sources_intensity}, respectively), and is therefore our most complete bispectrum.



\subsection{Scalar modes}
\label{sec:results_scalar}

We compute the contamination $\fnlcon$ induced by the intrinsic bispectra for the three models of primordial non-Gaussianity described in \sref{sec:templates}: local, equilateral and orthogonal.
Our results are shown in \tref{tab:fnl_results_planck}, where we assume a Planck-like experiment with the noise model described in \sref{sec:bispectrum_to_fnl}, and in \tref{tab:fnl_results_ideal}, where we assume an ideal experiment with $\lmax = 2000\,$.
These numbers do not include the non-scalar contributions, that is they have been computed using only the $m=0$ contribution to the sum in \eref{eq:intrinsic_angular_bispectrum_schematic_formula}. Therefore, for the equilateral and orthogonal models, they only represent the dominant contribution to the signal. On the other hand, we expect our local model results to be accurate, as vector and tensor modes are negligible in the squeezed configurations where the local template peaks.

\begin{table}[t]
  \caption[Observability of the intrinsic bispectrum for Planck]{
   Correlations between the primordial templates and the intrinsic bispectra, computed as $\fnlcon = \F{I}{T}/\F{T}{T}$, for a Planck-like experiment characterised by noise parameters in \tref{tab:planck_noise_and_beam}. The signal-to-noise \SN is given by the square root of the autocorrelation.}
  \label{tab:fnl_results_planck}
	\centering
   \begin{tabular}{l || *{3}c || r}
       Model       & ~~~~~~~$B^R$~~~~~~~~ & ~~~~~$B^{R+Z}$~~~~~ & ~~~~$B^{R+Z+M}$~~~~ & ~~~~~ \SN  \\[0.05cm]
       \hline
       \hline
       Local       &         2.3                  & 0.40                & 0.33           & 0.19   \\[0.05cm]
       Equilateral &         6.4                  & 4.2                 & 3.7            & 0.016  \\[0.05cm]
       Orthogonal  &         -4.3                 & -0.80               & -0.82          & 0.031  \\[0.05cm]
			\hline                                                         
			\hline                                                         
			\SN          &         0.57                 & 0.34                & 0.34           & ---    \\[0.05cm]
   \end{tabular}
\end{table}

\begin{figure}[p]
	\centering
	\includegraphics[width=0.7\linewidth]{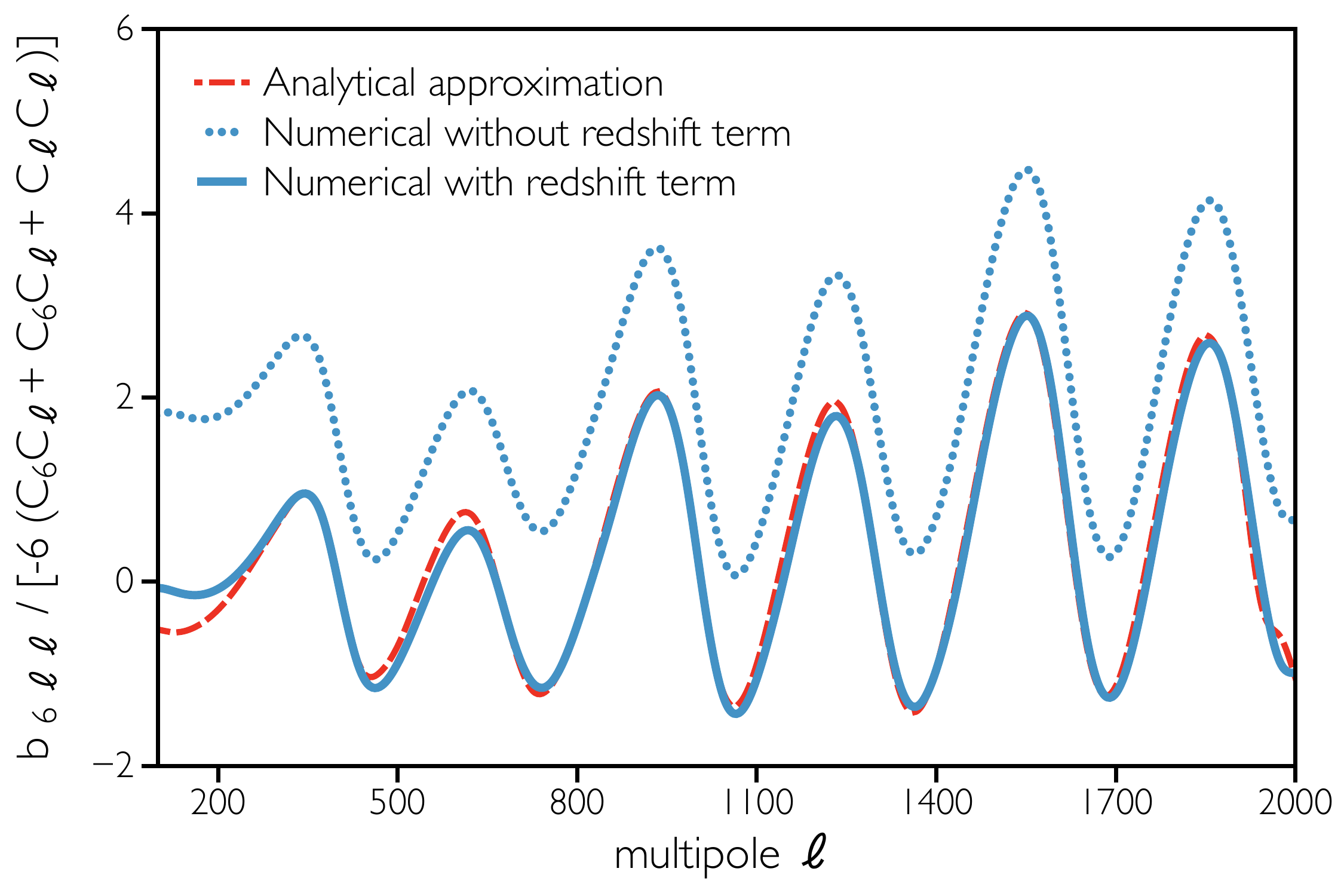}
	\caption[Squeezed limit of the intrinsic bispectrum]{Numerical temperature bispectra $B^R$ and $B^{R+Z}$, together with the squeezed-limit approximation in \eref{eq:squeezed_limit_bispectrum} for a WMAP7 cosmology \cite{komatsu:2011a}, where $\lone=6$ and $\ltwo=\ltre=\ell\,$.  We normalise the curves with respect to the ultra-squeezed limit for a local-type bispectrum with $\fnl^\Phi=1$ \cite{gangui:1994a, komatsu:2001a}, so that the primordial curve would appear as a constant horizontal line with amplitude close to unity.
  Plot taken from \citet{pettinari:2013a}, page 10. \textcopyright\xspace SISSA Medialab Srl. Reproduced by permission of IOP Publishing. All rights reserved.
	}
	\label{fig:squeezed_limit_bispectrum}
  \vspace{1.5cm}
	\includegraphics[width=0.7\linewidth]{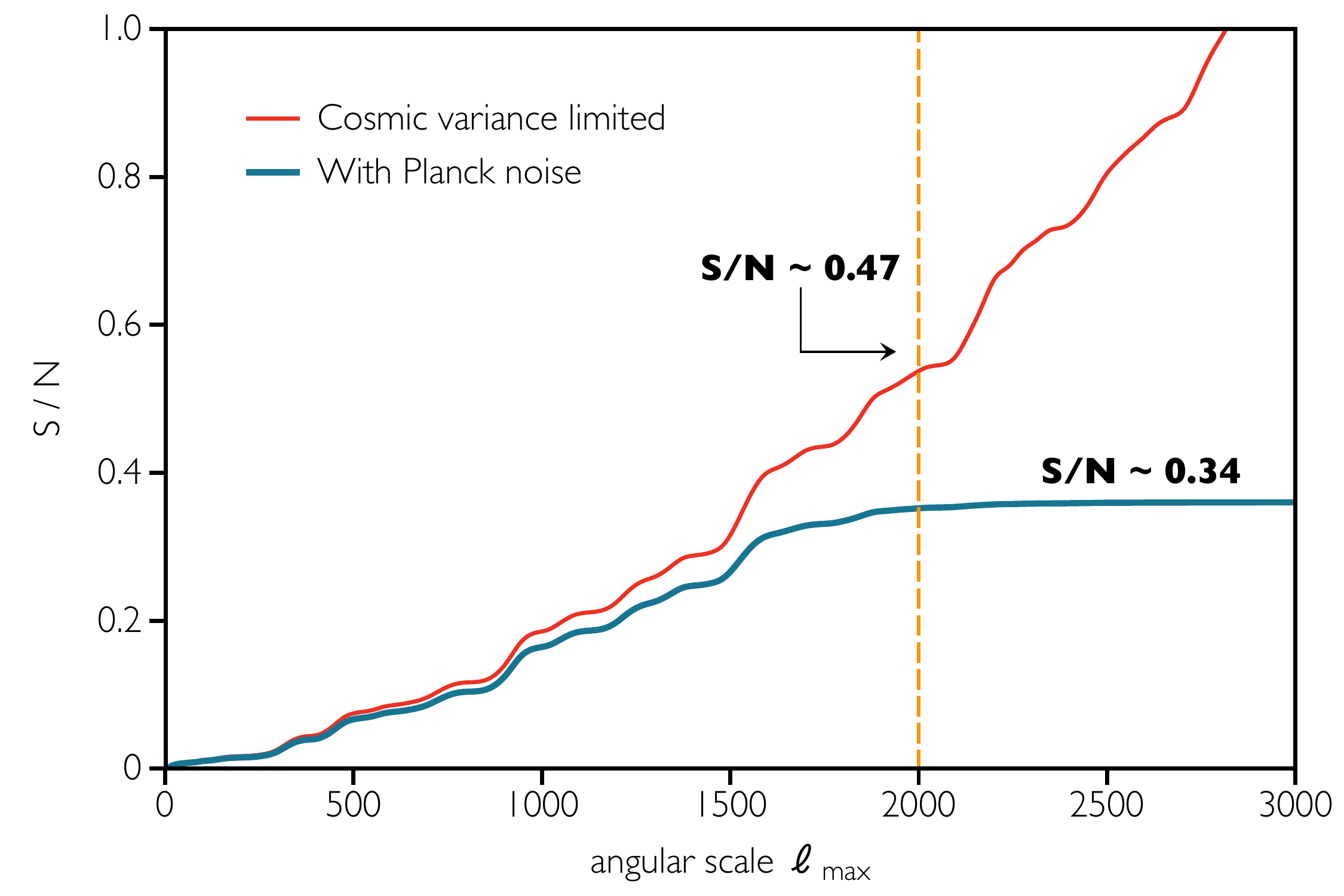}
	\caption[\SN of the intrinsic bispectrum]{Signal-to-noise ratio of the $B^{R+Z+M}$ bispectrum, which includes all effects apart from time-delay and lensing. The S/N saturates at $\sim0.34$ for $\L>2000\,$. A cosmic-variance limited experiment with a resolution of $\lmax=2000$ would yield $\SN\simeq0.47$; for the same ideal experiment, the $S/N$ reaches unity only at $\lmax\simeq3000$.}
	\label{fig:signal_to_noise}
\end{figure}

The most striking feature of Tables \ref{tab:fnl_results_planck} and \ref{tab:fnl_results_ideal} is the difference between the $B^R$ and $B^{R+Z}$ bispectra, with the former yielding a larger \fnl contamination.
This effect is clear also from \fref{fig:squeezed_limit_bispectrum}, where we plot $B^R$ and $B^{R+Z}$ for a squeezed configuration. The recombination-only curve exhibits a positive offset with respect to the integrated one showing the importance of the integrated effects which include $\Delta^{(1)}\,$.
On the other hand, the time-integrated effects given by the metric affect $\fnlcon$ only marginally, and do not seem to affect the signal-to-noise. This can be seen by comparing the $B^{R+Z}$ and $B^{R+Z+M}$ columns of \tref{tab:fnl_results_ideal}.

The last column of \tref{tab:fnl_results_ideal} can be computed by using a first-order Boltzmann code. Our value of $\SN = 0.24$ for the local-template agrees with the one obtained using the first-order code CAMB \cite{lewis:2000a} and with Ref.~\cite{komatsu:2001a}.

\begin{table}[t]
  \caption[Observability of the intrinsic bispectrum for an $\lmax=2000$ experiment.]{The same as \tref{tab:fnl_results_ideal}, but considering a cosmic variance limited CMB survey with $\lmax=2000\,$.}
  \label{tab:fnl_results_ideal}
	\centering
   \begin{tabular}{l || *{3}c || r}
       Model       & ~~~~~~~$B^R$~~~~~~~~ & ~~~~~$B^{R+Z}$~~~~~ & ~~~~$B^{R+Z+M}$~~~~ & ~~~~~ $\SN$  \\[0.05cm]
       \hline
       \hline
       Local       &         2.5                 & 0.58                & 0.51              & 0.24   \\[0.05cm]
       Equilateral &         6.7                 & 4.7                 & 4.2               & 0.018  \\[0.05cm]
       Orthogonal  &         -5.1                & -1.38               & -1.35             & 0.035  \\[0.05cm]
			\hline                                                                         
			\hline                                                                         
			\SN          &         0.77                & 0.47                & 0.47              & ---    \\[0.05cm]
   \end{tabular}
\end{table}

In \fref{fig:signal_to_noise}, we show the signal-to-noise ratio of the $B^{R+Z+M}$ bispectrum as a function of \lmax, which is the angular resolution of the considered experiment.
We find that, adopting the noise model of a Planck-like experiment, the signal to noise saturates at $\SN\simeq0.34\,$. For an ideal experiment which is limited only by cosmic variance, the signal-to-noise ratio reaches unity only for $\lmax\simeq3000$.

\subsubsection{Reproducing Pitrou's results} Pitrou et al.~(2010) \cite{pitrou:2010a} found $\fnlcon \sim 5$ and $\SN(\lmax=2000) \sim 1$ by using the Boltzmann code \emph{CMBquick} \cite{pitrou:2011a} and assuming a cosmic variance limited experiment.
In that code, the bispectrum was computed by including all line of sight sources in \eref{eq:line_of_sight_sources_compact}, including lensing and time-delay, and integrating them until shortly after recombination. This is perfectly achievable since lensing and time-delay pose numerical problems only at later times, when small-scale multipoles get excited. However, the choice of the cutoff time is arbitrary as the time-integrated effects are important throughout cosmic evolution.

We ran \SONG with the same parameters and cutoff time as \emph{CMBquick}, and we obtained similar values: $\fnlcon = 3.7$ and $\SN(\lmax=2000) = 1.1$. As pointed out in \sref{sec:convergence_tests}, the remaining discrepancy might be due to a lack of numerical convergence in \emph{CMBquick}. Furthermore, the most recent version of \emph{CMBquick} yields a value of $\fnlcon \sim 3$ which is more in line with what we find\footnote{Cyril Pitrou, private communication (2013).}.

\subsection{Non-scalar modes}
\label{sec:results_non_scalar_modes}

The results that we have discussed above were published in \citet{pettinari:2013a}. Since then, we have updated \SONG to implement the $m\neq0$ modes and produced the intrinsic bispectrum including the vector and tensor modes. That is, we have computed the bispectrum formula in \eref{eq:intrinsic_angular_bispectrum_final_formula} considering the elements of the azimuthal sum from $-2$ to $+2\,$.
Before showing our results, let us remark that we have not yet performed extensive convergence tests on the non-scalar modes; we cannot therefore guarantee their accuracy to more than the $10\%$ level.

The Fisher matrix that we obtain when we include the scalar, vector and tensor modes considering a cosmic variance limited experiment with $\lmax=2000$ is:
\begin{align}
  \label{eq:fisher_matrix_vector_tensor}
  F \;=\;
  \begin{pmatrix}
    614    \,(590) &  8.98    \,(8.98) &    -39.8  \,(-39.4) &    267    \,(299)   \\
    8.98   \,(8.98) &  3.18    \,(3.18) &  -0.44   \,(-0.45) &   13.9     \,(13.5)   \\
    -39.8  \,(-39.4) &  -0.44  \,(-0.45) &   12.6    \,(12.5) &   -6.84    \,(-16.9)  \\
    267  \,(299) &     13.9  \,(13.5) &   -6.84    \,(-16.9) &   2530    \,(2170)  \\
  \end{pmatrix} \times 10^{-4} \;.
\end{align}
The ordering of the rows and columns is local, equilateral, orthogonal and intrinsic. The values in parentheses correspond to the scalar contribution to the intrinsic bispectrum\footnote{Note that the inclusion of the non-scalar modes should not affect the \SN of the primordial templates, because we assume that the vector and tensor modes vanish at first order. However, we can see from  the Fisher matrix in \eref{eq:fisher_matrix_vector_tensor} that there are differences of the order $5\%$ for the local template. The reason for this discrepancy is purely numerical: in order to compute the intrinsic bispectrum for the $m\neq0$ modes we have adopted a different $\L$-grid that contains only configurations where $\lone+\ltwo+\ltre$ is even, as the bispectrum formula (\eref{eq:intrinsic_angular_bispectrum_final_formula}) vanishes otherwise. The local template is the most affected one by this slightly worse grid because it is very peaked for squeezed configurations.}.
The Fisher matrix elements translate to a signal-to-noise ratio of the intrinsic bispectrum of $\,\SN=0.50\,\,(0.47)$ and to biases on the primordial measurements of
\begin{align}
  \fnl^\local \;=\; 0.44 \,(0.51) \;, \quad\quad
  \fnl^\equil \;=\; 4.4 \,(4.2) \;, \quad\quad
  \fnl^\orth \;=\;  -0.54 \,(-1.35) \;.
\end{align}

Neither the signal-to-noise nor the bias to the primordial signal are significantly affected by the inclusion of the vector and tensor modes, with the exception of $\fnl^\orth$ which is small in both cases.
In principle, we should include in our analysis also the $|m|>2$ modes; however, we do not expect them to make a difference because they correspond to multipoles that are tight-coupling suppressed during recombination.




\section{Robustness of \SONG's bispectra}
\label{sec:bispectrum_numerical_checks}

The computation of the intrinsic bispectrum via \eref{eq:intrinsic_angular_bispectrum_final_formula} involves estimating a four-dimensional integral over six oscillatory functions;
one of them is the second-order transfer function, which is obtained by solving a large differential system (\sref{sec:diff_system}) and an oscillating integration (\sref{sec:line_of_sight}) for $\sim10^6$ configurations of the wavemodes.
The resulting bispectrum is then summed over $\sim10^9$ multipoles using a novel interpolation method to obtain the Fisher matrix (\sref{sec:the_estimator}).

\SONG implements all these steps in an efficient way, so that a Fisher matrix for a given cosmological model is produced to $5\%$ precision in about $4$ CPU-hours.
The point, however, is not only speed but accuracy: how can we trust \SONG's results after so much numerical processing? To answer the question, we have run several tests on \SONG's final products, that is the intrinsic bispectrum and its signal-to-noise ratio; these numerical and analytical checks are complementary to those involving the differential system, which we have discussed in \sref{sec:evolution_checks_of_robustness}.

\subsection{Convergence tests}
\label{sec:convergence_tests}

\begin{figure}[t]
	\centering
		\includegraphics[width=1\linewidth]{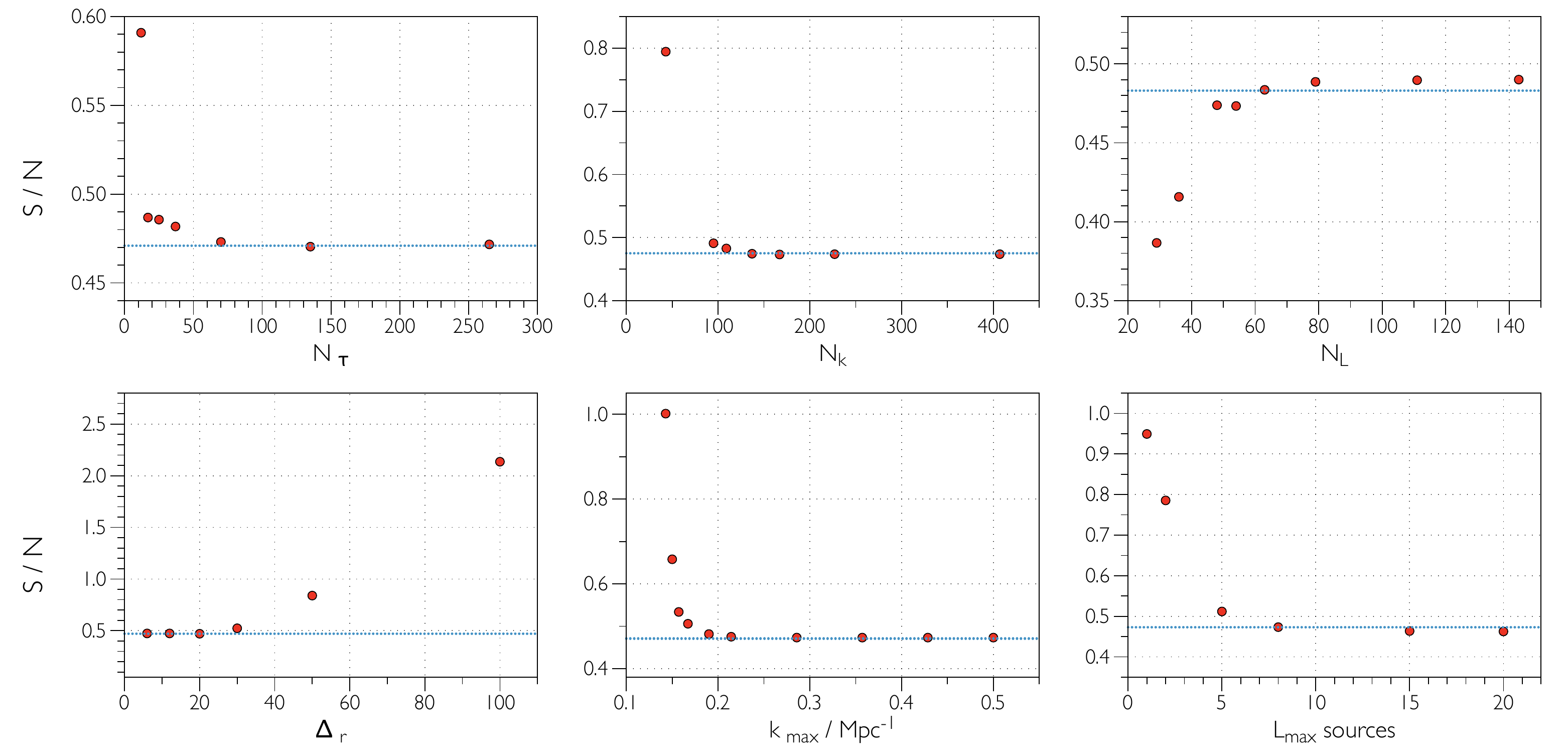}
	\caption[Convergence of the \SN of the intrinsic bispectrum]{Convergence of the signal-to-noise ratio for $B^{R+Z+M}$, our most complete bispectrum, for a cosmic variance limited experiment with $\lmax=2000\,$.
  The horizontal blue line in each panel represents the value obtained using the default parameters (\ie the typical run of \SONG).
  Refer to the text for details on the tested parameters.
  Plot taken from \citet{pettinari:2013a}, page 14. \textcopyright\xspace SISSA Medialab Srl. Reproduced by permission of IOP Publishing. All rights reserved.}
	\label{fig:convergence_tests}
\end{figure}

We have checked the numerical robustness of our bispectrum results by varying the most relevant numerical parameters in \SONG:
\begin{itemize}
	\item $N_\tau\!\,$, number of sampling points in conformal time for the line of sight sources (\sref{sec:sampling_strategies}).
	\item $N_k\!\,$, number of sampling points per direction of $k$-space $(k_1,k_1,k_3)$ for the transfer functions (\sref{sec:sampling_strategies}).
	\item $N_L\!\,$, number of sampling points per direction of $\L$-space $(\lone,\ltwo,\ltre)$ for the bispectrum (\sref{sec:the_estimator}).
	\item $\Delta_r\!\,$, step size of the $r$-grid in the bispectrum integrals in \eref{eq:intrinsic_angular_bispectrum_final_formula} and \ref{eq:final_linear_angular_bispectrum_k1k2k3_literature}.
	\item $k_\text{max}\!\,$, maximum value of $k$ for which we compute the transfer functions (\sref{sec:sampling_strategies}).
	\item $\Lmax\!\,$, highest multipole source considered in the line of sight integral in \eref{eq:los_integral_intensity}.
\end{itemize}
In \fref{fig:convergence_tests}, we show how quickly the signal-to-noise of the intrinsic bispectrum converges for all the tested parameters. (Note that the convergence of $\fnlcon = \F{B}{T}/\F{T}{T}$ is even faster than the convergence of $\SN=\sqrt{\F{B}{B}}$ as numerical errors tend to cancel when taking ratios.)

We find that the signal strongly depends on the number of multipoles included in the line of sight integration, $\,\Lmax\,$, as shown in the bottom-right panel of \fref{fig:convergence_tests}. While at first order there are no line of sight sources higher than the quadrupole (\eref{eq:gamma_coupling}), at second order the sum $\,J_{L\L m}\,S_{Lm}\,$ has to be cut at a suitable \Lmax ~-- see \eref{eq:los_integral_intensity} and the discussion in \sref{sec:los_source_function}. We obtain a convergence only for $\Lmax>8\,$, with lower values yielding a larger signal. This behaviour might partly explain the large value of \fnlcon found by Pitrou et al.~(2010) \cite{pitrou:2010a}, who used $\Lmax=4\,$.

As illustrated in \sref{sec:the_estimator}, we compute the Fisher matrix elements in \eref{eq:estimator} by interpolating the bispectra on a mesh. The top-right panel of \fref{fig:convergence_tests} shows how our interpolation technique yields percent-level precision with just $60$ points out of $2000$ in each $\ell$-direction. We also tested the interpolation against known results, such as the signal-to-noise of the local model, and obtained the same level of agreement.

\subsection{Squeezed limit}
\label{sec:squeezed_bispectrum_analytical}

For squeezed triangles, where the small-$k$ side is within the horizon today but was not at recombination, the intrinsic bispectrum is known  approximately \cite{creminelli:2004a}. In this configuration, the long-wavelength mode acts as a perturbation of the background that alters the observed angular scale of the short wavelength modes. The reduced bispectrum for the bolometric temperature then takes the following form \cite{lewis:2012a,bartolo:2012a,creminelli:2011a}:
\begin{align}
	\label{eq:squeezed_limit_bispectrum}
	b_{\lone \ltwo \ltre} [\,\Theta\,] \;&=\; 
    C_{\lone}C_{\ltwo} \,+\, C_{\lone}C_{\ltre} \,+\, C_{\ltwo}C_{\ltre}
    \msk&
    -\, C_{\lone}^{T\zeta} \;\frac{1}{2}\;\left(%
    C_{\ltwo} \frac{\dd\,\text{ln}\,(\ltwo^2\,C_{\ltwo})}{\dd\,\text{ln}\,\ltwo} \,+\,%
    C_{\ltre} \frac{\dd\,\text{ln}\,(\ltre^2\,C_{\ltre})}{\dd\,\text{ln}\,\ltre}%
    \right) \;, \notag
\end{align}
where $\,C_{\lone}^{T\zeta}\,$ is the correlation between the photon temperature and the super-horizon curvature perturbation $\,\zeta = \Delta/4 - \Phi\,$ at first order, and $\lone$ is the long-wavelength mode. The derivative term encodes the shift in the observed angular scales, known as Ricci focussing, while the first three terms represent the smaller effect due to anisotropic redshifting, known as redshift modulation \cite{lewis:2012a}. A quick comparison with \eref{eq:bispectrum_theta_to_delta} shows that the bispectrum induced by Ricci focussing corresponds to the bispectrum of $\DeltaT\,$.


In \fref{fig:squeezed_limit_bispectrum} we show two temperature bispectra obtained with \SONG compared to the analytical approximation for a squeezed configuration where the large-scale mode is fixed. The bispectrum computed using $\DeltaT$ (labelled $B^{R+Z}$ in \sref{sec:results}), which includes both the scattering sources and the time-integrated effect arising from the redshift term, matches the analytical curve to a precision of a few percent. On the other hand, the bispectrum computed using the standard brightness $\Delta\,$ (labelled $B^{R}$ in \sref{sec:results}), which does not include the redshift term, presents a nearly constant positive offset with respect to the analytical approximation.

\subsection{Local limit}
\label{sec:local_limit}

In \SONG, the initial conditions for the non-linear transfer functions are set using the gauge-invariant perturbation $\zeta\,$, as discussed in \sref{sec:initial_non_gaussianity}. Therefore, one can recreate any kind of initial non-Gaussianity by choosing an appropriate initial value for $\pert{T}{2}_\zeta(k_1,k_2,k_3)$.
If we choose for $\pert{T}{2}_\zeta(k_1,k_2,k_3)$ a local shape with a non-vanishing value for $\fnl$, and run \SONG with the quadratic sources deactivated, we expect to obtain an intrinsic bispectrum that perfectly matches the local template with an amplitude of $\fnl\,$; we call this the \emph{local limit}\index{local limit of the intrinsic bispectrum}.
This happens because deactivating the quadratic sources in the second-order Boltzmann-Einstein system is equivalent to solving the linear system, so that the resulting intrinsic bispectrum corresponds to the linearly propagated one.

\begin{figure}[t]
	\centering
		\includegraphics[width=1\linewidth]{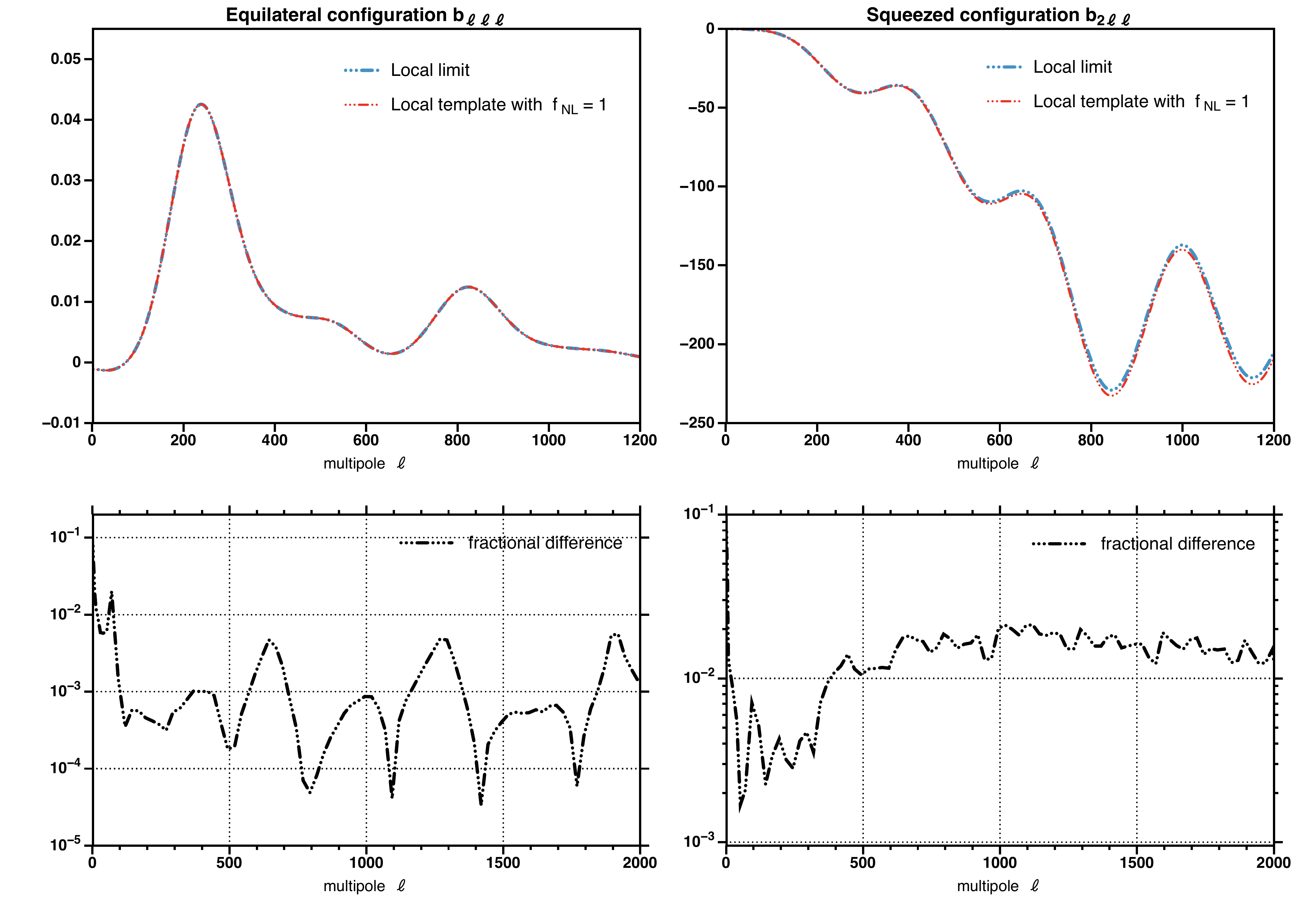}
	\caption[Local limit of the intrinsic bispectrum]{Local limit of the intrinsic bispectrum for an equilateral (left panels) and a squeezed (right panels) configuration. The red curve is the local template with $\fnl=1\,$; the blue curve is the intrinsic bispectrum with an equal amount of local NG and with the quadratic sources deactivated. The match between the two bispectra, which are shown multiplied by a factor $10^{16}\,\L^2\,(\L+1)^2/(2\,\pi)^2$ \cite{komatsu:2002a}, is always at the percent level or better.}
	\label{fig:local_limit}
\end{figure}

In \fref{fig:local_limit} we show that, for a typical run of \SONG, the intrinsic bispectrum in the local limit matches the linearly propagated bispectrum of the local template with percent-level accuracy.
By applying the $\fnl$ estimator (\eref{eq:estimator}) on the local-limit intrinsic bispectrum with $\fnl=1$, we recover $\fnl=1$ to $10^{-3}$ accuracy. (Note that the match in $\fnl$ is better than the one in the bispectrum because the former is obtained as a sum over all the bispectra configurations, which cancels the random error.) 
This is an important test on \SONG's implementation of the differential system, on the way the transfer functions are computed, on the bispectrum formula in \eref{eq:intrinsic_angular_bispectrum_final_formula} for the $m=0$ case, on the Fisher module and, in particular, on the mesh interpolation technique that we have discussed in \sref{sec:bispectrum_to_fnl}.


\chapterbib


\chapter{Conclusions}
\label{ch:conclusions}

\section{The intrinsic bispectrum}

In this thesis we have presented results from a new, efficient numerical code, \SONG, designed to calculate the cosmic microwave background anisotropies up to second order.  We have exploited it to find the temperature bispectrum which arises even for purely Gaussian initial density perturbations. This intrinsic non-Gaussianity will necessarily bias attempts to estimate different types of primordial non-Gaussianity from the CMB bispectrum. 
The efficiency of \SONG has allowed us to demonstrate convergence of our results with respect to several different numerical parameters. We have also demonstrated percent-level agreement with analytical estimates in the squeezed limit, and we believe our answers are robust. 

The contamination from the intrinsic bispectrum generated by the second-order Einstein-Boltzmann equations generally leads to a small bias in the estimates of non-Gaussianity, which is good news for the prospect of using CMB data to probe primordial non-Gaussianity.  
While the precise answer depends on the terms included, the biases for local templates of non-Gaussianity are below the level of primordial \fnl detectable by the Planck satellite. The biases from the intrinsic bispectrum for other primordial templates, equilateral and orthogonal, also appear to be small. (These results are summarised in \tref{tab:fnl_results_planck}.)
The intrinsic non-Gaussianity can be searched for directly, using the predicted signal as a template; our calculations suggest this signal is just beyond what is possible with Planck, with a signal-to-noise rising to unity only for $\lmax=3000$ (\fref{fig:signal_to_noise}.) 

In comparing to recent calculations, we find good agreement with the results of Huang and Vernizzi \cite{huang:2012a} when we include the integrated redshift term with the recombination contribution.  The signal-to-noise for the intrinsic signal matches well, while our bias to \fnlcon $ \simeq 0.5$ is slightly different, which appears to be due to differences in the implementation of the local template. 
Excluding the integrated redshift term yields a significantly higher answer, with \fnlcon $= 2.5$.  This is much more similar to the results of Pitrou et al \cite{pitrou:2010a}, which  focussed on the contributions on the recombination surface alone.  We have also found that the number of multipole sources in the line of sight  integral required for numerical convergence is $\Lmax\geq8$, and we find larger values of \fnlcon are obtained for $\Lmax=4$ as used in Ref.~\cite{pitrou:2010a}. Su et al.~\cite{su:2012a} find similar numerical values to Huang and Vernizzi \cite{huang:2012a} for the bias, but disagree on the signal-to-noise of the intrinsic signal.  We are unable to directly compare our numerical results with theirs, since they use integration by parts which leads to different line of sight source terms.

We have shown how the redshift terms along the line of sight  lead to a change in the value of the local-type \fnlcon bias of approximately 2. We interpret this as the evidence that effects which are not at recombination are important, and should be all included in order to obtain a complete result. We plan to  further develop our numerical code to include the time-delay and lensing contributions. The time-delay effect was studied in Ref.~\cite{hu:2001a} and is expected to be small. The lensing term, on the other hand, is known to strongly correlate with the linear integrated Sachs-Wolfe effect and thus yields a strong squeezed signal that contaminates the local measurement of Planck with a bias of $\fnlcon \sim 7$ \cite{lewis:2012a, hanson:2009a, smith:2011a, serra:2008a, lewis:2011a, lewis:2006a}.

We have calculated the intrinsic bispectrum from the scalar ($m=0$), vector ($m=\pm1$) and tensor ($m=\pm2$) modes, neglecting higher moments. This should give a reliable estimate of local-type \fnl since higher moments are suppressed for squeezed configurations.
We expect also the prediction on the signal-to-noise of the intrinsic bispectrum and on the bias on the equilateral and orthogonal templates to be robust. In fact, the higher moments that we are neglecting only exist for the multipoles with $\L\geq2\,$, which are suppressed by tight coupling during recombination.

\section{Current and future research}
\label{sec:future_prospects}

As we have seen in \cref{ch:perturbation_theory} and \ref{ch:evolution}, several non-linear effects in addition to the intrinsic bispectrum arise at second order that can be computed by \SONG.
In the following we give a brief outlook of these effects and, in general, of possible applications of \SONG.

\runinhead{B polarisation}\index{B polarisation}
Measuring the tensor-to-scalar ratio parameter, $\,r\,$, would shed light on the physics of the early Universe and provide an indirect detection of gravitational waves (\sref{sec:primordial_fluctuations}).
The $B$ polarisation of the cosmic microwave background is sourced by the tensor part of the metric and is therefore a promising probe for measuring $r\,$ \cite{kamionkowski:1997a, seljak:1997a}.
The $B$ polarisation, however, is also generated by the non-linear dynamics either via the conversion from $E$ to $B$-modes due to the propagation of light in an inhomogeneous Universe (either from lensing \cite{zaldarriaga:1998c, lewis:2006a} or time-delay terms \cite{hu:2001a}), by the vector and tensor modes in the metric \cite{mollerach:2004a} or by kinematic effects in the scattering term \cite{beneke:2011a}.
These effects are clearly recognisable in the second-order Boltzmann equation, as discussed in \sref{sec:final_Boltzmann_equation}.
We have implemented in \SONG a module to quantify the contribution to the power spectrum of the $B$-modes, $\,C_{\L}^{BB}\,$, induced by the second-order metric, scattering and propagation effects, excluding time-delay and lensing.
Our analysis \cite{fidler:2014a} indicates that these intrinsic $B$-modes from non-linear dynamics are comparable to a primordial signal of order $r\sim10^{-7}\,$ and, therefore, will not bias future CMB survey such as LiteBIRD \cite{hazumi:2012a,matsumura:2013a}, PIXIE
\cite{kogut:2011a} and Prism \cite{prism-collaboration:2013b}.


\runinhead{Spectral distortions}\index{spectral distortions}
When they collide through Compton scattering during recombination and reionisation, photons and electrons exchange a tiny amount of energy (\sref{sec:collision_energy_transfer}) that, at second order, needs to be taken into account.
This introduces a momentum dependence in the the CMB collision term that ultimately spoils its blackbody spectrum (see \sref{sec:photon_distribution_function} and \ref{sec:collision_contributions}).
This spectral distortion can be characterised using the Compton $y$ parameter \cite{pitrou:2010b} and has a signature similar to that of the thermal Sunyaev-Zeldovich effect \cite{sunyaev:1970a}.
The evolution of $y$ obeys the Boltzmann equation and is computed by solving an extra hierarchy that is sourced by the difference between the photon and electron velocities \cite{pitrou:2010b}.
Because the electrons' velocity grows after recombination ($v_b\propto k\tau$) and the photons' does not, the largest contribution to this type of spectral distortion comes from the time of reionisation.
Using \SONG, we have computed these spectral distortions both for temperature and polarisation and studied their dependence on the details of reionisation \cite{renaux-petel:2013a}. This is of interest in view of the proposed experiments Prism \cite{prism-collaboration:2013a} and Pixie \cite{kogut:2011a}, which are expected to measure the CMB frequency spectrum with unprecedented accuracy.
\annotate{From \cite{pitrou:2010b}: The spectral distortion arises mostly at the last scattering surface (LSS) and during reionization, from the electron flows irrespectively of their thermal velocity dispersion. We are thus dealing with a non-linear kinetic SZ effect, physically different from a thermal SZ effect. However, as we shall see, it has the same spectral signature.}


\runinhead{Magneto-genesis}\index{magnetic fields}
At second order, the electron and photon velocities are vortical even in the absence of primordial vector fluctuations.
During recombination, when the tight coupling between the two fluids breaks down, this vorticity translates into currents that unavoidably source a magnetic field.
The amplitude of this intrinsic magnetic field has been estimated in several limits and with varying accuracy; see for example Refs.~\cite{fenu:2011a, nalson:2014a, ichiki:2007a, saga:2015a}. 
In Ref.~\cite{fidler:2015a} we have numerically verified and extended these results down to cluster scales with a significantly higher degree of accuracy, by implementing in \SONG the cosmological Maxwell equations.
In particular we have solved a tension in the literature by confirming the $k^{3.5}$ slope on large scales of the magnetic field power spectrum.

\runinhead{Modified gravity}\index{modified gravity}
\citet{gao:2011a} has recently studied the dependence of the intrinsic bispectrum of the CMB on the theory of gravity.
By assuming an $f(R)$ model and considering only the Sachs-Wolfe effect, the author found that the intrinsic bispectrum depends strongly on the non-linear structure of the $f(R)$ function. In particular, he found that the existence of the second, third or fourth derivatives in $f(R)$ results in a bispectrum larger than the one produced for standard general relativity.
It would be interesting to explore this dependence in detail in view of constraining the $f(R)$ models using the observed CMB bispectrum.
We plan to do so by implementing an appropriate parametrisation of modified gravity into \SONG.

\vspace{0.5cm}

As mentioned in the preface, since I obtained my Ph. D. in 2013, my collaborators and I have carried out further research on the non-linearities of the CMB, extending the work in my thesis. In particular, we have found the polarised intrinsic bispectrum to be strongly enhanced with respect to the temperature one \cite{pettinari:2014b}; developed a formalism to treat all propagation effects, including lensing, at second order \cite{fidler:2014b}; computed the power spectrum of the second-order B-modes \cite{fidler:2014a}; quantified the intrinsic spectral distortions in the CMB \cite{renaux-petel:2013a}; provided the most precise numerical computation of the intrinsic magnetic field generated around and after recombination \cite{fidler:2015a}. These works can be freely accessed as preprints at this link: \url{http://arxiv.org/find/astro-ph/1/au:+Pettinari_G/0/1/0/all/0/1}. Furthermore, the code \SONG is available in the open-source format on the website \url{https://github.com/coccoinomane/song}.

\chapterbib

\appendix

\chapter{Projection on the sphere}
\label{app:sphere_projection}

In this Appendix we shall discuss how to treat the directional dependence in the Boltzmann and Einstein equations, in view of solving them numerically. The topic is also treated in Sec.~IIIB of \citet{beneke:2010a}, in Sec.~8.2 of \citet{pitrou:2009a} and in Sec.~C of \citet{hu:1997b}.


To characterise the angular dependence of the equations, we adopt a spherical coordinate system where the direction of propagation of a particle, \n, is parametrised by a polar angle $\theta$ (or colatitude) and an azimuthal angle $\phi$ (or longitude).
The polar angle is defined with respect to an arbitrary axis, the \keyword{zenith} or \keyword{polar axis}, and has the range $0\leq\theta\leq\pi$, the value $\pi/2$ corresponding to the equator. The azimuthal angle is the direction of $\n$ projected to the plane perpendicular to the zenith, and has the range $0\leq\phi<2\pi$, with the $y$ axis at $\phi=\pi/2$.
In a Cartesian coordinate system where the $z$-axis is aligned with the zenith, the coordinates of the particle's direction \n are given by:
\begin{align}
  &n_x \,=\, \sin\theta\,\cos\phi \;, \notag\\
  &n_y \,=\, \sin\theta\,\sin\phi \;, \notag\\
  &n_z \,=\, \cos\theta \;.
\end{align}

We expand the angular dependence of the distribution function, $f(\n)$, in spherical harmonics\index{spherical harmonics decomposition},
\begin{align}
  f(\n) \;=\; \sum\limits_{\L=0}^\infty\,\sum\limits_{m=-\L}^{\L}\,(-i)^\L\,\sqrt{\frac{4\pi}{2\L+1}}
  \,f_{\lm}\,Y_\lm(\n) \;.
  \label{eq:spherical_harmonics_expansion}
\end{align}
The coefficient $f_\lm$ are called the \keyword{multipoles} of $f$ and do not depend on the direction \n. The spherical harmonics $Y_{\lm}$ are defined as
\begin{align}
  Y_\lm(\theta,\phi) \;=\;
  \sqrt{\frac{2\L+1}{4\pi}\,\frac{(l-m)!}{(l+m)!}}\;\,P_\lm(\cos\theta)\;e^{i\,m\,\phi} \;,
  \label{eq:spherical_harmonics_definition}
\end{align}
where the $P_\lm$ are the associated Legendre polynomials of degree \L and order $m$ \cite{abramowitz:1977a}.
Note that we shall follow the literature and define two multipole expansions for the temperature perturbation of the CMB, $\,\Theta=(T-\overline{T})/\overline{T}\,$:
\begin{align}
  \Theta(\n) \;
  =\; \sum\limits_{\lm}\;a_{\lm}\,Y_\lm(\n) \;
  =\; \sum\limits_{\lm}\;(-i)^\L\,\sqrt{\frac{4\pi}{2\L+1}}\,\Theta_{\lm}\,Y_\lm(\n) \;.
  \label{eq:alm_vs_thetalm_1}
\end{align}
The $a_\lm$ are used to define the observables, such as the power spectrum $\,\avg{a_\lm,a_\lmp}\,$ and the bispectrum $\,\avg{a_\lmone,a_\lmtwo,a_\lmtwo}\,$; they are related to the $\Theta_\lm$ by
\begin{align}
  \Theta_{\lm} \;=\; i^\L\,\sqrt{\frac{2\L+1}{4\pi}}\;a_\lm\;.
  \label{eq:alm_vs_thetalm_2}
\end{align}
The extra coefficients in the definition of $\Theta_\lm$ and $f_\lm$ serve the purpose of simplifying the Boltzmann equation, and is a convention normally adopted in the literature.

The Legendre polynomials oscillate in the $\theta$ direction with a wavelength that is roughly inversely proportional to \L:
\begin{align}
  \lambda\,\sim\,2\pi/\L \;.
\end{align}
For example, for $\L=180$ the peaks of $P_{\L0}$ are separated by about $2\deg$.
Therefore, the multipole $f_\lm$ quantifies the autocorrelation of $f$ on angular scales $\sim2\pi/\L$; the larger \L is, the smaller are the scales being considered. For this reason, we shall often refer to $\L$ as the angular scale.

The \keyword{azimuthal mode} $m$ influences the $Y_\lm$ in two ways. First, it enters the associated Legendre polynomials as
\begin{align}
  P_\lm(\cos\theta) \;\propto\; (\sin\theta)^{\abs{m}} \;,
\end{align}
thus penalising $Y_\lm(\theta,\phi)$ for directions that are too close to the zenith ($\theta=0$).
Increasing $m$ makes $Y_\lm$ smaller at the zenith and larger at the equator; every spherical harmonics with $m\neq0$ vanishes at the zenith. For $m=l$, the spherical harmonic is peaked at the equator. Secondly, $m$ enters as a rotation parameter
\begin{align}
  Y_\lm \;\propto\; P_\lm \; e^{i\,m\,\phi} \;.
\end{align}

The normalisation factor of the spherical harmonics is chosen so that the $Y_\lm$ are orthonormal:
\begin{align}
  \int\dd\Omega(\n)\,Y_\lm(\n)\,Y^*_{\L'm'}(\n) \;=\; \delta_{\L'\L}\,\delta_{m'm}\;,
\end{align}
where
\begin{align}
  \int\dd\Omega(\n) \;=\; \int\limits_0^\pi\dd\theta\,\sin\theta\,\int\limits_0^{2\pi}\,\dd\phi 
\end{align}
denotes an integral over all possible directions. Because of the orthogonality of the spherical harmonics, the $(\L,m)$ multipole of the distribution function can be extracted using the relation
\begin{align}
  f_{\lm} \;=\; i^\L\,\sqrt{\frac{2\L+1}{4\pi}}\,\int\dd\Omega\:Y^*_\lm(\n)\:f(\n) \;.
  \label{eq:spherical_harmonics_projection}
\end{align}
In general, we define the projection operator $L$ as 
\begin{align}
  L_\lm[F] \;=\; i^\L\,\sqrt{\frac{2\L+1}{4\pi}}\,\int\dd\Omega\:Y^*_\lm(\n)\:F(\n) \;.
  \label{eq:L_operator}
\end{align}

We shall project the Boltzmann equation to harmonic space by applying the $L$ operator to both of its sides. This eliminates the angular dependence of the distribution function, at the cost of introducing two discrete indices, \L and $m$. The Boltzmann equation thus reduces to a hierarchy of ordinary differential equations in $(\L,m)$, which is numerically tractable. The hierarchy is in principle infinite, but it can be truncated at a suitable angular scale, $L_\text{max}$, as we detail in \cref{ch:boltzmann}. Therefore, the angular projection operator, $L$, is analogous to the Fourier projection operator, $\mathcal{F}$ (\sref{sec:fourier_formalism}), because it turns a partial differential equation into a system of ordinary differential equations by integrating out a functional dependence.

\section{Properties of the spherical harmonics}
\label{sec:spherical_harmonics}

The spherical harmonics have a number of important properties. We have already mentioned the orthonormality relation,
\begin{align}
  \int\dd\Omega(\n)\,Y_\lm(\n)\,Y^*_{\L'm'}(\n) \;=\; \delta_{\L'\L}\,\delta_{m'm}\;,
\end{align}
which allows to extract the multipole of a function by the simple projection in \eref{eq:spherical_harmonics_projection}. They also satisfy the conjugation relation,
\begin{align}
  Y_{\L-m}(\n) \;=\; (-1)^m\,Y^*_\lm(\n) \;,
\end{align}
the parity relation,
\begin{align}
  Y_{\lm}(-\n) \,=\, (-1)^\L\,Y_{\lm}(\n) \;,
\end{align}
where $-\n$ is characterised by the angles $(\pi-\theta, \phi+\pi)$, the completeness relation,
\begin{align}
  \sum\limits_{\L,m} \, Y_\lm(\theta,\phi) \, Y^*_\lm(\theta',\phi')
  \,=\, \delta(\cos\theta-\cos\theta') \, \delta(\phi-\phi') \;,
\end{align}
and the addition theorem \cite{abramowitz:1977a},
\begin{align}
  \sum\limits_m\,Y_{\lm}(\n)\,Y^*_{\lm}(\n')  \,=\,
  \frac{2\L+1}{4\pi}\,P_\L\,(\scalarP{\n}{\n'}) \,
  \label{eq:addition_theorem}
\end{align}
where $\n$ and $\n'$ are arbitrary unit vectors.

The product of two spherical harmonics can be itself expanded in spherical harmonics to yield a relation involving two Wigner 3$j$ symbols,
\begin{align}
  Y_{\lmone}(\n)\,Y_{\lmtwo}(\n) \,&=\, \sum\limits_{\lm}\,
  \sqrt{\frac{(2\lone+1)(2\ltwo+1)(2\L+1)}{4\pi}} \nmsk
  &\,\threej{\lone}{\ltwo}{\L}{0}{0}{0}
  \threej{\lone}{\ltwo}{\L}{m_1}{m_2}{m} \,Y^*_\lm (\n)\;.
  \label{eq:product_Ylm}
\end{align}
Integrating the above expression on the sphere yields the so-called \keyword{Gaunt relation}:
\begin{align}
  \label{eq:gaunt_integral}
  \int\dd\Omega\:&Y_{\lmone}(\n)\,Y_{\lmtwo}(\n)\,Y_{\lmtre}(\n) \,=\, \msk
  &\sqrt{\frac{(2\lone+1)(2\ltwo+1)(2\ltre+1)}{4\pi}}
  \,\threej{\lone}{\ltwo}{\ltre}{0}{0}{0}
  \threej{\lone}{\ltwo}{\ltre}{m_1}{m_2}{m_3} \;. \notag
\end{align}
In \cref{ch:intrinsic}, where we deal with three-dimensional integrals of the type $\int\dd\kone\dd\ktwo\dd\ktre$, the Gaunt relation will prove useful to integrate out analytically the angular dependence of the transfer functions. Sometimes, we shall also use the following shorthands:
\begin{align}
  h_{\lone\ltwo\ltre} \,=\, \sqrt{\frac{(2\lone+1)(2\ltwo+1)(2\ltre+1)}{4\pi}}
  \,\threej{\lone}{\ltwo}{\ltre}{0}{0}{0} \;,
  \label{eq:gaunt_factor_I}
\end{align}
and
\begin{align}
  \gaunt{\lone}{\ltwo}{\ltre}{m_1}{m_2}{m_3} \,=\, h_{\lone\ltwo\ltre}
  \,\threej{\lone}{\ltwo}{\ltre}{m_1}{m_2}{m_3} \;.
\end{align}

Finally, we list two properties of the associated Legendre polynomials \cite{abramowitz:1977a}
\begin{align}
  P_{\nu-\nu}(\cos\theta) \;=\; \frac{\sin(\theta)^\nu}{2^\nu\nu!}
  \qquad\text{and}\qquad P_{\nu-\nu}\;=\;(-1)^\nu\,\frac{1}{(2\nu)!}\,P_{\nu\nu}
\end{align}
that, together with the definition of the spherical harmonics in \eref{eq:spherical_harmonics_definition}, make it possible to derive a closed form for the spherical harmonics with $\L=m$ ,
\begin{align}
  \label{eq:appendix_Y_mm}
  &Y_{|m|m}(\theta,\phi) \;=\; (-1)^m\;\sqrt{\frac{2m+1}{4\,\pi}}\;
  \frac{\sqrt{(2m)!}}{2^m\,m!}\;\sin^m\theta\;e^{\,im\phi}\;
  &&\text{for $m\geq0$}\;,\msk
  &Y_{|m|m}(\theta,\phi) \;=\; (-1)^m\;Y^*_{|m||m|}(\theta,\phi) &&\text{for $m<0$}\;.
\end{align}
The formula will be useful in \sref{sec:intrinsic_bispectrum_derivation} to characterise the azimuthal dependence of the second-order transfer functions, and thus derive a numerically tractable expression for the intrinsic bispectrum.

\section{Properties of the 3$j$ symbols}

The Wigner 3$j$ symbol,
\begin{align}
  \threej{\lone}{\ltwo}{\ltre}{m_1}{m_2}{m_3} \;,
\end{align}
encodes the geometrical properties of a system of three vectors that form a triangle, $\vec{\lone}+\vec{\ltwo}+\vec{\ltre}=0$; the elements of the first line, ($\lone, \ltwo, \ltre$), must be positive and represent the magnitudes of the three vectors, while those of the second line, ($m_1, m_2, m_3$), must satisfy $-\L_i\,\leq\,m_i\,\leq\,\L_i$ and represent the projections of the three vectors on the zenith. The 3$j$ symbol is different from zero only for the configurations that respect the triangular inequality,
\begin{align}
  |\L_i\,-\,\L_j| \;\leq\; \L_k \;\leq\; \L_i\,+\,\L_j
\end{align}
and for those whereby ${\lone}_z+{\ltwo}_z+{\ltre}_z=0$, that is
\begin{align}
  m_1\;+\;m_2\;+\;m_3 \;=\; 0 \;.
\end{align}

The 3$j$ symbol is related to the Clebsch-Gordan coefficients, which are often used in quantum mechanics to describe the coupling of two angular momentum states, by the following relation
\begin{align}
  \threej{\lone}{\ltwo}{\ltre}{m_1}{m_2}{m_3} \;=\; 
  \frac{(-1)^{\lone-\ltwo-m_3}}{\sqrt{2\,\ltre\,+\,1}} \; \avg{\lone\,m_1\,\ltwo\,m_2|\ltre\,m_3} \;.
\end{align}

\subsection{Symmetries of the 3$j$ symbols}
\label{sec:wigner_3j_symmetries}

The 3$j$ symbols are symmetric under even permutations of their columns,
\begin{align}
  \threej{\lone}{\ltwo}{\ltre}{m_1}{m_2}{m_3} \;=\;
  \threej{\ltwo}{\ltre}{\lone}{m_2}{m_3}{m_1} \;=\;
  \threej{\ltre}{\lone}{\ltwo}{m_3}{m_1}{m_2} \;,
\end{align}
and they gain an alternating sign factor after an odd permutation,
\begin{align}
  \threej{\lone}{\ltre}{\ltwo}{m_1}{m_3}{m_2} \;&=\;
  \threej{\ltwo}{\lone}{\ltre}{m_2}{m_1}{m_3} \;=\;
  \threej{\ltre}{\ltwo}{\lone}{m_3}{m_2}{m_1} \nmsk
  &=\;(-1)^{\lone+\ltwo+\ltre}\;\threej{\lone}{\ltwo}{\ltre}{m_1}{m_2}{m_3} \;.
\end{align}
Changing the sign of the second line yields a phase factor, too,
\begin{align}
  \threej{\lone}{\ltwo}{\ltre}{-m_1}{-m_2}{-m_3} \;=\;
  (-1)^{\lone+\ltwo+\ltre}\;\threej{\lone}{\ltwo}{\ltre}{m_1}{m_2}{m_3} \;,
\end{align}
which implies that
\begin{align}
  \threej{\lone}{\ltwo}{\ltre}{0}{0}{0} \;=\; 0 \quad\text{if}\quad \lone+\ltwo+\ltre \quad \text{is odd} \;.
\end{align}
This property will be important in understanding the structure of the intrinsic bispectrum in \cref{ch:intrinsic}.

As we have anticipated in the previous section, the Gaunt integral can be expressed in terms of the product of two 3$j$ symbols (see \eref{eq:gaunt_integral}),
\begin{align}
  \gaunt{\lone}{\ltwo}{\ltre}{m_1}{m_2}{m_3} \;=\;
  \sqrt{\frac{(2\lone+1)(2\ltwo+1)(2\ltre+1)}{4\pi}}
  \,\threej{\lone}{\ltwo}{\ltre}{0}{0}{0}
  \threej{\lone}{\ltwo}{\ltre}{m_1}{m_2}{m_3} \;.
\end{align}
The Gaunt coefficients possess more symmetries than the 3$j$ symbols; in particular,
\begin{itemize}
  \item they are symmetric with respect to any permutation of their columns;
  \item they vanish for $\lone+\ltwo+\ltre$ odd, and
  \item they are invariant under sign flip of the $m$, that is $\gaunt{\lone}{\ltwo}{\ltre}{m_1}{m_2}{m_3} \;=\; \gaunt{\lone}{\ltwo}{\ltre}{-m_1}{-m_2}{-m_3}$.
\end{itemize}

\subsection{Orthogonality of the 3$j$ symbols}
\label{sec:wigner_3j_properties}

The 3$j$ symbols are orthogonal with respect to the summation over one column,
\begin{align}
  \sum\limits_{\ltre m_3}\;(2\,\ltre+1)\;
  \threej{\lone}{\ltwo}{\ltre}{m_1}{m_2}{m_3}\;
  \threej{\lone}{\ltwo}{\ltre}{M_1}{M_2}{m_3}\;=\;\delta_{m_1M_1}\:\delta_{m_2M_2} \;,
\end{align}
and with respect to the summation over two azimuthal numbers,
\begin{align}
  (2\,\ltre+1)\:\sum\limits_{m_1m_2}\;
  \threej{\lone}{\ltwo}{\ltre}{m_1}{m_2}{m_3}\;
  \threej{\lone}{\ltwo}{L_3}{m_1}{m_2}{M_3}\;=\;\delta_{\ltre L_3}\:\delta_{m_3M_3} \;.
\end{align}
The last identity implies also that
\begin{align}
  \sum\limits_{m_1m_2m_3}\;
  \threej{\lone}{\ltwo}{\ltre}{m_1}{m_2}{m_3}^2 \;=\; 1 \;,
  \label{eq:3j_symbols_squared}
\end{align}
a result that will be useful in defining the angle-averaged bispectrum.

\section{Projecting tensors}
\label{sec:projection_of_tensors}
To project the Einstein equation to spherical space, we need a prescription to extract the $(\L,m)$ multipoles out of a tensor. In this section we show how to do so by employing a set of projection vectors, $\xi$, and matrices, $\chi$. 


\subsection{The projection vectors $\xi$}
\label{sec:the_projection_vectors_xi}

We start by choosing a direction, \n, and noticing that it can be recast as
\begin{align}
  n^i \;=\; \sqrt{\frac{4\pi}{3}}\,\sum\limits_{m=-1}^{1}\,\xivector{m}{i}\,Y_{1m} \;,
  \label{eq:xi_vectors_definition}
\end{align}
where we have used the fact that
\begin{align}
  &\left\{\begin{aligned}
    &\vphantom{\sqrt{\nicefrac{4\pi}{3}}}n_x \,=\, \sin\theta\,\cos\phi \notag\\
    &\vphantom{\sqrt{\nicefrac{4\pi}{3}}}n_y \,=\, \sin\theta\,\sin\phi \notag\\
    &\vphantom{\sqrt{\nicefrac{4\pi}{3}}}n_z \,=\, \cos\theta
  \end{aligned}\right.
  &&\text{and}
  &&\left\{\begin{aligned}
    &\sqrt{\nicefrac{4\pi}{3}} \; Y_{1,-1} &&=\; \sqrt{\nicefrac{1}{2}}\;\sin\theta\,(\cos\phi-i\sin\phi) \\
    &\sqrt{\nicefrac{4\pi}{3}} \; Y_{1,+1} &&=\; \sqrt{\nicefrac{1}{2}}\;\sin\theta\,(-\cos\phi+i\sin\phi) \\
    &\sqrt{\nicefrac{4\pi}{3}} \; Y_{1,0}  &&=\; \cos\theta
  \end{aligned} \right.\quad\;.
\end{align}
We shall refer to the $\,\xi^{\,i}_{[\,m]}\,$ vectors as our \keyword{spherical basis}. They are a set of three unit vectors defined by \eref{eq:xi_vectors_definition}. Their cartesian coordinates are
\begin{align}
  \xi_{[0]}  \;=\; \threev{\,0\,}{0}{1}\;,  \quad\;
  \xi_{[+1]} \;=\; \sqrt{\frac{1}{2}}\;\threev{-1}{i}{0}\;, \quad\;
  \xi_{[-1]} \;=\; \sqrt{\frac{1}{2}}\;\threev{+1}{i}{0}\;,
\end{align}
and their indices are lowered and raised respectively with the Euclidean metric $\delta_{ij}$ and its inverse $\delta^{ij}$.
Since \n is real-valued, under complex conjugation the $\xi$ vectors transform like the spherical harmonics:
\begin{align}
  \xivector{-m}{i} \;=\; (-1)^m\,\xi^{*i}_{[m]} \;.
\end{align}
By using the orthogonality property of the spherical harmonics, we immediately see that the $\xi$ vectors are the coefficients for the spherical transformation of $n^i$, that is
\begin{align}
  L_{\lm}\,[\,n^i\,] \;=\; \delta_{\L 1}\;i\,\xivector{m}{i} \;,
\end{align}
where the operator $L$ is defined in \eref{eq:L_operator}.

\runinhead{Orthogonality}
It is straightforward to verify that the $\xi$ vectors are orthogonal with respect to both indices:
\begin{align}
  \sum\limits_{m=-1}^1\,\xivector{m}{i}\,\xi^{*\,j}_{\,[m]}
  \;&=\; \sum\limits_{m=-1}^1\,(-1)^m\,\xivector{m}{i}\:\xivector{-m}{j} \;=\; \delta^{ij} \;,\nmsk
  \sum\limits_{i=1}^3\,\xivector{m'}{i}\,\xi^{*\,i}_{\,[m]}
  \;&=\; \sum\limits_{i=1}^3\,(-1)^m\,\xivector{m'}{i}\:\xivector{-m}{i} \;=\; \delta_{mm'} \;.
  \label{eq:orthogonality_xi}
\end{align}
This property makes them suitable to be used as projection operators. We define the \index{spherical components of a vector}spherical components, $V_{[m]}$, of a real 3-vector, $V^i$, as
\footnote{Note that \citet{beneke:2010a} (Sec.~IIIB) define the spherical components so that $V^\text{BF}_{[m]}=i\,V_{[m]}$, while \citet{pitrou:2010a} (Sec.~7.2) use a notation whereby $V^P_{[m]}=-V_{[m]}$.\label{ftn:spherical_vector}}
\begin{align}
  V_{[m]} \;=\; \sum\limits_{i=1}^3\,\xivector{m}{i}\,V^i \;,
\end{align}
where $V^i$ are the vector's cartesian coordinates. The explicit form of the spherical components is given by
\begin{align}
  V_{[0]}  \;=\; V_z\;,  \quad\;
  V_{[+1]} \;=\; \sqrt{\frac{1}{2}}\,(-V_x\,+\,i\,V_y)\;, \quad\;
  V_{[-1]} \;=\; \sqrt{\frac{1}{2}}\,(+V_x\,+\,i\,V_y)\;,
  \label{eq:vector_spherical_decomposition}
\end{align}
and, like the $\xi$ vectors, they satisfy the relation
\begin{align}
  V_{[-m]} \;=\; (-1)^m\,V^*_{[m]} \;,
\end{align}
where $V^*_{[m]} \equiv \xivector{m}{i}\,V^*_i$.
The inverse relation is found by exploiting the orthogonality of $\xi$:
\begin{align}
  V^i \;=\; \sum\limits_{m=-1}^1\,\xi^{*\,i}_{\,[m]}\,V_{[m]}
  \;=\; \sum\limits_{m=-1}^1\,(-1)^m\,\xivector{-m}{i}\,V_{[m]} \;.
\end{align}
It should be noted the the spherical components of the reference direction, \n, are the azimuthal modes of the spherical harmonic $Y_{1m}$,
\begin{align}
  n_{[m]} \;=\; \xivector{m}{i}\,n_i \;=\; \sqrt{\frac{4\,\pi}{3}}\;Y_{1m}^* \;,
  \label{eq:spherical_components_n}
\end{align}
a property that can be proven by making use of the second orthogonality relation in \eref{eq:orthogonality_xi}.

\runinhead{Azimuthal modes}
By applying the spherical projection operator $L$ in \eref{eq:L_operator} to $n^i\,V_i$, it follows that the spherical components $V_{[m]}$ are the only non-vanishing multipoles of $n^i\,V_i$,
\begin{align}
  L_\lm\,[\,n^i\,V_i\,] \;=\; \delta_{\ell1}\; i\,V_{[m]} \;.
\end{align}
Due to this property, we shall refer to $V_{[0]}$ and $V_{[\pm1]}$ as the scalar and vector components of $V^i$, respectively.

\runinhead{Scalar product}
The scalar product of two real vectors, $U^i\,V_i$, has a simple form in terms of the spherical components,
\begin{align}
  \sum\limits_{i=1}^3\,V^i\,U^i \;&=\;
  \sum\limits_{m=-1}^1\,\sum\limits_{m'=-1}^1\,\left(\,\sum\limits_{i=1}^3\,
  \xi^{*\,i}_{\,[m]}\,\xi^{*\,i}_{\,[m']}\right)\,U_{[m]}\,V_{[m']} \nmsk
  &=\; \sum\limits_{m=-1}^1\,(-1)^m\,U_{[m]}\,V_{[-m]} \;,
\end{align}
which follows from the orthogonality relation in \eref{eq:orthogonality_xi}.
The scalar product is obviously a scalar quantity; however, it is given by the sum of scalar and vector quantities. This is a simple example of how the different azimuthal modes couple when considering the product of vectors.


\subsection{The projection matrices $\chi$}

Given a direction $\n$, the simplest rank-2 tensor that can be constructed is $n^in^j$. Using the expression for $n^i$ in \eref{eq:xi_vectors_definition}, $n^in^j$ is given by
\begin{align}
  n^in^j \,=\, \sum\limits_{m_1}\,\sum\limits_{m_2}\,\frac{4\pi}{3}\,
  \xivector{m_1}{i}\,\xivector{m_2}{j}\,Y_{1m_1}(\n)\,Y_{1m_2}(\n) \;.
\end{align}
The product of spherical harmonics can be expanded using \eref{eq:product_Ylm} into
\begin{align*}
  Y_{1m_1}(\n)\,Y_{1m_2}(\n) \,=\, \sum\limits_{\lm}\,
  \sqrt{\frac{9\,(2\L+1)}{4\pi}}
  \,\threej{1}{1}{\L}{0}{0}{0}
  \threej{1}{1}{\L}{m_1}{m_2}{m} \,Y^*_\lm \;.
\end{align*}
Because of the properties of the 3$j$ symbol, the sum over \L reduces to two terms: a monopole ($\L=0$) and a quadrupole ($\L=2$). The expansion of $n^in^j$ is then given by
\begin{align}
  n^in^j \;=\; \frac{\delta^{ij}}{3} \;+\;
  \sqrt{\frac{4\pi}{5}}\,\sum\limits_{m=-2}^{2}\,\chimatrix{2}{m}{ij}\,\,Y_{2m} \;,
  \label{eq:chi2m_matrices_ninj}
\end{align}
where we have used $Y_{00}=\sqrt{1/(4\pi)}$ and we have defined the symmetric and traceless $\chi$ matrices as
\begin{align}
  \chimatrix{2}{m}{ij} \;=\; (-1)^m\sum\limits_{m_1=-1}^1\,\sum\limits_{m_2=-1}^1\:
  \sqrt{\frac{10}{3}}\:\threej{1}{1}{2}{m_1}{m_2}{-m}\,\xivector{m_1}{i}\,\xivector{m_2}{j} \;.
  \label{eq:chi2m_matrices}
\end{align}
Their explicit form can be determined from \eref{eq:chi2m_matrices} and are given by
\begin{align}
  &\chi_{\,2,0} \;=\; \frac{1}{3}\;\begin{pmatrix}-1&0&0\\0&-1&0\\0&0&2\end{pmatrix} \;,
  &&\chi_{\,2,\pm1} \;=\; \sqrt{\frac{1}{6}}\;\begin{pmatrix}0&0&\mp1\\0&0&i\\\mp1&i&0\end{pmatrix} \;,
  &&\chi_{\,2,\pm2} \;=\; \sqrt{\frac{1}{6}}\;\begin{pmatrix}1&\mp i&0\\\mp i&-1&0\\0&0&0\end{pmatrix} \;.
  \label{eq:chi2m_matrices_explicit}
\end{align}

The Kronecker delta and the $\chi$ matrices are the $\L=0$ and $\L=2$ multipoles of the tensor $n^in^j$, respectively. All the other multipoles identically vanish; this is easily seen by applying the $L$ operator (\eref{eq:L_operator}) to the expansion of $n^in^j$ in terms of the $\chi$ matrices (\eref{eq:chi2m_matrices_ninj}):
\begin{align}
 &L_{00}\,[\,n^in^j\,] \;=\; \frac{\delta^{ij}}{3} \;, \nmsk
 &L_{2m}\,[\,n^in^j\,] \;=\; -\chimatrix{2}{m}{ij} \;.
\end{align}
Similarly, the contraction of an arbitrary tensor $E^{ij}$ with the tensor $n^in^j$ only has a monopole and a quadrupole contribution:
\begin{align}
  &L_{00}\,[\,n^in^jE_{ij}\,] \;=\; \frac{E^i_{\,i}}{3} \;, \nmsk
  &L_{2m}\,[\,n^in^jE_{ij}\,] \;=\; -\chimatrix{2}{m}{ij}\,E_{ij} \;=\; -E_{[m]}\;,
\end{align}
where in the last equality we have defined the azimuthal components of the symmetric tensor as\footnote{Note that \citet{beneke:2010a} (Sec.~IIIB) define the spherical components so that $E^\text{BF}_{[m]}=-E_{[m]}/\alpha_m$, with $\alpha_0=2/3$, $\alpha_{\pm1}=1/\sqrt{3}$ and $\alpha_{\pm2}=1$.\label{ftn:spherical_tensor}}
$E_{[m]}\equiv\chimatrix{2}{m}{ij}\,E_{ij}$.
Therefore, the $\chi$ matrices provide an easy way to extract from a symmetric three-tensor $E^{ij}$ its scalar ($E_{[0]}=\chimatrix{2}{0}{ij}\,E_{ij}$), vector ($E_{[\pm1]}=\chimatrix{2}{\pm1}{ij}\,E_{ij}$) and tensor ($E_{[\pm2]}=\chimatrix{2}{\pm2}{ij}\,E_{ij}$) parts.

\runinhead{Orthogonality}
The $\chi$ matrices are symmetric and traceless by construction. They satisfy
\begin{align}
  \chimatrix{2}{m}{ij*} \;=\; (-1)^m\,\chimatrix{2}{-m}{ij} \;
\end{align}
and are orthogonal with respect to summation over the spatial indices,
\begin{align}
  \sum\limits_{ij}\,\chimatrix{2}{m}{ij*}\,\chimatrix{2}{m'}{ij} \;=\; \frac{2}{3}\,\delta_{m\,m'} \;.
\end{align}
The orthogonality property can be used to extract the spherical components of $n^in^j$,
\begin{align}
  \chimatrix{2}{m}{ij}\,n_i\,n_j \;=\; \frac{2}{3}\,\sqrt{\frac{4\,\pi}{5}}\,Y_{2m}^* \;.
  \label{eq:spherical_components_ninj}
\end{align}


\section{Projecting functions}
\label{sec:spherical_projection_of_functions}

The most common direction-dependent term in the Boltzmann equation has the form
\begin{align}
  n^i\,V_i\,f(\n) \;,
\end{align}
where $V_i$ can be either a wavemode (in the Liouville term) or the electron bulk velocity (in the collision term). In both cases, the multipole space projection is obtained through the $L$ operator:
\begin{align}
  L_\lm\,[\,n^i\,V_i\,f(\n)\,] \;=\; \int\dd\Omega\;Y^*_\lm\;n^i\,V_i\:f(\n) \;.
\end{align}
Both $n^i$ and $f(\n)$ are further expanded in spherical harmonics according to Eq.~\ref{eq:xi_vectors_definition} and \ref{eq:spherical_harmonics_expansion}, respectively, to yield
\begin{align*}
  L_\lm\,[\,n^i\,V_i\,f(\n)\,] \;=\;& 
  \sum\limits_{m_2=-1}^{1}\,\sqrt{\frac{4\pi}{3}}\,\xivector{m_2}{i}\,
  \sum\limits_{\lone=0}^\infty\,\sum\limits_{m_1=-{\lone}}^{\lone}\,
  (-i)^{\lone}\,\sqrt{\frac{4\pi}{2\lone+1}}\,f_{\lone m_1} \nmsk
  &\times\;\int\dd\Omega\:Y^*_\lm(\n)\,Y_{1m_2}(\n)\,Y_{\lone m_1}(\n) \;.
\end{align*}
After noting that $Y^*_\lm=(-1)^m Y_\lm$, we take care of the angular integration using the Gaunt relation (\eref{eq:gaunt_integral}), and obtain
\begin{align}
  L_\lm\,[\,n^i\,V_i\,f(\n)\,] \;=\;&(-1)^m\,(2\L+1)\,
  \sum\limits_{\lone=0}^\infty\,\sum\limits_{m_1=-{\lone}}^{\lone}\,\sum\limits_{m_2=-1}^{1}\,
  \,V_{[m_2]}\,f_{\lone m_1} \nmsk
  &\times\;i^{\L-\lone}\,\GAUNT{1}{\lone}{\L}{m_2}{m_1}{-m} \;.
  \label{eq:nVf_expansion_raw}
\end{align}
The sum over $\lone$ is infinite but, due to the symmetries of the 3$j$ symbols, it has support only for triangular configurations; since one of the sides has length $1$, the sum consists of three terms:
\begin{align}
  \sum\limits_{\lone=0}^\infty \quad\rightarrow\quad \sum\limits_{\lone=\abs{\L-1}}^{\L+1} \;.
\end{align}
The first 3$j$ symbol also enforces that $1+\lone+\L$ is even, thus excluding the contribution with $\lone=\L$. Similarly, the second 3$j$ symbol enforces $m_1=m-m_2$, so that only the azimuthal modes of $f$ with $m_1=m$ and $m_1=m\pm1$ contribute to the sum. For example, the $(100,0)$ multipole of $n^i\,V_i\,f(\n)$ picks up contributions of the following types:
\begin{align*}
  L_{100,0}[n^i\,V_i\,f(\n)] \,\supset\,\Bigl\{\,
  f_{99,-1}\,V_{[1]},\:f_{99,0}\,\,V_{[0]},\:f_{99,1}\,V_{[-1]},\:
  f_{101,-1}\,V_{[1]},\:f_{101,0}\,V_{[0]},\:f_{101,1}\,V_{[-1]}\,\Bigr\} \;.
\end{align*}

In any gauge, the free-streaming term of the linearised Boltzmann equation is given by $\,n^i\,k_i\,f(\n,\k)\,$. Since we choose to align the zenith with the \k vector, the latter only has a scalar part, $k_{[m]}=\delta_{m0}\,k$ (see \eref{eq:vector_spherical_decomposition}). Thus, the sum over $m_2$ in \eref{eq:nVf_expansion_raw} reduces to only one term:
\begin{align*}
  L_\lm\,[\,n^i\,k_i\,f(\n)\,] \;=\;&(-1)^m\,(2\L+1)\,
  \sum\limits_{\lone=0}^\infty\,\sum\limits_{m_1=-{\lone}}^{\lone}\,
  \,k\,f_{\lone m_1} \; i^{\L-\lone}\,\GAUNT{1}{\lone}{\L}{0}{m_1}{-m} \;.
\end{align*}
The elements in the second line of a 3$j$ symbol must add up to zero; hence, the sum over $m_1$ only has support for $m_1=m$:
\begin{align}
  L_\lm\,[\,n^i\,k_i\,f(\n)\,] \;=\;&(-1)^m\,(2\L+1)\;k\,
  \sum\limits_{\lone=0}^\infty\,
  \,f_{\lone m} \; i^{\L-\lone}\,\GAUNT{1}{\lone}{\L}{0}{m}{-m} \;.
  \label{eq:nkf_expansion}
\end{align}
This is a manifestation of the decomposition theorem: when the zenith is aligned with $\k$, all the sums over the different azimuthal modes collapse and there is no coupling between the modes. As a result, the only contribution to the $(\L,m)$-th multipole of $n^i\,k_i\,f$ comes from the multipoles of $f$ with azimuthal mode $m$. On the other hand, the different angular scales $\L$ still couple, in analogy with the mode coupling of the Fourier modes that we have explored in \sref{sec:perturbations_mode_coupling}. For example, the $(100,0)$ multipole of $n^i\,k_i\,f(\n)$ picks up only two contributions:
\begin{align*}
  L_{100,0}[n^i\,k_i\,f(\n)] \,\supset\,\Bigl\{\,
  f_{99,0}\,\,k_{[0]},\:f_{101,0}\,k_{[0]}\,\Bigr\} \;.
\end{align*}

The Boltzmann equation at second order also contains the terms $\,n_ik^i_1f(\n)\,$ and $\,n_ik^i_2f(\n)\,$. Having aligned the zenith with \k, the wavemodes \kone and \ktwo are arbitrary vectors for which ${k_1}_{[\pm1]}$ and ${k_2}_{[\pm1]}$ do not need not vanish. Therefore, the sum over $m'$ in \eref{eq:nVf_expansion_raw} also includes the azimuthal modes of $f$ with $m_1=m\pm1$, meaning that the decomposition theorem does not apply for the quadratic part of the second-order equations.

\subsection{The coupling coefficients}
\label{sec:coupling_coefficients}

After enforcing the triangular inequality and setting $m_2=m-m_1$, the general multipole expansion of $\,n^i\,V_i\,f(\n)\,$ in \eref{eq:nVf_expansion_raw} takes the form
\begin{align}
  L_\lm\,[\,n^i\,V_i\,f(\n)\,] \;=\;&(-1)^m\,(2\L+1)\,
  \sum\limits_{\lone=\abs{\L-1}}^{\L+1}\,\sum\limits_{m_1=-{\lone}}^{\lone}\,
  \,V_{[m-m_1]}\,f_{\lone m_1} \nmsk
  &i^{\L-\lone}\,\GAUNT{1}{\lone}{\L}{m-m_1}{m_1}{-m} \;.
  \label{eq:nVf_expansion}
\end{align}
This type of term appears in the free-streaming and redshift part of the Liouville operator, where $V^i$ is one of \k, \kone or \ktwo, as well as in the collision term, where $V^i$ is the electron velocity. Thus, to express the Boltzmann equation in a compact way, we follow \citet{beneke:2010a} and introduce the coupling coefficients $C^{\pm}$,
\begin{align}
  C^{\pm,\L}_{m_1\,m} \;\equiv\; (-1)^m\,(2\L+1)\,\GAUNT{1}{\L\pm1}{\L}{m-m_1}{m_1}{-m} \;,
  \label{eq:C_coupling_coefficients}
\end{align}
so that \eref{eq:nVf_expansion} can be rewritten as
\begin{align}
  L_\lm\,[\,n^i\,V_i\,f(\n)\,] \;=\; -i\;\sum\limits_\pm\;\sum\limits_{m_1=m-1}^{m+1}\;
  \pm\;V_{[m-m_1]}\;f_{\L\pm1,m_1}\;C^{\pm,\L}_{m_1\,m} \;,
\end{align}
with the caveat that $C^{-,0}$ should be set to zero. For the polarisation hierarchies, a class of terms slightly different than \eref{eq:nVf_expansion} appear where the first 3$j$ symbol has $(0,2,-2)$ in the second line; in that case, we define the $D^{\pm}$ and $D^0$ coupling coefficients as
\begin{align}
  &D^{\pm,\L}_{m_1\,m} \;\equiv\; (-1)^m\,(2\L+1)\,
  \threej{1}{\L\pm1}{\L}{0}{2}{-2}\,\threej{1}{\L\pm1}{\L}{m-m_1}{m_1}{-m} \;, \nmsk
  &D^{0,\L}_{m_1\,m} \;\equiv\; (-1)^m\,(2\L+1)\,
  \threej{1}{\L}{\L}{0}{2}{-2}\,\threej{1}{\L}{\L}{m-m_1}{m_1}{-m} \;,
  \label{eq:D_coupling_coefficients}
\end{align}
The $D^0$ coefficients encode the mixing between the $E$ and $B$ modes. Note that there is no thing such as a $C^0$ coefficient because the 3$j$ symbol
\begin{align*}
  \threej{1}{\L}{\L}{0}{0}{0}
\end{align*}
would vanish. The explicit form of the $C$ and $D$ coupling coefficients is
\begin{align}
  &C^{+,\L}_{m\pm 1,m} \;=\; -\frac{\sqrt{(\L+1\pm m)\:(\L+2\pm m)}}{\sqrt{2}(2\L+3)}\;,
  &&C^{+,\L}_{m\,m} \;=\; \frac{\sqrt{(\L+1)^2-m^2}}{2\L+3} \;,
  \nmsk
  &C^{-,\L}_{m\pm 1,m} \;=\; \frac{\sqrt{(\L-1\mp m)\:(\L\mp m)}}{\sqrt{2}(2\L-1)} \;,
  &&C^{-,\L}_{m,m} \;=\; \frac{\sqrt{\L^2-m^2}}{2\L-1} \;,
  \nmsk
  &D^{+,\L}_{m_1 m} \;=\; \frac{\sqrt{(\L-1)\:(\L+3)}}{\L+1}\;C^{+,\L}_{m_1 m} \;,
  &&D^{-,\L}_{m_1 m} \;=\; \frac{\sqrt{\L^2-4}}{\L}\;C^{-,\L}_{m_1 m} \;,
  \nmsk
  &D^{0,\L}_{m\pm 1,m} \;=\; \mp \frac{\sqrt{2(\L+1\pm m)\:(\L \mp m)}}{\L (\L+1)} \;,
  &&D^{0,\L}_{m\,m} \;=\; -\frac{2m}{\L (\L+1)} \;.
  \label{eq:coupling_coefficients_explicit_CD}
\end{align}

The multipole expansion of the lensing term in the Liuoville equation is different from the others, because it includes the derivative of the distribution function with respect to the direction of propagation, $\pfrac{f}{n^i}$. We thus define another set coefficients, the $R^{\pm}$,
\begin{align}
  &R^{\pm,\L}_{m_1\,m} \;\equiv\; (-1)^m\,(2\L+1)\,\sqrt{2\,(\L\pm1)(\L\pm1+1)}\;
  \threej{1}{\L\pm1}{\L}{1}{-1}{0}\,\threej{1}{\L\pm1}{\L}{m-m_1}{m_1}{-m} \;,
  \label{eq:R_coupling_coefficients}
\end{align}
so that
\begin{align}
  L_\lm\,\left[\;(\KronUD{i}{j}-n^in_j)\;\pfrac{f(\n)}{n^i}\;V^j\;\right] \;=\;
  i\;\sum\limits_\pm\;\sum\limits_{m_1=m-1}^{m+1}\;
  \pm\;V_{[m-m_1]}\;f_{\L\pm1,m_1}\;R^{\pm,\L}_{m_1\,m} \;.
\end{align}
Their explicit form is given by
\begin{align}
  &R^{+,\L}_{m_1 m} \;=\; -(l+2)\;C^{+,\L}_{m_1 m}  \;,
  &&R^{-,\L}_{m_1 m} \;=\;(l-1)\;C^{-,\L}_{m_1 m} \;,
  \nmsk
  &K^{+,l}_{m_1 m} \;=\; -(l+2)\;D^{+,\L}_{m_1 m} \;,
  &&K^{-,l}_{m_1 m} \;=\; (l-1)\;D^{-,\L}_{m_1 m} \;,
  &&K^{0,l}_{m_1 m} \;=\; - D^{0,\L}_{m_1 m} \;,
  \label{eq:coupling_coefficients_explicit_RK}
\end{align}
where the $K$ coefficients are the equivalent of the $R$ coefficients but for the polarisation hierarchies.

\chapterbib


\chapter{Geometry of the wavemodes}
\label{app:perturbations_geometry}

The non-linear transfer functions are defined inside a convolution integral over two dummy wavemodes, $\kone$ and $\ktwo\,$:
\begin{align}
  &X_\lm(\k) \;=\;\, \pert{\T_\lm}{1}(\k)\;\Phi(\k) \msk
  &\qquad+\;\,\int\,\frac{\dd\kone\,\dd\ktwo}{(2\pi)^3}\;\,\delta(\kone+\ktwo-\k)\;\,
  \pert{\T_\lm}{2}(\kone,\ktwo,\k) \;\, \Phi(\kone)\;\Phi(\ktwo) \;, \notag
\end{align}
where the $(\ell,m)$ indices come from the decomposition in spherical harmonics of the directional dependence of $X$, as explained in Appendix~\ref{app:sphere_projection}.
In principle, $\,\pert{\T}{2}_\lm\,$ depends on the 9 coordinates of the wavemodes: the magnitudes $k_1$, $k_2$ and $k\,$; the polar angles $\theta_1$, $\theta_2$ and $\theta\,$; the azimuthal angles $\phi_1$, $\phi_2$ and $\phi\,$.
In solving the Boltzmann-Einstein differential system for $\,\pert{\T}{2}_\lm\,$, however, we adopt the following simplifying assumptions that reduce the number of independent parameters to 3, which we choose to be the three magnitudes; we shall denote the resulting transfer function as $\pert{\widetilde{\T}}{2}_\lm(k_1,k_2,k)\,$.

First, we solve the system only for those configurations where the polar axis is aligned with $\k\,$. That is, we always take
\begin{align}
  \theta \;=\; \phi \;=\; 0 \;,
\end{align}
which also implies $k_x = k_y = k_{[\pm1]}=0\,$.
The statistical isotropy of the Universe ensures that $\pert{\T}{2}_\lm$ can be obtained in the other configurations by performing a rotation of the polar axis, as we will detail in \sref{sec:intrinsic_bispectrum_derivation} where we compute the intrinsic bispectrum.

Secondly, we note that the Dirac delta function enforces $\k=\kone+\ktwo\,$. This allows to express $\theta_1$, $\theta_2$ and $\phi_2\,$ as functions of the other variables,
\begin{align}
  \label{eq:costheta1_costheta2_phi_2}
  &\cos\theta_1 \;=\; \frac{k^2\,+\,k_1^2\,-\,k_2^2}{2\,k\,k_1} \;,
  &&\cos\theta_2 \;=\; \frac{k^2\,-\,k_1^2\,+\,k_2^2}{2\,k\,k_2} \;,
  &&\phi_2 \;=\; \phi_1 \,+\,\pi \;,
\end{align}
so that only 6 independent parameters are left.
Together with the alignment of the polar axis, the Dirac delta condition allows us to set $\,k_{1x}=-k_{2x}\,$, which implies
\begin{align}
  \label{eq:sin_theta}
  k_1\,\sin\theta_1 \;=\; k_2\,\sin\theta_2 \;,
\end{align}
an expression that will be useful in \sref{sec:intrinsic_bispectrum_numerical} to optimise the bispectrum computation.

Finally, we only compute the transfer functions in $\phi_1=0\,$ and $\,\phi_2=\pi\,$, so that the $\kone$ and $\ktwo$ wavevectors both lie in the $zx$ plane. Again, thanks to the statistical isotropy, the value of $\pert{\T}{2}_\lm$ in the general case is obtained with the simple rotation
\begin{align}
  \pert{\widetilde{\T}}{2}_\lm(k_1,k_2,k,\phi_1) \;=\;
  e^{\,im\phi_1}\;\pert{\widetilde{\T}}{2}_\lm(k_1,k_2,k,0) \;.
\end{align}
We shall use this property in \eref{eq:azimuthal_rotation} to analytically solve the $\phi_1$ dependence in the bispectrum integral.
To sum up, the second-order transfer function computed by \SONG, $\,\pert{\widetilde{\T}}{2}_\lm(k_1,k_2,k_3)\,$, is related to the general one by
\begin{align}
  \pert{\widetilde{\T}}{2}_\lm(k_1,k_2,k) \;=\;
  \pert{\T}{2}_\lm \bigl(
  k_1,\theta_1(k_1,k_2,k),\phi_1=0,\;
  k_2,\theta_2(k_1,k_2,k),\phi_2=\pi,\;
  k, \theta=0,\phi=0\bigr) \;.
\end{align}

\section{Rotation}
The second-order equations are sourced by terms quadratic in the linear transfer functions; because the Fourier transform of a product in real space is a convolution in Fourier space (\eref{eq:fourier_convolution_dirac}), these quadratic sources are evaluated in the dummy wavemodes $\kone$ and $\ktwo\,$.
For example, the equation for the photon dipole transfer function, $\,\pert{{\T}_{1m}}{2}\,$, at second order includes the term
\begin{align}
  & \pert{\dot{\T}_{1m}}{2} \;\supset\;
  4\,\pert{\T}{1}_{1m}(\kone)\;\pert{\dot{\T}}{1}_\Phi(\ktwo) \;.
  \label{eq:quadratic_term_example}
\end{align}
In \SONG, we compute the linear transfer functions only in the direction of the polar axis,
\begin{align}
  \pert{\widetilde{\T}}{1}_\lm(k_1) \;=\; \pert{\T}{1}_\lm(k_1,\theta_1=0,\phi_1=0) \;.
\end{align}
The $\,\pert{\widetilde{\T}}{1}(k_1)\,$'s cannot be inserted directly in the quadratic sources of the second-order system, like \eref{eq:quadratic_term_example}, which instead involve the transfer functions in the general direction $\ktwo\,$.
Thanks to statistical isotropy, however, the two are related by a Wigner rotation,
\begin{align}
  \label{eq:rotation}
  \pert{\T}{1}_{\lm}(\kone) \;=\;
  \sqrt{\frac{4\,\pi}{2\,\L+1}}\;
  Y_\lm(\vec{\hat{k}_1})\;\,\pert{\widetilde{\T}}{1}_{\L\,0}(k_1) \;.
\end{align}
Here we have implicitly used the fact that only the scalar mode exists at first order, since we assume vanishing initial conditions for the vector and tensor modes: $\,\pert{\widetilde{\T}}{1}_\lm(k_1)\propto\delta_{m0}\,$.

All the quadratic sources in \SONG are expressed using \eref{eq:rotation}, including the baryon velocity,
\begin{align}
  v_{b[m]}(\kone) \;=\; \sqrt{\frac{4\,\pi}{3}}\;
  Y_{1m}(\vec{\hat{k}_1})\;\,\pert{\widetilde{v}}{1}_{b[0]}(k_1) \;.
\end{align}
It should be noted that, having chosen the azimuthal angle of \kone to be $\phi_1=0$ and $\phi_1=\phi_1+\pi\,$, the $\,Y_\lm\,$ function is always real-valued. This is a favorable property because it is numerically simpler to evolve a system of real-valued differential equations. (See also Eq.~A.37 of \cite{pitrou:2010a} and Eq.~A.6 of \cite{beneke:2011a}.)

\section{Symmetrisation}

The second-order transfer functions are defined inside a convolution integral (\eref{eq:transfer_function_definition}) where $\kone$ and $\ktwo$ are the integration variables.
This reflects the structure of the Boltzmann and Einstein equations, which, in Fourier space, include the same convolution over the quadratic sources (\sref{sec:perturbations_mode_coupling}).
Because $\kone$ and $\ktwo$ are dummy variables, the quadratic sources can be arranged to be symmetric with respect to their exchange:
\begin{align}
  \pert{\T}{2}_\lm(\kone,\ktwo,\k) \;=\; \pert{\T}{2}_\lm(\ktwo,\kone,\k) \;.
\end{align}

The $\kone\leftrightarrow\ktwo$ symmetry is exploited in \SONG to reduce the computation time of the transfer functions by half.
We do so by building quadratic sources that are symmetric with respect to the exchange of the magnitudes $k_1$ and $k_2$.
Since $\theta_1$ and $\theta_2$ are determined by $k_1$, $k_2$ and $k$ via \eref{eq:costheta1_costheta2_phi_2}, this choice also ensures that the quadratic sources are symmetric with respect to $\theta_1\leftrightarrow\theta_2\,$.
The azimuthal angles of the convolution wavemodes, on the other hand, are independent from the magnitudes and satisfy $\phi_2=\phi_1+\pi\,$.
Then, the identity $\,\pert{\T}{2}_\lm(\kone,\ktwo,\k)=\pert{\T}{2}_\lm(\ktwo,\kone,\k)\,$ implies
\begin{align}
  \label{eq:alternating_sign_exchange_k1_k2}
  \pert{\widetilde{\T}}{2}_\lm(k_2,k_1,k) \;=\; e^{\,im\pi}\;
  \pert{\widetilde{\T}}{2}_\lm(k_1,k_2,k) \;=\;
  (-1)^m\,\pert{\widetilde{\T}}{2}_\lm(k_1,k_2,k) \;.
\end{align}
Thus, by symmetrising the quadratic sources with respect to $k_1\leftrightarrow k_2\,$ we only need to evolve the transfer functions with $k_2\geq k_1$; the other configurations are obtained by multiplication with the $(-1)^m$ factor.
We shall use this fact in \sref{sec:intrinsic_bispectrum_numerical} to perform the bispectrum integral.



\chapterbib

\printindex

\end{document}